\documentclass[seceqn]{elsart}
\usepackage{graphicx}
\usepackage{times,latexsym,euscript,amstext,amssymb,amsbsy,amsmath,amsfonts,bm}
\usepackage{slashed}

\hoffset   =- .5 cm
\def\be{\begin{equation}}
\def\ee{\end{equation}}
\def\bea{\begin{eqnarray}}
\def\eea{\end{eqnarray}}
\def\beqa{\begin{equation}\begin{array}{l}}
\def\eeqa{\end{array}\end{equation}}
\def\bsub{\begin{subequations}}
\def\esub{\end{subequations}}
\def\bce{\begin{center}}
\def\ece{\end{center}}
\def\eqlab#1{\label{eq:#1}}
\def\figlab#1{\label{fig:#1}}

\def\seclab#1{\label{sec:#1}}
\def\barr{\left(\begin{array}{c}}
\def\earr{\end{array}\right)}
\def\bmat{\left(\begin{array}{cc}}
\def\emat{\end{array}\right)}

\def\Eqref#1{Eq.~(\ref{eq:#1})}

\def\fref#1{\ref{fig:#1}}
\def\Figref#1{Fig.~\ref{fig:#1}}

\def\secref#1{Section~\ref{sec:#1}}
\def\appref#1{Appendix~\ref{sec:#1}}

\newcommand{\ksm}{k\hspace{-0.24cm}\slash}

\newcommand{\psm}{p\hspace{-0.18cm}\slash}

\newcommand{\ks}{\slashed{k}}
\newcommand{\ps}{\slashed{p}}

\newcommand{\Rxi}{$R_\xi$ }
\def\gb{\bm{\Gamma}}
\def\kb{\bm{\mathrm K}}
\def\l{\bm{\mathrm L}}
\def\g{\widetilde\gb}
\def\k{\widetilde\kb}
\newcommand{\sw}{s_{w}}
\newcommand{\cw}{c_{w}}
\newcommand{\gw}{g_{w}}
\newcommand{\tw}{\theta_{w}}
\newcommand{\Mw}{M_\chic{W}}
\newcommand{\Mz}{M_\chic{Z}}

\newcommand{\Sq}{{\sigma_{\mu\nu}q^{\nu}}}
\newcommand{\Sm}{{\frac{i\Sq}{2m_{t}}}}
\newcommand{\Sme}{{\frac{i\Sq}{2m_{e}}}}
\def\chic#1{{\scriptscriptstyle #1}}
\newcommand{\D}{\displaystyle}
\def\ie{{{\it i.e.}, }}
\def\eg{{{\it e.g.}, }}
\newcommand{\cal}{\mathcal}
\newcommand{\xiQ}{{\xi_Q}}

\begin{document}
\begin{frontmatter}

\title{Pinch Technique: Theory and Applications}

\author[ECT]{Daniele~Binosi},
\author[UV]{Joannis~Papavassiliou}

\address[ECT]{ECT*, Villa Tambosi, Str.\ delle Tabarelle 286,
I-38100 Villazzano (Trento), Italy}

\address[UV]{Departamento de F\'\i sica Te\'orica and IFIC, Centro Mixto, 
Universidad de Valencia-CSIC, E-46100, Burjassot, Valencia, Spain}

\begin{abstract}
We review  the theoretical  foundations and
the most important physical  applications of the Pinch Technique (PT).
This  general  method  allows  the construction of  off-shell  Green's
functions in  non-Abelian gauge theories  that are independent  of the
gauge-fixing  parameter and  satisfy ghost-free  Ward  identities.  We
first present  the diagrammatic formulation  of the technique  in QCD,
deriving  at  one  loop   the  gauge  independent  gluon  self-energy,
quark-gluon vertex, and  three-gluon vertex, together with their
Abelian  Ward identities.  The  generalization of  the PT  to theories
with spontaneous symmetry  breaking is carried out in  detail, and the
profound  connection  with  the  optical theorem  and  the  dispersion
relations are explained within  the electroweak sector of the Standard
Model.  The  equivalence between the PT  and the Feynman  gauge of the
Background  Field   Method  (BFM)  is  elaborated,   and  the  crucial
differences  between the  two methods  are critically  scrutinized.  A
variety of field theoretic techniques needed for the generalization of
the PT to  all orders are introduced, with  particular emphasis on the
Batalin-Vilkovisky  quantization method and  the general  formalism of
algebraic renormalization.   The main conceptual  and technical issues
related  to  the  extension  of  the technique  beyond  one  loop  are
described, using the two-loop construction as a concrete example.  Then
the all-order generalization  is thoroughly examined, making extensive
use of the field theoretic machinery previously introduced; of central
importance  in this  analysis  is the  demonstration  that the  PT-BFM
correspondence  persists to  all orders  in perturbation  theory.  The
extension  of  the  PT  to  the non-perturbative  domain  of  the  QCD
Schwinger-Dyson  equations is presented  systematically, and  the main
advantages  of  the resulting  self-consistent  truncation scheme  are
discussed.  A plethora of physical  applications relying on the PT are
finally  reviewed,   with  special  emphasis  on   the  definition  of
gauge-independent   off-shell   form-factors,   the  construction   of
non-Abelian  effective  charges,   the  gauge-invariant  treatment  of
resonant transition amplitudes and unstable particles, and finally the
dynamical generation of an effective gluon mass.
\end{abstract}

\begin{keyword}
Non-Abelian gauge theories \sep 
Gluons, gauge bosons \sep 
Gauge-invariance \sep 
Schwinger-Dyson equations \sep 
GreenÕs functions \sep 
Dynamical mass generation 
\PACS 
12.38.Aw \sep 
14.70.Dj \sep 
12.38.Bx \sep 
12.38.Lg
\end{keyword}

\end{frontmatter}

\begin{flushright}
{\it prepared for Physics Reports}
\end{flushright}

\newpage
\tableofcontents
\newpage


\section{\seclab{intro}Introduction}
\noindent
When quantizing  gauge theories in  the continuum 
one usually resorts to an appropriate gauge-fixing procedure in
order   to  remove  redundant   (non-dynamical)  degrees   of  freedom
originating    from    the   gauge    invariance    of   the    theory
\cite{Abers:1973qs}.    
Thus,  one  adds   to  the   gauge  invariant
(classical) Lagrangian,  ${\mathcal L}_{\rm I}$, a  gauge-fixing term, 
${\mathcal L}_{\rm GF}$, which allows for the consistent derivation of
Feynman  rules.  At  this point  a new  type of  redundancy  makes its
appearance, this time at the level of the building blocks defining the
perturbative  expansion. In  particular, individual  off-shell Green's
functions  ($n$-point  functions) carry  a  great  deal of  unphysical
information,  which disappears when  physical observables  are formed.
$S$-matrix elements, for example,  are independent of the gauge-fixing
scheme  and  parameters  chosen  to  quantize  the  theory are
unitary and  
well-behaved at high energies.  Green's functions, on  the other hand, depend
explicitly (and, in general, non-trivially) on the gauge-fixing parameter (gfp)
entering  in the  definition of  ${\mathcal L}_{\rm  GF}$, 
contain unphysical thresholds, and grow much
faster than  physical amplitudes at  high energies (\eg  they grossly
violate   the   Froissart-Martin   bound \cite{Froissart:1961ux}).
Evidently, in  going from  unphysical Green's
functions to physical  amplitudes, subtle field-theoretic mechanisms
are at work, enforcing vast cancellations among the various Green's
functions. While it is clear that the realization of these
cancellations  mixes non-trivially contributions stemming from Feynman diagrams of
different  kinematic nature (propagators, vertices, boxes), 
the prevailing attitude is to condense all this down to the 
standard statement that the Becchi-Rouet-Stora-Tyutin (BRST) symmetry \cite{Becchi:1975nq,Tyutin:1975qk} 
guarantees eventually 
the gauge-independence of physical observables, and nothing more. 
\newline
\indent
It turns  out, however,  that all aforementioned  cancellations inside
physical amplitudes  (such as $S$-matrix elements,  Wilson loops, etc)
take place in  a very particular way: not only  is the entire physical
amplitude   gauge-independent,   but  it   may   be  decomposed   into
kinematically  distinct  subamplitudes   that  are  themselves  {\it
individually}    gauge-independent.      In    addition    to    being
gauge-independent,  these  subamplitudes   are  endowed  with  further
properties, such as analyticity and a profound connection with the optical
theorem.  The precise field-theoretic method that exposes 
this  particular  stronger  version  of gauge  independence and  enforces all
ensuing   physical   properties is   the   Pinch   Technique   (PT)
\cite{Cornwall:1976hg,Cornwall:1981ru,Cornwall:1981zr,Cornwall:1989gv,Papavassiliou:1989zd}.        
The basic observation is that all relevant cancellations are realized when
a very  particular subset of longitudinal  momenta, circulating inside
vertex  and   box  diagrams,  extracts   out  of  them,   through  the
``pinching'' of  internal lines, structures  that are in  all respects
propagator-like, and should therefore be reassigned to the conventional
self-energy Feynman graphs.  This  particular reshuffling of terms has
far-reaching consequences, giving rise to effective Green's functions,
which,  in contradistinction  to the  conventional  unphysical Green's
functions,   have  properties   generally  associated   with  physical
observables.  In particular, the  PT Green's functions are independent
of  the gauge-fixing  scheme  and parameters  chosen  to quantize  the
theory  ($\xi$ in covariant  gauges, $n_\mu$  in axial  gauges, etc.)
are gauge-invariant,  \ie they  satisfy  the all-order
simple  tree-level  Ward Identities (WIs), associated  with  the  gauge  symmetry of  the
classical Lagrangian  ${\mathcal L}_{\rm  I}$, instead of the ghost-infested 
Slavnov-Taylor identities (STIs),
they display  only physical
thresholds, and they are well-behaved at high energies.
\newline
\indent
But why should one worry at all about the gauge-dependence or 
other unphysical properties that individual 
Green's functions may have?
After all, when one uses them to 
construct observables, they do conspire to furnish the right answer, 
which is all that really matters. Things are not so simple, however; 
in fact, as we will explain in detail in this report, 
there are  considerable theoretical and  phenomenological advantages in
reformulating the perturbative expansion in terms of off-shell Green's
functions with improved properties.
\newline
\indent 
Even within 
a fixed order perturbative calculation,  
the sharp  difference between observables
and  Green's functions  suggests a  great  deal of  redundancy in  the
conventional diagrammatic formulation of  gauge theories, in the sense
that extensive underlying cancellations beg to be made manifest and be
explicitly  exploited  as  early  within a  calculation  as  possible.
Implementing these  cancellations at an  early stage renders
the book-keeping aspects more tractable \cite{Feng:1995vg}. 
Moreover, there is an unpleasant mismatch between our intuition based on Quantum Electrodynamics
(QED) and the way non-Abelian theories seem to work; however, very often  
this mismatch is not due to inherent properties of the non-Abelian physics,  
but is rather an artifact of the 
quantization procedure, and of the way 
this affects individual Green's functions. 
For example, the text-book concept of the 
effective charge, so familiar in QED, becomes completely obscured 
in a non-Abelian setting, because of the gauge-dependence of the 
vector meson's self-energy, a complication that is automatically resolved 
in the  PT context.
\newline
\indent
The main reason that clearly  favors 
employing the PT Green's functions, however, is the fact that 
a variety of important physical problems cannot be addressed within
the framework of fixed-order perturbation theory, \ie by simply 
computing all Feynman diagrams contributing to a given process 
at a given order.
This is often the case within
Quantum Chromodynamics (QCD), where, 
due to the large disparities of the physical
scales involved,  
a complicated interplay between perturbative
and non-perturbative effects takes place.  Similar limitations appear when
physical kinematic singularities, such as resonances, render the
perturbative expansion divergent at any finite order, or when
perturbatively exact symmetries prohibit the appearance of certain
phenomena, such as chiral symmetry breaking or gluon mass generation.
In such cases one often resorts to various reorganizations of the
perturbative expansion, or to completely non-perturbative techniques 
such as the Schwinger-Dyson equations (SDEs). 
One of the main difficulties encountered when dealing with the
problems mentioned above is the fact that several physical properties,
which are automatically preserved in fixed-order perturbative
calculations by virtue of powerful field-theoretic principles, may
be easily compromised when rearrangements of the perturbative series,
such as resummations, are carried out.  These complications may, in
turn, be traced down to the fundamental fact that we have emphasized from the 
outset: in non-Abelian gauge theories
individual off-shell Green's functions are unphysical.
\newline
\indent
We now take a closer look at some of the aforementioned issues, in order 
to fully appreciate the usefulness of the PT formalism.

\begin{itemize}
 
\item[$\ast$] 
{\it  Non-Abelian effective    charges.}
The unambiguous extension  of the  concept of  the 
gauge-independent, renormalization
group invariant,  and process-independent effective charge  
from QED to
QCD~\cite{Cornwall:1981zr,Cornwall:1976ii}
is of special  interest for several reasons~\cite{Watson:1996fg}.  
The PT  construction of this quantity 
 accomplishes
the explicit identification  of the conformally-variant  
and conformally-invariant
subsets of
QCD  graphs~\cite{Brodsky:1982gc},   
usually assumed in the field of renormalon calculus~\cite{Mueller:1992xz}.  
Moreover, the PT effective charge  can serve as the natural scheme for
defining the  coupling in the proposed  ``event amplitude generators''
based on  the the light-cone formulation  of QCD~\cite{Brodsky:2001ha}.
In addition, the electroweak effective charges constructed  
with the PT are used to define  
the physical  renormalization schemes~\cite{Binger:2003by}, which 
provide a superior framework for the study of gauge coupling unification. 
\newline
\item[$\ast$] 
{\it Off-shell form-factors.}
In non-Abelian theories their proper definition poses in general 
problems related to the gauge invariance \cite{Fujikawa:1972fe}.
Specifically, if one attempts to define the form-factors 
from the conventional vertices, for off-shell momentum transfers, 
one is invariably faced with residual 
gauge-dependences, together with the various pathologies that these imply.
Some representative cases are the 
magnetic dipole and electric 
quadrupole moments of the $W$~\cite{Papavassiliou:1993ex}, 
the top-quark magnetic moment~\cite{Papavassiliou:1993qe}, 
and the neutrino charge radius~\cite{Bernabeu:2000hf}. 
The PT allows for an unambiguous   
definition of such quantities, without any additional assumptions whatsoever:
one must simply extract the corresponding physical off-shell form-factors
from the corresponding gauge-independent PT vertex. A celebrated  
example of such a successful construction has been the 
neutrino charge radius;  
the gauge-independent, renormalization-group-invariant, 
and target-independent neutrino charge radius obtained from the 
corresponding PT vertex  
constitutes a genuine {\it physical} observable,   
since it can be 
extracted (at least in principle) from an appropriate combination of 
scattering experiments~\cite{Bernabeu:2002nw}.
\newline
\item[$\ast$] 
{\it   Resonant   transition  amplitudes.} 
The Breit-Wigner  procedure used  for regulating
the  physical  singularity appearing  in  the  vicinity of  resonances
($\sqrt{s}\sim  M$) is  equivalent to  a {\it  reorganization}  of the
perturbative  series  \cite{Veltman:1963th}.   In  particular,  the  Dyson
summation of the  self-energy, which is the standard  way for treating
resonant  amplitudes,  effectively amounts  to  removing a  particular
term from  each order of  the perturbative expansion, since  from all
the Feynman  graphs contributing to a  given order one only keeps 
the  part  that contains  self-energy   bubbles.   Given  that
non-trivial  cancellations involving the  various Green's  function
generally  take  place  at  any  given  order  of  this
expansion, the  act of  removing one  of them  from each
order  may distort those  cancellations; this  is indeed  what happens
when constructing  non-Abelian {\it running  widths}.  The way  the PT
solves this  problem is by ensuring that  all unphysical contributions
contained inside  the conventional self-energies  have been identified
and  properly  discarded,  {\it before}  
any  resummations  are  carried  out~\cite{Papavassiliou:1995fq}.
\newline
\item[$\ast$] 
{\it  Schwinger-Dyson  equations.}  
The most widely  used framework for studying in  the continuum various
dynamical  questions  that  lie  beyond perturbation  theory  are  the
Schwinger-Dyson  equations (SDE)~\cite{Dyson:1949ha,Schwinger:1951ex}.
This infinite system of  coupled non-linear integral equations for all
Green's  functions of  the theory  is inherently  non-perturbative, and
captures the  full content of  the quantum equations of  motion.  Even
though  these  equations  are   derived  by  an  expansion  about  the
free-field  vacuum,  they finally  make  no  reference  to it,  or  to
perturbation theory,  and can be  used to address problems  related to
chiral  symmetry  breaking, dynamical  mass  generation, formation  of
bound      states,     and     other      non-perturbative     effects
\cite{Cornwall:1974vz,Marciano:1977su}.   Since  this  system involves  
an infinite hierarchy  of equations, in practice one  is severely limited
in their use, and the  need for a self-consistent truncation scheme is
evident.  Devising such  a scheme, however, is far  from trivial; 
the crux of the matter is that the  SDEs, 
in their conventional formulation,
are  built out  of unphysical  Green's functions. Thus, the
extraction  of  reliable  physical  information depends  crucially  on
delicate all-order cancellations, which may be inadvertently distorted
in the  process of the truncation.   
The PT addresses  this problem at  its root, by introducing 
a drastic modification  already at the  level  of  the building  
blocks  of  the  SD series,  namely  the off-shell  
Green's functions  themselves.   
\end{itemize}
\indent
Let us emphasize from the beginning  that, to date, there is no formal
definition of the PT procedure at the level of the functional integral
defining  the theory.   In particular,  let  us assume  that the  path
integral has  been defined  using an arbitrary  gauge-fixing procedure
(\eg   linear covariant  gauges); then,  there is  no known  a priori
procedure (such  as, \eg functional differentiation  with respect to
some combination of appropriately  defined sources) that would furnish
directly the  gauge-independent PT Green's  functions.  The definition
of the  PT procedure is operational,  and is intimately  linked to the
diagrammatic expansion of the theory  (\ie  one must know the Feynman
rules).  In fact, the starting point of the PT construction can be any
gauge-fixing scheme that furnishes a set of well-defined Feynman rules
and   gauge-independent  physical   observables.    Specifically,  one
operates  at  a  certain  well-defined  subset of  diagrams,  and  the
subsequent  rearrangements give  rise to  the same  gfp-independent PT
answer, regardless of the  gauge-fixing scheme chosen for deriving the
Feynman rules.  However, as we will  see in the last  sections of this
report, the PT in its ultimate formulation is not diagrammatic, in the
sense  that one  does not  need to  operate on  individual  graphs but
rather on  a handful of classes  of diagrams (each  one containing an
infinite number of individual graphs).
\newline
\indent
Today's  distilled wisdom on the structure of the PT can 
be essentially captured by the profound connection between the PT and the
the well-known quantization scheme known as the 
Background Field Method (BFM)~
\cite{Dewitt:1967ub,Honerkamp:1972fd,Kallosh:1974yh,KlubergStern:1974xv,Arefeva:1974jv,Hooft:1975vy,Abbott:1980hw,Weinberg:1980wa,Shore:1981mj,Abbott:1983zw,Hart:1984jy}.
The  BFM  is  a  special gauge-fixing 
procedure, implemented  at the  level of  the generating functional.  In 
particular, it preserves  the symmetry of  the action under ordinary  gauge
transformations  with respect to  the background (classical) gauge field
$\widehat{A}_{\mu}$,  while the quantum gauge fields $A_{\mu}$ appearing in
the  loops transform  homogeneously under  the gauge  group,  \ie
as ordinary matter  fields which happened  
to be assigned to  the adjoint representation 
\cite{Weinberg:1996kr}.   As a  result  of the  background gauge symmetry, the BFM
$n$-point  functions  $\langle 0  | T \left[ \widehat{A}_{\mu_1}(x_1)
\widehat{A}_{\mu_2}(x_2)\cdots  \widehat{A}_{\mu_n}(x_n) \right] |0 \rangle$
satisfy naive QED-like Ward-identities, but they do depend  explicitly on
the quantum gauge-fixing parameter
$\xi_Q$ used  to define the tree-level  propagators of  the 
quantum gluons.   It  turns out that, to all orders in perturbation theory,
the gauge-fixing parameter-independent effective 
$n$-point functions constructed by means of the PT (starting from 
any   gauge-fixing  scheme)  {\it coincide}  
with  the  corresponding background 
$n$-point functions when  the latter  are computed  at the special   value 
$\xi_Q =1$   
(BFM Feynman  gauge, BFG in short)~\cite{Denner:1994nn,Hashimoto:1994ct,Pilaftsis:1996fh} . 
Some important conceptual issues 
related to this correspondence  
will be discussed extensively in the corresponding sections.
\newline
\indent
We now turn to a somewhat more technical issue, and discuss briefly 
the formal machinery necessary for the implementation of the PT.
Evidently, there is a gradual increase in the sophistication 
of the field-theoretic tools employed when going from the 
one-loop construction, presented in the early articles,  
all the way to the recently derived new SD series.
\newline
\indent
The original one-loop \cite{Cornwall:1981zr} and 
two-loop \cite{Papavassiliou:1999az}
PT calculations consist in carrying out algebraic manipulations 
  inside individual box- and vertex-diagrams, 
following well-defined rules. In particular,  
one tracks down the rearrangements 
induced when the action of 
(virtual) longitudinal momenta ($k$) on the bare vertices of 
diagrams trigger elementary WIs.
The  longitudinal momenta  responsible  for these
rearrangements stem either from the bare gluon propagators or from 
a very characteristic  decomposition  of the  tree-level
(bare) three-gluon   vertex.
Eventually, a WI of the form 
$ k_{\mu}\gamma^{\mu} = S^{-1}(\ksm + \psm)- \,S^{-1}(\psm)$
gives rise to 
propagator-like parts, by removing (pinching out) the 
internal bare fermion propagator $S(\ksm + \psm)$.  
Depending on the order and topology 
of the diagram under consideration, the  
final  WI may be activated 
immediately, as happens at one loop, 
or as the final outcome of a sequential 
triggering of intermediate WIs, as happens at two loops. 
The propagator-like contributions so obtained 
are next reassigned to the usual gluon self-energies,  
giving rise to the PT  gluon self-energy.  
\newline
\indent
The direct diagram-by-diagram treatment 
followed up until the two-loops 
cannot be possibly used to generalize the PT to all orders.
Indeed, the resulting logistic complexity clearly 
advocates for  
the use of a non-diagrammatic approach, 
\ie a method that treats at once entire subsets of  
diagrams. 
The  non-diagrammatic formulation  of the
PT  introduced in~\cite{Binosi:2002ft} accomplishes this, 
by recognizing that 
the  aforementioned one- and two-loop 
rearrangements  are  but lower-order  manifestations  of a  more fundamental
cancellation. This cancellation  
takes place
when computing the divergence (STI) of
a special
Green's function, which serves as  a common kernel to all higher order
self-energy  and  vertex  diagrams. 
In addition, and most importantly, 
the parts of the Feynman diagrams that are
shuffled around  during the pinching process are expressed 
in  terms of well-defined
field-theoretic  objects,  namely   the  ghost   Green's  functions
appearing  as a standard ingredient in the  STI satisfied 
by the three-gluon vertex~\cite{Ball:1980ax}. 
These ghost 
Green's  functions involve  composite operators,  such as  $\langle 0
\vert T[s\Phi(x)\cdots]\vert0\rangle$,  where $s$ is  the BRST operator
and  $\Phi$ is  a  generic QCD  field.   It turns  out  that the  most
efficient  framework for  dealing with  these type  of objects  is the
Batalin-Vilkovisky formalism~\cite{Batalin:1984jr}. In this framework, 
one adds to the original
gauge-invariant  Lagrangian  ${\mathcal  L}_\mathrm{I}$  the  term  ${\mathcal
L}_\mathrm{BRST}=\sum_\Phi\Phi^*s\Phi$,   thus coupling    the   composite
operators  $s\Phi$ to  the  BRST invariant  external sources (usually  called anti-fields)  $\Phi^*$, 
to  obtain the  new Lagrangian  ${\mathcal
L}_\mathrm{BV}={\mathcal   L}_\mathrm{I}+{\mathcal   L}_\mathrm{BRST}$.    One
advantage of  this formulation  is that it allows one to express the
STIs  of  the theory  in  terms of  auxiliary  functions,  which can  be
constructed using  a well-defined  set of Feynman rules  (derived from
${\mathcal L}_\mathrm{BRST}$). The Batalin-Vilkovisky formalism, and in 
particular  
a multitude of useful identities derived from it, is 
used extensively in the derivation of the new series of 
gauge-invariant SDEs.
\newline
\indent
We conclude by presenting a roadmap 
of the topics discussed in this report. 
\begin{itemize}

\item[{\bf \secref{QCD_one-loop}}.]  This section contains  a detailed introduction
to the one-loop  PT in the context of a theory  like QCD, \ie without
tree-level symmetry breaking.  The  method is implemented at the level
of every single one-loop Feynman diagram contributing to a quark-quark
scattering  amplitude.  The PT  two-point  functions  at one-loop  are
derived, with particular emphasis on  the PT gluon self-energy and the
quark-gluon  vertex.   The QED-like  WI  satisfied  by  the latter  is
derived  in detail.   A similar  construction is  carried out  for the
one-loop   three-gluon  vertex,  the   corresponding  Abelian   WI  is
presented,  and the  supersymmetric  structure of  its form-factors  is
discussed.   We  dedicate  a  large  part  of  the  first  section  in
establishing the precise connection between the imaginary parts of the
one-loop PT  Green's functions and the optical  theorem, together with
the corresponding dispersion relations.
\newline
\item[{\bf \secref{PTBFM}.}]
Here we review the formal aspects of the BFM, 
and derive the corresponding set of Feynman rules, 
emphasizing the dependence of the {\it bare} three- and four-gluon 
vertices on the gfp, and the characteristic ghost sector
containing a symmetric ${\widehat A} {\bar c} c$ vertex, and 
a new ${\widehat A}{\widehat A} {\bar c} c$ four-field vertex. 
We next establish the correspondence between the 
PT and the BFG at the one-loop level, and clarify various 
conceptual issues regarding this correspondence 
and its correct interpretation.
The final item in this section 
is the introduction to the ``generalized'' PT, which is 
a diagrammatic procedure that permits one to start out with any 
arbitrary conventional gauge and be dynamically projected to the 
corresponding BFM gauge.
\newline
\item[{\bf \secref{SM_one-loop}.}] 
In this section, the one-loop PT construction 
for the electroweak sector of the Standard Model (SM) is 
presented. This exercise is significantly more involved 
than in the case of QCD, mainly due to the book-keeping 
complications introduced by the proliferation of particles.
We pay particular attention to the modifications 
introduced to the PT procedure due to the spontaneous breaking of the symmetry 
through the Higgs mechanism. We first present the technically  
simpler situation of massless external test fermions, 
an assumption that considerably simplifies  the algebra. 
The absorptive construction of the first section is repeated, 
and the same underlying principles and patterns are recovered. 
The generalization of the method to the case of massive external fermions is 
then discussed, and the central role of the would-be Goldstone bosons 
for maintaining gauge-invariance is elucidated. 
We demonstrate how in this latter case the 
requirement of the complete gauge-independence of the PT-rearranged 
scattering amplitude furnishes non-trivial WIs relating the various PT Green's 
functions.
\newline
\item[{\bf \secref{Applications - I}.}] 
We present some of the 
most characteristic applications of the PT, that can be worked 
out based on the material presented in the previous three sections.
We focus on four particular subjects.
First, we study in detail the construction of non-Abelian effective charges
that satisfy the same properties as the prototype QED effective 
charge. The analysis includes the QCD effective charge, as well as the 
those appearing in the electroweak sector, most notably the effective  
electroweak mixing angle. We demonstrate how the unitarity and analyticity 
properties built into these charges allow (at least in principle) 
their reconstruction from experiments. As a particularly 
interesting phenomenological application 
of the PT effective charges, we focus on the so-called ``physical renormalization schemes'', 
relevant for the correct quantitative study of the unification 
of the gauge couplings. Second, we explain how to 
define gauge-independent off-shell form-factors with the PT.
Particular emphasis is placed on the more recent case of the neutrino 
charge radius, which is shown to be endowed with a plethora of physical
properties, and to constitute a genuine physical observable.
The third application is related to the gauge-independent definition of 
some important electroweak parameters, such as the $S$, $T$, and $U$, 
and the universal part of the $\rho$ parameter. 
The fourth main application is the gauge-invariant framework for 
treating self-consistently resonant transition amplitudes. 
The intricate nature of this problem 
requires an elaborate synthesis of practically all the material 
that has been presented in the first three sections. 
In the corresponding subsections the reader may fully appreciate 
how tightly intertwined  the various physical principles really are,  
and eventually recognize the superiority of the 
PT-based resonant transition formalism 
over any other similar attempt that has appeared in the literature to date.   
\newline
\item[{\bf \secref{beyond_1l}.}]  
The application of the PT beyond one loop is presented.
We start with the explicit two-loop construction, 
which still proceeds by applying the PT algorithm on individual graphs.
The upshot of the analysis 
is that all the PT properties known from the one-loop 
construction  are replicated at two loops, 
without any additional assumptions;   
most notably, it is established that the PT-BFG correspondence
persists at two loops. Next, we shift gears and 
turn into the non-diagrammatic formulation of the PT:
the all-order construction is carried out by recognizing that 
all crucial PT cancellations are encoded into the STI satisfied 
by a special Green's function, and the PT-BFG correspondence 
is proven to be valid to all orders.
\newline
\item[{\bf \secref{PTBV}.}]  
We introduce the powerful quantization formalism 
of  Batalin and Vilkovisky, which will allow us to streamline 
elegantly the entire PT procedure, in a way especially
suited for accomplishing the important task of the next section. 
After introducing the basic formalism, we revisit the 
one- and two-loop cases, and show how the various terms participating 
in the construction are expressed in terms of the auxiliary 
Green's functions characteristic of the Batalin-Vilkovisky formalism.
In addition, we derive a set of identities relating the 
conventional and BFM Green's functions, which will turn out to 
be of paramount importance for the SD analysis that follows. 
\newline
\item[{\bf \secref{PT_SDEs}.}]  
This section contains the holy grail of the PT.  We
first  explain that  the naive truncation  of  the conventional  SD
series is bound  to introduce artifacts, such as  the violation of the
transversality  of  the gluon  self-energy.   We  then  show that  the
application  of  the  PT  to   the  conventional  SDE  for  the  gluon
propagator and three-gluon vertex gives rise to
new  SDEs endowed  with  special properties.   The fully  dressed
vertices appearing in this new SD series satisfy Abelian all-order WIs
instead of the STIs  satisfied by their conventional counterparts.  As
a result,  and contrary to  the standard case,  the new series  can be
truncated  {\it gauge-invariantly}  at   any  order  in  the  dressed  loop
expansion, and {\it separately} for gluonic and ghost contributions. 
\newline
\item[{\bf \secref{Applications-II}.}]  
Here we present a highly non-trivial application 
of the new SD formalism derived in the previous section. In particular,  
after truncating the SD series gauge-invariantly, we  
solve the resulting system of coupled integral equations, 
and determine the infrared behavior of the gluon and ghost propagator 
(in the Landau gauge). We explain that, under very special 
assumptions for the three-gluon vertex entering into the SDE, 
one can obtain an infrared finite gluon propagator. The 
physics behind this behavior is 
associated with the phenomenon of dynamical gluon mass generation, 
which is the $4-d$ analogue of the $2-d$ Schwinger mechanism. 
In addition, the numerical treatment of the SD system reveals that 
the dressing function of the ghost propagator is   
also finite in the infrared. These results are then 
compared with several recent large-volume lattice simulations, 
and are found to be in good qualitative agreement. 
\end{itemize}

The review ends with some concluding remarks in \secref{conclrem}, and three appendices collecting material used in the main text.

\newpage


\section{\seclab{QCD_one-loop}The one-loop pinch technique in QCD}
\noindent
In this section, we present in detail the PT construction at one-loop
for a non-Abelian gauge theory like QCD, where there is no tree-level 
symmetry breaking (no Higgs mechanism). The analysis we present here applies 
to any gauge group [$SU(N)$, exceptional groups, etc], but for concreteness 
we will adopt the QCD terminology (thus talking about quarks, gluons, etc). 
The calculations presented in this section are purposefully very detailed,  
and aim to provide a completely self-contained guide to the one-loop PT.    

\subsection{\label{prolegomena}The QCD Lagrangian, gauge-fixing, and BRST symmetry}
\noindent
Throughout this report we will adopt the conventions of the book by Peskin \& Schr\"oder~\cite{Peskin:1995ev}.
The QCD Lagrangian density is given by
\begin{equation}
{\mathcal L} = {\mathcal L}_{\mathrm I} + {\mathcal L}_{\mathrm{GF}} + {\mathcal L}_{\mathrm{FPG}}.
\label{QCD_lag}
\end{equation}
${\mathcal L}_{\mathrm I}$ represents the gauge invariant $SU(3)$ Lagrangian, namely
\begin{equation}
{\mathcal L}_{\mathrm I} = -\frac14 F_a^{\mu\nu}F^a_{\mu\nu}+\bar{\psi}^i_\mathrm{f}
\left(i\gamma^\mu{\mathcal D}_\mu-m\right)_{ij}\psi^j_\mathrm{f} ,
\label{Linv}
\end{equation}
where $a=1,\dots,8$ (respectively $i,j=1,2,3$) is the color index for the adjoint (respectively fundamental) 
representation, while ``f'' is the flavor index. 
The field strength is
\begin{equation}
F^a_{\mu\nu}=\partial_\mu A^a_\nu-\partial_\nu A^a_\mu+gf^{abc}A^b_\mu A^c_\nu,
\end{equation}
and the covariant derivative is defined as 
\begin{equation}
({\mathcal D}_\mu)_{ij}=\partial_\mu (\mathbb{I})_{ij}-ig A^a_\mu (t^a)_{ij},
\end{equation}
with $g$ the (strong) coupling constant. Finally, the $SU(N)$ generators $t^a$ satisfy the commutation relations
\begin{equation}
[t^a,t^b]=if^{abc}t^c,
\label{SU(N)_gen_comm_rel}
\end{equation}
with $f^{abc}$ the totally antisymmetric $SU(N)$ structure constants. Useful formulas 
involving the $SU(N)$ structure constants are reported in \appref{SU(N)ids}. 
\newline
\indent
${\mathcal L}_{\mathrm I}$ is invariant under the (infinitesimal) local gauge transformations
\be
\delta A^a_\mu=-\frac1g\partial_\mu\theta^a+f^{abc}\theta^bA^c_\mu\qquad
\delta_{\theta}\psi^i_\mathrm{f}=-i \theta^a(t^a)_{ij}\psi^j_\mathrm{f} \qquad 
\delta_{\theta}\bar\psi^i_\mathrm{f}=i \theta^a\bar\psi^j_\mathrm{f} (t^a)_{ji}, 
\label{QCD_gauge_trans}
\ee
where $\theta^a(x)$ are the local infinitesimal parameters corresponding to the $SU(N)$ generators $t^a$.
\newline
\indent
In order to quantize the theory, the gauge invariance needs to be broken; this is achieved through a (covariant) gauge fixing function ${\mathcal F}^a$, giving rise to the  (covariant) gauge fixing Lagrangian ${\mathcal L}_{\mathrm{GF}}$ and its associated Faddeev-Popov ghost term ${\mathcal L}_{\mathrm{FPG}}$. The most general way of writing these 
terms is through the BRST operator $s$~\cite{Becchi:1974md,Becchi:1975nq} and the Nakanishi-Lautrup multiplier $B^a$ \cite{Nakanishi:1966cq,Lautrup:1966cq} which represents an auxiliary, non-dynamical field, that can be eliminated through 
its (trivial) equation of motion. Then
\begin{eqnarray}
{\mathcal L}_{\mathrm{GF}}&=&-\frac\xi2(B^a)^2+B^a{\mathcal F}^a,\nonumber\\
{\mathcal L}_{\mathrm{FPG}} &=&-\bar c^a s{\mathcal F}^a,
\label{FPG_Lag}
\end{eqnarray}
where 
\be
\delta_{\mathrm{BRST}}\Phi=\epsilon s\Phi,
\ee
with $\epsilon$ a Grassmann constant parameter, and $s$ the BRST operator acting on the QCD fields as 
\begin{eqnarray}
sA^a_\mu=\partial_\mu c^a+gf^{abc}A^b_\mu c^c
&\qquad& s c^a=-\frac12g f^{abc}c^bc^c, 
\nonumber \\
s\psi^i_\mathrm{f}=ig c^a(t^a)_{ij}\psi^j_\mathrm{f} &\qquad& s\bar c^a=B^a, \nonumber \\
s\bar\psi^i_\mathrm{f}=-ig c^a\bar\psi^j_\mathrm{f} (t^a)_{ji}  &\qquad& sB^a=0.
\label{BRSTtrans}
\end{eqnarray}
We thus see that the sum of the gauge fixing and Faddev-Popov terms can be written as  a total BRST variation
\begin{eqnarray}
{\mathcal L}_{\mathrm{GF}}+{\mathcal L}_{\mathrm{FPG}}=s\left(\bar c^a\mathcal{F}^a-\frac\xi2\bar c^a B^a\right).
\label{gffer}
\end{eqnarray}
This is of course expected, since it is well known that total BRST variations cannot appear in the physical spectrum of the theory, implying, in turn, the gfp independence of the $S$-matrix elements and physical observables.
\newline
\indent
As far as the gauge fixing function is concerned, there are several possible choices.
The usual linear $R_\xi$ gauges, correspond to the covariant choice 
\be
{\mathcal F}^a_{R_\xi}=\partial^\mu A^a_\mu.
\label{Rxi_ch2}
\ee
In this case one has
\begin{eqnarray}
{\mathcal L}_{\mathrm{GF}}&=&\frac1{2\xi}(\partial^\mu A^a_\mu)^2,\nonumber\\
{\mathcal L}_{\mathrm{FPG}} &=&\partial^\mu\bar c^a \partial_\mu c^a+gf^{abc}(\partial^\mu\bar c^a)A^b_\mu c^c;
\end{eqnarray}
the Feynman rules corresponding to such gauge are reported in \appref{Frules}.
One can also consider non-covariant gauge fixing functions, such as~\cite{Delbourgo:1974pa,Kummer:1974ze,Konetschny:1975he,Frenkel:1976zk,Dokshitzer:1978hw,Leibbrandt:1987qv,Capper:1981rd,Andrasi:1981rr}
\be
{\mathcal F}^a_\eta=\frac{\eta^\mu\eta^\nu}{\eta^2}\partial_\mu A^a_\nu,
\label{axial_gauge}
\ee
where $\eta^\mu$ is an arbitrary but constant four-vector.  In general, we can classify these gauges from the different value of $\eta^2$, \ie $\eta^2<0$ (axial gauges),  $\eta^2=0$ (light-cone gauge) and, finally, $\eta^2>0$ (Hamilton or time-like gauge).  In this case
\begin{eqnarray}
{\mathcal L}_{\mathrm{GF}}&=&\frac1{2\xi(\eta^2)^2}(\eta^\mu\eta^\nu\partial_\mu A^a_\nu)^2,\nonumber\\
{\mathcal L}_{\mathrm{FPG}} &=&\frac{\eta^\mu\eta^\nu}{\eta^2}\left[\partial_\mu\bar c^a \partial_\nu c^a+gf^{abc}(\partial^\mu\bar c^a)A^b_\nu c^c\right].
\end{eqnarray}
Notice that these non-covariant gauges are ghost-free, since it can be shown
 that,  in dimensional regularization, the ghosts decouple completely from the $S$-matrix \cite{Capper:1981rd,Andrasi:1981rr}. 
Another non-covariant gauge fixing function is the one determining the Coulomb gauge, which arises from choosing
\be
{\mathcal F}^a=\left(g_{\mu\nu}-\frac{\eta^\mu\eta^\nu}{\eta^2}\right)\partial_\mu A^a_\nu.
\ee
Finally, due to their central importance for the PT,  
the particular class of gauges known as background field gauges~\cite{Abbott:1981ke,Abbott:1980hw}
will be described in detail in \secref{PTBFM}.
\newline
\indent
Throughout this report we will use dimensional regularization to regulate loop integrals.
We will employ the short-hand notation
\be
\int_{k}\equiv\mu^{2\varepsilon}(2\pi)^{-d}\int\!d^d k,
\ee
where $d=4-\epsilon$ is the dimension of space-time and $\mu$ the 't Hooft mass-scale, 
introduced to guarantee that the coupling constant remains dimensionless in $d$ dimensions.
In addition, the standard result 
\be
\int_k \frac{1}{k^{2}} = 0,
\label{dreg}
\ee
will be often used to set to zero various terms appearing in the PT procedure.

\subsection{Gauge cancellations in the $S$-matrix and the origin of the pinch technique}
\label{sec:Ti}
\noindent
Consider the $S$-matrix
element $T$ for the elastic scattering of two fermions of masses
$m_{1}$ and $m_{2}$. To any order in perturbation theory $T$ is independent
of the gfp  $\xi$. 
On the other hand, 
the conventionally defined proper box, vertex, and self-energy, collectively
depicted in \Figref{s-t-u_channels} $(a)$, $(b)$, and $(c)$, respectively, 
depend on {\it explicitly} on the gfp $\xi$ already at one-loop level.
Specifically, since $s+t+u = 2(m_1^2+m_2^2)$, we have that 
\be
T(s,t,m_i)= T_{1}(t,\xi) + T_{2}(t,m_i,\xi)+T_{3}(t,s,m_i,\xi) \,.
\label{S-matrix}
\ee
where the gfp-dependent subamplitudes $T_{1}$, $T_{2}$, and $T_{3}$ 
are composed of self-energy, vertex, and box diagrams, respectively;  
for example,  $T_{1}(t,\xi)$ corresponds to the 
standard propagator, depending kinematically only on $t=(r_1 - r_2)^2 = (p_1 - p_2)^2$, 
but not on $s=(r_1+p_1)^2 = (r_2+p_2)^2$, nor on the external masses. 
\newline
\indent
The central observation of the PT is that 
the $\xi$-dependence  of the proper self-energy will cancel against 
contributions from the vertex- and box-graphs, 
which, at first glance, do not seem to contain propagator-like parts.
In turn, this cancellation can be employed to define 
{\it gfp-independent} subamplitudes with distinct kinematic properties. 
Indeed, given that the total sum
$T(s,t,m_{i})$ is gfp-independent, it is relatively easy to show that
Eq.~(\ref{S-matrix}) can be recast in the form
\be
T(s,t,m_i)=
{\widehat{T}}_1 (t) + {\widehat{T}}_2 (t,m_i)+
{\widehat{T}}_3 (t,s,m_i) ,
\label{S2-matrix}
\ee
where the ${\widehat{T}}_{i}$ ($i=1,2,3$) are {\it individually} $\xi$-independent.
An immediate way to see this is by differentiating both sides of (\ref{S-matrix})
with respect to $\xi$ and $s$; the rhs vanishes because \linebreak$dT(s,t,m_i)/d\xi =0$; on the lhs we have that 
$dT_{1}(t,\xi)/ds = dT_{2}(t,m_i,\xi)/ds =0$. Thus, 
\be
\frac{d^2 T_{3}(t,s,m_i,\xi)}{ds\, d\xi}=0,
\ee
from which it follows that  $T_{3}$ can be written as a sum of two functions, 
one independent of $\xi$ and one independent of $s$, \ie
\be
T_{3}(t,s,m_i,\xi) = {\widehat{T}}_3 (t,s,m_i) + h(t,m_i,\xi).
\ee
So, we have 
\be
T(s,t,m_i)= T_{1}(t,\xi) + {\tilde T_{2}}(t,m_i,\xi) + {\widehat{T}}_3 (t,s,m_i),
\label{S3-matrix}
\ee
where ${\tilde T_{2}}(t,m_i,\xi)\equiv T_{2}(t,m_i,\xi)+h(t,m_i,\xi)$.
The argument may be continued by differentiating both sides of Eq.~(\ref{S3-matrix})
with respect to $\xi$ and $m_i$, now obtaining   
\be
\frac{d^2 {\tilde T_{2}}(t,m_i,\xi)}{dm_i\, d\xi}=0,
\ee
and thus 
\be
{\tilde T_{2}}(t,m_i,\xi) = {\widehat{T}}_2 (t,m_i) + f(t,\xi).
\ee
The last step is to write ${\widehat{T}}_1 (t,\xi) \equiv T_{1}(t,\xi) + f(t,\xi)$;
clearly, since $dT(s,t,m_i)/d\xi =0$, we must have that  $d{\widehat{T}}_1 (t,\xi)/d\xi =0$, 
and therefore ${\widehat{T}}_1 (t,\xi) = {\widehat{T}}_1 (t)$, thus 
arriving at Eq.~(\ref{S2-matrix}).
\begin{figure}[!t]
\bce
\includegraphics[width=13cm]{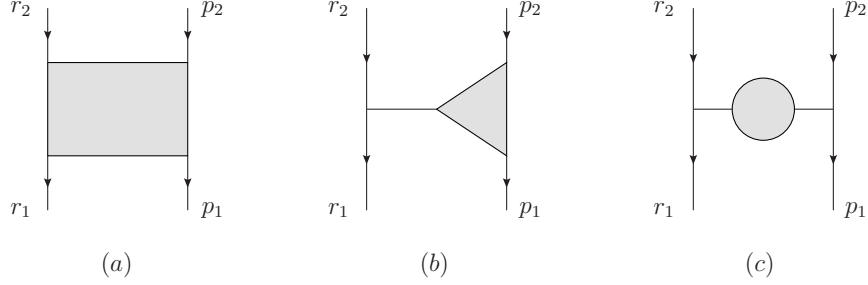}
\ece
\caption{\figlab{s-t-u_channels} The diagrams contributing to the $S$-matrix, grouped according to their topologies
and their dependence on the  Mandelstam variables $s$, $t,$ and $u$, with $s=(r_1+p_1)^2 = (r_2+p_2)^2$, 
$t=(r_1 - r_2)^2 = (p_1 - p_2)^2$,  and $u=(r_1 - p_2)^2 = (p_1 - r_2)^2$,
with  $s+t+u = \Sigma_i m_i^2$. Evidently,     
box-diagrams $(a)$ depend on $s,t,m_i^2$, vertex-diagrams $(b)$ depend on $t,m_i^2$, and self-energy diagrams $(c)$ 
depend only on $t$.}
\end{figure}
\newline
\indent
The above proof is meant to demonstrate the possibility of decomposing 
$T(s,t,m_i)$ in terms of individually gfp-independent subamplitudes, as  
in (\ref{S2-matrix}), but does not specify how this decomposition is realized operationally, 
nor whether it is physically unique. 
To be sure, at the level presented above, the decomposition is not mathematically unique, since 
one can always add an arbitrary function $g(t)$ to ${\widehat{T}}_1 (t)$ and subtract it from 
${\widehat{T}}_2 (t,m_i)$; this changes the 
definition of what the individual 
subamplitudes are, without changing the value of 
the full $T(s,t,m_i)$. However, when the ${\widehat{T}}_i$ are endowed with 
physical properties, such as unitarity and analyticity, Dyson resummability, and invariance under the 
renormalization group (RG), to name a few, the above arbitrariness disappears.  
As we will see in the rest of this review, 
the PT Green's functions, which, by construction, have all the aforementioned 
physical properties built in, provide the  field-theoretically and physically unique  way  
of realizing the decomposition of Eq.~(\ref{S2-matrix}).

\subsection{\label{PT1} The pinch technique mechanism of gauge fixing parameter cancellations at one loop}
\noindent
Let us start by considering the  $S$-matrix
element  for the  quark-quark elastic scattering  process 
\mbox{$q(p_1) q(r_1)\to q(p_2)q(r_2)$} in  QCD.
We have that $p_1+ r_1= p_2 +r_2$, and 
set $q=r_2-r_1 = p_1-p_2$,
with $t=q^2$ the  square of the momentum transfer.
The longitudinal momenta responsible  for triggering the
kinematical rearrangements characteristic of the PT stem either
from   the  bare  gluon   propagator, $\Delta_{\alpha\beta}^{(0)}(k)$,
or   from  the  {\it external} bare
(tree-level) three-gluon vertices, {\it i.e.}, the vertices where the
physical momentum transfer $q$ is entering. 
\newline
\indent
To study the origin of the longitudinal momenta in detail, consider first the gluon propagator 
$\Delta_{\alpha\beta}(k)$; after factoring out the trivial color factor $\delta^{ab}$, in the \Rxi gauges it has the form\footnote{In the definition of the gluon propagator $\Delta_{\alpha\beta}$ we explicitly pull out an $i$ factor on the lhs, which accounts for the slightly unusual (but totally equivalent) form of writing Eq.~(\ref{sluninv}). This is done in order to be consistent with the definition of the Green's functions in terms of functional differentiation of the generating functional introduced later on (Section 7).} 
\begin{equation}
i\Delta_{\alpha\beta}(q,\xi)= -i\left[P_{\alpha\beta}(q)\Delta(q^2,\xi) + 
\xi\frac{q_{\alpha}q_{\beta}}{q^4}\right],
\label{prop_cov}
\end{equation}
with $P_{\alpha\beta}(q)$ the dimensionless transverse projector defined as 
\be
P_{\alpha\beta}(q)= \ g_{\alpha\beta} - \frac{q_\alpha q_\beta}{q^2}.
\label{projector}
\ee
The scalar function $\Delta(q^2,\xi)$ is related to the 
all-order gluon self-energy 
\be
\Pi_{\alpha\beta}(q,\xi)=P_{\alpha\beta}(q)\Pi(q^2,\xi),
\ee
through
\be
\Delta(q^2,\xi) = \frac{1}{q^2 + i\Pi(q^2,\xi)}.
\label{fprog}
\ee
Since $\Pi_{\alpha\beta}$ has been defined in (\ref{fprog}) 
with the imaginary  factor $i$ factored out in front, it is simply 
given by the corresponding Feynman diagrams in Minkowski space.
The inverse of $\Delta_{\alpha\beta}$ can be found by requiring that
\begin{equation}
i\Delta^{am}_{\alpha\mu}(q,\xi)(\Delta^{-1})^{\mu\beta}_{mb}(q,\xi)=\delta^{ab}g_\alpha^\beta,
\label{sluninv}
\end{equation}
and it is given by
\begin{equation}
\Delta^{-1}_{\alpha\beta}(q,\xi)=
 iP_{\alpha\beta}(q) \Delta^{-1}(q^2,\xi) + \frac i\xi q_{\alpha}q_{\beta}.
\label{inv_prog}
\end{equation}
\indent
At tree-level we have that 
\bea
i\Delta_{\alpha\beta}^{(0)}(q,\xi) &=& 
-id(q^2) \left[\ g_{\alpha\beta} - (1-\xi) \frac{\D q_\alpha
q_\beta}{q^2}\right]\,,\label{GluProp-0}\nonumber\\
d(q^2)&=&\frac1{q^2}\,.
\label{GluProp}
\eea 
Evidently, the longitudinal (pinching) momenta 
are proportional to $(1-\xi)$, and vanish for 
the particular choice $\xi=1$, to be referred to as the ``Feynman gauge''; in that gauge the propagator 
is simply proportional to $g_{\alpha\beta}d(q^2)$. 
The case $\xi=0$, known as the ``Landau gauge'', gives rise to a transverse
$\Delta_{\alpha\beta}^{(0)}(k)$, but does not eliminate the pinching momenta.
\newline
\indent
In order to gradually build up the concepts, and at the same time introduce some useful notation,  
let us see what happens to the pinching momenta at tree-level.
Defining 
\be
{\mathcal V}^{a\alpha}(p_1,p_2) = \bar{u}(p_1) gt^{a}\gamma^{\alpha} u(p_2), 
\label{defV}
\ee
the tree-level amplitude reads
\be
{\mathcal T}^{(0)} = 
i{\mathcal V}^{a\alpha}(r_1,r_2)
i\Delta^{(0)}_{\alpha\beta}(q)
i{\mathcal V}^{a\beta}(p_1,p_2).
\label{treeamp}
\ee
Then, since the on-shell spinors satisfy the equations of motion
\be
\bar{u}(p)(\not\! p - m) = 0 = (\not\! p - m)u(p), 
\label{spinors}
\ee
the longitudinal part coming from $\Delta_{\alpha\beta}^{(0)}$ vanishes, and we obtain
\be
{\mathcal T}^{(0)} = 
i{\mathcal V}^{a\alpha}(r_1,r_2) d(q^2) {\mathcal V}^{a}_{\alpha}(p_1,p_2).
\ee
\indent
Let us next consider the conventional three-gluon vertex, to be denoted by
$\Gamma^{amn}_{\alpha\mu\nu}(q,k_1,k_2)$; of course, in the case of the specific process we consider   
this vertex appears for the first time at one loop.
It is given by the following manifestly Bose-symmetric expression
(all momenta are incoming, {\it i.e.}, $q+k_1+k_2 = 0$)
\bea
& &i\Gamma^{amn}_{\alpha\mu\nu}(q,k_1,k_2)=g f^{amn}
\Gamma_{\alpha \mu \nu}(q,k_1,k_2), \nonumber \\
& & \Gamma_{\alpha \mu \nu}(q,k_1,k_2)= g_{\mu\nu}(k_1-k_2)_{\alpha}+g_{\alpha\nu} (k_2-q)_{\mu}+
g_{\alpha\mu}(q-k_1)_{\nu}.
\label{tgv_ch2}
\eea
It is elementary to verify that the vertex satisfies the following WIs:
\bea 
q^{\alpha} \Gamma_{\alpha \mu \nu}(q,k_1,k_2) &=& 
k_2^2 {P}_{\mu\nu}(k_2) - k_1^2 {P}_{\mu\nu}(k_1),
\label{3gWI-1} \nonumber\\
k_1^{\mu} \Gamma_{\alpha \mu \nu}(q,k_1,k_2) &=& 
q^2 { P}_{\alpha\nu}(q) - k_2^2 {P}_{\alpha\nu}(k_2),
\label{3gWI-2} \nonumber\\
k_2^{\nu} \Gamma_{\alpha \mu \nu}(q,k_1,k_2)&=&
k_1^2 {P}_{\alpha\mu}(k_1) - q^2 {P}_{\alpha\mu}(q).
\label{3gWI-3}
\eea
\indent
To show how the relevant pinching momenta are identified in  
the conventional three-gluon vertex, we split $\Gamma_{\alpha \mu \nu}(q,k_1,k_2)$ into two parts, 
\be
\Gamma_{\alpha \mu \nu}(q,k_1,k_2) 
=\Gamma_{\alpha \mu \nu}^{{\rm F}}(q,k_1,k_2)+
\Gamma_{\alpha \mu \nu}^{{\rm P}}(q,k_1,k_2),
\label{decomp}
\ee
with 
\bea
\Gamma_{\alpha \mu \nu}^{{\rm F}}(q,k_1,k_2) &=& 
(k_1-k_2)_{\alpha} g_{\mu\nu} + 2q_{\nu}g_{\alpha\mu} 
- 2q_{\mu}g_{\alpha\nu}, \label{GF}\nonumber\\
\Gamma_{\alpha \mu \nu}^{{\rm P}}(q,k_1,k_2) &=& 
 k_{2\nu} g_{\alpha\mu} - k_{1\mu}g_{\alpha\nu}.  
\label{GP}
\eea
The vertex $\Gamma_{\alpha \mu \nu}^{{\rm F}}(q,k_1,k_2)$ 
is Bose-symmetric only with respect to the
$\mu$ and $\nu$ legs. 
Evidently the above decomposition assigns a special role 
to the $q$-leg,
and allows $\Gamma_{\alpha \mu \nu}^{{\rm F}}(q,k_1,k_2)$ to satisfy the WI
\be 
q^{\alpha} \Gamma_{\alpha \mu \nu}^{{\rm F}}(q,k_1,k_2) = 
(k_2^2 - k_1^2)g_{\mu\nu}.
\label{WI2B}
\ee
where the rhs is the difference of two inverse tree-level 
propagators in the Feynman gauge.  
The term $\Gamma_{\alpha \mu \nu}^{{\rm P}}(q,k_1,k_2)$, which 
in configuration space corresponds to a pure divergence, 
contains the longitudinal momenta that will pinch. 
\newline
\indent
When considering a vertex or a box diagram,
the effect of the pinching momenta, regardless of their origin (gluon propagator or three-gluon vertex), 
is to trigger the elementary WI   
\bea
k_{\nu}\gamma^{\nu} &=& (\slashed{k} + \slashed{p} - m) -  (\slashed{p} -m)
\nonumber\\
&=&  -i[S^{-1}_{(0)}(k + p)  - S^{-1}_{(0)}(p)],
\label{BasicWI}
\eea
where the rhs is the difference of two inverse tree-level quark propagators. 
The first of these terms removes (pinches out) the internal tree-level fermion propagator $S^{(0)}(k+p)$, whereas 
the second term on the rhs vanishes when hitting the on-shell external leg, i.e. using 
the appropriate Dirac equation of \ref{spinors}. 
Diagrammatically, what appears in the place where the $S^{(0)}(k + p)$ was is an unphysical effective vertex, \ie a
vertex that does not appear in the original Lagrangian; as we will see, 
all such vertices  cancel in the full, gauge-invariant amplitude.
\begin{figure}[!t]
\bce
\includegraphics[width=14cm]{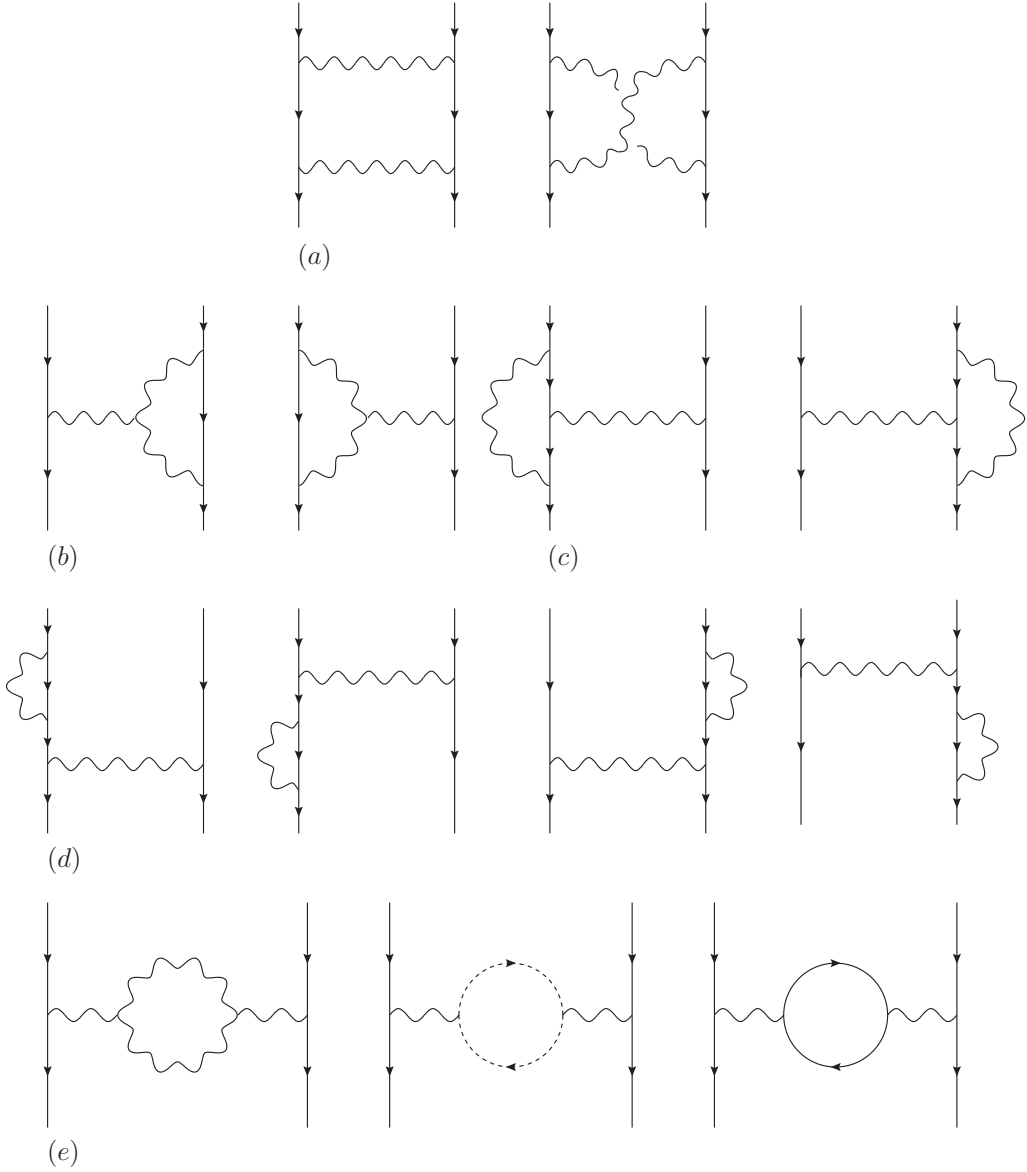}
\ece
\caption{\figlab{1l_S-matrix}The diagrams contributing to the one-loop quark elastic scattering $S$-matrix element. $(a)$ box contributions, $(b)$ non-Abelian and $(c)$ Abelian vertex contributions, $(d)$ quark self-energy corrections, and $(e)$ gluon self-energy contributions.}
\end{figure}
\newline
\indent
We next consider all one-loop graphs contributing to the $S$-matrix element
shown in \Figref{1l_S-matrix}, and isolate their 
gfp-dependent parts using the PT procedure;
what we will find is that all gfp-dependent parts, irrespectively of whether they come from 
box- or vertex-diagrams, are effectively propagator-like (we emphasize that no integration over virtual 
momenta is necessary for carrying out the pinching procedure). 
\newline
\indent

\subsubsection{\label{ptbox}The box}
\begin{figure}[t]
\bce
\includegraphics[width=11cm]{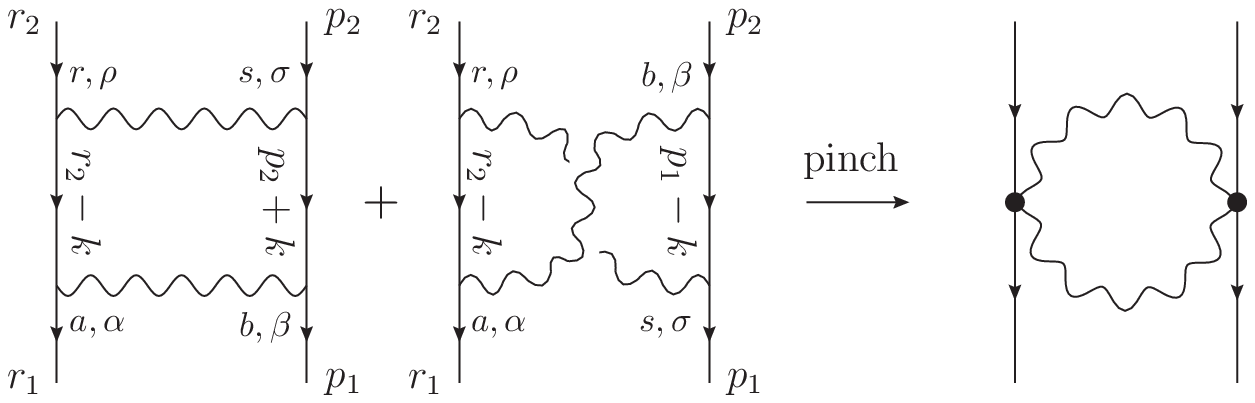}
\ece
\caption{\figlab{1l_box_pinch_contrib} Schematic representation of the propagator-like parts extracted from the boxes for a general $\xi$. Black dots indicate effective vertices that do not exist in the original theory.}
\end{figure}

We start our one-loop analysis from the two box diagrams, direct and crossed,
shown in graphs $(a)$ of \Figref{1l_S-matrix} (for the kinematics used see~\Figref{1l_box_pinch_contrib}).
For the sum of the two graphs we have
\bea
(a) &=&g^2\int_k\!\bar u(r_1)\gamma^\alpha t^aS^{(0)}(r_2-k)\gamma^\rho t^ru(r_2)\Delta^{(0)}_{\alpha\beta}(k-q)\Delta^{(0)}_{\rho\sigma}(k)\times\nonumber \\
&\times&g^2\bar u(p_1)\left\{\gamma^\beta t^aS^{(0)}(p_2+k)\gamma^\sigma t^r+\gamma^\sigma t^rS^{(0)}(p_1-k)\gamma^\beta t^a\right\}u(p_2).
\eqlab{box}
\eea
To see how the PT works, we must now 
study the action of the longitudinal momenta 
appearing in the product $\Delta^{(0)}_{\alpha\beta}(k-q)\Delta^{(0)}_{\rho\sigma}(k)$.
Therefore, let us, for concreteness, see what happens to the  
term $k_{\rho}k_{\sigma}$ coming from $\Delta^{(0)}_{\rho\sigma}(k)$.
Using  Eqs~(\ref{BasicWI}) and~(\ref{spinors}), we find that 
the contraction of $k_{\sigma}$ with the term contained in the 
brackets in the second line on the rhs of \Eqref{box} gives rise to the expression 
\bea
g^2\bar{u}(p_1) k_{\sigma}\left\{\cdots\right\}^{\beta\sigma}u(p_2) &=& 
g^2\bar{u}(p_1) \gamma^{\beta} \left\{t^a t^r - t^r t^a\right\}u(p_2)
\nonumber\\
&=&
i g^2f^{arn}\bar{u}(p_1) \gamma^{\beta}  t^nu(p_2)
\nonumber\\
&=& g f^{arn}P^\beta_\nu(q) \bar{u}(p_1)ig\gamma^{\nu} t^n  u(p_2) 
\nonumber\\
&=& \left[g f^{arn}P^{\beta}_{\nu}(q)\right] i{\mathcal V}^{n\nu}(p_1,p_2).
\eqlab{box2}
\eea
Notice that in the second step we have used the commutation relation 
of Eq.~(\ref{SU(N)_gen_comm_rel}), while in the third step  
we have used the fact that, for the on-shell process we consider,
longitudinal pieces proportional to 
$q_{\beta}q_{\nu}$ may be added for free (since they vanish anyway due to current conservation), 
thus converting $g^\beta_\nu$ to 
$P^\beta_\nu(q)$.
The term in the last line of \Eqref{box2} couples to 
the external on-shell quarks 
as a propagator; evidently all reference to the internal (off-shell) quarks 
inside the brackets has disappeared. 
To continue the calculation, ({\it i}) multiply the result
by $k_{\rho}$, ({\it ii}) let 
$k_{\rho}$ get contracted with the $\gamma^{\rho}$ in the first line of \Eqref{box},
({\it iii}) employ again the WI of Eq.~(\ref{BasicWI}), and ({\it iv}) 
use that $if^{abc}t^{a}t^{b}=-\frac{1}{2} C_{A} t^{c}$,  
where $C_{A}$ is the Casimir eigenvalues of the adjoint
representation, defined in \appref{SU(N)ids}.
The final result is a purely propagator-like term, \ie
a term that only depends on $q$ (even though it originates from a box diagram),
and couples to the external on-shell quarks as a propagator
(see \Figref{1l_box_pinch_contrib}). 
Armed with these observations, 
it is relatively easy to track down the action of all 
terms proportional to $(1-\xi)$; setting $\lambda \equiv (1-\xi)$, 
we can write the two boxes as follows,
\be
(a)= (a)_{\xi =1} + {\mathcal V}^{a}_{\alpha}(r_1,r_2)d(q^2)
\Pi_{{ {\rm box}}}^{\alpha\beta}(q,\lambda)d(q^2)  {\mathcal V}^{a}_{\beta}(p_1,p_2),
\label{box1}
\ee
where the gfp-dependent propagator-like term $\Pi^{\alpha\beta}_{{ {\rm box}}}$ is given by
\be
\Pi_{{ {\rm box}}}^{\alpha\beta}(q,\lambda) =  
\lambda g^2 C_{A} q^4\left[
\frac{\lambda}{2} 
{P}^{\alpha\mu}(q){P}^{\beta\nu}(q)
\int_k\,\frac{k_{\mu} k_{\nu}}{k^4 (k+q)^4}-
 {P}^{\alpha\beta}(q) \int_k \frac{1}{k^4 (k+q)^2}\right].
\label{box2}
\ee

\subsubsection{\label{qgv} The quark-gluon vertex}
\noindent
We next turn to the two vertex graphs,  
the non-Abelian graphs $(b)$ and the
Abelian graphs $(c)$,
shown in \Figref{1l_S-matrix}. We will analyze only one graph per subgroup, since the mirror graphs are 
to be treated in exactly the same way.
As in the case of the boxes, we want to isolate the gfp-dependent pieces
coming from the internal gluon propagators. The action of the 
corresponding longitudinal momenta is determined following the 
PT procedure; again, they give rise to effectively propagator-like 
terms, as shown schematically in \Figref{1l_vertex_pinch_contrib}, 
where the kinematics used are explicitly shown. 
Note that we do {\it not} yet split  the three gluon vertex 
as described in Eq.~(\ref{decomp}); for the moment
we simply collect the terms proportional to different powers of $\lambda$.
After a straightforward calculation, we find for the corresponding results
\bea
(b) &=& (b)_{\xi =1} + {\mathcal V}^{a}_{\alpha}(r_1,r_2) d(q^2)
\Pi_{{ {\rm nav}}}^{\alpha\beta}(q,\lambda)d(q^2)  {\mathcal V}^{a}_{\beta}(p_1,p_2),
\nonumber\\
(c) &=& (c)_{\xi =1} + {\mathcal V}^{a}_{\alpha}(r_1,r_2) d(q^2)
\Pi_{{ {\rm av}}}^{\alpha\beta}(q,\lambda)d(q^2)  {\mathcal V}^{a}_{\beta}(p_1,p_2),
\eea
with the propagator-like pieces given by
\bea
\Pi_{{ {\rm nav}}}^{\alpha\beta}(q,\lambda) &=& -\frac{\lambda^2}{2} g^2 C_{A} q^4 
P^{\alpha\mu}(q)P^{\beta\nu}(q)\!\int_k\!\frac{k_{\mu} k_{\nu}}{k^4 (k+q)^4} \nonumber \\
&+&\lambda  g^2 C_{A} q^2\! \left[q^2 P^{\alpha\beta}(q) 
\int_k\frac{1}{k^2 (k+q)^4} 
+P^{\beta\mu} (q)\! \int_k\frac{k^{\alpha}k_{\mu}}{k^4 (k+q)^2}
- P^{\alpha\beta}(q)\! \int_k\! \frac{1}{k^4}\right]\!,
\nonumber\\
\nonumber\\
\Pi_{{ {\rm av}}}^{\alpha\beta}(q,\lambda) &=&
\lambda g^2 \left(\frac{C_{A}}{2} -  C_{f}\right) q^2 {P}^{\alpha\beta}(q)\int_k\!\frac{1}{k^4},
\label{severt}
\eea
where $C_{f}$ is the Casimir eigenvalues in the 
fundamental representation, see again \appref{SU(N)ids}.
\begin{figure}[t]
\bce
\includegraphics[width=9cm]{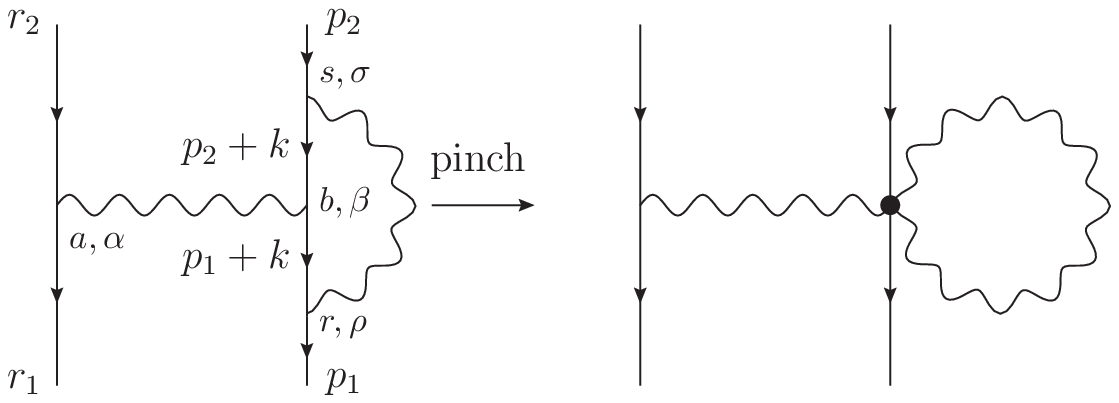}
\includegraphics[width=11cm]{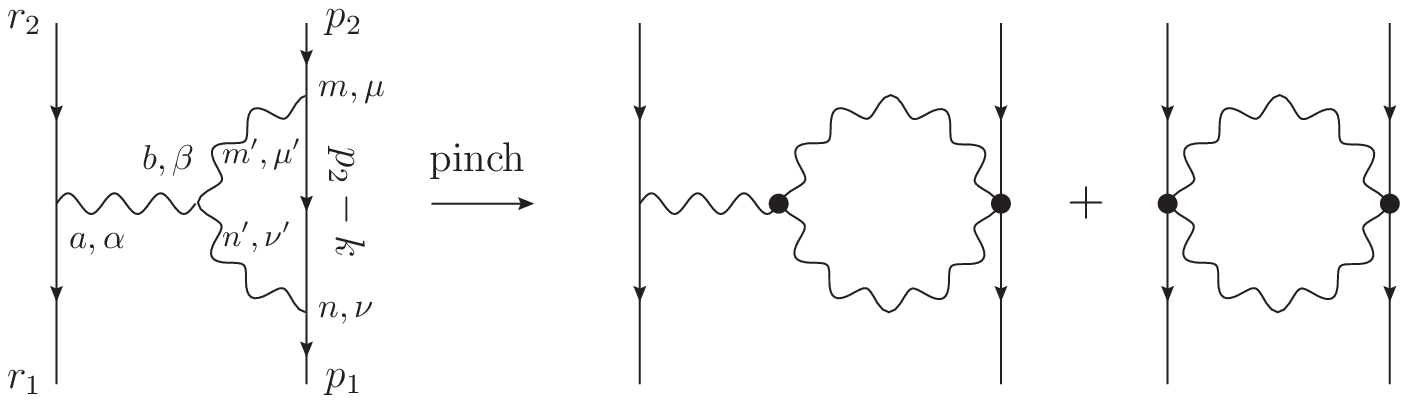}
\ece
\caption{\figlab{1l_vertex_pinch_contrib} Schematic representation of the propagator-like parts extracted from the 
one-loop vertex graphs for general $\xi$.}
\end{figure}

\subsubsection{The quark self-energy}
\noindent
Let us now turn to the one-loop corrections to 
the self-energy of the on-shell test quarks shown in the group $(d)$ of \Figref{1l_S-matrix}; notice that these four graphs are multiplied by a factor of $\frac{1}{2}$. Let us then concentrate on one of these graphs, shown in \Figref{1l_wavef_pinch_contrib}. It reads
\be
(d)=\frac12{\mathcal V}^{a\alpha}(r_1,r_2)\Delta^{(0)}_{\alpha\beta}(q)\bar u(p_1)g\gamma^\beta t^aS^{(0)}(p_2)\Sigma_\xi(p_2)u(p_2),
\eqlab{d}
\ee
where
\be
\Sigma_\xi(p_2)=g^2C_{f}\int_k\!\gamma^\rho\Delta^{(0)}_{\rho\sigma}(k)S^{(0)}(p_2+k)\gamma^\sigma.
\label{Sigma_xi}
\ee
Using the same methodology employed so far, it is easy to show that 
\bea
\Sigma_\xi(p_2)&=&\Sigma_{\xi=1}(p_2)-\lambda g^2C_{f}(\ps_2-m)\int_k\!\frac1{k^4}S^{(0)}(p_2+k)\ks\nonumber \\
&=&\Sigma_{\xi=1}(p_2)-\lambda g^2C_{f}\left\{(\ps_2-m)\int_k\!\frac1{k^4}-(\ps_2-m)\int_k\!\frac1{k^4}S^{(0)}(p_2+k)(\ps_2-m)\right\}.\nonumber \\
\label{STOT}
\eea
We next insert the rhs of (\ref{STOT}) back into \Eqref{d}. 
Clearly, the second term in the brackets vanish on-shell as the second fermion inverse propagator will trigger the Dirac equation;
also the term $\Sigma_{\xi=1}(p_2)$ gives simply $(d)_{\xi=1}$. Thus the only term furnishing a propagator part will be the first one in the brackets and we will have
\be
(d)=(d)_{\xi=1}+{\mathcal V}^{a}_{\alpha}(r_1,r_2)d(q^2)\Pi^{\alpha\beta}_\mathrm{qse}(q,\lambda)d(q^2){\mathcal V}^{a}_{\beta}(p_1,p_2),
\ee
where
\be
\Pi^{\alpha\beta}_\mathrm{qse}(q,\lambda)=\frac 12\lambda g^2C_{f}q^2P^{\alpha\beta}(q)\int_k\!\frac1{k^4}.
\ee
Notice that $\Pi^{\alpha\beta}_{{ {\rm qse}}}$ is proportional to 
 $C_{f}$ instead of $C_{A}$, and is in that sense of Abelian nature. 
Indeed, after multiplying it by a factor of 2 (accounting for both quark fields), 
$\Pi_{{ {\rm qse}}}$ 
cancels exactly against the part of $\Pi_{{ {\rm av}}}$
proportional to $C_{f}$ in Eq.~(\ref{severt}). For  $C_{f}=1$ this is simply the 
standard QED gfp-cancellation between the elector-photon vertex and the 
electron wave-function.  
\newline
\indent
Note that in obtaining the rhs of (\ref{STOT}) we have not assumed that  
$\Sigma(p_2)$ is actually sandwiched between on-shell spinors, as indicated
in \Eqref{d}. Thus, the gfp-independent quark self-energy should be 
identified with the first term on the rhs of (\ref{STOT}), \ie
\be
\widehat\Sigma(p) =  \Sigma_{\xi=1}(p).
\label{Sigmahat}
\ee
A more thorough analysis~\cite{Papavassiliou:1994yi,Binosi:2001hy}, 
where an off-shell $\Sigma(p)$ is embedded  
into a quark-gluon scattering process [$g(k_1) q(k_2)\to g(k_3) q(k_4)$, with 
$p=k_1+k_2=k_3+k_4$] and the pinching procedure is 
repeated, shows that 
Eq.~(\ref{Sigmahat}) is absolutely general: the gfp-independent 
off-shell quark self-energy {\it coincides} with the conventional 
quark self-energy calculated in the Feynman gauge. 
\begin{figure}[t]
\bce
\includegraphics[width=9.5cm]{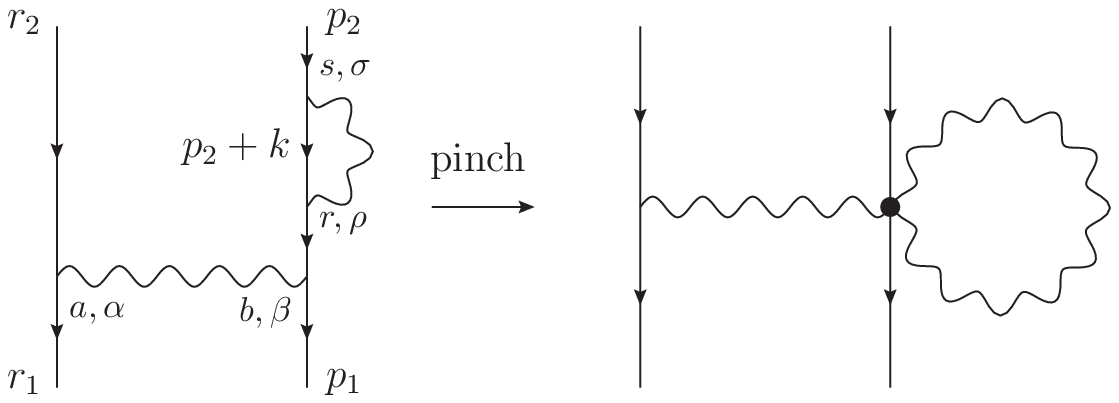}
\ece
\caption{\figlab{1l_wavef_pinch_contrib} Schematic representation of the propagator-like parts extracted from (one of) the 
quark self-energy corrections graphs for general $\xi$.}
\end{figure}

\subsubsection{\label{PT2} Final cancellation of all gauge fixing parameter dependence}
\begin{figure}[t]
\bce
\includegraphics[width=11.5cm]{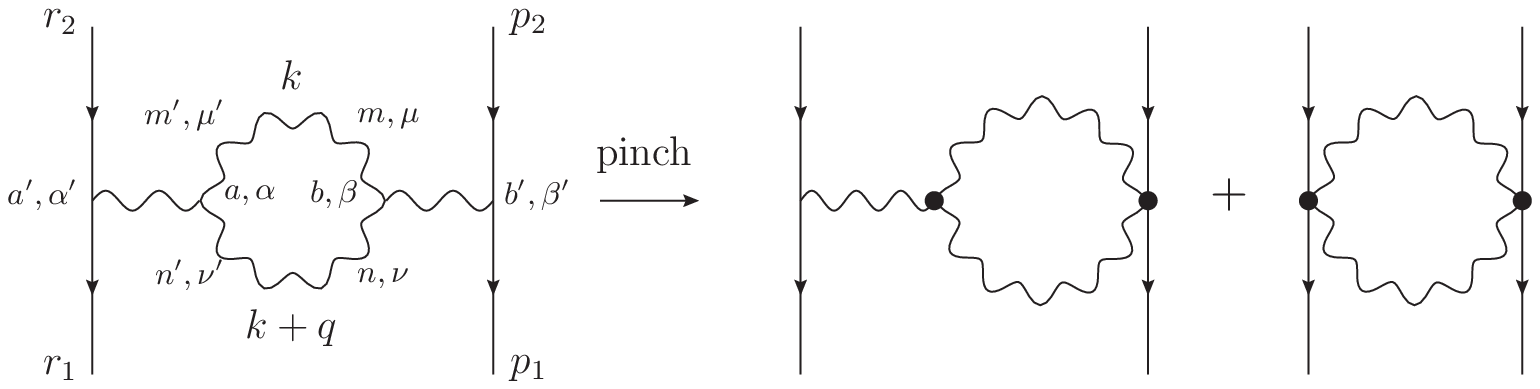}
\ece
\caption{\figlab{1l_selferg_pinch_contrib} Schematic representation of the propagator-like parts extracted from the 
gluon self-energy corrections graph for general $\xi$.}
\end{figure}
\noindent
We will now show that 
the propagator-like parts extracted from all the 
previous diagrams cancel exactly against 
analogous terms contained in the conventional self-energy graphs $(e)$ of \Figref{1l_S-matrix}.
Of course this cancellation is guaranteed
to take place, regardless of how one may choose to organize the 
calculation, given that it amounts to the gfp-independence 
of the one-loop amplitude.
It is conceptually important, however,  
to establish a systematic way for extracting the 
relevant terms from the conventional one-loop self-energy
using nothing but tree-level WIs.
\newline
\indent
To that end, we concentrate only on the graph containing the three-gluon vertices (see \Figref{1l_selferg_pinch_contrib}); the ghost and fermion graphs have no pinching momenta and thus will be inert. We have
\be
(e)={\mathcal V}^a_\alpha(r_1,r_2)d(q^2)\Pi^{\alpha\beta}(q,\lambda)d(q^2){\mathcal V}^a_\beta(p_1,p_2).
\ee
Then, we let the longitudinal momenta coming from the tree-level propagators
act on the two bare three-gluon vertices,
triggering the two WIs of  Eqs~(\ref{3gWI-2}). It turns out that only the 
terms proportional to the transverse projector 
$P_{\alpha\beta}(q)$ survive, furnishing  
\be
\Pi^{\alpha\beta}(q,\lambda)  = 
\Pi^{\alpha\beta}_{\xi=1}(q) + \Pi^{\alpha\beta}_{{ {\rm gse}}}(q,\lambda),
\ee
with
\bea
\Pi^{\alpha\beta}_{{ {\rm gse}}}(q,\lambda) &=& \frac{\lambda^2}{2}g^2 C_{A} q^4 
{P}^{\alpha\mu}(q){P}^{\beta\nu}(q) \int_k\!
\frac{k_{\mu} k_{\nu}}{k^4 (k+q)^4} \nonumber \\
 &-&\lambda g^2  C_{A} q^2\!\left[q^2 {P}^{\alpha\beta}(q)\! \int_k \frac{1}{k^2 (k+q)^4} + 
2 {P}^{\beta\mu} (q) 
\int_k \frac{k^{\alpha}k_{\mu} }{k^4 (k+q)^2}
- {P}^{\alpha\beta}(q)\! \int_k \frac{1}{k^4} \right]\!.\nonumber \\
\label{gse}
\eea
\indent
We are now in the position of showing the cancellation of the gfp-dependent pieces; in fact,
adding all the terms we have been isolating, we find 
\be
\Pi^{\alpha\beta}_{{ {\rm gse}}}(q,\lambda) +\Pi_{{ {\rm box}}}^{\alpha\beta}(q,\lambda ) +
2 \left[\Pi_{{ {\rm av}}}^{\alpha\beta}(q,\lambda) 
+\Pi_{{ {\rm nav}}}^{\alpha\beta}(q,\lambda)\right]
+4 \Pi_{{ {\rm qse}}}^{\alpha\beta}(q,\lambda)
= 0.
\label{tgc}
\ee
In the above formula 
the multiplicative factor of 2 comes from the mirror vertex graphs 
and the 4 from the four external quarks. The contributions of each term to the different gfp-dependent structures appearing in the PT process is shown in Table~\ref{gfp_canc_details}.
\begin{table}
\centering
\begin{tabular}{c||c|c|c|c|}
 & $\lambda^2\int_k\!\frac{k_\mu k_\nu}{k^4(k+q)^4}$ & $\lambda\int_k\!\frac{k_\mu k_\nu}{k^4(k+q)^2}$ & $\lambda\int_k\!\frac{1}{k^2(k+q)^4}$ & $\lambda\int_k\!\frac{1}{k^4}$\\
 \hline\hline
 $\Pi_{{ {\rm box}}}$ & $\frac12C_{A}$ & 0 &  $-C_{A}$ & 0 \\
 \hline
 $2\Pi_{{ {\rm av}}}$ & 0 & 0  &  0& $C_{A}-2C_{f}$ \\
 \hline
 $2\Pi_{{ {\rm nav}}}$ & $-C_{A}$ & $2C_{A}$ & $2C_{A}$ & $-2C_{A}$ \\
 \hline
 $4\Pi_{{ {\rm qse}}}$ & 0 &  0 & 0 & $2C_{f}$ \\
 \hline
 $\Pi_{{ {\rm gse}}}$ & $\frac12 C_{A}$ & $-2C_{A}$ & $-C_{A}$ & $C_{A}$\\
 \hline\hline
 Total & 0 & 0 & 0 & 0 
\end{tabular}
\caption{\label{gfp_canc_details}Contributions of the box, vertex and self-energy diagrams to the different $\xi$-dependent structures appearing in the PT process. The sum of each column is zero, showing the well-known property of the gfp-independence of the $S$-matrix elements.}
\end{table}
\newline
\indent
In summary,  all gfp-dependent  terms have been  eliminated in  a very
particular way.  Specifically, due  to the PT procedure employed, all
gfp-dependent pieces  turned out to be propagator-like.   As a result,
all  gfp-dependence has  canceled  giving rise  to subamplitudes  that
maintain  their  original  kinematic  identity (boxes,  vertices,  and
self-energies),  and are,  in addition,  individually gfp-independent.
It is important to appreciate  the fact that the explicit cancellation
carried  out  amounts  effectively  to  choosing  the  Feynman  gauge,
$\xi=1$, from the  beginning.  Of course, there is  no doubt that this
can be done for the entire physical amplitude considered; the point is
that,  thanks to  the  PT, one  may  move from  general  $\xi$ to  the
specific  $\xi=1$  without   compromising  the  notion  of  individual
topologies.  Such a notion would  have been lost if, for instance, the
demonstration  of the gfp-independence  involved the  integration over
virtual momenta;  had one  opted for this  latter approach,  one would
have eventually succeeded to  demonstrate the $\xi$-independence of the
entire $S$-matrix element, but would have missed out on the ability to
identify gfp-independent subamplitudes, as we did.  In addition, this
result  indicates that  there  is  no loss  of  generality in  choosing
$\xi=1$  from  the  beginning,  thus  eliminating a  major  source  of
longitudinal  pieces, that  are bound  to cancel  anyway,  through the
special pinching procedure outlined above.
\newline
\indent
It would be tempting at this point to identify the gfp-independent 
subamplitudes obtained here with the 
${\widehat T}_i$ $(i=1,2,3)$ introduced in Eq.~(\ref{S2-matrix}). While this identification 
would be justified, as far as the gfp-independence is concerned, 
it will be postponed until the end of the next two subsections,
in order to endow the ${\widehat T}_i$
with one additional powerful ingredient: QED-like WIs. 

\subsection{The one-loop pinch technique Green's functions}

\subsubsection{\label{PT3}The one-loop pinch technique quark-gluon vertex and its Ward identity}
\noindent
Let us now turn to the longitudinal terms contained in the 
pinching part 
$\Gamma_{\alpha \mu \nu}^{\rm P}$ of the three-gluon vertex [see Eq.~(\ref{GP})]
appearing in the non-Abelian vertex graph $(b)$ (first line of \Figref{1l_vertex_pinch_contrib}),
and the two such vertices inside the gluon self-energy graph (\Figref{1l_selferg_pinch_contrib}).
One may ask at this point what is the purpose of
carrying the PT decomposition of the vertex given that one 
has already achieved $\xi$-independent
structures. The answer is that the
effect of the pinching 
momenta of $\Gamma_{\alpha \mu \nu}^{\rm P}$ 
is to make the effective  $\xi$-independent Green's functions  
satisfy, in addition, QED-like WIs instead of the usual STIs.
\newline
\indent
This is best seen in the case of one-loop quark-gluon vertex 
$\Gamma_{\alpha}^{a}(p_1,p_2)$, composed by graphs $(b)$ and $(c)$ of \Figref{1l_S-matrix}
now written (after the $\xi$-cancellations described above) 
in the Feynman gauge. It is well known that the QED counterpart of 
$\Gamma_{\alpha}^{a}(p_1,p_2)$, namely the photon-electron vertex 
$\Gamma_{\alpha}(p_1,p_2)$, satisfies to all orders (and for every gfp) the WI
\be
q^{\alpha} \Gamma_{\alpha}(p_1,p_2) =  ie\left\{S_e^{-1} (p_1) -  S_e^{-1} (p_2)\right\},
\label{vertQED}
\ee
where $S_e$ is the (all-order) electron propagator; Eq.~(\ref{vertQED}) is the naive, 
all-order generalization of the tree-level WI of (\ref{BasicWI}).  
\newline
\indent
The quark-gluon vertex $\Gamma_{\alpha}^{a}(p_1,p_2)$ also 
obeys the  WI of (\ref{BasicWI}) at tree-level (multiplied by $t^{a}$):
\be
q^{\alpha} \Gamma^a_{\alpha}(p_1,p_2)=igt^a\left\{S_e^{-1} (p_1) -  S_e^{-1} (p_2)\right\}.
\label{tl_WI}
\ee
However, at  higher orders it obeys an STI that is not the naive generalization of this
tree-level WI. Instead,  $\Gamma^{\alpha}_a(p_1,p_2)$ satisfies the STI~\cite{Pascual:1984zb}
\be
q^{\alpha} \Gamma_{\alpha}^a(p_1,p_2) = 
\left[q^2 D^{aa'}(q)\right] \left\{S^{-1} (p_2) H^{a'}(q,p_1)
+\bar H^{a'}(p_1,q)S^{-1}(p_2)\right],
\label{STIqg}
\ee
where $D^{aa'}(q)$ and $S(p)$ represents the full ghost and quark propagator respectively, and  
$H^{a}$ is a composite operator defined as (see also \Figref{H_aux_fer})
\be
iS(p)iD^{aa'}(q) iH^{a}(p, q) = 
-gt^d \int\! d^4x\int\!d^4y\,\e^{ip\cdot x} \,e^{iq\cdot y} 
\left\langle 0 \left\vert\ T \left\{\bar q(x)\bar c^{a'}(y)\left[c^d(0)q(0)\right]\right\} \right\vert0 \right\rangle, 
\label{compopH}
\ee
where $T$ denotes the time-ordered product of fields, and
${\bar H}$ is the hermitian conjugate of $H$. At tree-level,  
$H^{a}_{ij}$ reduces to $H^{(0)a}_{ij}= t^{a}_{ij}$. 
\begin{figure}[!t]
\bce
\includegraphics[width=9cm]{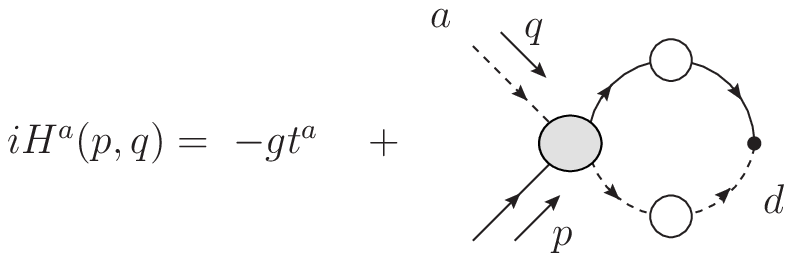}
\ece
\caption{The auxiliary function $H$ appearing in the quark-gluon vertex STI. The gray blob represents the (connected) ghost-fermion kernel appearing in the usual QCD skeleton expansion. }
\figlab{H_aux_fer}
\end{figure}
\newline
\indent
After these general considerations, 
let us carry out the decomposition of Eq.~(\ref{decomp}) to the  non-Abelian vertex 
of graph $(b)$ in \Figref{1l_S-matrix}. Then, let us write, suppressing again the color indices, 
\be
(b)_{\xi=1}=i{\mathcal V}^{\alpha}_aid(q^2)\bar u(p_1)i\widetilde{\Gamma}^a_\alpha(p_1,p_2) u(p_2),
\ee
and concentrate on the (one-loop) non-Abelian contribution to the quark-gluon vertex $\widetilde{\Gamma}^a_\alpha$. We have
\bea
i\widetilde{\Gamma}^a_\alpha(p_1,p_2)&=&\frac12g^3C_{A}t^a\int_k\!\frac{\Gamma_{\alpha\mu\nu}\gamma^\nu S^{(0)}(p_2-k)\gamma^\mu}{k^2(k+q)^2}\nonumber\\
&=&\frac12g^3C_{A}t^a\left\{\int_k\!\frac{\Gamma^{\mathrm{F}}_{\alpha\mu\nu}\gamma^\nu S^{(0)}(p_2-k)\gamma^\mu}{k^2(k+q)^2}+\int_k\!\frac{\Gamma^{\mathrm{P}}_{\alpha\mu\nu}\gamma^\nu S^{(0)}(p_2-k)\gamma^\mu}{k^2(k+q)^2}\right\},
\eea
where in this case
\bea
\Gamma_{\alpha \mu \nu}^{{\rm F}}&=& 
 g_{\mu\nu}(2k+q)_{\alpha} + 2q_{\nu}g_{\alpha\mu} 
- 2q_{\mu}g_{\alpha\nu},  
\label{GFapp}\nonumber\\
\Gamma_{\alpha \mu \nu}^{{\rm P}}&=& 
 - (k+q)_{\nu} g_{\alpha\mu} - k_{\mu}g_{\alpha\nu}. 
\label{GPapp}
\eea
Despite appearances, if we use
that $\bar{u}(p_2)(\not\! p_2 - m) = 0$ and  $(\not\! p_1 - m)u(p_1)=0$,
the part of the vertex graph containing
$\Gamma^{{\rm P}}$ is in fact 
purely propagator-like:  
\be
\int_k\!\frac{\Gamma^{\mathrm{P}}_{\alpha\mu\nu}\gamma^\nu S^{(0)}(p_2-k)\gamma^\mu}{k^2(k+q)^2}
\stackrel{\stackrel{{\rm PT}}{\rm Dirac\, Eq.}}{\longrightarrow}
2\gamma_\alpha\int_k \frac{1}{k^2(k+q)^2}.
\ee
\begin{figure}[t]
\bce
\includegraphics[width=9cm]{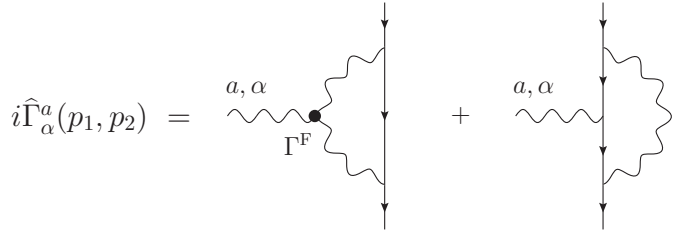}
\ece
\caption{\figlab{1l_PT_qgvert} Diagrammatic representation of 
the PT quark-gluon vertex at one-loop.}
\end{figure}
\indent
Thus, using the by now familiar methodology employed before,
one obtains from the one-loop quark-gluon vertex a propagator-like 
contribution, to be denoted by $\Pi_{\mu\nu}^{{\rm P}}(q)$,
given by
\be
\Pi_{\mu\nu}^{{\rm P}}(q) = g^2C_{A}q^2P_{\mu\nu}(q)
\int_k \frac{1}{k^2(k+q)^2}.
\label{proplikevert}
\ee
This term, together with an identical one 
coming from the mirror vertex, will be reassigned to the PT self-energy, soon to be constructed; for the moment let us concentrate on the remaining terms in the vertex.
In fact, the part of the vertex graph containing 
$\Gamma^{{\rm F}}$ remains unchanged, 
since it has no longitudinal momenta. Adding it to the usual 
Abelian-like graph, we obtain the one-loop 
PT quark-gluon vertex,
to be denoted by $\widehat{\Gamma}_{\alpha}^{a}$, given by  (see \Figref{1l_PT_qgvert})
\bea
i\widehat{\Gamma}_{\alpha}^{a}(p_1,p_2) &=& g^3 t^{a} \left\{\frac12C_{A}
\int_k\!\frac{\Gamma^{\mathrm{F}}_{\alpha\mu\nu}\gamma^\nu S^{(0)}(p_2-k)\gamma^\mu}{k^2(k+q)^2}\right.
\nonumber \\
&+&  \left.\left(C_{f}-\frac{C_{A}}{2}\right)\int_k\!\frac{\gamma^\mu S^{(0)}(p_1+k)\gamma_\alpha S^{(0)}(p_2+k)\gamma_\mu}{k^2}\right\}.
\label{PTvert}
\eea 
Now it is easy to derive the WI that the 
$\widehat\Gamma_{\alpha}^{a}(p_1,p_2)$ satisfies, simply by 
contracting the rhs of (\ref{PTvert}); this will trigger inside the integrands the 
corresponding tree-level WIs. Thus, 
using  Eqs~(\ref{BasicWI}) and (\ref{WI2B}), together with the definitions~(\ref{Sigma_xi}) and~(\ref{Sigmahat}), 
we have that 
\bea
q^{\alpha} \widehat\Gamma_{\alpha}^{a}(p_1,p_2) &=&-i  gt^{a}   
\left\{g^2C_{f} 
\int_k \frac{\gamma^{\mu} S^{(0)}(p_2+k)\gamma_{\mu} }{k^2} -  
g^2C_{f} \int_k \frac{\gamma^{\mu} S^{(0)}(p_1+k)\gamma_{\mu} }{k^2} 
\right\}
\nonumber\\
&=& igt^{a}\left\{\widehat\Sigma (p_1) -  \widehat\Sigma (p_2)\right\}.
\label{WIvert}
\eea 
Clearly, Eq.~(\ref{WIvert}) is the naive generalization 
of (\ref{tl_WI}) at one-loop, \ie the WI satisfied by $\Gamma_{\alpha}^{a}$ at tree-level; this 
makes the analogy with Eq.~(\ref{vertQED}) fully explicit.
An immediate consequence of Eq.~(\ref{WIvert}) is that the 
renormalization constants of $\widehat\Gamma_{\alpha}^{a}$ and $\widehat\Sigma$, 
to be denoted by ${\widehat Z}_1$ and ${\widehat Z}_2$, respectively, 
are related by the relation ${\widehat Z}_1={\widehat Z}_2$, which is 
none other than the textbook relation $Z_1=Z_2$ of QED, but now  
realized in a non-Abelian context.
\newline
\indent
A direct comparison of the STI of Eq.~(\ref{STIqg}), 
obeyed by the conventional vertex $\Gamma_{\alpha}^{a}$,  
with the WI of Eq.~(\ref{WIvert}), satisfied by the 
PT vertex $\widehat\Gamma_{\alpha}^{a}$, suggests  
a connection between 
the terms removed from  $\Gamma_{\alpha}^{a}$ 
during the process of pinching and the ghost-related quantities 
$D^{ab}$ and $H^{a}_{ij}$. As we will see in detail in the next chapter,  
such a connection indeed exists, and is, in fact, of central importance 
for the generalization of the PT to all orders. 

\subsubsection{The pinch technique gluon self-energy at one loop}
\noindent
Next, we construct the PT gluon self-energy, to be denoted by $\widehat\Pi_{\alpha\beta}(q)$.
It is given by the sum  of the conventional self-energy graphs and the self-energy-like parts
extracted from the two vertices, as shown schematically in \Figref{1l_PT_gprop},  \ie
\be
\widehat\Pi_{\alpha\beta}(q) = \Pi_{\alpha\beta}(q) + 2 \Pi_{\alpha\beta}^{{\rm P}}(q).
\ee
Specifically, in a closed form~\cite{Cornwall:1989gv},  
\be
\widehat\Pi_{\alpha\beta}(q) =
\frac12g^2C_{A}\left\{
\int_k \frac{\Gamma_{\alpha\mu\nu} 
\Gamma_{\beta}^{\mu\nu}}{k^2 (k+q)^2}-\int_k\!\frac{
k_{\alpha}(k+q)_{\beta}+k_{\beta}(k+q)_{\alpha}}{k^2 (k+q)^2}\right\}
+ 
2g^2C_{A} \int_k\!\frac{q^2 P_{\alpha\beta}(q)}{k^2 (k+q)^2},
\label{PTprop}
\ee
where we have symmetrized the ghost contribution [graph $(b)$ in \Figref{1l_PT_qgvert}] for later convenience, and neglected the fermion contribution [graph $(c)$ of the same figure].
\begin{figure}[!t]
\bce
\includegraphics[width=16cm]{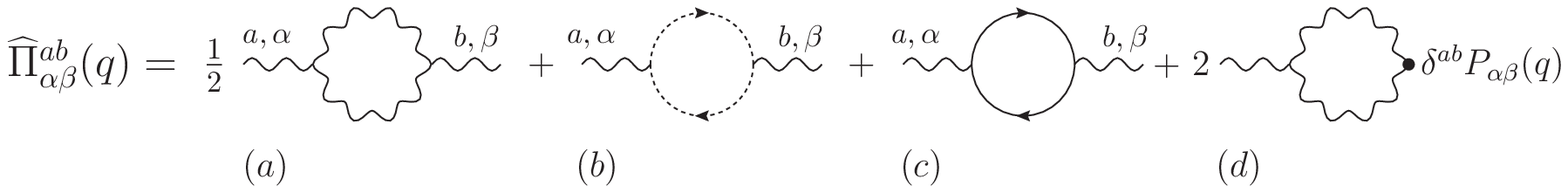}
\ece
\caption{\figlab{1l_PT_gprop} Diagrammatic representation of the one-loop PT gluon
 self-energy $\widehat\Pi_{\alpha\beta}$ as the sum of the
 conventional gluon self-energy terms and the pinch contributions coming
 from the vertex.}
\end{figure}
\newline
\indent
It would be elementary to compute $\widehat\Pi_{\alpha\beta}$ directly from the rhs of (\ref{PTprop}).
It is very instructive, however, to identify exactly the 
parts of the conventional $\Pi_{\alpha\beta}$ 
that combine with (and eventually cancel against)
the term $\Pi_{\alpha\beta}^{{\rm P}}$.  
To make this cancellation manifest, 
one carries out the following  
rearrangement of the two elementary three-gluon vertices
appearing in graph $(a)$ of \Figref{1l_PT_qgvert}
\bea
\Gamma_{\alpha \mu \nu}\Gamma^{\,\mu \nu}_{\beta}  &=&
\left[\Gamma^{{\rm F}}_{\alpha \mu \nu} + \Gamma^{{\rm P}}_{\alpha \mu \nu}\right]\,
\left[\Gamma^{{\rm F}\,\mu \nu}_{\beta} + \Gamma^{{\rm P}\,\mu\nu}_{\beta}\right]
\nonumber\\
&=&\Gamma^{{\rm F}}_{\alpha \mu \nu}\Gamma^{{\rm F}\,\mu \nu}_{\beta}
+\Gamma^{{\rm P}}_{\alpha \mu \nu}\Gamma^{\mu \nu}_{\beta}
+\Gamma_{\alpha \mu \nu}\Gamma^{{\rm P}\,\mu \nu}_{\beta}
-\Gamma^{{\rm P}}_{\alpha \mu \nu} \Gamma^{{\rm P}\,\mu\nu}_{\beta}.
\label{INPTDEC1}
\eea
Then, using the elementary WIs of  Eqs~(\ref{3gWI-1}) we have  
\bea
\Gamma^{{\rm P}}_{\alpha \mu \nu}\Gamma_{\beta}^{\mu \nu}+
\Gamma_{\alpha\mu\nu}\Gamma^{{\rm P}\,\mu\nu}_{\beta}
&=& - 4 q^2 P_{\alpha\beta}(q) - 
2 k_{\alpha} k_{\beta} - 2 (k+q)_{\alpha}(k+q)_{\beta}, \label{GPG+GGP}\nonumber\\
\Gamma^{{\rm P}}_{\alpha \mu \nu}\Gamma^{{\rm P}\,\mu\nu}_{\beta}  
&=& 2 k_{\alpha}k_{\beta}+ (k_{\alpha}q_{\beta}+q_{\alpha}k_{\beta}),
\label{GPGP}
\eea
where several terms have been set to zero by virtue of Eq.~(\ref{dreg}).
Thus we obtain~\cite{Cornwall:1989gv}
\be
\widehat{\Pi}_{\alpha\beta}(q) = \frac12g^2C_{A}\left\{ 
\int_k\!\frac{\Gamma^{{\rm F}}_{\alpha \mu \nu}\Gamma^{{\rm F}\,\mu \nu}_{\beta}}{k^2 (k+q)^2}-\int_k\!\frac{2 (2k+q)_{\alpha}(2k+q)_{\beta}}{k^2 (k+q)^2}\right\},
\label{propexp}
\ee
which may be further evaluated, using 
\be
\Gamma^{{\rm F}}_{\alpha \mu\nu}\Gamma^{{\rm F}\,\mu\nu}_{\beta}
= d (2k+q)_{\alpha} (2k+q)_{\beta} + 8 q^2 P_{\alpha\beta}(q),
\ee
and 
\be
\int_k \frac{(2k+q)_{\alpha} (2k+q)_{\beta}}{k^2 (k+q)^2} = - \left(\frac{1}{d-1}\right) q^2 P_{\alpha \beta}(q)
\int_k \frac{1}{k^2 (k+q)^2}, 
\ee
to finally cast $\widehat{\Pi}_{\alpha \beta}(q)$ in the simple form
\be
\widehat{\Pi}_{\alpha \beta}(q) = \left(\frac{7d-6}{d-1}\right) g^{2}\frac{C_A}{2} 
q^2 P_{\alpha \beta}(q) \int_k \frac{1}{k^2 (k+q)^2}.
\label{Pid}
\ee
Writing 
\be
\widehat{\Pi}_{\alpha \beta}(q)= P_{\alpha \beta}(q)\widehat\Pi(q^2), 
\label{Pid1}
\ee
and 
following the standard integration rules for the Feynman integral, we obtain for 
the unrenormalized $\widehat\Pi$ 
\be
\widehat\Pi(q^2) = ib g^2 q^2 
\left[\frac{2}{\epsilon} + \ln 4\pi -\gamma_{ E}  - \ln \frac{q^2}{\mu^2} + \frac{67}{33}\right],
\label{1l_PT_prop_res}
\ee
where $\gamma_{ E}$ is the Euler-Mascheroni constant ($\gamma_{E}\approx0.57721$) and
\be
b= \frac{11C_{A}}{48\pi^2},
\ee
is the one-loop coefficient of the $\beta$ function of QCD ($\beta = -bg^3$) in the absence of quark loops.
\newline
\indent
The appearance of $b$ in front of the logarithm is not accidental, and  
is exactly what happens with the vacuum polarization of QED. In the latter case the 
corresponding coefficient is $-\alpha/3 \pi$;
of course, the difference in the sign is related to the fact that 
QCD is asymptotically free, whereas QED is not.
The fact that the PT gluon propagator 
captures the leading RG logarithms  is a direct consequence of the 
WI of Eq.~(\ref{WIvert}) and the corresponding relation ${\widehat Z}_1={\widehat Z}_2$.
Indeed, if ${\widehat Z}_1={\widehat Z}_2$, then  
the charge renormalization constant, $Z_g$, and the wave-function renormalization of the 
PT gluon self-energy, ${\widehat Z}_A$, are related by 
$Z_{g}={\widehat Z}_{A}^{-1/2}$, exactly as in QED. 

\subsubsection{Process-independence of the pinch technique}

\begin{figure}[!t]
\bce
\includegraphics[width=4cm]{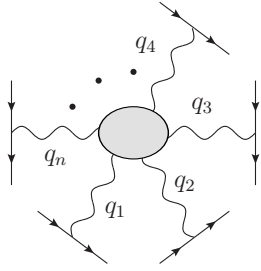}
\ece
\caption{\figlab{n-point_embed} $S$-matrix embedding necessary for
constructing  a  gfp-independent,  fully off-shell  gluonic  $n$-point
function.}
\end{figure}
\noindent
It  is  important  to stress, at this point,
that  the  only  completely
off-shell Green's function involved in the previous 
construction was the gluon
self-energy; instead,  the quark-gluon  vertex has the  incoming gluon
off-shell and  the two  quarks on  shell, while the  box has  all four
incoming  quarks on  shell.  These  latter quantities  were  also made
gfp-independent  in the  process of  constructing the  fully off-shell
gfp-independent gluonic two-point function. Similarly, 
as already mentioned after Eq.~(\ref{Sigmahat}),  the
construction of  a fully off-shell  PT quark self-energy  requires its
embedding in  a process such  as quark-gluon elastic  scattering.  The
generalization  of the  methodology  is now  clear;  for example,  for
constructing  a  gfp-independent,  fully off-shell  gluonic  $n$-point
function (\ie with $n$ off-shell gluons) one must consider the entire
gfp-independent   process   consisting   of   $n$-pairs   of   quarks,
\mbox{$q(p_1)q(k_1)$, $q(p_2)q(k_2)$, $\cdots$,  $q(p_n)q(k_n)$} and  hook each
gluon  $A_i$  to one  pair  of  test  quarks; the  off-shell  momentum
transfer $q_i$ of  the $i^\mathrm{th}$ gluonic leg will  be $q_i= p_i-k_i$ (see
\Figref{n-point_embed}).  Note, however, that one may equally well 
use gluons as external test
particles, or  even (not observed) fundamental  scalars carrying color.
Provided  that the  embedding process  is gfp-independent,  the answer
that the PT  furnishes for a given fully  off-shell $n$-point function
is  unique, \ie  it is  independent  of the  embedding process.  This
property is usually referred to as the process-independence of the PT,
and the  PT Green's  functions are said  to be  process-independent or
universal.   The universality  of the  one-loop gluon  self-energy has
been demonstrated  through explicit  computations, using a  variety of
external  test  particles~\cite{Watson:1994tn}.  For  example,  when
gluons  are used  as external  test particles,  the  pinching isolates
propagator-like  pieces  that  are  attached to  the  external  gluons
through  a tree-level three-gluon  vertex (see  \Figref{1l_vert_glemb_pinch}). In
this    case    the     analogue    of    the    quark-gluon    vertex
$\widehat\Gamma_{\alpha}^{a}$  is  a  gfp  one-loop  vertex  with  one
off-shell and two on-shell gluons, which, as we will see in a later
section, is the one-loop generalization of $\Gamma^{{\rm F}}$. This
latter vertex  should not be  confused with the PT  three-gluon vertex
with all three gluons off-shell,  that can be constructed by embedding
it into a  six quark process (one pair for each  leg), to be discussed
in the next subsection.  The distinction between these two three-gluon
vertices is crucial, and will be made more explicit later on; in addition, 
a more precise field-theoretic  notation  will  be adopted, 
that will allow us  to distinguish  them unambiguously.
\newline
\indent
We emphasize that the PT construction is not restricted to the 
use of on-shell $S$-matrix amplitudes,
and works equally well inside, for example,  
a gauge-invariant current correlation function or a Wilson loop.
This fact is particularly relevant for 
the correct interpretation of the correspondence between PT and 
BFM, which will be discussed in the \secref{PTBFM}.
Actually, in the first PT calculation ever~\cite{Cornwall:1981zr}, 
Cornwall studied the 
set of one-loop Feynman diagrams contributing to 
the  {\it gauge-invariant} Green's function   $G(x,y)  =  \left\langle 0 \left\vert
T\left\{\mathrm{Tr}\left[\Phi(x)\Phi^{\dagger}(x)\right] \mathrm{Tr}\left[\Phi(y)\Phi^{\dagger}(y)\right]\right\}\right\vert 0
\right\rangle$, where $\Phi(x)$ is a matrix describing a set of scalar test
particles in  an appropriate representation of  the gauge  group.
In this case, the special momentum, 
with respect to which the vertex decomposition of Eq.~(\ref{decomp}) 
should be carried out (\ie the equivalent of $q$ in that same equation), 
is the momentum transfer  between the two sides  of the scalar loop (\ie
one should count  loops as if the $\Phi$ loop had been opened at $x$ and $y$).
The advantage of using an $S$-matrix amplitude is purely operational: 
the PT construction becomes more expeditious, because several terms 
can be set to zero directly due to the equation of motion of the on-shell 
test particles. Instead, in the case of a  Wilson loop, one 
would have to carry out the additional step of demonstrating 
explicitly their cancellation against other similar terms.

\begin{figure}[t]
\bce
\includegraphics[width=12cm]{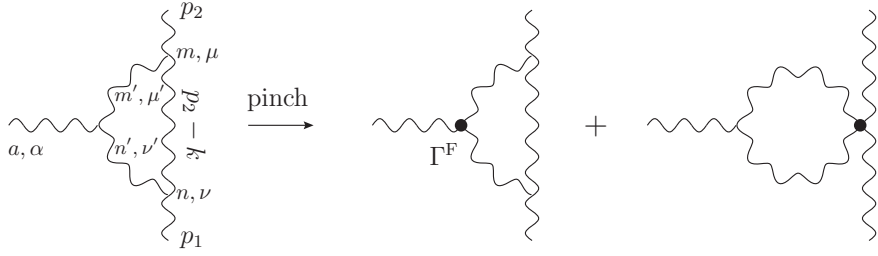}
\ece
\caption{\figlab{1l_vert_glemb_pinch}The pinching procedure when the embedding particles are ``on-shell'' gluons. Despite appearances, the vertex to which the pinching contribution is connected to the external gluons is a three-gluon vertex.}
\end{figure}

\subsubsection{Intrinsic pinch technique and the gauge-independent three-gluon vertex at one loop}
\begin{figure}[!t]
\bce
\includegraphics[width=9cm]{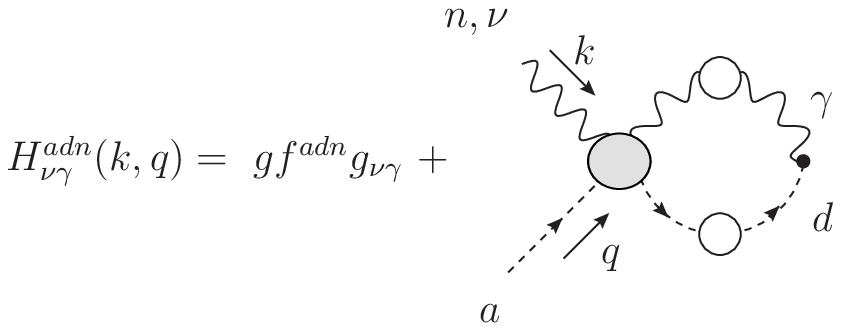}
\ece
\caption{The auxiliary function $H$ appearing in the three-gluon vertex STI. The gray blob represents the (connected) ghost-gluon kernel ${\mathcal K}$ appearing in the usual QCD skeleton expansion. }
\label{fig:H_aux}
\end{figure}

\noindent
The  central achievement  of  the previous  subsections  has been  the
construction  of  the  gfp-independent  off-shell  gluon  self-energy,
$\widehat{\Pi}_{\mu  \nu}$, through its  embedding into  a physical
$S$-matrix element,  corresponding to quark-quark  elastic scattering.
This was  accomplished by  identifying propagator-like pieces  from the
vertices  and the  boxes contributing  to the  embedding  process, and
reassigning  them to the  conventional gluon  self-energy, ${\Pi}_{\mu\nu}$.   
This procedure has been carried out for a general value of the gfp, leading 
to a unique answer, which is most economically reached by choosing 
the Feynman gauge from the beginning. Thus, $\widehat{\Pi}_{\mu  \nu}$ 
is obtained by adding  to ${\Pi}_{\mu\nu}$ the propagator-like pieces 
$2 \Pi_{\mu\nu}^{{\rm P}}$ extracted from the vertices, 
as shown in Eq.~(\ref{PTprop}). 
In the analysis following Eq.~(\ref{PTprop}) it became clear that 
these latter terms cancel very precise terms of the  
conventional self-energy ${\Pi}_{\mu\nu}$, furnishing finally $\widehat{\Pi}_{\mu  \nu}$.
Specifically, after the vertex
decomposition of Eq.~(\ref{INPTDEC1}), the terms $\Gamma^{{\rm P}}$ 
acted on the corresponding $\Gamma$, triggering  
the WIs of  Eqs~(\ref{3gWI-1}): the term $2 \Pi_{\mu\nu}^{{\rm P}}$ 
cancels against the terms of the  WIs that are proportional to 
$q^2 P_{\mu\nu}$. This observation motivates the 
following more expeditious course of action: instead of identifying 
the propagator-like pieces from the various graphs, focus on 
${\Pi}_{\mu\nu}$, carry out the 
decomposition of Eq.~(\ref{INPTDEC1}), and discard the terms 
coming from the  WIs that are   
proportional to $q^2 P_{\mu\nu}$; what is left is then the PT answer.
\newline
\indent
This alternative, and completely equivalent, 
approach to pinching was first introduced in~\cite{Cornwall:1989gv} 
and is known as ``intrinsic'' PT. Its main virtue is that 
it avoids as much as possible
the embedding of the Green's function under construction into a physical amplitude. 
As we will see later on, the intrinsic approach is particularly suited for 
extending the PT construction at the level of the SDE of the theory. 
\begin{figure}[!t]
\bce
\includegraphics[width=15.8cm]{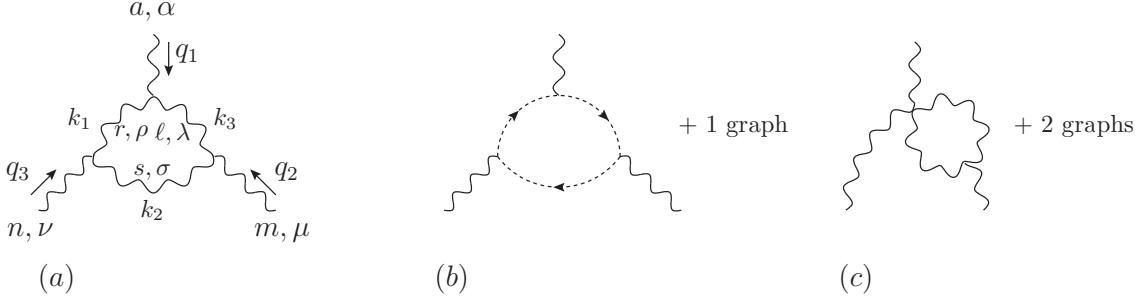}
\ece
\caption{\figlab{1l_3gvert_Rxi_diag} \Rxi diagrams contributing to the one-loop three-gluon vertex. Diagrams $(c)$ carry a $\frac12$ symmetry factor. Fermion diagrams are not shown.}
\end{figure}
\newline
\indent
As an application of the intrinsic PT algorithm, we will construct the one-loop PT three gluon vertex~\cite{Cornwall:1989gv}. 
The conventional \Rxi diagrams are shown in \Figref{1l_3gvert_Rxi_diag} and read
\be
\Gamma^{amn}_{\alpha\mu \nu}(q_1, q_2, q_3) =- \frac 12 g^{3} C_{A}f^{amn}\left\{ 
\int _{k_1}\!
\frac1{k^2_1 k^2_2 k^2_3}N_{\alpha\mu \nu}+B_{\alpha\mu \nu}\right\},
\label{conv3glvert}
\ee
with
\bea
 N_{\alpha\mu\nu} &=& 
\Gamma_{\alpha\lambda\rho}(q_1,k_3, -k_1)
\Gamma_{\mu\sigma\lambda}(q_2,k_2,-k_3)
\Gamma_{\nu\rho\sigma}(q_3,k_1, -k_2) 
- k_{1\alpha}k_{2\nu}k_{3\mu}-k_{1\alpha}k_{2\nu}k_{3\mu},
\label{Nterm}\nonumber \\
\nonumber\\
B_{\alpha\mu\nu}&=&\frac92\left(g_{\alpha\mu}q_{1\nu}-g_{\alpha\nu}q_{1\mu}\right)\int_{k}\!\frac1{k^2(k+q_1)^2}
+\frac92\left(g_{\alpha\mu}q_{2\nu}-g_{\mu\nu}q_{2\alpha}\right)\int_{k}\!\frac1{k^2(k+q_2)^2}\nonumber \\
&+&\frac92\left(g_{\mu\nu}q_{3\alpha}-g_{\alpha\nu}q_{3\mu}\right)\int_{k}\!\frac1{k^2(k+q_3)^2}.
\label{Bterm}
\eea
Let us then introduce the short-hand notation $\Gamma_1\Gamma_2\Gamma_3$ for the product of (bare) three gluon vertices appearing in Eq.~(\ref{Nterm}). 
In this notation all the Lorentz indices are suppressed and the number appearing in each vertex is the one corresponding to its external momentum $q_i$. Then, decomposing each of the $\Gamma_i$ into $\Gamma^{\mathrm{F}}_i+\Gamma^{\mathrm{P}}_i$, we obtain the analogue of (\ref{INPTDEC1}), namely 
\bea
\Gamma_1\Gamma_2\Gamma_3&=&\Gamma^{\mathrm{F}}_1\Gamma^{\mathrm{F}}_2\Gamma^{\mathrm{F}}_3
+\Gamma^{\mathrm{P}}_1\Gamma_2\Gamma_3+\Gamma_1\Gamma^{\mathrm{P}}_2\Gamma_3+\Gamma_1\Gamma_2\Gamma^{\mathrm{P}}_3-\Gamma^{\mathrm{P}}_1\Gamma^{\mathrm{P}}_2\Gamma_3-\Gamma_1^{\mathrm{P}}\Gamma_2\Gamma^{\mathrm{P}}_3-\Gamma_1\Gamma^{\mathrm{P}}_2\Gamma^{\mathrm{P}}_3\nonumber \\
&+&\Gamma_1^{\mathrm{P}}\Gamma^{\mathrm{P}}_2\Gamma^{\mathrm{P}}_3.
\label{3v_dec}
\eea
Now, the first term contains no pinching momenta, and therefore will be kept in the PT answer, giving rise to the term
\be
(\widehat{a})=- \frac i2 g^{3} C_{A}f^{amn}
\int _{k_1}\!\frac1{k^2_1 k^2_2 k^2_3}\Gamma^{\mathrm{F}}_{\alpha\lambda\rho}(q_1,k_3, -k_1)
\Gamma^{\mathrm{F}}_{\mu\sigma\lambda}(q_2,k_2,-k_3)
\Gamma^{\mathrm{F}}_{\nu\rho\sigma}(q_3,k_1, -k_2). 
\ee
\indent
Each of the next six terms  gives rise to pinching contributions, generated
when $\Gamma^{\mathrm{P}}_i$ acts on the full $\Gamma$'s, thus triggering the WIs of (\ref{3gWI-3}).
Some of the terms so generated will be proportional to $d^{-1}(q_i^2)$, i.e. inverse {\it external} gluon propagators; 
according to the rules of the intrinsic pinch, we simply discard them.  
However, all other terms generated from the  WIs of (\ref{3gWI-3}) must be kept; as we will see, they are crucial for furnishing the correct final answer.
For example, collectively denoting the terms discarded with ellipses, one has
\bea
\Gamma^{\mathrm{P}}_1\Gamma_2\Gamma_3&=&
d^{-1}(k_3^2)\left[\Gamma_{\nu\alpha\mu}(k_1,-k_2)+\Gamma_{\mu\nu\alpha}(k_2,-k_3)\right]+k_{2\mu}\left[d^{-1}(k_1^2)g_{\alpha\nu}-k_{1\alpha}k_{1\nu}\right]\nonumber\\
&+&k_{2\nu}\left[d^{-1}(k_3^2)g_{\alpha\mu}-k_{3\alpha}k_{3\mu}\right]+\cdots,\nonumber\\
\Gamma_1^{\mathrm{P}}\Gamma^{\mathrm{P}}_2\Gamma_3&=&d^{-1}(k_3^2)\Gamma_{\nu\alpha\mu}(k_1,-k_2)-
k_{3\alpha}\left[d^{-1}(k_2^2)g_{\mu\nu}-k_{2\mu}k_{2\nu}\right]\nonumber\\
&-&k_{3\mu}\left[d^{-1}(k_1^2)g_{\nu\alpha}-k_{1\nu}k_{1\alpha}\right]+\cdots, 
\eea 
with similar expressions for the other such terms on on the rhs of Eq.~(\ref{3v_dec}). 
The last term on the rhs of Eq.~(\ref{3v_dec}) does not have terms proportional to $d^{-1}(q_i^2)$, so there is nothing to discard; 
it must be kept in its entirety.
Specifically,  
\bea
\Gamma_1^{\mathrm{P}}\Gamma^{\mathrm{P}}_2\Gamma^{\mathrm{P}}_3&=&-d^{-1}(k_1^2)\left(g_{\mu\nu}k_{3\alpha}+g_{\alpha\mu}k_{1\alpha}\right)-d^{-1}(k_2^2)\left(g_{\alpha\mu}k_{1\nu}+g_{\alpha\nu}k_{3\mu}\right)\nonumber\\
&-&d^{-1}(k_3^2)\left(g_{\alpha\nu}k_{2\mu}+g_{\mu\nu}k_{1\alpha}\right)-k_{1\alpha}k_{2\mu}k_{3\nu}-k_{1\nu}k_{2\mu}k_{3\alpha}.
\eea
\indent
Isolating all terms that are not proportional to  a $d^{-1}(k_i^2)$, 
and adding them to the conventional ghost graph $(b)$ of \Figref{1l_3gvert_Rxi_diag}, we get the result
\be
(\widehat{b}) =\frac i2g^3C_{A}f^{amn}\int_{k_1}\frac1{k_1^2k_2^2k_3^2}2(k_1+k_3)_\alpha(k_2+k_3)_\mu(k_1+k_2)_\nu \,.
\ee
Evidently, all terms proportional to $d^{-1}(k_i^2)$ will cancel against one internal gluon propagator, giving rise to 
integrands with only two such propagators, i.e. 
\bea
(\widetilde{c})&=&- \frac i2 g^{3} C_{A}f^{amn}
\int _{k_2}\!\frac1{k^2_2 k^2_3}\left[g_{\alpha\mu}(k_1-q_3)_\nu+2g_{\alpha\nu}(q_3-q_1)_\mu+g_{\mu\nu}(k_1+q_1)_\alpha\right]\nonumber\\
&-&\frac i2 g^{3} C_{A}f^{amn}\int _{k_1}\!\frac1{k^2_1 k^2_3}\left[g_{\alpha\mu}(k_2+q_3)_\nu+g_{\alpha\nu}(k_2-q_2)_\mu+2g_{\mu\nu}(q_2-q_3)_\alpha\right]\nonumber\\
&-&\frac i2 g^{3} C_{A}f^{amn}\int _{k_1}\!\frac1{k^2_1 k^2_2}\left[2g_{\alpha\mu}(q_1-q_2)_\nu+g_{\alpha\nu}(k_3+q_2)_\mu+g_{\mu\nu}(k_3-q_1)_\alpha\right].
\label{ctilde}
\eea
\indent
This is, however, not the end of the story. As we have seen, 
in the presence of longitudinal momenta the topology of a Feynman diagram is 
not a well-defined property, since longitudinal momenta will pinch out internal 
propagators, turning $t$-channel diagrams into $s$-channel ones. 
This same caveat 
applies also to the notion of one particle reducibility. Remember that a diagram is called one-particle irreducible (1PI) if it cannot be split into two disjoined pieces by cutting a single internal line; otherwise it is called one-particle reducible (1PR).
Now, it turns out that by pinching out internal propagators, 
one can effectively convert 1PR diagrams into 1PI ones (see \Figref{1PR_pinch_contrib});  
of course the opposite cannot happen. Evidently the notion of a 1PR diagram is gauge-dependent!  
Thus, when constructing the purely (1PI) gauge-invariant three-gluon vertex at one-loop, 
one has to take into account possible 1PI pinching contribution coming from seemingly 1PR diagrams,  
such as those shown in \Figref{1PR_pinch_contrib}. How do we actually obtain these terms ?
Simply by carrying out the intrinsic PT construction inside the self-energy graph $(d_1)$:   
in doing so, the terms that will remove the ``internal'' gluon propagator will furnish the 
(effectively 1PI) diagram $(d_1)^\mathrm{P}$, 
while those proportional to an inverse ``external'' gluon propagator [diagram $(d_1')^\mathrm{P}$] ought to be discarded,
in full accordance with the rules 
of the intrinsic PT. Indeed, as the reader should be able to verify, in the $S$-matrix PT implementation
these latter terms will cancel anyway against analogous contribution coming 
from non-Abelian vertices attached to the external test-quark.
\begin{figure}[!t]
\bce
\includegraphics[width=12cm]{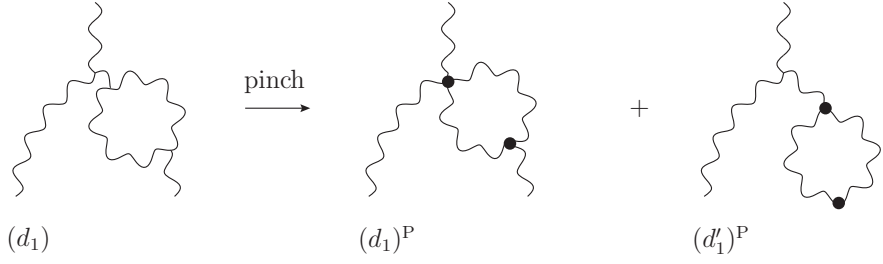}
\ece
\caption{\figlab{1PR_pinch_contrib} 1PR diagram giving to effectively 1PI pinching contributions [diagram $(d_1)^{\mathrm{P}}$]. Two more diagrams (corresponding to having the gluon self-energy correction on the remaining legs) that give rise to similar terms are not shown.}
\end{figure}
\newline
\indent
Let us see in detail what happens in the case shown in  \Figref{1PR_pinch_contrib}. One has
\bea
(d_1)&=&-\frac{\mathrm{i}}2g^2C_A\Gamma_{\alpha\mu'\nu}(q_1,q_2,q_3)d(q_2^2)g^{\mu'\nu'}\times\nonumber \\
&\times&\int_k\!\frac1{k^2(k+q_2)^2}\Gamma_{\nu'\rho\sigma}(-q_2,k+q_2,-k)\Gamma_\mu^{\rho\sigma}(-q_2,k+q_2,-k).
\eea
As explained above, of all the possible pinching contributions appearing after the splitting of the two three gluon vertices inside this gluon self-energy 
according to (\ref{INPTDEC1}) and (\ref{GPGP}), 
one needs to retain only half of the first term appearing in the rhs of Eq.~(\ref{GPG+GGP}), the other half removing instead the external propagator, thus generating diagram $(d'_1)^\mathrm{P}$ of \Figref{1PR_pinch_contrib}. Therefore one has
\be
(d_1)^{\mathrm{P}}=ig^3C_{A}f^{amn}\Gamma_{\alpha\mu\nu}(q_1,q_2,q_3)\int_k\!\frac1{k^2(k+q_2)^2},
\ee
where we kept only the $g^{\mu\sigma}$ part of the $P^{\mu\sigma}$ appearing in the pinching term, since the $q_2^\mu q_2^\sigma$ term will remove the external propagator and thus ought to be discarded. Adding this to the first term on the rhs of Eq.~(\ref{ctilde}), to be denoted by $(\widetilde{c}_1)$, we find  
\be
(\widetilde{c}_1)+(d_1)^{\mathrm{P}}=-i\frac74g^3C_{A}f^{amn}(g_{\alpha\mu}q_{2\nu}-g_{\mu\nu}q_{2\alpha})\int_k\frac1{k^2(k+q_2)^2}.
\ee 
The same procedure can be repeated for the diagrams $(d_2)$ and $(d_3)$; 
after adding them to the corresponding contributions, $(\widetilde{c}_2)$ and $(\widetilde{c}_3)$, we obtain 
\bea
(\widetilde{c}_2)+(d_2)^{\mathrm{P}}&=&-i\frac74g^{3}C_{A}f^{amn}(g_{\mu\nu}q_{3\alpha}-g_{\alpha\nu}q_{3\mu})\int_k\frac1{k^2(k+q_3)^2},\nonumber\\
(\widetilde{c}_3)+(d_3)^{\mathrm{P}}&=&-i\frac74g^{3}C_{A}f^{amn}(g_{\alpha\mu}q_{1\nu}-g_{\alpha\nu}q_{1\mu})\int_k\frac1{k^2(k+q_1)^2}.
\eea
Notice that these terms have exactly the same structure as the conventional $(c)$ diagrams, to which they can be added.
\newline
\indent
Thus, the PT one-loop three-gluon vertex is finally given by 
\be
i\widehat{\Gamma}^{amn}_{\alpha\mu \nu}(q_1, q_2, q_3) =- \frac i2 g^{3} C_{A}f^{amn}\left\{ 
\int _{k_1}\!
\frac1{k^2_1 k^2_2 k^2_3} \widehat{N}_{\alpha\mu \nu}+ \widehat{B}_{\alpha\mu \nu}\right\},
\label{c3glvert-PT}
\ee
where
\bea
 \widehat{N}_{\alpha\mu\nu} &=& 
\Gamma^{\mathrm{F}}_{\alpha\lambda\rho}(q_1,k_3, -k_1)
\Gamma^{\mathrm{F}}_{\mu\sigma\lambda}(q_2,k_2,-k_3)
\Gamma^{\mathrm{F}}_{\nu\rho\sigma}(q_3,k_1, -k_2)\nonumber\\ 
&-&2(k_1+k_3)_\alpha(k_2+k_3)_\mu(k_1+k_2)_\nu,
\label{PTNterm}\nonumber \\
\nonumber\\
 \widehat{B}_{\alpha\mu\nu}&=&8\left(g_{\alpha\mu}q_{1\nu}-g_{\alpha\nu}q_{1\mu}\right)\int_{k}\!\frac1{k^2(k+q_1)^2}
+8\left(g_{\alpha\mu}q_{2\nu}-g_{\mu\nu}q_{2\alpha}\right)\int_{k}\!\frac1{k^2(k+q_2)^2}\nonumber \\
&+&8\left(g_{\mu\nu}q_{3\alpha}-g_{\alpha\nu}q_{3\mu}\right)\int_{k}\!\frac1{k^2(k+q_3)^2}.
\label{PTBterm}
\eea
Note that $\widehat{\Gamma}^{amn}_{\alpha\mu \nu}(q_1, q_2, q_3)$ is manifestly Bose-symmetric with respect to all three of its legs. 
\newline
\indent
It is now of central importance 
to recognize that, unlike the conventional three gluon vertex that satisfies an STI, the $\widehat{\Gamma}^{amn}_{\alpha\mu \nu}(q_1, q_2, q_3)$ 
constructed above satisfies a simple Abelian-like WI.
Specifically, the conventional three-gluon vertex (in the\Rxi gauges) satisfies at all orders the STI~\cite{Ball:1980ax}
\bea
q_1^\alpha \Gamma^{amn}_{\alpha\mu\nu}(q_1,q_2,q_3) &=&
\left[q_1^2D^{aa'}(q_1)\right]\left\{
\Delta^{-1}(q_2^2) P^{\gamma}_{\mu}(q_2) H^{a'nm}_{\nu \gamma}(q_3,q_2)\right. \nonumber \\
&+&\left.\Delta^{-1}(q_3^2) P^{\gamma}_{\nu}(q_3) H^{a'mn}_{\mu\gamma}(q_2,q_3)
\right\},
\label{sti3gv}
\eea
where  the auxiliary function $H_{\alpha\beta}$ is the 1PI part of the composite operator
\bea
i\Delta^{nn'}_{\nu\nu'}(k)iD^{aa'}iH^{adn}_{\nu\gamma}(k,q)&=&-igf^{eds}\int\!d^4x\int\!d^4y\e^{iq\cdot x}\e^{ik\cdot y}\times
\nonumber\\
&\times&\left\langle0\left\vert T\left\{\bar c^{a'}(x)A^{n'}_{\nu'}(y)[c^e(0)A^s_\gamma(0)]\right\}\right\vert0\right\rangle^\mathrm{1PI}.
\eea
and it is defined in Fig.~\ref{fig:H_aux}. Notice that the kernel appearing in this auxiliary function is the conventional connected
ghost-ghost-gluon-gluon kernel appearing in the usual
QCD skeleton expansion~\cite{Marciano:1977su,BarGadda:1979cz}. 
Also, $H_{\alpha\beta}(k,q)$ is related to the conventional gluon-ghost
vertex ${\Gamma}_{\beta}(k,q)$ (with $k$ the gluon and $q$ the anti-ghost momentum) 
by~\cite{Pascual:1984zb,Ball:1980ax,Marciano:1977su,BarGadda:1979cz}.
\be
q^{\alpha} H_{\alpha\beta}(k,q) = {\Gamma}_{\beta}(k,q).
\label{H_ghost_rel}
\ee
Of course, the STI of (\ref{sti3gv}) reduces at tree-level to that of (\ref{3gWI-1}).
\newline
\indent
On the other hand, contracting Eq.~(\ref{c3glvert-PT}) with $q_1^\alpha$,  one obtains easily the result~\cite{Cornwall:1989gv}
\be
q_1^\alpha \widehat{\Gamma}^{amn}_{\alpha\mu\nu}(q_1,q_2,q_3) =gf^{amn}\left\{\widehat{\Delta}^{-1}(q_2)P_{\mu\nu}(q_2)-\widehat{\Delta}^{-1}(q_3)P_{\mu\nu}(q_3)\right\},
\label{WI3g_PT}
\ee
with $\widehat{\Delta}^{-1}(q)=q^2+i\widehat{\Pi}(q^2)$; of course, contracting with respect $q_2^\mu$ and $q_3^\nu$ 
furnishes the expected Bose-symmetric analogues  of (\ref{WI3g_PT}).  
Thus, rather remarkably, we find the naive one-loop generalization of the tree-level identity of Eq.~(\ref{WI2B}). 
Note, in particular, that any reference to auxiliary ghost Green's functions has disappeared: Eq.~(\ref{WI3g_PT}) is completely gauge-invariant.
\newline
\indent
Let us now focus on a very interesting property of the one-loop PT three-gluon vertex, 
discovered recently by Binger and Brodsky ~\cite{Binger:2006sj}. 
These authors have first added quark and scalar loops to $\widehat{\Gamma}^{amn}_{\alpha\mu\nu}(q_1,q_2,q_3)$;
this is straightforward, from the point of view of gauge-independence and gauge-invariance, 
since these loops are automatically gfp-independent and satisfy (\ref{WI3g_PT}).
Then, all resulting one-loop integrals, including those of  
(\ref{PTNterm}) and (\ref{PTBterm}), were evaluated for the first time, 
thus determining the precise tensorial decomposition  
 of $\widehat{\Gamma}^{amn}_{\alpha\mu\nu}(q_1,q_2,q_3)$.
Then, 
after choosing a convenient tensor basis, 
$\widehat{\Gamma}^{amn}_{\alpha\mu\nu}(q_1,q_2,q_3)$ was  expressed as a linear combination
of fourteen independent tensors, each one multiplied by its own scalar form-factor. 
Every form-factor receives, in general, contributions from   
gluons(G), quarks(Q), and scalars(S). It turns out that these three types of contributions 
satisfy very characteristic  
relations, that are closely linked to
supersymmetry and conformal symmetry, and in particular the ${\mathcal N}=4$
non-renormalization theorems. Specifically, for all form-factors $F$ (in $d$-dimensions) it was shown that
\be
 F_G+4F_Q+(10-d)F_S=0,
 \label{GQSsum}
\ee
which encodes the vanishing contribution of the ${\mathcal N}=4$ supermultiplet
in four dimensions. Similar relations have been found in the context
of supersymmetric scattering amplitudes
\cite{Bern:1994zx,Bern:1994cg}.
\newline
\indent
It should be emphasized that relations such as Eq.(\ref{GQSsum}) do {\it not} exist
 for the gauge-dependent three-gluon vertex \cite{Davydychev:1996pb}, since the
 gluon contributions depend on the gfp, while the
 quarks and scalars do not. Indeed, it is
 uniquely the PT (or equivalently BFM in the BFG $\xi_Q = 1$, see next section)
 Green's function that satisfies this homogeneous sum rule.
 Most importantly, calculating in the BFM with $\xi_Q \neq 1$ leads to
 a nonzero rhs of Eq.~(\ref{GQSsum}).

\subsubsection{The pinch technique four-gluon vertex at one loop}
\noindent
The construction of the gauge-invariant four-gluon vertex has been outlined in detail in~\cite{Papavassiliou:1992ia}, in the context 
of the $S$-matrix PT. The actual derivation is technically rather cumbersome  
because of ({\it i}) the large number of graphs and ({\it ii}) certain subtle exchange of pinching contributions between 1PI and 1PR diagrams (some of which have been seen in the construction carried out in the previous subsection). The exact closed form of this vertex (see, \eg~\cite{Hashimoto:1994ct})
is too lengthy to be reported here. Far more interesting is the WI that this vertex satisfies:
it is simply the naive one-loop generalization of the tree-level result, as can be easily confirmed 
using the explicit expressions for the bare three- and four-gluon vertices. 
Specifically, we have~\cite{Papavassiliou:1992ia}
\bea
q^{\alpha}{\Gamma}_{\alpha\mu\nu\rho}^{amnr}(q,k_1,k_2,k_3) &=&
g f^{adm} \widehat{\Gamma}_{\mu\nu\rho}^{dnr}(q+k_1,k_2,k_3)+
g f^{adr} \widehat{\Gamma}_{\nu\rho\mu}^{drm}(q+k_2,k_3,k_1)
\nonumber\\
 &+&
g f^{adn} \widehat{\Gamma}_{\nu\mu\rho}^{dmr}(q+k_3,k_1,k_2),
\label{WI4gl_PT}
\eea
where the three-gluon vertices appearing are the PT ones constructed in the previous subsection.
Again we find a fully gauge invariant, Abelian-like WI, that makes no reference to ghost Green's functions.

\subsection{\label{abspt}The absorptive  pinch technique construction}
\noindent
In the previous subsection, we  worked at the level of 
one-loop perturbation theory, and constructed non-Abelian Green's functions 
that are gfp-independent and satisfy QED-like WIs. The analogy between the 
PT Green's functions and those of QED is best exemplified by comparing the 
PT gluon self-energy with that of the photon (vacuum polarization): 
they are both gfp-independent and capture the leading RG logarithms.  
It is therefore natural to want to explore until what point     
this analogy with QED may persist.   
Specifically, in QED knowledge of the vacuum polarization spectral
function determined from the tree level $e^{+}e^{-}\to\mu^{+} \mu^{-}$
cross sections, together with a single low energy measurement of
the fine structure constant $\alpha$, enables the construction of the 
one-loop vacuum polarization, and the 
corresponding effective charge, $\alpha_{\rm eff}(q^2)$, for all $q^2$. 
What makes this possible in the case of QED 
is the unitarity of the $S$-matrix, expressed 
in the form of the optical theorem, and the requirement of 
the analyticity of Green's functions, as captured by the so-called 
dispersion relations (see, \eg \cite{BjDr}). 
In this subsection we will study in detail 
how the above crucial properties are encoded into the Green's 
functions constructed by the PT. Specifically, we will see that, 
in a non-Abelian context,  
the PT construction enforces {\it at the level of individual 
Green's functions} properties of unitarity and analyticity that 
are completely analogous to those of QED. 

\subsubsection{\label{sec:OT_anal}Optical theorem and analyticity}
\noindent
The $T$-matrix element of a reaction $i\to f$ is defined via the relation
\begin{equation}
\langle f | S | i \rangle\ =\ \delta_{fi}\ +\ i(2\pi )^4
\delta^{(4)}(P_f - P_i)\langle f | T|i\rangle ,\label{tmatrix}
\end{equation}
where $P_i$ ($P_f$) is the sum of all initial (final) momenta of the
$|i\rangle$ ($| f \rangle$) state. Furthermore, imposing the unitarity
relation $S^\dagger S = 1$ leads to the generalized optical theorem: 
\begin{equation}
\langle f|T|i\rangle - \langle i |T|f\rangle^*\ =\
i\sum_{j} (2\pi )^4\delta^{(4)}(P_{j} - P_i)\langle j | T | f \rangle^*
\langle j | T | i \rangle. \label{optical}
\end{equation}
In Eq.~(\ref{optical}), the sum $\sum_{j}$ should be understood to be over
the entire phase space and spins of all possible on-shell intermediate
particles $j$. 
\newline
\indent
An important corollary of this theorem is obtained if $f=i$, 
corresponding to the case of the so-called ``forward scattering''.
For this particular kinematic choice, setting  on the lhs
\mbox{$T^{ii} \equiv \langle i|T|i\rangle$} and on the rhs 
${\mathcal T}^{ij}\equiv \langle j | T | i \rangle $
and 
${\mathcal M}^{ij}\equiv |\langle j | T | i \rangle |^2 = |{\mathcal T}^{ij}|^2$ 
we have 
\begin{equation}
\Im m \{ T^{ii}\} =\ \frac{1}{2}
\sum_j (2\pi )^4\delta^{(4)}(P_j - P_i){\mathcal M}^{ij} .
\label{absorptive}
\end{equation}
For the rest of this review we will be referring to 
the relation given in Eq.~(\ref{absorptive}) as the optical theorem (OT).
\newline
\indent
The rhs of the OT consists of the sum 
of the (squared) amplitudes, ${\mathcal M}^{ij}$,   
of all kinematically allowed  elementary processes 
connecting the initial and final states.
Note, in particular, that only {\it physical particles} 
may appear as intermediate $\vert j\rangle$ states. 
If the particles involved are fermions and/or gauge bosons,
when calculating ${\mathcal M}^{ij}$ one averages 
over the initial state polarizations and sums 
over the final state polarizations. In addition, 
the  integration over all available phase-space,
implicit in the sum $\sum_j$,  
must be carried out. 
The lhs of the OT is given by the imaginary part of the {\it entire}
amplitude, \ie including all Feynman diagrams contributing to it.
For example, in the case of non-Abelian gauge theories 
to obtain the lhs of the OT 
one must calculate the imaginary part of all diagrams, regardless of 
whether they contain physical (gluons, quarks) or unphysical 
(ghosts or would-be Goldstone bosons) fields inside their loops. 
The way how these imaginary parts will be actually computed is   
a mathematics rather than a physics question. For instance, in the simple 
case of one-loop graphs one may carry out the integration over 
virtual momenta, and then determine where the resulting expressions
develop imaginary parts.
Equivalently, one can use a set of rules known as  ``Cutkosky rules'' 
or ``cutting rules''. One has to first  
cut through all diagrams on the lhs of the OT in all possible ways
such that the cut propagators can be put simultaneously on-shell. Note that 
one cuts through physical and unphysical particles 
(given that we are operating on the lhs of the OT), and that 
higher order diagrams have, in general, multiple (two-particle, three-particle, etc.)
cuts. Then, for each cut propagator one must substitute
\be
(k^2-m^2 + i \epsilon)^{-1} \to -2i \pi \delta_+(k^2-m^2),
\ee 
where 
\be
\delta_+(k^2-m^2)\equiv \theta(k^0)\delta(k^2-m^2),
\ee
and carry out the 
resulting integral. Finally, one must sum up the contributions of all cuts.  
This approach has the advantage of 
casting the lhs of the OT into a form that, for certain simple theories
such as scalar field theories or QED,  makes the equality 
with the rhs manifest. 
\newline
\indent
An issue of central importance for what follows is the way that the 
OT is realized at the level 
of the conventional diagrammatic expansion, or equivalently, at the level of the 
\mbox{propagator-,} vertex-, and box-like amplitudes,  
$T_1$, $T_2$, and $T_3$, respectively, introduced in subsection~\ref{sec:Ti}. 
Specifically, in its general formulation of Eq.~(\ref{absorptive}), the OT
is a statement at the level of {\it entire} amplitudes
and not of individual Feynman graphs, nor of the corresponding 
subamplitudes.
Thus, the imaginary part of a 
given diagram appearing on the rhs does {\it not} necessarily 
correspond to an easily identifiable diagrammatic (or kinematic) piece on the rhs. 
For example, in QCD the {\it conventional} propagator-like pieces of the two sides, \ie
those defined from the {\it standard} (as opposed to the ``pinched'') 
diagrammatic expansion, do not have to coincide in general. 
Of course, there are theories where the OT holds also at the 
level of individual graphs and kinematic subamplitudes. 
This stronger version of the OT is realized in scalar theories, 
but fails in non-Abelian gauge theories, 
such as QCD and the electroweak model. 
A crucial advantage of the PT is that it permits 
the realization of the OT at the level of kinematically distinct, well-defined 
subamplitudes, even in the context of non-Abelian gauge-theories; these   
privileged subamplitudes are, of course, none other than the 
$\widehat{T}_1$, $\widehat{T}_2$, and $\widehat{T}_3$.
\begin{figure}[t]
\bce
\includegraphics[width=15cm]{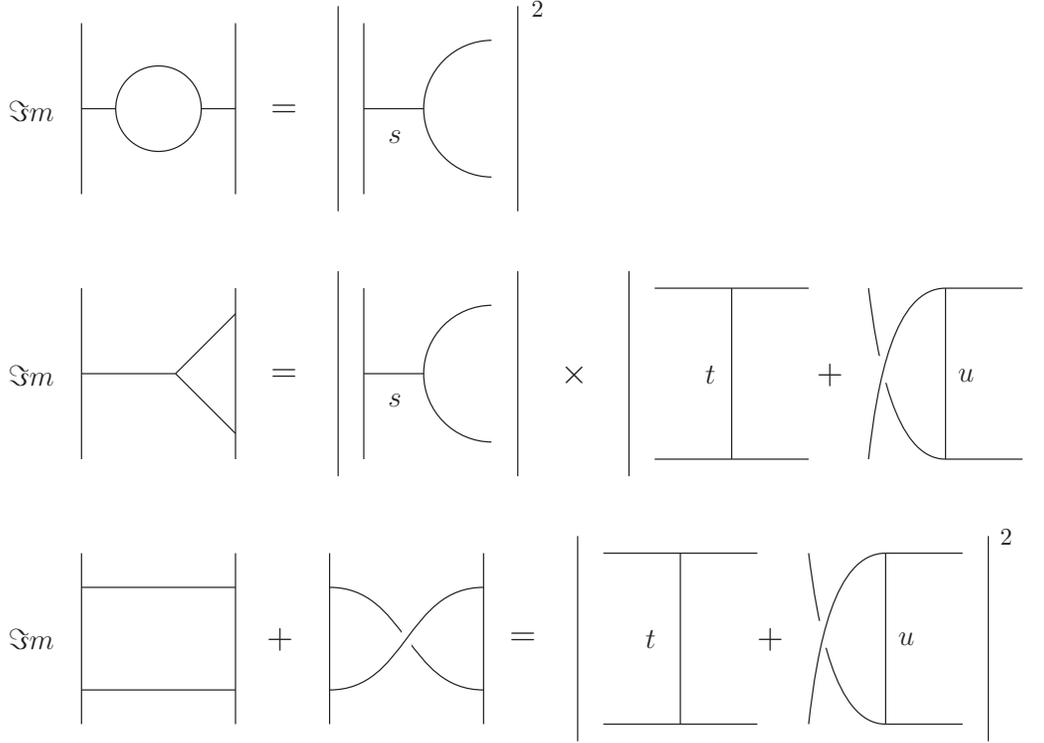}
\ece
\caption{\figlab{strong_OT} The stronger version of the OT in the case of a scalar theory.}
\end{figure}
\newline
\indent
In order to see 
a concrete example where the OT holds at the level of individual subamplitudes,
let us turn to a scalar, $\lambda \phi^3$ theory, where pinching is impossible (no WIs),
and therefore the topological structures 
given by the Feynman graphs cannot be modified.
We will consider the 
process $\phi(p_1)\phi(p_2) \to \phi(p_1)\phi(p_2)$; thus, 
$| i \rangle = |\phi(p_1) \phi (p_2)\rangle$, 
and, at lowest order, the only intermediate 
state possible is $| j \rangle = |\phi(k_1) \phi (k_2)\rangle$ (see \Figref{strong_OT}). 
In such a case, the OT
assumes its stronger version 
\begin{equation}
\Im m \{ T^{ii}_{\ell}\} =\ \frac{1}{2}
\sum_j (2\pi )^4\delta^{(4)}(P_j - P_i){\mathcal M}^{ij}_{\ell} ,
\label{OTstrong}
\end{equation}
where $\ell=1,2,3$ denotes, respectively, the 
propagator-, the vertex-, 
and box-like parts of either side (to recover the full OT, one 
simply sums both sides over $\ell$). 
\newline
\indent
The way the ${\mathcal M}^{ij}_{\ell}$ are determined is simply through the dependence 
on the Mandelstam variables $s$, $t$, and $u$; the latter are defined in this case as
$s=(p_1+p_2)^2= (k_1+k_2)^2$, $t= (p_1-k_1)^2 = (p_2-k_2)^2$, and $u= (p_1-k_2)^2 = (p_2-k_1)^2$.
Specifically (suppressing the superscript $ij$),  
\be
{\mathcal T}  = \lambda^2 
\left[\frac{1}{s-m^2} + \frac{1}{t-m^2} +  \frac{1}{u-m^2} \right],
\ee
and so
\bea
{\mathcal M}_{1} &=& \lambda^4 \left(\frac{1}{s-m^2}\right)^2,
\label{M1}\nonumber\\
{\mathcal M}_{2} &=& 2 \lambda^4 
\left[\frac{1}{t-m^2} + \frac{1}{u-m^2} \right]\frac{1}{s-m^2},
\label{M2}\nonumber\\
{\mathcal M}_{3} &=& \lambda^4 \left[\frac{1}{t-m^2} + \frac{1}{u-m^2} \right]^2.
\label{M3}
\eea
As we will see later on, this simple identification 
fails in the case of non-Abelian theories.
\newline
\indent
Let us now verify Eq.~(\ref{OTstrong})
for the propagator-like parts of the amplitude ($\ell=1$) and 
at the lowest non-trivial order in $\lambda$.
We have 
that 
\bea
T^{ii}_{1}&=&-\lambda^2 i\Delta(q^2,m^2)\nonumber \\
&=&-\lambda^2 \frac i{q^2-m^2-i\Pi(q^2,m^2)}\nonumber \\
&=&-\lambda^2 iD^{(0)}(q^2,m^2)+\lambda^2D^{(0)}(q^2,m^2)\Pi(q^2,m^2)D^{(0)}(q^2,m^2),\nonumber\\
{\mathcal M}^{ij}_{1} &=& \lambda^4 [D^{(0)}(q^2,m^2)]^2
\eea
with 
$D^{(0)}(q^2,m^2)= (q^2-m^2)^{-1}$ the tree-level scalar propagator and 
$\Pi(q^2,m^2)$ its one-loop self-energy, given by  
\be
i\Pi(q^2,m^2)= \frac{\lambda^2}{2} \int_k \frac1{(k^2-m^2)[(k+q)^2-m^2]}.
\label{Iint}
\ee
Then, denoting the two sides of the OT by $(\rm lhs)_{1}$ and $(\rm rhs)_{1}$,  
we have 
\bea
(\rm lhs)_{1} &=& \lambda^2 \Im m \{T^{ii}_1\},
\label{qwer2-a}\nonumber\\
(\rm rhs)_{1}  &=& \frac{1}{2}\times\frac{1}{2}\lambda^4 [D^{(0)}(q^2)]^2 \int_\mathrm{PS},
 \label{qwer2-b}
\eea
where the additional combinatorial $\frac12$  factor 
accounts for having two identical particles in the final state. The integral 
$\int_\mathrm{PS}$ is the two-body  phase-space integral, given by
\be
\int_\mathrm{PS}= \frac{1}{(2\pi)^2}\, 
\int d^{4}k_1 \int d^{4}k_2\, \delta_+ (k^2_1-m^2_1)\delta_+ (k^2_2-m^2_2)
\delta^{(4)}(q-k_1-k_2),
\label{2BPS}
\ee
where $m_1$ and $m_2$ are the masses of the intermediate 
particles produced (in the case at hand \mbox{$m_1=m_2=m$}). 
To demonstrate the equality $(\rm lhs)_{1}=(\rm rhs)_{1}$, use
for $(\rm rhs)_{1}$
the standard result
\be
\int_\mathrm{PS} =  
 \theta(q^0)\theta [q^2-{(m_1+m_2)}^2] \frac{1}{8\pi q^2}
\lambda^{1/2} (q^2,m_1^2,m_2^2) ,
\label{LIPS1}
\ee
where $\lambda (x,y,z)= (x-y-z)^2-4yz$, and for $(\rm lhs)_{1}$ that 
\bea
\Im m \{\Pi(q^2)\} 
&=&
- \frac{\lambda^2}{32\pi^2}\Im m\left\{\int^1_{0} dx
\ln[m^2 -q^2x(1-x)]\right\}
\nonumber\\
&=& \frac{\lambda^2}{32 \pi} \frac{\theta (q^2- 4 m^2 )}{q^2}  
\lambda^{1/2}(q^2,m^2,m^2)\nonumber\\
&=& \frac{\lambda^2}{4} \int_\mathrm{PS},
\label{impi}
\eea
obtained from Eq.~(\ref{Iint}) after the Feynman parametrization 
and standard integration over $k$. It is 
relatively straightforward to show, to lowest order, the validity of  
(\ref{OTstrong}) for $\ell=2,3$, 
especially if the Cutkosky rules are employed to determine the rhs.
Notice that, in QCD, the stronger version of the OT holds for the quark loop but fails when the virtual particles circulating in the loop are gluons (see \Figref{strong_OT_QCD}). 
\begin{figure}[t]
\bce
\includegraphics[width=9.5cm]{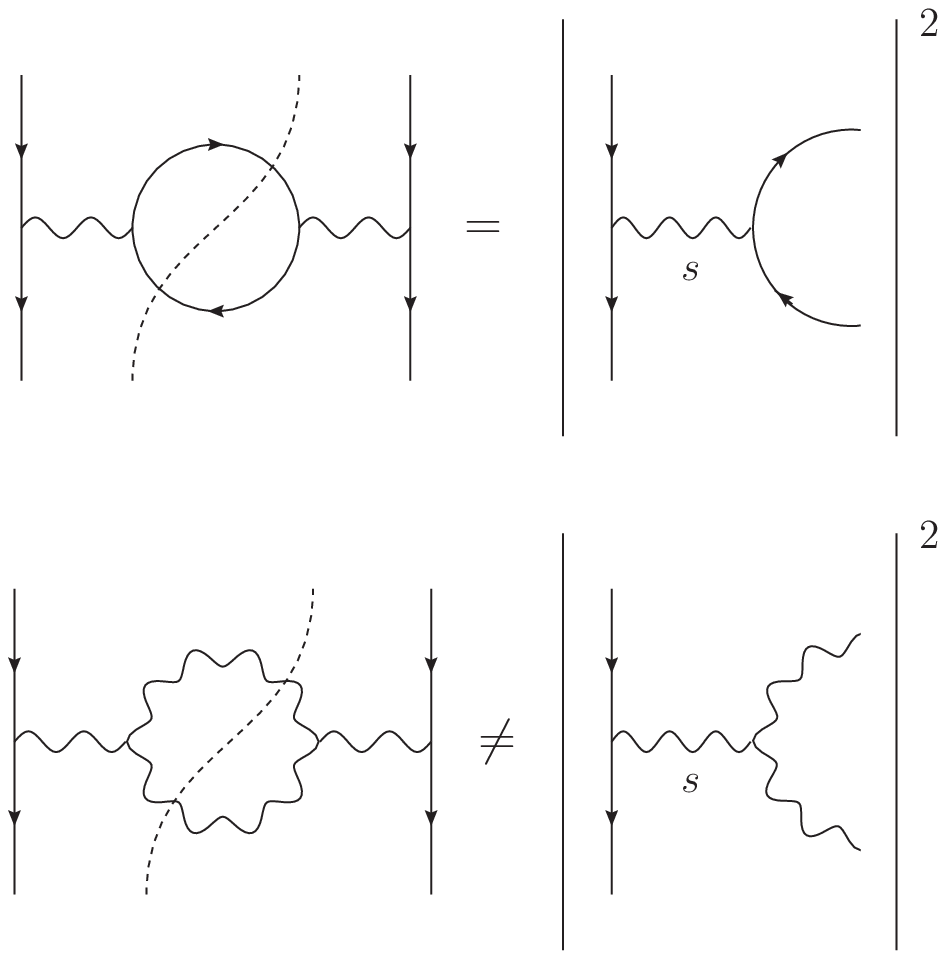}
\ece
\caption{\figlab{strong_OT_QCD} The stronger version of the OT in QCD: it holds for
the quark loop but fails for the gluon loop.}
\end{figure}
\newline
\indent
An additional important ingredient that accompanies the OT is that of analyticity;
together they provide a powerful framework that restricts severely the allowed  
structure of Green's functions.
Specifically, Green's functions are considered 
to be  analytic functions of their kinematic variables; 
this property, in turn, relates their real and imaginary parts  
by the so-called dispersion relations.
\newline
\indent
In particular, let us recall that if a complex function $f(z)$ is analytic in the interior of and upon a closed
curve, say $C_\uparrow$  in \Figref{disprel}, and $x+i\varepsilon$ (with $x,\varepsilon
\in\mathbb{R}$ and $\varepsilon >0$) is a point within the closed curve
$C_\uparrow$, we then have the Cauchy's integral form, 
\begin{equation}
f(x+i\varepsilon)\ =\ \frac{1}{2\pi i} \oint_{C_\uparrow} dz\,
\frac{f(z)}{z-x-i\varepsilon}\ ,
\end{equation} 
where $\oint$ denotes that the path $C_\uparrow$ is singly wound. Using
Schwartz's reflection principle, one also obtains
\begin{equation}
f(x-i\varepsilon)\ =\ -\, \frac{1}{2\pi i} \oint_{C_\downarrow} dz\,
\frac{f(z)}{z-x+i\varepsilon}\ .
\end{equation} 
Note that $C_\uparrow^*=C_\downarrow$. Sometimes, an analytic function is 
called holomorphic; both terms are equivalent for complex functions.
\begin{figure}[t]
\bce
\includegraphics[width=6cm]{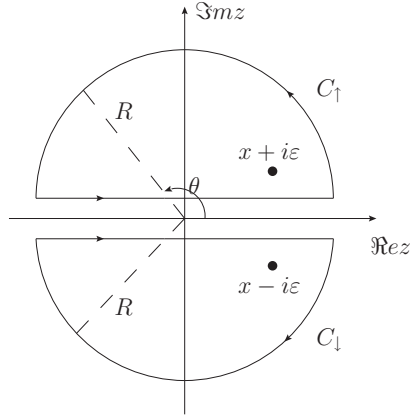}
\ece
\caption{Contours of complex integration for the dispersion relations}
\figlab{disprel}
\end{figure}
\newline
\indent
Then, let us assume 
that the analytic function
$f(z)$ has the asymptotic behavior, $|f(z)|\le C/R^k$, for large radii $R$ 
with  $C$ a real non-negative constant and $k>0$. Taking
the limit $\varepsilon\to 0$, it is easy to evaluate $\Re e f(x)$ through 
\begin{equation} 
2\Re e f(x)\ =\ `\lim_{\varepsilon\to 0}\mbox{'}\Big[ f(x+i\varepsilon)
+ f^*(x-i\varepsilon)\Big]\ =\ `\lim_{\varepsilon\to 0}\mbox{'}\, 
\frac{1}{\pi}\int\limits_{-\infty}^{+\infty}dx'\, 
\Im m \left( \frac{f(x')}{x'-x-i\varepsilon}\right)\, +\, \Gamma_{\infty}.
\end{equation}
Here, $`\lim_{\varepsilon\to 0}\mbox{'}$ means that the limit
should be taken {\em after} the integration has been performed, and 
\begin{equation}
\label{orio}
\Gamma_\infty\ =\ \frac{1}{\pi}\lim_{R\to \infty}\, \Re e\,
\int_0^\pi d\theta\, f(Re^{i\theta})\, .
\end{equation}
Because of the assumed asymptotic behavior of $f(z)$ at infinity, the
integral over the upper infinite semicircle in \Figref{disprel} can
be easily shown to vanish:  $\Gamma_{\infty}=0$. Then, employing the well-known identity for
distributions, 
\be
`\lim_{\varepsilon\to 0}\mbox{'}\frac{1}{x'-x-i\varepsilon} =
\mbox{P}\frac{1}{x'-x} + i\pi\delta (x'-x),
\ee
we arrive at the unsubtracted dispersion relation,
\begin{equation}
\label{DR1}
\Re e f(x)=\frac{1}{\pi} \mbox{P}\int\limits^{+\infty}_{-\infty}
dx'\frac{\Im m f(x')}{x'-x}. 
\end{equation} 
where the ``P'' denotes the 
principle value of the integral. Following similar arguments, one can
express the imaginary part of $f(x)$ as an integral over $\Re e f(x)$. 
\newline
\indent
In the previous derivation, the assumption that $|f(z)|$ approaches zero
sufficiently fast at infinity has been crucial, since it guarantees that
$\Gamma_{\infty}\to 0$. However, if this assumption does not hold,
additional subtractions need be included in order to arrive at a finite
expression. For instance, for $|f(z)|\le CR^k$ with $k<1$, it is sufficient to
carry out a single subtraction at a point $x=a$. In this way, one has 
\begin{equation}
\label{DR2}
\Re e f(x) = \Re e f(a) + \frac{(x-a)}{\pi} \mbox{P} 
\int\limits_{-\infty}^{+\infty}dx'\, \frac{\Im m f(x')}{(x'-a)(x'-x)}.
\end{equation}
From Eq.\ (\ref{DR2}), it is obvious that $\Re e f(x)$ can be
obtained from $\Im m f(x)$, up to an unknown, real constant $\Re e f(a)$.
Usually, the point $a$ is chosen in a way such that $\Re e f(a)$ takes a
specific value on account of some physical 
requirement or normalization condition.
\newline
\indent
To see how analyticity works in a simple case, let us return to  
the scalar self-energy $\Pi(s)$ of Eq.\ (\ref{Iint}).
Setting $q^2=s$, and defining 
the ``velocity''
\be
\beta(s,m^2)\equiv (1-4m^2/s)^{1/2} = s^{-1} \lambda^{1/2}(s,m^2,m^2),
\ee
a standard integration yields 
\be
\Pi(s,m^2)= \frac{\lambda^2}{32\pi^2} \left[ \frac{2}{\epsilon}
- \gamma_{ E} + \ln\frac{4\pi\mu^2}{m^2} + 2 - 
\beta(s) \ln\frac{\beta(s,m^2) + 1}{\beta(s,m^2) - 1}\right],
\label{RePi}
\ee
where $s$ should be analytically continued to $ s + i\varepsilon $. In fact,
for $s >4m^2$, the logarithmic function in Eq.~(\ref{RePi}) assumes the form
\be
\ln\frac{1+\beta(s,m^2)}{1-\beta(s,m^2)}- i\pi\theta (s-4m^2). 
\ee
Evidently, the absorptive part of $\Pi (s)$ obtained from 
Eq.~(\ref{RePi}) is equal to the $\Im m \Pi(s)$ appearing in Eq.~(\ref{impi}).
Furthermore, one can verify the validity of the dispersion 
relation of Eq.~(\ref{DR2}), singly
subtracted at $s=0$. Since
\begin{equation}
\Pi(0,m^2) = \frac{\lambda^2}{32\pi^2}\left[ \frac{2}{\epsilon}
- \gamma_E + \ln\frac{4\pi\mu^2}{m^2} \right],
\end{equation}
the renormalized $\Pi (s)$ is obtained simply as 
$\Pi_\mathrm{ R}(s,m^2) = \Pi(s,m^2)-\Pi(0,m^2)$. 
Noting that $\Im m \Pi(s,m^2) = \Im m \Pi_\mathrm{ R}(s,m^2)$ 
(the divergent parts have to be real in order 
for the hermiticity of the Lagrangian to be preserved),
it is elementary to demonstrate that indeed (principle value implied),  
\be
\Re e \Pi_\mathrm{ R}(s,m^2) =
\frac{s}{\pi}\, \int\limits_{4m^2}^{\infty}ds' 
\frac{\Im m \Pi(s',m^2)}{s'(s'-s)} .
\ee
\newline
\indent
The synergy between unitarity (OT) and analyticity (dispersion relations)
constitutes the basis of the dynamical framework  known from the sixties as ``$S$-matrix theory'' (see, \eg\cite{ELOP}).
In the context relevant to our purposes, one may resort to 
this framework in order to (re)construct dynamically (at least in principle)
a given Green's function.
A possible procedure one may adopt to accomplish this is the following.
The lhs of the OT is an experimentally measurable quantity: up to simple 
kinematic ingredients (flux-factors, etc) it can be identified with a 
physical cross-section. In fact, in the case of scalar theories or QED, the contributions  
of the {\it individual} subamplitudes ${\mathcal M}^{ij}_{\ell}$ to this cross-section  
may be projected out; for example, in the 
center-of-mass frame the ${\mathcal M}^{ij}_{\ell}$ display a different
dependence on the scattering angle $\theta$, which, in turn, allows their 
extraction  from the entire cross-section. 
Then, through the lhs of the OT,  the measured  ${\mathcal M}^{ij}_{\ell}$ is identified 
with the imaginary part 
of the Green's function under construction, 
namely (up to trivial factors) 
the  $T^{ii}_{\ell}$ (for example, the propagator, for $\ell=1$).
Having determined the $\Im m  T^{ii}_{\ell}$ from the OT, 
the dispersion relation can finally furnish 
(up to subtractions) the real part of $T^{ii}_{\ell}$.

\subsubsection{The fundamental $s$-$t$ cancellation}
\noindent
As already alluded to in the previous subsection, the 
strong version of the OT, expressed in \linebreak Eq.~(\ref{OTstrong}), does 
not hold in general in the case of non-Abelian theories. 
This is so because, with the exception of certain gauges, the naive (diagrammatic) 
propagator-, vertex-, and box-like subamplitudes 
of each side are totally different.
For example, 
in the case of the forward QCD process  
$q(p_1) {\bar q}(p_2)\!\to\! q(p_1){\bar q}(p_2)$ 
the propagator-like part of the lhs, computed in the renormalizable gauges, 
is determined by cutting 
 through one-loop graphs  containing  
$\xi$-dependent gluon propagators and unphysical ghosts (omit quark-loops), while 
the propagator-like part of the rhs contains the polarization tensors 
corresponding to 
physical massless  particles of spin 1 (two physical polarizations). 
This profound difference complicates the 
diagrammatic verification of the OT, and invalidates, at the same time, its stronger version.
\newline
\indent
As we will demonstrate in this subsection, the application of the 
PT on the rhs (the physical side) of the OT is tantamount 
to the explicit use of  an  underlying fundamental cancellation between $s$-channel
and  $t$-channel  graphs~\cite{Papavassiliou:1996zn,Papavassiliou:1996fn}. 
 This  cancellation is  exposed  after  the
judicious  combination of two  fundamental WIs,  one operating  on the
$s$-channel and one on the $t$-channel amplitude.
This  cancellation results in a non-trivial reshuffling of terms, 
which, in turn, allows for the definition of 
kinematically 
distinct contributions, to be denoted by $\widehat{{\mathcal M}}^{ij}_{\ell}$;
interestingly enough, they 
correspond to the imaginary parts of the one-loop PT subamplitudes 
constructed in the previous section. 
Specifically, the  
PT subamplitudes satisfy the strongest version of the OT, \ie 
\begin{equation}
\Im m  {\widehat T}^{ii}_{\ell} = \frac{1}{2}
\sum_j (2\pi )^4\delta^{(4)}(P_j - P_i) \widehat{{\mathcal M}}^{ij}_{\ell},
\label{OTPTstrong}
\end{equation}
In other words, the strong version of the OT holds iff the 
identification of the subamplitudes on each
side occurs {\it after} the application of the PT. 
\newline
\indent
To see all this in detail, 
we   consider  the  forward  scattering   process  
$q(p_1)  {\bar q}(p_2)\!\to\! q(p_1){\bar q}(p_2)$, 
and  concentrate   on  the  OT   to  lowest  order.    Obviously,  the
intermediate  states  appearing  on  the  rhs may  involve  quarks  or
gluons. The  quarks can be treated  essentially as in QED,  and are, in  
that  sense,  completely straightforward.   
We will therefore focus on the part of the OT
where the intermediate states are
two gluons; we have that 
\begin{equation}
\Im m \langle q\bar{q}|T|q\bar{q}\rangle = \frac12\times
\frac12\int_{\mathrm{PS}_{gg}}  
\langle q\bar{q}|T|gg\rangle \langle gg|T|q\bar{q}\rangle^{*}.
\label{OTgg}
\end{equation}
The extra $\frac12$ factor is statistical and arises from
the fact that the final on-shell gluons should be considered as identical
particles in the total rate. 
\begin{figure}[!t]
\bce
\includegraphics[width=16cm]{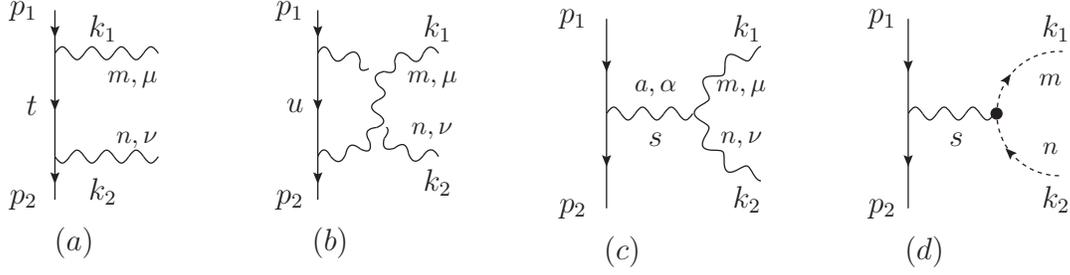}
\ece
\caption{\figlab{absorpt_PT} Diagrams defining the amplitudes ${\mathcal T}_t$ [graphs $(a)$ and $(b)$] and ${\mathcal T}_s$ [graph $(c)$]. Diagram $(d)$ will contribute to the amplitude ${\mathcal S}$ defined in Eq.~(\ref{S_def}).}
\end{figure}
\newline
\indent
In Eq.~(\ref{OTgg}) we set for the lhs  $T\equiv \langle q\bar{q}|T|q\bar{q}\rangle $ 
and for the rhs ${\mathcal T} \equiv\langle q\bar{q}|T|gg\rangle$ and  
${\mathcal M}\equiv {\mathcal T}{\mathcal T}^{*}$. 
Let us now focus on the rhs of Eq.\ (\ref{OTgg}). Diagrammatically, 
the tree-level amplitude ${\mathcal T}$ 
consists of two distinct parts: $t$ and $u$-channel graphs that contain an
internal quark propagator, ${{\mathcal T}_{t}}^{mn}_{\mu\nu}$, as shown in diagrams $(a)$ and $(b)$ of \Figref{absorpt_PT},
 and an $s$-channel amplitude, ${{\mathcal T}_{s}}^{mn}_{\mu\nu}$,
given in diagram $(c)$ of that same figure. The subscript $s$ and $t$ refers as usual to the
corresponding Mandelstam variables, \ie $s=q^2=
(p_1+p_2)^2=(k_1+k_2)^2$, and $t=(p_1-k_1)^2=(p_2-k_2)^2$. 
\newline
\indent
Defining the analogue of Eq.~(\ref{defV}) for these kinematics, namely 
\begin{equation}
i{\mathcal V}_{\rho}^{a}(p_2,p_1)= \bar{v}(p_2)ig t^a\gamma_{\rho}u(p_1) ,
\end{equation}
we have that
\begin{equation}
{\mathcal T}^{mn}_{\mu\nu}={{\mathcal T}_{s}}^{mn}_{\mu\nu}+
{{\mathcal T}_{t}}^{mn}_{\mu\nu},
\label{DefT}
\end{equation}
with
\begin{eqnarray}
{{\mathcal T}_{s}}^{mn}_{\mu\nu} & =&
-gf^{amn}{\mathcal V}^{a\rho}\Delta^{(0)}_{\rho\alpha}(q,\xi)
\Gamma^{\alpha}_{\mu\nu}(q,-k_1,-k_2) ,
\label{Ts}\nonumber\\
{{\mathcal T}_{t}}^{mn}_{\mu\nu} &=& -ig^2\bar{v}(p_2)\left[ 
t^n\gamma_{\nu} S^{(0)}(p_1-k_1)t^m \gamma_{\mu}\ + t^m \gamma_{\mu}
S^{(0)}(p_1-k_2) \gamma_{\nu} t^n\right]u(p_1).
\label{Tt}
\end{eqnarray}
We then have
\begin{eqnarray}
{\mathcal M} &=&  {\mathcal T}^{mn}_{\mu\nu} L^{\mu\mu^{\prime}}
(k_1) L^{\nu\nu^{\prime}}(k_2)\, {\mathcal T}^{mn*}_{\mu^{\prime}\nu^{\prime}}
\nonumber\\
&=& \left[ {\mathcal T}_{s} +
{\mathcal T}_{t}\right]_{\mu\nu}^{mn}L^{\mu\mu^{\prime}}(k_1)
L^{\nu\nu^{\prime}}(k_2)\left[ {{\mathcal T}_{s}}^{*}
+{{\mathcal T}_{t}}^{*}\right]_{\mu^{\prime}\nu^{\prime}}^{mn},
\label{MM}
\end{eqnarray}
where the polarization tensor $L^{\mu\nu}(k)$ corresponding to a 
massless spin one particle is given by
\begin{equation}
L_{\mu\nu}(k)\ =\ -g_{\mu\nu}+ \frac{n_{\mu}k_{\nu}
+n_{\nu}k_{\mu} }{n\cdot k} + 
\eta^2 \frac{k_{\mu}k_{\nu}}{{(n\cdot k)}^2}\, .
\label{PhotPol}
\end{equation}
Notice that in accordance with  Eq.~(\ref{OTgg}),
the rhs of Eq.~(\ref{MM}) is inside the 
phase space integral \mbox{$\frac{1}{4}\int_{\mathrm{PS}_{gg}}$}, which we 
will suppress for simplicity.
Also, the auxiliary four-vectors $n_{\mu}$ are unphysical, and any 
dependence on them must cancel in a physical process. 
The same holds for the parameter $\eta^2$, 
which acts as a gauge-fixing term. 
\newline
\indent
It is clear now that if we were to apply at this point 
the same criterion for determining the ${\mathcal M}_{\ell}$
as in the scalar case [see  Eqs~(\ref{M1})] 
we would get (we suppress Lorentz and color indices)
\bea
{\mathcal M}_{1} &=& {\mathcal T}_{s}L(k_1)L(k_2){\mathcal T}_{s}^{*},
\label{M1naive}\nonumber\\
{\mathcal M}_{2} &=&  {\mathcal T}_{s} L(k_1)L(k_2){\mathcal T}_{t}^{*} + 
{\mathcal T}_{t} L(k_1)L(k_2){\mathcal T}_{s}^{*},
\label{M2naive}\nonumber\\
{\mathcal M}_{3} &=& {\mathcal T}_{t} L(k_1)L(k_2)\,{\mathcal T}_{t}^{*}.
\label{M3naive}
\eea
However, the  ${\mathcal M}_{\ell}$ appearing in the equations above cannot play the role of 
the subamplitudes 
appearing on the rhs of the strong version of OT,
for at least one obvious reason:  they depend explicitly on the  unphysical  $n_{\mu}$ 
and  $\eta^2$. This is the crux of the matter:
the cancellation of $n_{\mu}$ and $\eta^2$ 
at the level of the 
entire ${\mathcal M}$ becomes possible only 
by combining contributions between the $s$- and the 
 $t$-channel graphs. It is only {\it after} this cancellation has taken place that the 
criterion of  Eqs~(\ref{M1}) may be used to define the 
$\widehat{{\mathcal M}}^{ij}_{\ell}$ that will appear on the rhs of Eq.~(\ref{OTPTstrong}).
\newline
\indent
We now study the above points concretely. Before turning to the 
$n_{\mu}$ and $\eta^2$ cancellations, 
let us  demonstrate the cancellation of gfp inside the 
off-shell tree-level gluon propagator,  $\Delta^{(0)}_{\mu\nu}(q,\xi)$, 
appearing in the $s$-channel graph.
To that end note 
that the external quark current 
is conserved, namely $q^{\rho} {\mathcal V}_{\rho}^{a} =0$.
In addition, we have that for ``on-shell'' gluons, 
\ie for $k^{2}=0$, $k^{\mu} L_{\mu\nu}(k) = 0$. 
By virtue of  this last property, we see immediately
that if we carry out the PT decomposition of Eq.~(\ref{decomp})
to the three-gluon vertex $\Gamma$, 
the term $\Gamma^{\mathrm{P}}$ vanishes after being contracted 
with the polarization vectors,
and only the $\Gamma^{\mathrm{F}}$ piece of the vertex survives.
Then, it is immediate to verify that
the longitudinal (gfp-dependent)  parts of
$\Delta^{(0)}$ either vanish because of current conservation,
or because they trigger the WI
\begin{equation}
q^{\alpha}\Gamma^{\mathrm{F}}_{\alpha\mu\nu}(q, -k_1, -k_2)=g_{\mu\nu}
(k_1^2 - k_2^2),
\label{FWI}
\end{equation}
which vanishes on-shell.  This last WI is crucial because, in general,
current conservation alone is not sufficient to guarantee the gfp-independence
of the final answer. Clearly, in the covariant gauges the gauge fixing term
is proportional to $q^{\mu}q^{\nu}$ and therefore current conservation ensures that such a term vanishes. 
However, had we chosen an axial gauge instead, the gluon propagator would be of the form
\begin{equation}
\label{Deleta}
\Delta^{(0)}_{\mu\nu}(q,\tilde{\eta}) = \frac{L_{\mu\nu}
(q,\tilde{n},\tilde{\eta})}{q^2}, 
\end{equation}
where $\tilde{n} \neq n$ in general; then only the term
${\tilde{\eta}_{\nu}}q_{\mu}$ vanishes because of current conservation,
whereas the term ${\tilde{n}_{\nu}}q_{\mu}$  can only disappear if 
Eq.~(\ref{FWI}) holds.  Thus, either way, Eq.~(\ref{MM}) finally becomes 
\begin{equation}
{\mathcal M}=\frac{1}{4}
({\mathcal T}_{s}^{\mathrm{F}}+{\mathcal T}_{t})^{mn}_{\mu\nu}
L^{\mu\mu^{\prime}}(k_1)\, L^{\nu\nu^{\prime}}(k_2)
({\mathcal T}_{s}^{\mathrm{F}}+{\mathcal T}_{t})^{mn*}_{\mu^{\prime}\nu^{\prime}} ,
\label{MM2}
\end{equation}
where the gfp-independent quantity ${\mathcal T}_{s}^{\mathrm{F}}$ is given by
\begin{equation}
{{\mathcal T}_{s}}^{{\mathrm{F}},mn}_{\mu\nu} =
gf^{amn} g^{\rho\alpha} d(q^2)
\Gamma^{\mathrm{F}}_{\alpha\mu\nu}(q,-k_1,-k_2) {\mathcal V}_{\rho}^{a}.
\label{TsF}
\end{equation}
\indent
We now want to show that the dependence on 
the unphysical quantities $n_{\mu}$ and $\eta^2$,
coming from the polarization vectors, disappears. 
The exact way this happens is very instructive,  
and can be  traced to a very particular cancellation
operating between the $s$- and $t$- channel components.
This cancellation, in turn, is crucial for the PT construction (and its all-order generalization, see \secref{beyond_1l}), 
because it captures precisely the mechanism that enforces 
the corresponding cancellations inside Feynman 
diagrams (where the gluons are off-shell).
\newline
\indent
To see this  $s$-$t$ cancellation
in detail, let us first define the quantities 
${\mathcal S}^{mn}$ and ${\mathcal R}^{mn}_{\mu}$ as follows:
\begin{eqnarray}
{\mathcal S}^{mn}&=& 
\frac 12g f^{amn}d(q^2)(k_1-k_2)^\mu {\mathcal V}_{\mu}^{a}, 
\label{S_def}\nonumber\\
{\mathcal R}_{\mu}^{mn} &=&  gf^{amn}{\mathcal V}_{\mu}^{a},
\end{eqnarray}
which are related by
\begin{equation}
k_1^{\mu}{\mathcal R}_{\mu}^{mn} =
-k_2^{\mu}{\mathcal R}_{\mu}^{mn}= q^2 {\mathcal S}^{mn}.
\label{D2}
\end{equation}
\newline
\indent
Second, using the 
conditions $k_1^2=k_2^2=0$, together with current conservation, 
$q^{\rho} {\mathcal V}_{\rho}^{a} =0$, we obtain the elementary WIs
\begin{eqnarray}
k_1^{\mu}\Gamma^{{\mathrm{F}}}_{\alpha\mu\nu}(q,-k_1,-k_2)& = &
-q^2 g_{\alpha\nu}+(k_1-k_2)_{\alpha} k_{2\nu},
\label{WIGFLR-1}\nonumber\\
k_2^{\nu} \Gamma^{{\mathrm{F}}}_{\alpha\mu\nu}(q,-k_1,-k_2)& = &
q^2 g_{\alpha\mu}+(k_1-k_2)_{\alpha} k_{1\mu}.
\label{WIGFLR-2}
\end{eqnarray}
Now the crucial point is that the $q^2$ term on the rhs of 
the above WIs will cancel against the $d(q^2)$ inside 
${{\mathcal T}_{s}}^{{\mathrm{F}}}$, allowing the communication of this 
part with the (contracted) $t$-channel graph. Specifically, we will have
\begin{eqnarray}
k_1^{\mu}{{\mathcal T}_{s}}^{{\mathrm{F}},mn}_{\mu\nu} & = &  
2 k_{2\nu}{\mathcal S}^{mn} - {\mathcal R}_{\nu}^{mn} ,
\label{w4-1}\nonumber\\
k_2^\nu {{\mathcal T}_{s}}^{{\mathrm{F}},mn}_{\mu\nu}  & = &  2 k_{1\mu}{\mathcal S}^{mn} 
 + {\mathcal R}_{\mu}^{mn},
\label{w4-2}\nonumber\\
k_1^{\mu}{{\mathcal T}_{t}}^{mn}_{\mu\nu} & = & {\mathcal R}_{\nu}^{mn},
\label{w4-3}\nonumber\\
k_2^{\nu}{{\mathcal T}_{t}}^{mn}_{\mu\nu} & = & -{\mathcal R}_{\mu}^{mn} 
\label{w4-4},
\end{eqnarray}
so that, evidently, 
\begin{eqnarray}
k_1^{\mu} [{{\mathcal T}_{s}}^{{\mathrm{F}}} + {{\mathcal T}_{t}}]^{mn}_{\mu\nu} 
 & = & 2 k_{2\nu}{\mathcal S}^{mn},
\label{stcan-1}\nonumber\\
k_2^\nu [{{\mathcal T}_{s}}^{{\mathrm{F}}} +{{\mathcal T}_{t}}]^{mn}_{\mu\nu}  
& = & 2 k_{1\mu}{\mathcal S}^{mn}.
\label{stcan-2}
\end{eqnarray}
This is the $s$-$t$ cancellation~\cite{Papavassiliou:1996zn,Papavassiliou:1996fn}:   
the term ${\mathcal R}$ comes with opposite sign, 
and drops out in the sum.
In addition, notice also that
\begin{eqnarray}
k_1^{\mu}k_2^{\nu}{{\mathcal T}_{s}}^{{\mathrm{F}},mn}_{\mu\nu}& = & q^2{\mathcal S}^{mn},
\label{w5} \nonumber\\
k_1^{\mu}k_2^{\nu}{{\mathcal T}_{t}}^{mn}_{\mu\nu} &=& -q^2{\mathcal S}^{mn}.
\label{w6}
\end{eqnarray}
\indent
Using the above results, it is now easy to check that indeed, all dependence on
both $n_{\mu}$ and $\eta^2$ cancels in Eq.~(\ref{MM2}), as it should, and
we are finally left with (omitting the fully contracted color and Lorentz indices)
\be
{\mathcal M} =  
\left( {\mathcal T}_{s}^{{\mathrm{F}}}{{\mathcal T}_{s}^{{\mathrm{F}}}}^{*} -8 {\mathcal S}{\mathcal S}^{*}\right) 
+ \left( {\mathcal T}_{s}^{{\mathrm{F}}}{\mathcal T}_{t}^{*} + {{\mathcal T}_{s}^{{\mathrm{F}}}}^{*}
{\mathcal T}_{t} \right) + {\mathcal T}_{t}{\mathcal T}_{t}^{*} 
\label{MM3}
\ee
At this point we can naturally define the genuine propagator-like, vertex-like,
and  box-like subamplitudes, as in the scalar case, \ie
\bea
\widehat{{\mathcal M}}_1 &=& {\mathcal T}_{s}^{{\mathrm{F}}}{{\mathcal T}_{s}^{{\mathrm{F}}}}^{*} -8 {\mathcal S}{\mathcal S}^{*}, 
\label{M1PT}\nonumber\\
\widehat{{\mathcal M}}_2 &=& {\mathcal T}_{s}^{{\mathrm{F}}}{\mathcal T}_{t}^{*} + {{\mathcal T}_{s}^{{\mathrm{F}}}}^{*}{\mathcal T}_{t}, 
\label{M2PT}\nonumber\\
\widehat{{\mathcal M}}_3 &=& {\mathcal T}_{t}{\mathcal T}_{t}^{*}.
\label{M3PT}
\eea
\indent
Let us now focus on $\widehat{{\mathcal M}}_1$; employing the relation
\begin{equation}
\Gamma^{{\mathrm{F}}}_{\rho\mu\nu}\Gamma^{{\mathrm{F}},\mu\nu}_{\alpha} = 
8q^2 P_{\alpha\rho}(q) +
4{(k_1-k_2)}_{\alpha}{(k_1-k_2)}_{\rho},
\label{FRE}
\end{equation}
and 
\be
{\mathcal S}{\mathcal S }^{*} =
\frac{1}{4}g^2 C_{A} {\mathcal V}^a_{\alpha}d(q^2) \left[(k_1-k_2)^\alpha
(k_1-k_2)^\rho \right] d(q^2) {\mathcal V}^{a}_{\rho},
\ee
we obtain for $\widehat{{\mathcal M}}_1$
\begin{equation}
\widehat{{\mathcal M}}_1=g^2 C_{A}
{\mathcal V}^{a}_{\mu}d(q^2)\left[ 8q^2 P^{\mu\nu}(q)+
2 {(k_1-k_2)}^{\mu}{(k_1-k_2)}^{\nu}\right] d(q^2){\mathcal V}^{a}_{\nu},
\end{equation} 
and the propagator-like part of the rhs of the OT reads 
\be
({\rm rhs})_{1}  = \frac{1}{2}\times\frac{1}{2}
\int_{\mathrm{PS}_{gg}}\!\widehat{{\mathcal M}}_1.
\label{rhs1a}
\ee
Using the result of Eq.~(\ref{LIPS1}), together with the additional general formula  
\bea
\int_{\mathrm{PS}}\!{(k_1-k_2)}_{\mu}{(k_1-k_2)}_{\nu} &=&
- \frac{\lambda (q^2,m_1^2,m_2^2)}{3q^2}{P}_{\mu\nu}(q)\int_{\mathrm{PS}}\!
\nonumber\\
&+&\left[ \frac{\lambda (q^2,m_1^2,m_2^2)}{q^2} - q^2 + 2 (m^2_1+m^2_2) \right]
 \frac{q_{\mu}q_{\nu}}{q^2}\int_{\mathrm{PS}},
\label{LIPS2}
\eea
we obtain for the case of two massless gluons in the final state
\begin{eqnarray}
&&\int_{\mathrm{PS}_{gg}} =  
 \frac{1}{8\pi},
 \label{LIPS2g-1} \nonumber\\
&&\int_{\mathrm{PS}_{gg}}\! {(k_1-k_2)}_{\mu}{(k_1-k_2)}_{\nu} = - \frac{1}{8\pi} 
\frac{1}{3}q^2P_{\mu\nu}(q),
\label{LIPS2g-2}
\end{eqnarray}
and thus (\ref{rhs1a}) becomes 
\be
({\rm rhs})_{1} = {\mathcal V}^\mu_a d(q^2)\pi  b g^2 q^2 P_{\mu\nu}(q)d(q^2){\mathcal V}_a^\nu.
\label{rhs1b}
\ee
On the other hand, for the propagator-like part of the lhs of the OT we have 
\be
({\rm lhs})_{1} = \Im m \widehat{T}_1=
{\mathcal V}^a_\mu d(q^2) \Im m \widehat{\Pi}^{\mu\nu}(q)d(q^2) {\mathcal V}^a_\nu,
\label{lhs1a}
\ee
Equating (\ref{rhs1b}) and (\ref{lhs1a}) we finally obtain 
\begin{equation}
\Im m \widehat{\Pi}_{\mu\nu}(q) = \pi b g^2  q^2 P_{\mu\nu}(q).
\label{IMQCD}
\end{equation}
\indent
Let us now write the (dimensionful)
$\widehat{\Pi}(q^2)$ introduced in Eq.~(\ref{Pid1})
as $\widehat{\Pi}(q^2) = q^2 {\bf \widehat{\Pi}}(q^2)$.
From (\ref{IMQCD}) we have that 
$\Im m  {\bf \widehat{\Pi}}(q^2) =\pi bg^2$,
with the renormalized ${\bf \widehat{\Pi}}_{ R}(q^2)$ given 
by 
\be
{\bf \widehat{\Pi}}_{{\mathrm{R}}}(q^2) 
= {\bf \widehat{\Pi}}(q^2) - {\bf \widehat{\Pi}}(\mu^2),
\ee
and from the corresponding  single subtraction dispersion relation
\bea
\Re e {\bf \widehat{\Pi}}_{{\mathrm{R}}}(q^2) &=&  \int_{0}^{\infty}\!
ds \left[\frac{1}{s-q^2}- \frac{1}{s-\mu^2}\right]
\frac{\Im m {\bf \widehat{\Pi}}(s)}{\pi}
\nonumber\\
&=& - bg^2 \ln \frac{q^2}{\mu^2}.
\eea
\indent
We emphasize that the above procedure furnishes an 
alternative way for constructing 
the gfp-independent PT Green's functions at one-loop, 
for {\it every} gauge-fixing scheme. 
Indeed, in our derivation we have solely
relied on the rhs of the OT, which we have rearranged in a well-defined way,
{\it after} having explicitly demonstrated its gfp-independence. The proof of
the gfp-independence of the rhs presented here is, of course, expected on
physical grounds, and it only relies on the use of WIs, triggered by the
longitudinal parts of the tree-level gluon propagators. 
Since the gfp-dependence at the level of the Feynman rules is 
carried entirely by the longitudinal parts of the gluon tree-level propagator,
its cancellation at the level of $\mathcal M$ proceeds exactly as described 
before Eq.~(\ref{TsF}). Obviously, the final step of reconstructing the real part 
from the imaginary by means of a (once subtracted)  
dispersion relation does not introduce any new gauge-dependences. 
\newline
\indent
In addition, in the context of theories with tree-level (spontaneous) 
symmetry breaking (such as the electroweak theory)
the lhs of the OT contains, in general, 
would-be Goldstone bosons or ghost fields, with gfp-dependent masses. 
Such contributions manifest themselves as 
unphysical cuts, \eg at $q^2=\xi M_{ W}^2$ for a $W$ propagator in 
the renormalizable $R_{\xi}$ gauges, and, eventually, as  
unphysical thresholds, at \eg $s=4\xi M_{ W}^2$. 
However, unitarity requires that these unphysical contributions
should vanish, as can be read off from the rhs of Eq.~(\ref{absorptive}). 
As we will see in detail in \secref{Applications - I}, this observation is of 
central importance when devising a gauge-invariant resummation formalism 
for resonant transition amplitudes~\cite{Papavassiliou:1995fq,Papavassiliou:1995gs,Papavassiliou:1997fn,Papavassiliou:1998pb}.

\newpage


\section{\seclab{PTBFM}The background field method and its correspondence with the PT}
\noindent
As  we  have  seen  in  the  previous  section,  in  the  conventional
formulation of gauge theories the Lagrangian $\mathcal L$-- consisting of
the  classical  term plus  the  gauge-fixing  and Faddeev-Popov  ghost
terms-- is no longer gauge  invariant, but rather BRST invariant. As a
consequence, off-shell  Green's functions satisfy complicated STIs
reflecting  BRST invariance.
In this context  we have seen how the PT implements
a rearrangement  of the  perturbative series, that  allows the
construction  of  {\it effective}  one-loop  Green's functions  which,
among several other important properties, satisfy naive, QED-like WIs.
\newline
\indent
It turns out that there exists a formal framework, 
known as the  background field method (BFM), which gives rise to  
Green's functions that satisfy automatically 
this last property (but not all others).
\newline
\indent
The BFM was initially introduced at the one-loop level
\cite{Dewitt:1967ub,Honerkamp:1972fd,KlubergStern:1974xv,Arefeva:1974jv,Dewitt:1967uc,Honerkamp:1971sh,Sarkar:1974db,Sarkar:1974ni,KlubergStern:1975hc,Hooft:1973us,Grisaru:1975ei}, 
and was generalized soon afterwards to higher orders~\cite{Hooft:1975vy,Abbott:1980hw,DeWitt:1980jv,Boulware:1980av,Capper:1982tf}. 
In the BFM one arranges 
things such that the explicit gauge invariance present at the level 
of the classical Lagrangian is retained even after the gauge-fixing 
and ghost terms have been added. 
Thus, the (formally defined) off-shell Green's functions of the BFM obey the naive 
WIs dictated by gauge invariance, 
exactly as the (diagrammatically defined) PT Green's functions. 
Notice however two important points: The  BFM Green's functions ({\it i}) depend explicitly on the (quantum) gfp, denoted by $\xi_{Q}$, and  
({\it ii}) obey the aforementioned WIs for every value of  $\xi_{Q}$.  Thus one encounters a situation very similar to 
that of QED: the photon-electron vertex and the electron self-energy depend 
explicitly on the gfp, and for every value of the 
gfp they satisfy the text-book WI of Eq.~(\ref{vertQED}).
The reason why the BFM has become so relevant in the study 
of the PT was the observation ~\cite{Denner:1994nn,Hashimoto:1994ct,Papavassiliou:1994yi} 
that for a very simple choice of  $\xi_{Q}$ the BFM Green's functions become identical to those of the PT. 
This particular value is the BFG, $\xi_{Q}=1$. 
\newline
\indent
The purpose of this section is twofold. First, 
we introduce the BFM quantization, discussing its advantages 
over the conventional quantization formalism. Second, we will establish (at the one-loop level) the important correspondence
between the PT and BFM Green's functions mentioned above.
Since this correspondence will accompany us for the rest of this report, 
we will pay particular attention in explaining the conceptual differences that distinguish the PT from the BFM 
(for a related general discussion see also \secref{intro}, as well as subsection~\ref{resum} for further physical arguments sharpening this distinction). 

\subsection{The background field method}
\noindent
Let us start by considering a theory describing a (scalar) field $\phi$ with an associated classical action 
\be
S[\phi]=\int\!d^4x{\mathcal L}[\phi].
\label{class_act}
\ee
The $S$-matrix of this theory can be derived from the knowledge of its corresponding Green's functions through the LSZ reduction formula (see below). The Green's functions of the theory can, in turn, be obtained by taking functional derivatives with respect to a suitable source $J$ of the generating functional
\be
Z[J]=\int[d\phi]\e^{i\left\{S[\phi]+J\cdot\phi\right\}},\qquad J\cdot\phi=\int\!d^4xJ(x)\phi(x).
\label{gen_func}
\ee 
The functional integral appearing in the equation above is performed over all the possible configurations of the field $\phi$.
The disconnected Green's functions of the theory are then defined through the relation
\be
\langle0\vert T(\phi\cdots\phi)\vert0\rangle=\int[d\phi](\phi\cdots\phi)\e^{iS[\phi]}=\left.\left(\frac1i\frac{\delta}{\delta J}\right)^nZ[j]\right|_{J=0}.
\ee
Since the disjoined pieces appearing in the Green's function  do not contribute to the $S$-matrix, it is more convenient to work with the connected Green's functions which are generated by taking functional derivatives with respect to the source $J$ of the generating functional $W[J]$ defined as
\be
\e^{iW[J]}=Z[J].
\ee
Finally, yet another simplification can be achieved by expressing connected Green's functions in terms of 1PI pieces, which are  generated by the effective action defined through the Legendre transform
\be
\Gamma[\overline{\phi}]=W[J]-J\cdot\overline{\phi},
\label{eff_act}
\ee
where
\be
\overline{\phi}=\frac{\delta W}{\delta J}.
\label{phi_bar}
\ee
Notice the difference between $\phi$ and $\overline{\phi}$; since
\be
\overline{\phi}=-i\frac1{Z[J]}\frac{\delta}{\delta J}Z[J]=\frac{\langle0\vert\phi\vert0\rangle_J}{\langle0\vert0\rangle_J},
\ee
we see that $\overline{\phi}$ corresponds to the vacuum expectation value of the field $\phi$ in the presence of a source $J$. In particular notice that the equation
\be
\frac{\delta\Gamma}{\delta\overline{\phi}}=-J,
\ee
corresponds to the quantum mechanical field equation for $\overline{\phi}$, which replaces the classical field equation
\be
\frac{\delta S}{\delta\phi}=-J,
\ee
in the quantized theory.
\newline
\indent
Thus, the key quantity to calculate in field theories 
is the effective action (\ref{eff_act}), because,  
once it is known, the $S$-matrix can be constructed through the 
LSZ reduction formula, \ie by stringing together trees of 
1PI Green's functions (thus generating the connected ones), 
amputating external propagators, putting external momenta on-shell,  
and adding appropriate external wave-functions renormalization factors.
\newline
\indent
At this stage the BFM can be seen as a convenient way of computing the effective action. 
Let us begin by defining a new generating functional $\widetilde{Z}$ by shifting the argument of 
the classical action $S[\phi]$ appearing in Eq.~(\ref{class_act}) by 
an arbitrary background field $\varphi$ independent of $J$:
\be
\widetilde{Z}[J,\varphi]=\int[d\phi]\e^{i\left\{S[\phi+\varphi]+J\cdot\phi\right\}}.
\label{gen_func_bck}
\ee
By analogy we can now go on and define $\widetilde{W}$ as
\be
\e^{i\widetilde{W}[J,\varphi]}=\widetilde{Z}[J,\varphi],
\ee
and, finally, the corresponding effective action
\be
\widetilde{\Gamma}[J,\varphi]=\widetilde{W}[J,\varphi]-J\cdot\widetilde{\phi},
\label{eff_act_bck}
\ee
with
\be
\widetilde{\phi}=\frac{\delta\widetilde{W}}{\delta J}.
\label{phi_tilde}
\ee
Therefore the background effective action constructed in Eq.~(\ref{eff_act_bck}) is a conventional effective action computed in the presence of a background field $\varphi$, 
and, as such, will give rise to 1PI Green's functions in the presence 
of this background field $\varphi$.
\newline
\indent
Let us now look for a connection between the generating functionals $\widetilde{Z}$ and $Z$. Shifting the variable of the functional integration in Eq.~(\ref{gen_func_bck}) through $\phi\to\phi-\varphi$, we arrive immediately at the relation
\be
\widetilde{Z}[J,\varphi]=Z[J]\e^{-iJ\cdot\varphi},
\ee
and thus we are led to
\be
\widetilde{W}[J,\varphi]=W[J]-J\cdot\varphi.
\ee
Differentiating this last equation with respect to $J$ and taking into account the definitions of  Eqs~(\ref{phi_bar}) and~(\ref{phi_tilde}) we find
\be
\widetilde{\phi}=\overline{\phi}-\varphi.
\ee
Finally, making use of   Eqs~(\ref{eff_act}) and~(\ref{eff_act_bck}) we find 
\be
\widetilde{\Gamma}[\widetilde{\phi},\varphi]=W[J]-J\cdot(\widetilde{\phi}+\varphi)=\Gamma[\overline{\phi}]=\Gamma[\widetilde{\phi}+\varphi].
\ee
As a special case of the above equation we can choose $\widetilde{\phi}=0$ to get
\be
\widetilde{\Gamma}[0,\varphi]=\Gamma[\varphi].
\label{vac_eff_act}
\ee   \indent   Notice    that   the   background   effective   action
$\widetilde{\Gamma}[0,\varphi]$      has     no      dependence     on
$\widetilde{\phi}$, so that it can  only generate 1PI vacuum graphs in
the  presence of  the background  field $\varphi$;  on the  other hand
Eq.~(\ref{vac_eff_act})  tells us  that  the effective  action of  the
theory  can be  obtained  precisely  by summing  up  all these  vacuum
diagrams. However, unless $\varphi$  is a very simple background field
(\eg a  constant), the calculation  of $\widetilde{\Gamma}[0,\varphi]$,
treating $\varphi$ exactly, is not possible. Thus, one needs to resort to
a perturbative  treatment of the background field  $\varphi$, in which
case the background field appearing in external lines is arbitrary, and
does not need to be specified at all.
\newline
\indent To pursue this  latter approach, one starts generating Feynman
rules from ${\mathcal L}[\phi+\varphi]$.  From this process there will
be two type of interaction  vertex emerging: ({\it i}) those involving
only $\phi$ fields, which must be used inside diagrams only, and ({\it
ii}) those  involving $\phi$ and  $\varphi$ fields, which ought  to be
used in order  to generate external lines. This, in  turn, means that in
the  BFM  one   will  calculate  the  same  diagrams   needed  in  the
conventional  formulation,   with  the  only   proviso  that  vertices
appearing inside loops might  have different Feynman rules compared to
those connecting external legs.

\subsection{Background field gauges}
\noindent
All the results discussed above for the simplified setting of a scalar field theory have analogs 
in non-Abelian gauge theories, with the obvious complications coming from the fact that, in the latter cases,  
one must choose a gauge fixing term; this, in turn, implies the appearance of the corresponding Faddeev-Popov ghost determinant. 
The generating functional for pure gluodynamics (since fermions play no role in the BFM construction they will be neglected in what follows) can be written as
\be
Z[J]=\int[dA]\mathrm{Det}\left[\frac{\delta{\mathcal F}^a}{\delta\theta^b}\right]\e^{i\left\{S_{\mathrm{I}}[A]+S_{\mathrm{GF}}[A;\xi]+J\cdot A\right\}}, \qquad J\cdot A=\int\!d^4xJ^a_\mu A^\mu_a.
\ee
The gauge invariant and gauge-fixing Lagrangian have been introduced in \secref{QCD_one-loop} and read
\be
S_{\mathrm{I}}[A]=\int\!d^4x\,{\mathcal L}^{A}_{\mathrm{I}}=-\frac14\int\!d^4x\,F^a_{\mu\nu}F^{\mu\nu}_a,
\qquad S_{\mathrm{GF}}[A;\xi]=\frac1{2\xi}\int\!d^4x\,{\mathcal F}^a{\mathcal F}^a,
\ee  
with ${\mathcal F}^a$ the gauge fixing function; finally, $\delta{\mathcal F}^a/\delta\theta^b$ represents the derivative of the gauge fixing function with respect to the infinitesimal gauge transformation of the gluon field, see Eq.~(\ref{QCD_gauge_trans}), with the determinant of this term expressing the result of the integral over the ghost and anti-ghost field variables of the Faddeev-Popov Lagrangian term, introduced in Eq.~(\ref{FPG_Lag}). 
\newline
\indent
The background field generating functional can be defined in an way completely analogous to the scalar case [see Eq.~(\ref{gen_func_bck})], namely\footnote{From now on we will indicate background fields with a caret. Notice however that we will indicate Green's functions containing background fields with a tilde, to distinguish them from the PT ones.}
\be
\widetilde{Z}[J,\widehat{A}]=\int[dA]\mathrm{Det}\left[\frac{\delta\widehat{\mathcal F}^a}{\delta\theta^b}\right]e^{i\left\{S_{\mathrm{I}}[A+\widehat{A}]+S_{\mathrm{GF}}[A;\xiQ]+J\cdot A\right\}},
\label{gen_func_bck_gauge}
\ee
where now $\widehat{\mathcal F}=\widehat{\mathcal F}(A,\widehat{A})$, with $\delta\widehat{\mathcal F}^a/\delta\theta^b$ represents the derivative of the gauge fixing function with respect to an infinitesimal gauge transformation
\be
\delta A^a_\mu=-\frac1g\partial_\mu\theta^a+f^{abc}\theta^b\left(\widehat{A}^c_\mu + A^c_\mu\right).
\label{gauge_trans_A}
\ee
At this point, all quantities defined in the scalar case can be also defined here. Moreover, since the relation 
between normal and background quantities can be found, as before, by shifting the integration variable in the path integral 
through \mbox{$A\to A-\widehat{A}$}, one finds in analogy with Eq.~(\ref{vac_eff_act}),
\be
\widetilde{\Gamma}[0,\widehat{A}]=\left.\Gamma[\overline{A}]\right|_{\overline{A}=\widehat{A}}.
\label{gauge_eff_act}
\ee
Notice that for arriving at this equation one has to write $\widetilde{Z}[J,\widehat{A}]$ in terms of $Z[J]$; this, as already said, can be easily done by shifting the integration variable. Thus, for Eq.(\ref{gauge_eff_act}) to be valid,  
if on the lhs the effective action is calculated using the gauge fixing function $\widehat{\mathcal F}(A,\widehat{A})$, 
then on the rhs we must use the gauge fixing function ${\mathcal F}= \widehat{\mathcal F}(A-\widehat{A},\widehat{A})$ (and the gauge fixing parameter $\xi=\xiQ$).
Then, while it is clear that the BFM will give rise to different Green's functions with respect to 
those appearing in the conventional formalism, the gauge independence of 
the physical observables, together with Eq.~(\ref{gauge_eff_act}), guarantee that one will finally obtain the same $S$-matrix.
\newline
\indent
The crucial feature that makes the BFM such an advantageous way of quantizing gauge theories is the following.
There exist special choices of the gauge fixing function $\widehat{\mathcal F}$ 
for which the form of the BFM effective action $\widetilde{\Gamma}[0,\widehat{A}]$ 
is severely restricted, being a gauge invariant functional of $\widehat{A}$, \ie invariant under the gauge transformations
\bea
\delta \widehat{A}^a_\mu&=& -\frac1g\partial_\mu\widehat{\theta}^a+f^{abc}\widehat{\theta}^b\widehat{A}^c_\mu,\label{gauge_trans_Ahat}\nonumber\\
\delta J^a_\mu&=&-f^{abc}\widehat{\theta}^bJ^c_\mu.
\label{gauge_trans_J}
\eea
Notice that the (infinitesimal) parameter of the gauge transformations above 
has been denoted by $\widehat\theta$, because it is different  
from the one appearing in the gauge transformations of the quantum field $A$, shown \eg in Eq.~(\ref{gauge_trans_A}). 
The (covariant) gauge fixing function that enforces the background gauge invariance is
\bea
\widehat{\mathcal F}^a&=&(\widehat{\mathcal D}^\mu A_\mu)^a\nonumber \\
&=&\partial^\mu A_\mu^a+gf^{abc}\widehat{A}^b_\mu A_c^\mu,
\label{BFMgff}
\eea
where $\widehat{\mathcal D}^\mu$ is the covariant background field derivative.
\newline
\indent
In order to prove the invariance of the effective action under the transformations~(\ref{gauge_trans_Ahat}) and~(\ref{gauge_trans_J}), 
we carry out the following change of variables in the functional integral appearing in Eq.~(\ref{gen_func_bck_gauge})
\be
A^a_\mu\to A^{'a}_\mu=A^a_\mu-f^{abc}\widehat{\theta}^b A^c_\mu=O^{ab}(\widehat{\theta})A^b_\mu,
\label{ch_of_var}
\ee
where
\be
O^{ab}(\widehat{\theta})=\delta^{ab}-f^{abc}\widehat{\theta}^c
\ee
is an orthogonal matrix representing a rotation by an infinitesimal amount $\widehat{\theta}$ 
in the vector space spanned by the generators of the $SU(N)$ gauge group. Thus it is clear that 
this change of variables leaves the integral measure invariant.
Also, on the one hand, since both  Eqs~(\ref{gauge_trans_J}) and~(\ref{ch_of_var}) represent adjoint group rotations, 
the term $J\cdot A$ is clearly invariant. On the other hand, one has that
\be
\int[dA]\e^{iS[A+\widehat{A}]}\to\int[dA']\e^{iS[A'+\widehat{A}+\delta(A+\widehat{A})]},
\ee
with
\be
\delta(A^a_\mu+\widehat{A}^a_\mu)=-\frac1g\partial_\mu\widehat{\theta}^a+f^{abc}\widehat{\theta}^b\left(\widehat{A}^c_\mu + A^c_\mu\right).
\ee
The latter represents a gauge transformation (with parameter $\widehat{\theta}$) of the $A+\widehat{A}$ field, 
so that also this part of the generating functional is invariant.
\newline 
As far as the gauge fixing term is concerned, one has that it transforms as follows
\bea
\widehat{\mathcal F}(\widehat{A}+\delta\widehat{A},A)&=&\widehat{\mathcal F}^a(\widehat{A},A)-f^{abc}\partial^\mu\left(\widehat{\theta}^bA^c_\mu\right)+gf^{abc}\widehat{\theta}^b\partial^\mu A^c_\mu+gf^{abc}f^{bde}\widehat{\theta}^d\widehat{A}^e_\mu A^\mu_c\nonumber \\
&=&\widehat{\mathcal F}^a(\widehat{A},A')+f^{abc}\widehat{\theta}^b\widehat{\mathcal F}^c(\widehat{A},A'),
\label{BFM_gff_trans}
\eea
where in the last step the Jacobi identity has been used, and $\widehat{\theta}^2$ terms have been discarded. Thus we can write
\be
\widehat{\mathcal F}^a(\widehat{A}+\delta\widehat{A},A)=O^\mathrm{T}_{ac}\widehat{\mathcal F}^c(\widehat{A},A'),
\ee
and therefore since $({\mathcal F}^a)^2$ is manifestly invariant under orthogonal rotations, the background gauge invariance of the gauge fixing term is evident. Finally for the derivative term $\delta\widehat{\mathcal F}^a/\delta\theta^b$ one has the relation~\cite{Pilaftsis:1996fh}
\bea
\frac{\delta}{\delta\theta^b}{\mathcal F}^a(\widehat{A}+\delta\widehat{A},A^\theta(\widehat{A}+\delta\widehat{A}))=
O^{T,ac}(\widehat{\theta})\frac{\delta}{\delta\widetilde{\theta}^b}{\mathcal F}^c(\widehat{A},A'^{\widetilde{\theta}}(\widehat{A}))O^{db}(\widehat{\theta}),
\label{BFM_FPdet}
\eea
where $\widetilde{\theta}^a=O^{ab}(\widehat{\theta})\theta^b$, and we have explicitly indicated the dependence of the variation 
of the quantum field $A$ on the background field $\widehat{A}$ and the infinitesimal parameter $\theta$.  
From the above identity follows that the determinant is also invariant, since the determinant of an orthogonal matrix is equal to 1. 
Thus, this concludes our proof of the gauge invariance of the background effective action. 
\newline
\indent 
Summarizing, the general idea behind the BFM is to first make a linear decomposition of the gauge field appearing 
in the classical action in terms of a background field, $\widehat{A}$, and the quantum field, $A$, which is the variable of integration in the path integral.
Then, using the Faddeev-Popov quantization method, one eliminates the unphysical degrees of freedom of the gauge field by breaking 
the gauge invariance of the classical Lagrangian through a gauge fixing condition, which is 
usually taken to be of covariant form, even if such a choice may not be unique 
(see the next subsection). Most importantly, 
it is possible to choose a gauge fixing condition that is invariant under the gauge transformations of the background field  $\widehat{A}$, 
so that the whole Lagrangian retains (background) gauge invariance with respect to the latter field, 
which only appears in external lines. The gauge symmetry is, however, explicitly broken by the quantum field $A$, which enters only in loops.

\subsubsection{\label{gbg}Generalized background gauges}
\noindent
An interesting question to ask is whether the gauge fixing function of Eq.~(\ref{BFMgff}) is unique. To address this issue, 
we start by noticing that, in order to get from the conventional $R_\xi$ gauge fixing function $\partial^\mu A^a_\mu$ 
to the covariant background equivalent  given in Eq.~(\ref{BFMgff}), it has been  sufficient to make the replacement
\be
\delta^{ab}\partial_\mu\to\widehat{D}^{ab}_\mu,
\label{BFM_replace}
\ee 
where $\widehat{D}^{ab}_\mu=\widehat{D}^{ab}_\mu(\widehat{A})=\delta^{ab}\partial_\mu+gf^{amb}\widehat{A}^m_\mu$. 
On the other hand, Eq.~(\ref{BFM_gff_trans}) 
tells us that the background covariant derivative transforms under the background gauge transformation~(\ref{gauge_trans_Ahat}) as
\be
\widehat{D}^{ab}_\mu(\widehat{A}+\delta\widehat{A})A^b_\nu=O^\mathrm{T}_{ab}\widehat{D}^{bc}_\mu(\widehat{A})A'^b_\nu, 
\label{bck_cov_dev_trans}
\ee
which, in turn, ensures that the corresponding Faddeev-Popov determinant transforms as in Eq.~(\ref{BFM_FPdet}).
In fact, we see that Eq.~(\ref{bck_cov_dev_trans}), together with the fact that the gauge fixing Lagrangian is given by the group space
scalar product of the corresponding function, ensures that $\widehat{\mathcal F}^a$ leaves $Z[J,\widehat{A}]$ invariant under background field gauge transformations. 
Thus, using Eq.~(\ref{BFM_replace}), all the non-covariant gauge fixing functions introduced in subsection~\ref{prolegomena} can be converted into background 
gauge fixing functions that preserve the background gauge invariance of the effective action.

\subsection{Advantages over the conventional formalism}

\subsubsection{Preliminaries: Green's function and $S$-matrix calculation in the BFM}
\noindent
The background gauge invariance of the effective action $\widetilde{\Gamma}[0,\widehat{A}]$, imposes  
a drastic restriction on the form of the 1PI Green's functions generated by taking functional 
derivative of the background effective action 
with respect to the background fields $\widehat{A}$. In fact, 
exactly as happens in the Abelian (QED) case, 
these functions are forced to satisfy naive WIs rather 
than the usual STIs associated with the non-Abelian character of the theory.    
\newline
\indent
BFM Green's functions are calculated starting from the shifted Lagrangian ${\mathcal L}_\mathrm{I}[A,\widehat{A}]$, 
the gauge fixing term ${\mathcal L}^{\mathrm{BFM}}_{\mathrm{GF}}=\frac1{2\xiQ}\widehat{\mathcal F}^a\widehat{\mathcal F}^a$, 
and the Faddeev-Popov determinant, which can be written in terms of an anti-commuting scalar field $c$, 
giving rise to the Faddeev-Popov ghost Lagrangian
\bea
{\mathcal L}^{\mathrm{BFM}}_{{\mathrm{FPG}}}&=&\partial^\mu\bar c^a \partial_\mu c^a+gf^{abc}(\partial^\mu\bar c^a)A^b_\mu c^c+gf^{abc}(\partial^\mu\bar c^a)\widehat{A}^b_\mu c^c-gf^{abc}\bar c^a\widehat{A}^b_\mu(\partial^\mu c^c)\nonumber \\
&-&g^2f^{abe}f^{cde}\bar c^a\widehat{A}^b_\mu(A_c^\mu+\widehat{A}^c_\mu)c^d.
\label{BFM_FPG}
\eea
Notice the appearance of a modified ghost sector: the interactions between ghosts and background gluons are very characteristic,  
consisting of a symmetric $\widehat{A}c\bar c$ ghost vertex and a completely new, four particle vertex, $\widehat{A}Ac\bar c$.
\newline
\indent
As discussed in the scalar case, vertices appearing inside loops 
contain only quantum fields $A$, while vertices involving the background field $\widehat{A}$ 
connect external lines. 
Notice that all the propagators ought to be those of the quantum fields, 
since gauge invariance is unbroken in the background sector, and therefore the $\widehat{A}$ propagator is not defined.
The complete set of Feynman rules for QCD in the BFM gauge are reported in \appref{Frules}. 
As a rule of thumb to remember what vertices we expect to have different Feynman rules,  
notice that the BFM covariant gauge fixing term is linear in the quantum fields $A$;
therefore, apart from vertices involving ghost fields, 
only vertices containing exactly two quantum fields can differ from the conventional ones. 
Thus, for example, the vertex $\widehat{A}AAA$ will have to lowest order the same Feynman rule as the conventional vertex $AAAA$. 
Despite the distinction between background and quantum fields, calculations in the BFM are, in general, simpler. 
This is particularly so in the BFG, where many vertices simplify considerably (see again \appref{Frules}).  
\newline
\indent
When calculating 
$S$-matrix elements from the BFM Feynman rules remember that 
fields inside loops (\ie fields irrigated by virtual momenta) are always quantum, 
while those irrigated by physical momentum transfers are background.
As a result, box diagrams are exactly the same 
as in the conventional case, \ie all fields are quantum. 
On the other hand,  self-energy and vertex diagrams 
are attached to the on-shell particles by background gluons, having quantum 
gluons and quarks inside their loops. Note also that, eventually,   
one must choose a gauge for the background fields $\widehat{A}$ as well, which is completely 
unrelated to the gauge used for the quantum fields. For example, the propagators inside the loops 
may be in the BFG, while the background propagators in the axial gauge. 
After the background gauge is fixed, the background field propagator is well-defined, and one can use 
it to build up strings of 1PI functions, thus generating the connected Green's functions. 
Finally, the $S$-matrix will be determined from the LSZ reduction formula. 

\subsubsection{Special transversality properties of the BFM\label{2lBFM}}

\begin{figure}[!t]
\begin{center}
\includegraphics[width=9cm]{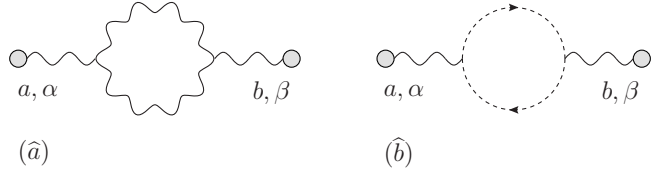}
\end{center}
\caption{\figlab{1loop_BFM}Feynman diagrams contributing to the one-loop background gluon self-energy. 
Gray circles on external lines represent background fields. 
Seagull contribution are not shown since, perturbatively, they do not contribute to the self-energy.}
\end{figure}
\noindent
To show what are the simplifications that the explicit preservation of gauge invariance implies at the  practical level, let us consider the calculation of the gluon two-point function.  The diagrams contributing to the one-loop background gluon self-energy are shown in \Figref{1loop_BFM} (seagull terms do not contribute due to the dimensional regularization result~\mbox{$\int_kk^{-2}=0$}). 
Choosing the background Feynman gauge (the $\beta$ function being, of course, a $\xi$-independent quantity) 
and using the BFM Feynman rules reported in \appref{Frules} one has
\bea
(\widehat{a})_{\alpha\beta}&=&g^2\frac{C_{A}}2\int_k\!\frac1{k^2(k+q)^2}
\widehat{\Gamma}_{\alpha\mu\nu}(q,-k-q,k)\widehat{\Gamma}_\beta^{\mu\nu}(q,-k-q,k),
\label{BFM_1}\nonumber\\
(\widehat{b})_{\alpha\beta}&=&-g^2C_{A}\int_k\!\frac1{k^2(k+q)^2}(2k+q)_\alpha(2k+q)_\beta,
\label{BFM_2}
\eea
Introducing, then, the function 
\be
f(q^2,\epsilon)=i\frac{C_{A}}3\frac{g^2}{(4\pi)^2}\Gamma\left(\frac\epsilon2\right)\left(\frac{q^2}{\mu^2}\right)^{-\frac\epsilon2},
\label{f_qsquare}
\ee
($\Gamma$ being the Euler gamma function) we get\footnote{Notice that from this result, and the fact that in the BFM $Z_g=Z_{\widehat{A}}^{-\frac12}$ one immediately obtains that $\beta^{(1)}=-\frac{11}3C_{A}\frac{g^3_{\mathrm{R}}}{(4\pi)^2}$. This has to be contrasted with the conventional formalism (\eg $R_\xi$ gauges), where in order to get $Z_g$ 
one must calculate the divergent parts of the gauge and ghost self-energies, as well as the ghost-gauge vertex~\cite{Politzer:1973fx,Gross:1973id,Jones:1974mm}.}
\be
\left\{
\begin{array}{l} 
({\widehat a})_{\alpha\beta} = 10 f(q^2,\epsilon)  q^2 P_{\alpha\beta}(q),
\\
(\widehat{b})_{\alpha\beta} =  f(q^2,\epsilon) q^2 P_{\alpha\beta}(q),\\
\end{array}\right.\qquad\Rightarrow\qquad
\widetilde\Pi_{\alpha\beta}^{(1)}(q)  = 11  f(q^2,\epsilon)  q^2P_{\alpha\beta}(q).
\ee
Notice that each of the contributions above is individually transverse, as a result of background gauge invariance. 
This is to be contrasted with \eg the $R_\xi$ result, where
\bea
(a)_{\alpha\beta}&=&g^2\frac{C_{A}}2\int_k\!\frac1{k^2(k+q)^2}\Gamma_{\alpha\mu\nu}(q,-k-q,k)\Gamma_\beta^{\mu\nu}(q,-k-q,k),
\nonumber\\
(b)_{\alpha\beta}&=&-g^2C_{A}\int_k\!\frac1{k^2(k+q)^2}k_\alpha(k+q)_\beta,
\eea
and $(a)$ and ($b$) are the diagrams corresponding to those shown in \Figref{1loop_BFM}, when the external gluons are quantum gluons. 
Carrying out the integrals, we obtain
\be
\left\{
\begin{array}{l}
(a)_{\alpha\beta} = 
 \frac{1}{4}f(q^2,\epsilon) \left( 19 q^2 g_{\alpha\beta} - 22  q_{\alpha}q_{\beta}\right)
\nonumber\\
(b)_{\alpha\beta} =
  \frac{1}{4}f(q^2,\epsilon) \left(q^2 g_{\alpha\beta} + 2   q_{\alpha}q_{\beta}\right)
\end{array}
\right.
\qquad\Rightarrow\qquad  
 \Pi_{\alpha\beta}^{(1)}(q) = 
5 f(q^2,\epsilon) q^2P_{\alpha\beta}(q).
\ee
Thus, while the sum of the two diagrams 
results in a transverse one-loop gluon self-energy  (as it should), 
the individual diagrams are not transverse. 
\begin{figure}[!t]
\bce
\includegraphics[width=12cm]{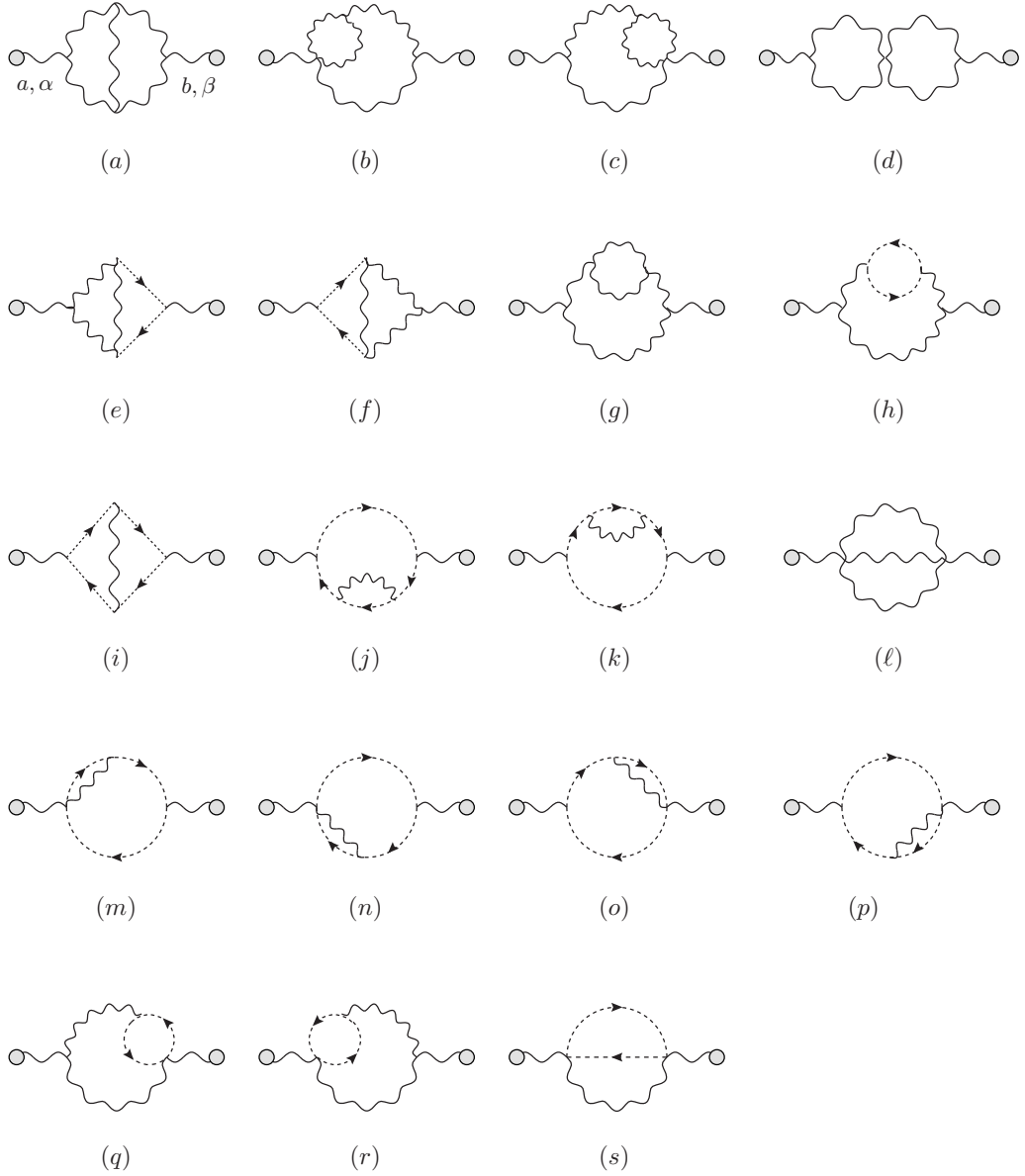}
\ece
\caption{\figlab{BFM_2l}The Feynman diagrams contributing to the  BFM
two-loop gluon self-energy ${\widetilde\Pi}_{\alpha\beta}^{(2)}$.}
\end{figure}
\newline
\indent
At two loops the diagrams to be calculated are shown in \Figref{BFM_2l}.
 Let us start by noticing that 
out of these diagrams one can form combinations 
that corresponds to the one-loop dressed gluonic and ghost contributions, 
and two-loop (dressed) gluonic and ghost contributions. 
Using the notation of~\Figref{ghatghat_SDE}, one has respectively
\bea
& &\left[(d_1)+(d_2)\right]^{(2)}=(a)+\frac12[(b)+(c)]+(d)+(g)+(h)+\frac12[(q)+(r)],\nonumber\\
& &\left[(d_3)+(d_4)\right]^{(2)}=(e)+(f)+(i)+(j)+(k)+\frac12[(m)+(n)+(o)+(p)],\nonumber\\
& & \left[(d_5)+(d_6) \right]^{(2)}=\frac12[(b)+(c)]+(\ell),\nonumber\\
& & \left[(d_7)+(d_8)+(d_9)+(d_{10}) \right]^{(2)}=\frac12[(q)+(r)]+(s)+\frac12[(m)+(n)+(o)+(p)],
\eea
and focussing on the divergent parts of the diagrams, we obtain~\cite{Abbott:1980hw}
\bea
& & \left[(d_1)+(d_2)\right]_{\alpha\beta}^{(2)}=i\frac{g^4C^2_A}{(4\pi)^4} \frac7{2\epsilon^2}\left(1+\frac{43}{12}\epsilon-\rho\epsilon\right)q^2P_{\alpha\beta}(q),\label{1ld_g}\nonumber\\
& & \left[(d_3)+(d_4)\right]_{\alpha\beta}^{(2)}=-i\frac{g^4C^2_A}{(4\pi)^4} \frac1{2\epsilon^2}\left(1+\frac{3}{4}\epsilon-\rho\epsilon\right)q^2P_{\alpha\beta}(q),\label{1ld_c}\nonumber\\
& & \left[(d_5)+(d_6) \right]_{\alpha\beta}^{(2)}=-i\frac{g^4C^2_A}{(4\pi)^4} \frac9{4\epsilon^2}\left(1+\frac{31}{12}\epsilon-\rho\epsilon\right)q^2P_{\alpha\beta}(q),\label{2ld_gl}\nonumber\\
& & \left[(d_7)+(d_8)+(d_9)+(d_{10}) \right]_{\alpha\beta}^{(2)}=-i\frac{g^4C^2_A}{(4\pi)^4} \frac3{4\epsilon^2}\left(1+\frac{9}{4}\epsilon-\rho\epsilon\right)q^2P_{\alpha\beta}(q),\label{2ld_c}
\eea
with $\rho=\gamma_{\mathrm{E}}-\ln4\pi+\ln(q^2/\mu^2)$. Then, adding up all the contributions, we get
\be
\widetilde\Pi^{(2)}_{\alpha\beta}(q)=\sum_{i=1}^{10}(d_i)^{(2)}_{\alpha\beta}=i\frac{g^4C^2_A}{(4\pi)^4}\frac{14}{3\epsilon}q^2P_{\alpha\beta}(q).
\ee
Notice that, as happened in the one-loop case, the contribution of each of the four subgroups is {\it individually} transverse. 
This property is certainly not accidental; in fact,  
as we will see in \secref{PT_SDEs}, it is valid to all orders. Its validity has been established for the first time in 
\cite{Aguilar:2006gr}, and is a direct consequence of the 
 linear WIs satisfied by the (fully-dressed) vertices entering in the SD expansion of the background gluon self-energy.
As we will explain there in detail, this grouping of diagrams into individually transverse subsets  
has profound implications, constituting the cornerstone of the gauge-invariant truncation scheme implemented by the PT.

\subsection{The pinch technique/background Feynman gauge correspondence}
\noindent
The key observation~\cite{Denner:1994nn,Hashimoto:1994ct} that immediately suggests a connection 
between the PT and BFM Green's functions is related to the form of 
the $\xiQ$-dependent
tree-level BFM vertex \linebreak ${\widehat A}_{\alpha} (q) A_{\mu}(k_1) A_{\nu}(k_2)$, 
denoted by $\tilde{\Gamma}_{\alpha\mu\nu}^{\xiQ}(q,k_1,k_2)$ (see the BFM Feynman rules of \appref{Frules}).  
Specifically, using the PT decomposition of Eq.~(\ref{decomp}) for
the standard tree-level three-gluon vertex $\Gamma_{\alpha\mu\nu}(q,k_1,k_2)$, we find that   
\bea
\tilde{\Gamma}_{\alpha\mu\nu}^{(\xiQ)}(q,k_1,k_2) &=& \Gamma^{\mathrm{F}}_{\alpha\mu\nu}(q,k_1,k_2)
+\left(\frac{\xiQ-1}\xiQ\right)\Gamma^{\mathrm{P}}_{\alpha\mu\nu}(q,k_1,k_2),
\label{BFM_PTdec}
\nonumber\\
&=&  \Gamma_{\alpha\mu\nu}(q,k_1,k_2) - \frac{1}{\xiQ} \Gamma^{\mathrm{P}}_{\alpha\mu\nu}(q,k_1,k_2).
\label{BFM_PTdec2}
\eea
Evidently, at $\xiQ=1$ we have that 
\be
\tilde{\Gamma}_{\alpha\mu\nu}^{(\xiQ=1)}(q,k_1,k_2)\equiv\Gamma^{\mathrm{F}}_{\alpha\mu\nu}(q,k_1,k_2).
\ee
Given that, in addition, at $\xiQ=1$ the 
longitudinal parts of the gluon propagator vanish also, one realizes that at this point there is nothing there 
that could pinch. Thus, ultimately, the BFG is singled out because of 
the total absence, in this particular gauge, of any pinching 
momenta. 
\newline
\indent
In what follows we will only prove the PT/BFG correspondence by means of  
explicit calculations at the one-loop level, following the aforementioned original articles. 
An all-order, more profound proof 
of this correspondence will be  postponed until \secref{beyond_1l}; there  
we will show that, as a result of the BRST symmetry, the PT/BFG correspondence persists 
to all orders in perturbation theory. Finally, in \secref{PT_SDEs}, the proof will be generalized to 
the non-perturbative case of the SDEs.
\newline
\indent
The case of the gluon self-energy is almost immediate: 
simply compare the two terms on the rhs of  Eq.~(\ref{propexp})   
with the two terms given  
in  Eqs~(\ref{BFM_1}). Evidently the PT and BFG gluon self-energies 
are identical at one loop.  
\begin{figure}[!t]
\bce
\includegraphics[width=12cm]{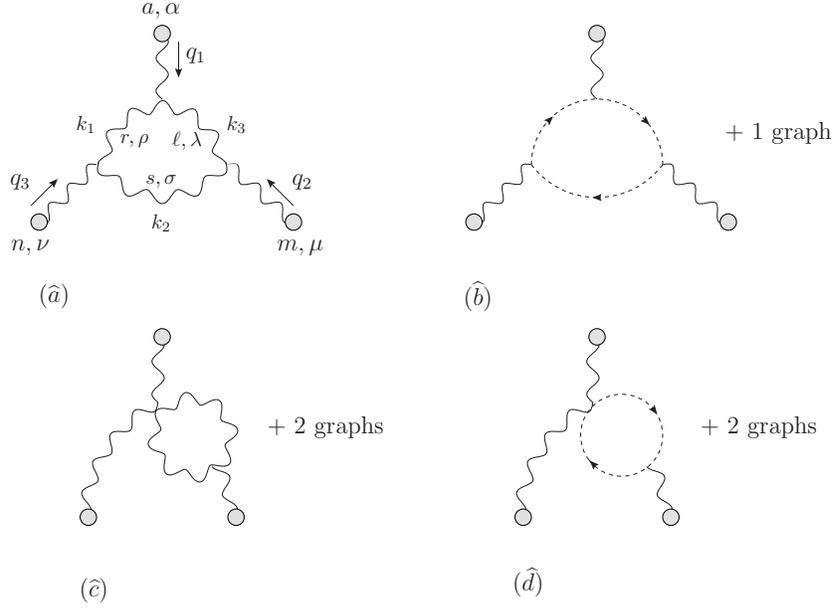}
\ece
\caption{\figlab{1l_3gvert_BFM_diag}One-loop diagrams contributing to the 
three-gluon vertex in the BFM. Diagrams $(\widehat{c})$ carry a $\frac12$ symmetry factor.}
\end{figure}
\newline
\indent 
Then, let us consider the case of the three-gluon vertex~\cite{Hashimoto:1994ct}. 
The one-loop diagrams contributing to the BFM three-gluon vertex 
(i.e., with three incoming background gluons)  are shown in \Figref{1l_3gvert_BFM_diag}; there  
we also fix the conventions for momenta, Lorentz, and color indices, used in what follows.   
From diagram $(\widehat{a})$ one has
\begin{equation}
(\widehat{a})=- \frac i2 g^{3} C_{A}f^{amn}
\int _{k_1}\!\frac1{k^2_1 k^2_2 k^2_3}\tilde{\Gamma}^{(\xiQ=1)}_{\alpha\lambda\rho}(q_1,k_3, -k_1)
\tilde{\Gamma}^{(\xiQ=1)}_{\mu\sigma\lambda}(q_2,k_2,-k_3)
\tilde{\Gamma}^{(\xiQ=1)}_{\nu\rho\sigma}(q_3,k_1, -k_2).
\end{equation}
For diagram $(\widehat{b})$, and the one with the ghost charge running in the opposite direction, we have instead
\begin{equation}
(\widehat{b}) =
ig^{3}C_{A}f^{amn}\int\!\frac1{k^2_1 k^2_2 k^2_3}(k_1+k_3)_\alpha(k_2+k_3)_\mu(k_1+k_2)_\nu.
\end{equation}
For diagrams $(\widehat{c})$, and the two other possible diagrams of the same type, 
we get (using the BFG for the four-gluon vertex with two background legs)
\bea
(\widehat{c})&=&4g^3C_{A}f^{amn}\left(g_{\alpha\nu}q_{1\mu}-g_{\alpha\mu}q_{1\nu}\right)\int_{k}\!\frac1{k^2(k+q_1)^2}
\nonumber\\
&+&4g^3C_{A}f^{amn}\left(g_{\mu\nu}q_{2\alpha}-g_{\alpha\mu}q_{2\nu}\right)\int_{k}\!\frac1{k^2(k+q_2)^2}\nonumber \\
&+&4g^3C_{A}f^{amn}\left(g_{\alpha\nu}q_{3\mu}-g_{\mu\nu}q_{3\alpha}\right)\int_{k}\!\frac1{k^2(k+q_3)^2}.
\eea
Finally, diagram $(\widehat{d})$ (and the other two similar diagrams) 
turns out to be zero at this order, due to group-theoretical identities for the structure constants such as
\begin{equation}
 f^{ead} ( f^{dbx} f^{xce}+f^{dcx}f^{xbe}) = 0.
\end{equation}
Adding the contributions found,  notice that the sum $(\widehat{a})+(\widehat{b})$ 
coincides with the term $\widehat{N}_{\alpha\mu\nu}$ of Eq.~(\ref{PTNterm}), while $(\widehat{c})$ 
coincides exactly with $\widehat{B}_{\alpha\mu\nu}$ of Eq.~(\ref{PTBterm}). 
Thus the BFM result matches the expression of Eq.(\ref{c3glvert-PT}), obtained by the (intrinsic) PT. 
\newline
\indent
The calculation of the one-loop four-gluon vertex in the BFM Feynman gauge has been carried out in~\cite{Hashimoto:1994ct}; 
again it was found to coincide with the one constructed through the PT algorithm.

\subsection{\label{PT/BFG:conceptual}The pinch technique/background Feynman gauge correspondence: conceptual issues}
\noindent
While it is a remarkable and extremely useful fact that the one-loop PT Green's functions 
can be calculated in the BFG, particular care is needed for the correct interpretation 
of this correspondence. 
\newline
\indent
First of all,  
the PT is a way of enforcing gauge independence (and
several  other physical properties, such as unitarity and analyticity)  
on off-shell  Green's functions, 
whereas the BFM,  in a general gauge, is  not.  This is reflected  in the fact
that  the BFM $n$-point  functions are  gauge-invariant, in  the sense
that they  satisfy (by construction)  QED-like WIs, but are  {\it not}
gauge-independent, \ie  they depend explicitly on  $\xi_{Q}$.  
For example, the BFM gluon self-energy at one loop is given by 
\be
\tilde{\Pi}^{(\xiQ)}_{\alpha\beta}(q) = \tilde{\Pi}^{(\xiQ=1)}_{\alpha\beta}(q) + \frac{i}{4(4\pi)^2}g^2C_A
(1-\xiQ)(7+\xiQ) q^2{P}^{\alpha\beta}(q).
\label{gfo}
\ee
Had the
BFM  $n$-point functions  been $\xi_{Q}$-independent,  in  addition to
being  gauge-invariant,  there  would   be  no  need  for  introducing,
independently, the PT. 
\newline
\indent
We emphasize that the objective of  the PT
construction is not to derive  diagrammatically the BFG, but 
rather to exploit the underlying BRST symmetry in order to 
expose a large number of cancellations, and eventually define 
gauge-independent Green's functions satisfying Abelian WIs.
Thus, that the PT Green's functions can also be calculated in the
BFG always needs  a very extensive demonstration.  
Therefore, the correspondence must be  
verified at the end of the PT construction and should not be  
assumed beforehand. 
\newline
\indent
Moreover, the  $\xi_{Q}$-dependent BFM Green's functions are 
{\it not} physically equivalent. This is best seen  
in theories with 
spontaneous symmetry breaking: the dependence of the 
BFM Green's functions on $\xi_Q$ gives rise 
to {\it unphysical} thresholds inside these Green's functions 
for $\xi_Q \neq 1$, a fact
which limits their usefulness for resummation purposes
(this point will be studied in detail in subsection 5.4).
Only the case of the BFG
is free from unphysical poles; that is because  
then (and only then) the BFM results collapse to the physical 
PT Green's functions.
\newline
\indent
It is also important to realize that the PT construction  goes through
unaltered under  circumstances  where  the BFM
Feynman rules cannot even be  applied.  Specifically, if instead of an
$S$-matrix element  one were to consider a  different observable, such
as a current correlation function  or a Wilson loop (as was in
fact    done    by    Cornwall    in    the    original    formulation
\cite{Cornwall:1981zr}, and more recently in \cite{Binosi:2001hy}), one
could not start  out using the background Feynman  rules, because {\it
all} fields  appearing inside the  first non-trivial loop  are quantum
ones. Instead, by following the PT rearrangement inside these physical
amplitudes the unique PT answer emerges again.
\newline
\indent
Perhaps the most compelling fact 
that demonstrates that the PT and the BFM 
are intrinsically two completely disparate  methods is 
that one can apply the PT within the BFM.   
Operationally, this is 
easy to understand: away from $\xi_Q =1$ even in the BFM there 
are longitudinal (pinching momenta) that will initiate the 
pinching procedure. Thus,  
one starts out with the $S$-matrix written with the BFM Feynman rules using a general $\xiQ$, and  
applies the PT algorithm as in any other gauge-fixing scheme; one will recover again the 
unique PT answer for all Green's functions involved (\ie will get projected dynamically to  $\xiQ=1$).

\subsubsection{\label{PTinBFM} Pinching within the background field method}
\noindent
Let us study in some detail how  pinching works  within the BFM.
It is expeditious to organize this calculation 
using as reference point 
the corresponding one-loop construction 
presented in the the previous section, in the context of the $R_{\xi}$ gauges.  
The reason is that a great deal of the results needed here can be recovered directly 
from the analysis of the one-loop pinching in subsection~\ref{PT1}, simply 
by setting $\xi \to \xiQ$. 
\newline
\indent
To begin with, the box diagrams in the BFM are identical to those of the $R_{\xi}$, 
shown in \Figref{1l_box_pinch_contrib}, with the trivial 
replacement $\xi \to \xiQ$. Therefore, their pinching proceeds  
exactly as described in  \ref{PT1}, and the results are precisely those  
found in  subsection~\ref{ptbox},  Eqs~(\ref{box1}) and~(\ref{box2}), with $\xi \to \xiQ$.
\newline
\indent
We then turn to the vertex graphs. The topologies are the same as in the  $R_{\xi}$ gauges.
The Abelian graph, together with the external leg 
corrections are again identical, with the replacement $\xi \to \xiQ$.
The non-Abelian graph of \Figref{1l_vertex_pinch_contrib}, however, needs particular 
treatment, because  
the $\xi_Q$ dependent BFM three-gluon vertex does not coincide with the 
conventional three-gluon vertex (of the $R_{\xi}$ gauge). 
In fact, as we can see from Eq.~(\ref{BFM_PTdec2}), 
the two vertices differ 
by an amount proportional to $\Gamma^{\mathrm{P}}$. 
So, we will use Eq.~(\ref{BFM_PTdec2}) inside the non-Abelian graph: the first term 
converts the graph into its conventional counterpart 
(with  $\xi \to \xiQ$ for the gluon propagators in the loop); the second term 
is  purely pinching in nature, and we will track down its effect 
[Note that one could equally well employ (\ref{BFM_PTdec}) directly, as was done in~\cite{Papavassiliou:1994yi}, 
but then one could not use the results of the $R_{\xi}$ so straightforwardly].
\begin{figure}[!t]
\begin{center}
\includegraphics[width=10cm]{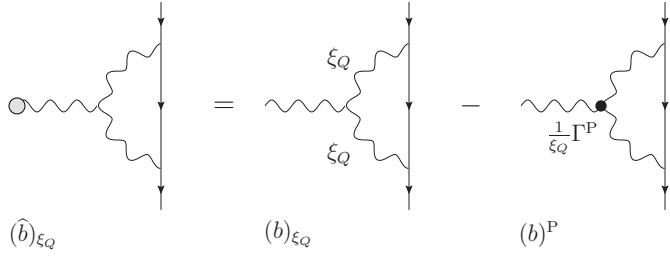}
\end{center}
\caption{\figlab{pinch_gen_xiQ}Relation between the one loop non-Abelian BFM vertex, 
and the standard non-Abelian graph in the $R_\xi$ gauge (with the substitution $\xi\to\xiQ$), for a general value of the gfp parameter $\xi_{Q}$. The difference is a pinching contribution.}
\end{figure}
\newline
\indent
The term $\Gamma^{\mathrm{P}}_{\alpha\mu\nu}(q,k_1,k_2)$, 
when multiplied by the two gluons inside the non-Abelian graph, leads to  
the expression
\be
\frac{1}{\xiQ}\Gamma_{\alpha}^{\mathrm{P}\,\mu\nu}(q,k_1,k_2)\Delta_{\mu\rho}^{(\xiQ)}(k_1) \Delta_{\nu\sigma}^{(\xiQ)}(k_2)
=
\frac{k_{1\rho}}{k_1^2}\Delta_{\alpha\sigma}^{(\xiQ)}(k_2)-\frac{k_{2\sigma}}{k_2^2}\Delta_{\alpha\rho}^{(\xiQ)}(k_1),
\label{GPT_struct}
\ee 
which, after pinching, generates propagator-like terms. In particular, 
the original non-Abelian vertex graph, $(\widehat{b})_{\xiQ}$, can be written as  (\Figref{pinch_gen_xiQ})
\bea
(\widehat{b})_{\xiQ} &=&(b)_{\xiQ}+(b)^{\mathrm{P}}\nonumber \\
&=&(b)_{\xiQ} + g^2C_{A}\delta^{ab}\int_k\left[-\frac{g_{\alpha\beta}}{k^2(k+q)^2}+
\lambda_{Q}\frac{k_\alpha k_\beta}{k^4(k+q)^2}\right] [g\gamma^\beta t^b],
\label{vbrd1}
\eea
where $(b)_{\xiQ}$ denotes the standard non-Abelian graph in the $R_\xi$ gauge when $\xi\to\xiQ$. 
Then, we can convert $(b)_{\xiQ}$ directly to $(\widehat{b})_{\xiQ=1}$, \ie 
the PT answer,  following  exactly the procedure described in subsections~\ref{qgv} and~\ref{PT3}; 
in this way the total pinching contribution coming from the BFM non-Abelian vertex is
\be
(\widehat{b})_{\xiQ} \to \Pi^{\alpha\beta}_\mathrm{nav}(q,\lambda_Q)+
 \lambda_{Q} g^2C_{A}q^2P^{\beta\mu}(q)\int_k 
\frac{k^\alpha k_\mu}{k^4(k+q)^2}. 
\label{vbrd2}
\ee
After these observations, it is easy to determine 
the total propagator-like pinching contribution, $\tilde\Pi^{\alpha\beta}_{{\rm P}}(q,\lambda_Q)$, 
that should be added to the BFM self-energy  
$\tilde{\Pi}^{\xiQ}_{\alpha\beta}(q)$ (following the universal PT rules): one has
\bea
\tilde\Pi^{\alpha\beta}_{{\rm P}}(q,\lambda_Q) &=&
\Pi_{{ {\mathrm box}}}^{\alpha\beta}(q,\lambda_{Q}) +
2 \left[\Pi_{{ {\rm nav}}}^{\alpha\beta}(q,\lambda_{Q}) 
+\Pi_{{ {\rm nav}}}^{\alpha\beta}(q,\lambda_{Q})\right]
+4 \Pi_{{ {\rm qse}}}^{\alpha\beta}(q,\lambda_{Q}) \nonumber \\
&+&2\lambda_{Q} g^2C_{A}q^2P^{\beta\mu}(q)\int_k 
\frac{k^\alpha k_\mu}{k^4(k+q)^2}\nonumber \\
&=& -\Pi^{\alpha\beta}_{{ {\rm gse}}}(q,\lambda_Q)+2\lambda_{Q} g^2C_{A}q^2P^{\beta\mu}(q)\int_k 
\frac{k^\alpha k_\mu}{k^4(k+q)^2},
\label{vbrd3}
\eea
where the last step is  by virtue of Eq.~(\ref{tgc}). Thus, 
\begin{eqnarray}
\tilde\Pi^{\alpha\beta}_{{\mathrm P}}(q,\lambda_Q)&=&\lambda_Q g^2  C_{A} q^2\left\{
-\frac{\lambda_Q}{2}q^2 
{P}^{\alpha\mu}(q){P}^{\beta\nu}(q) \int_k\!
\frac{k_{\mu} k_{\nu}}{k^4 (k+q)^4}\right. \nonumber \\
 &+&\left.\left[q^2 {P}^{\alpha\beta}(q)\! \int_k \frac{1}{k^2 (k+q)^4} + 
4 {P}^{\beta\mu} (q) 
\int_k \frac{k^{\alpha}k_{\mu} }{k^4 (k+q)^2}
- {P}^{\alpha\beta}(q)\! \int_k \frac{1}{k^4} \right]\right\}.\nonumber \\
\label{vbrd4}
\end{eqnarray}
Using the exact relation
\be
{P}^{\beta\mu} (q) \int_k  \frac{4 k^{\alpha}k_{\mu} }{k^4 (k+q)^2} 
= {P}^{\alpha\beta} (q) \int_k \frac{1}{k^2 (k+q)^2},
\label{vbrd5}
\ee
we finally find that 
\be
\tilde\Pi^{\alpha\beta}_{{\rm P}}(q,\lambda_Q) = 
- \lambda_Q g^2 C_{A} q^2 \left[
\frac{\lambda_Q}{2} q^2 {P}^{\alpha\mu}(q){P}^{\beta\nu}(q) \int_k\!
\frac{k_{\mu} k_{\nu}}{k^4 (k+q)^4} 
+ {P}^{\alpha\beta}(q) \int_k\! \frac{2q\cdot k}{k^4 (k+q)^2}\right]. 
\label{rc}
\ee
Note that the result is ultraviolet (UV) finite, as expected. 
After carrying out the integration, we obtain  
\be
\tilde\Pi^{\alpha\beta}_{{\rm P}}(q,\lambda_Q) = -\frac{i}{4(4\pi)^2}g^2C_A
(1-\xiQ)(7+\xiQ) q^2{P}^{\alpha\beta}(q),
\label{vbrd7}
\ee
which, when added to $\tilde{\Pi}^{(\xiQ)}_{\alpha\beta}(q)$ given in Eq.~(\ref{gfo}) will give 
$\tilde{\Pi}^{(\xiQ=1)}_{\alpha\beta}(q)$, as announced. 

\subsection{Generalized pinch technique\label{GPT}}
\noindent
As we have seen in detail, 
the PT is an algorithm that gives rise to the 
same unique Green's functions, regardless of the 
gauge-fixing scheme one starts out from. 
Thus, irrespective of the starting point, the  
PT projects one dynamically to the BFG. 
A question one may naturally ask at this point is 
the following: could we devise a PT-like procedure 
that would project us to some other 
value of the background gfp  $\xiQ$? 
Or, going one step further, could one rearrange the Feynman 
graphs is such a way as to be projected 
to the generalized background gauges introduced in subsection~\ref{gbg}?
As was shown by Pilaftsis~\cite{Pilaftsis:1996fh}, such a 
construction is indeed possible; the 
systematic algorithm that accomplishes  
this is known as the  {\it generalized}  pinch  technique (GPT).
The GPT essentially  
modifies the  starting point of the PT algorithm, namely  Eq.~(\ref{decomp}), distributing  
differently the longitudinal momenta between (the now modified) $\Gamma_{\alpha \mu \nu}^{{\rm F}}$
and $\Gamma_{\alpha \mu \nu}^{{\rm P}}$ type of terms. 
\newline
\indent
As explained by the author of ~\cite{Pilaftsis:1996fh},  
the GPT represents a  fundamental departure from  the primary aim of  the PT, 
which is to construct gfp-independent off-shell Green's functions. The GPT, instead,  
deals exclusively with gfp-dependent Green's functions, 
with all the pathologies that this dependence entails. 
Nonetheless, it is certainly useful
to have a method that allows us to move systematically from 
one gauge-fixing scheme to another, at the level of individual Green's functions.
In addition to the possible applications mentioned in ~\cite{Pilaftsis:1996fh}, 
we would like to emphasize the usefulness of the GPT in truncating 
gauge-invariantly (\ie maintaining transversality) sets of 
SDEs written in gauges other than the Feynman gauge
(see \secref{PT_SDEs}). This possibility becomes particularly relevant, 
for example,  when one attempts to compare SDE predictions with 
lattice simulations, carried out usually in the Landau gauge.
\newline
\indent
So, let us suppose, for starters, that we want to devise an algorithm that 
will take us from the $R_{\xi}$ gluon self-energy, calculated  
at   $\xi = \xi_0$,  to  the corresponding  BFM self-energy, calculated at the same $\xiQ =  \xi_0$ 
(for example, say we want to go from the normal Yennie gauge, $\xi=3$, 
to the BFM  Yennie gauge,  now at $\xiQ=3$). 
\newline
\indent
It is clear that in this case the box diagrams in both schemes are automatically identical. 
So, what one should focus on is the one-loop vertex. What we want to do is get from the 
conventional vertex graph (with a normal ($\xi$-independent!)
three-gluon vertex, and gluon propagators written  at $\xi = \xi_0$) to the 
corresponding BFM graph (with the $\xiQ$-dependent three-gluon vertex, and 
 gluon propagators written  at $\xiQ = \xi_0$), and hope that the remainder is a purely 
propagator-like piece. Then, the solution is almost obvious: one must 
simply use Eq.~(\ref{BFM_PTdec2}), with the term proportional to 
$\Gamma^{\mathrm{P}}$ moved to the lhs, \ie 
\be
\Gamma_{\alpha\mu\nu}(q,k_1,k_2) = \tilde{\Gamma}_{\alpha\mu\nu}^{(\xi_0)}(q,k_1,k_2) + 
\frac{1}{\xi_0} \Gamma^{\mathrm{P}}_{\alpha\mu\nu}(q,k_1,k_2).
\label{GPTdecomp}
\ee
In fact, this particular decomposition predates the GPT by two decades;
it appears for the first time in Eq.(4.4) of \cite{Cornwall:1976ii}, 
and has also been employed by Haeri in \cite{Haeri:1988af}. 
It was essentially motivated by the observation that the first term, corresponding to $\Gamma^{{\rm F}}$ in the usual PT procedure, 
satisfies precisely the correct generalization of the WI given in (\ref{WI2B}), namely   
\be
q^\alpha\Gamma_{\alpha\mu\nu}^{(\xi_0)}(q,k_1,k_2)=i\left\{\Delta_{(\xi_0)}^{(0)-1}(k_1)
-\Delta_{(\xi_0)}^{(0)-1}(k_2)\right\}_{\mu\nu},
\label{new_WI3g_PT}
\ee
with the inverse propagator given in Eq.~(\ref{inv_prog}). This property, in turn, guarantees 
that the resulting one-loop vertex satisfies the correct Abelian WI (see below).  
Evidently, Eq.~(\ref{GPTdecomp}) reduces to the usual PT decomposition of Eq.~(\ref{decomp}) when $\xi_0=1$.
\newline
\indent
Now, the first term on the rhs, when inserted into the original non-Abelian graph,  
gives exactly the corresponding graph in  the BFM (at $\xiQ =  \xi_0$); of course,  
the Abelian graphs and the external leg corrections are identical. Thus, one  eventually
obtains  the BFM one-loop gluon-quark vertex, 
$\tilde{\Gamma}_\alpha^{(\xiQ)}(p_1,p_2)$,  at $\xiQ =  \xi_0$.
As is known from the general discussion on the formal properties of the 
BFM Green's function, or as can be demonstrated explicitly (at one-loop) using Eq.~(\ref{new_WI3g_PT}), 
$\tilde{\Gamma}_\alpha^{(\xi_0)}(p_1,p_2)$ satisfies a 
naive QED-like WI, namely 
\be
q^\alpha\tilde{\Gamma}_\alpha^{(\xi_0)}(p_1,p_2)=g\left\{\widehat\Sigma^{(\xi_0)}(p_1) -  \widehat\Sigma^{(\xi_0)}(p_2)\right\},
\label{GPT_qg_WI}
\ee
where $\widehat\Sigma^{(\xi_0)}$ is the GPT 
quark self-energy coinciding with the usual quark self-energy when evaluated 
in the gauge $\xiQ=\xi_0$.
\newline
\indent
The final step is to determine the (exclusively  propagator-like)
contributions of the remaining term  
$\Gamma^{\mathrm{P}}$, coming from Eq.~(\ref{GPTdecomp}). 
In fact, $\Gamma^{\mathrm{P}}$ will trigger first 
Eq.~(\ref{GPT_struct}), and then pinch, as usual. Once the 
propagator-like contributions have been alloted to $\Pi_{\alpha\beta}^{(\xi_0)}$, the conventional 
$R_{\xi}$ one-loop self-energy at $\xi=\xi_0$, we will obtain 
$\tilde\Pi_{\alpha\beta}^{(\xi_0)}$, namely the BFM one-loop self-energy at $\xiQ= \xi_0$, thus 
concluding the construction. 
\newline
\indent
The method can be systematically generalized to more complicated situations~\cite{Pilaftsis:1996fh}.  
For instance, one 
may be projected from the $R_{\xi}$ gauges to one of the generalized BFM gauges, such as the BFM axial gauge; 
this would lead to a proliferation of pinching momenta. The resulting construction is therefore 
more cumbersome, but remains conceptually rather straightforward.

\newpage


\section{\seclab{SM_one-loop}The Pinch Technique  one-loop construction in the electroweak sector of the Standard Model}
\noindent
In this section we give a general overview of how the PT construction is 
modified in a case of a theory with spontaneous (tree-level) symmetry breaking 
(Higgs mechanism)~\cite{Papavassiliou:1989zd,Degrassi:1992ue,Papavassiliou:1994pr}, 
using the electroweak sector of the  SM as the reference theory.

\subsection{The electroweak lagrangian}
\noindent
In order to define the relevant quantities and set up the
notation used throughout this section, we begin by writing the classical
(gauge invariant) SM Lagrangian as
\be
{\mathcal L}^{\rm cl}_{{\rm SM}}={\mathcal L}_{{\rm YM}}+{\mathcal L}_{{\rm H}}+{\mathcal
L}_{{\rm F}}.
\ee
The gauge invariant 
${SU}(2)_{{W}}
\otimes{U}(1)_{{Y}}$ 
Yang-Mills
part ${\mathcal L}_{\rm YM}$ consists of an isotriplet $W^a_\mu$ (with
$a={1,2,3}$) associated with the weak isospin generators
$T^a_w$, and 
an isosinglet $W^4_\mu$ with weak hypercharge $Y_w$
associated to the group factor $U(1)_{{Y}}$; it reads
\bea
{\mathcal L}_{\rm YM}&=&-\frac14F^a_{\mu\nu}F^{a\,\mu\nu} \nonumber \\
&=& -\frac14\left(\partial_\mu W^a_\mu-\partial_\nu W^a_\mu+\gw
f^{abc}W^b_\mu W^c_\nu\right)^2-\frac14\left(\partial_\mu W^4_\nu
-\partial_\nu W^4_\mu\right)^2. 
\eea 
The Higgs-boson part ${\mathcal L}_{\rm H}$ involves a complex 
${SU}(2)_{{W}}$
scalar doublet field $\Phi$ and its complex (charge) conjugate
$\widetilde\Phi$, given by
\bea
\Phi=\left(
\begin{array}{c} 
\phi^+\\
\frac1{\sqrt2}\left(H+i\chi\right)
\end{array}
\right),
&\hspace{2.0cm}&
\widetilde\Phi\equiv i\sigma^2\Phi^*=\left(
\begin{array}{c} 
\frac1{\sqrt2}\left(H-i\chi\right)\\
-\phi^-
\end{array}
\right).
\eea
Here $\sigma$ denotes the Pauli matrices, $H$ denotes the physical Higgs field, while $\phi^\pm$ and $\chi$
represents, respectively, the charged and neutral unphysical
degrees of freedom (would-be Goldstone bosons) .
Then ${\mathcal L}_{\rm H}$ takes the form
\be
{\mathcal L}_{\rm
H}=\left({\mathcal D}_\mu\Phi\right)^\dagger\left({\mathcal D}^\mu\Phi\right)-V(\Phi),
\ee
with the covariant derivative ${\mathcal D}_\mu$ defined as
\be
{\mathcal D}_\mu=\partial_\mu-i\gw T^a_wW^a_\mu+ig_1\frac{Y_w}2W^4_\mu,
\ee
and the Higgs potential as
\be
V(\Phi)=\frac\lambda4\left(\Phi^\dagger\Phi\right)^2
-\mu^2\left(\Phi^\dagger\Phi\right).
\ee
The SM leptons (we neglect the quark sector in what follows) 
are grouped into left-handed doublets
\be
\Psi^{L}_i=P_{L}\Psi_i=\left(
\begin{array}{c} 
\nu_i^{L}\\
\ell_i^{L}
\end{array}
\right), 
\ee
which transform
under the fundamental representation of ${SU}(2)_{{W}}
\otimes{U}(1)_{{Y}}$, and right-handed singlets
(which comprise only the charged leptons)
\be
\psi^{R}_i=P_{R}\psi_i=\ell^{R}_i
\ee
transforming with
respect to the Abelian subgroup ${U}(1)_{{Y}}$ only.
In the previous formulas, $i$ is the generation index, and the chirality
projection operators are defined according to\linebreak $P_{{L,R}}=(1\mp\gamma_5)/2$.
In this way the leptonic part of ${\mathcal L}_{\rm{ F}}$ reads
\be
{\mathcal L}_{\rm{ F}}=\sum_i\left(i\overline\Psi^{L}_i\gamma^\mu
{\mathcal D}_\mu\Psi^{L}_i+i\overline\psi^{R}_i\gamma^\mu
{\mathcal D}_\mu\psi^{R}_i-\overline\Psi^{L}_iG^\ell_i\psi^{R}_i\Phi
+{\rm h.c.}\right),
\ee
with $G^\ell_i$ the Yukawa coupling.
\newline
\indent
The Higgs field $H$ will give mass to all the SM
fields, by acquiring
a vacuum expectation value (vev) 
$v$; in particular, the masses of the gauge fields are generated
after absorbing the massless would-be Goldstone bosons $\phi^\pm$ and
$\chi$. The physical massive gauge-bosons $W^\pm,\,Z$ and the
(massless) photon $A$ 
are then obtained by diagonalizing the mass matrix; they are given by 
\be
W^\pm_\mu=\frac1{\sqrt2}\left(W^1_\mu\mp iW^2_\mu\right),
\hspace{1.0cm}
\left( 
\begin{array}{c}
Z_\mu \\
A_\mu
\end{array}
\right)=
\left(
\begin{array}{cc}
\cw & \sw \\
-\sw & \cw
\end{array}\right)
\left(
\begin{array}{c}
W^3_\mu \\
W^4_\mu
\end{array}
\right),
\ee
where 
\be
\cw=\cos\tw=\frac{\gw}{\sqrt{g_1^2+\gw^2}}, \hspace{1.5cm}
\sw=\sin\tw=\sqrt{1-\cw^2},
\ee
with $\tw$ the weak mixing angle. The resulting gauge-boson masses are
\be
M_{W}=\frac12\gw v,\qquad M_{Z}=\frac12\sqrt{g_1^2+\gw^2}v,\qquad M_{A}=0,
\label{gb_masses}
\ee 
which for the weak mixing angle give the relation
\be
\cw=\frac{\Mw}{\Mz}.
\ee
Finally, identifying the photon-electron coupling constant with 
the usual electrical charge $e$, we find 
\be
e=\frac{g_1\gw}{\sqrt{g_1^2+\gw^2}},\qquad g_1=\frac e{\cw},\qquad \gw=\frac e{\sw}.
\ee
\newline
\indent
For quantizing the theory, a gauge fixing term must be added to the
classical Lagrangian ${\mathcal L}^{\rm cl}_{{\rm SM}}$. 
To avoid tree-level mixing between gauge and
scalar fields, a renormalizable $R_\xi$ gauge of the 't Hooft type is
most commonly chosen \cite{Fujikawa:1972fe}. This latter gauge 
is specified by introducing a different gauge parameter for
each gauge-boson, and is defined through the linear gauge fixing (no sum over the color index $a$)
functions
\bea
{\mathcal F}^a&=&\partial^\mu W^a_\mu-\frac i2\gw\xi_a\left[v^\dagger_i\sigma^a_{ij}\phi_j-\phi^\dagger_i\sigma^a_{ij}v_j\right], \nonumber\\
{\mathcal F}^4&=&\partial^\mu W^4_\mu+\frac i2g_1\xi_4\left[v^\dagger_i\phi_i-\phi^\dagger_iv_i\right],
\eea
with $v_1=0$, $v_2=v$, and the gfp parameters given by $\xi_1=\xi_2=\xi_{W}$, $\xi_3=\xi_4=\xi_{Z}$. In terms of the mass eigenstates these translate into the gauge fixing functions 
\bea
{\mathcal F}^\pm
&=&\partial^\mu W^\pm_\mu\mp i\xi_{ W}\Mw\phi^\pm, \nonumber\\
{\mathcal F}^{Z}&=&
\partial^\mu Z_\mu- \xi_{Z}\Mz\chi,\nonumber\\
{\mathcal F}^{A}&=&
\partial^\mu A_\mu, 
\eea
finally yielding to the $R_\xi$ gauge fixing Lagrangian
\be
{\mathcal L}_{{\rm GF}}=-\frac1{\xi_{W}}{\mathcal
F}^+{\mathcal F}^-
-\frac1{2\xi_{Z}}
\left({\mathcal F}^{Z}\right)^2
-\frac1{2\xi_{A}}\left({\mathcal F}^{
A}\right)^2.
\label{SM_GF_Lag}
\ee
\indent
The Faddeev-Popov ghost sector corresponding to the above gauge fixing
Lagrangian reads 
\be
{\mathcal L}_{{\rm FPG}}=-\bar u^+ s{\mathcal F}^+-\bar u^- s{\mathcal F}^-
-\bar u^{Z} s{\mathcal F}^{Z}-\bar u^{A} s{\mathcal F}^{A},
\label{SM_FPG_Lag}
\ee
with $s$ the BRST operator for the SM fields (for the full set of the BRST transformation see \eg~\cite{Binosi:2004qe}).
Notice that the ghost Lagrangian contains kinetic terms for the Faddeev-Popov
fields, allowing one to introduce them as dynamical fields of the
theory.
\newline
\indent
In the case of the BFM gauge-fixing one replaces the Higgs vev  by the background scalar field
\be
\widehat{\Phi}=\left(
\begin{array}{c} 
\widehat{\phi}^+\\
\frac1{\sqrt2}\left(v+\widehat{H}+i\widehat{\chi}\right)
\end{array}
\right),
\ee
and adds to the derivative term the background $SU(2)_{W}$ triplet field
$\widehat{W}^a_\mu$. Thus we get
\bea
{\mathcal F}^a&=&\left(\delta^{ac}\partial^\mu+\gw f^{abc}\widehat{W}^b_\mu\right)W^{c\mu}-\frac i2\gw\xi_{Q}^{W}\left[\widehat{\phi}^\dagger_i\sigma^a_{ij}\phi_j-\phi^\dagger_i\sigma^a_{ij}\widehat{\phi}_j\right], 
\nonumber\\
{\mathcal F}^4&=&\partial^\mu W^4_\mu+\frac i2g_1\xi_{Q}^4\left[\widehat{\phi}^\dagger_i\phi_i-\phi^\dagger_i\widehat{\phi}_j\right].
\eea
Notice that: ({\it i}) the background scalar doublet $\widehat{\Phi}$ field has the usual 
non-vanishing vev $v$, while that of the quantum field $\Phi$ is 
zero, and ({\it ii}) the background field gauge 
invariance restricts the number of quantum gauge parameters to two, one for $SU(2)_{W}$ and 
one for $U(1)_{Y}$. Setting $\xi_{Q}^{W}=\xi_{Q}^4=\xi_{Q}$, to avoid tree-level mixing between the photon and the $Z$ boson, we get
\bea
{\mathcal F}^\pm
&=&\partial^\mu W^\pm_\mu\pm i\gw\left(\sw\widehat{A}^\mu-\cw\widehat{Z}^\mu\right)W^\pm_\mu\mp i\gw\left(\sw A^\mu-\cw Z^\mu\right)\widehat{W}^\pm_\mu\nonumber\\
&\mp&\frac i2\gw\xi_{Q}\left[\left(v+\widehat{H}\mp i\widehat{\chi}\right)\phi^\pm-\left(H\mp i\chi\right)\widehat{\phi}^\pm\right],\nonumber\\
{\mathcal F}^{Z}&=&\partial^\mu Z_\mu-i\gw\cw\left(\widehat{W}^+_\mu W^{-\mu}-W^+_\mu\widehat{W}^{-\mu}\right)-\frac i2\gw\frac{\cw^2-\sw^2}{2\cw}\xi_{Q}\left(\widehat\phi^-\phi^+-\widehat\phi^+\phi^-\right)\nonumber \\
&+&\frac1{2\cw}\gw\xi_{Q}\left(\widehat{\chi}H-\widehat{H}\chi-v\chi\right),\nonumber\\
{\mathcal F}^{A}&=&\partial^\mu A_\mu+i\gw\sw\left(\widehat{W}^+_\mu W^{-\mu}-W^+_\mu\widehat{W}^{-\mu}\right)+i\gw\sw\xi_{Q}\left(\widehat\phi^-\phi^+-\widehat\phi^+\phi^-\right).
\eea

After setting the gauge parameters all equal to $\xiQ$,
the corresponding gauge-fixing and \linebreak Faddeev-Popov terms are still given by  Eqs~(\ref{SM_GF_Lag}) and~(\ref{SM_FPG_Lag}). \newline
\indent
Summarizing, the complete electroweak sector of the SM Lagrangian in the $R_\xi$/BFM
gauges is given by
\be
{\mathcal L}_{{\rm SM}}={\mathcal L}^{\rm cl}_{{\rm SM}}+{\mathcal L}_{{\rm F}}+
{\mathcal L}_{{\rm GF}}+{\mathcal L}_{{\rm FPG}}.
\label{SMlag}
\ee
The full set of Feynman rules derived from this Lagrangian can be found in~\cite{Denner:1994xt}, and will be used
throughout this section. 

\subsection{\label{sec:gencons} Pinch technique  with Higgs mechanism: general considerations} 
\noindent
Before proceeding with the general discussion, we report some useful ingredients.
In particular, in the $R_{\xi}$ gauges the tree-level gauge boson propagators--
three massive gauge bosons ($W^{\pm}$ and $Z$), and a massless photon ($A$)--
are given by [notice the $i$ factor difference with respect to our definitions 
of Eq.~(\ref{GluProp-0}) and (\ref{GluProp}) in the previous section] 
\bea
\Delta^{\mu\nu}_{i} (q) &=& \left [g^{\mu\nu}-
\frac{(1-\xi_{i})q^{\mu}q^{\nu}}{ q^2 -\xi_{i} M_{i}^2}\right]d_{i}(q^2)
\label{GenProp-1}\nonumber\\
d_{i}(q^2) &=& \frac{-i}{q^2 -M_{i}^2}
\label{GenProp-2}
\eea
where $i=W,Z,A$, and $M^2_{ A}=0$.  
In general, the gfps $\xi_{ W}$, $\xi_{ Z}$, and $\xi_{ A}$ 
will be considered to be different from one another.  
The inverse of the  gauge boson propagators (\ref{GenProp-1}), to be denoted by  
$\Delta^{-1}_{i,\mu\nu}$, is given by
\be
\Delta^{-1}_{i,\mu\nu}(q) = i\left[(q^{2}-M^{2}_i)g_{\mu\nu}
- q_{\mu}q_{\nu}+ \frac{1}{\xi_i}q_{\mu}q_{\nu}\right].
\label{InvKsi}
\ee
\newline
\indent
There are three unphysical (would-be) Goldstone bosons  
associated with the three massive gauge boson, 
to be denoted by $\phi^{\pm}$ and $\chi$.
Their tree-level propagators are $\xi$-dependent, and are given by 
\be
D_i(q)=\frac{i}{q^{2}-\xi_{i} M^{2}_{i}},
\label{Gold}
\ee
with $i=W,Z$ (of course, there should be no Goldstone boson associated with the photon).
In addition, the ghost propagators are also given by 
$D_i(q)$, with $i=W,Z,A$ (there is, however, a massless ghost associated with the photon).
Finally, the bare propagator of the Higgs-boson is given by  
\be
\Delta_{ H}(q)=\frac{i}{q^{2}-M^{2}_{ H}},
\label{Higgs}
\ee
\newline
\indent
For the massive gauge bosons ($i=W,Z$)
the following identities, valid for any value of $\xi_i$,
will be used frequently:
\be
\Delta^{\mu\nu}_i(q)=
U^{\mu\nu}_i (q)-\frac{q^{\mu}q^{\nu}}{M^{2}_i}D_i(q),
\label{Id1}
\ee
where
\be
U^{\mu\nu}_i(q)= \left(g^{\mu\nu} - \frac{q^{\mu}q^{\nu}}{M^2_i}\right) d_i (q^2),
\label{UnitaryProp}
\ee
is the corresponding propagator in the so-called {\it unitary gauge} 
($\xi_i\rightarrow\infty$).
In addition, in order to rearrange various expressions appearing in the computations, we will often employ the algebraic identity
\be
\frac{1}{k^2 -\xi_{i} M_{i}^2} = \frac{1}{k^2 -M_{i}^2}
-\frac{(1-\xi_{i}) M_{i}^2}{ (k^2 -M_{i}^2)
(k^2 -\xi_{i} M_{i}^2)} .
\label{AlgId}
\ee
Finally, when dealing with the case where the fermions are considered to be massive,
we will use extensively the identities ~\cite{Papavassiliou:1994pr}
\bea
ig^{\nu}_{\alpha}&=& i\Delta^{\nu\mu}_i(q) \Delta^{-1}_{i\, \mu\alpha}(q)\nonumber \\
&=& q^{\nu}q_{\alpha}D_i(q)
-\Delta^{\nu\mu}_i(q)\left[(q^{2}-M^{2}_i)g_{\mu\alpha}
- q_{\mu}q_{\alpha}\right], 
\label{Id12}\nonumber\\
iq^{\mu} &=& q^{2} D_i(q) q^{\mu} + M^{2}_i q_{\nu}\Delta_{i}^{\mu\nu}(q).
\label{Id13}
\eea
\indent
Now, the application of the PT in the electroweak sector 
is significantly more involved than in the QCD case; in addition 
to the general proliferation of graphs intrinsic to the electroweak sector,
there are three PT-specific reasons that complicate the 
construction~\cite{Papavassiliou:1989zd,Papavassiliou:1994pr}. 

\begin{itemize}

\item[{\it i}.] In addition to the longitudinal momenta coming from 
the propagators of the gauge bosons (proportional to $\lambda_i=1-\xi_{i}$)
and  the PT decomposition of the vertices involving three gauge bosons,  
a new source of pinching momenta appears, originating 
from graphs having an external (\ie carrying the physical momentum $q$)
would-be Goldstone boson.
Specifically, interaction vertices such as $\Gamma_{A_\alpha\phi^{\pm}\phi^{\mp}}$,
$\Gamma_{Z_\alpha\phi^{\pm}\phi^{\mp}}$, $\Gamma_{W^{\pm}_{\alpha}\phi^{\mp} \chi}$, and $\Gamma_{W^{\pm}_{\alpha}\phi^{\mp} H}$ 
also furnish pinching momenta, when the 
gauge boson is inside the loop carrying (virtual) momentum $k$. 
Such a vertex will then be decomposed as (see \Figref{PT_svert_deco})
\be
\Gamma^{(0)}_\alpha(q,k,-q-k)=
\Gamma^{\mathrm F}_\alpha(q,k,-q-k)+\Gamma^{\mathrm P}_\alpha(q,k,-q-k), 
\label{PTsplit1}
\ee
 with
\begin{eqnarray}
\Gamma^{(0)}_\alpha(q,k,-q-k) &=& (2q+k)_\alpha,\label{PTsplit2-1} \nonumber\\
\Gamma^{\mathrm F}_\alpha(q,k,-q-k) &=& 2 q_{\alpha},\label{PTsplit2-2}\nonumber\\
\Gamma^{\mathrm P}_\alpha(q,k,-q-k) &=& k_\alpha,
\label{PTsplit2-3}
\end{eqnarray}
which is the scalar case analogue of (\ref{decomp}), (\ref{GF}) and (\ref{GP}).
\begin{figure}[!t]
\bce
\includegraphics[width=12cm]{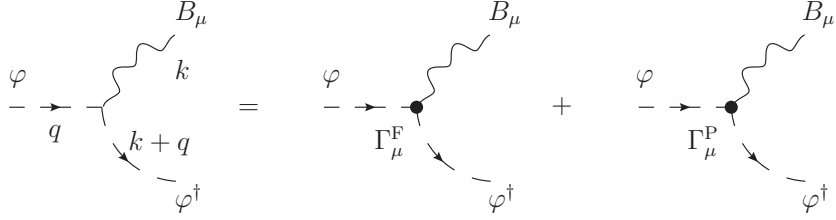}
\ece
\caption{\figlab{PT_svert_deco}The PT decomposition of the generic elementary gauge-boson-scalar vertex $\Gamma_{B_\mu\varphi\varphi^\dagger}$.}
\end{figure}
\newline
\item[{\it ii}.] When the fermions involved (external or inside loops)
are massive,
the WI of Eq.~(\ref{BasicWI}) receives additional contributions, 
which correspond precisely 
to the tree-level coupling of the 
would-be Goldstone bosons to the fermions. 
To see this concretely, let us consider the 
analogue of the fundamental pinching WI of Eq.~(\ref{BasicWI}) \eg
in the case where the incoming boson is a $W$. Contracting with $k^{\mu}$
the  $\Gamma_{W_{\mu}^{+}{\bar u}d}$ vertex 
(the fermions $u$ and $d$ are isodoublet partners), 
we have (we omit a factor $\gw/\sqrt{2}$)
\be
\ks  P_{ L} = 
 P_{ R} S_{d}^{-1}(k + p) - 
S_{u}^{-1}(p) P_{ L} + 
[m_{d} P_{ R}  -  m_{u} P_{ L}].
\label{EWI}
\ee
The first two terms will pinch and vanish on-shell,
respectively, as they did in the case of Eq.~(\ref{BasicWI}); the leftover 
term in the square bracket 
corresponds precisely to the coupling $\phi^{+} {\bar u}d$
(the case involving the $\Gamma_{W_{\mu}^{-}{\bar d} u}$ is identical).
A completely analogous WI is obtained when the incoming boson is a $Z$. 
Again, contraction with the vertex  $\Gamma_{Z {\bar f} f}$ furnishes 
a WI completely analogous to (\ref{EWI}), with 
the additional term proportional to $ m_{f} \gamma_5$,  
which corresponds to the coupling $\Gamma_{\chi{\bar f} f}$. 
\newline
\item[{\it iii}.] After the various pinch contributions have been identified, 
particular care is needed when allotting them among the
PT quantities that one is constructing.
So, unlike the QCD case 
where all propagator-like pinch contributions
were added to the only available self-energy, ${\Pi}_{\alpha\beta}$ 
(in order to construct ${\widehat{\Pi}}_{\alpha \beta}$), in the electroweak case 
such pinch contributions must, in general,
be split among various propagators. Thus, in the case of 
the charged channel, they will be shared, in general, between 
the self-energies $\Pi_{ W_\alpha  W_\beta}$, 
$\Pi_{ W_\alpha \phi}$, $\Pi_{\phi  W_{\beta}}$, 
and $\Pi_{\phi\phi}$.
This is equivalent to saying that, when forming the
inverse of the $W$ self-energy, in general 
the longitudinal parts may no longer be
discarded from the four-fermion amplitude, since the external current
is not conserved, up to terms proportional to the fermion masses.
As we will see in detail, the correct way of treating the longitudinal 
pieces is provided by  
the identities~(\ref{Id12}).
The neutral channel is even more involved; 
one has to split the 
propagator-like pinch contributions 
among the self-energies 
$\Pi_{ Z_\alpha  Z_\beta}$,
$\Pi_{ A_\alpha  A_\beta}$,
$\Pi_{ Z_\alpha  A_\beta}$,
$\Pi_{ A_\alpha  Z_\beta}$,
$\Pi_{ Z_\alpha \chi}$,
$\Pi_{\chi Z_\beta}$,
$\Pi_{\chi  \chi}$, and $\Pi_{ H   H}$.
\end{itemize}
\indent  
We emphasize that points ({\it i}), ({\it ii}), and ({\it iii}) above are tightly intertwined.
The extra terms appearing in the WI are precisely needed to cancel 
the gauge-dependence of the corresponding graph where the 
gauge boson is replaced by its associated Goldstone boson.
In addition, as we will see later, 
when the external currents are not conserved, 
the appearance of the scalar--scalar or  scalar--gauge-boson
self-energies is crucial for enforcing the gfp-independence  
of the physical amplitude.
\newline
\indent
We close this general discussion by briefly presenting an alternative
approach to the PT, known as the ``current algebra formulation of the PT'', 
introduced in~\cite{Degrassi:1992ue}.
In this
approach the interaction of gauge bosons with external fermions
is expressed in terms of
current correlation functions,
\ie matrix elements of Fourier transforms
of time-ordered products of current operators.
This is particularly economical because these amplitudes automatically
include several closely related Feynman diagrams. When one of the current
operators is contracted with the appropriate four-momentum, a WI
is triggered. The pinch part is then identified with the contributions
involving the equal-time commutators in the WIs, and therefore
involve amplitudes where the number of current operators has been
reduced by one or more. A basic ingredient in this formulation are the
following equal-time commutators
\begin{eqnarray}
\delta(x_0-y_0) [J^{0}_{ W}(x),J^{\mu}_{ Z}(y)] &=&
 \cw^{2}J^{\mu}_{ W}(x)\delta^{4}(x-y),\label{Commut2-1}\nonumber\\
\delta(x_0-y_0)[J^{0}_{ W}(x),J^{\mu\dagger}_{ W}(y)] &=&
 - J^{\mu}_{3}(x)\delta^{4}(x-y),\label{Commut2-2}\nonumber\\
\delta(x_0-y_0)[J^{0}_{ W}(x),J^{\mu}_{ A}(y)] &=&
 J^{\mu}_{ W}(x)\delta^{4}(x-y),\label{Commut2-3}\nonumber\\
\delta(x_0-y_0)[J^{0}_{ V}(x),J^{\mu}_{ V^{'}}(y)] &=& 0.
\label{Commut2-4}
\end{eqnarray}
where $J_{3}^{\mu}\equiv 2(J_{ Z}^{\mu}+\sw^{2}J_{ A}^{\mu})$
and $V,V^{'} \in \{ A,Z \}$.
To demonstrate the method with an example, consider
the one-loop vertex $\Gamma_{\mu}$, where the gauge
particles in the loop are $W$s,
and the incoming ($\vert \psi_i\rangle$) and outgoing ($\vert \psi_f\rangle$) fermions are massless.
It can be written as follows (with $\xi=1$):
\begin{equation}
\Gamma_{\mu}=\int_k
\Gamma_{\mu\alpha\beta}(q,k,-k-q)\int\!d^{4}x \, \e^{ik\cdot x}
\left\langle \psi_f\left\vert T[J^{\alpha\dagger}_ {W}(x)J^{\beta}_{ W}(0)]\right\vert \psi_i \right\rangle.
\label{Papous}
\end{equation}
When an appropriate momentum, say $k_{\alpha}$,
 from the vertex is pushed into the integral over
$dx$, it gets transformed into a covariant derivative
 $d/dx_{\alpha}$ acting on the time ordered product
$\left\langle\psi_f\left\vert T[J^{\alpha\dagger}_{ W}(x)J^{\beta}_{ W}(0)]\right\vert \psi_i \right\rangle$.
Then, after using current
conservation and differentiating the
$\theta$-function terms, implicit in the definition of
the $T^{*}$ product, we end up with the lhs of  Eq.~(\ref{Commut2-2}).
So, the contribution of each such term is proportional to the
matrix element of a single current operator,
namely $\left\langle \psi_f\left\vert J_{3}^{\mu}\right\vert \psi_i \right\rangle$; 
this is precisely the pinch part. 

\subsection{The case of massless fermions}
\noindent
We will now study the application of the PT in the case
where all fermions involved are massless.
This simplification facilitates the PT procedure 
considerably, 
because no scalar particles can be attached to the massless fermions. 
As a result ({\it i}) 
the scalars can appear only  
{\it inside} the self-energy graphs, 
where they obviously cannot pinch; ({\it ii}) Eq.(\ref{EWI}) is practically 
reduced to its QCD equivalent; ({\it iii}) there are no 
self-energies with incoming scalars 
(\ie no $\Pi_{ W_\alpha \phi}$, $\Pi_{ Z_\alpha \chi}$, 
$\Pi_{\phi\phi}$, etc).
\newline
\indent
Let us now focus, for simplicity, on the  
the process \mbox{$f_1(p_1){\bar f}_1(p_2)\to f_2(r_1) {\bar f}_2(r_2)$},   
mediated at tree-level by a $Z$-boson  and a photon.
At one-loop order, the box and vertex graphs furnish 
propagator-like contributions, every time the WI of Eq.(\ref{EWI}) 
is triggered by a pinching momentum. 
Specifically, the term in  Eq.~(\ref{EWI}) proportional to the inverse 
of the internal fermion propagator gives rise to a propagator-like term, 
whose coupling to the external fermions $f$ and ${\bar f}$ 
(with $f=f_1,f_2$) is proportional
to an effective vertex $C_{W_{\alpha} { f} {\bar f}}$ 
given by (see also \Figref{basic_SM_pinch})
\be
C_{W_\alpha { f} {\bar f}}
= -i\left(\frac{g_w }{2}\right) \gamma_\alpha P_{ L}.
\label{unphvert}
\ee
Note that this effective vertex is unphysical, in the sense that 
it does not correspond to any of the elementary vertices 
appearing in the electroweak Lagrangian.
However, it can be 
written as a linear combination of the two {\it physical} tree-level
vertices $\Gamma_{{ A}_\alpha { f} \bar{ f}}$ and
$\Gamma_{{ Z}_\alpha { f} \bar{ f}}$ given by
\bea
\Gamma_{{ A}_\alpha { f} {\bar f}} &=& -i\gw\sw 
Q_{ f} \gamma_{\alpha}, \label{GenVer-1}\nonumber\\ 
\Gamma_{{ Z}_\alpha { f} {\bar f}} &=&
-i \left(\frac{\gw}{\cw}\right) \gamma_{\alpha}
[ (\sw^2 Q_{ f} - T^{ f}_z) P_{ L} + \sw^2 Q_{ f} 
P_{ R}],
\label{GenVer-2}
\eea
as follows:
\be 
C_{W _{\alpha}{ f} {\bar f}}= \left(\frac{\sw}{2 T^{ f}_z}\right)
\Gamma_{{ A}_{\alpha}{ f} {\bar f}} - 
\left(\frac{\cw}{2 T^{ f}_z}\right)
\Gamma_{{ Z}_{\alpha}{ f}{\bar f}}.
\label{SepVer}
\ee
In the above formulas,
$Q_{ f}$ is the electric charge of the fermion $f$, and
$T^{ f}_z$ its $z$-component of the weak isospin.
The identity established in Eq.~(\ref{SepVer}) above,
allows one to combine the propagator-like
parts with the conventional 
self-energy graphs by writing $1=d_{i} (q^2) d_{i}^{-1}(q^2)$.
\begin{figure}[t]
\bce
\includegraphics[width=10cm]{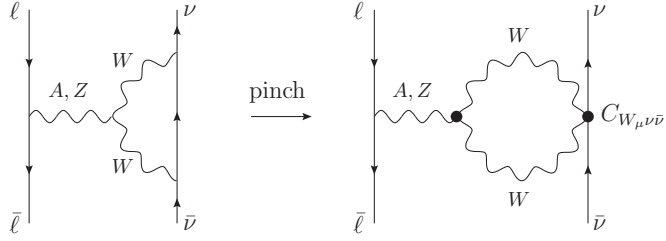}
\ece
\caption{\figlab{basic_SM_pinch}The basic pinching and one of the unphysical vertex produced in the process $\ell\bar\ell\to \nu \nu$ with $\ell$ a lepton.}
\end{figure}

\subsubsection{Gauge fixing parameter cancellations}
\noindent
Next, we will describe how the cancellation of
the gfp proceeds at one-loop level for the simple case where 
$f_1$ is a lepton, to be denoted by $\ell$, and $f_2$ is a
neutrino, denoted by $\nu$. 
Of course, based on general field-theoretic principles,  one knows in advance  that the
entire amplitude  will be gfp-independent.  What is important to recognize, however,  
is that this cancellation  goes through without having
to carry  out  any    of    the integrations  over   virtual  loop
momenta, exactly as happened in the case of QCD.
From the practical point of view,
the extensive gauge cancellations that are implemented through the PT finally amount 
to the statement that one may start out in the Feynman gauge, 
\ie set directly $\xi_{ W}=1$ and $\xi_{ Z}=1$,
with no loss of generality.
\begin{figure}[t]
\bce
\includegraphics[width=13cm]{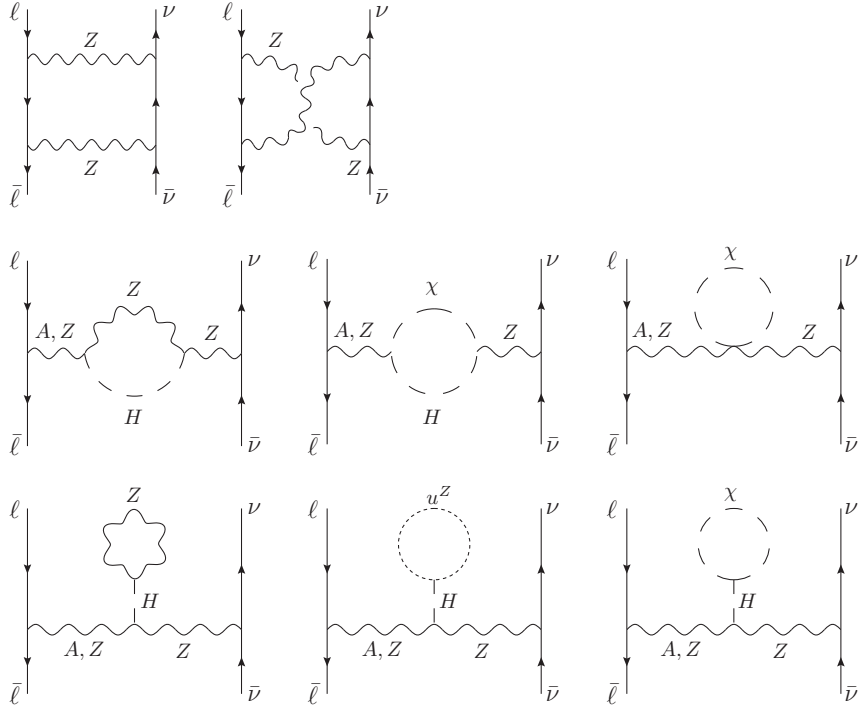}
\ece
\caption{\figlab{xiz_gfp_dep_diag}The subset of diagrams of the process that depends on the gfp~$\xi_{ Z}$.}
\end{figure}
\newline
\indent
Demonstrating the cancellation of $\xi_{ Z}$ is rather easy.
First of all, it is straightforward to verify that the 
box diagrams containing two $Z$-bosons (direct and crossed, see \Figref{xiz_gfp_dep_diag}) 
form a $\xi_{ Z}$-independent subset. 
The way this works is completely analogous to 
the QED case, where the two boxes contain photons: 
the $\xi_{ Z}$-dependence of the 
direct box cancels exactly against the gfp-dependence of the crossed. 
\newline
\indent
The only other graphs with a $\xi_{ Z}$ dependence 
are the self-energy graphs shown in \Figref{xiz_gfp_dep_diag};
it is easy to show by employing 
Eq.~(\ref{AlgId}) that their sum is independent of $\xi_{ Z}$, separately for 
$ZZ$ and $AZ$.
In fact, from the PT point of view it is clear why this must be so:
at one-loop there are no vertex graphs containing  $Z$, $\chi$, or $H$, that could possibly furnish pinch contributions 
which might mix with (and cancel against) the self-energy graphs.
Therefore, the $\xi_{ Z}$ dependent contributions are isolated in the 
self-energy, and must cancel completely, since the $S$-matrix is gfp independent.
\begin{figure}[!t]
\bce
\includegraphics[width=11cm]{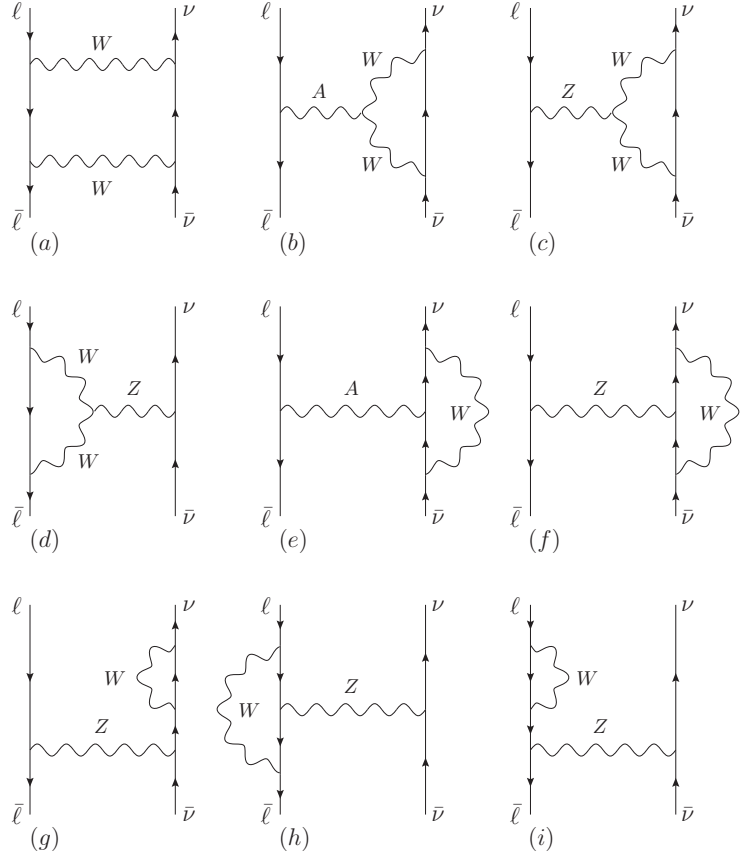}
\ece
\caption{\figlab{massless_box-vertex_diag}The box and vertex diagrams that depends on the gfp~$\xi_{ W}$.}
\end{figure}
\newline
\indent
Proving the cancellation of $\xi_{ W}$ is significantly more 
involved. In what follows we set  $\lambda_{ W} \equiv 1-\xi_{ W}$, and suppress 
a factor $g_w^2 \int_k$. We also define
\bea
I_3 & \equiv & \left[(k^2 -\xi_{ W} M_{ W}^2)(k^2 -M_{ W}^2) 
[(k+q)^2-M_{ W}^2]\right]^{-1},
\nonumber\\ 
I_4 & \equiv & \left[(k^2 -\xi_{ W} M_{ W}^2)
[(k+q)^2-\xi_{ W} M_{ W}^2](k^2 -M_{ W}^2) 
[(k+q)^2-M_{ W}^2]\right]^{-1},
\label{DefIs}
\eea
Note that terms proportional to $q_{\mu}$ or $q_{\nu}$ may be dropped directly, 
because the external currents are conserved (massless fermions). 
\begin{figure}[t]
\bce
\includegraphics[width=12cm]{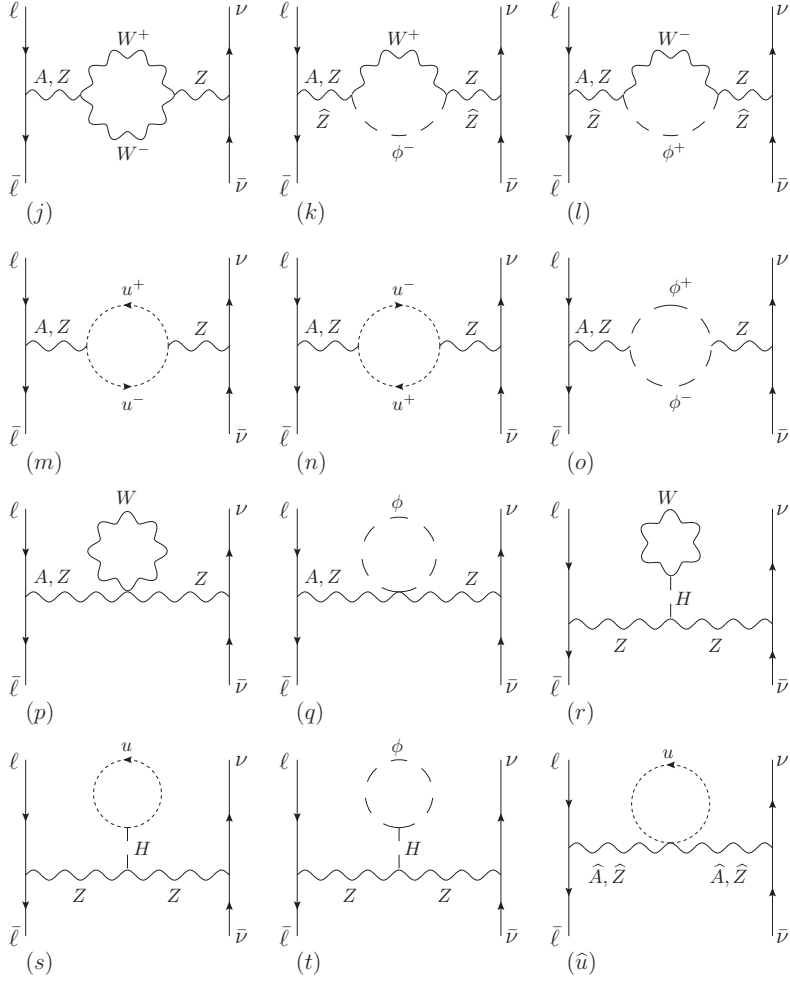}
\ece
\caption{\figlab{massless_self-nrg_diag}The self-energy diagrams that depends on the gfp~$\xi_{ W}$.}
\end{figure}
\begin{figure}[t]
\bce
\includegraphics[width=16cm]{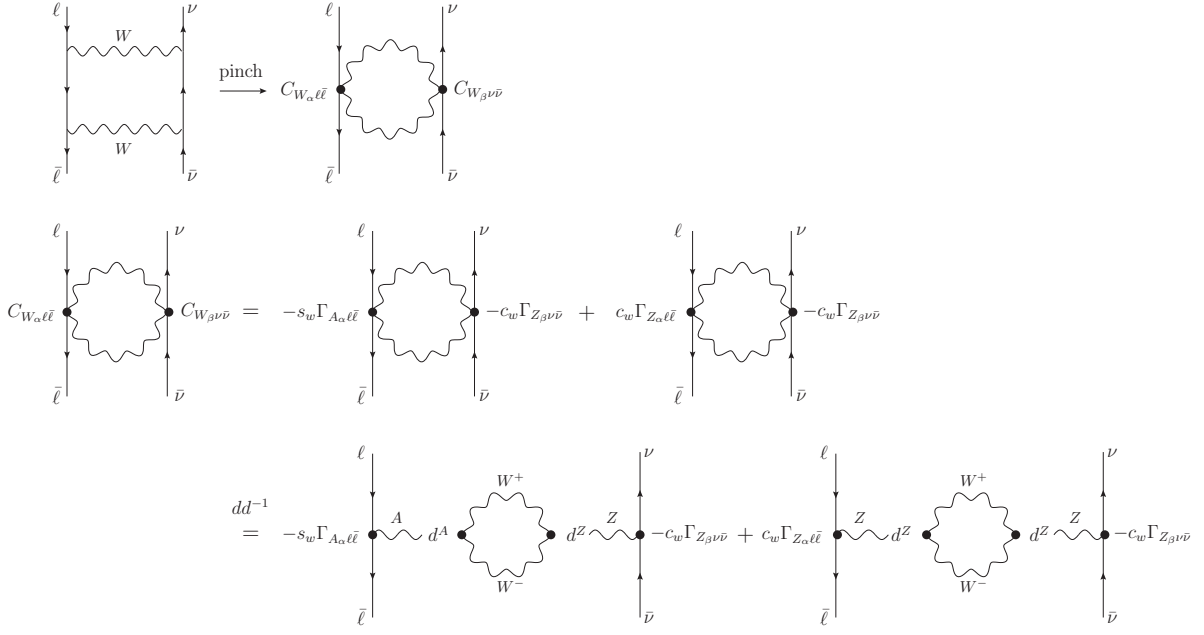}
\ece
\caption{\figlab{massless_box_PTsplit}The procedure needed for splitting the propagator-like 
pieces coming from the $WW$ box among the different $AZ$ and $ZZ$ self-energies.}
\end{figure}
\newline
\indent
To get a feel of how the PT organizes the various gauge-dependent terms, 
consider the box graphs shown in \Figref{massless_box-vertex_diag}. We have:
\be
(a) = (a)_{\xi_{ W} = 1} +  
{\mathcal V}_{{W}^{\alpha}\ell\bar\ell}\left( 
\lambda_{ W}^2 I_4 k_{\alpha}k_{\beta}
-2\lambda_{ W} I_3 g_{\alpha\beta} \right){\mathcal V}_{{ W}^{\beta} 
\nu\bar{\nu}},
\label{box}
\ee
where the vertices ${\mathcal V}$ are defined according to 
\bea
{\mathcal V}_{W_\alpha f\bar f}&=&\bar v_f C_{W_\alpha f\bar f} u_f,\nonumber\\
{\mathcal V}_{V_\alpha f\bar f}&=&\bar v_f \Gamma_{V_\alpha f\bar f} u_f, \qquad V=A,\ Z.
\eea
The first term on the  rhs of (\ref{box}) is the ``pure'' box, 
\ie the part that does not contain any propagator-like 
structures, whereas the 
second term is the propagator-like contribution that must 
be combined with the conventional propagator graphs of \Figref{massless_self-nrg_diag}.
To accomplish this, we employ Eq.~(\ref{SepVer}), 
in order to write the unphysical vertices 
${\mathcal V}_{{ W}\ell\bar\ell}$ and 
${\mathcal V}_{{ W} \nu\bar{\nu}}$ in terms of the physical 
ones, ${\mathcal V}_{{ A}\ell\bar\ell}$,
${\mathcal V}_{{ Z}\ell\bar\ell}$, and 
${\mathcal V}_{{ Z} \nu \bar{\nu}}$.
Specifically, using that 
in our case $T_z^\ell= -\frac{1}{2}$ and  $T_z^{\nu} = \frac{1}{2}$, 
we have 
\bea
C_{\ell\bar\ell}^{\alpha}  &=& -\sw 
\Gamma_{{ A}^{\alpha}\ell\bar\ell}  
+ \cw \Gamma_{{ Z}^{\alpha}\ell\bar\ell},
\label{abc-1}\nonumber\\  
C_{\nu\bar{\nu}}^{\alpha}  &=& - \cw 
\Gamma_{{ Z}^{\alpha} \nu \bar{\nu}}.
\label{abc-2}
\eea
 The equations above determine, unambiguously,
the parts that must be appended to  
$\Pi_{Z_\alpha Z_\beta}$ and 
$\Pi_{A_\alpha Z_\beta}$ self-energies.
To make this separation manifest, one must take the extra step of writing   
$d_{ Z}(q^2) d_{ Z}^{-1}(q^2) =d_{ A}(q^2) d_{ A}^{-1}(q^2) =1$, 
in order to 
force the external tree-level propagators to appear explicitly 
[see \Figref{massless_box_PTsplit}]. 
Thus, from the propagator-like part of the box we finally obtain
\bea
(a)_{A_\alpha Z_\beta} &=&  
\sw  \cw  q^2(q^2-M_{ Z}^2)
 \left( \lambda_{ W}^2 I_4 k_{\alpha}k_{\beta}
-2\lambda_{ W} I_3 g_{\alpha\beta}\right), 
\label{nGZnZZ-1}\nonumber\\
(a)_{Z_\alpha Z_\beta} &=& - 
\cw ^2 (q^2-M_{ Z}^2)^2
 \left( \lambda_{ W}^2 I_4 k_{\alpha}k_{\beta}
-2\lambda_{ W} I_3 g_{\alpha\beta} \right).
\label{nGZnZZ-2}
\eea
\indent
A similar procedure must be followed 
for the vertex graphs 
shown in \Figref{massless_box-vertex_diag}.
In doing so, recall that 
there is a relative minus sign between the 
$ZW^{+}W^{-}$ and $A W^{+}W^{-}$ vertices, namely
$\Gamma_{A_\alpha W_\mu W_\nu}(q,k_1,k_2) = i\gw \sw 
\Gamma_{\alpha\mu\nu} (q,k_1,k_2)$,
while $\Gamma_{Z_\alpha W_\mu W_\nu}(q,k_1,k_2)= -i\gw \cw \Gamma_{\alpha\mu\nu} (q,k_1,k_2)$.
Then, all propagator-like terms identified 
from the boxes and the vertex-graphs must be added 
to the conventional self-energy diagrams, 
given in \Figref{massless_self-nrg_diag}.
At this point, it would be a matter of straightforward algebra to verify that 
all $\xi_{ W}$-dependent terms cancel. Of course, this 
cancellation proceeds completely independently for the 
$ZZ$ and $A Z$ contributions.
To make the cancellation explicit 
({\it i.e.}, identify exactly the parts of the 
conventional  self-energy diagrams that will cancel 
against those coming from the boxes and the vertex-graphs)
we can repeat  what we did in the case of QCD. 
Thus, employing the  WIs of  Eqs~(\ref{3gWI-2}) 
triggered by the action of the longitudinal parts of the
internal $W$ propagators on the three-boson vertices
we can rearrange diagram $(j)$, 
exposing a large part of the underlying gfp-cancellation.
For the rest of the diagrams in \Figref{massless_self-nrg_diag}
one must instead use the identity of Eq.~(\ref{AlgId}).
Note that the inclusion of the tadpole 
graphs, namely $(r)$, $(s)$, and $(t)$, is crucial for the final cancellation 
of the gfp-dependent contributions that do not depend on $q^2$. 

\subsubsection{Final rearrangement and connection with the background Feynman gauge}
\noindent
Exactly as happened in the QCD case, 
the  gfp-cancellations described in the previous subsection 
amount effectively to choosing the Feynman gauge,  $\xi_{ W} =1$ 
(the self-energy diagrams of  \Figref{xiz_gfp_dep_diag} are also in 
$\xi_{ Z}=1$ and we will suppress them in what follows).
The next step is to consider the action of the remaining 
pinching momenta
stemming from the three-gauge-boson vertices 
inside the non-Abelian diagrams
$(o)$, $(p)$, and $(q)$, exposed  
after employing the PT decomposition of  Eqs~(\ref{decomp}) and~(\ref{GP}). 
The propagator-like contributions that will emerge from 
the action of $\Gamma^{\rm P}$
must be then reassigned to the conventional self-energy graphs,
thus giving rise to the one-loop PT self-energies, 
in this case $\widehat{\Pi}_{Z_\alpha Z_\beta}$
and $\widehat{\Pi}_{A_\alpha Z_\beta}$.
The part of the vertex graph containing the $\Gamma^{\rm F}$,
 together with the Abelian graph which in the Feynman gauge remains 
unchanged, constitute the one-loop PT vertices
$A \nu \bar{\nu}$,  $Z \nu \bar{\nu}$, and $Z \ell\bar\ell$, 
to be denoted by 
$\widehat\Gamma_{{ A} \nu \bar{\nu}}$,
$\widehat\Gamma_{{ Z} \nu \bar{\nu}}$, and  
$\widehat\Gamma_{{ Z} \ell\bar\ell}$,
respectively.
\newline
\indent
Let us see this in detail. Setting
\be
I_{ W W}(q) = \int_k \frac{1}{(k^2 - M_{ W}^2)[(k+q)^2- M_{ W}^2]},
\label{iww}
\ee
we obtain from the non-Abelian vertex graphs (now in the Feynman gauge):
\bea
(b)_{\xi_{W} =1} &=&  
(b)^{\rm F} 
-2 {\mathcal V}_{{ A} ^{\alpha}\ell\bar\ell}\ d_{ A} (q^2)
\left[\sw \cw (q^2-M^2_{{ Z}}) I_{{ W}{ W}}(q) g_{\alpha\beta} \right]
d_{ Z}(q^2)\, {\mathcal V}_{{ Z}^{\beta} \nu \bar{\nu}},
\label{hatopq-1}\nonumber\\
(c)_{\xi_{ W} =1} &=&
(c)^{\rm F}
 + 2 {\mathcal V}_{{ Z}^{\alpha} \ell\bar\ell}\ d_{ Z}(q^2)
\left[\cw^2 (q^2-M^2_{{ Z}})I_{{ W}{ W}}(q) g_{\alpha \beta} \right]
d_{ Z}(q^2){\mathcal V}_{{ Z}^{\beta} \nu \bar{\nu}},
\label{hatopq-2}\nonumber\\
(d)_{\xi_{ W} =1}  &=&
(d)^{\rm F}
-2 {\mathcal V}_{{ A}^{\alpha}\ell\bar\ell}\ d_{ A} (q^2)
\left[\sw \cw q^2 I_{{ W}{ W}} g_{\alpha \beta} \right]
d_{ Z}(q^2) {\mathcal V}_{{ Z}^{\beta} \nu \bar{\nu}},
\nonumber\nonumber\\
&+&
2 {\mathcal V}_{{ Z}^{\alpha} \ell\bar\ell}\ d_{ Z}(q^2)
\left[c_w^2 (q^2-M^2_{{ Z}}) I_{{ W}{ W}}(q) g_{\alpha \beta} \right]
d_{ Z}(q^2) {\mathcal V}_{{ Z}^{\beta} \nu \bar{\nu}}.
\label{hatopq-3}
\eea
The one-loop PT vertices 
$\widehat\Gamma_{{ A} \nu \bar{\nu}}$,
$\widehat\Gamma_{{ Z} \nu \bar{\nu}}$, and  
$\widehat\Gamma_{{ Z} \ell\bar\ell}$,
are given by 
\bea
(e)+ (b)^{\rm F}  &=&
{\mathcal V}_{{ A}^{\alpha} \ell\bar\ell} \, d_{ A}(q^2)\,
\widehat{\Gamma}_{{ A}^{\alpha} \nu \bar{\nu}}, 
\label{nuevo-1}\nonumber\\
(f)+ (c)^{\rm F}  &=& 
{\mathcal V}_{{ Z}^{\alpha}  \ell\bar\ell} \, d_{ Z}(q^2)\,
\widehat{\Gamma}_{{ Z}^{\alpha}\nu \bar{\nu}}, 
\label{nuevo-2}\nonumber\\
(h)+ (d)^{\rm F} &=&
\widehat{\Gamma}_{{ Z}^{\alpha} \ell\bar\ell} \, d_{ Z}(q^2)\,
{\mathcal V}_{{ Z}^{\alpha} \nu \bar{\nu}},
\label{nuevo-3}
\eea
whereas the PT self-energies
$\widehat{\Pi}_{Z_\alpha Z_\beta}$ and 
$\widehat{\Pi}_{Z_\alpha Z_\beta}$ are simply 
the sum of all propagator-like contributions, namely 
\bea
\widehat{\Pi}_{Z_\alpha Z_\beta}(q)
&=& 
\Pi_{Z_\alpha Z_\beta}^{(\xi_{ W}=1)}(q) +  
4 \gw^2 \cw^2 (q^2-M^2_{{ Z}})g_{\alpha\beta}  I_{{ W}{ W}}(q) ,
\label{PTZZ}
\nonumber\\
\widehat{\Pi}_{A_\alpha Z_\beta}(q)
&=& 
\Pi_{A_\alpha Z_\beta}^{(\xi_{ W}=1)}(q)
-2 \gw^2 \sw \cw (2 q^2-M^2_{{ Z}}) g_{\alpha\beta}  I_{{ W}{ W}}(q).
\label{PTGZ}
\eea
\indent
It is now relatively straightforward to prove that the 
$\xi_{ W}$-independent 
PT self-energies constructed in  Eqs~(\ref{PTZZ})
coincide with their BFM counterparts computed at  $\xi_{ W}^{ Q} =1$, \ie
\bea
\widehat{\Pi}_{Z_\alpha Z_\beta}(q) &=& 
\tilde{\Pi}_{Z_\alpha Z_\beta}^{(\xi_{ W}^{ Q} =1)}(q), 
\label{hatZZ}\nonumber\\
\widehat{\Pi}_{A_\alpha Z_\beta}(q) &=& 
\tilde{\Pi}_{A_\alpha Z_\beta}^{(\xi_{ W}^{ Q} =1)}(q). 
\label{hatGZ}
\eea
To see this explicitly, we will 
start from the rhs of  (\ref{PTZZ})
and reorganize, appropriately, the individual 
Feynman diagrams contributing to the $ZZ$ and $AZ$ self-energies.
Specifically, we will cast all diagrams in \Figref{massless_self-nrg_diag}, computed at $\xi_{ W}=1$, 
into the form of the corresponding diagrams 
in the BFM at $\xi_{ W}^{ Q} =1$, plus the leftover contributions.   
\newline
\indent
Let us then start with  diagrams $(j)_{Z_\alpha Z_\beta}$ 
and $(j)_{A_\alpha Z_\beta}$, and employ Eq.~(\ref{INPTDEC1}), together with  Eqs~(\ref{GPG+GGP}),
to write them in the form
\bea
({j})_{Z_\alpha Z_\beta} &=& 
(\widehat{j})_{Z_\alpha Z_\beta}
-  2\gw^2 \cw^2 \left[2q^2 I_{ W W}(q) g_{\alpha\beta}+ 
\int_k \frac{3k_{\alpha}k_{\beta}-k^2 g_{\alpha\beta}}
{(k^2 - M_{ W}^2)[(k+q)^2- M_{ W}^2]}\right],
\label{xi1aZZ-1}\nonumber\\
({j})_{A_\alpha Z_\beta} &=& 
(\widehat{j})_{A_\alpha Z_\beta}
+ 2 \gw^2 \sw \cw \left[ 2q^2 I_{ W W}(q) g_{\alpha\beta} 
\,+\int_k \frac{3k_{\alpha}k_{\beta}-k^2 g_{\alpha\beta}}
{(k^2 - M_{ W}^2)[(k+q)^2- M_{ W}^2]}\right].\nonumber \\
\label{xi1aZZ-2}
\eea
Notice that the terms  $(\widehat{j})_{Z_\alpha Z_\beta}$ and 
$(\widehat{j})_{A_\alpha Z_\beta}$
on the rhs come from the $\Gamma^{{\rm F}}\Gamma^{{\rm F}}$ part,
while the remainders come from  the expressions given in  Eqs~(\ref{GPG+GGP}), when the 
terms proportional to  $q_{\alpha}$ and $q_{\beta}$ are set equal to zero 
(due to current conservation).
\newline
\indent
For the remaining diagrams, simple algebra yields:
\bea
(k)_{Z_\alpha Z_\beta}
+ (l)_{Z_\alpha Z_\beta}
&=& 
 (\widehat{k})_{Z_\alpha Z_\beta} + 
(\widehat{l})_{Z_\alpha Z_\beta} 
+ 
2\gw^2\cw^2 (2 M_{ Z}^{2} - M_{ W}^{2}) I_{ W W}(q) g_{\alpha\beta},
\label{detil-1}\nonumber\\
(m)_{Z_\alpha Z_\beta}
+ (n)_{Z_\alpha Z_\beta}
&=& (\widehat{m}_{Z_\alpha Z_\beta}
+ (\widehat{n})_{Z_\alpha Z_\beta}
+  6 \gw^2 \cw^2 \int_k \frac{k_{\alpha}k_{\beta}}{(k^2 - M_{ W}^2)[(k+q)^2- M_{ W}^2]},\nonumber \\
\label{detil-2}\nonumber\\
(p)_{Z_\alpha Z_\beta}   &=& 
(\widehat{p})_{Z_\alpha Z_\beta}  +  (\widehat{u})_{Z_\alpha Z_\beta}
- 2 \gw^2 \cw^2 \int_k \frac{g_{\alpha\beta}}{k^2 - M_{ W}^2},
\label{detil-3}
\eea
and
\bea
(k)_{A_\alpha Z_\beta}
+ (l)_{A_\alpha Z_\beta}
&=& - 2 \gw^2 \sw \cw (M_{ Z}^{2} - M_{ W}^{2}) I_{ W W}(q)g_{\alpha\beta},
\label{gdtil-1}\nonumber\\
(m)_{A_\alpha Z_\beta}
+ (n)_{A_\alpha Z_\beta} 
&=& (\widehat{d})_{A_\alpha Z_\beta}  + 
(\widehat{e})_{A_\alpha Z_\beta}
- 6 \gw^2 \sw \cw \int_k \frac{k_{\alpha}k_{\beta}}{(k^2 - M_{ W}^2)[(k+q)^2- M_{ W}^2]},\nonumber \\
\label{gdtil-2}\nonumber\\
(p)_{A_\alpha Z_\beta} &=& 
(\widehat{p})_{A_\alpha Z_\beta} +   (\widehat{u})_{A_\alpha Z_\beta}
+ 2 \gw^2 \sw \cw \int_k \frac{g_{\alpha\beta}}{k^2 - M_{ W}^2}. 
\label{gdtil-3}
\eea
Then, adding by parts all the terms above, we obtain
\bea
\Pi_{Z_\alpha Z_\beta}^{(\xi_{ W}=1)}
&=& \tilde{\Pi}_{Z_\alpha Z_\beta}^{(\xi_{ W}^{ Q} =1)}
 -  4 \gw^2 \cw^2 (q^2-M^2_{{ Z}})g_{\alpha\beta} \, I_{{ W}{ W}}(q) ,
\label{ZZIP}\nonumber\\
\Pi_{A_\alpha Z_\beta}^{(\xi_{ W}=1)}
&=& 
\tilde{\Pi}_{A_\alpha Z_\beta}^{(\xi_{ W}^{ Q} =1)}
+ 2 \gw^2 \sw \cw (2 q^2-M^2_{{ Z}}) g_{\alpha\beta} \, I_{{ W}{ W}}(q).
\label{GZIP}
\eea
Substituting   Eqs~(\ref{ZZIP}) into the rhs of 
 Eqs~(\ref{PTZZ}) we obtain immediately   Eqs~(\ref{hatZZ}), 
as announced. In particular, notice that: ({\it i}) in the BFM there is no 
 ${\widehat A} W^{\pm} \phi^{\mp}$ interaction, and therefore 
graphs $(k)$ and $(l)$ are absent in $\tilde\Pi_{A_\alpha Z_\beta}$, and ({\it ii}) diagram ($\hat{u}$), 
corresponding to the characteristic BFM four-field coupling $\widehat{V}\widehat{V}uu$, has been
generated dynamically from the simple rearrangement of terms. 
\newline
\indent
With a small extra effort we can now obtain the closed expressions for the    
$\widehat{\Pi}_{Z_\alpha Z_\beta}$ and 
$\widehat{\Pi}_{A_\alpha Z_\beta}$ in terms of the Passarino-Veltman functions~\cite{Passarino:1978jh}.
We will only focus 
on the parts of the self-energies 
originating from Feynman graphs containing $W$ propagators, together with 
the associated Goldstone boson and ghosts. 
The contributions coming form the rest of the diagrams
(\eg containing loops with fermions, or $Z$- and $H$-bosons)
are common to the conventional and PT self-energies, 
\ie $\Pi_{{ A}{ Z}}^{(\bar{f}f)}= \widehat\Pi_{{ A}{ Z}}^{(\bar{f}f)}$
and $\Pi_{{ Z}{ Z}}^{(\bar{f}f)}= \widehat\Pi_{{ Z}{ Z}}^{(\bar{f}f)}$,
and we do not report them here. Therefore the only  Passarino-Veltman function
that will appear is $B_{0}(q^2,M^2_{{ W}},M^2_{{ W}})$.
\newline
\indent
To that end, one may use the closed expressions for 
$\Pi_{Z_\alpha Z_\beta}^{(\xi_{ W}=1)}$
and $\Pi_{A_\alpha Z_\beta}^{(\xi_{ W}=1)}$ given in~\cite{Denner:1991kt}, 
and add to them the pinch terms given in  Eqs~(\ref{PTZZ}). 
Equivalently, one may 
employ the correspondence established in  Eqs~(\ref{hatZZ}), and calculate 
directly the graphs contributing to  
$\tilde{\Pi}_{A_\alpha Z_\beta}^{(\xi_{ W}^{ Q} =1)}$
and $\tilde{\Pi}_{A_\alpha Z_\beta}^{(\xi_{ W}^{ Q} =1)}$ 
using the Feynman rules of~\cite{Denner:1994xt}. 
Opting for the former procedure, and concentrating on the part proportional to $g_{\alpha\beta}$ (which we factor out), from~\cite{Denner:1991kt} we have that 
\bea
\left.\Pi_{{ A}{ Z}}^{({ W}{ W})}(q^2)\right|_{\xi_{ W}=1} &=&
\frac{\alpha}{4\pi}\frac1{3 \sw \cw}
\left\{\left[\left(9 \cw^2 + \frac{1}{2}\right)q^2 + (12\cw^2+4)M^2_{{ W}} \right] B_{0}(q^2,M^2_{{ W}},M^2_{{ W}})
\right.\nonumber\\
&-&\left.(12 \cw^2 -2)M^2_{{ W}}B_{0}(0,M^2_{{ W}},M^2_{{ W}})+\frac{1}{3}q^2 \right\} ,
\label{ZZAZconv-1}\nonumber\\
\left.\Pi_{{ Z}{ Z}}^{({ W}{ W})}(q^2)\right|_{\xi_{ W}=1} &=&
- \frac{\alpha}{4\pi}\frac1{6 \sw^2 \cw^2}
\left\{\left[\left(18 \cw^4 + 2 \cw^2 -\frac{1}{2}\right)q^2 + 
(24c_w^4 + 16 c^2_w -10)M^2_{{ W}} \right] \times\right.\nonumber \\
&\times&B_{0}(q^2,M^2_{{ W}},M^2_{{ W}})
-(24 \cw^4 - 8\cw^2 +2)M^2_{{ W}}
B_{0}(0,M^2_{{ W}},M^2_{{ W}})\nonumber \\
&+&\left.\frac{1}{3}(4\cw^2-1)q^2 \right\}.
\label{ZZAZconv-2}
\eea
Adding to the above expressions the pinch terms given in  Eqs~(\ref{PTZZ}), 
and using the identity
$i B_{0}(q^2,M^2_{{ W}},M^2_{{ W}}) = 16 \pi^2 \, I_{{ W}{ W}}(q)$,
we finally obtain
\bea
\widehat{\Pi}_{{ A}{ Z}}^{({ W}{ W})}(q^2) &=&
\frac{\alpha}{4\pi}\frac1{3 \sw \cw}
\left\{\left[\left(21 \cw^2 + \frac{1}{2}\right)q^2 + (12\cw^2-2)M^2_{{ W}} \right] B_{0}(q^2,M^2_{{ W}},M^2_{{ W}})
\right.\nonumber\\
&-&\left.(12 \cw^2 -2)M^2_{{ W}}B_{0}(0,M^2_{{ W}},M^2_{{ W}})+\frac{1}{3}\,q^2 \right\} ,
\label{ZZDDW-1}\nonumber\\
\widehat{\Pi}_{{ Z}{ Z}}^{({ W}{ W})}(q^2) &=&
- \frac{\alpha}{4\pi}\frac1{6 \sw^2 \cw^2}
\left\{\left[\left(42 \cw^4 + 2 \cw^2 -\frac{1}{2}\right)q^2 + 
(24c_w^4 -8 c^2_w -10)M^2_{{ W}} \right]\times\right. 
\nonumber\\
&\times&B_{0}(q^2,M^2_{{ W}},M^2_{{ W}}) -(24 \cw^4 - 8\cw^2 +2)M^2_{{ W}}
B_{0}(0,M^2_{{ W}},M^2_{{ W}})\nonumber \\
&+&\left.\frac{1}{3}(4\cw^2-1) \,q^2 \right\}.
\label{ZZDDW-2}
\eea
\indent
It is easy to establish from the closed expressions reported in~\cite{Denner:1991kt} that 
$\Pi_{{ A}{ Z}}^{(f\bar{f})}(0) =0$ and $\widehat\Pi_{{ A}{ Z}}^{(f\bar{f})}(0)=0$. 
On the other hand, 
from  Eqs~(\ref{ZZAZconv-2})  we see that 
${\Pi}_{{ A} { Z}}^{({ W} { W})}(0)\neq 0$, 
while $\widehat{\Pi}_{{ A} { Z}}^{ (W  W)}(0)=0$.
Evidently, as a result of the PT rearrangement, bosonic and fermionic 
radiative corrections are treated on the same footing. 
As we will see in the next section, this last 
property is of great importance for phenomenological applications, 
such as the
self-consistent generalization of the universal part of the $\rho$-parameter,    
the unambiguous definition of the physical charge radius of the neutrinos, 
and the gauge-invariant formalism for treating resonant 
transition amplitudes.

\subsubsection{A very special case: the unitary gauge}
\noindent
In the previous subsections we have applied the PT in 
the framework of the linear
renormalizable $R_{\xi}$ gauges, and we have obtained 
$\xi$-independent
one-loop self-energies for the gauge bosons. 
What would happen, however, if one were to work 
{\it directly} in the unitary gauge? 
The unitary gauge is reached after gauging away the would-be Goldstone 
bosons, through an appropriate field redefinition 
(which, at the same time, corresponds to a gauge transformation) 
$\phi(x) \to \phi^{\prime}(x) = \phi(x) \exp{(-i \zeta(x)/v})$, where 
$\zeta(x)$ denotes, generically, the Goldstone fields. 
Note that the unitary gauge is defined completely independently of the 
$R_{\xi}$ gauges; of course, operationally, it  
is identical to the 
$\xi_{W} , \xi_{Z} \rightarrow\infty$ limit of the latter.
In particular,  in the unitary gauge 
the $W$ and $Z$ propagators
are given by (\ref{UnitaryProp}), where $i=W,Z$. 
\newline
\indent
Given that 
the contributions of unphysical scalars and ghosts cancel in this gauge, 
the unitarity of the theory becomes 
{\it manifest} [and hence its name]. 
In the language employed in subsection \ref{abspt}, 
``manifest unitarity'' 
means that, in the unitary gauge, the OT (a direct consequence of unitarity) 
holds in its strong version. 
The most immediate way to realize this is by noticing 
that the unitary gauge propagators, (\ref{UnitaryProp}),     
and the expression for the sum 
over the polarization vectors 
of a massive spin one vector boson 
[see (\ref{WPol}) in the following subsection] are practically identical.
\newline
\indent
Since the early days of spontaneously broken non-Abelian gauge theories,
the unitary gauge has been known to give rise to
non-renormalizable Green's functions, in 
the sense that their divergent parts cannot
be removed 
by the usual mass and field-renormalization counter-terms.
It is easy to deduce from the 
tree-level expressions of the 
gauge-boson propagators why this happens:
the longitudinal contribution in (\ref{UnitaryProp})
is divided by a squared mass 
instead of a squared momentum, 
i.e. $q^{\mu}q^{\nu}/{M^2_i}$ instead of $q^{\mu}q^{\nu}/{q^2}$, and therefore, 
${U}_{\mu\nu}^{i}(q)\sim 1 $ as $q\rightarrow\infty$. 
As a consequence, when ${U}_{\mu\nu}^{i}(q)$ is inserted inside quantum loops 
(and $q$ is the virtual momentum that is being integrated over), 
it gives rise to highly divergent integrals.
If dimensional regularization is applied, this hard short-distance behavior
manifests itself in the occurrence of divergences proportional to high
powers of $q^{2}$. 
Thus, at one loop, the  
divergent part of the $W$ or $Z$ self-energies
proportional to $g_{\mu\nu}$ has the general form 
\be
{\Pi}_{WW}^{\rm div}(q^{2}) = \frac{1}{\epsilon}
(c_1 q^{6} + c_2 q^{4} + c_3 q^{2} + c_4 )\,,
\label{undiv}
\ee
where the coefficients $c_i$, of appropriate dimensionality, 
depend on the gauge coupling and combinations of 
$M^{2}_{W}$ and $M^{2}_{Z}$. 
The important point is that, whereas 
the last two terms on the rhs of (\ref{undiv}) can be absorbed into 
mass and wave-function renormalization as usual, 
the first two {\it cannot} be absorbed into 
a redefinition of the parameters in the original Lagrangian, 
because they are proportional to $q^{6}$ and  $q^{4}$.  
\newline
\indent
As was shown in a series of papers \cite{Weinberg:1971fb,Lee:1973xp,Appelquist:1972tn},
when one puts together the individual Green's functions  
to form $S$-matrix elements, an extensive cancellation 
of all non-renormalizable divergent terms takes place, and 
the resulting $S$-matrix element can be rendered finite 
through the usual mass and gauge coupling renormalization.
Actually, in retrospect, this cancellation is nothing 
but another manifestation of the PT (of course, the 
papers mentioned above predate the PT).  
Even though this situation may be considered acceptable from the
practical point of view, in the sense that $S$-matrix elements may 
be still computed consistently, 
the inability to define renormalizable Green's functions has
always been a theoretical shortcoming of the unitary gauge.
\newline
\indent
The actual demonstration of how to construct 
renormalizable Green's functions at one-loop
starting from the unitary gauge was given in~\cite{Papavassiliou:1994fp}.
The methodology is identical to that used in the context of the $R_{\xi}$ gauges:
the propagator-like parts of vertices and boxes are identified and 
subsequently redistributed among the various gauge-boson self-energies.
Evidently, the pinch contributions contain, themselves, 
divergent terms proportional
to $q^{6}$ and  $q^{4}$, which, when added to the analogous contributions 
contained in the conventional propagators, cancel exactly. 
After this cancellation, the remaining terms reorganize themselves 
in such a way as give rise {\it exactly} to the unique 
PT gauge boson self-energies, {\it viz.}   Eqs~(\ref{ZZAZconv-1}).

\subsubsection{Pinch technique absorptive construction in the electroweak sector}
\noindent
We will now study exactly how the PT subamplitudes
of the  electroweak theory satisfy the OT~\cite{Papavassiliou:1996zn,Papavassiliou:1996fn}.
The upshot of this section is 
that the conclusions drawn from the corresponding QCD 
analysis, in particular the $s$-$t$ cancellations and  
the validity of the strong OT version for the PT Green's functions, 
persist in the case of tree-level symmetry breaking. 
The richness of the electroweak spectrum and the complexity of the 
corresponding Feynman rules make the actual demonstrations slightly more cumbersome, 
but the underlying philosophy of the construction is very similar to that of QCD.
\newline
\indent
We consider the forward process $f(p_1)\bar{f}(p_2)\to f(p_1)\bar{f}(p_2)$
and study both sides of the OT to lowest order. The PT rearrangement 
of the one-loop amplitude of this process proceeds as described 
in the previous section, giving rise to one-loop PT Green's functions, such as 
the PT self-energies
$\widehat{\Pi}_{Z_\alpha Z_\beta}(q)$, 
$\widehat{\Pi}_{A_\alpha Z_\beta}(q)$, 
$\widehat{\Pi}_{A_\alpha A_\beta}(q)$,
the PT vertices 
 $\widehat\Gamma_{{ A_\alpha} \nu \bar{\nu}}$,
$\widehat\Gamma_{{ Z_\alpha} f \bar{f}}$, 
$\widehat\Gamma_{{ Z_\alpha} \nu \bar{\nu}}$, and the PT boxes.
From them the corresponding subamplitudes may be straightforwardly constructed; 
their imaginary parts will determine the propagator-, vertex-, and box-like
parts of the rhs of the OT, to be denoted by \mbox{$(\rm rhs)_i$, $i=1,2,3$} as usual. 
\newline
\indent
Let us now focus on the lhs of the OT. 
Evidently, there are several intermediate states  $| j \rangle$ that may appear;
specifically, depending on the available center-of-mass energy,
all fermionic pairs (quarks and leptons)
$|f_i \bar{f}_i \rangle$ (with $i$ the flavor index), together with the bosonic channels 
$|W^{+} W^{-}\rangle$ and $|Z H\rangle$, may enter in principle.
Notice, however, that the energy thresholds for the appearance of 
all these intermediate states are different, being given by $s_{\rm th}= (m_1+m_2)^2$; for 
instance, the intermediate state $|W^{+} W^{-}\rangle$ will 
appear on the rhs of the OT when $s \geq 4 M^2_{ W}$, while 
 the $|Z H\rangle$ channel opens up when $s \geq  (M_{ Z}+M_{ H})^2$.
This clear kinematic separation of the various possible channels indicates that 
the pertinent PT cancellations (\ie the $s$-$t$ cancellation) 
take place {\it independently} within each intermediate state; indeed, there is no 
way that the  $|W^{+} W^{-}\rangle$ and $|Z H\rangle$ can talk to each other 
(unless $M_{ H} = 2 M_{ W} - M_{ Z}$, which is experimentally excluded). 
Therefore, on the rhs of the OT we will keep only the contribution of the $S$-matrix element
$ \langle f\bar{f}|T|W^{+} W^{-}\rangle$, \ie
\begin{equation}
\Im m \langle f\bar{f}|T|f\bar{f}\rangle_{{WW}} =\ \frac{1}{2} 
\int_{\mathrm{PS}_{{ W}{ W}}}\hspace{-.5cm} 
\langle f\bar{f}|T| W^{+}W^{-} \rangle \langle W^{+}W^{-}|T|f\bar{f}\rangle^{*}.
\label{OTWW}
\end{equation}
Notice that, unlike Eq.~(\ref{OTgg}), 
now there is no additional 
statistical factor, since the two (positively and negatively charged) $W$'s are 
distinguishable particles. 
The two-body phase space integral is given by Eq.~(\ref{2BPS}) with $m_1=m_2 = M_{{ W}}$.
As in the QCD case, in what follows we set 
\mbox{$T=\langle f\bar{f}|T|f\bar{f}\rangle_{{WW}} $},  
\mbox{${\mathcal T}=\langle f\bar{f}|T| W^{+}W^{-}\rangle$}, and ${\mathcal M}= |{\mathcal T}|^2$.
\begin{figure}[!t]
\bce
\includegraphics[width=12cm]{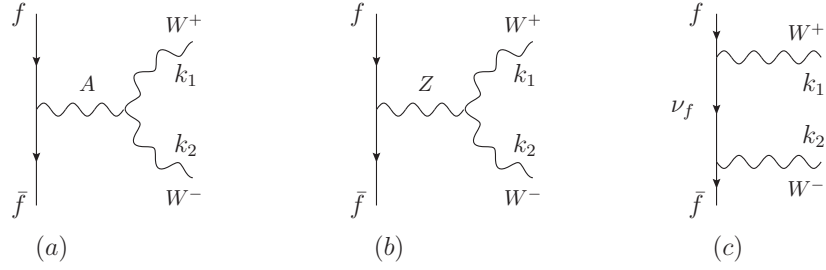}
\ece
\caption{\figlab{SM_absorpt_PT}The two $s$-channel and one $t$-channel graphs contributing to the tree-level process  $f(p_1)\bar{f}(p_2) \to W^{+}(k_1) W^{-}(k_2)$.}
\end{figure}
\newline
\indent
Let us now focus on the rhs of (\ref{OTWW}),
considering the process $f(p_1)\bar{f}(p_2)\! \to\! W^{+}(k_1) W^{-}(k_2)$,
with $q=p_1+p_2 = k_1+k_2$, and $s= q^{2} = (p_1+p_2)^{2}=(k_1+k_2)^2 > 4 M_{{ W}}^2$.
In this case, we have that ${\mathcal T}^{\mu\nu}$ is given by two $s$-channel graphs,
one mediated by a photon and the other by a $Z$-boson, to be denoted by 
 ${\mathcal T}_{{ A}}^{\mu\nu}$ and ${\mathcal T}_{{ Z}}^{\mu\nu}$, respectively,
and one $t$-channel graph, to be denoted by ${\mathcal T}_{t}^{\mu\nu}$, \ie (see also \Figref{SM_absorpt_PT})
\be
{\mathcal T}^{\mu\nu} =  {\mathcal T}_{s,{ A}}^{\mu\nu} + {\mathcal T}_{s,{ Z}}^{\mu\nu} + 
{\mathcal T}_{t}^{\mu\nu},
\label{k4m}
\ee
where
\bea
{\mathcal T}_{s,{ A}}^{\mu\nu}
&=& - {\mathcal V}_{{ A}^{\alpha} {{f\bar f}}}
  d_{ A}(q^2) \gw \sw \Gamma_{\alpha}^{\mu\nu}(q,k_1,k_2),
\label{k3m-1} \nonumber\\  
{\mathcal T}_{s,{ Z}}^{\mu\nu}  &=& {\mathcal V}_{{ Z}^{\alpha} 
{{f\bar f}}}
  d_{ Z}(q^2)\gw \cw \Gamma_{\alpha}^{\mu\nu}
(q,k_1,k_2),
\label{k3m-2}\nonumber\\
{\mathcal T}_{t}^{\mu\nu} &=& -\frac{g_w^2}{2} \bar{v}_{f} (p_2) 
\gamma^\mu P_{ L}\
  S^{(0)}_{ f'}(p_1-k_1)  \gamma^\nu P_{ L}u_{ f}(p_1) .
\label{k3m-3}
\eea
Note that 
we have already used current conservation to eliminate the (gfp-dependent)  longitudinal 
parts of the tree-level photon and $Z$-boson propagators.   
Then,
\be
{\mathcal M} =  \left[ {\mathcal T}_{{s, A}} + {\mathcal T}_{{s, Z}} + {\mathcal T}_{t}
\right]^{\mu\nu} L_{\mu\mu^{\prime}}(k_1)
L_{\nu\nu^{\prime}}(k_2)
\left[{\mathcal T}_{{s, A}}^{*} + {\mathcal T}_{{s, Z}}^{*} + {\mathcal T}_{t}^{*}\right]^{\mu^{\prime}\nu^{\prime}},
\label{MM_SM}
\ee
where now the polarization tensor $L^{\mu\nu}(k)$ corresponds
to a massive gauge boson (and thus with {\it three} polarization states), \ie
\begin{equation}
L_{\mu\nu}(k)= \sum_{\lambda=1}^{3} 
\varepsilon_{\mu}^{\lambda}(k)\varepsilon_{\nu}^{\lambda}(k)  
=-g_{\mu\nu}+ \frac{k_{\mu}k_{\nu}}{M^2_{ W}}.
\label{WPol}
\end{equation}
On shell ($k^2 = M^2_{ W}$) we have that $k^{\mu} L_{\mu\nu}(k)=0$.
Therefore, as in the QCD case, 
when the two non-Abelian vertices are decomposed as in Eq.~(\ref{decomp}),
the $\Gamma^{\rm P}$ parts vanish, and only the $\Gamma^{\rm F}$ parts contribute
in the $s$-channel graphs; we denote them by ${\mathcal T}_{s,{ A}}^{{\rm F},\mu\nu}$ and 
${\mathcal T}_{s,{ Z}}^{{\rm F},\mu\nu}$, respectively.
\begin{figure}[!t]
\bce
\includegraphics[width=13cm]{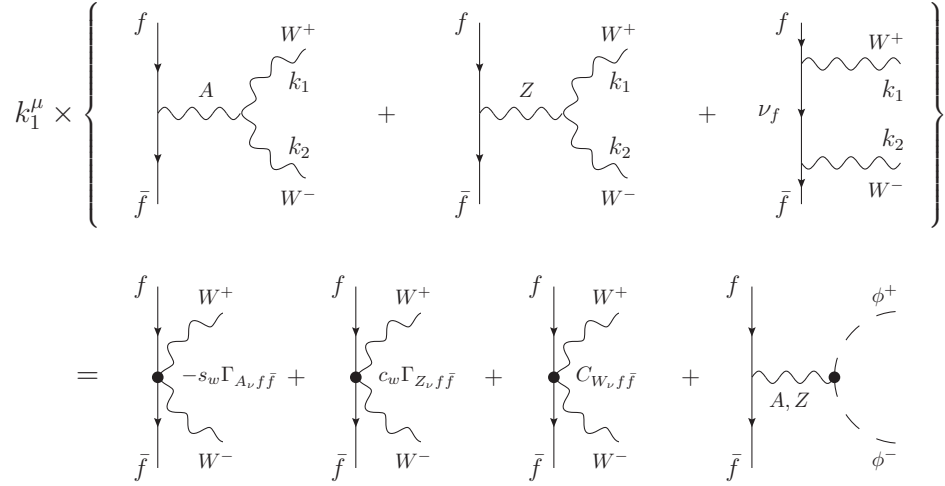}
\ece
\caption{\figlab{SM_st_canc}The fundamental $s$-$t$ cancellation in the SM case.}
\end{figure}
\newline
\indent
Let us then study what happens when ${\mathcal T}_{\mu\nu}$ is contracted
by a longitudinal momentum, $k_{1}^{\mu}$ or $k_{2}^{\nu}$, 
coming from the polarization tensors. 
The WIs of  Eqs~(\ref{3gWI-2}) will operate at the two $s$-channel graphs,
whereas that of  Eq.~(\ref{EWI}) at the  $t$-channel graph (\Figref{SM_st_canc}), yielding 
\bea
k_{1 \mu} {\mathcal T}_{{s, A}}^{F,\mu\nu}&=& 
- s_w {\mathcal V}_{{ A}{^{\nu}{f\bar f}}}
 + {\mathcal S}_{{ A}}^{\nu},
\label{k1m-1} \nonumber\\  
k_{1\mu}  {\mathcal T}_{{s, Z}}^{F,\mu\nu} &=& c_w 
{\mathcal V}_{{ Z}^{\nu}{{f\bar f}}}
+  {\mathcal S}_{{ Z}}^{\nu},
\label{k1m-2}\nonumber\\
k_{1 \mu} {\mathcal T}_{t}^{\mu\nu} &=& {\mathcal V}_{W{^{\nu}f\bar f}},
\label{k1m-3} 
\eea
with
\bea
{\mathcal S}_{{ A}}^{\nu} &=&  - {\mathcal V}_{{ A}^{\alpha} {{f\bar f}}}d_{ A}(q^2)  
\gw \sw (k_1- k_2)_{\alpha}k_{2}^{\nu},\nonumber\\
{\mathcal S}_{{ Z}}^{\nu} &=&  {\mathcal V}_{{ Z}^{\alpha} {{f\bar f}}}d_{ Z}(q^2) 
\gw \cw \left[(k_1- k_2)_{\alpha}k_{2}^{\nu} -M_{ Z}^2 g_{\alpha\nu} \right] .
\eea
Adding by parts both sides of  Eqs~(\ref{k1m-3})
we see that a major cancellation takes place:
the pieces containing the vertices ${\mathcal V}_{{ A}^{\nu}{{f\bar f}}}$ and 
${\mathcal V}_{{ Z}^{\nu}{{f\bar f}}}$
cancel against ${\mathcal V}_{W{^{\nu}f\bar f}}$
by virtue of Eq.~(\ref{abc-1}), and
one is left on the rhs with purely $s$-channel contribution, namely
\be
k_{1 \mu} {\mathcal T}^{\mu\nu} = {\mathcal S}_{{ A}}^{\nu} +  {\mathcal S}_{{ Z}}^{\nu}. 
\label{k6m} 
\ee
An exactly analogous cancellation takes place when one contracts with $k_2^{\nu}$.
Of course, Eq.~(\ref{k6m}) is nothing more than the manifestation of the  
$s$-$t$ cancellation already encountered in QCD, in a slightly  more involved context.
\newline
\indent
It is important to   recognize~\cite{Bernabeu:2000hf} that the cancellation   described above
goes  through,  even when the  initial  fermions  are right-handedly
polarized, regardless of the fact that, in that particular  case, there is no  $t$-channel graph, 
since the fermions do not couple to the $W$ (\Figref{SM_st_canc_rh}). What happens, then, is that  
the  elementary  vertices given in  Eqs~(\ref{GenVer-1}),
together   with  the  corresponding   ${\mathcal V}_{ A_{\alpha} {{f\bar f}}}$ and 
${\mathcal   V}_{ Z_\alpha{{f\bar f}}}$    are   appropriately    modified.  Specifically,
\bea
\Gamma_{ A_{\alpha} { f}_{ R} {\bar f }_{ R}} &=& 
-i\gw \sw Q_{ f} \gamma_{\alpha},
\label{GenVerPol-1}\nonumber\\ 
\Gamma_{  Z_{\alpha}{ f}_{ R} {\bar f}_{ R}} 
&=& -i \frac{\gw}{\cw}\sw^2 Q_{ f}  \gamma_{\alpha},
\label{GenVerPol-2}
\eea
and
\be
{\mathcal V}_{ Z_{\alpha} { f}_{ R} {\bar f }_{ R}} = 
\frac{\sw}{\cw}{\mathcal V}_{ { A}_{\alpha} { f}_{ R} {\bar f }_{ R}}.
\label{xyz}
\ee
Clearly, in that case, from  Eqs~(\ref{abc-1}) it follows immediately that 
\be
{\mathcal V}_{W{_{\alpha} f}_{ R} {\bar f }_{ R}} = 0 ,
\label{Vwfr}
\ee
so that Eq.~(\ref{k6m}) is still satisfied.  
\begin{figure}[!t]
\bce
\includegraphics[width=11cm]{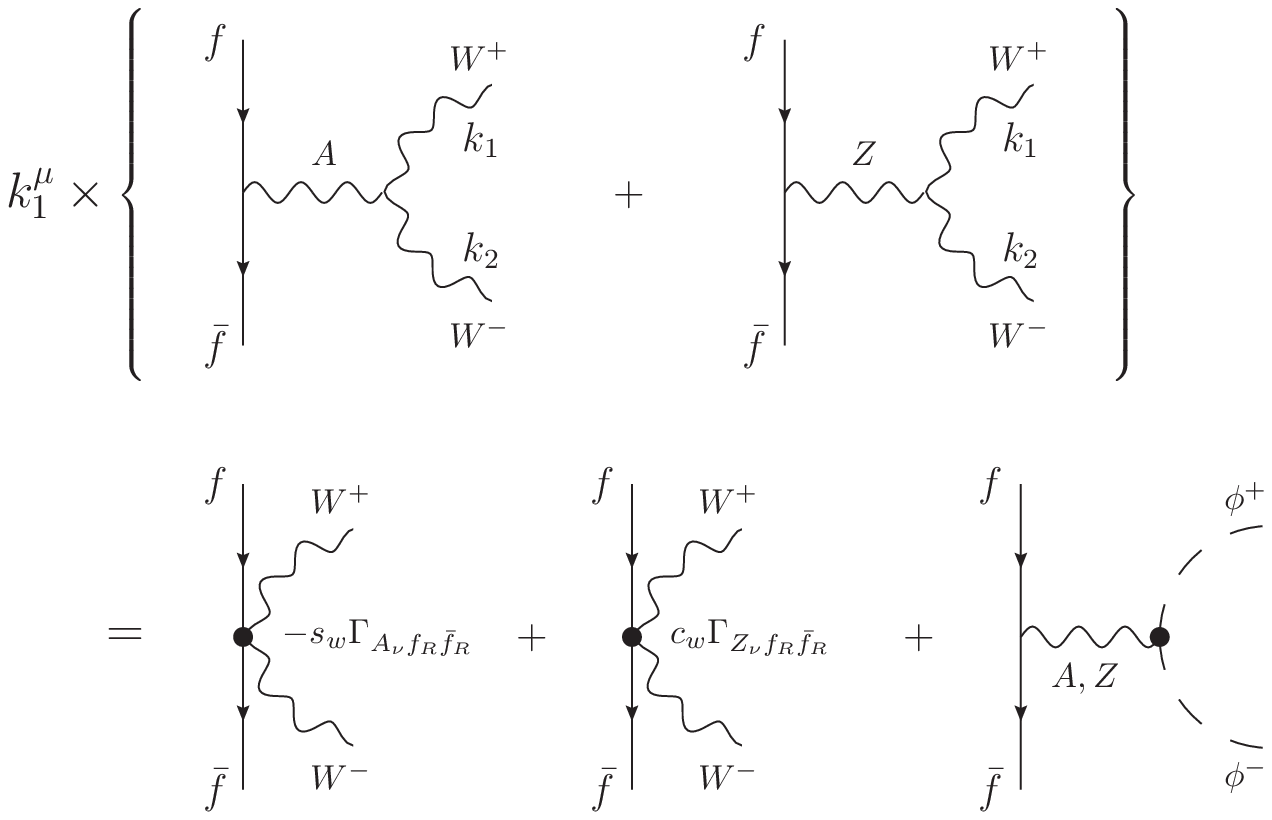}
\ece
\caption{\figlab{SM_st_canc_rh}The fundamental cancellation in the case of right-handed fermions; due to the absence of the $t$-channel graph, it is implemented through the two remaining $s$-channel graphs.}
\end{figure}
\indent
Let us now simplify the algebra, by choosing the  
initial fermions to be neutrinos, \ie let us consider the  
process  $\nu \bar{\nu} \to W^{+} W^{-}$. This choice 
eliminates all terms mediated by a photon, and one has 
\be
{\mathcal M} =  \left[{\mathcal T}_{{s, Z}}^{{ F}} + {\mathcal T}_{t}
\right]^{\mu\nu} L_{\mu\mu^{\prime}}(k_1)
L_{\nu\nu^{\prime}}(k_2)
\left[{\mathcal T}_{{s, Z}}^{{ F}*} + {\mathcal T}_{t}^{*}\right]^{\mu^{\prime}\nu^{\prime}},
\label{MMF}
\ee
Then,  Eqs~(\ref{k1m-1}) simplify to  
\bea
k_{1 \mu}  [{\mathcal T}_{{s, Z}}^{{ F}} + {\mathcal T}_{t}]^{\mu\nu} &=& 
{\mathcal S}_{ Z}^{\nu}, 
\label{k10-1} \nonumber\\
k_{2 \nu}  [{\mathcal T}_{{s, Z}}^{{ F}} + {\mathcal T}_{t}]^{\mu\nu}&=& 
{\bar{\mathcal S}}_{ Z}^{\mu}. 
\label{k10-2} 
\eea
with 
\bea
{\mathcal S}_{{ Z}}^{\nu} &=& 
{\mathcal V}_{{ Z}^{\alpha} {\nu} \bar{\nu}} d_{ Z}(q^2)\, 
g_w c_w \left[(k_1- k_2)_{\alpha}k_{2}^{\nu} -M_{ Z}^2 g_{\alpha}^{\nu} \right],
\label{brd1-1}\nonumber\\
{\bar {\mathcal S}}_{{ Z}}^{\mu} &=&  
{\mathcal V}_{{ Z}^{\alpha} {\nu} \bar{\nu}} d_{ Z}(q^2)\, g_w c_w \left[
(k_1- k_2)_{\alpha}k_{1}^{\mu} + M_{ Z}^2 g_{\alpha}^{\mu} \right],
\label{brd1-2}
\eea
In addition,  
\bea
k_{2 \nu} {\mathcal S}_{{ Z}}^{\nu} &=& k_{1 \mu} {\bar {\mathcal S}}_{{ Z}}^{\mu}\nonumber\\
&=& {\mathcal V}_{{ Z}^{\alpha} {\nu} \bar{\nu}} d_{ Z}(q^2)
\gw \cw \left[M_{ W}^2 + \frac{1}{2} M_{ Z}^2 \right]  (k_1- k_2)_{\alpha}.
\label{brd12}
\eea
Now we will isolate from (\ref{MMF}) the part that is purely $s$-channel 
(or, equivalently, purely propagator-like),
to be denoted by $\widehat{{\mathcal M}}_1$. It is 
composed by the sum of the following terms
\be
\widehat{{\mathcal M}}_1 = {\mathcal T}_{{ Z}}^{{ F}}\cdot {\mathcal T}_{{ Z}}^{{ F}*}
- \frac{{\mathcal S}_{{ Z}}\cdot {\mathcal S}_{{ Z}}^{*}}{M_{ W}^2}
-\frac{{\bar {\mathcal S}}_{{ Z}}\cdot {\bar {\mathcal S}}_{{ Z}}^{*}}{M_{ W}^2}
+ \frac{(k_{2}\cdot {\mathcal S}_{{ Z}})\cdot (k_{2}\cdot {\mathcal S}_{{ Z}}^{*})}{M_{ W}^4}\,.
\label{fbr1}
\ee
Using (\ref{FRE}), (\ref{brd1-1}),  (\ref{brd1-2}) and~(\ref{brd12}), 
we find
\be
\widehat{{\mathcal M}}_1  = {\mathcal V}_{{ Z}^{\alpha} {\nu} \bar{\nu}} d_{ Z}(q^2) 
K_{\alpha\beta}
d_{ Z}(q^2) {\mathcal V}_{{ Z}^{\beta} {\nu} \bar{\nu}}
\label{fbr3}
\ee
with 
\be
K_{\alpha\beta} = 
-\frac{g^2_w}{c_w^2}\left[ \left(8q^2 c_w^4 -2 M_{ W}^2\right)g_{\alpha\beta}
+ 
\left(3 c_w^4 -c_w^2 + \frac{1}{4}\right){(k_1-k_2)}_{\alpha}{(k_1-k_2)}_{\beta}\right].
\ee
Thus, the propagator-like part of the rhs of the OT becomes 
\be
({\rm rhs})_{1}  = \frac{1}{2}\int_ {{\rm PS}_{WW}}\hspace{-0.5cm} \widehat{{\mathcal M}}_1.
\label{rhs1a_SM}
\ee
Now, from Eq.~(\ref{LIPS2}), we have that 
\mbox{$\lambda (q^2,M_{ W}^2,M_{ W}^2) = q^2(q^2-4M_{ W}^2)$}, and therefore
\be
\int_ {{\rm PS}_{WW}}\hspace{-0.5cm}
{(k_1-k_2)}_{\alpha}{(k_1-k_2)}_{\beta} = -\frac{1}{3}(q^2-4M_{ W}^2)g_{\alpha\beta}
\int_ {{\rm PS}_{WW}}+\cdots,
\ee
where the ellipses stand for terms proportional to $q_\alpha q_\beta$. 
Then, using the elementary result
\be
8\pi^2\int_ {{\rm PS}_{WW}} \hspace{-0.5cm}= \Im m  B_{0}(q^2,M^2_{{ W}},M^2_{{ W}}),
\label{imww}
\ee
Eq.~(\ref{rhs1a_SM}) becomes 
\be
({\rm rhs})_{1} = {\mathcal V}_{{ Z}^{\alpha}{\nu} \bar{\nu}} d_{ Z}(q^2)
Kd_{ Z}(q^2) {\mathcal V}_{ Z_\alpha {\nu} \bar{\nu}},
\label{fbr4}
\ee
with 
\be 
K = 
-\frac{\alpha}{4\pi}\frac1{6 \sw^2 \cw^2}\left[\left(42 \cw^4 + 2  \cw^2 - \frac{1}{2}\right)q^2 
+ (24 \cw^4 -8 \cw^2 -10) M_{ W}^2 \right]\! 
\Im m  B_{0}(q^2,M^2_{{ W}},M^2_{{ W}}).
\label{fbr5}
\ee
\indent
Let us now compare Eq.~(\ref{rhs1a_SM}) with
the propagator-like part of the lhs of the OT, given by 
\be
({\rm lhs})_{1} = {\mathcal V}_{{ Z}^{\alpha} {\nu} \bar{\nu}} d_{ Z}(q^2) 
\left[\Im m \widehat{\Pi}_{ Z  Z}^{(WW)}(q) \right] 
d_{ Z}(q^2) {\mathcal V}_{{ Z}_\alpha {\nu} \bar{\nu}}.
\label{fbr2}
\ee
The equality between Eq.~(\ref{fbr2}) and~(\ref{fbr4}) requires that 
\bea
\Im m \widehat{\Pi}_{ Z  Z}^{(WW)}(q) &=& 
-\frac{\alpha}{4\pi}\frac1{6 s_w^2 c_w^2}\left[\left(42 \cw^4 + 2  \cw^2 - \frac{1}{2}\right)q^2 
+ (24 \cw^4 -8 \cw^2 -10) M_{ W}^2 \right] \times\nonumber \\
&\times& \Im m  B_{0}(q^2,M^2_{{ W}},M^2_{{ W}}).
\label{leftright}
\eea
Taking the imaginary part of $\widehat{\Pi}_{ Z  Z}^{(WW)}$ given in Eq.~(\ref{ZZDDW-2})
we see that Eq.~(\ref{leftright}) is indeed fulfilled.
\newline
\indent
At this point one could go one step further, 
and employ a twice subtracted dispersion relation, in order 
to reconstruct from (\ref{leftright}) the real part of the 
$\widehat{\Pi}_{ Z  Z}^{(WW)}(q)$. The end result 
of this procedure 
will coincide with the corresponding 
expression obtained from Eq.~(\ref{ZZDDW-2}) after renormalization.
(for a detailed derivation, see~\cite{Papavassiliou:1996fn}).   
\newline
\indent
Finally, let us return to the non-renormalizability of the unitary gauges, now seen 
from the absorptive point of view. As mentioned in the previous subsection, in the unitary 
gauge the strong version of the OT is satisfied; to make contact with this section, what this 
means is that the OT is satisfied diagram-by-diagram, 
{\it without} having to resort explicitly to the $s$-$t$ cancellation.
For example, the imaginary part of the conventional self-energy ${\Pi}_{ Z  Z}^{(WW)}(s)$
in the unitary gauge is 
\be
\Im m {\Pi}_{ Z  Z}^{(WW)}(s) \sim (s-M_Z)^2 \int_ {{\rm PS}_{WW}} 
{\mathcal T}_{{ Z}}^{\mu\nu} L_{\mu\mu^{\prime}}(k_1) 
L_{\nu\nu^{\prime}}(k_2){\mathcal T}_{{ Z}}^{*\mu^{\prime}\nu^{\prime}}\,. 
\label{ungcon}
\ee
What is the price one pays for {\it not} implementing the $s-t$ cancellation? 
Simply, the conventional subamplitudes, such as the one given above, contain terms that  
grow as $s^2$ or as $s^3$ [see, e.g.,~\cite{Alles:1976qv,Papavassiliou:1996fn}]; 
indeed, the $s$-$t$ cancellation eliminates
precisely terms of this type. Consequently,  
if one were to substitute the $\Im m {\Pi}_{ Z  Z}^{(WW)}(s)$
obtained from the rhs of (\ref{ungcon})
into a twice subtracted dispersion relation --the maximum number of subtractions allowed by renormalizability--
one would encounter UV divergent real parts proportional to $q^4$ or as $q^6$. 
Of course, these are precisely the non-renormalizable terms encountered in (\ref{undiv}), 
now obtained not from a direct one-loop calculation but rather 
from the combined use of unitarity and analyticity 
(and with a hard UV cutoff instead of $1/\epsilon$).

\subsubsection{Background field method away from $\xi_{Q}=1$: physical versus unphysical thresholds} 
\noindent
As we have seen in the previous section, 
from the fact that the BFM Green's functions satisfy the same QED-like WIs for every 
value of the quantum gfp  $\xi_{Q}$ one should {\it not} conclude that the 
PT Green's functions, reproduced from the BFM at $\xi_{Q}=1$, 
are simply one among an infinity of physically equivalent choices, parametrized 
by  $\xi_{Q}$. This interpretation is not correct:
the BFM Green's functions obtained  away from $\xi_{Q}=1$ 
are {\it not} physically equivalent to the privileged case of $\xi_{Q}=1$.
\newline
\indent
In addition to the reasons 
outlined in \secref{PTBFM}, when dealing with the SM the following    
crucial observation clarifies the above point beyond any doubt:
for $\xi_{Q}\neq 1$
the imaginary parts of the BFM electroweak self-energies
include terms with {\it unphysical thresholds}~\cite{Papavassiliou:1996fn,Papavassiliou:1996zn}.
For example, for the one-loop contributions of the $W$ and its associated
would-be Goldstone boson and ghost to 
$\tilde\Pi_{ Z  Z}^{(WW)}(\xi_{ Q},s)$
one obtains
\be
\Im m \tilde\Pi_{ Z  Z}^{(WW)}(s,\xi_{ Q})
= \Im m \widehat{\Pi}_{ Z  Z}^{(W  W)}(s) 
+ \frac{\alpha}{24 \sw^{2} \cw^{2}} 
\left(\frac{s - M_{ Z}^{2}}{s M_{ Z}^{4}}\right) 
\left[W_1(s)+W_2(s,\xi_{ Q})+W_3(s,\xi_{ Q})\right],
\label{Imunph}
\ee
with
\bea
W_1(s) &=& f_1(s) \theta(s - 4M_{ W}^{2}),
\label{W-1}\nonumber\\
W_2(s,\xi_{ Q}) &=&  f_2(s,\xi_{ Q})
\lambda^{1/2}(s,\xi_{ Q}M_{ W}^{2},\xi_{ Q} M_{ W}^{2})
\theta(s - 4\xi_{ Q}M_{ W}^{2}),
\label{W-2}\nonumber\\
W_3(s,\xi_{ Q}) &=& f_3(s,\xi_{ Q}) \lambda^{1/2}(s,M_{ W}^{2}, \xi_{ Q}M_{ W}^{2})
\theta(s - M_{ W}^{2}(1 + \sqrt{\xi_{ Q}})^{\!2}),
\label{W-3}
\eea
and
\bea
f_1(s) &=& \left(8 M_{ W}^{2} + s\right) \left(M_{ Z}^{2} + s \right)
+ 4M_{ W}^{2} \left(4M_{ W}^{2} + 3M_{ Z}^{2} + 2s \right),  
\label{f1}\nonumber\\
f_2(s,\xi_{ Q})  &=& f_1(s) -4 \left(\xi_{ Q}-1 \right)M_{ W}^{2} \left(4M_{ W}^{2} + M_{ Z}^{2} + s \right), 
\label{f2}\nonumber\\
f_3(s,\xi_{ Q}) &=& -2\left[8M_{ W}^{2} + s -2 \left(\xi_{Q}-1 \right)M_{ W}^{2}
+ \left(\xi_{ Q}-1 \right)^{2}M_{ W}^{4}s^{-1}\right] \left(M_{ Z}^{2} + s\right). 
\label{f3}
\eea
These gauge-dependent unphysical thresholds (see the arguments of the $\theta$ functions)  
are artifacts of the
BFM gauge fixing procedure,
and exactly cancel in the calculation of any physical process
against unphysical contributions from the imaginary parts of the 
one-loop vertices and boxes.
After these cancellations have been implemented
one is left just with the contribution proportional to
the tree level cross section for the on-shell physical process 
$\nu\bar\nu \to W^{+}W^{-}$, given in Eq.~(\ref{leftright}),
with thresholds only at $q^{2} = 4M_{W}^{2}$.
In fact, by obtaining in the previous subsection the full $W$-related contribution 
to the PT self-energy, namely  $\widehat{\Pi}^{ Z  Z}_{WW}(s)$, 
directly from the on-shell physical
process $\nu\bar\nu \to W^{+}W^{-}$,
we have shown explicitly that, in the BFM at $\xi_{Q} = 1$,
the thresholds that occur at $q^{2} = 4M_{W}^{2}$ 
are due {\it solely} to the {\it physical} $W^{+}W^{-}$ pair.
\newline
\indent
We therefore conclude that the particular value $\xi_{Q} = 1$ in the BFM
{\it is} distinguished on physical grounds from
all other values of $\xi_{Q}$. In the next section we will further elaborate on this point,
by exposing various  pathologies resulting in from the Dyson
summation of self-energies with unphysical thresholds.

\subsection {PT with massive fermions: an explicit example}
\begin{figure}[!t]
\bce
\includegraphics[width=11cm]{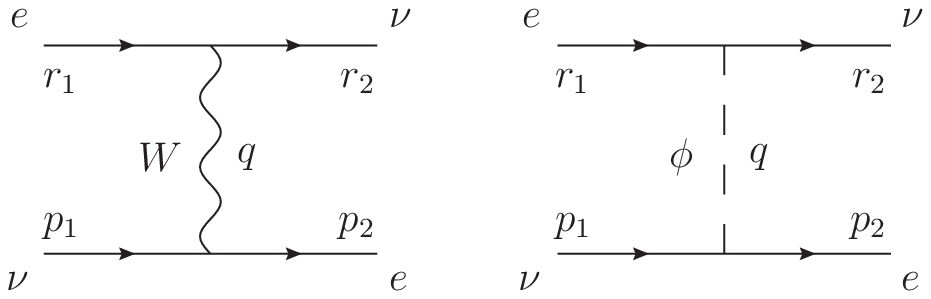}
\ece
\caption{\figlab{e_nu_tree-level}The process $e \nu_e\to e \nu_e$ at tree level in the SM.}
\end{figure}
\noindent
In this section, we discuss the technical subtleties encountered
in the application of the PT
when the fermions are massive. 
Consider the elastic process 
$e^{-}(r_1) \nu_{e}(p_1) \to e^{-}(p_2) \nu_{e}(r_2)$, and concentrate 
on the charged channel which, at tree-level, is shown in \Figref{e_nu_tree-level}. 
The momentum transfer $q$ is defined as $q=p_1-p_2= r_2-r_1$.
We will consider the electrons to be massive, with a mass $m_e$, while the 
neutrinos will be treated for simplicity as if they were massless. 
The tree-level propagators of the $W$ and the corresponding Goldstone 
boson are those given in Eq.~(\ref{GenProp-1}) and  Eq.~(\ref{Gold})
(for $i=W$); the index ``$W$'' will be suppressed in what follows. 
The elementary vertices describing the coupling of the charged bosons  
with the external fermions are 
\mbox{$\Gamma_{\alpha} \equiv\Gamma_{W^{+}_{\alpha}\bar{\nu}_e e} = \Gamma_{W^{-}_{\alpha}\bar{e} {\nu}_e }$}, 
\mbox{$\Gamma_{+}\equiv\Gamma_{\phi^{+}\bar{\nu}_e e }$}, and 
\mbox{$\Gamma_{-} \equiv \Gamma_{\phi^{-}\bar{e} {\nu}_e} $}, and are given by
\be
\Gamma_{\alpha} = \frac{ig_w}{\sqrt{2}} \gamma_{\alpha} P_{ L}\,, 
\,\,\,\,\,\,\,\,
\Gamma_{+(-)} = -\frac{ig_w}{\sqrt{2}} \frac{m_e}{M_{ W}}  P_{R (L)}\,. 
\label{ECV-3}
\ee
We also define the corresponding vertices sandwiched between the external spinors, \ie
\bea
\Gamma_{1}^{\alpha} &=& 
\bar{u}_{\nu_{e}}(r_2)\,\Gamma^{\alpha} u_{e}(r_1), 
\,\,\,\,\,\,\,\,\,
\Gamma_{2}^{\alpha} = 
\bar{u}_{e}(p_2)\,\Gamma^{\alpha}\, u_{\nu_{e}}(p_1), 
\label{GL0-2}\nonumber\\
\Gamma_{1} &=& 
\bar{u}_{\nu_{e}}(r_2)\, \Gamma_{+} \,u_{e}(r_1),
\,\,\,\,\,\,\,\,
\Gamma_{2} =
\bar{u}_{e}(p_2)\, \Gamma_{-}\, u_{\nu_{e}}(p_1). 
\label{GL0-4}
\eea
Note that both $\Gamma_{1}^{\alpha}$ and $\Gamma_{2}^{\alpha}$ contain 
a $P_{ L}$, whereas $\Gamma_{1}$ and $\Gamma_{2}$ 
a  $P_{ R}$ and a $P_{ L}$, respectively.
The subscripts $(1,2)$ are related to the 
electric charge carried by the $W^{\pm}$ entering into the 
corresponding tree-level vertices 
by setting $+\to 1$, $-\to 2$.
The following elementary identities
\bea
q_{\alpha}\Gamma_{1,2}^{\alpha} &=&  M_{ W}\Gamma_{1,2},  
\label{idaa}
\nonumber\\
i\Gamma_{1,2}  & = &   
 M_{ W} q^{\beta} \Delta_{\beta\alpha}(q)\Gamma_{1,2}^{\alpha} 
+ q^2 D(q) \Gamma_{1,2} , 
\label{idab}
\eea
valid for every $\xi_{ W}$ (which will be indicated simply as $\xi$ in what follows), will be  
frequently used [in deriving (\ref{idab}) we have used Eq.~(\ref{Id13})].
\newline
\indent
We will start by considering the $S$-matrix at tree-level (\Figref{e_nu_tree-level}),
to be denoted by $T_0$, given by
\be 
T_0 = \Gamma_{1}^{\alpha}\,\Delta_{\alpha\beta}(q) \,\Gamma_{2}^{\beta}
+ \Gamma_{1}\, D(q)\,\Gamma_{2}\,.
\label{T0}
\ee
Of course, $T_0$ must be $\xi$-independent, and it is easy to 
demonstrate that this is indeed so. 
There are three, algebraically equivalent but physically 
rather distinct, ways of writing the $\xi$-independent expression for $T_0$.

\begin{itemize}

\item[{\it i}.]  Using  Eqs~(\ref{AlgId}) and~(\ref{idaa}) we can 
see immediately that all dependence on $\xi$ cancels, and 
one can cast $T_0$ in terms of $\Delta_{\beta\alpha}^{\xi=1}(q)$ and 
$D^{\xi=1}(q)$ as follows 
\be 
T_0 = \Gamma_{1}^{\alpha} \Delta_{\alpha\beta}^{\xi=1}(q) \Gamma_{2}^{\beta}
+ \Gamma_{1} D^{\xi=1}(q)\Gamma_{2}.
\label{T0a}
\ee
The physical amplitude 
is the sum a massive gauge boson 
and a massive (would-be) Goldstone boson. The fact that the 
Goldstone boson is massive is, of course,  a consequence of the 
gauge-fixing used, namely the $R_{\xi}$-gauges.
\newline
\item[{\it ii}.] Using  Eqs~(\ref{Id1}) and~(\ref{idaa}), it is elementary 
to verify that $T_0$ can also be written as  
\be 
T_0 =  \Gamma_{1}^{\alpha} U_{\alpha\beta}(q) \Gamma_{2}^{\beta}.
\label{T0b}
\ee
Thus, even though one works in the $R_{\xi}$ gauge,  
making no assumption on the value of $\xi$ (in particular,
not taking the limit $\xi\to\infty$) one is led {\it effectively} 
to the unitary gauge, with no (unphysical) would-be Golstone bosons
present.  
\newline
\item[{\it iii}.] The third way of writing $T_0$ is slightly more subtle,
as far as its physical interpretation is concerned.
It is well-known (but often underemphasized) that  
the so-called ``spontaneous symmetry breaking'' is not actually ``breaking'' 
the local gauge symmetry, but simply realizing it in a different way.
Specifically, the WIs or STIs of the theory, which encompass the gauge symmetry 
at the level of Green's functions, maintain their form, at the expense of introducing 
massless longitudinal poles. The
role of these massless poles is obscured by the fact that,  
through the process of gauge fixing, they can be changed to poles of arbitrary mass
(as explained above).
These massless poles do not appear in the $S$-matrix, to the 
extent that they are absorbed by gauge bosons. 
However, simple algebra can recast the tree-level amplitude  
into a form where the presence of the massless poles becomes manifest. 
Using the algebraic identity 
\be
\frac{1}{M^{2}}= \frac{1}{q^{2}}+ \frac{q^{2}-M^{2}}{q^{2}M^{2}},
\label{AnotherId}
\ee
we can write $U_{\alpha\beta}(q)$ as 
\be
U_{\alpha\beta}(q) = {P}_{\alpha\beta}(q) d_W(q^{2}) 
+ \frac{q_{\alpha}q_{\beta}}{M^{2}_{ W}}\frac{i}{q^{2}},
\label{Upoles}
\ee
where we have used the transverse projector 
${P}_{\alpha\beta}(q)$ defined in Eq.~(\ref{projector}). 
Then Eq.~(\ref{T0b}) can be rewritten as
\be
T_0 = \Gamma_{1}^{\alpha} {P}_{\alpha\beta}(q)d_W(q^2) \Gamma_{2}^{\beta} + 
\Gamma_{1} \frac{i}{q^{2}}\Gamma_{2}.
\label{T0c}
\ee
\end{itemize}

As we will see later on, thanks to the PT,  the ways of writing the $S$-matrix given in Eqs~(\ref{T0b}) and~(\ref{T0c}) 
go through at one-loop, and eventually at all orders.  
\begin{figure}[!t]
\bce
\includegraphics[width=14cm]{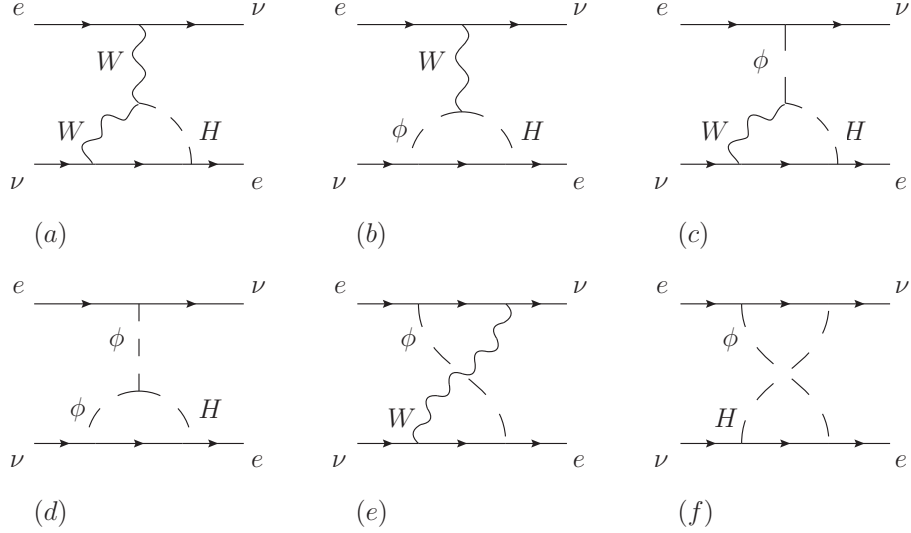}
\ece
\caption{\figlab{WH_box_diag}The subset of box and vertex diagrams containing a $W$ gauge boson and a Higgs field $H$.}
\end{figure}

\subsubsection{Gauge fixing parameter cancellations} 
\noindent
Let us now turn to the one-loop PT construction.
The main motivation is to construct via the PT 
the gfp-independent self-energies 
$\Pi_{W_\alpha W_\beta}$, 
$\Pi_{W_\alpha\phi}$, $\Pi_{\phi  W_\beta}$, 
and $\Pi_{\phi\phi}$, to be denoted by ${\widehat{\Pi}}_{\alpha\beta}$,
 ${\widehat{\Theta}}_{\alpha}$,
${\widehat{\Theta}}_{\beta}$, and $\widehat{\Omega}$, respectively, as well as
gfp-independent $W{f_{1}}\bar f_{2}$ and $\phi{f_{1}}\bar f_{2}$
vertices, which we denote by ${\widehat{\Gamma}}_{\alpha}$ and $\widehat{\Gamma}_{\pm}$,
respectively.
\newline
\indent
We will show the PT construction for a characteristic subset of diagrams
contributing to the amplitude $e^{-}(r_1) \nu_{e}(p_1) \to e^{-}(p_2) \nu_{e}(r_2)$.
Specifically, we will consider the 
subset of all Feynman graphs 
that contain, inside the loop, a $W$- and a $H$-propagator.
The relevant vertex and box diagrams are shown in \Figref{WH_box_diag} and
the self-energy diagrams in \Figref{WH_self-nrg_diag} .   
It is relatively easy to understand why this subset must be 
gfp-independent by itself: The dependence on the Higgs mass 
forces the gfp-cancellation to take place within this subset (as we will see, up to seagull-like terms). 
From the absorptive point of view,  
the graphs we consider display  
a threshold (\ie they develop imaginary parts if cut) at 
$q^2 \geq (M_{ W}+ M_{ H})^2$; therefore they should 
form a gfp-independent subset, since they cannot communicate 
with the rest (this absorptive argument does not apply to seagulls and tadpoles, 
since they do not have imaginary parts, but the PT construction takes care of them as well).
In what follows, we will use the sub- or super-script ``$W\!H$'' for the aforementioned subset; 
for example, ${\widehat{\Pi}}_{\alpha\beta}^{(W\!H)}$ denotes the subset of graphs contributing to 
the $WW$ self-energy that contain, in their loop, a $W$  or $\phi$ propagator 
and a Higgs-boson propagator (see \Figref{WH_self-nrg_diag}). 
\begin{figure}[t]
\bce
\includegraphics[width=16cm]{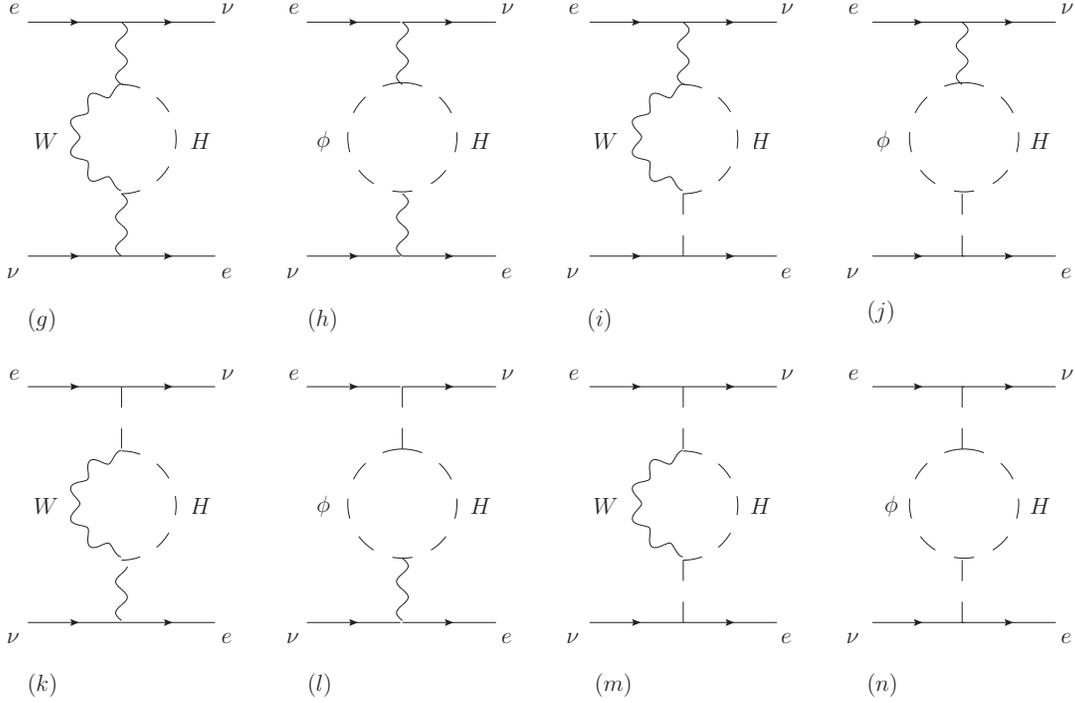}
\ece
\caption{\figlab{WH_self-nrg_diag}The self-energy diagrams containing a $WH$  or  $\phi H$ loop. The various SM self-energies are given by the diagrams' combinations $(g)+(h)=\Pi_{\alpha\beta}^{(W\!H)}$,  $(i)+(j)=\Theta_{\alpha}^{(W\!H)}$, $(k)+(l)=\Theta_{\beta}^{(W\!H)}$ and, finally, $(m)+(n)=\Omega^{(W\!H)}$.}
\end{figure}
\newline
\indent
We will introduce the following ingredients appearing in the intermediate steps 
of our demonstration:
\begin{itemize}

\item[\it{i}.] The coupling of the Higgs boson to the electrons 
$\Gamma_{ H}\equiv \Gamma_{H {\bar e} e}$ is given by
\be
\Gamma_{ H}= -i\frac{\gw}{2} \frac{m_e}{M_{ W}},
\ee
and we have that 
\be
\Gamma_{ H} P_{R (L)} = \frac{1}{\sqrt{2}}  \Gamma_{+(-)}\,. 
\label{HtoW-2}
\ee
\newline
\item[\it{ii}.] We set 
\be
A_{ W\! H}^{\xi} (q,k) =\left[(k^2-M_{ W}^2) (k^2-\xi M_{ W}^2)\left((k+q)^2 -  M_{ H}^2\right)\right]^{-1},
\ee 
and define the propagator-like structures
\bea
F_{ W\! H}(q) &=& (1-\xi) \gw^2 \int_k\! A_{ W\!  H}^{\xi} (q,k), 
\nonumber\\
J_{ W\! H}^{\alpha}(q) &=& (1-\xi)\gw^2 \int_k\!k^{\alpha} 
A_{ W\! H}^{\xi} (q,k),
\nonumber\\
K_{ W\! H}^{\alpha\beta}(q) &= & (1-\xi)\gw^2 \int_k  
k^{\alpha} k^{\beta} A_{ W\! H}^{\xi} (q,k),
\eea
the vertex-like structures
\bea
L_{ W\! H} (q,p_1) &= &  (1-\xi)\gw 
\int_k\! \left[\Gamma_{{ H}} S^{(0)}_e (p_1 + k)\Gamma_{-} \right] A_{ W\! H}^{\xi} (q,k), 
\nonumber\\
{\bar L}_{ W\! H} (q,r_2) &= &  (1-\xi)\gw 
\int_k\!\left[\Gamma_{+} S^{(0)}_e (r_2 + k)\Gamma_{{ H}}\right] 
A_{ W\! H}^{\xi} (q,k), 
\nonumber\\
N_{ W\! H}^{\alpha}(q,p_1)  &= & (1-\xi) \gw 
\int_k\! \left[\Gamma_{{ H}} S^{(0)}_e (p_1 + k)\Gamma_{-}\right]  
k^{\alpha} A_{ W\! H}^{\xi} (q,k),
\eea
and, finally, the box-like structure
\be
R_{ W\! H}(q,p_1,r_2) = (1-\xi)\,g_w \int_k\,
\left[\Gamma_{{ H}} S^{(0)}_e (p_1 + k)\Gamma_{-} \right]
\left[\Gamma_{+} S^{(0)}_e (r_2 + k) \Gamma_{{ H}} \right] A_{ W\! H}^{(\xi)} (q,k).
\ee
\end{itemize}
\indent
Then, we start with the vertex diagrams $(a)$ and $(c)$ in \Figref{WH_box_diag}
and we let the $\xi$-dependent longitudinal parts appearing in the tree-level $W$ trigger 
the WI of Eq.~(\ref{EWI}). Now that the electrons are considered to be massive, 
the term on the rhs in the square brackets of  Eq.~(\ref{EWI}) is turned on; as a result, and for the first time until now, 
the outcome of the pinching action is {\it not} only  propagator-like contributions: 
in addition, we obtain a vertex-like contribution, precisely due to the 
additional term in Eq.~(\ref{EWI}) proportional to the electron mass. 
As we will see, this vertex-like term will mix with the graphs $(b)$ and $(d)$, 
and will combine to form a $\xi$-independent vertex-like structure.  
\newline
\indent
Specifically, we have (suppressing a common factor $\Gamma^{\beta}_{1}\Delta_{\beta\alpha}(q)$ in front)
\bea
(a)^{\alpha} &=& 
(a)^{\alpha}_{\xi=1} + \left\{\frac{1}{2} M_{ W}  J_{ W\! H}^{\alpha}(q)\right\} 
\left(i\Gamma_{2}\right) 
- i M_{ W}^2 N_{ W\! H}^{\alpha}(q,p_1),
\label{ab-1}\nonumber\\
(b)^{\alpha} &=& 
(b)^{\alpha}_{\xi=1} + \frac{i}{2} q^{\alpha} M_{ W}^2 L_{ W\! H} (q,p_1) 
+ iM_{ W}^2 N_{ W\! H}^{\alpha}(q,p_1),
\label{ab-2}
\eea
We next turn to graphs $(c)$ and $(d)$.
As far as diagram $(c)$ is concerned, one of the
longitudinal momenta coming from the $W$-propagator will pinch as before; however, in addition,
we will use the  identity \mbox{$(2q+k)\cdot k = [(k+q)^2 - M_{ H}^2] + M_{ H}^2 - q^2$},
triggered when the second longitudinal momentum is contracted with the $\Gamma_{\phi W H}$ vertex.
The first term in this identity will cancel the $M_{H}$-dependent part appearing in 
$A_{ W\! H}^{\xi} (q,k)$, thus   
generating a term that is independent of $M_{ H}$. These terms cancel against other
similar terms, coming from the $M_{ H}$-independent 
Feynman graphs that are not considered, and will be discarded. Therefore, keeping only $M_{ H}$-dependent terms, we have 
[suppressing a  common factor $\Gamma_{1} D(q)$ in front]
\bea
(c) &=& (c)_{\xi=1} + \left\{-\frac{1}{4}(q^2 - M_{ H}^2) 
F_{ W H}(q)\right\} \left(i\Gamma_{2}\right) 
+ \frac{i}{2}(q^2 - M_{ H}^2) M_{ W} L_{ W H} (q,p_1),
\label{cd-1}\nonumber\\
(d) &=& (d)_{\xi=1} + \frac{i}{2} M_{ H}^2  M_{ W} L_{ W H} (q,p_1).
\label{cd-2}
\eea
Finally, following a similar methodology for the boxes $(e)$ and $(f)$, we have  
\bea
(e) &=& (e)_{\xi=1} + \left(i\Gamma_{1}\right)\left\{\frac{1}{4}F_{ W\! H}(q)\right\} 
\left(i\Gamma_{2}\right)
- R_{ W\! H}(q,p_1,r_2)
\nonumber \\
& +& \frac{1}{2}\left[ \Gamma_{1} M_{ W} L_{ W\! H} (q,p_1) 
+ \Gamma_{2} M_{ W} {\bar L}_{ W\! H} (q,r_2)\right],
\label{ef-1}\nonumber\\
(f) &=& (f)_{\xi=1} +  R_{ W\! H}(q,p_1,r_2).
\label{ef-2}
\eea
\indent
It is now straightforward to verify that: 
\begin{itemize}
\item[\it{i}.] the vertex-like $N_{ W\! H}^{\alpha}$ in~$(a)^{\alpha}$ and~$(b)^{\alpha}$, 
and the box-like $R_{ W\! H}$ in~$(e)$ and~$(f)$ 
cancel directly;
\newline
\item[\it{ii}.] the vertex-like terms 
proportional to $L_{ W H}$ in $(b)^{\alpha}$, $(c)$, $(d)$ and~$(e)$
cancel (after restoring the suppressed factors in front) by evoking the identity 
of Eq~(\ref{idab}); 
\newline
\item[{\it iii}.] the term proportional to $\bar L_{ W\! H}$ will 
cancel, in exactly the same way described above, against the contributions coming from the mirror vertex
graphs, not shown. 
\end{itemize}
\indent
Thus one is left only with $\xi$-dependent propagator-like pieces, contained in the 
curly brackets in  Eqs~(\ref{ab-1}), (\ref{cd-1}) and~(\ref{ef-1}), 
together with the contributions coming from the mirror vertex graphs;
the latter are identical to those already identified, up to 
trivial adjustments. 
All aforementioned terms will cancel exactly against the  $\xi$-dependent parts
of the conventional self-energy graphs, shown in \Figref{WH_self-nrg_diag}. 
\newline
\indent
Let us now turn to this remaining cancellation. Separating out the 
contributions at $\xi=1$ from the rest, we have for the self-energy graphs of  \Figref{WH_self-nrg_diag} 
\bea
(g)^{\alpha\beta}+ (h)^{\alpha\beta} &=& (g)^{\alpha\beta}_{\xi=1} + (h)^{\alpha\beta}_{\xi=1}\nonumber \\
&-&\frac{1}{4} q^{\alpha} q^{\beta} M_{ W}^2 F_{ W\! H}(q) 
- \frac{1}{2} M_{ W}^2 
\left[q^{\alpha} J_{ W\! H}^{\beta}(q)+ q^{\beta} J_{ W\! H}^{\alpha}(q)\right],
\label{ppp-1}\nonumber\\
(i)^{\alpha}+ (j)^{\alpha} &=& (i)^{\alpha}_{\xi=1} + (j)^{\alpha}_{\xi=1} 
- \frac{1}{2} q^2 M_{ W}  J_{ W\! H}^{\alpha}(q) - \frac{1}{4} q^{\alpha} 
M_{ W} M_{ H}^2 F_{ W\! H}(q),
\label{ppp-2}\nonumber\\
(k)^{\beta}+ (l)^{\beta} &=& (k)^{\beta}_{\xi=1} + (l)^{\beta}_{\xi=1}
- \frac{1}{2} q^2 M_{ W}  J_{ W\! H}^{\beta}(q) - \frac{1}{4} q^{\beta} 
M_{ W} M_{ H}^2 F_{ W\! H}(q),
\label{ppp-3}\nonumber\\
(m)+(n) &=& (m)_{\xi=1}+(n)_{\xi=1} + 
\frac{1}{4} q^{2} (q^{2} - 2 M_{ H}^2) F_{ W\! H}(q) .
\label{ppp-4}
\eea
Using again the identity of Eq.~(\ref{idab}) 
one may separate, unambiguously, the $\xi$-dependent propagator like pieces 
from  Eqs~(\ref{ab-1}), (\ref{cd-2}) and~(\ref{ef-2}) into 
$WW$, $W \phi$, $\phi W$, and $\phi\phi$ structures, and add them 
to the corresponding contributions in the equation above. 
It is then straightforward to verify that 
a complete cancellation of all $\xi$-dependent terms takes place; 
in fact, the terms
proportional to $F_{ W\! H}$  and $J_{ W\! H}^{\alpha}$ cancel separately.
Even though we have restricted ourselves to the subset of Feynman diagrams 
that depend explicitly on  $M_{ H}$, the methodology presented goes 
through, unchanged, also for the remaining graphs. 
We emphasize again that, as in all previous examples,  
the conceptual and technical advantage  of this demonstration lies  precisely in the fact that 
all cancellations take place systematically, by identifying the 
appropriate kinematic structures, with no need to carry out any integrations.  
\newline
\indent
Notice, finally, the following important point: all aforementioned cancellations take place 
{\it inside} the loops, {\it without} touching the  $\xi$-dependence of the (external) bare propagators
attached to the external fermions; indeed, all we have used, in addition to pinching,
is the algebraic identity of Eq~(\ref{idab}), which is valid for every   $\xi$.
As we will see shortly, after the completion of the PT procedure at one-loop,  
the requirement that this residual $\xi$-dependence 
also cancels imposes Abelian-like WIs on the PT Green's functions. 

\subsubsection{Final rearrangement and comparison with the background Feynman gauge}
\noindent
Let us now turn again to the subset of graphs considered above.
As we have demonstrated, inside the loops all propagators have been dynamically reduced   
to the Feynman gauge, $\xi=1$.
At this point the genuine box contributions have been isolated; thus, the 
$W\!H$-part of the one-loop PT box is given simply by $(e)_{\xi=1}+ (f)_{\xi=1}$.
\newline
\indent
To get the corresponding part of the one-loop PT vertices and  
self-energies, an additional step is required: we must extract 
from the vertex graphs in the Feynman gauge possible propagator-like
pieces generated by the momentum-dependent vertices. 
For the case at hand, the only graph that can furnish such a contribution 
is ($c$); the propagator-like piece is generated when 
the longitudinal momentum $k_{\mu}$, coming from the 
elementary vertex $\Gamma_{\phi W H} \propto (2q+k)_{\mu}$ 
in ($c$) is allowed to pinch, according to our earlier general discussion [subsection~\ref{sec:gencons} point ({\it i})]
(Of course, had we considered the entire set of vertex diagrams
then the three-boson vertices $\Gamma_{ZWW}$ and $\Gamma_{AWW}$ should also undergo 
the standard PT splitting).
\newline
\indent
Then, separate $(c)_{\xi=1}$ into the purely vertex-like part,  denoted by $(c)_{\xi=1}^{\rm v}$, 
and the propagator-like part, $(c)_{\xi=1}^{\rm se}$, 
\be
(c)_{\xi=1} = (c)_{\xi=1}^{\rm v} + (c)_{\xi=1}^{\rm se}
\label{cvertcprop1}
\ee
with
\bea
(c)_{\xi=1}^{\rm v} &=&  \frac{\gw}{2}   
\int_k\! \left[\Gamma_{{ H}} S^{(0)}_e (p_1 + k)\Gamma_{\mu} (2 q^{\mu})\right] 
d_{ W}(k^2) \Delta_{ H}(k+q)
\nonumber\\
&+& 
\frac{\gw}{2} M_{ W} \int_k\!\left[\Gamma_{{ H}} S^{(0)}_e (p_1 + k) \Gamma_{-}\right] 
d_{ W}(k^2) \Delta_{ H}(k+q) 
\label{cvertcprop2-1}\nonumber\\
(c)_{\xi=1}^{\rm se} &=& \frac{g_w^2}{4}
  I_{ W  H}(q) \left(i\Gamma_{2}\right).
\label{cvertcprop2-2}
\eea
and
\be
I_{ W\! H}(q) = \int_k d_{ W}(k^2) \Delta_{ H}(k+q) 
= \int_k \frac{1}{(k^2 - M_{ W}^2)[(k+q)^2- M_{ H}^2]}.
\label{iWH}
\ee
Then, the  $W\!H$-parts of the 
one-loop $\Gamma_{W_{\mu}^{-}\bar{e} {\nu}_e}$ and $\Gamma_{\phi^{-}\bar{e} {\nu}_e}$ PT vertices,
to be denoted by ${\widehat \Gamma}_{\alpha}^{(W\!H)}$ and  
${\widehat \Gamma}_{-}^{(W\!H)}$, respectively,  
are given schematically by  
\bea 
{\widehat \Gamma}_{\alpha}^{(W\!H)} &=&  (a)_{\alpha}^{\xi=1} + (b)_{\alpha}^{\xi=1},
\label{vertpt-1}\nonumber\\ 
{\widehat \Gamma}_{-}^{(W\!H)}&=&  (c)_{\xi=1}^{\rm v} + (d)_{\xi=1}.
\label{vertpt-2}
\eea 
\newline
\indent
Finally, to obtain the corresponding parts of the one-loop PT self-energies,
to be denoted by $\widehat{\Pi}_{\alpha\beta}^{({W\!H})}$, 
$\widehat{\Theta}_{\alpha}^{({W\!H})}$,
$\widehat{\Theta}_{\beta}^{({W\!H})}$, and $\widehat{\Omega}^{({W\!H})}$, we must use 
 Eq.~(\ref{idab}) in order to distribute among the different self-energies the contributions coming from 
 $(c)_{\xi=1}^{\rm{se}}$ and its mirror graph. 
Evidently, these terms will not contribute anything to 
$\widehat{\Pi}_{\alpha\beta}^{({W\!H})}$, and therefore
$\widehat{\Pi}_{\alpha\beta}^{({W\!H})}$ will 
be identical to the conventional 
one-loop $\Pi_{\alpha\beta}^{(W\!H)}$ (of course, the non $W\!H$-parts will differ).
Then, we will have
\bea 
\widehat{\Pi}_{\alpha\beta}^{(W\!H)}(q) &=& (g)_{\alpha\beta}^{\xi=1} + (h)_{\alpha\beta}^{\xi=1},
\label{propspt-1}\nonumber\\ 
\widehat{\Theta}_{\alpha}^{(W\!H)}(q) &=& (i)_{\alpha}^{\xi=1} + (j)_{\alpha}^{\xi=1}
+ \frac{\gw^2}{4}  M_{ W}  I_{{W\!H}}(q)  q_{\alpha},
\label{propspt-2}\nonumber\\ 
\widehat{\Theta}_{\beta}^{(W\!H)}(q) &=& (k)_{\beta}^{\xi=1} + (l)_{\beta}^{\xi=1}
+ \frac{\gw^2}{4}  M_{ W}   I_{{W\!H}}(q) q_{\beta},
\label{propspt-3}\nonumber\\ 
\widehat{\Omega}^{(W\!H)}(q)&=& (m)^{\xi=1}+(n)^{\xi=1} + 
\frac{\gw^2}{2} q^2  I_{{W\!H}}(q) .
\label{propspt-4}
\eea 
\indent
It is a relatively straightforward exercise to verify that the 
parts of the PT Green's functions constructed 
above {\it coincide} with the corresponding quantities calculated in the BFG; 
of course, this coincidence holds for the entire Green's functions, 
and it is not restricted to the $W\!H$-parts analyzed here.
\newline
\indent
In the case of ${\widehat \Gamma}_{\alpha}^{({W\!H})}$ 
given in (\ref{vertpt-1}), the coincidence with the BFG is obvious;
the external $W$'s may be converted into $\widehat W$'s for free,
since the corresponding (lowest-order) vertices are identical
in the $R_{\xi}$ and the BFM gauges.  
The case of ${\widehat \Gamma}_{-}^{({W\!H})}$ is more interesting;
the coincidence with the BFG takes place because of the 
extraction of the propagator-like piece $(c)_{\xi=1}^{\rm se}$ 
from $(c)_{\xi=1}$, as described in Eq.~(\ref{cvertcprop1})
and (\ref{cvertcprop2-2}). 
Specifically, the purely vertex-like piece 
$(c)_{\xi=1}^{\rm v}$ in Eq.~(\ref{cvertcprop2-1})
consists of two parts,
the first one corresponds to graph $(c)$ computed in the BFG [indeed, the 
factor $(2 q^{\mu})$ is proportional to the bare BFM vertex  $\Gamma_{\widehat\phi W\!H}$],     
while the second part, when added to $(d)_{\xi=1}$, furnishes the 
graph $(d)$ computed in the BFG, given that the BFM vertex $\Gamma_{\widehat\phi \phi H}$
and the $R_{\xi}$ vertex $\Gamma_{\phi \phi H}$ are related by 
$\Gamma_{\widehat\phi \phi H} = \Gamma_{\phi \phi H} + i\gw \xi_{Q} M_{ W}/2$. 
\newline
\indent
Turning to the self-energies of Eq.~(\ref{propspt-3}), 
$\widehat{\Pi}_{\alpha\beta}^{({W\!H})}$ coincides with the corresponding 
BFM quantity, for the same reason as in the case of the ${\widehat \Gamma}_{\alpha}^{({W\!H})}$ vertex:
the elementary vertices appearing in $(g)$ and $(h)$ coincide in both 
gauge-fixing schemes, $R_{\xi}$ and BFM. 
As for $\widehat{\Theta}_{\alpha}^{({W\!H})}$, the term 
$(\gw^2/4)  M_{ W}   I_{ W \! H}(q) q_{\beta}$ accounts
precisely for the difference between $\Gamma_{\widehat\phi \phi H}$ and $\Gamma_{\phi \phi H}$.
Finally, the case of $\widehat{\Omega}^{(W\! H)}$ is slightly more involved:
to demonstrate the equality one must write the $(2q+k)^2$ appearing in $(m)$ in the form
\be
(2q+k)^2 = 2 q^2 + 2 [(k+q)^2 - M_{ H}^2] - (k^2-M_{ W}^2)
+ (2  M_{ H}^2 -M_{ W}^2)\,.
\ee
Then for the terms on the rhs we have that 
the first is added to $(\gw^2/2) q^2  I_{ W \! H}$ and   
furnishes graph $(m)$ in the BFG; 
the second 
combines with other $M_{ H}$-independent parts 
(note that in the BFM we have also the coupling 
$\Gamma_{{\widehat\phi}^{+}{\widehat\phi}^{-} u^{\pm} {\bar u}^{\mp}} \propto -\xiQ/2$; 
the third cancels with the 
$(\gw^2/2) q^2  I_{ W \! H}$; the third converts the seagull graph containing a Higgs-boson 
propagator (not shown) 
to the same graph in the BFM, given that
$\Gamma_{{\widehat\phi}^{+}{\widehat\phi}^{-} HH} = \Gamma_{{\phi}^{+}{\phi}^{-} HH} - i\gw^2 \xiQ/2$;
finally, the last term converts $(n)$ into its BFM counterpart. 

\subsubsection {Deriving Ward identities from the gfp-independence of the $S$-matrix.}

\begin{figure}[t]
\includegraphics[width=16cm]{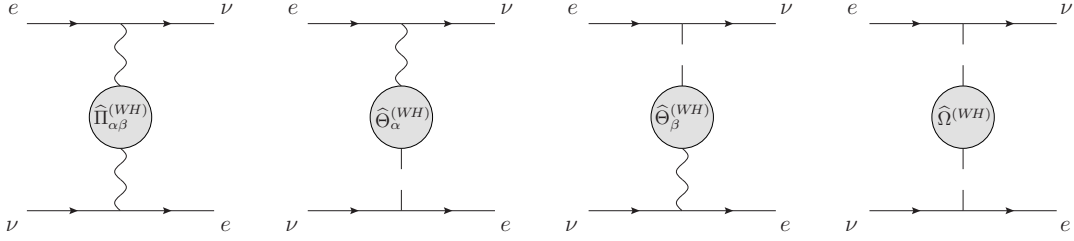}
\caption{\figlab{PT_self-nrg}The (one-loop) $\xi$-independent PT self-energies (gray blobs); 
the tree-level propagators are still $\xi$-dependent. By requiring that any gfp-dependence coming from  these  tree-level propagators 
must cancel, imposes a set of non-trivial WIs on the one-loop PT self-energies (and vertices).}
\end{figure}
\noindent
In the previous subsections we showed explicitly how the application of the PT
gives rise to gfp-independent self energies, vertices, and boxes.
As already emphasized there, the gfp-cancellation proceeded without reference 
to the tree-level propagators connecting the one-loop graphs to the external fermions. 
Any gfp-dependence coming from  these  tree-level propagators [see  Eqs~(\ref{GenProp-1}) and~(\ref{Gold})]
must also cancel, in order to obtain fully gfp-independent subamplitudes
${\widehat{T}}_{1}$ and ${\widehat{T}}_{2}$ 
(${\widehat{T}}_{3}$ being box-like does not have external propagators and is already fully gfp-independent).
It turns out that, quite remarkably, 
the requirement of this final gfp-cancellation imposes 
a set of non-trivial WIs on the one-loop PT self-energies and 
vertices~\cite{Papavassiliou:1989zd,Papavassiliou:1994pr}.
These 
WIs are identical to those obtained some years later within the BFM~\cite{Denner:1994xt},  
but are derived through a procedure that has no apparent connection 
with the BFM; all that one evokes really is the  full gfp-independence of the $S$-matrix.
Actually, this $S$-matrix derivation 
could be considered as an 
all-order proof of the above WIs, assuming that 
the various Green's functions [the gray blobs in \Figref{PT_self-nrg})]
can be made gfp-independent to all orders.
\newline
\indent
Let us see how the WIs for the self-energies are derived 
from the gfp-independence of ${\widehat{T}}_{1}$. 
Neglecting tadpole contributions from the external fermions, 
we have that ${\widehat{T}}_{1}$ is given by 
\bea
{\widehat{T}}_{1} &=& \Gamma^{\mu}_{1}
\Delta_{\mu\alpha}(q){\widehat{\Pi}}^{\alpha\beta}(q)
\Delta_{\beta\nu}(q) \Gamma^{\nu}_{2}
+ \Gamma_{1} D(q)\, \widehat{\Omega}(q)\, D(q)\Gamma_{2}
\nonumber\\
&+& \Gamma^{\mu}_{1}
\Delta_{\mu\alpha}(q) {\widehat{\Theta}}^{\alpha}(q) D(q) \Gamma_{2}
+ \Gamma_{1}D(q) {\widehat{\Theta}}^{\beta}(q)\Delta_{\beta\nu}(q) 
\Gamma^{\nu}_{2},
\label{T1}
\eea
or, after using Eq.~(\ref{Id1}),
\bea
{\widehat{T}}_{1}&=& \Gamma^{\mu}_{1}
\left[U_{\mu\alpha}(q) - \frac{q_{\mu}q_{\alpha}}{M^{2}_{ W}}D(q)\right]
{\widehat{\Pi}}^{\alpha\beta}(q)
\left[U_{\beta\nu}(q) - \frac{q_{\beta}q_{\nu}}{M^{2}_{ W}}D(q)\right ]
\Gamma^{\nu}_{2}
\nonumber\\
&+& \Gamma^{\mu}_{1}
\left[U_{\mu\alpha}(q) - \frac{q_{\mu}q_{\alpha}}{M^{2}_{ W}}D(q)\right]
{\widehat{\Theta}}^{\alpha}(q) D(q) \Gamma_{2}\nonumber \\
&+&\Gamma_{1}D(q){\,\widehat{\Theta}}^{\beta}(q)
\left[U_{\beta\nu}(q) - \frac{q_{\beta}q_{\nu}}{M^{2}_{ W}}D(q)\right]
\Gamma^{\nu}_{2}+ \Gamma_{1} D(q)\widehat{\Omega}(q) D(q) \Gamma_{2}.
\label{PCon}
\eea
This way of writing ${\widehat{T}}_{1}$ has the advantage of isolating all residual 
$\xi$-dependence inside the propagators $D(q)$.
Demanding that ${\widehat{T}}_{1}$ should be  $\xi$-independent, we obtain
as a condition for the cancellation of the terms quadratic in $D(q)$ 
\be
q^{\beta}q^{\alpha}{\widehat{\Pi}}_{\alpha\beta}(q)
-2 M_{ W} q^{\alpha}{\widehat{\Theta}}_{\alpha}(q) 
+ M^{2}_{ W} \widehat{\Omega}(q)=0,
\label{PWI}
\ee
while for the cancellation of the linear terms we must have 
\be
q^{\alpha}{\widehat{\Pi}}_{\alpha\beta}(q) - M_{ W}{\widehat{\Theta}}_{\beta}(q)=0.
\label{PWI1}
\ee
From  Eqs~(\ref{PWI}) and~(\ref{PWI1}) it follows that
\be
q^{\beta}q^{\alpha}{\widehat{\Pi}}_{\alpha\beta}(q)= M^{2}_{ W} \widehat{\Omega}(q),
\label{PWI2}
\ee
and
\be
q^{\alpha}{\widehat{\Theta}}_{\alpha}(q) = M_{ W} \widehat{\Omega}(q).
\label{PWI3}
\ee
 Eqs~(\ref{PWI}) and~(\ref{PWI3}) are the announced WIs. 
Applying an identical procedure for  ${\widehat{T}}_{2}$ one obtains the corresponding WI 
relating the one-loop PT vertices ${\widehat{\Gamma}}_{\alpha}$ and $\widehat{\Gamma}_{\pm}$.
\newline
\indent
Finally, the gfp-independent ${\widehat{T}}_{1}$ is given by
\be
{\widehat{T}}_{1}= \Gamma^{\mu}_{1}
U_{\mu\alpha}(q) \,{\widehat{\Pi}}^{\alpha\beta}(q)
U_{\beta\nu}(q)\Gamma^{\nu}_{2}.
\label{RRB}
\ee
Notice that Eq.~(\ref{RRB}) is the one-loop generalization of  Eq.~(\ref{T0b}). 
\newline
\indent
We can now use the WIs derived above in order 
to reformulate the $S$-matrix in a very particular way; 
specifically, we will show that the higher-order physical amplitude given above  
may be cast in the tree-level form of Eq.~(\ref{T0c}).
Such a reformulation gives rise to a
new transverse gfp-independent $W$ self-energy
${\widehat{\Pi}}_{\alpha\beta}^{t}$ with a gfp-independent longitudinal part, 
exactly as in  Eq.~(\ref{T0c}).
To be sure, the cost of such a reformulation
is the appearance of {\it massless} Goldstone poles in our expressions. However,
since both the old and the new quantities originate from the same
{\it unique} $S$-matrix, all poles introduced by this reformulation will cancel
against each other, because the $S$-matrix contains no massless poles to
begin with.
\newline
\indent
To see how this works out, write ${\widehat{\Theta}}_{\alpha}$ in the form
\be
{\widehat{\Theta}}_{\alpha}(q)=q_{\alpha}\widehat{\Theta}(q)\,;
\label{TensProp}
\ee
from (\ref{PWI3}) follows that 
\be
\widehat{\Theta}(q)= \frac{M_{ W}}{q^{2}} \widehat{\Omega}(q)
\label{TTT}
\ee
Then, we can define 
${\widehat{\Pi}}_{\alpha\beta}^{t}(q)$ in terms of
${\widehat{\Pi}}^{\alpha\beta}(q)$ and $\widehat{\Theta}(q)$ as follows:
\be
{\widehat{\Pi}}_{\alpha\beta}^{\rm t}(q) =
{\widehat{\Pi}}_{\alpha\beta}(q)- \frac{q_{\alpha}q_{\beta}}{q^{2}}M_{ W} \widehat{\Theta}(q)
\label{TransProp}
\ee
Evidently ${\widehat{\Pi}}_{\alpha\beta}^{\rm t}(q)$ is transverse, \eg
$q^{\alpha}{\widehat{\Pi}}_{\alpha\beta}^{t}(q)=q^{\beta}{\widehat{\Pi}}_{\alpha\beta}^{\rm t}(q)=0$.
Moreover, using  Eqs~(\ref{PWI1}) and~(\ref{TTT}),
\be
{\widehat{\Pi}}_{\alpha\beta}^{\rm t}(q) =
P_{\alpha\mu}(q) {\widehat{\Pi}}^{\mu\nu}(q) P_{\beta\nu}(q).
\label{TransProp2}
\ee
\newline
\indent
We may now re-express ${\widehat{T}}_{1}$ of (\ref{RRB}) in terms of
${\widehat{\Pi}}_{\alpha\beta}^{t}$ and $\widehat{\Omega}$; using (\ref{Upoles}) and (\ref{idaa}) we have 
\be
{\widehat{T}}_{1} = 
{\Gamma}^{\alpha}_{1}d_W(q^2){\widehat{\Pi}}_{\alpha\beta}^{\rm t}(q)d_W(q^2) {\Gamma}^{\beta}_{2} +
\Gamma_{1}\frac{i}{q^{2}}\widehat{\Omega}(q)\frac{i}{q^{2}}\Gamma_{2}.
\label{T1Ref}
\ee
\indent
Eq.~(\ref{T1Ref}) is the generalization of Eq.~(\ref{T0c}):
${\widehat{T}}_{1}$ is the sum of two self-energies, one
corresponding to a {\it transverse massive vector field} and one 
to a {\it massless Goldstone boson}. 
It is interesting to notice that the above 
rearrangements have removed the mixing terms ${\widehat{\Theta}}_{\alpha}$ and
${\widehat{\Theta}}_{\beta}$ between $W$ and $\phi$, thus leading to the  
generalization of the well known tree-level property of the $R_{\xi}$ gauges to higher orders.
It is important to emphasize again that the massless poles in the above
expressions would not have appeared had we not insisted on the 
transversality of the $W$ self-energy (or the vertex); notice in particular that they are 
{\it not} related to any particular gauge choice, such as the Landau gauge
($\xi=0$). 
A completely analogous procedure may be followed for 
the one-loop (and beyond) vertex \cite{Papavassiliou:1989zd}, yielding the 
corresponding Abelian-like WI; as in the case of te self-energy studied above, 
the WI of the vertex is realized by means  of massless Goldstone bosons.
\newline
\indent
The rearrangement of the $S$-matrix carried out above,
in additional to the conceptual transparency that it provides, 
brings about considerable calculational simplifications, since it organizes the
transverse and longitudinal pieces in individually gauge-invariant blocks.
As was first recognized in~\cite{Papavassiliou:1995rj},
this is particularly economical if one is only interested in
gauge-invariant longitudinal
contributions, \eg in the context of resonantly enhanced $CP$ violation.

\newpage


\section{\seclab{Applications - I}Applications - I}
\noindent
In this section we will present a variety
of phenomenological applications of the PT. 
Specifically, we will focus on the following representative topics: 

\begin{itemize}

\item[{\it i}.] 
The field-theoretic construction 
and the observable nature of the PT effective charges, as well as   
the conceptual and practical advantages of 
the physical renormalization schemes, 
which use these  effective charges,   
over the unphysical schemes, such as the popular $\overline{MS}$.
\newline
\item[{\it ii}.] The definition and measurement 
of gauge-invariant off-shell form-factors, with particular emphasis 
on the neutrino charge radius.
\newline
\item[{\it iii}.] The gauge-invariant definition of basic electroweak 
parameters, such as the S,T, and U, 
and the universal part of the $\rho$ parameter. 
\newline
\item[{\it iv}.] The gauge-invariant resummation formalism for resonant transition amplitudes. 

\end{itemize}

\subsection{Non-Abelian effective charges}
\noindent
The  possibility  of extending  the  concept  of  an effective  charge~\cite{GellMann:1954fq}
from QED to non-Abelian gauge
theories  is  of fundamental  interest  for  at  least three  reasons.
First, in QCD, the existence of an effective charge
analogous to that of QED  is explicitly assumed in renormalon studies
~\cite{Hooft:1979aa,David:1985xj,Beneke:1994sw}.  
However, in the absence of a concrete 
guiding principle (such as the PT), the diagrammatic identification of the 
subset of  (conformally-variant) corrections that should be resummed is rather obscure.
Second,  in theories
involving unstable  particles (\eg in the electroweak SM)
the Dyson summation of (appropriately defined) self-energies is needed, in order 
to  regulate   the
kinematic 
singularities of the corresponding  tree-level
propagators in the vicinity of resonances
~\cite{Pilaftsis:1989zt,Sirlin:1991fd,Sirlin:1991rt,Baur:1995aa}. 
Third,  in theories  involving  disparate energy  scales (\eg  grand
unified  theories) the extraction  of accurate  low-energy predictions
requires  an  exact  treatment  of  threshold  effects  due  to  heavy
particles~\cite{Ellis:1990wk,Amaldi:1991cn,Langacker:1993xb}.  
The construction of  
effective  charges, valid  for all  momenta $q^{2}$  and not  just the
asymptotic    regime   governed    by   the  
$\beta$-functions,  constitutes the  natural way  to
account  for such threshold  effects.  In  all cases,  the fundamental
problem is  the gfp-dependence of  the conventionally defined gauge
boson  self-energies.
The PT cures this  problem and leads to the definition of
physical effective charges, both in QCD and the electroweak sector of
the SM.

\subsubsection{QED effective charge: the prototype}
\noindent
The quantity that serves as the field-theoretic prototype 
for guiding our analysis 
is the effective charge of QED. In the rest of this section, a `0' super or subscript will indicate (bare) unrenormalized quantities.
\newline
\indent
In QED consider the unrenormalized photon self-energy   $\Pi_0^{\mu\nu}(q) =  P^{\mu\nu}(q)\Pi_0(q^2)$, 
where  $\Pi_0 (q^2)$ has dimensions of mass squared, 
and is gfp-independent to all  orders in perturbation theory.
One usually sets $\Pi(q^2) = q^2 {\bf \Pi}(q^2)$, 
where the dimensionless quantity ${\bf \Pi}(q^2)$ is referred to 
as the ``vacuum polarization''.
Carrying out the  standard  Dyson   summation,  
we obtain   the  dressed photon propagator between conserved external currents,
\be
\Delta_0^{\mu\nu}(q^2)\ = \frac{g^{\mu\nu}}{q^2[1+{\bf \Pi}_0(q^2)]}.
\label{qed0}
\ee
$\Delta_0^{\mu\nu}(q^2)$ is renormalized multiplicatively according to 
$\Delta_0^{\mu\nu}(q^2) = Z_{ A} \Delta^{\mu\nu}(q^2)$, where 
$Z_{ A}$ is the wave-function renormalization of the photon 
($A_0 = Z_{ A}^{1/2} A$). Imposing the on-shell renormalization condition 
for the photon we obtain 
\be 
1+ {\bf \Pi}(q^2)= Z_{ A} [1+{\bf \Pi}_0(q^2)], 
\ee
where $Z_{ A} = 1-{\bf \Pi}_0(0)$, and ${\bf \Pi}(q^2) = {\bf \Pi}_0(q^2)-{\bf \Pi}_0(0)$; 
clearly ${\bf \Pi}(0) = 0$.
\newline
\indent
The renormalization procedure introduces, in  addition, 
the standard relations  between  renormalized and unrenormalized electric charge,   
\be
e = Z_{e}^{-1} e^0 = Z_f Z_{ A}^{1/2} Z_1^{-1} e^0,
\label{qed1}
\ee
where $Z_{e}$ is the charge renormalization constant,  
$Z_f$ the wave-function renormalization 
constant of the fermion, and $Z_1$ the vertex renormalization.
\newline
\indent
The  Abelian symmetry  of the   theory   gives  rise   to  the well-known  
WI [given also in (\ref{vertQED})]
\be
q^{\mu}\Gamma^0_{\mu}(p,p+q)= S_0^{-1}(p+q)-S_0^{-1}(p), 
\label{qed2}
\ee
where
$\Gamma^0_{\mu}$ and  $S_0 (k)$ are the  unrenormalized one-loop 
photon-electron vertex and electron propagator, respectively.
The requirement  that     the   renormalized vertex       $\Gamma_{\mu}  =
Z_{1}\Gamma^0_{\mu}$   and  the  renormalized  self-energy   $S  =
Z_{f}^{-1} S_0$ should satisfy the same WI imposes the equality
\be
Z_{1}=Z_{f}, 
\label{qed3}
\ee
from which it immediately follows that 
\be
Z_{e}\ =\ Z_{ A}^{-1/2}. 
\label{qed4}
\ee
\newline
\indent
Given these relations between the renormalization constants,
we can now form the following renormalization group (RG) invariant combination:
\begin{equation}
{R}_0^{\mu\nu}(q^2) =
 \frac{(e^0)^2}{4\pi} \Delta_0^{\mu\nu}(q^2)
 =\frac{e^2}{4\pi} \Delta^{\mu\nu}(q^2) = {R}^{\mu\nu}(q^2).
\label{qed5}
\end{equation}   
From ${R}_{\mu\nu}(q^2)$, after pulling out a 
the trivial kinematic factor $(1/q^2)$, one may define 
the  QED {\it effective charge}  ${\alpha}_{\rm eff} (q^2)$, namely 
\be
{\alpha}_{\rm eff} (q^2) = \frac{\alpha}{[1+{\bf \Pi}(q^2)]},
\label{alphaqed}
\ee
where $\alpha$ is the fine-structure constant.  
\newline
\indent
The  QED effective charge of (\ref{alphaqed}) has the following crucial properties:
\begin{itemize}

\item[{\it i}.] It is gfp-independent, to all orders in perturbation theory.
\newline
\item[{\it ii}.] It is RG-invariant by virtue of the WI of \ref{qed2} and the resulting 
relation (\ref{qed4}).
\newline
\item[{\it iii}.] Given that ${\bf \Pi}(0)=0$, 
at low energies the effective charge matches the
fine structure constant: $\alpha_{\rm eff}(0) = \alpha = 1/137.036\cdots$.
\newline
\item[{\it iv}.] For asymptotically large values of $q^2$, \ie for  $q^2\gg  m^2_f$, where  $m_f$ denotes 
the masses of the fermions contributing to the vacuum polarization loop ($f=e,\mu,\tau$,...),
${\alpha} (q^2)$ matches the {\em running coupling}
$\bar\alpha(q^{2})$ defined from the RG: at the one-loop level,
\be
\alpha_{\rm eff}(q^2) 
\stackrel{q^2\gg  m^2_f}{\longrightarrow} 
\bar\alpha(q^{2}) = 
\frac{\alpha}{1 - (\alpha\beta_{1}/2\pi)\log(q^{2}/m_{f}^{2})},
\ee
where $\beta_{1} = \frac{2}{3}n_{f}$ is the coefficient of the QED
$\beta$ function for $n_{f}$ fermion species.
\newline
\item[{\it v}.]  The effective charge has a non-trivial dependence on the masses $m_f$, 
which   allows its   reconstruction  from  physical amplitudes 
by resorting to the OT and analyticity, \ie dispersion 
relations.  Specifically, given a particular contribution to the spectral function $\Im m {\bf \Pi}(s)$,
the corresponding contribution to 
${\bf \Pi}(q^2)$ can be reconstructed via a {\it once-subtracted dispersion relation} (see, e.g.~\cite{deRafael:1976di}).
For example, for the one-loop contribution
of the fermion $f$, choosing the on-shell renormalization scheme,
\be
{\bf \Pi}_{f\bar f}(q^2) = \frac{1}{\pi}
q^{2}\,\int_{4m_f^{2}}^{\infty}ds\frac{\Im m{\bf \Pi}_{f\bar f}(s)}{s(s-q^2)}.
\label{qeddisprel}
\ee
For $f\neq e$, $\Im m{\bf \Pi}_{f\bar f}(s)$  
is {\it measured directly} in the tree-level cross-section for
\mbox{$e^{+}e^{-} \rightarrow f^{+}f^{-}$}, see \Figref{QED_OT_epem}.
For $f = e$, it is necessary to isolate the self-energy-like
component of the tree-level Bhabha cross-section, see \Figref{Strong_QED_OT_epem}. 
This is indeed possible because the self-energy-, vertex- and box-like components
of the Bhabha {\it differential}  cross-section are 
{\it linearly independent functions} of $\cos\theta$;
they may therefore  be projected out by convoluting the {\it differential}  cross-section
with appropriately chosen polynomials in $\cos\theta$.
\end{itemize}
\indent
Thus, in QED, knowledge of the spectral function $\Im m{\bf \Pi}_{f\bar f}(s)$,
determined from the tree-level $e^{+}e^{-} \rightarrow f^{+} f^{-}$ cross sections,
together with a single low energy measurement of the fine structure constant
$\alpha$ (obtained \eg from the Josephson effect and the quantized Hall effect, 
or from the anomalous magnetic moment of the electron~\cite{Cohen:1990aa}), 
enables the construction of the one-loop effective charge
$\alpha_{\rm eff}(q^2)$ for all $q^2$.

\begin{figure}[!t]
\bce
\includegraphics[width=10.5cm]{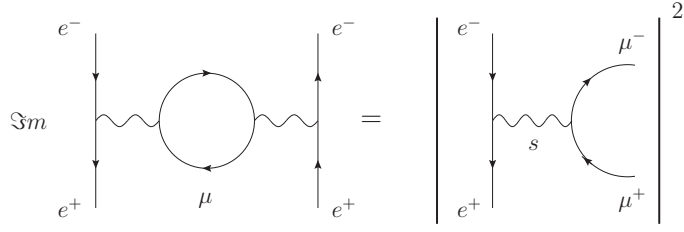}
\ece
\caption{\figlab{QED_OT_epem} The OT relation between the imaginary part of the
muon contribution to ${\bf \Pi}(s)$ and the
tree-level cross section $\sigma(e^{+}e^{-}\rightarrow \mu^{+}\mu^{-})$ in QED.}
\end{figure}

\subsubsection{QCD effective charge\label{QCDech}}
\noindent
In non-Abelian gauge theories  the crucial equality $Z_1=Z_f$ does not
hold in  general.  Furthermore, in contrast to   the photon  case, the
gluon vacuum  polarization depends on  the gfp, already  at one-loop
order.  These  facts make the  non-Abelian  generalization  of the QED
concept of  the  effective  charge  non-trivial.  The  possibility  of
defining an effective  charge for QCD in  the framework of the PT  was
established first by Cornwall \cite{Cornwall:1981zr}, and was further investigated 
in a series of papers~\cite{Watson:1996fg,Papavassiliou:1995fq,Papavassiliou:1995gs,Papavassiliou:1996zn,Watson:1995ad,Binosi:2002vk}.
\newline
\indent
As we have shown in detail in \secref{QCD_one-loop}
the PT   rearrangement  of   physical amplitudes  gives   rise   to  a
gfp-independent effective gluon self-energy,  restoring, at the same
time, the equalities
\begin{equation}
\widehat Z_1 = \widehat Z_f,\qquad \widehat Z_{g_s} = 
\widehat Z_{ A}^{-1/2},
\end{equation}
where $g_s$ is the QCD coupling. Then, using the additional fact  that  the   
PT self-energy is process-independent~\cite{Watson:1994tn}
and can be Dyson-resummed to all orders~\cite{Papavassiliou:1995fq,Papavassiliou:1995gs,Papavassiliou:1996zn,Watson:1996fg},
the  construction of
the  universal RG-invariant combination  and the  corresponding QCD  effective
charge is immediate.  We have 
\begin{equation}
\widehat{R}_0^{\mu\nu}(q^2) =
\frac{(g_s^0)^2}{4\pi}
\widehat{\Delta}^0_{\mu\nu}(q) = \frac{g_s^2}{4\pi}  \widehat{\Delta}_{\mu\nu}(q) = \widehat{R}^{\mu\nu}(q^2).
\label{Rqcd}
\end{equation} 
and after pulling out a $1/q^2$ factor we arrive at the QCD effective charge,  
\begin{equation}
\alpha_{s, \rm eff}(q^2) = 
 \frac{\alpha_{s}}{1 + {\bf \widehat{\Pi}}(q^2)},
 \label{alphaqcd}
\end{equation}
where $\alpha_{s}= g_s^2/4\pi$. 

\begin{figure}[!t]
\bce
\includegraphics[width=15cm]{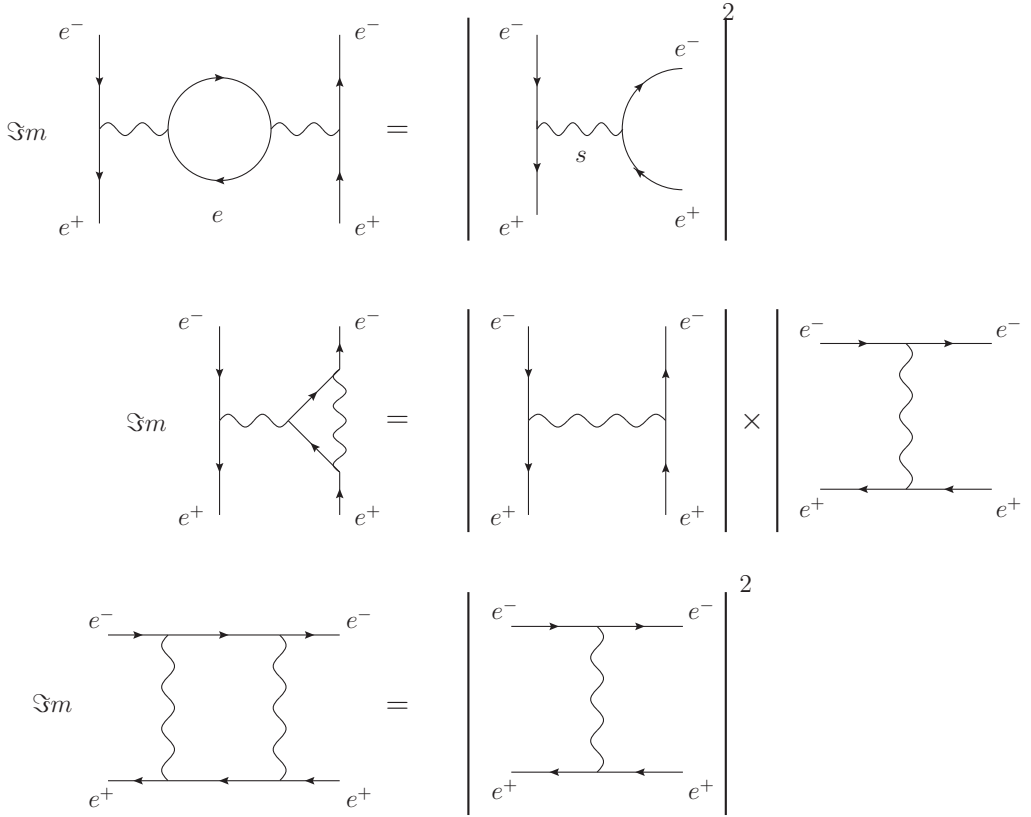}
\ece
\caption{\figlab{Strong_QED_OT_epem} 
The OT relation between the imaginary parts of the
electron contribution to the one-loop vacuum polarization,
vertex and box diagrams and the components of the
tree-level cross section
$\sigma(e^{+}e^{-}\rightarrow e^{+}e^{-})$ in QED}
\end{figure}

\indent
Let us now turn to properties ({\it iii})--({\it v}) of the QED effective 
charge, and see how they are modified due to the fact that 
the low-energy sector of QCD is strongly coupled and must be 
treated non-perturbatively. 
\newline
\indent
Evidently point  ({\it iii}) must be replaced by a measurement 
where perturbation theory holds, such as the $\alpha(M_{ Z})$.  
Point ({\it iv}) remains true; actually, due to asymptotic freedom,  
the high energy limit is where the QCD charge is completely 
unambiguous. Finally, point ({\it v}) is trickier: clearly, the 
absorptive analysis of \secref{QCD_one-loop}
demonstrates that, perturbatively, 
the imaginary part of ${\bf \widehat{\Pi}(s)}$ may be identified 
with a well-defined part of the $q \bar{q}\rightarrow g g$ 
tree-level cross-section. These parts may be, in principle, 
extracted out of the full differential cross-section through the 
convolution with an appropriately constructed function of the scattering angle
[see subsection~\ref{EW_eff_ch} on the electroweak effective charges], and then be fed into 
the dispersion relation (\ref{DR2}) to furnish ${\bf \widehat{\Pi}(q^2)}$. 
The problem is that, in QCD, such a procedure is not 
feasible, even at the level of a thought-experiment, because 
in the limit of low $s$ QCD is strongly coupled, and non-perturbative  effects become significant. 
\newline
\indent
An additional important point related to the non-perturbative nature of QCD is the following.
As has been argued long ago by Cornwall~\cite{Cornwall:1981zr}, and has been corroborated 
by a large number of lattice simulations and SDE studies,  
the gluons generate dynamically an effective mass,  
which cures the Landau singularity and makes 
the gluon propagator finite in the infrared [we will revisit this issue in much more detail in \secref{Applications-II}]. 
In such a case, it would  of course be 
 wrong to define the effective charge as in Eq.~(\ref{alphaqcd}), \ie
by  forcing out a factor of $1/q^2$ from Eq.~(\ref{Rqcd}).
Such a procedure would furnish a completely unphysical strong QCD coupling, 
namely one that would vanish in the deep infrared(!) where 
QCD is supposed to be strongly coupled. 
The correct treatment (see again \secref{Applications-II}) yields, instead, an 
effective coupling that in the deep infrared ``freezes'' at a  {\it finite value}, 
in complete agreement with a plethora of phenomenological and theoretical works.

\subsubsection{\label{sec:RGI}Effective mixing (Weinberg) angle}
\noindent
We now turn to the  electroweak sector of the SM, and consider the 
subset of neutral gauge boson self-energies,  
$\widehat{\Pi}_{\mu\nu}^{{ A} { A}}(q)$, 
and  $\widehat{\Pi}_{\mu\nu}^{{ Z} { Z}}(q)$, 
 together with the mixed self-energies  $\widehat{\Pi}_{\mu\nu}^{{ A} { Z}}(q)$
and  $\widehat{\Pi}_{\mu\nu}^{{ Z} { A}}(q)$.
Let us now see how the above self-energies organize
themselves into  RG-invariant combinations. 
We will assume that we are working between massless fermions (conserved currents) and we will therefore 
retain only the parts of the self-energies proportional to $g_{\mu\nu}$.
The general framework presented in this
subsection has been established in~\cite{Denner:1994xt} and~\cite{Papavassiliou:1997fn};  
here we will adopt the notation and philosophy of the latter article.
\newline
\indent
We begin   by listing the relations between the
bare and renormalized parameters for the neutral part of the electroweak sector.
For the masses we have
\begin{equation}
  \label{massren}
(M^0_{ W})^2 = M_{ W}^2+ \delta M_{ W}^2 , \qquad
(M^0_{ Z})^2 = M_{ Z}^2+ \delta M_{ Z}^2 .
\end{equation}
The wave-function renormalizations for the neutral sector are defined as
\be
\left( \begin{array}{c} Z_{0} \\ A_{0}
\end{array} \right)\ =
\left( \begin{array}{cc}
\widehat{Z}_{{Z} {Z}}^{1/2} & 
\widehat{Z}_{ {Z} {A}}^{1/2} \\
\widehat{Z}_{{A} {Z}}^{1/2}        
& \widehat{Z}_{{A}{A}}^{1/2}
\end{array} \right)\,
\left( \begin{array}{c} Z \\ A
\end{array} \right)
\ = 
\left( \begin{array}{cc}
1+\frac{1}{2}\delta \widehat{Z}_{{Z} {Z}} & 
\frac{1}{2}\delta \widehat{Z}_{ {Z} {A}} \\
\frac{1}{2}\delta \widehat{Z}_{{A} {Z}}        
&1+\frac{1}{2}\delta \widehat{Z}_{{A}{A}}
\end{array} \right)\,
\left( \begin{array}{c} Z \\ A
\end{array} \right).
\label{GammaNCINV}
\ee
In addition, the coupling renormalization constants are defined by
\be
e_0 = \widehat{Z}_{e} e = 
(1 + \delta \widehat{Z}_{e}) e,\qquad
g^0_w = \widehat{Z}_{g_w}g_w = 
(1 + \delta \widehat{Z}_{g_w})g_w,\qquad
c^0_w= \widehat{Z}_{c_{w}}c_{w},
\ee
with
\begin{equation}
\widehat{Z}_{c_w}\ =\left(1+ \frac{\delta M_W^2}{M_W^2}\right)^{1/2}
\left(1+ \frac{\delta M_Z^2}{M_Z^2}\right)^{-1/2}.
\end{equation}
If we expand $\widehat{Z}_{c_{w}}$ perturbatively, we have
that
$\widehat{Z}_{c_{w}}= 1+ \frac{1}{2} (\delta c^2_w / c^2_w) + \cdots,$
with
\begin{equation}
\frac{\delta c^2_w}{c^2_w} =\ \frac{\delta M_W^2}{M_W^2}-
\frac{\delta M_Z^2}{M_Z^2},
\end{equation}
which is the usual one-loop result.  
\newline
\indent
Imposing the requirement that the  PT Green's functions should respect
the  same  WI's  before and   after renormalization  we  arrive at the
following relations:
\be
\widehat{Z}_{{A}{A}} =  \widehat{Z}_{e}^{-2},\qquad
\widehat{Z}_{{Z}{Z}} = {\widehat{Z}}_{g_w}^{-2}
\widehat{Z}_{c_{w}}^2,
\label{W3}
\ee
or, equivalently, at the level of the counter-terms
\begin{eqnarray}
\delta\widehat{Z}_{{A}{A}} &=& -2 \delta \widehat{Z}_{e},
\nonumber\\
\delta\widehat{Z}_{{Z}{Z}} &=& -2 \delta \widehat{Z}_{e}
- \frac{c^2_w - s^2_w}{s^2_w} \left(\frac{\delta c^2_w}{c^2_w}\right),
\nonumber\\ 
\delta \widehat{Z}_{{A}{Z} } &=& 
2 \frac{c_w}{s_w} \left(\frac{\delta c^2_w}{c^2_w}\right),
\nonumber\\
\delta \widehat{Z}_{{Z}{A} } &=& 0.
\end{eqnarray}
\indent
The corresponding propagators relevant for the neutral sector
may be obtained by inverting the matrix $\widehat{L}$, whose 
entries are the PT self-energies, \ie
\be
\widehat{L} =
\left( \begin{array}{cc}
q^2 + \widehat{\Pi}_{{A} {A}}(q^2) & 
\widehat{\Pi}_{{A}{Z}}(q^2) \\
\widehat{\Pi}_{{A} {Z} }(q^2)         
& q^{2} - M_{Z}^{2} + \widehat{\Pi}_{{Z}{Z}}(q^2)
\end{array} \right).
\label{GammaNC}
\ee
Casting the inverse in the form
\be
\widehat{L}^{-1} =
\left( \begin{array}{cc}
\widehat{\Delta}_{{A} {A}}(q^2) & 
\widehat{\Delta}_{ {A} {Z}}(q^2) \\
\widehat{\Delta}_{{A} {Z}}(q^2)        
& \widehat{\Delta}_{{Z} {Z}}(q^2)
\end{array} \right) 
\label{GammaNCI}
\ee
one finds that
\bea
\widehat{\Delta}_{{A} {A}}(q^2) &=& 
\frac{- [q^{2} -M_{ Z}^{2} +
\widehat{\Pi}_{{Z}{Z}}(q^2)]}
{\widehat{\Pi}_{{A}{Z}}^{2}(q^2) -
[q^{2} -M_{ Z}^{2} + \widehat{\Pi}_{{Z}{Z} }(q^2)]
[q^{2} + \widehat{\Pi}_{{A}{A}}(q^2)]},\nonumber\\
\widehat{\Delta}_{{Z}{Z}}(q^2)      &=&
\frac{- [q^{2} + \widehat{\Pi}_{{A}{A}}(q^2)]}
{\widehat{\Pi}_{{A}{Z}}^{2}(q^2) -
[q^{2} -M_{ Z}^{2} + \widehat{\Pi}_{{Z}{Z} }(q^2)]
[q^{2} + \widehat{\Pi}_{{A}{A}}(q^2)]},\nonumber\\
\widehat{\Delta}_{{A} {Z}}(q^2) &=& 
\frac{-\widehat{\Pi}_{{A}{Z}}(q^2)}
{\widehat{\Pi}_{{A}{Z}}^{2}(q^2) -
[q^{2} -M_{ Z}^{2} + \widehat{\Pi}_{{Z}{Z} }(q^2)]
[q^{2} + \widehat{\Pi}_{{A}{A}}(q^2)]}.
\eea
The above expressions at one-loop reduce to
\bea
\widehat{\Delta}_{{A} {A}} (q^2) &=& 
\frac{1}
{q^{2} + \widehat{\Pi}_{{A}{A}}(q^2)},\nonumber\\
\widehat{\Delta}_{{Z}{Z}}(q^2) &=&
\frac{1}
{q^{2} -M_{ Z}^{2} + \widehat{\Pi}_{{Z}{Z}}(q^2)},\nonumber\\
\widehat{\Delta}_{{A} {Z}}(q^2) &=& 
\frac{\widehat{\Pi}_{{A}{Z}}(q^2)}
{q^{2} (q^{2} -M_{Z}^{2})}.
\eea
\indent
The standard re-diagonalization procedure of the neutral sector~\cite{Baulieu:1981ux,Kennedy:1988sn,Philippides:1995gf}
may then be followed,  
for the PT self-energies; it will finally amount to 
introducing the effective (running) weak mixing angle.
In particular,  
after the PT rearrangement, 
the propagator-like part $\widehat{\mathcal D}_{ff'}$ 
of the neutral current
amplitude for the interaction between fermions
with charges $Q$, $Q'$ and isospins $T^{f}_z $,  $T^{f'}_z $,  
is given
in terms of the inverse of the matrix $\widehat{L}$ by the expression
\bea
\widehat{\mathcal D}_{ff'}\, &=&\, 
\lefteqn{
\left( \begin{array}{cc}
e Q_{f}, & {\displaystyle \frac{g_w}{c_w}}
\left[ s^2_w Q_{f} - T^{f}_z P_{ L} \right]
\end{array} \right)
\widehat{L}^{-1}
\left( \begin{array}{c}
e Q_{f'} \\ {\displaystyle \frac{g_w}{c_w}}
(s^2_w Q_{f'} - T^{f'}_z P_{ L} )
\end{array} \right)}\nonumber \\
&=& \left( \begin{array}{cc}
e Q_{f},  & {\displaystyle \frac{g_w}{c_w}}
\left[ \bar{s}^2_w(q^{2}) Q_{f} - T^{f}_z P_{ L}\right] 
\end{array} \right)
\widehat{L}^{-1}_{{D}}
\left( \begin{array}{c}
e Q_{f'} \\ {\displaystyle \frac{g_w}{c_w}}
\left[ \bar{s}^2_w(q^{2})Q_{f'} - T^{f'}_z P_{ L}\right]
\end{array} \right), 
\label{born}
\eea
where
\be
\widehat{L}^{-1}_{{D}} = 
\left( \begin{array}{cc}
\hat{\Delta}_{{A} {A}}(q^2)             & 0 \\ 
\phantom{\displaystyle \frac{e}{s_w}}0
\phantom{\displaystyle \frac{e}{s_w}}  & 
\hat{\Delta}_{{Z} {Z}}(q^2)
\end{array} \right).
\ee
The rhs of this equation, where the neutral current interaction
between the fermions has been written in diagonal
(\ie Born-like) form, defines the diagonal propagator functions
$\widehat{\Delta}_{{A}{A}}$ 
and $\widehat{\Delta}_{{Z}{Z}}$ and
the effective weak mixing angle $\bar{s}_w^{2}(q^2)$
\be
\label{sweff}
\bar{s}_w^{2}(q^2) = ({s_w^{0}})^{2}\left[1 -
\left(\frac{c_w^{0}}{s_w^{0}}\right)\frac{\widehat{\Pi}_{{A} {Z}}^0(q^2)}
{q^{2} + \widehat{\Pi}_{{A}{A}}^0(q^2)} \right]
= s_w^{2}\left[1 -
\left(\frac{c_w}{s_w}\right)\frac{\widehat{\Pi}_{{A}{Z}}(q^2)}
{q^{2} + \widehat{\Pi}_{{A}{A}}(q^2)} \right].
\ee
It is easy to show that,  
by virtue of the special relations of 
Eq.~(\ref{W3}), $\bar{s}_w^{2}(q^2)$ is an RG-invariant quantity. 
\newline
\indent
Using the fact that 
$\widehat{\Pi}_{{A}{Z}}(0)=0$, we may write 
$\widehat{\Pi}_{ A  Z}(q^2) = q^2 {\bf \widehat{\Pi}}_{ A  Z}(q^2)$; 
then, Eq.~(\ref{W3}) yields 
\bea
\bar{s}_w^{2}(q^2) &=& 
s_w^{2}\left[1 -
\left(\frac{c_w}{s_w}\right)\frac{{\bf \widehat{\Pi}}_{{A}{Z}}(q^2)}
{1 +{\bf \widehat{\Pi}}_{{A}{A}}(q^2)} \right]
\nonumber\\
&=&  s_w^{2} \left[1 - \left(\frac{c_w}{\alpha \,s_w}\right) 
{\alpha}_{\rm eff} (q^2) 
{\bf \widehat{\Pi}}_{{A}{Z}}(q^2) \right], 
\label{sweff2}
\eea
where in the last step we used Eq.~(\ref{alphaqed}).
At one-loop level,
$\bar{s}_w^{2}(q^2)$ reduces to 
\be
\label{sweff1}
\bar{s}_w^{2}(q^2) = s_w^{2}\left[1 - \left(\frac{c_w}{s_w}\right) 
{\bf \widehat{\Pi}}_{ A  Z}(q^2)\right].
\ee
Notice that in the case where the fermion $f'$ is a neutrino 
($f'=\nu$, with $Q_{\nu}=0$ and $T^{\nu}_z = 1/2$), Eq.~(\ref{born}) assumes the form
\be
\widehat{\mathcal D}_{f\nu } = 
\left( \begin{array}{cc}
e Q_{f},  & {\displaystyle \frac{g_w}{c_w}}
\left[ \bar{s}^2_w(q^{2}) Q_{f} - T^{f}_z P_{ L}\right] 
\end{array} \right)
\widehat{L}^{-1}_{{D}}
\left( \begin{array}{c}
0 \\ {\displaystyle - \frac{g_w}{2c_w}} P_{ L}
\end{array} \right)  
\label{born2}
\ee
Evidently, $\bar{s}_w^{2}(q^2)$ constitutes a universal   
modification to the effective vertex of the charged fermion.
  
\subsubsection{\label{EW_eff_ch}Electroweak effective charges}
\noindent
The analogue of Eq.~(\ref{alphaqed}) 
may be defined for 
the $Z$- and $W$-boson propagators. 
In particular, 
the  bare and renormalized PT resummed
$Z$-boson  propagators,      
$\widehat{\Delta}_{{Z}{Z},0}^{\mu\nu}(q)$
and $\widehat{\Delta}_{{Z}{Z}}^{\mu\nu}(q)$
respectively,     satisfy     the
following relation
\begin{equation}
  \label{renmult}
\widehat{\Delta}_{{Z}{Z}}^{0,\,\mu\nu}(q) = 
\widehat{Z}_{{Z}{Z}}
\widehat{\Delta}_{{Z}{Z}}^{\mu\nu}(q).
\end{equation}
In what follows we only consider the cofactors of $g^{\mu\nu}$, \ie
$\widehat{\Delta}_{{Z}{Z}}^{0,\,\mu\nu}(q) =
\widehat{\Delta}_{{Z}{Z}}^{0}(q) g^{\mu\nu}$ and
\linebreak 
$\widehat{\Delta}_{{Z}{Z}}^{\mu\nu}(q) =
\widehat{\Delta}_{{Z}{Z}}(q) g^{\mu\nu}$,   
since the longitudinal parts vanish when contracted with the
conserved external currents of massless fermions. 
The standard renormalization procedure  is to define the wave function
renormalization, 
$\widehat{Z}_{{Z}{Z}}$, by means of  the transverse part  of
the resummed $Z$-boson propagator:
\begin{equation}
  \label{TransRen}
\widehat{Z}_{{Z}{Z}} [ q^2 - (M^0_{ Z})^2 +
\widehat{\Pi}_{{Z}{Z}}^0(q^2) ]=q^2 - M_{ Z}^2 +
\widehat{\Pi}_{{Z}{Z}}(q^2).
\end{equation}
It is then straightforward to verify  that the universal 
RG-invariant quantity for the $Z$ boson, which constitutes a common part of all 
neutral current processes, is given by 
(we omit a factor  $g^{\mu\nu}$):
\be
  \label{RW}
\bar{R}_{{Z}}^0(q^2) =  
\frac{1}{4 \pi} \left(\frac{g^0_w}{c^0_w}\right)^2
\widehat{\Delta}_{{Z}{Z}}^0(q^2)  = \frac{1}{4 \pi} \left(\frac{g_w}{c_w}\right)^2
\widehat{\Delta}_{{Z}{Z}}(q^2)  
= \bar{R}_{{Z}}(q^2).
\ee
A completely analogous analysis holds also for the 
$\widehat{\Delta}_{{W}{W}}(q^2)$ propagator; in this 
case the corresponding RG-invariant combination is 
\be
\bar{R}_{{W}}(q^2) = \left(\frac{g_w^2}{4 \pi}\right) 
\widehat{\Delta}_{{W}{W}}(q^2).
\label{Wrgi}
\ee
\indent
If one retains only the real parts 
in the above equation,   
one may define from  $\bar{R}_{i}(q^2)$ ($i=W,Z$) a  
dimensionless quantity, corresponding to    
an effective charge by  casting 
$\Re e \widehat{\Pi}(q^2)_{ii}$ in the form 
$\Re e \widehat{\Pi}_{ii}(q^2) = \Re e \widehat{\Pi}_{ii}(M^2_{i}) + ( q^2\, -\, M_{i}^2) 
\Re e {\bf\widehat{\Pi}}_{ii}(M^2_{i})$ 
and then pulling out a common factor $( q^2\, -\, M_{i}^2)$.
In that case, 
\be
\bar{R}_{i}(q^2) = 
\frac{{\alpha}_{i,\rm{eff }}(q^2)}{q^2\, -\, M_{i}^2}\,, \qquad i=W,Z
\ee 
with 
\bea
{\alpha}_{w,\rm{eff}}(q^2) &=& \frac{{\alpha}_{w}}{1+\Re e {\bf \widehat{\Pi}}_{{W}{W}}(q^2)}\,,
\label{aw}
\nonumber\\
{\alpha}_{z,\rm{eff}}(q^2) &=& \frac{{\alpha}_{z}}{1+\Re e {\bf \widehat{\Pi}}_{{Z}{Z}}(q^2)}\,,
\label{az}
\eea
where 
\be
{\bf \widehat{\Pi}}_{{i}{i}}(q^2) = 
\frac{\widehat{\Pi}_{ii}(q^2)- \widehat{\Pi}_{ii}(M^2_{i})}{ q^2\, -\, M_{i}^2}\,, \qquad  i=W,Z\,.
\ee
and ${\alpha}_{w}=g^2_w/4\pi$ and ${\alpha}_{z}={\alpha}_{w}/c_w^2$.
Notice, however, that, whereas  Eqs~(\ref{RW}) and~(\ref{Wrgi}) remain valid in the presence of imaginary 
parts (\ie when $\widehat{\Pi}_{ii}(q^2)$ develops 
physical thresholds)  the above separation into 
a dimensionful and a dimensionless part is ambiguous and should be avoided~\cite{Papavassiliou:1998pb} . 

\subsubsection{Electroweak effective charges and their relation to physical cross-sections}

\begin{figure}[!t]
\bce
\includegraphics[width=11cm]{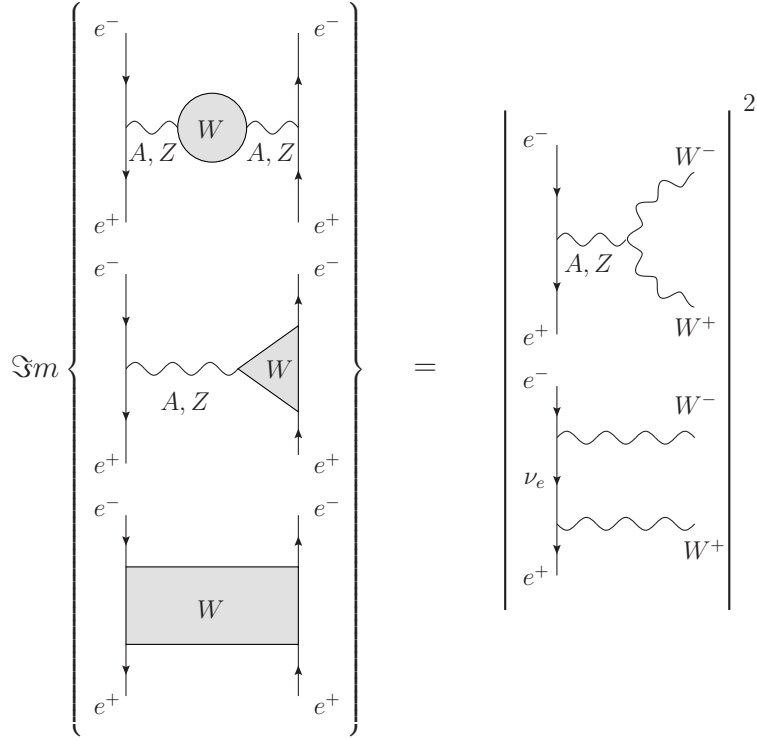}
\ece
\caption{\figlab{SM_OT_epem} 
The OT  relation between the imaginary parts of the subset of the 
$W$-related  one-loop corrections to  $e^{+}e^{-}\rightarrow e^{+}e^{-}$ 
and the tree-level process $e^{+}e^{-}\rightarrow W^{+}W^{-}$.}
\end{figure}
\noindent
Let us now turn to the relation of the RG-invariant and universal 
quantity $\bar{R}_{{Z}}(q^2)$ to physical cross-sections, and 
the procedure that would allow, at least in principle, its extraction 
from experiment~\cite{Papavassiliou:1996fn}.
In general, the renormalization of $\widehat\Pi_{{Z}{Z}}$ 
requires two subtractions, for mass and field renormalization. If we denote the 
subtraction point by $s_{0}$, then  
the 
twice-subtracted dispersion relation corresponding to the  $W^{+}W^{-}$ contributions reads
\be
\widehat\Pi_{{Z}{Z}}^{\scriptscriptstyle (WW)}(q^2)
=  \frac{1}{\pi} (q^2-s_0)^{2}  \int_{4M_{ W}^2}^{\infty}
ds \frac{\Im m \widehat\Pi_{{Z}{Z}}^{\scriptscriptstyle (WW)}(s)}{(s- q^{2})(s- s_0)^{2}}.
\label{sigmadisprel}
\ee
The property that is instrumental for the observability of  
$\bar{R}_{{Z}}^{\scriptscriptstyle (WW)}(q^2)$ 
is that, in contrast to the conventional gauge-dependent self-energies,  
the absorptive parts 
of the PT self-energies appearing on the rhs of Eq.~(\ref{sigmadisprel})
are directly related to components of the physical cross-section  $e^{+}e^{-}\rightarrow W^{+}W^{-}$ 
which are {\it experimentally observable} (see \Figref{SM_OT_epem}).
Indeed, as we have already seen in \secref{SM_one-loop}, 
the characteristic $s$-$t$ cancellation, triggered by the longitudinal momenta 
of the on-shell polarization tensors, rearranges the  
tree-level cross-section $e^{+}e^{-}\rightarrow W^{+}W^{-}$
into subamplitudes, which,  through the use of the OT, can be   
connected unambiguously with the absorptive parts of the one-loop PT Green's functions. 
\newline
\indent
To simplify the algebra without compromising the principle, 
let us consider the limit of $e^{+}e^{-}\rightarrow W^{+}W^{-}$ 
when the electroweak mixing angle vanishes, $s_w^{2}=0$. In this limit  
all photon related contributions are switched off , and the two massive gauge bosons become 
degenerate ($M_{ Z}=M_{ W} \equiv M$). 
Let us denote by 
$\theta$ the center-of-mass scattering angle, 
and set $x = \cos\theta$, $\beta = \sqrt{1-4M^{2}/s}$, and $z = (1 + \beta^{2})/2\beta$.  
Then, it is relatively straightforward to show that the 
differential tree-level cross-section for $e^{+}e^{-}\rightarrow W^{+}W^{-}$ 
can be cast in the form~\cite{Papavassiliou:1996fn}
\be
\label{obs}
(z - x)^{2}
\left(\frac{d\sigma}{dx}\right)_{s_w = 0}
=
\frac{g^{4}}{64\pi} \frac{s \beta}{(s-M^{2})^{2}}\theta(s-4M^{2})
\sum_{i=1}^{5}A_{i}(s)F_{i}(s,x)
\ee  
where 
\bea
& & F_{1}(s,x) = (z-x)^{2} \hspace{3cm}
A_{1}(s)   = \frac{5}{32}\left(\beta^{2}-12\right),\nonumber\\
& & F_{2}(s,x) = (z-x)^{2}\,x^{2}\hspace{2.5cm}
A_{2}(s)   = -\frac{9}{32}\beta^{2},\nonumber\\ 
&  & F_{3}(s,x) = (z-x)(1-x^{2})\qquad \qquad
A_{3}(s)   = -\frac{\beta}{2}\left(\frac{s-M^{2}}{s}\right),\nonumber\\
&  & F_{4}(s,x) = (z-x)(1-\beta x)\hspace{1.6cm}
A_{4}(s)   = \frac{2}{\beta}\left(\frac{s-M^{2}}{s}\right),\nonumber\\ 
&  & F_{5}(s,x) = 1-x^{2}\hspace{3.3cm}
A_{5}(s)   =\frac{1}{2}\left(\frac{s-M^{2}}{s}\right)^{2}.
\eea
Let us emphasize  an important point: for $s \to\infty$, the $A_{i}(s)$ 
reported above go asymptotically to constant values.   
This good high-energy behavior is to be contrasted with that of 
the conventional subamplitudes, corresponding to the $A_{i}(s)$, 
obtained in ~\cite{Alles:1976qv} (no $s$-$t$ cancellation explicitly carried out): they grow  
rapidly as functions of $s$,  violating individually unitarity 
(see also comments at the end of subsection 4.3.4). 
Notice also that the five polynomials $F_{i}(s,x)$, $i = 1,2\ldots 5$   
are  {\it linearly independent}, and also that  
the coefficients $A_{1}(s)$ and $A_{2}(s)$ contribute only to the
self-energy-like component of the cross-section, 
being related to $\Im m\,\widehat{\Pi}_{ Z  Z}^{\scriptscriptstyle (WW)}(s)$ by 
\be
\label{imSigmaA1A2}
\left.\Im m\,\widehat{\Pi}_{ Z  Z}^{\scriptscriptstyle (WW)}(s)
\right|_{s_w = 0} =
\frac{g^{2}}{4\pi}\beta s\left( A_{1}(s) + \frac{1}{3}A_{2}(s)\right).
\ee
\indent
To project out the functions $A_{i}(s)$, we construct a further set of five polynomials
$\tilde{F}_{i}(s,x)$ satisfying the orthogonality conditions
\be
\int_{-1}^{1}dx\,F_{i}(s,x)\tilde{F}_{j}(s,x) = \delta_{ij}.
\ee
The explicit expressions for the $\tilde{F}_{i}(s,x)$ can be found in~\cite{Papavassiliou:1996fn}. The coefficient functions
$A_{i}(s)$ may then be projected out from the observable formed by taking the
product of the differential cross section with the kinematic factor
$(z-x)^{2}$:
\be
\label{proj}
\int_{-1}^{1}dx\,\tilde{F}_{i}(s,x)\,
(z-x)^{2}\left(\frac{d\sigma}{dx}\right)_{s_w = 0}
=
\frac{g^{4}}{64\pi}\frac{s \beta}{(s-M^{2})^{2}}A_{i}(s).
\ee
Thus, it is is possible to extract 
$\Im m\,\widehat{\Pi}_{ Z  Z}^{\scriptscriptstyle (WW)}(s)|_{s_w = 0}$
directly from
$d\sigma(e^{+}e^{-}\!\to\! W^{+}W^{-})/dx |_{s_w = 0}$.
The general case with $s_w^{2}\neq 0$ requires, in addition, the observation of spin density matrices~\cite{Bilenky:1993ms};
though technically more involved, the procedure is, in principle, the same. 

Finally notice the following:
\begin{itemize}
\item[{\it i}.] In order to use the dispersion relation of (\ref{sigmadisprel}) to compute 
$\widehat{\Pi}_{ Z  Z}^{(WW)}(q^2)$ , 
one needs to integrate the spectral density
$\Im m\,\widehat{\Pi}_{ Z  Z}^{ (WW)}(s)$
over a large number of values of $s$. This, in turn, means 
that one needs experimental data for the process $e^{+}e^{-}\rightarrow W^{+}W^{-}$ 
for a variety of center-of-mass energies $s$, and for each value of $s$ one must repeat the 
procedure described above. Regardless of whatever practical difficulties this might entail, 
it does not constitute a problem of principle.  
\newline
\item[{\it ii}.] The experimental extraction of the contributions  
$\widehat{\Pi}_{ Z  Z}^{(ZH)}(q^2)$ 
is conceptually far more straightforward, given that 
it involves the {\it entire} cross-section 
of the process $e^{+}e^{-}\rightarrow ZH$ (known as ``Bjorken process'' or 
``Higgsstrahlung''); specifically, we have that  
\be
\label{imsigmaZHZZ}
e^{2}\frac{(a^{2} + b^{2})}{s_w^{2} c_w^{2}}
\frac{\Im m \widehat{\Pi}_{ZZ}^{(ZH)}(s)}{(s-M_{ Z}^{2})^{2}}=
\sigma(e^{+}e^{-}\rightarrow ZH).
\ee
\end{itemize}

\subsubsection{The effective charge of the Higgs boson}
\noindent
As has been shown in~\cite{Papavassiliou:1997fn}, there is a universal RG-invariant quantity related to the Higgs boson 
that  leads to the {\it novel concept} of the ``Higgs boson effective charge''.
Specifically, after applying the PT algorithm, the linearity of the WI satisfied by the PT Green's functions provides us with the following 
relations 
\begin{equation}
  \label{ZWZH}
\widehat{Z}_{ W}\ =\ \widehat{Z}_{g_w}^{-2},\qquad
\widehat{Z}_{ H} = \widehat{Z}_{\chi} = \widehat{Z}_{\phi}
= \widehat{Z}_{ W}(1+ \delta M^2_{ W}/M^2_{ W}),
\end{equation}
where $\widehat{Z}_{ W}$ and $\widehat{Z}_{ H}$  are the
wave-function renormalizations of  the   $W$ and  $H$    fields,
respectively, $\widehat{Z}_{g_w}$ is the coupling renormalization.
The  renormalization of the bare  resummed Higgs-boson
propagator $\hat{\Delta}^0_{ H}(s)$ proceeds, then, as follows:
\begin{equation}
  \label{DeltH}
\widehat{\Delta}^0_{ H}(s)=[s - (M^0_{ H})^2 +
\widehat{\Pi}^0_{{HH}}(s) ]^{-1} = \widehat{Z}_{ H} [\,s - M_{ H}^2 +
\widehat{\Pi}_{{HH}}(s)]^{-1} = \widehat{Z}_{ H}
\widehat{\Delta}^{ H}(s), 
\end{equation}
with $(M^0_{ H})^2 = M_{ H}^2 +  \delta M_{ H}^2$. The renormalized Higgs-boson
mass $M_{ H}^2$   may be defined  as the  real part  of  the complex pole
position    of  $\hat{\Delta}^H(s)$.   
Employing the relations of Eq.~(\ref{ZWZH}), together with the relation between 
${(M^0_{ W})^2}$ and $M^2_{ W}$ given in (\ref{massren}), 
we observe that the universal (process-independent) quantity
\begin{equation}
  \label{RGIC}
  \bar{R}_{ H}^0(s) = \frac{(g^0_w)^2}{(M^0_{ W})^2}
  \widehat{\Delta}^0_{ H}(s) =  \frac{g^2_w}{M_{ W}^2} \widehat{\Delta}_{ H}
  (s) = \bar{R}_{ H}(s)
\end{equation}
constitutes, in fact, a RG-invariant. Interestingly enough, $\bar{R}^{ H}(s)$
provides a natural extension of the notion of the QED effective charge
for the SM Higgs boson:  $H$  couples universally to matter
with an effective ``charge'' inversely proportional to its vev. 

\subsubsection{Physical renormalization schemes vs  $\overline{MS}$}
\noindent
It is well-known, but often overlooked, that the widely used 
(unphysical) renormalization\linebreak schemes, such as  
$\overline{\rm  MS}$ and $\overline{\rm DR}$, are plagued with   
persistent threshold and matching errors.
The origin of these errors can be
understood by noting that the aforementioned unphysical schemes implicitly integrate out
all masses heavier than the physical energy scale until they are
crossed, and then they turn them back on abruptly, by means of a step function.
Integrating out heavy fields, however, is only valid for energies
well below their masses. 
This procedure is conceptually problematic, since it
does not correctly incorporate the finite probability that the
uncertainty principle gives for a particle to be pair produced
below threshold~\cite{Binger:2003by}. In addition,
complicated matching conditions must be applied when crossing
thresholds to maintain consistency for such desert scenarios. In
principle, these schemes are only valid for theories where all
particles have zero or infinite mass, or if one knows the full
field content of the underlying physical theory.
\newline
\indent
Instead, in  the physical  renormalization scheme  defined with  the  PT, gauge
couplings are defined directly in terms of {\it physical observables},
namely  the effective charges.  The latter  
run smoothly over space-like momenta and have non-analytic
behavior only at the expected physical thresholds for time-like momenta;
as a result, the  thresholds  associated  with  heavy
particles  are treated  with  their correct  analytic dependence.   In
particular, particles will contribute  to physical predictions even at
energies below their threshold~\cite{Binger:2003by}.  
\newline
\indent
Historically, the gauge-invariant parametrization of physics 
offered by the PT has been first 
systematized by Hagiwara, Haidt, Kim, and Matsumoto~\cite{Hagiwara:1994pw}, 
and has led to an alternative 
framework for confronting the precision 
electroweak data with the theoretical predictions. This approach resorts to the PT in order  
to separate the one-loop corrections into gfp-independent 
universal (process-independent) and process-specific pieces; the former 
are parametrized using  the PT effective charges,  
${\alpha}_{\rm eff} (q^2)$, $\bar{s}_w^{2}(q^2)$, ${\alpha}_{w,\rm{eff}}(q^2)$, and  
${\alpha}_{z,\rm{eff}}(q^2)$, defined earlier. 
There is a total of  nine electroweak parameters that must be determined in this approach:
the eight universal parameters 
$M_{ W}$, $M_{ Z}$, ${\alpha}_{\rm eff} (0)$,  $\bar{s}_w^{2}(0)$, 
${\alpha}_{w,\rm{eff}}(0)$, ${\alpha}_{z,\rm{eff}}(0)$,  $\bar{s}_w^{2}(M^2_{ Z})$,
${\alpha}_{z,\rm{eff}}(M^2_{ Z})$, and one process-dependent parameter $\delta_b (M^2_{ Z})$,
related to the form-factor of the $Z \bar{b}_{ L} b_{ L}$ vertex.
\newline
\indent
The authors of ~\cite{Hagiwara:1994pw} explain 
in detail the advantage of their approach over the   $\overline{MS}$ scheme.
In particular they emphasize that the non-decoupling nature of the $\overline{MS}$ 
forces one to adopt an effective field theory approach, where the heavy particles are 
integrated out of the action. The couplings of the effective theories are then related to each other 
by matching conditions ensuring 
that all effective theories give identical results
at zero momentum transfer, since the effects of heavy particles in the
effective light field theory
must be proportional to $q^2/M^2$, where $M$ is the heavy mass scale. 
This procedure, however, is not only impractical in the presence of many quark and lepton mass scales, 
but it introduces errors due to the mistreatment of the 
threshold effects. In addition, the direct use of the  $\overline{MS}$  couplings 
leads to expressions where the masses used for the light quarks are affected by  
sizable non-perturbative QCD effects.
\newline
\indent
The relevance of the 
effective charges in the quantitative study of threshold corrections due to heavy particles 
in Grand Unified Theories (GUTs) was already  
recognized in ~\cite{Hagiwara:1994pw}, 
but it was not until a decade later when 
this was actually accomplished by Binger and Brodsky~\cite{Binger:2003by}.
As was shown by these authors,  
the effective charges defined with the PT furnish a 
conceptually superior and calculationally more accurate
framework for studying the important issue of gauge coupling unification. 
The main advantage of the  effective charge formalism 
is that it provides a template for calculating all 
mass threshold effects for any given grand unified theory;
such threshold corrections may be instrumental in making 
the measured values of the gauge couplings consistent with unification. 
\newline
\indent
The analysis  presented in~\cite{Binger:2003by} in the context of a toy model 
makes a most compelling case in favor 
of  the {\it physical renormalization schemes}; here we reproduce it practically  
unchanged (up to minor notational modifications).
\newline
\indent
Consider QED with three fermions $e$, $\mu$, and $\tau$, and focus 
on the process $e^-\mu^-{\rightarrow}e^-\mu^-$. 
The corresponding 
amplitude can be written as a dressed skeleton expansion, \ie
the dressed tree-level graph plus the dressed box diagram plus the dressed
double box, etc. 
It is easy to see (\eg, by carrying out the standard Dyson summation) 
that the tree-level diagram, dressed to all orders in perturbation theory, is
equal to the tree-level diagram with one modification: the QED coupling
(fine structure constant) $\alpha$
is replaced by the effective charge $\alpha_{\rm eff}(q^2)$ given in Eq.~(\ref{alphaqed}).
\newline
\indent
Let us now consider two rather disparate scales, denoted by $q_h$ (for ``high'') 
and $q_{\ell}$ (for ``low'').
From measurements of the cross-section, one can actually extract  the
the value of the effective charge at these two scales,  
$\alpha_{\rm eff}(q^2_h)$ and $\alpha_{\rm eff}(q^2_{\ell})$.   
Let us suppose for a moment that 
the electron charge is not known, and we are trying to test the predictions
of QED.
The way to proceed is to use one measurement, say at the low scale $q_{\ell}$,
as an input
to determine $e$. Once  $e$ is known the prediction at the high scale $q_h$ is well
defined, and represents a test of the theory. 
More directly, we could just write
$\alpha_{\rm eff}(q^2_h)$ in terms of $\alpha_{\rm eff}(q^2_{\ell})$, 
which would lead us to the same prediction.
\newline
\indent
Since the cross-section $\sigma_{e^-\mu^-{\rightarrow}e^-\mu^-}(q^2)$ is
proportional to
$\alpha^2_{\rm eff}(q^2)$, we are clearly relating one observable to another.
Such a
scheme is referred to as an {\it effective charge scheme}, 
since a given observable,
here just $\sigma_{e^-\mu^-{\rightarrow}e^-\mu^-}(q_h^2)$ (or $\alpha_{\rm eff}(q_h^2)$), 
is expressed in terms of
an effective charge, $\alpha_{\rm eff}(q_{\ell}^2)$, defined from a measurement of
the cross-section at the scale $q_{\ell}$. One could equally well express any
observable in terms
of this effective charge. Note that this approach to renormalization works
for arbitrary
scales, even if the low scale lies below some threshold, say $q_{\ell}<m_{\tau}$,
while $q_h>m_{\tau}$. Thus, the decoupling 
and the smooth (not through a $\theta$ function!) ``turning on'' of the
$\tau$ is manifest, due to the intrinsic analyticity and unitarity properties of the 
vacuum polarization (see subsection~\ref{sec:OT_anal}).
\newline
\indent
Let us now consider the results obtained by using the
conventional implementation of $\overline{MS}$. In this case one begins
by calculating the cross-section at $q_{\ell}$ using the rules of
$\overline{MS}$, which stipulate that 
{\it only} the electrons and muons are allowed to propagate in loops, 
since $q_{\ell}<m_{\tau}$. Comparing the observed cross
section to this result will fix the value of the $\overline{MS}$
coupling for two flavors, $\hat{\alpha}_2(q_{\ell})$. To predict the result
of the same experiment at scale $q_h>m_{\tau}$, we need to evolve
$\hat{\alpha}_2$ to the $\tau$ threshold using the {\it two-flavor} $\beta$
function, match with a three-flavor coupling, $\hat{\alpha}_3$,
through the relation $\hat{\alpha}_2(m_{\tau})=\hat{\alpha}_3(m_{\tau})$,
and then evolve $\hat{\alpha}_3(m_{\tau})$ to $q_h$ using the {\it three-flavor}
$\beta$ function. We will now have a prediction for
$\sigma_{e^-\mu^-{\rightarrow}e^-\mu^-}(q_h^2) \propto \alpha^2(q_h^2)$.
\newline
\indent
Now, one might expect, from the general principle of RG-invariance of 
physical predictions, that this result should be the same as the
prediction derived using the physical effective charge scheme
above. However, there is a discrepancy arising from the incorrect
treatment of the threshold effects in the $\overline{MS}$. 
One finds that the ratio of the cross
section derived using $\overline{MS}$ with the cross section derived
using effective charges, to first order in perturbation theory, is
given by~\cite{Binger:2003by} 
\be
{\hat{\sigma}(q_h^2)\over \sigma(q_h^2)} =
1+ {2 \alpha_{\rm eff}(q^2_{\ell})\over 3\pi}\left[ L\left(\frac{q_{\ell}}{m_{\tau}}\right) - 5/3 \right],
\ee
with 
\begin{eqnarray}
\label{Ltau}
L_{\tau}\left(\frac qm\right) &=&\int_0^1 dx\, 6x(1-x)
\log{\left(1+{q^2\over m^2}x(1-x) \right)}+ \frac{5}{3} 
\nonumber\\
&=&[\beta {\rm tanh}^{-1}(\beta^{-1}) - 1] ( 3-\beta^2 ) + 2,
\end{eqnarray}
where $\beta = \sqrt{ 1 + 4m^2/q^2}$.
Since $L_{\tau}(0)=5/3$,
there is no discrepancy when the low reference scale $q_{\ell}$ is much
lower than the $\tau$ mass threshold. This reflects the important fact 
that unphysical schemes, such as  $\overline{MS}$, are formally consistent only
in {\it desert regions} where particle masses can be neglected. 
Notice that in this example there is an error only for $q_l<m_{\tau}$.
However, in the more general case of multiple flavor thresholds, 
as well as in the analysis of grand unification, 
there are errors from both high and low scales.
\newline
\indent
An important difference between the physical effective charges and
the unphysical $\overline{MS}$, intimately related to the analyticity 
properties built into the former,
is the distinction between time-like and space-like momenta. 
As emphasized in~\cite{Binger:2003by}, in conventional approaches the 
thresholds are treated using a step function approximation [$\theta$ functions], and hence, 
the running is always logarithmic. The analytic continuation from
space-like to time-like momenta is trivial, yielding $i\pi$
imaginary terms on the time-like side; thus, the real parts of such
couplings are the same [modulo three loop $(i\pi)^2$ corrections].
In contrast, the PT charges on time-like and space-like
sides have considerable differences at one-loop. 
Specifically, the relevant quantities to consider are 
functions such as the $L_{\tau}(q/m)$ defined above. 
The analytic continuation  to time-like momenta below
threshold, $ 0 < q^2 = -Q^2 < 4m^2$, is obtained by replacing
\begin{eqnarray}
\beta {\rightarrow} i\bar{\beta}, \quad {\rm where} \quad
\bar{\beta} = \sqrt{{4m^2\over q^2}- 1 }, \quad {\rm and} \quad
{\rm tanh}^{-1}(\beta^{-1}) {\rightarrow} -i{\rm tan}^{-1}(\bar{\beta}^{-1}).
\end{eqnarray}
Above threshold, $q^2 > 4m^2$, one should replace
\begin{eqnarray}
{\rm tanh}^{-1}(\beta^{-1}) {\rightarrow} {\rm tanh}^{-1}(\beta) + i{\pi\over 2},
\quad {\rm where}\quad \beta = \sqrt{ 1 - {4m^2\over q^2} }.
\end{eqnarray}
From these results, it is clear
that significant differences will arise between the space-like  and time-like
couplings evaluated at scale $M_{ Z}^2$, mainly due to the $W^{\pm}$ boson
threshold asymmetry.
\newline
\indent
In~\cite{Binger:2003by} the effective charges  
$\alpha_{\rm eff}(q^2)$, $\alpha_{s, \rm eff}(q^2)$, and the effective mixing angle $\bar{s}_w^{2}(q^2)$
[defined in  (\ref{alphaqed}), (\ref{alphaqcd}), and (\ref{sweff1}), respectively], 
were used to define new effective charges, 
$\tilde{\alpha_1}(q^2)$, $\tilde{\alpha_2}(q^2)$, and $\tilde{\alpha_3}(q^2)$,  
which correspond to the standard combinations of gauge couplings 
used to study gauge-coupling unification. Specifically,   
\be
\tilde{\alpha}_1(q^2) = \left({5\over 3}\right){\alpha_{\rm eff}(q^2)\over 1-\bar{s}_w^{2}(q^2)}\qquad
\tilde{\alpha}_2(q^2) = {\alpha_{\rm eff}(q^2)\over \bar{s}_w^{2}(q^2)}\qquad
\tilde{\alpha}_3 (q^2) = \alpha_{s, \rm eff}(q^2).
\ee
\newline
\indent
The above couplings were  used to obtain novel heavy and light 
threshold corrections, and 
the resulting impact on the unification 
predictions for a general GUT model were studied.
Notice that 
even in the absence of new physics (\ie using only the known SM spectrum)
there are appreciable numerical discrepancies between the values of 
of the conventional and PT couplings at $M_{ Z}$ (see Table I in \cite{Binger:2003by}).
 Given that these values are used as initial conditions for the 
evolution of the couplings to the GUT scale, 
these differences alone may affect the unification properties of the couplings
(\ie even if no additional threshold effects due to new particles are considered). 

\subsection{Gauge-independent off-shell form-factors: general considerations}
\noindent
It is well-known that renormalizability and gauge-invariance severely  restricts
the type of interaction vertices that can appear at the 
level of the fundamental Lagrangian. Thus, the 
tensorial possibilities allowed 
by Lorentz invariance are drastically reduced 
down to relatively simple tree-level vertices.  
Beyond tree-level, the tensorial structures that have been 
so excluded appear due to quantum corrections, \ie 
they are generated from loops. This fact is not in 
contradiction with renormalizability and gauge-invariance,
provided that the tensorial structures generated, not present 
at the level of the original Lagrangian, are {\it UV finite}, \ie
no counterterms need be introduced to the fundamental Lagrangian 
proportional to the forbidden structures.
\newline
\indent
In order to fix the ideas, let us 
consider a concrete, text-book example. 
In standard QED the tree-level photon-electron vertex
is simply proportional to $\gamma_{\mu}$, 
while 
kinematically one may have, in addition, (for massive on-shell electrons, using
the Gordon decomposition) a term proportional to 
$\sigma_{\mu\nu}q^{\nu}$,  that would correspond to a non-renormalizable 
interaction. 
Of course, the one-loop photon-electron vertex 
generates such a term; specifically, one has 
\be
\Gamma_{\mu}(q) = \gamma_{\mu} F_{1}(q^2)+ \Sme F_{2}(q^2)\,,
\ee
where the scalar cofactors multiplying the two tensorial structures 
are the corresponding form-factors; they are, 
 in general, non-trivial functions of the photon momentum transfer
(the photon ``off-shellness''). $F_{1}(q^2)$ is the 
electric form-factor, whereas $F_2(q^2)$ is the magnetic form-factor.
$F_{1}(q^2)$ is UV divergent, and becomes finite after 
carrying out the standard vertex renormalization. On the other 
hand,  $F_2(q^2)$ comes out UV finite, as it should, given that 
there is no term proportional to  
$\sigma_{\mu\nu}q^{\nu}$ 
(in configuration space) in the original Lagrangian, where 
a potential UV divergence could be absorbed. 
Of course, in the limit of  $q^2 \to 0$ the 
magnetic form-factor $F_2(q^2)$ reduces to the 
famous Schwinger anomaly~\cite{Schwinger:1948iu} 
~(see, e.g., \cite{Peskin:1995ev}).
\newline
\indent
At the level of an Abelian theory, such as QED, the 
above discussion exhausts more or less the theoretical 
complications associated with the calculation of off-shell 
form-factors. However, in non-Abelian theories, such as the 
electroweak sector of the SM, there is an additional important
complication: The  off-shell form-factors obtained from the 
conventional one-loop vertex (and beyond)
depend explicitly on the gfp. This dependence 
disappears when going to the on-shell limit of the incoming 
gauge boson ($q^2 \to 0$ for a photon,  $q^2 \to M^2_{ Z}$ for a $Z$ boson, etc)
but is present for any other value of $q^2$. This fact becomes 
phenomenologically relevant, because one often wants to 
study the various form-factors of particles that are produced
in high-energy collisions, where the 
gauge boson mediating the interaction is far off-shell.
In the case of $e^{+}e^{-}$ annihilation into heavy fermions, 
the value of $q^{2}$ must be above the heavy fermion threshold.
For example, top quarks may be pair-produced
through the reaction $e^{+}e^{-}\rightarrow t\bar{t}$, 
with center-of-mass energy $s=q^{2}\ge 4m_{t}^{2}$.
Due to their large masses,
the produced top quarks are expected to
decay weakly
($t\bar{t} \rightarrow bW^{+}\bar{b} W^{-}$, with subsequent leptonic
decays of the $W$), before hadronization takes place; therefore
electroweak properties of the top can be studied in detail, and QCD
corrections can be reliably evaluated in the context of perturbation theory,
when the energy of the collider is well above the threshold for $t\bar{t}$
production.
The problem is that, in such a case, the intermediate
photon and $Z$ are far off-shell, and therefore, the 
form-factors  $F_{i}^{V}$, appearing in the standard decompositions
\begin{equation}
\Gamma_{\mu}^{V}
(q^2) = \gamma_{\mu}F_{1}^{V}(q^2,\xi)+ \Sm F_{2}^{V}(q^2,\xi)
+\gamma_{\mu}\gamma_{5} F_{3}^{V}(q^2,\xi)
+\Sm\gamma_{5} F_{4}^{V}(q^2,\xi)~,
\label{GeneralForm}
\end{equation}
depend explicitly on $\xi$, which stands collectively for 
$\xi_{ W}$, $\xi_{ Z}$, $\xi_{ A}$,  and $V=A,Z$.
\newline
\indent
The  situation described  above  is rather  general  and affects  most
form-factors; very often the residual gauge-dependences have serious physical
consequences.   For  example,   the  form-factors  display  unphysical
thresholds, bad high-energy  behavior, and sometimes they  are UV and
infrared (IR) divergent.  The way out is  to use the PT construction, and extract
the  physical, gauge-independent  form-factors from  the corresponding
off-shell one-loop PT vertex (and beyond). Applying the PT to the case
of  the  form-factors  amounts  to saying  that  one  has to  identify
vertex-like  contributions (with  the appropriate  tensorial structure
corresponding  to   the  form-factor  considered)   contained  in  box
diagrams, as shown in \Figref{anomal_mag_mom}. The latter, 
when added to the usual vertex graphs, render
all form-factors $\xi$-independent and well-behaved in all respects.
\newline
\indent
In what follows, we will present certain characteristic examples,
in order for the reader to appreciate the nature of the  
pathologies encountered in the conventional formulation,  
and see how the PT resolves all of them at once.

\subsubsection{Anomalous gauge boson couplings}

\begin{figure}[!t]
\bce
\includegraphics[width=16cm]{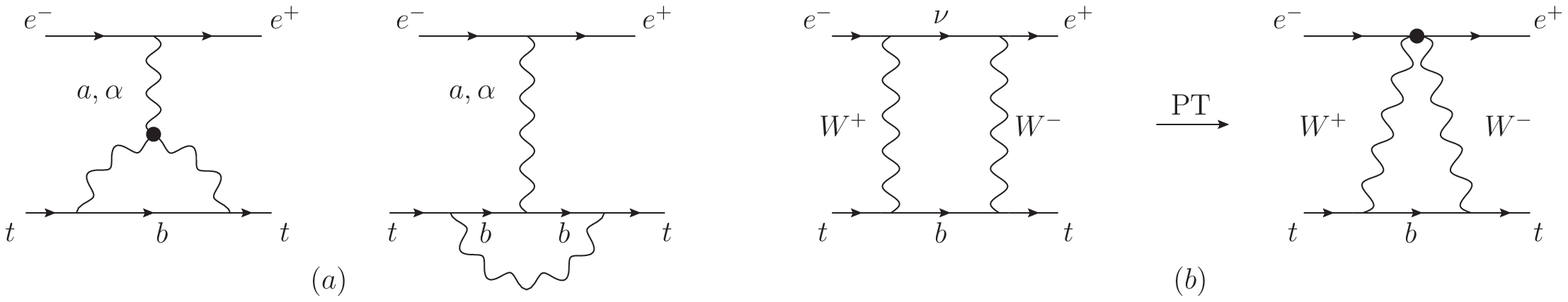}
\ece
\caption{\figlab{anomal_mag_mom} The conventional one-loop vertex (a) and the 
vertex-like piece extracted from the box for $\xi\neq 1$}
\end{figure}

\noindent
A significant amount of research activity has been devoted to the 
study of the three-boson vertices 
$A W^{+}W^{-}$ and $Z W^{+}W^{-}$, with the neutral gauge bosons
{\it off-shell} and the $W$ pair on-shell, or off-shell and subsequently  
decaying to on-shell particles.
Historically, the main motivation for exploring their properties 
was the fact that they were going to be tested 
at LEP2 by direct $W$-pair production, proceeding through 
the process $e^{+}e^{-}\rightarrow W^{+}W^{-}$; 
their experimental scrutiny could provide 
invaluable information on 
non-Abelian nature of the electroweak sector of the SM.
Particularly appealing in this quest has been
the possibility of measuring anomalous gauge boson couplings,
\ie the appearance of contributions to $A W^{+}W^{-}$ and $Z W^{+}W^{-}$
not encoded in the fundamental Lagrangian of the SM.
Such contributions may originate from two sources: 
({\it i}) from radiative corrections within the SM, and/or ({\it ii}) from 
physics beyond the SM. Therefore the first theoretical 
task one is faced with is to carry out the necessary calculations 
for completing part ({\it i}).  
\newline
\indent
The most general parametrization of the trilinear gauge boson vertex for
on-shell $W$s and off-shell  $V=A,Z$ is given by (see \Figref{anomal_gauge_bos_def})
\bea
\Gamma_{\mu\alpha\beta}^{V} &= & -ig_{V}\Big\{
 f \left[ 2g_{\alpha\beta}\Delta_{\mu}+ 4(g_{\alpha\mu}Q_{\beta}-
g_{\beta\mu}Q_{\alpha})\right]
 +2\Delta\kappa_{V}\left(g_{\alpha\mu}Q_{\beta}-g_{\beta\mu}Q_{\alpha}\right)
\nonumber \\
& + &  4\frac{\Delta Q_V}{M_W^2}
\left(\Delta_{\mu}Q_{\alpha}Q_{\beta}-
\frac{1}{2}Q^{2}g_{\alpha\beta}\Delta_{\mu}\right)\Big\}+\cdots,
\eea
with $g_{A}= g_w s_w$, $g_{Z}= g_w c_w$, 
 and the ellipses denote omission of $C$, $P$, or $T$ violating terms.
 The four-momenta $Q$ and $\Delta$
 are related to the incoming momenta $q$, $p_{1}$ and $p_{2}$ of
the gauge bosons $V,~W^-$and $W^+$ respectively, by
$q=2Q$, $p_{1}=\Delta -Q$ and $p_{2}=-\Delta - Q$ ~\cite{Bardeen:1972vi}.
\begin{figure}[!t]
\bce
\includegraphics[width=6cm]{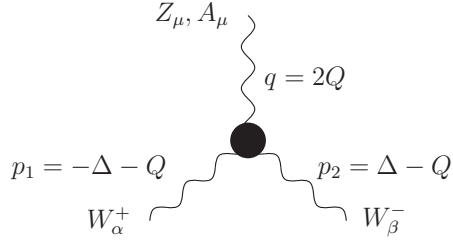}
\ece
\caption{\figlab{anomal_gauge_bos_def}The kinematics of the trilinear gauge boson vertex $VW^+W^-$ (all momenta incoming)  
}
\end{figure}
The {\it off-shell} form-factors $\Delta\kappa_{V}$ and $\Delta Q_{V}$,
also defined as
$\Delta\kappa_{V}= \kappa_{V} + \lambda_{V} - 1$,
and $\Delta Q_{V}= -2\lambda_{V}$,
are compatible with $C$, $P$,
and $T$ invariance, and
are  related to the magnetic dipole moment $\mu_{W}$ and the electric
quadrupole moment $Q_{W}$, by the following expressions:
\be
\mu_{W} = \frac{e}{2M_{ W}}(2+ \Delta\kappa_{A}), \qquad
Q_{W}= -\frac{e}{M^{2}_{ W}}(1+\Delta\kappa_{A}+\Delta Q_{A}).
\ee
In the context of the SM
 their canonical, tree-level values are
 $f=1$ and  $\Delta\kappa_{V}=\Delta Q_{V}=0$.
To determine the radiative corrections to these quantities
one must cast the resulting one-loop expressions in the  form
\be
\Gamma_{\mu\alpha\beta}^{V}= -ig_{V}[
 a_{1}^{V}g_{\alpha\beta}\Delta_{\mu}+ a_{2}^{V}(g_{\alpha\mu}Q_{\beta}-
g_{\beta\mu}Q_{\alpha})
 + a_{3}^{V}\Delta_{\mu}Q_{\alpha}Q_{\beta}],
\label{1loopParametrization}
\ee
where $a_{1}^{V}$, $a_{2}^{V}$, and $a_{3}^{V}$ are
 complicated functions of the
momentum transfer $Q^2$, and the masses of the particles
appearing in the loops.
It then follows that  $\Delta\kappa_{V}$ and $\Delta Q_{V}$
are given by 
\be
\Delta\kappa_{V}=\frac{1}{2}(a_{2}^{V}-2a_{1}^{V}-Q^{2}a_{3}^{V}), \qquad
\Delta Q_{V}= \frac{M^{2}_{W}}{4}a_{3}^{V}~.
\label{1loopdeltakappaandQ}
\ee

\begin{figure}[!t]
\bce
\includegraphics[width=10cm]{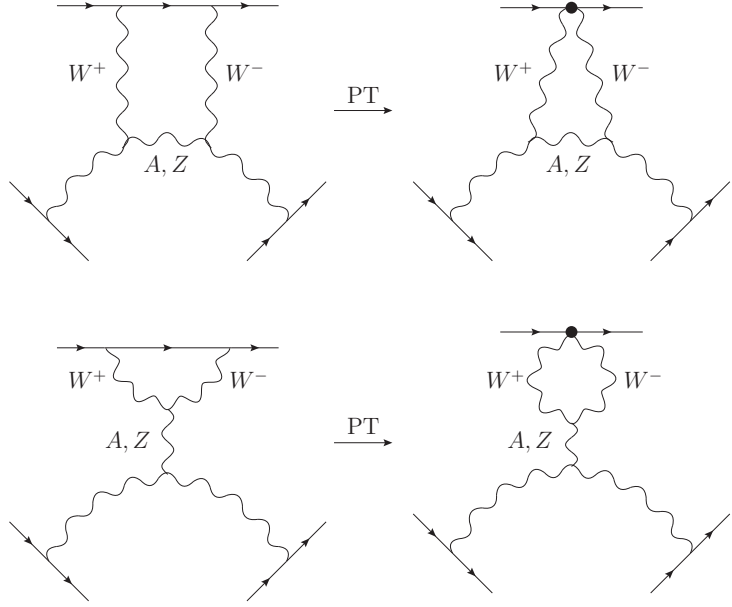}
\ece
\caption{\figlab{anomal_gauge_bos_PTterms}Two of the graphs contributing pinching parts to the gauge independent $VW^+W^-$ vertex.}
\end{figure}
Calculating the one-loop expressions for $\Delta\kappa_{V}$ and
 $\Delta Q_{V}$ is a non-trivial task, both from
the technical and the conceptual point of view.
Let us focus, for concreteness, on the case $V=A$. If one
calculates
just the Feynman diagrams contributing to the $A W^{+}W^{-}$
vertex and then extracts from them the contributions to
$\Delta\kappa_{A}$ and $\Delta Q_{A}$,
one arrives at
expressions that are
plagued with several pathologies, gfp-dependence being one of them.
Indeed, even if the two $W$s are considered to be on-shell ($p_1^{2}=p_2^{2}=M_{ W}^{2}$)
since the incoming
photon is not, there is no {\it a priori}
 reason why a gfp-independent answer
should emerge. Indeed, in the context of the renormalizable $R_{\xi}$ gauges
the final answer depends on the
choice of the gfp $\xi$, which enters into the one-loop
calculations through the gauge-boson propagators
( $W$, $Z$,$A$, and unphysical ``would-be'' Goldstone bosons).
In addition, as shown by an explicit calculation
performed
in the Feynman gauge ($\xi=1$), the answer for
$\Delta\kappa_{A}$
is {\it infrared divergent} and
violates perturbative unitarity,
\eg it grows
monotonically for $Q^2 \rightarrow \infty$~\cite{Argyres:1992vv}.
\newline
\indent
All the above pathologies may be circumvented
if one uses the PT definition of the relevant (off-shell) 
gauge boson vertices~\cite{Papavassiliou:1993ex}. 
The application of the PT 
identifies vertex-like contributions from the box graphs, as shown in \Figref{anomal_gauge_bos_PTterms},
which are subsequently distributed, in a unique way, among the  
various form-factors. The final outcome is that 
one arrives at
new expressions, to be denoted by 
$\widehat{\Delta}\kappa_{A}$ and $\widehat{\Delta} Q_{A}$,
which are
gauge fixing parameter ($\xi$) independent,
ultraviolet {\it and} infrared finite, and
monotonically decreasing for large momentum transfers $Q^{2}$.
\newline
\indent
Using ``hats'' to denote the gfp-independent one-loop contributions,
we have
\bea
\widehat{\Delta}\kappa_{A} &=& \Delta\kappa_{A}^{(\xi=1)}
+ \Delta\kappa_{A}^{P},\nonumber\\
\widehat{\Delta}Q_{A} &=& \Delta Q_{A}^{(\xi=1)}+  \Delta Q_{A}^{P},
\eea
where $\Delta Q_{A}^{(\xi=1)}$ and $\Delta Q_{A}^{(\xi=1)}$
are the contributions of the usual vertex diagrams in
the Feynman gauge~\cite{Argyres:1992vv}, 
whereas
$\Delta Q_{A}^P$ and $\Delta Q_{A}^P$ are the analogous
contributions from the pinch parts.
A straightforward calculation yields
\be
{\Delta\kappa}_{A}^{P}= - \frac{Q^{2}}{M^{2}_{ W}}
\sum_{V} \frac{\alpha_{V}}{\pi} \int_{0}^{1}da \int_{0}^{1}dt\,
\frac{t(at-1)}{L^{2}_{V}},
\ee
where
\be
L^{2}_{V} = t^{2}-t^{2}a(1-a)\left(\frac{4Q^{2}}{M^{2}_{ W}}\right) +
 (1-t)\frac{M^{2}_{ V}}{M^{2}_{ W}},
\ee
and
\be
{\Delta Q}_{A}^{P} = 0.
\ee
Notice an important point: $\Delta\kappa_{A}^{P}$ contains an
infrared divergent term, stemming from the double
integral shown above, when $V=A$. This term cancels {\it exactly}
against a similar infrared divergent piece
contained in $\Delta\kappa_{A}^{(\xi = 1)}$,
thus rendering
$\widehat{\Delta}\kappa_{A}$ infrared finite.
After the infrared pieces have been canceled, one notices that
the remaining contribution of $\Delta\kappa_{A}^{P}$
decreases monotonically
as $Q^2 \rightarrow \pm \infty$; due to the difference
in relative signs, this contribution cancels
asymptotically against the monotonically increasing
contribution from
$\Delta\kappa_{A}^{(\xi = 1)}$.
Thus, by including the pinch parts the
unitarity of $\widehat{\Delta}\kappa_{A}$ is restored and
$\widehat{\Delta}\kappa_{A} \rightarrow 0$ for large values of $Q^2$.

\subsubsection{Neutrino charge radius}
\noindent
The neutrino electromagnetic form-factor and the 
neutrino charge radius (NCR) have constituted an important   
theoretical puzzle for over three decades.
Since the dawn of the SM it was 
pointed out that 
radiative   corrections will induce 
an effective  
one-loop $A^{*}(q^2)\nu \nu$ vertex, 
to be denoted by ${\Gamma}^{\mu}_{A \nu \bar{\nu}}$, 
with $A^{*}(q^2)$ an off-shell photon. 
Such a vertex  would, in turn, give rise  to   
a small but non-vanishing NCR. 
Traditionally (and, of course, non-relativistically and rather heuristically) 
the NCR has been interpreted as 
 a measure of the ``size'' of the neutrino $\nu_i$ when probed 
electromagnetically, owing to  
its classical definition  
(in the static limit) as the second moment 
of the  spatial neutrino charge density $\rho_{\nu}(\bf{r})$, \ie 
\be
\left\langle r^2_{\nu}\right\rangle\sim \int\! d {\bf{r}}\, r^2 \rho_{\nu}({\bf{r}}).
\ee 
\indent
From the quantum field theory point of view, the NCR is defined as follows.
If we write ${\Gamma}^{\mu}_{A \nu \bar{\nu}}$ in the form
\be
{\Gamma}^{\mu}_{A \nu \bar{\nu}}(q^2)
=  \gamma_{\mu}(1-\gamma_{5})F_{ D}(q^2),
\ee
where $F_{D}(q^2)$ is the (dimensionless) Dirac electromagnetic form-factor,
then the NCR is given by
\be
\left\langle r^2_\nu\right\rangle   =  6   
\left.\frac{\partial  F_{ D}(q^2)}{\partial q^2}\right\vert_{q^2 = 0}.  
\ee
Gauge invariance (if not compromised) requires that, 
in the limit $q^2\to 0$,
$F_{ D}(q^2)$ must be proportional to $q^2$, 
\ie that it can be cast in the form 
$F_{ D}(q^2) = q^2 F (q^2)$, with the dimensionful 
form-factor $F(q^2)$  being regular as $q^2\to 0$.
As a result, the $q^2$ contained in $F_{ D}(q^2)$ cancels against 
the $(1/q^2)$ coming from the propagator of the off-shell photon, 
and one effectively obtains  a contact interaction between 
the neutrino and the sources of the (background) photon, as 
one would expect from classical considerations.
\newline
\indent
Even though in the SM the one-loop
computation of the 
{\it entire} $S$-matrix element 
describing the  electron-neutrino scattering, shown in \Figref{NCR_diagrams},  
is conceptually straightforward, the identification
of a {\it subamplitude},  
which would serve as the  effective 
${\Gamma}^{\mu}_{A \nu \bar{\nu}}$ has
been   faced   with  serious   complications,   associated  with   the
simultaneous   reconciliation   of   crucial  requirements   such   as
gauge-invariance,       finiteness,       and      target-independence. 
Specifically,
various  attempts to define the value of the NCR within the SM from the
one-loop  ${\Gamma}^{\mu}_{A \nu \bar{\nu}}$ 
vertex   calculated  in the   renormalizable
($R_{\xi}$) gauges   reveal  that  the   corresponding electromagnetic
form-factor depends  explicitly  on  
the  gauge-fixing parameter $\xi$ in a prohibiting  way.  
In particular, even though  in  the static limit of
zero momentum transfer,  $q^2 \to 0$,  the Dirac form-factor 
becomes independent of  $\xi$,  its first  derivative with respect  to
$q^2$, which corresponds to the definition  of the NCR, 
continues to depend on it. 
Similar (and sometimes worse) problems occur in the context of other
gauges (\eg unitary gauge).
These complications have obscured the 
entire concept of an NCR, and have casted serious doubts 
on whether it can be regarded as a genuine physical observable.   
\indent
\subsubsection{The physical NCR}
\begin{figure}[!t]
\bce
\includegraphics[width=14cm]{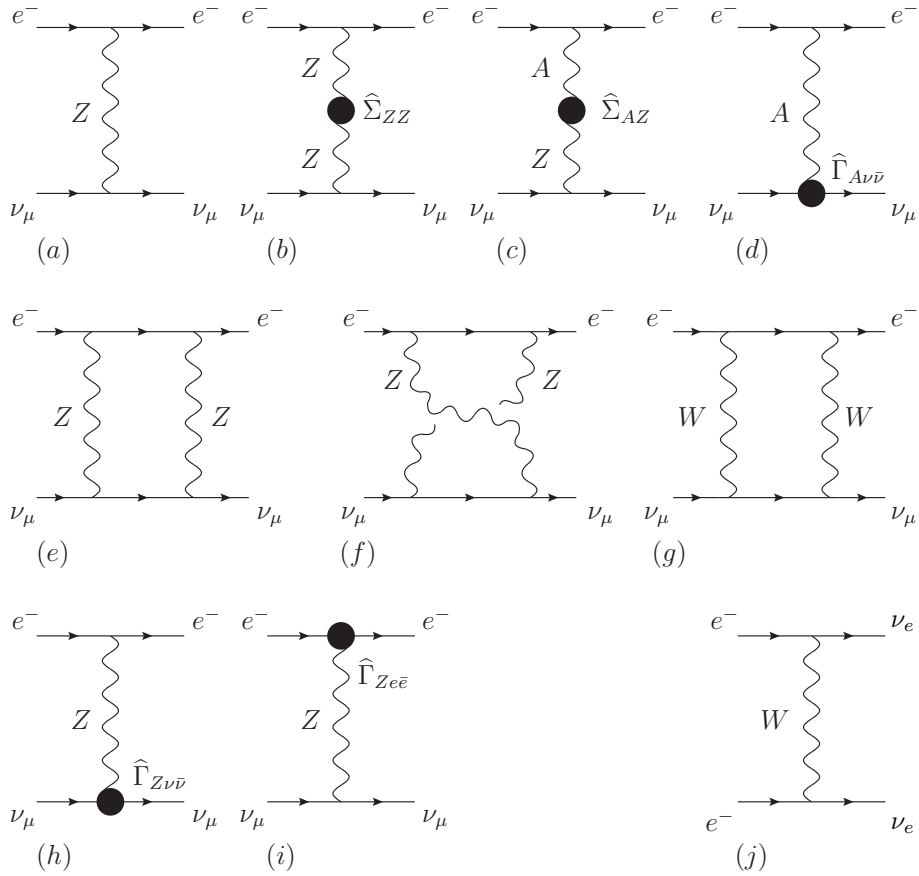}
\ece
\caption{\figlab{NCR_diagrams}The electroweak diagrams contributing to the entire 
electron-neutrino scattering process at one-loop. The last diagram (and all of its dressing) is absent when the neutrino species is muonic.}
\end{figure}
\noindent
Of course, if a quantity  is gauge-dependent it is not physical. But the fact that 
the off-shell vertex is gauge-dependent only means that it just does not 
serve as a physical definition of the NCR. It does not mean that 
an {\it effective} NCR cannot be encountered which satisfies {\it all} necessary  
physical properties, gauge-independence being one of them.  
Indeed, several authors have attempted to find a 
{\it modified} vertex-like amplitude, that would lead 
to a consistent definition of the electromagnetic 
NCR (see~\cite{Bernabeu:2000hf,Bernabeu:2002nw,Bernabeu:2002pd} for an extended list of references). 
The common underlying idea in all these works is 
to rearrange, somehow, the Feynman graphs contributing to the 
scattering amplitude 
of neutrinos with charged particles, 
in an attempt to find a vertex-like   
combination that would satisfy all desirable properties. 
What makes this exercise so difficult is that, in addition to 
gauge-independence, a multitude of 
other crucial physical requirements  
need to be satisfied as well. For example, one should  
not enforce gauge-independence at the expense of 
introducing target-dependence.  
Therefore, a definite guiding-principle is needed, allowing 
for the construction of physical subamplitudes with definite kinematic 
structure (\ie self-energies, vertices, boxes).
\newline
\indent
The guiding-principle in question has been provided by the PT.
As was shown for the first time in ~\cite{Papavassiliou:1989zd}, the 
rearrangement of the physical amplitude 
$f^{\pm}\nu \to f^{\pm} \nu$, where $f^{\pm}$ are the target fermions,
into PT self-energies, vertices, and boxes
conclusively settles the issue: the {\it proper} PT vertex 
with an off-shell photon 
and two on-shell neutrinos (see \Figref{NCR_PT_vertex}),  denoted by 
$\widehat{\Gamma}^{\mu}_{A \nu_i \bar{\nu}_i}$,   
{\it furnishes unambiguously and uniquely the physical NCR}.
As we know from the general discussion of the previous section,   
$\widehat{\Gamma}^{\mu}_{A \nu_i \bar{\nu}_i}$ 
{\it coincides} with the BFM vertex involving an off-shell
background photon and two on-shell neutrinos, calculated  
by putting the quantum $W$-bosons inside the loops in the Feynman gauge (the BFG). 
\newline
\indent 
Several years after its original resolution within the PT ~\cite{Papavassiliou:1989zd}, 
the NCR issue was revisited in ~\cite{Bernabeu:2000hf}.
There, in addition to an exhaustive demonstration of the various 
gauge cancellations, two important conceptual points have been conclusively settled:
\begin{itemize}
\item[{\it i}.] 
As already explained in ~\cite{Papavassiliou:1989zd}, 
the box diagrams furnish gauge-dependent (propagator-like) 
 contributions that are crucial for the gauge-cancellations, 
but once these contributions have been identified and extracted,
the remaining ``pure'' box cannot form part of the NCR, because 
it would introduce process-dependence (due to its non-trivial  
dependence on the target-fermion masses, for one thing). 
The most convincing way to understand why the pure box could not possibly  
enter into the NCR definition  
is to consider the case of right-handedly
polarized target fermions, which do not couple to the $W$s: in that 
case, the box diagram is not even there! (the gauge-cancellations 
proceed now differently, since the coupling of the $Z$ boson to the 
target fermions is also modified)~\cite{Bernabeu:2000hf}.
\newline
\item[{\it ii}.] The mixing self-energy  ${\widehat{\Pi}}_{{ A} { Z}} (q^2)$ 
should {\it not} be included in the definition of the NCR either.
The reason is more subtle (and had not been recognized in ~\cite{Papavassiliou:1989zd}): 
${\widehat{\Pi}}_{{ A} { Z}} (q^2)$ is {\it not} an RG-invariant 
quantity; adding it to the finite contribution coming from the 
proper vertex would convert the resulting NCR to a $\mu$-dependent, 
and therefore unphysical quantity. 
Instead,
${\widehat{\Pi}}^{ {A} { Z}}  (q^2)$ must be combined with the
appropriate $Z$-mediated  {\it tree-level} contributions 
(which  evidently  do not  enter into  the
definition of the NCR) in order to form, with them, the RG-invariant combination 
$\bar{s}_w^{2}(q^2)$ of Eq.(\ref{sweff1}),
whereas the ultraviolet-finite 
NCR will be determined from the {\it proper} vertex only.
\end{itemize}
\indent
\begin{figure}[!t]
\bce
\includegraphics[width=14cm]{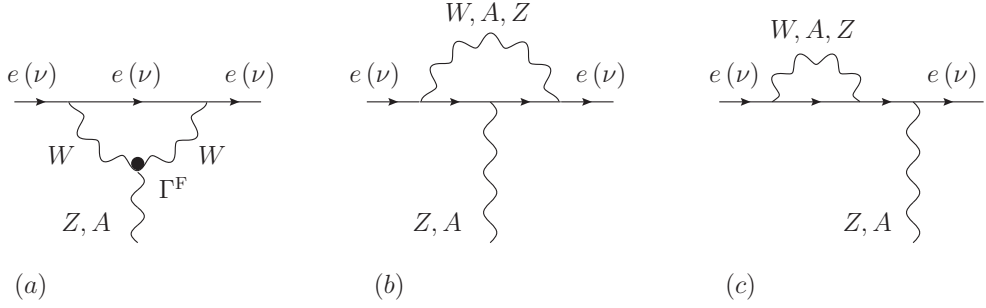}
\ece
\caption{\figlab{NCR_PT_vertex} The PT vertex with an off-shell gauge-boson ($A$,$Z$)
and two on-shell fermions ($\nu$, $e$)}
\end{figure}
Writing 
$\widehat{\Gamma}^{\mu}_{A \nu_i \bar{\nu}_i}
= q^2  \widehat{F}_{i}(q^2) \gamma_{\mu}(1-\gamma_{5})$,
the physical NCR is  
then defined as   
$\left\langle r^2_{\nu_i}\right\rangle = 6 \widehat{F}_{i}(0)$, and the
explicit calculation yields  
\be
\left\langle r^2_{\nu_i}\right\rangle =
\frac{G_{ F}}{4\, {\sqrt 2 }\pi^2} 
\left[3 
- 2\log \left(\frac{m_{ i}^2}{M_{ W}^2} \right) \right],
\label{ncr}
\ee
where $i= e,\mu,\tau$, 
$m_i$ denotes the mass of the charged isodoublet
partner of the neutrino under consideration, and $G_{ F}$
is the Fermi constant. Note that the logarithmic term on the rhs of 
Eq.~(\ref{ncr}) originates from the Abelian graph, and  is not 
affected by the PT procedure, \ie it is, from the beginning, gauge-independent.   
This term may be obtained directly if one considers the problem 
from an effective field theory point of view, in the limit of a very heavy $W$-boson. 
\newline
\indent
The observable nature of the
NCR was established in~\cite{Bernabeu:2002nw}, by
demonstrating that, at least in principle,  
the probe-independent NCR  
may be extracted from a judicious combination of scattering experiments involving neutrinos
and anti-neutrinos. 
\newline
\indent
Consider, in fact, the elastic processes 
$ f(k_1) \nu(p_1)\!  \to\! f(k_2) \nu(p_2) $ and 
$f(k_1) \bar{\nu}(p_1)\!  \to\! f(k_2) \bar{\nu}(p_2) $,
where $f$ denotes an electrically charged 
fermion belonging to a different
isodoublet than the neutrino $\nu$, in order to eliminate 
the diagrams mediated by a charged $W$-boson.
The Mandelstam variables are defined as
$s=(k_1+p_1)^2 = (k_2+p_2)^2$, 
$t= q^2 = (p_1-p_2)^2 = (k_1-k_2)^2$, 
$u = (k_1-p_2)^2 = (k_2-p_1)^2$, 
and $s+t+u=0$. 
In what follows, we will restrict ourselves to the 
limit $t=q^2 \to 0$ of the above amplitudes,
assuming that all external (on-shell) fermions are massless.
As a result of this special kinematic situation we have the
following relations:
$p_1^2 = p_2^2 = k_1^2 = k_2^2 = p_1 \cdot p_2 = k_1 \cdot k_2 = 0$
and 
$p_1 \cdot k_1  = p_1 \cdot k_2 = p_2 \cdot k_1 = p_2 \cdot k_2 = s/2 $.
In the center-of-mass system, we have that 
$t=-2 E_{\nu}E_{\nu}'(1-x)\leq 0 $, 
where $E_{\nu}$ and $E_{\nu}'$
are the energies of the neutrino before and after the
scattering, respectively, and 
$x \equiv \cos\theta_{cm}$, where
$\theta_{cm}$
is the scattering angle. Clearly, the condition $t=0$ 
corresponds to the exactly forward amplitude, 
with $\theta_{cm}=0$, \, $x=1$.
\newline
\indent
At tree-level the amplitude  $ f \nu  \to f \nu $ is 
mediated by an off-shell $Z$-boson.
At one-loop, the relevant  
contributions are determined  
through the PT 
rearrangement of the amplitude, giving rise to 
gauge-independent subamplitudes. 
In particular, $\widehat{\Pi}_{{A}{Z}}^{\mu\nu}(q^2)$
obtained is transverse, for {\it both} 
the fermionic and the bosonic contributions,
\ie $\widehat{\Sigma}_{{A}{Z}}^{\mu\nu}(q^2)
= (q^2 g^{\mu\nu}  - q^{\mu} q^{\nu}) 
{\bf \widehat{\Pi}}_{ {A} { Z}} (q^2)$, 
and we only keep the $g^{\mu\nu}$ part of $\widehat{\Pi}_{{Z}{Z}}^{\mu\nu}(q^2)$. 
The one-loop vertex 
$\widehat\Gamma_{{ Z} { F} \bar{ F}}^{\mu}(q,p_1,p_2)$, 
with $F = f$  or $F = \nu$, 
satisfies a QED-like WI, relating it to the one-loop  
inverse fermion propagators $\widehat\Sigma_{ F}$,
\ie  
$ q_{\mu} 
\widehat\Gamma_{{ Z} { f} \bar{ F}}^{\mu}(q,p_1,p_2)
= 
\widehat\Sigma_{ F} (p_1) - \widehat\Sigma_{ F} (p_2)$.
In the limit of 
$q^2 \to 0$,  
$\widehat\Gamma_{{ Z} { F} \bar{ F}}^{\mu} 
\sim q^2 \gamma^{\mu}(c_1 + c_2 \gamma_5)$; 
since it is multiplied by a
massive $Z$ boson propagator $(q^2 - M_{ Z})^{-1}$, its 
contribution to the amplitude vanishes when 
$q^2 \to 0$. This is to be contrasted with the 
$\widehat{\Gamma}^{\mu}_{A \nu_i \bar{\nu}_i}$,  
which is accompanied by a 
$(1/q^2)$ photon-propagator, thus giving rise 
to a contact interaction between the target-fermion and the neutrino,
described by the NCR. 
\newline
\indent 
In order to experimentally isolate from the amplitude the contribution due to the NCR, 
we first eliminate the box-contributions.
The basic observation 
is that the tree-level amplitudes 
${\mathcal M}_{\nu f}^{(0)}$,  
as well as the
part of the one-loop amplitude ${\mathcal M}_{\nu f}^{(B)}$
consisting of 
the propagator and vertex corrections 
(namely the ``Born-improved'' amplitude),
are proportional to 
\be
[\bar{u}_{f}(k_2)\gamma_{\mu}( v_f + a_f \gamma_5 ) 
u_{f}(k_1)]
[\bar{v}(p_1)\gamma_{\mu} P_{ L}  v(p_2)],
\ee
and  
therefore transform differently than the boxes 
under the replacement $\nu \to \bar{\nu}$, since~\cite{Sarantakos:1982bp}
\be
\bar{u}(p_2)\gamma_{\mu} P_{ L} u(p_1) \to 
- \bar{v}(p_1)\gamma_{\mu} P_{ L}  v(p_2)
= - \bar{u}(p_2)\gamma_{\mu} 
P_{ R} u(p_1).
\label{sar}
\ee
Thus, under the above transformation, 
${\mathcal M}_{\nu f}^{(0)} + {\mathcal M}_{\nu f}^{(B)}$ reverse 
sign once, 
whereas the box contributions reverse sign twice.
These distinct transformation properties 
allow for the isolation of
the box contributions 
when the forward differential cross-sections
$(d\sigma_{\nu f}/dx)_{x=1}$ and  
$(d\sigma_{\bar{\nu} f}/dx)_{x=1}$ 
are appropriately combined. 
In particular, the combination
\be
\sigma^{\pm}_{\nu f} \equiv 
\left.\frac{d\sigma_{\nu f}}{dx}\right\vert_{x=1}
\pm \left.\frac{d\sigma_{\bar{\nu} f}}{dx}\right\vert_{x=1}
\ee 
either does not contain 
boxes (when choosing the plus sign), or precisely   
isolates the contribution of the boxes (when choosing the minus sign).
\newline
\indent
Finally, a detailed analysis shows 
that in the kinematic limit considered, 
the Bremsstrahlung contribution vanishes, 
due to a completely destructive interference 
between the two relevant  diagrams corresponding to the
processes $f A \nu (\bar{\nu}) \to f \nu (\bar{\nu})$ and 
$f \nu (\bar{\nu}) \to f A \nu (\bar{\nu})$. 
The absence of such corrections is consistent with the
fact that there are no infrared divergent contributions 
from the (vanishing) vertex
$\widehat\Gamma_{{ Z} { F} \bar{ F}}^{\mu}$, 
to be canceled against.  
\newline
\indent 
$\sigma^{+}_{\nu f}$ receives contributions from the 
tree-level exchange of a $Z$-boson, the one-loop contributions
from the ultraviolet divergent quantities 
$\widehat{\Sigma}_{{Z}{Z} }(0)$ and  
${\widehat{\Pi}}^{ {A}  {  Z}} (0)$, 
and the (finite) NCR, coming from the proper vertex 
$\widehat{\Gamma}^{\mu}_{A \nu_i \bar{\nu}_i}$.
The first three contributions are universal (\ie common to all 
neutrino species) whereas that of the proper vertex 
$\widehat{\Gamma}^{\mu}_{A \nu_i \bar{\nu}_i}$
is flavor-dependent. 
After organizing the one-loop corrections of $\sigma^{(+)}_{\nu f}$ 
in terms of the RG-invariant quantities $\bar{R}_{ Z}$ and $\bar{s}_w^{2}(q^2)$,
one may  fix $\nu = \nu_{\mu}$, 
and then consider three different choices for $f$: ({\it i}) 
right-handed electrons, $e_{ R}$; 
({\it ii}) left-handed electrons, $e_{ L}$, and ({\it iii}) neutrinos, 
$\nu_{i}$ 
other than  $\nu_{\mu}$, \ie $i=e,\tau$. 
Thus, we arrive at the system 
\bea
\sigma^{+}_{\nu_{\mu} \nu_i} &=& s \pi \bar{R}^2(0),\nonumber\\
\sigma^{+}_{\nu_{\mu} \,e_{ R}}  &=& 
s \pi \bar{R}^2(0)\bar{s}_w^{4}(0) 
- 2 \lambda s_w^{2} 
\left\langle r^2_{\nu_{\mu}}\right\rangle,\nonumber\\
\sigma^{+}_{\nu_{\mu} e_{ L}} &=& 
s \pi \bar{R}^2(0) 
\left(\frac{1}{2} - \bar{s}_w^{2}(0)\right)^{2} 
+ \lambda (1-2 s_w^{2}) 
\left\langle r^2_{\nu_{\mu}}\right\rangle,
\eea
where $\lambda \equiv (2\sqrt{2}/3) s \alpha \,G_{ F}$.
Now $\bar{R}^2(0)$, $\bar{s}_w^{2}(0)$, and  
$\left\langle r^2_{\nu_{\mu}}\right\rangle$ are treated as three unknown 
quantities, to be determined from the above algebraic equations. 
Substituting  $s \pi \bar{R}^2(0) \to \sigma^{+}_{\nu_{\mu} \nu_i}$
into the equations above, we arrive at 
a system which is linear in the unknown quantity 
$\left\langle r^2_{\nu_{\mu}}\right\rangle$, and 
quadratic in $\bar{s}_w^{2}(0)$. 
The corresponding solutions are given by
\bea
\bar{s}_{w}^{2}(0) &=&  s_w^{2} \pm \Omega^{1/2}
\nonumber\\
\left\langle r^2_{\nu_{\mu}}\right\rangle &=&  
\lambda^{-1}
\left[\left(s_w^{2}-\frac{1}{4} \pm \Omega^{1/2}\right)
\sigma^{+}_{\nu_{\mu} \nu_i}  
+ \sigma^{+}_{\nu_{\mu} e_{ L}} - 
\sigma^{+}_{\nu_{\mu} e_{ R}}
\right],
\label{sol2}
\eea
where the discriminant $\Omega$ is given by
\be
\Omega = (1- 2 s_w^{2}) \left(\frac{\sigma^{+}_{\nu_{\mu} e_{ R}}}
{\sigma^{+}_{\nu_{\mu} \,\nu_i}} -  
\frac{1}{2} s_w^{2}  
\right)
+  2 s_w^{2} 
\frac{\sigma^{+}_{\nu_{\mu} e_{ L}}}
{\sigma^{+}_{\nu_{\mu} \nu_i}}
\label{disc}
\ee
and must satisfy $\Omega > 0$. 
The actual sign in front of $\Omega$ may be chosen by requiring that 
it correctly accounts for the sign of the shift of $\bar{s}_{w}^{2}(0)$
with respect to $ s_w^{2}$ predicted by the theory \cite{Hagiwara:1994pw}. 
\newline
\indent
To extract the experimental values of the quantities 
$\bar{R}^2(0)$, $\bar{s}_w^{2}(0)$, and $\left\langle r^2_{\nu_{\mu}}\right\rangle$,
one must substitute  
in the above equations the experimentally
measured values for the differential cross-sections 
$\sigma^{+}_{\nu_{\mu} e_{ R}}$, 
$\sigma^{+}_{\nu_{\mu} e_{ L}}$,
and $\sigma^{+}_{\nu_{\mu} \nu_i}$. 
Thus, one would 
have to carry out three different pairs of experiments. 

\subsubsection{Neutrino-Nuclear coherent scattering and the NCR}
\noindent
The above analysis establishes the observable nature of the NCR 
in terms of Gedanken-type of experiments. 
In practice, however, one needs to resort to a more feasible procedure, 
even at the expense of using as an input the theoretical SM values for certain 
parts of the process (\eg boxes).
\newline
\indent
One such proposal aims to extract the value of the NCR 
from the {\it coherent scattering} of a neutrino against a heavy nucleus~\cite{Papavassiliou:2005cs}.
 The notion of coherent nuclear scattering 
is well-known from electron scattering. In the neutrino case it was developed 
in connection with the discovery of weak neutral currents, with a component 
proportional to the number operator \cite{Bernabeu:1975tw}.
\newline
\indent
 When a
projectile (\eg a neutrino) scatters elastically from a composite system
(\eg a nucleus), the amplitude $F({\bf p'},{\bf p})$ for scattering from
an incoming momentum ${\bf p}$ to an outgoing momentum ${\bf p'}$ is given 
as the sum of the contributions from
each constituent,
\be
F({\bf p'},{\bf p})= \sum_{j=1}^{A} f_j({\bf p'},{\bf p}) 
e^{i {\bf q}\cdot {\bf x}_j} ,
\label{coh1}
\ee
where ${\bf q}={\bf p'}-{\bf p}$ is the momentum transfer and the individual
amplitudes $f_j({\bf p'},{\bf p})$ are added with a relative phase-factor, 
determined by the corresponding wave function.
The differential cross-section is then
\be 
\frac{d\sigma}{d\Omega} = |F({\bf
p'},{\bf p})|^2 = \sum_{j=1}^{A} |f_j({\bf p'},{\bf p})|^2 +
\sum_{j,i}^{i\neq j} f_i({\bf p'},{\bf p})f_j^{\dagger}({\bf p'},{\bf p})
e^{i {\bf q}\cdot ({\bf x}_j-{\bf x}_i)}.
\label{coh2}
\ee
In principle, due to the presence of the phase factors, major cancellations
may take place among the $A(A-1)$ terms in the second (non-diagonal) sum.
This happens for  $qR \gg 1$, where  $R$ is the size of the composite system,  
and the scattering would be incoherent.
On the contrary, under the condition that $qR \ll 1$, 
then all phase factors may be approximated by unity, and the terms
in (\ref{coh2}) add coherently.  If there were only one type of constituent,
\ie $f_j({\bf p'},{\bf p}) = f({\bf p'},{\bf p})$ for all $j$, then
(\ref{coh2}) would reduce to
\be
\frac{d\sigma}{d\Omega} = A^2 \left|f({\bf p'},{\bf p})\right|^2 
\ee
Evidently, in that case, the {\it coherent} scattering cross-section would
be enhanced by a factor of $A^2$ compared to that of a single constituent.
In the realistic case of a nucleus with $Z$ protons and $N$ neutrons,
and assuming zero nuclear spin, the corresponding
differential cross-section reads~\cite{Bernabeu:1975tw}
\be
\frac{d\sigma}{d\Omega} = \frac{G^2_{{ F}}}{4(2\pi)^2} E^2 
(1+\cos\theta) \left[(1-4 s_w^2)Z -  N\right]^2,
\label{dsdy}
\ee
where $s_w$ is the sine of the weak mixing angle, $d\Omega = d\phi
d(\cos\theta)$, and $\theta$ is the scattering angle. 
\newline
\indent
In such an experiment the relevant quantity to measure is the kinetic energy
distribution of the recoiling nucleus, which, in turn, may be directly
related to the shift in the value of $s_w^{2}$
produced by the NCR.  
Specifically, the UV-finite contribution from the NCR may be absorbed into an additional 
(flavor-dependent!) shift of $\bar{s}_w^{2}(q^2)$. In
fact, a detailed analysis based on the methodology developed in~\cite{Hagiwara:1994pw}, 
reveals that, in the kinematic range of interest,
the numerical impact of $\bar{R}_{{Z}}(q^2)$ and $\bar{s}_w^{2}(q^2)$ is
negligible, \ie these quantities do not run appreciably.  Instead, the
contribution from the NCR amounts to a correction of few percent
to $s_w^{2}$, given by an expression of the form
$s_w^{2} \longrightarrow s_w^{2} 
\left(1-\frac{2}{3} M_{ W}^{2}\left\langle r^2_{\nu_i}\right\rangle\right)$~~\cite{Papavassiliou:2005cs}.
The contributions of the boxes are of the order $g_{ W}^4/M_{ W}^{2}$,
and they may have to be subtracted out ``by hand''.
This type of experiment has been
proposed in order to observe the coherent elastic neutrino-nuclear
scattering for the first time, and it could also furnish the first {\it terrestrial}
measurement of the NCR.
\newline
\indent
Finally, it is interesting to mention that if one were to consider the
differences in the cross-sections between two {\it different} neutrino species
scattering coherently off the same nucleus, as proposed by Sehgal 
long ago \cite{Sehgal:1985iu}, one would eliminate all unwanted
contributions, such as boxes, thus measuring the {\it difference} between
the two corresponding charge radii. Such a difference would contribute 
to a difference for the neutrino index of refraction in nuclear matter~\cite{Botella:1986wy}.

\subsection{Gauge-independent definition of electroweak parameters}
\noindent
The PT offers the possibility to define a set of electroweak parameters 
that are completely gauge-independent. This is useful 
because one usually tends to place bounds on new physics 
by comparing with the most sensitive  electroweak parameters. Clearly, 
gauge artifacts may give 
misleading information on the viability and relevance of possible 
extensions of  the SM.   

\subsubsection{The $S$, $T$, and $U$ parameters}
\noindent
One  of  the most  frequently  used  parametrizations  of the  leading
contributions of electroweak radiative  corrections is in terms of the
$S$, $T$,  and $U$ parameters~\cite{Peskin:1995ev}.
The expressions for these parameters are suitable combinations 
of self-energies (usually referred to also as ``oblique corrections''); 
in terms of the conventional SM self-energies they are given by
\bea 
\hat{\alpha} S &=&
\frac{4 \hat{c}^2 \hat{s}^2}{M_{ Z}^2}
\Re e 
\Big\{
{\Pi}_{{Z}{Z}}(M_{ Z}^2) - {\Pi}_{{Z}{Z}}(0)\nonumber \\
&-& \frac{\hat{c}^2-\hat{s}^2}{\hat{c}\hat{s}}
\left[{\Pi}_{{A}{Z}}(M_{ Z}^2) - {\Pi}_{{A}{Z}}(0)\right]
- {\Pi}_{{A}{A}}(M_{ Z}^2)
\Big\},\label{STU-1}\nonumber\\
\hat{\alpha}  T &=& 
\frac{{\Pi}_{{W}{W}}(0)}{M_{ W}^2}
- \frac{\hat{c}^2}{M_{ W}^2}
\left[{\Pi}_{{Z}{Z}}(0) +
\frac{2\hat{s}}{\hat{c}}{\Pi}_{{A}{Z}}(0) \right],\label{STU-2}\nonumber\\
\hat{\alpha} U &=& 4\hat{s}^2 
\Re e \left\{
\frac{{\Pi}_{{W}{W}}(M_{ W}^2)-{\Pi}_{{W}{W}}(0)}{M_{ W}^2}
- \hat{c}^2 \frac{{\Pi}_{{Z}{Z}}(M_{ Z}^2)-{\Pi}_{{Z}{Z}}(0)}{M_{ Z}^2}\right.
\nonumber\\
&&\left. -2 \hat{c}\hat{s}\frac{{\Pi}_{{A}{Z}}(M_{ Z}^2)-{\Pi}_{{A}{Z}}(0)}{M_{ Z}^2}
-\hat{s}^2\frac{{\Pi}_{{A}{A}}(M_{ Z}^2)}{M_{ Z}^2}\right\},
\label{STU-3}
\eea
where $\Pi_{ W  W}$, $\Pi_{ Z  Z}$, $\Pi_{ A  Z}$, and 
$\Pi_{ A  A}$ are the  cofactors 
of $g^{\mu\nu}$ in  the $W W$,  $Z Z$,  $AZ$, and $AA$
self-energies,  respectively. 
The $\overline{\rm{MS}}$ values 
$\hat{e}^2 \equiv \hat{e}^2(M_{ Z})$, 
$\hat{s}^2 \equiv \sin^2\hat{\theta}_{w}(M_{ Z})$ are usually employed, 
 since they are considered well suited to describe 
physics at the $M_{ Z}$ scale.
\newline
\indent
The main practical function of these parameters is to furnish constraints for models 
of new physics; this is done by computing the 
contributions of the new physics to these parameters, and then comparing them against 
the SM values. However, a serious problem arises already at the level of the SM, \ie 
before any new physics has been put in: 
the above expressions for the $S$, $T$, and $U$ are {\it not} gfp-independent.
 Specifically, as was  shown by Degrassi,
Kniehl,  and  Sirlin~\cite{Degrassi:1993kn},   they
become infested  with  gauge-dependencies  as  soon as  the  one-loop  SM bosonic  
contributions are  taken into
account.  In addition,  these  quantities are,  in general,  ultraviolet
divergent, unless one  happens to work within a  very special class of
gauges, namely those satisfying the relation  
\begin{equation}
\xi_{ W}= \hat{c}^2 \xi_{ Z} + \hat{s}^2 \xi_{ A}.
\label{spg}
\end{equation}
The  above shortcomings may be  circumvented automatically if
one defines the $S$, $T$, and $U$ parameters through the corresponding
gfp-independent PT self-energies, \ie simply by replacing, in  Eqs~(\ref{STU-1}), all 
$\Pi$s by the corresponding $\widehat{\Pi}$s. 
If one restricts oneself only to the contributions within the SM (no new physics)  
one obtains the following relation between the conventional and gfp-independent
(hatted) quantities (note that the SM tadpoles cancel exactly)  
\bea 
\hat{\alpha} \widehat{S}_{\rm {SM}} &=& \hat{\alpha} {S}_{\rm {SM}} + 8 \hat{e}^2 \hat{c}^2
[I_{ W  W}(M_{ Z}^2) - I_{ W  W}(0)],
\label{PTSTU-1}\nonumber\\
\hat{\alpha}  \widehat{T}_{\rm {SM}} &=& \hat{\alpha}  {T}_{\rm {SM}}
+ 4 \hat{g}^2 [\hat{c}^2 I_{ Z  W}(0) +\hat{s}^2  I_{ A W}(0) - I_{ W  W}(0)],
\label{PTSTU-2}\nonumber\\
\hat{\alpha}\widehat{U}_{\rm {SM}} &=& \hat{\alpha} {U}_{\rm {SM}}
+ 16 \hat{e}^2 \left\{ \hat{c}^2\left[ I_{ W  W}(0)- I_{ Z  W}(0) \right] 
+ \hat{s}^2[I_{ W  W}(M_{ Z}^2) - I_{ A W}(0)] \right\},
\label{PTSTU-3}
\eea 
where   
\be
I_{ij}(q^2) = i\int_k \frac{1}{(k^2 - M_{i}^2)[(k+q)^2- M_{j}^2]}\,.
\label{ias}
\ee
Note that, since ${\widehat\Pi}_{{A}{Z}}(0)=0$, we have that 
\be
\hat{\alpha}  \widehat{T}_{\rm {SM}} =  
\frac{\widehat{\Pi}^{ W  W}(0)}{M^2_{ W}}-\frac{\widehat{\Pi}^{ Z  Z}(0)}{M^2_{ Z}}\,;
\label{Trho}
\ee
thus, $\hat{\alpha}  \widehat{T}_{\rm {SM}}$ serves as the gfp-independent definition of the 
universal part of the  $\rho$ parameter at one loop (see next topic).
\newline
\indent
It goes without saying that  
the additional contributions to the $\widehat{S}$, $\widehat{T}$, and $\widehat{U}$ 
parameters from new physics must also be cast in a PT form (unless they involve only fermion loops). 
Thus, contributions to the self-energies from 
new gauge bosons (such as, \eg the Kaluza-Klein modes in models with universal  
extra dimensions~\cite{Carone:1999nz,Appelquist:2000nn,Oliver:2002up}) 
must undergo the PT rearrangement, and be written in the form 
$\widehat{\Pi}_{\rm {NP}}$; for some recent applications of this methodology in various SM extensions,  
see, \eg~\cite{Dawson:2008as,Burdman:2008gm,Dawson:2007yk,Cacciapaglia:2006pk,Christensen:2005cb,Montano:2005gs}. 

\subsubsection{The universal part of the $\rho$ parameter beyond one loop}
\noindent
The $\rho$ parameter~\cite{Veltman:1977kh} is defined as  the  ratio  of  the relative  strength
between  neutral and  charged  current interactions,  at low  momentum
transfer, namely
\be
\rho = \frac{G_{ N  C}(0)}{G_{ C  C}(0)}=\frac{1}{1-\Delta\rho}
\label{rhofull}
\ee
where $G_{ N  C}$ and $G_{ C  C}$ 
are the corresponding full amplitudes, with all Feynman diagrams included.
The $\rho$ parameter displays  a strong dependence on $m_t$  and affects most
electroweak parameters such as $\Delta r$, 
$M_{ W}$, and $\sin^2 \theta_{\rm eff} (M_{ Z})$. The
$\rho$ parameter defined  above as the ratio of two  amplitudes is a gauge
independent and finite quantity. In addition, it is manifestly process
dependent, since  its value  depends  on  the  quantum numbers  of  the
external particles  chosen. To fully determine  the value of  $\rho$ for a
given neutral and  charged process, one must compute  the complete set
of Feynman diagrams  (self-energy, vertex, and box graphs) to a given
order  in  perturbation  theory.  However, traditionally  one  focuses
instead on the quantity $\Delta_{\rm un}$ , defined 
in terms of the subset of Feynman diagrams containing only 
the gauge-boson self-energies, \ie
\be
\Delta_{\rm un} = \frac{\Pi_{ W  W}(0)}{M^2_{ W}}- 
\frac{\Pi_{ Z  Z}(0) +(2 s_w/c_w) \Pi_{ A  Z}(0)}{M^2_{ Z}}.
\label{rhoconv}
\ee
The quantity $\Delta_{\rm un}$
is meant to capture the "universal" (\ie process-independent) part of $\rho$, since, by
definition, it does not  depend on the details of  the process. According
to the standard lore~\cite{Chanowitz:1978uj,Chanowitz:1978mv,Einhorn:1981cy,vanderBij:1986hy}, 
$\Delta_{\rm un}$ contains the dominant contributions to $\rho$.
\newline
\indent
From Eq.~(\ref{STU-2}) we see that, at one-loop, 
$\Delta_{\rm un}= \hat{\alpha}  {T}_{\rm {SM}}$. 
Given the discussion of the previous subsection, 
the problematic nature of this identification, as well as its one-loop remedy,  
should be clear by now.
Specifically, the leading one-loop $m_t$ contributions 
(of order $G_\mu m^2_t$) to $\Delta_{\rm un}$ 
are  trivially gauge-independent (since  the  gfp
does  not appear in the fermion loop),  and UV
finite. On the  other  hand,
the one-loop bosonic  contributions (subleading in $m^2_t$ , of order  $g^2 m^0_t$ ) 
to $\Delta_{\rm un}$ are gauge-dependent 
and, except when  formulated within  a restricted  
class of  gauges given in Eq.~(\ref{spg}),  UV divergent.
The remedy is, of course, to use, instead, the definition appearing on the rhs of Eq.~(\ref{Trho}).
\newline
\indent
As one may imagine, things do not get any better at two-loops. 
Thus, if one attempts to use Eq.~(\ref{rhoconv}) at two loops (a dubious proposition, 
given that it does not even work at one loop) one encounters more problems 
(compounded by the book-keeping complications  typical of the two loops)~\cite{Degrassi:1994kt,Degrassi:1994tf}.   
In particular, the leading  two-loop contributions (of order $G^2_\mu m^4_t$) 
to $\Delta_{\rm un}$
are also  gauge-independent and UV finite,  exactly as
their  one-loop counterparts.  On the  other hand,  
subleading  two-loop $m_t$  contributions (of  order
$G^2_\mu m^2_t M^2_Z$) are $\xi$-dependent in the context of the $R_{\xi}$ gauges.
In addition, even when computed in the Feynman gauge ($\xi_{ W} =\xi_{ Z} = 1$), 
which satisfies the (one-loop) relation  of (\ref{spg}), 
the answer turns out to be UV divergent. 
\newline
\indent
In order to understand the  origin of the problems associated with the
subleading  contributions, one  should first  establish the  mechanism
that  enforces the  good behavior  of the  leading  contributions-- in
particular their UV finiteness~\cite{Papavassiliou:1995hj}. 
If we denote the leading (fermionic) contributions
(both at one and two loops) to the $W W$ and $Z Z$ self-energies 
by $\Pi^{(\ell)\mu\nu}_{ W  W}(q)$ and $\Pi^{(\ell)\mu\nu}_{ Z  Z}(q)$,
respectively (the label $\ell$ stands for ``leading''), and use the important fact that 
\be
\Pi^{(\ell)}_{ A  Z}(0)=0,
\label{sk02}
\ee
(valid for fermionic contributions only!) we can write for $\Delta_{\rm un}^{(\ell)}$
\be
\Delta_{\rm un}^{(\ell)} = \frac{\Pi^{(\ell)}_{ W  W}(0)}{M^2_{ W}}
-\frac{\Pi^{(\ell)}_{ Z  Z}(0)}{M^2_{ Z}}.
\label{sk01}
\ee
\indent
The  finiteness  of  $\Delta_{\rm un}^{(\ell)}$ may be established 
as follows.
The  $WW$ and $ZZ$ self-energies appearing in this problem,  
denoted by $\Pi^{\mu\nu}_{ii}(q)$  ($i=W,Z$) 
may be written in the form 
\be
\Pi^{\mu\nu}_{ii}(q)= \Pi_{ii}(q^2)g_{\mu\nu}+ \Pi^{ L}_{ii}(q^2) q^{\mu} q^{\nu},
\label{sk1}
\ee
and the dimensionality of $\Pi_{ii}(q^2)$ will be saturated either by 
$q^2$ or by the masses appearing in the theory, the latter being all proportional
to the value $v$ of the vev. Thus, we have 
\be
\Pi_{ii}(q^2)= v^2 \Pi_{1\,ii}(q^2) +  q^2 \Pi_{2\,ii}(q^2), 
\label{sk2}
\ee
and therefore
\be
\Pi_{ii}(0)= v^2 \Pi_{1\,ii}(0).   
\label{sk3}
\ee
Then it is elementary to establish that 
\be
\Pi_{ii}(0) =\left.\frac{d}{dq^2}\left\{q_\mu q_\nu \Pi^{\mu\nu}_{ii}(q^2)\right\}\right|_{q^2=0}.
\ee
Next, combine this last result with the WIs (valid only for the fermionic loops, 
and, in particular, the leading contributions containing top-quark loops)
\bea
q_{\mu} q_{\nu} \Pi_{(\ell)\mu\nu}^{ W  W}(q) &=& 
M^2_{ W} \Pi^{(\ell)}_{\phi\phi}(q^2),
\\ \nonumber
q_{\mu} q_{\nu} \Pi^{(\ell)\mu\nu}_{ Z  Z}(q) &=& 
M^2_{ Z} \Pi^{(\ell)}_{\chi\chi}(q^2),
\label{sk0}
\eea
where $\Pi^{(\ell)}_{\phi\phi}$ and $\Pi^{(\ell)}_{\chi\chi}$  are the
leading contributions of the $\phi \phi$ and $\chi \chi$ self-energies,
respectively. We may then write $\Delta_{\rm un}^{(\ell)}$ of Eq.~(\ref{sk01}) as 
\be
\Delta_{\rm un}^{(\ell)} =\left.
\frac{d}{dq^2}\left\{\Pi^{(\ell)}_{\phi\phi}(q^2)-\Pi^{(\ell)}_{\chi\chi}(q^2)\right\}\right|_{q^2=0}.
\ee
\indent
The final ingredient that enforces the finiteness of $\Delta_{\rm un}^{(\ell)}$
is the equality of the divergent parts of $\Pi^{(\ell)}_{\phi\phi}(q^2)$ and 
$\Pi^{(\ell)}_{\chi\chi}(q^2)$, reflected in the corresponding equality 
\be
Z^{(\ell)}_{\phi\phi} = Z^{(\ell)}_{\chi\chi},
\label{sk4}
\ee
between the wave-function renormalization constants.
\newline
\indent
Notice, however,  that  the crucial relations  (\ref{sk02}), (\ref{sk0}), and (\ref{sk4})
are not longer valid when one includes the bosonic (subleading) parts of 
$\Pi^{\mu\nu}_{ W  W}(q)$ and $\Pi^{\mu\nu}_{ Z  Z}(q)$
in the  framework of  the $R_{\xi}$ gauges. 
Consequently,  since  the mechanism
enforcing  the finiteness  does not  operate any  more,  the resulting
expressions do not have to be UV finite, and indeed, 
as an explicit calculation showed, they are not.
\newline
\indent
The  easy way out of these complications would be to 
abandon  the notion of  a "universal"  part of  $\rho$,  
and adopt the conservative point of view that the  entire  process must
be considered in order to  restore the finiteness and gauge-independence
of  the  final  answer. In that case,  one would  introduce  vertex  and  box
corrections, which  would render the result gauge-independent  and finite, at
the   expense   of   making   it  process-dependent,   and   therefore
non-universal.
\newline
\indent
Of course, as the reader must have realized by now, 
this    unpleasant trade-off    between   gauge-independence    and
process-independence  is completely artificial, 
and can be easily 
avoided by defining $\widehat\Delta_{\rm un}$ beyond one loop 
in terms of the physical PT self-energies, namely~\cite{Papavassiliou:1995hj} 
\be
\widehat\Delta_{\rm un}
= \frac{\widehat{\Pi}_{ W  W}(0)}{M^2_W}-\frac{\widehat{\Pi}_{ Z  Z}(0)}{M^2_Z}, 
\label{sk5}
\ee
\ie use exactly the same definition [{\it viz.} (\ref{Trho})] as at one loop! 
Indeed, all  PT  self-energies  are gauge-independent, and due to the
Abelian WIs they satisfy, for {\it both fermionic and bosonic contributions}, 
all aforementioned conditions 
enforcing  the finiteness of $\Delta_{\rm un}^{(\ell)}$, 
and in particular (\ref{sk02}), (\ref{sk0}), and (\ref{sk4})
 are  valid {\it both  for leading and  subleading}
contributions. Evidently,  the  PT  restores the  mechanism  for  the
cancellation of  the UV divergences,  and  guarantees at the same time 
the gauge- and process-independence of the final answer.
\newline
\indent
Thus the $\widehat\Delta_{\rm un}$ defined in  (\ref{sk5}) in terms of the
PT  self-energies 
constitutes the 
natural  extension  of the universal part of the $\rho$-parameter,
that can accommodate
consistently  both leading and subleading contributions, at one and two loops.
$\widehat\Delta_{\rm un} $
is  endowed with three crucial
properties; it is ({\it i}) independent  of  the  gfp, ({\it ii}) UV finite, and
({\it iii}) process-independent (universal).  
In addition to the above important conceptual advantages, 
the calculation  of $\widehat\Delta_{\rm un} $ is
facilitated  enormously from the fact that no   vertex  or  box  diagrams  
need to be calculated, since the terms that restore the good properties are all propagator-like.
Therefore, the answer can be expressed in a closed analytic form up to two loops.
The actual calculation of the two-loop subleading corrections of order $G^2_\mu m^2_t M^2_Z$ 
was carried out in the 
typical limit of $s_w^2=0$, where $M_{ Z}= M_{ W}$ (custodial symmetry restored).
It turned out that their relative size is about 25$\%$ with respect to the leading ones~\cite{Papavassiliou:1995hj}; 
this result is in complete agreement with naive expectations, given that the 
expansion parameter employed is  $M^2_{ W}/ m^2_t \approx 1/4$.

\subsection{\label{resum}Self-consistent resummation formalism for resonant transition amplitudes}
\noindent
The physics of unstable particles   
 and the computation  of
resonant transition amplitudes 
has attracted significant
attention  in  recent  years, because  it   is both phenomenologically
relevant   and  theoretically  challenging.    
The practical interest in the problem is related to the resonant 
production of various particles in all sorts of accelerators,
most notably LEPI and  LEP2 in the past,  the TEVATRON at present, and, of course, 
the LHC in the very near future.  
From the theoretical point of view, the 
issue comes up every time fundamental resonances,
\ie unstable particles that appear as basic degrees   
of freedom in the original 
Lagrangian of the theory (as opposed to composite bound-states), 
can be produced resonantly.
The presence of such fundamental resonances makes it
impossible to compute physical amplitudes for arbitrary values of the
kinematic parameters, unless a resummation has taken place first. Simply
stated, perturbation theory breaks down in the vicinity of resonances, and
information about the dynamics to ``all orders'' needs be encoded already at
the level of Born amplitudes. 
The  difficulty arises from   the fact that  in the  context of non-Abelian
gauge theories the standard Breit-Wigner  resummation used for regulating
physical amplitudes near resonances is  at odds with gauge invariance,
unitarity, and  the equivalence theorem~\cite{Cornwall:1974km,Vayonakis:1976vz,Chanowitz:1985hj}.
Consequently,  the resulting
Born-improved  amplitudes, in general,   fail to capture faithfully  the
underlying dynamics. 
\newline
\indent
Whereas the need for a resummed propagator is evident when dealing
with  unstable  particles  within  the  framework  of  the  $S$-matrix
perturbation theory, its incorporation  to the amplitude of a resonant
process is non-trivial.  When this incorporation is done naively 
(\eg   by  simply replacing  the  bare  propagators  of a  tree-level
amplitude  by resummed  propagators) one  is often  unable  to satisfy
basic  field  theoretical requirements,  such  as the  gfp-independence   
of  the   resulting  $S$-matrix   element, the  $U(1)_{em}$
symmetry, high-energy  unitarity, and the OT.  This fact
is  perhaps not  so surprising,  since  the naive  resummation of  the
self-energy graphs  takes into  account higher order  corrections, for
{\em  only}  certain  parts   of  the  tree-level  amplitude.  Indeed,
resumming  the conventional  two-point function  of a  gauge  boson in
order to construct a Breit-Wigner  type of propagator does not include
properly  crucial  contributions   originating  from  box  and  vertex
diagrams.   Even  though  the  amplitude  possesses  all  the  desired
properties,  this  unequal  treatment  of its  parts  distorts  subtle
cancellations, resulting  in numerous pathologies,  that are artifacts
of the resummations method used.   It is therefore important to devise
a  self-consistent   calculational  scheme,  which   {\em  manifestly}
preserves  the aforementioned field  theoretical properties, 
{\it intrinsic} to every $S$-matrix  element.  In what follows we will
briefly review how this is accomplished using the PT; 
the presentation is almost exclusively 
based on a series of articles written on the subject
by A.~Pilaftsis and one of the authors~\cite{Papavassiliou:1995fq,Papavassiliou:1996zn,Papavassiliou:1995gs,Papavassiliou:1997fn,Papavassiliou:1998pb}.

\subsubsection{The Breit-Wigner Ansatz and the Dyson summation}
\noindent
The mathematical  expressions for computing  transition amplitudes are
ill-defined  in the  vicinity  of resonances,  because the  tree-level
propagator of  the particle  mediating the interaction (\ie $\Delta=
(s-M^2)^{-1}$),   becomes  singular   as   the  center-of-mass   energy
approaches the mass of the resonance, \ie as 
$\sqrt{s}\sim  M$.   The standard  way  for  regulating this  physical
kinematic  singularity is to  use a  Breit-Wigner type  of propagator;
thus, near the resonance, one carries out the substitution
\be
\frac{1}{s-M^2} \longrightarrow \frac{1}{s-M^2 + i M \Gamma},
\label{BWcw}
\ee
where  $\Gamma$ is  the width of  the unstable
(resonating)  particle. The presence of the  $i M \Gamma$ in the denominator prevents the
amplitude from being divergent, even at the physical resonance, 
\ie when $s=M^2$. 
\newline
\indent
This physically motivated Breit-Wigner  Ansatz may seem unjustified at
first,  considering the  fact  that the  width  of the  particle is  a
parameter that  does not appear in the  fundamental Lagrangian density
defining the theory; indeed, such a term would violate immediately the
hermiticity of ${\mathcal L}$, thus producing all sorts of complications.
The actual field-theoretic mechanism that justifies 
a replacement similar to that of (\ref{BWcw})   
is  the   Dyson  resummation  of   the 
self-energy  $\Pi(s)$ of  the unstable  particle. This 
bubble resummation amounts to the rigorous 
substitution 
\be
\frac{1}{s-M^2} \longrightarrow \frac{1}{s-M^2+\Pi(s)} . 
\label{BW}
\ee
The running width
of the particle is then defined as $M\Gamma(s) =\Im m \Pi(s)$, whereas
the usual (on-shell) width is simply its value at $s=M^2$; 
thus, (\ref{BWcw}) is a special case of (\ref{BW}). 
\newline
\indent
It is relatively  easy to realize now 
that  the Breit-Wigner  procedure, as described above, is tantamount 
to a reorganization  of  the  perturbative  series. Indeed,  resumming  the
self-energy  $\Pi(s)$ amounts to  removing a  particular piece  from each
order of the perturbative expansion, since, from all the Feynman graphs
contributing to a given order $n$  we only pick the part that contains
the corresponding string of self-energy bubbles $\Pi(s)$, and then take $n \to \infty$.  
Notice, however, that the off-shell Green's functions contributing to 
a physical quantity,  at  any  finite   order  of  the
conventional perturbative  expansion, participate in a subtle 
cancellation, which eliminates all unphysical terms. 
Therefore, the act of resummation, which treats unequally the various Green's functions,  
 is in general liable to distort these cancellations. To put it
differently, if  $\Pi(s)$ contains unphysical  contributions (which would
eventually cancel against other terms within a given order), resumming
it naively  would mean that  these unphysical contributions  have also
undergone infinite  summation (they now  appear in the  denominator of
the propagator $\Delta(s)$).  In order to remove them, one would have  to add the
remaining  perturbative  pieces  to  an  infinite  order,  clearly  an
impossible  task,  since  the  latter  (boxes  and  vertices)  do  not
constitute a resumable set.  Thus,  if the resumed $\Pi(s)$ happened to
contain such unphysical terms, one would finally arrive at predictions
for  the physical  amplitude close  to  the resonance  which would  be
plagued with unphysical artifacts.  It turns out that, while in scalar
field theories $\Pi(s)$ does not contain such
unphysical  contributions, this  ceases  to  be true  in  the case  of
non-Abelian gauge theories.
The crucial novelty introduced by the PT is that
the resummation of the (physical) self-energy graphs must take place only
{\em after} the amplitude of interest has been cast
via the PT algorithm into
manifestly physical subamplitudes, with distinct
kinematic properties, order by order in perturbation theory.
Put in the language employed earlier, the PT 
ensures that all unphysical contributions contained inside $\Pi(s)$ have
been identified and properly discarded, {\it before} $\Pi(s)$ undergoes 
resummation.

\subsubsection{The non-Abelian setting}

\begin{figure}[!t]
\bce
\includegraphics[width=14.5cm]{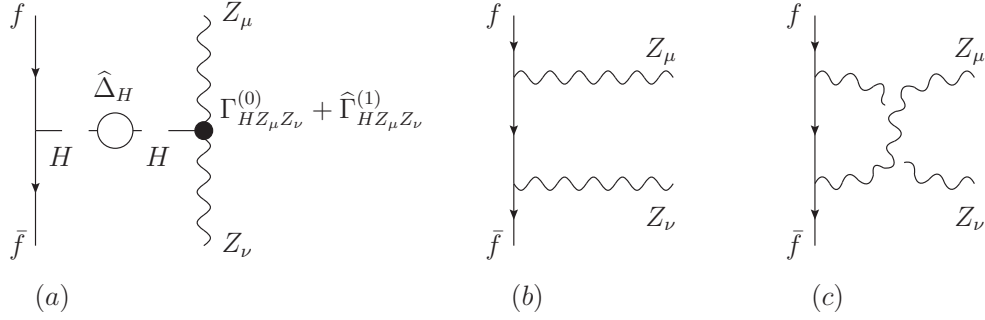}
\ece
\caption{\figlab{PT_resum_nunu-ZZ} The amplitude for the process 
$f \bar{f} \to ZZ$. The s-channel graph (a) may become resonant, 
and must be regulated by appropriate  
resummation of the Higgs propagator and dressing of the $HZZ$ vertex}
\end{figure}
\noindent
We now turn to the case of a non-Abelian gauge theory, 
such as the electroweak sector of the SM.
As has been discussed extensively in the relevant literature, 
the physical requirements that must be encoded 
into a properly regulated resonant amplitude 
are the following~\cite{Papavassiliou:1995fq,Papavassiliou:1996zn,Papavassiliou:1995gs,Papavassiliou:1997fn,Papavassiliou:1998pb}:

\begin{itemize}

\item[{\it i}.] The resonant amplitude must be gfp-independent.
\newline
\item[{\it ii}.] Unitarity (OT) and analyticity 
(dispersion relations) must hold.
\newline
\item[{\it iii}.] The position of the pole must be unchanged. 
\newline
\item[{\it iv}.] The external gauge-invariance must remain intact. 
\newline
\item[{\it v}.] The equivalence theorem must be satisfied.
\newline
\item[{\it vi}.] The resonant amplitude must be 
invariant under the renormalization group.  
\newline
\item[{\it vii}.] The amplitude must display 
good asymptotic (high-energy) behavior. 
\end{itemize}
Note that if all incoming and outgoing particles are fermions then points ({\it iv}) and ({\it v}) 
do not enter into the discussion.
\newline
\indent
In order to fully appreciate the subtle interplay between all these  
issues, let us consider a concrete example that has sufficient structure 
for all the above points to make their appearance. 
Specifically, we study the   process
$f(p_1)\bar{f}(p_2)\to  Z(k_1)Z(k_2)$, shown in  \Figref{PT_resum_nunu-ZZ},   
and $s=(p_1+p_2)^{2}=(k_1+k_2)^2$  is the   
c.m.\ energy squared. 
The tree-level amplitude of this process  
is the sum of an $s$- and a
$t$-  channel contribution, denoted   by ${\mathcal T}_s$ 
and ${\mathcal T}_t$, respectively, given by
\begin{eqnarray}
  \label{THsZZ}
{\mathcal T}_{s}^{\mu\nu} &=&\Gamma_{ H  Z  Z}^{\mu\nu}
\Delta_{ H}(s)\ \bar{v}(p_2) \Gamma_{{H} {f} {\bar{f}}} u(p_1),\nonumber\\
  \label{TtZZ}
{\mathcal T}_{t}^{\mu\nu}  &=& \bar{v}(p_2)\left[ 
\Gamma_{ Z {f} {\bar{f}}}^{\nu} S^{(0)}(\ps_1 + \ks_1) 
\Gamma_{ Z {f} {\bar{f}} }^{\mu} + 
\Gamma_{ Z {f} {\bar{f}}}^{\mu} S^{(0)} (\ps_1 + \ks_2)
\Gamma_{ Z {f} {\bar{f}} }^{\nu} \right]u(p_1),
\end{eqnarray}
where
\be
\Gamma_{ H  Z  Z}^{\mu\nu}   =  
ig_w \frac{M^2_{ Z}}{M_{ W}}     g^{\mu\nu},\qquad
\Gamma_{{H} {f} {\bar{f}}}  =    -i    g_w 
\frac{m_f}{2    M_{ W}},\qquad
\Gamma_{ Z {f} {\bar{f}}}^{\mu} = -i\frac{g_w}{c_w} 
\gamma_\mu\, (  T^f_z P_{ L} - Q_f s^2_w), 
\ee
are  the   tree-level  $HZZ$,  $Hf\bar{f}$  and $Zf\bar{f}$ couplings,
respectively.
\newline
\indent
Obviously the  $s$-channel contribution is mediated by the Higgs boson of mass $M_{ H}$
and becomes resonant if the kinematics are such that $\sqrt s$ lies in the vicinity of $M_{ H}$;
in that case the resonant amplitude must be properly regulated, as explained earlier. 
As we will see in detail in what follows, the minimal way for accomplishing this is by: (i) 
Dyson-resumming the one-loop PT self-energy of the (resonating) Higgs boson, and (ii)
 by appropriately 
``dressing'' the tree-level vertex $\Gamma_{ H  Z  Z}^{\mu\nu}$, \ie 
by replacing in the amplitude the vertex $\Gamma_{ H  Z  Z}^{\mu\nu}$ 
by the  one-loop PT vertex $\widehat\Gamma_{ H  Z  Z}^{\mu\nu}$. 

\noindent
{\it i}.\hspace{.1cm} {\it gfp-independence}

Let us first see what happens if one attempts to regulate the 
resonant amplitude by means of the conventional one-loop Higgs
self-energy in the $R_{\xi}$ gauges. 
A straightforward calculation yields (tadpole  and  seagull terms omitted)~\cite{Papavassiliou:1997fn,Papavassiliou:1998pb}:

\bea
\Pi_{ H  H}^{( W  W)}(s,\xi_{ W}) &=& \frac{\alpha_w}{4 \pi}\, \left[ \left(
\frac{s^2}{4 M^2_{ W}} - s + 3M^2_{ W}\right)B_0(s,M^2_{ W},M^2_{ W})\right.\nonumber \\
&+&\left. \frac{M_{ H}^4-s^2}{4M^2_{ W}}
B_0(s,\xi_{ W} M^2_{ W},\xi_{ W} M^2_{ W})
\right].
\label{DRKSI}
\eea
\indent
We see  that for $\xi_{ W}\neq 1$ the term growing as $s^2$  
survives and is proportional  to the  difference 
\mbox{$B_0(s,M^2_{ W},M^2_{ W}) -  
B_0(s,\xi_{ W} M^2_{ W},\xi_{ W} M^2_{ W})$}.  
For  any finite  value of  $\xi_{ W}$  this term
  vanishes for  sufficiently large  $s$, \ie  $s\gg  M^2_{ W}$
  and  $s\gg   \xi_{ W} M^2_{ W}$.   Therefore,  
 the   quantity  in  Eq.~(\ref{DRKSI}) 
displays good high  energy behavior in compliance with
  high energy unitarity.   Notice, however, that the onset  of this good
  behavior depends crucially on the choice of $\xi_{ W}$.  
Since $\xi_{ W}$ is a free  parameter, and may be chosen to  be arbitrarily large, but
  finite, the  restoration of unitarity may be  arbitrarily delayed as
  well.  This  fact poses no problem  as long as one  is restricted to
  the  computation  of  physical  amplitudes  at  a  finite  order  in
  perturbation theory.   However, if the  above self-energy were  to be
  resummed  in order  to regulate  resonant transition  amplitudes, it
  would  lead  to  an   artificial  delay  of  unitarity  restoration,
   which becomes numerically significant for large values  of $\xi_{ W}$. 
   In addition,
  a serious pathology occurs
  for any value of $\xi_{ W}\neq 1$,  namely the appearance of unphysical
  thresholds~\cite{Papavassiliou:1995fq,Papavassiliou:1995gs,Papavassiliou:1996zn}.   Such  thresholds   may be particularly
  misleading if $\xi_{ W}$  is  chosen in the  vicinity of  unity, giving
  rise to distortions in the lineshape of the unstable particle.
\newline
\indent
How does the situation change if instead we compute  the 
corresponding part of the Higgs-boson self-energy 
in the BFM, for an {\it arbitrary} value of $\xi_{ Q}$?
Denoting it 
by $\widetilde\Pi^{ H  H}_{( W  W)}(s,\xi_{ Q})$, 
and using the appropriate set of Feynman rules~\cite{Denner:1994xt}, we obtain   
\bea
\widetilde\Pi_{ H  H}^{( W  W)}(s,\xi_{ Q})  
&=& \frac{\alpha_w}{4\pi}
\left[
\left(\frac{s^2}{4M^2_{ W}} - s + 3 M^2_{ W} \right)
B_0(s,M^2_{ W},M^2_{ W})\right. \nonumber \\
&+&\left. \frac{M_{ H}^4-s^2}{4 M^2_{ W}} B_0(s,\xi_{ Q} M^2_{ W},\xi_{ Q} M^2_{ W})\right]
\nonumber\\
&-& \frac{\alpha_w}{4\pi}\xi_{ Q} (s-M_{ H}^2) 
B_0(s,\xi_{ Q} M^2_{ W},\xi_{ Q} M^2_{ W}), 
\label{DBFG}
\eea
or simply 
\be
\widetilde\Pi_{ H  H}^{( W  W)}(s,\xi_{ Q}) = 
\Pi_{ H  H}(s,\xi_{ W} \to \xi_{ Q}) - 
\frac{\alpha_w}{4\pi} \xi_{ Q} (s-M_{ H}^2) 
B_0(s,\xi_{ Q} M^2_{ W},\xi_{ Q} M^2_{ W})
\ee
Evidently, away from $\xi_{ Q}=1$, 
$\widetilde\Pi_{ H  H}^{( W  W)}(s,\xi_{ Q})$ 
displays the same unphysical characteristics mentioned above for 
$\Pi_{HH}^{(WW)}(s,\xi_{ W})$ ! Therefore, when it comes to 
the study of resonant amplitudes, calculating in the BFM 
for general $\xi_{ Q}$ is as pathological as calculating  
in the conventional $R_{\xi}$ gauges. 
\newline
\indent
To solve these problems one has to simply follow the PT procedure,
within either gauge-fixing scheme, $R_{\xi}$ or BFM, 
identify the corresponding Higgs-boson related pinch parts from the 
vertex and box diagrams, and add them to  (\ref{DRKSI}) or (\ref{DBFG}).
Then a unique answer emerges, 
the  PT     one-loop   Higgs   boson   self-energy, given by 
$\widehat{\Pi}_{HH}(q^2)$,
\begin{equation}
  \label{HPT}
\widehat{\Pi}_{ H  H}^{( W  W)}(s) = \frac{\alpha_w}{16\pi}\frac{M_{ H}^4}{M^2_{ W}}
\left[ 1+4\frac{M^2_{ W}}{M_{ H}^2}- 4\frac{M^2_{ W}}{M_{ H}^4}
(2s-3M^2_{ W}) \right] B_0(s,M^2_{ W},M^2_{ W}).
\end{equation}
Setting $\xi_{ Q}=1$  in the  expression of  Eq.~(\ref{DBFG}), we
  recover the full  PT answer of Eq.~(\ref{HPT}), as expected.
Clearly, $\widehat{\Pi}^{ H  H}_{( W  W)}(s)$ has none of the pathologies 
observed above.
\newline
\indent
At this point one may wonder why not use simply the 
$R_{\xi}$ expression for $\Pi_{ H  H}(s,\xi_{ W})$ at $\xi_{ W}=1$,
given that it too becomes free of the aforementioned problems. 
The answer is that if the external particles are gauge bosons 
then the vertices connecting them with the  
resonating Higgs boson will not satisfy an Abelian WI, but rather an STI, and this, in turn, will 
make it impossible to satisfy the external gauge invariance 
[see also subsection ({\it iv})]. 
\newline
\indent
Exactly the same arguments presented above hold for the part of the Higgs self-energy containing 
the $Z$-bosons, together with the associated would-be Golstone bosons and ghosts.
The corresponding one-loop PT result reads
\begin{equation}
  \label{HPTZ}
\widehat{\Pi}_{ H  H}^{( Z  Z)}(s)\ =\ \frac{\alpha_w}{32\pi}\frac{M_{ H}^4}{M^2_{ W}}
\left[ 1+ 4\frac{M^2_{ Z}}{M_{ H}^2}- 4\frac{M^2_{ Z}}{M_{ H}^4}
(2s - 3 M^2_{ Z})\right] B_0(s,M^2_{ Z},M^2_{ Z}) .
\end{equation}

\noindent
{\it ii}.\hspace{.1cm} {\it  Running width and the optical theorem.}

When the kinematic singularity of the resonant amplitude is regulated  
through the resummation of the self-energy of the resonating particle, 
then, in the Breit-Wigner language, one obtains 
automatically a running  ($s$-dependent) width.    
The main reason why 
a $s$-dependent instead of a constant width must be used 
comes from the OT.
To appreciate this in  a simpler context, we consider a toy model~\cite{Veltman:1963th}, with interaction 
Lagrangian ${\mathcal L}_{int}\ =\ \frac{\lambda}{2}\phi^2\Phi$,  
and assume that $M_\Phi > 2 M_\phi$, so that the decay of 
 $\Phi$ into a pair of  $\phi$s  is kinematically allowed.
For concreteness, let 
us consider the reaction $\phi\phi \to \Phi^{*}(s) \to\phi\phi$ 
at c.m.s.\ energies
$s\simeq M^2_\Phi$. There are three relevant graphs, one
resonant $s$-channel graph, and two non-resonant $t$ and $u$ graphs.
We next focus on the resonant channel, $T_\mathrm{res}(s)$, and  carry out 
the Dyson summation of the (irreducible) $\Phi\Phi$ self-energy, 
to be denoted by $\Pi(s)$,  obtaining for the transition amplitude
\be
T_\mathrm{res}(s) = - \frac{\lambda^2}{s-M^2_\Phi +\Re e\Pi(s)
+i\Im m\Pi(s)}.
\label{Tstu}
\ee
For the case at hand the OT states that 
\begin{equation}
\Im m\, T_\mathrm{res}(s) = \frac{1}{2}\, \int (d PS)_{\phi\phi}\left\vert T_s(s)\right\vert^2;
\label{AbsTs}
\end{equation}
on the other hand, the lhs of Eq.~(\ref{AbsTs}) is 
given simply by the imaginary part of Eq.~(\ref{Tstu}), namely
\begin{equation}
\Im m \,T_\mathrm{res}(s) = \frac{\lambda^2 \Im m\Pi(s)}{
\left[s-M^2_\Phi+\Re e \Pi(s)\right]^2 + \left[\Im m \Pi(s)\right]^2}.
\label{AbsTres}
\end{equation}
Eq.~(\ref{AbsTs}) is consistent
with Eq.~(\ref{AbsTres}) in a perturbative sense; if one is 
suffuciently away from the resonance, such that 
the perturbative expansion makes sense, then 
Eq.~(\ref{AbsTres}) expanded to first order 
reproduces Eq.~(\ref{AbsTs}).
Notice, however, that this becomes possible
{\it only}  when the resummation
involves an $s$-dependent two-point function and width for the
unstable scalar $\Phi$. If a constant
width
for $\Phi$ had been considered instead,
unitarity would have been violated, \ie Eq.~(\ref{AbsTres}) would not go over to 
Eq.~(\ref{AbsTs}), when $s\neq M^2_\Phi$. 
\newline
\indent
It is therefore
evident that the regulator of a resummed propagator in a scalar theory should
be $s$-dependent. Needless to say, 
a similar situation occurs if one
attempts to use a constant pole expansion in the context of
a gauge field theory; indeed, 
it would be unrealistic to expect that 
one could consistently describe gauge theories using a
resummation procedure that is defective even for scalar
theories.
The reorganization of the perturbative expansion implemented by the PT and,
in particular the resummation of the PT self-energies,  
strictly enforces the required unitarity relations. 
The reason for this is that 
the PT self-energies  satisfy the OT {\it  individually},
as  explained   in  Sections \ref{sec:QCD_one-loop} and \ref{sec:SM_one-loop}.     
\newline
\indent
To     verify    that
$\widehat{\Pi}_{ H  H}^{(Z  Z)}(s)$    
has indeed this   property, 
in complete analogy to the methodology developed in 
sections I and III, we must turn to the tree-level version of the 
process shown in  \Figref{PT_resum_nunu-ZZ},
[no dressing for graph $(a)$], and study  the quantity   
\begin{equation}
{\mathcal M}= \left[{\mathcal  T}_{s}^{\mu\nu} +  {\mathcal  T}_{t}^{\mu\nu}\right]  
L_{\mu\rho}(k_1) L_{\nu\sigma}(k_2) [{\mathcal T}_{s}^{\rho\sigma}
+{\mathcal     T}_{t}^{\rho\sigma}]^*,   
\end{equation}
where the $s$-channel contribution comes from graph (a) and the $t$-channel from graphs (b) and (c). 
$L_{\mu\nu}(k)$    is the usual   polarization
tensor introduced in Eq.~(\ref{WPol}). Then we must use the longitudinal 
momenta  coming  from $L_{\mu\rho}(k_1)$  and  $L_{\nu\sigma}(k_2)$
to extract from ${\mathcal T}_{t}^{\mu\nu}$ the effectively $s$-dependent, 
Higgs-boson mediated, part, to be denoted by ${\mathcal T}^{ P}_{s}$ (see \Figref{H_contrib_tch_ZZ})
   Specifically, 
\begin{equation}
  \label{TP}
\frac{k_{1\mu} k_{2\nu}}{M^2_{ Z}}\, {\mathcal T}_{t}^{\mu\nu} =
{\mathcal T}^{ P}_{s}+ \cdots = - \frac{ig_w}{2M_{ W}}
\bar{v}(p_2) \Gamma_{{H} {f} {\bar{f}}} u(p_1)+ \cdots,
\end{equation}
where   the   ellipses denote   genuine  $t$-channel  
(not Higgs-boson related)  contributions.  Then,  one    must append the   piece  
${\mathcal T}^{ P}_{s}{\mathcal T}^{ P *}_{s}$
to   the ``naive'' Higgs-dependent
part  \mbox{${\mathcal    T}_{s}^{\mu\nu} L_{\mu\rho}(k_1)
L_{\nu\sigma}(k_2){\mathcal   T}_{s}^{\rho\sigma}$}.
\newline
\indent
Integrating the expression obtained in   Eq.~(\ref{TP}) over the two-body phase space of two $Z$ bosons
we finally arrive at the imaginary  part of Eq.\ (\ref{HPTZ}), which is
the announced result. A completely analogous procedure must be applied to the 
process $f(p_1)\bar{f}(p_2)\to  W^{+}(k_1) W^{-}(k_2)$, in order to verify that 
the $\widehat{\Pi}_{ H  H}^{( W  W)}(s)$  of Eq.~(\ref{HPT}) has the same property.

\begin{figure}[!t]
\bce
\includegraphics[width=14.5cm]{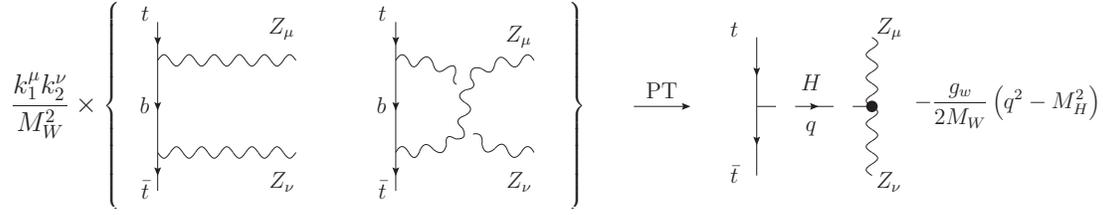}
\ece
\caption{\figlab{H_contrib_tch_ZZ} The Higgs-boson related contribution extracted from the boxes 
through pinching; to get it we must contract with both momenta.}
\end{figure}

\noindent
{\it iii}.\hspace{.1cm} {\it Position of the pole}
 
Since the position of the pole is the
only gauge-invariant quantity that one can extract from conventional
self-energies, any acceptable resummation procedure should give rise to
effective self-energies that do not shift the position of the pole. 
This requirement is rather stringent and constrains significantly 
any alternative resummation procedure. 
It is a non-trivial exercise to demonstrate that, indeed, the 
PT self-energies do not shift the position of the pole.
The easiest
way to see that at one-loop is by noticing that the 
pinch terms are always proportional to $(q^2-M^2)$, and therefore 
vanish at the pole. The all-order demonstration becomes 
far more involved,
and relies on a careful construction, where the contributions coming from 
the 1PR diagrams must be  properly taken into account.
\newline
\indent
Let us mention in passing that an early attempt 
towards a self-consistent resummation scheme 
has been based on the observation
that the position of the complex pole
is a gauge independent quantity~\cite{Willenbrock:1991hu,Stuart:1991cc,Stuart:1991xk,Stuart:1992jf,Sirlin:1991fd,Sirlin:1991rt}. Exploiting this
fundamental property of the $S$-matrix,
a perturbative
approach in terms of three gauge invariant quantities has been proposed:
the constant complex pole position of the resonant amplitude, the residue
of the pole, and a $s$-dependent non-resonant background term.
Even though this approach,
which finally boils down to a Laurent series expansion of the
resonant transition element~\cite{Stuart:1991cc,Stuart:1991xk,Stuart:1992jf}, 
furnishes a gauge invariant result, it clashes with point ({\it ii}) above:
the use of a constant instead of a running width leads to   
the violation of the OT. 
The perturbative treatment of these three gfp-independent 
quantities~\cite{Veltman:1992tm} introduces unavoidably residual space-like
threshold terms, which become more apparent in CP-violating
scenarios of new-physics~\cite{Pilaftsis:1989zt}. In fact, the precise $q^2$-dependent shape
of a resonance~\cite{Sirlin:1991fd,Sirlin:1991rt} is reproduced, to a given loop order, by considering
quantum corrections to the three gfp-independent quantities mentioned
above~\cite{Stuart:1991cc,Stuart:1991xk,Stuart:1992jf,Veltman:1992tm}, while the space-like threshold contributions,
even though shifted to higher orders, do not disappear completely.

\noindent
{\it iv}.\hspace{.1cm}{\it External gauge-invariance}

\begin{figure}[!t]
\bce
\includegraphics[width=17cm]{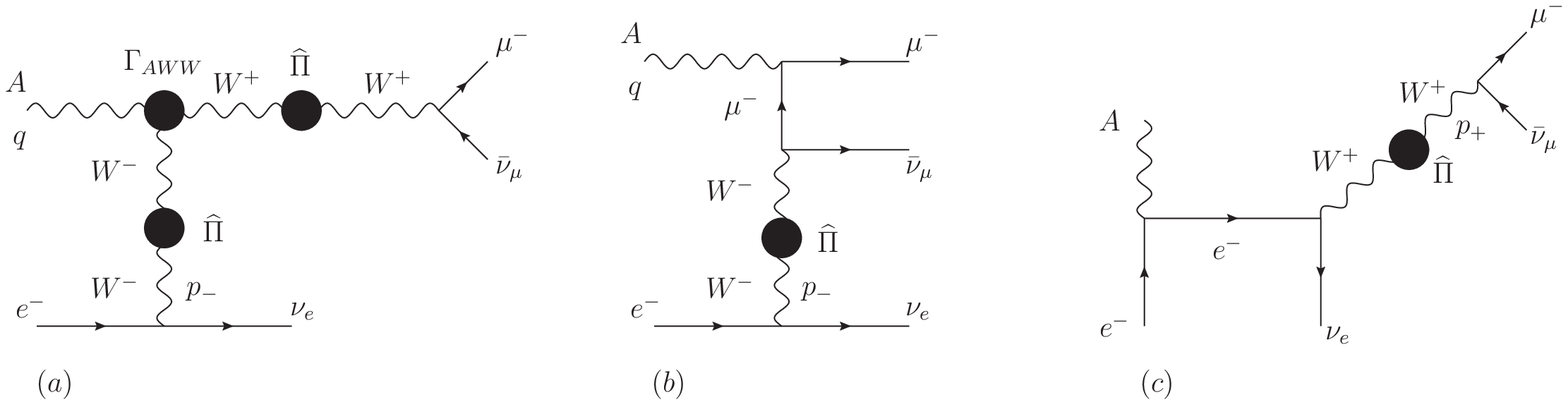}
\ece
\caption{\figlab{PT_Ae_mu-barnumu-barnue} The process $\gamma e^- \to \mu^-\bar{\nu}_\mu \nu_e$ 
appropriately dressed.}
\end{figure}

As is well-known already from the studies of QED, 
gauge-invariance imposes WIs on physical amplitudes. Let us    
consider a physical (on-shell) amplitude with 
$n$ incoming photons \linebreak $A^{\mu_i}(k_i)$, $i=1,\dots,n$
[$n$ must be even, otherwise the amplitude vanishes by  Furry's theorem].
Denoting the amplitude by 
$T_{\mu_1\mu_2 ...\mu_i ...\mu_n}(k_1,k_2, \dots, k_i, \dots, k_n,p)$, where $p$ 
stands collectively for the momenta of the incoming fermions,  
gauge invariance imposes the relation
\be  
k_i^{\mu_i} T_{\mu_1\mu_2\cdots\mu_i\cdots\mu_n}(k_1,k_2, \dots, k_i, \dots,k_n,p)= 0,
\label{QEDamp}
\ee
for every $i$.
This result is 
valid to all orders in perturbation theory 
as well as non-perturbatively.
In non-Abelian theories without tree-level symmetry breaking, 
such as QCD, the 
result mentioned above holds unchanged for the corresponding 
gauge bosons mediating the interaction (gluons). 
In the case of QCD such an example is given in  Eqs~(\ref{stcan-1}); 
note that the rhs of these equations vanish 
when contracted with the corresponding (on-shell) polarization tensors.
In the case of non-Abelian theories with spontaneous symmetry breaking 
(electroweak sector), the above result gets, in general, modified 
by symmetry breaking effects. 
When the amplitude  
is contracted by the momentum carried by a massive gauge 
boson ($W$ or $Z$), the result does not vanish up to terms related to the 
would-be Goldstone bosons and the corresponding gauge boson masses.
However, for the particular case of the photons, \ie the gauge boson corresponding to 
the unbroken $U(1)_{em}$, the vanishing of the contracted amplitude persists
(no mass nor would-be Goldstone boson associated with the photon). 
\newline
\indent
The way the fundamental property of Eq.~(\ref{QEDamp}) is realized 
diagrammatically at tree-level is through a number of delicate cancellations triggered  
by the elementary Abelian WIs satisfied by the bare photonic vertices. 
Beyond tree-level, to any finite order in perturbation theory, 
 Eq.~(\ref{QEDamp}) is still valid.
However, its diagrammatic demonstration 
is more involved, due to the proliferation of graphs and 
the fact that, in a non-Abelian context, 
the quantum corrections (concretely, the bosonic loops)
to the  conventional photonic vertex introduce new terms that convert the 
simple tree-level WI into a more complicated  STI. 
Nonetheless, at the conceptual level, the situation is straightforward; 
all one has to do is contract  with the relevant momentum 
all Feynman diagrams contributing to the amplitude in that order,  
and sum up the terms: the various contributions 
conspire in such a way as to enforce Eq.~(\ref{QEDamp}). 
The PT rearrangement of the amplitude 
facilitates the demonstration, for one thing because it restores 
the simple tree-level WIs to higher orders, but, strictly speaking, 
is not necessary.  
\newline
\indent
The PT becomes indispensable, however, when one attempts to maintain 
Eq.~(\ref{QEDamp}) near the resonance.   
Let us  
imagine that the kinematic configuration is such that a part of the amplitude 
becomes resonant and must be regulated through appropriate resummation 
of some of its parts. Since the Dyson resummation 
is not a fixed order calculation, it leads, in general, 
to the distortion of the cancellations enforcing the validity of 
Eq.~(\ref{QEDamp}) at any finite order. Moreover, near resonance a great 
number of graphs become numerically subleading (box diagrams are not 
resonant, etc), and the tendency is to omit them, even though, away 
from the resonance, when fixed order 
perturbation theory works, their inclusion is necessary for Eq.~(\ref{QEDamp}) to 
be valid. Evidently, the fixed-order wisdom  
does not carry over easily to the resonant case. 
The question that arises naturally, therefore, is under what conditions the resummed 
amplitude will be still gauge-invariant, \ie it will 
satisfy  the crucial Eq.~(\ref{QEDamp}). To put it differently, 
what is the {\it minimum number of graphs} that must be kept, and what is the 
{\it minimum amount of ``dressing''} they must undergo, in order to  maintain  Eq.~(\ref{QEDamp})
 in the vicinity of a resonance?
\newline
\indent
Let us focus on the concrete example of the 
process $\gamma e^- \to \mu^-\bar{\nu}_\mu \nu_e$, shown in  \Figref{PT_Ae_mu-barnumu-barnue}
(recall that Furry's theorem does not apply in non-Abelian theories).  
We assume, for simplicity, that the external fermions are all massless, 
and therefore we keep only the $g^{\mu\nu}$ parts of the propagators involved.
At tree-level, 
and away from the resonance, it is elementary to demonstrate that 
the amplitude $T_{\alpha}^{(0)}$ satisfies  
\be
q^\alpha T_{\alpha}^{(0)}=0 .
\label{zes1}
\ee
To that end, one must employ the 
the WI satisfied by $Ae^{+}e^{-}$ and $A\mu^{+}\mu^{-}$, \ie Eq.~(\ref{BasicWI}),
together with
the 
elementary WI of the tree-level  $AW^{+}W^{-}$ vertex, namely
($q+p_-+p_+=0$)  
\be
q_\alpha \Gamma_{{AWW}}^{\alpha\mu\nu} =
(p_{-}^2 - M_{ W}^2)g^{\mu\nu} - (p_{+}^2 - M_{ W}^2)g^{\mu\nu},  
\label{zes2}
\ee
where we have omitted longitudinal momenta, since they vanish when contracted with the 
conserved currents. 
\newline
\indent
It is clear that 
the negatively charged  $W$ (carrying momentum $p_{-}$) in graphs (a)
may become resonant, in which case   
the tree-level expression for its propagator 
must be regulated. It is important to notice, however, that if one were 
to simply replace, by hand, the tree-level propagator by a Dyson-resummed one, 
with no further modifications,
one would violate (\ref{zes2}) and, as a consequence, also  (\ref{zes1}).
In order to maintain (\ref{zes2}) one must: 
\begin{itemize}
\item[{\it i}.] Replace in {\it all three graphs} the tree-level $W$ propagators  
by the (one-loop) Dyson-resummed  PT propagators
$\widehat{\Delta}_{ W}^{\mu\nu}(p_{\pm})$,   
\begin{equation}
\widehat{\Delta}_{ W}^{\mu\nu}(p_{\pm}) = \frac{-ig^{\mu\nu}}
{p^2_{\pm} - M_{ W}^2 + \widehat{\Pi}_{ W  W} (p^2_{\pm})};
\label{DeltaPT}
\end{equation}
\item[{\it ii}.]  Add to the tree-level vertex $\Gamma_{{AWW}}^{\alpha\mu\nu}$ 
of graph (a) the one-loop PT 
vertex $\widehat{\Gamma}_{{AWW}}^{ \alpha\mu\nu}$, satisfying the~WI 
\be
q_\alpha \widehat{\Gamma}_{{AWW}}^{\alpha\mu\nu} =
\widehat{\Pi}_{ W  W}^{\mu\nu}(p_-) - 
\widehat{\Pi}_{ W  W }^{\mu\nu}(p_+).
\label{zes3}
\ee
\end{itemize}
\indent
Thus, one arrives at the following resonant transition amplitude,
\bea
\widehat{T}^{\alpha}_\mathrm{res} &=&
\Gamma_{{W e \nu_{e}}}
\widehat{\Delta}_{ W}(p_{+})
\left[\Gamma_{{AWW}}^{\alpha}+ \widehat{\Gamma}_{{AWW}}^\alpha\right]
\widehat{\Delta}_{ W}(p_{-}) \Gamma_{{W \mu \nu_{\mu}}}
\nonumber\\
&+& 
\Gamma^{{W e\nu_{ e}}}
S_{e}^{(0)} \Gamma_{{A e e}}^{\alpha}\widehat{\Delta}_{ W}(p_{-})\Gamma_{{W \mu \nu_{\mu}}} 
+ 
\Gamma_{{W e\nu_{ e}}}
\widehat{\Delta}_{ W}(p_{+}) \Gamma_{{A \mu \mu}}^{\alpha}
S_{\mu}^{(0)} \Gamma_{{W \mu \nu_{\mu}}},
\label{Twres}
\eea
where  contraction over all Lorentz indices except of the photonic one
is implied. 
\newline
\indent
Then, since by virtue of (\ref{zes3}) we have that 
\be
q_\alpha \left[\Gamma_{{AWW}}^{\alpha\mu\nu}+ \widehat{\Gamma}_{{AWW}}^{\alpha\mu\nu}\right] =
\widehat{\Delta}_{ W}^{-1}(p_{-}) g^{\mu\nu} 
- \widehat{\Delta}_{ W}^{-1}(p_{+}) g_{\mu\nu}, 
\ee
namely the generalization of Eq.~(\ref{zes2}), 
it is straightforward to verify that the $U(1)_{em}$ gauge invariance 
of this resonant process is maintained, \ie that 
\be
q_\alpha \widehat{T}^{\alpha}_\mathrm{res}=0.
\label{zes4}
\ee

\noindent
{\it v}.\hspace{.1cm}{\it  Equivalence theorem}

The  equivalence  theorem  states  that  at  very  high  energies 
($s\gg M^2_{ Z}$)
the
amplitude  for emission  or absorption  of a  longitudinally polarized
gauge boson becomes equal to the amplitude in which the gauge boson is
replaced  by the  corresponding would-be  Goldstone boson~\cite{Cornwall:1974km,Vayonakis:1976vz,Chanowitz:1985hj,Gounaris:1986cr}.   The above
statement is a consequence of the underlying local gauge invariance of
the SM, and has been known to  hold to all orders in perturbation theory for
multiple absorptions and emissions of massive vector bosons.
Compliance  with  this  theorem is  a  necessary  requirement  for any
resummation algorithm, since  any Born-improved  amplitude which fails
to satisfy  it is bound  to be missing important physical information.
The  reason why most resummation methods  are at odds  with  the equivalence  theorem is
that, in the usual diagrammatic analysis, the underlying symmetry of the
amplitudes is not manifest. Just as happens in the case of the OT, the
conventional subamplitudes,  defined  in terms of  Feynman diagrams, do
{\it  not}  satisfy the equivalence  theorem individually.    The resummation of  such a
subamplitude will, in turn, distort  several subtle cancellations, thus
giving rise to artifacts and unphysical effects.
 Instead, the PT subamplitudes satisfy the equivalence  theorem {\it individually}; 
as usual, the  only
non-trivial step for establishing this  is the proper exploitation of
elementary WIs.
\newline
\indent
To see an explicit example, let us return to the process 
$f(p_1)\bar{f}(p_2)\to  Z(k_1)Z(k_2)$. The equivalence  theorem states 
that the full amplitude 
${\mathcal T} ={\mathcal T}_s+{\mathcal T}_t$ satisfies 
\be
  \label{GETZZ}
{\mathcal T}(Z_{ L}Z_{ L}) = - {\mathcal T}(\chi\chi)  
-i{\mathcal T}(\chi z) -i {\mathcal T}(z \chi) + {\mathcal T}(\chi\chi),
\ee
where $Z_{ L}$ is  the longitudinal component  of the $Z$ boson, $\chi$ is
its  associated    would-be    Goldstone boson,  and     $z_\mu  (k) =
\varepsilon_\mu^{ L}(k) - k_\mu/M_{ Z}$ is the energetically suppressed part
of  the longitudinal polarization  vector  $\varepsilon_\mu^{ L}$.  It is
crucial  to observe,  however, that already    at the tree-level,  the
conventional $s$- and  $t$- channel subamplitudes  ${\mathcal T}_s$ and
${\mathcal T}_t$ fail  to satisfy the equivalence  theorem individually~\cite{Papavassiliou:1997fn,Papavassiliou:1998pb}.
\newline
\indent
To  verify that, one   has  to   calculate 
${\mathcal   T}_s (Z_{ L}Z_{ L})$,  using  explicit
expressions for the   longitudinal polarization vectors, and check  if
the  answer obtained is  equal to the Higgs-boson mediated $s$-channel
part  of the lhs of Eq.\ (\ref{GETZZ}).   In particular, in the center of mass (c.m.) 
system, we have 
\be
z_\mu (k_1) = \varepsilon_\mu^{ L}(k_1) - \frac{k_{1\mu}}{M_{ Z}} =
- 2M_{ Z} \frac{k_{2\mu}}{s}    +  {\mathcal  O}\left(\frac{M^4_{ Z}}{s^2}\right),  
\ee
and an exactly    analogous
expressions  for $z_\mu (k_2)$.  The  residual  vector $z^\mu (k)$ has
the properties    $k^\mu z_\mu   =  -M_{ Z}$  and  $z^2  = 0$.    
 After a straightforward calculation, we obtain 
\be
{\mathcal T}_s (Z_{ L}Z_{ L}) = -{\mathcal T}_s (\chi\chi) -i {\mathcal T}_s (z\chi) 
-i {\mathcal T}_s(\chi z) + {\mathcal T}_s (zz) - {\mathcal T}_s^{ P} , 
\ee
where
\begin{eqnarray}
{\mathcal T}_s (\chi\chi) &=&  \Gamma_{{H\chi\chi}} 
\Delta_{ H} (s) \bar{v}(p_2)\Gamma_{{Hf\bar{f}}}^{(0)} u(p_1),  \label{THsGG}\nonumber\\
{\mathcal T}_s (z\chi) + {\mathcal T}_s(\chi z)&=&
[z_\mu (k_1) \Gamma_{{HZ\chi}}^{\mu} +
z_\nu (k_2) \Gamma_{{H\chi Z}}^{\nu}] \Delta_{ H} (s)
\bar{v}(p_2)\Gamma_{{Hf\bar{f}}}u(p_1),  \label{THsZG}\nonumber\\ 
{\mathcal T}_s  (zz) &=& z_\mu (k_1)z_\nu (k_2) {\mathcal T}_{s}^{\mu\nu}(ZZ),
\end{eqnarray}
with  $\Gamma_{{H\chi\chi}}   =    -i  g_w   M^2_{ H}/(2M_{ W})$  and
$\Gamma_{{HZ\chi}}^{\mu}    = -  g_w  (k_1  +   2  k_2)_\mu /(2c_w)$.
Evidently,  the  presence of  the 
term ${\mathcal T}_s^{ P}$ prevents ${\mathcal  T}_s^H (Z_{ L}Z_{ L})$  
from satisfying the equivalence theorem.
This is, of course, not surprising, 
given that an  important Higgs-boson mediated $s$-channel part has
been  omitted.    Specifically,   the    momenta  $k_{1}^{\mu}$    and
$k_{2}^{\nu}$, stemming from  the  leading  parts of the   longitudinal
polarization        vectors          $\varepsilon^\mu_{ L}(k_1)$      and
$\varepsilon^\nu_{ L}(k_2)$,    extract    such  a   term    from   ${\mathcal
  T}_t(Z_{ L}Z_{ L})$ (see \Figref{H_contrib_tch_ZZ}).
Just as happens  in Eq.\  (\ref{TP}), this term  is
precisely  ${\mathcal T}_s^{ P}$,   and  must be    added  to  ${\mathcal  T}_s
(Z_{ L} Z_{ L})$, in  order  to form a  well-behaved  amplitude at high
energies.   In other  words,  the  amplitude 
\be
\widehat{{\mathcal   T}}_s (Z_{ L}Z_{ L})   = 
{\mathcal T}_s (Z_{ L}Z_{ L}) + {\mathcal T}_s^{ P}
\label{lm1}
\ee
satisfies the equivalence theorem 
by itself [see  Eq.~(\ref{GETZZ})].  
\newline
\indent
In fact, this crucial property   persists {\it after  resummation}, provided 
that one follows the same methodology for maintaining the external gauge-invariance, 
presented in the previous subsection.  
Indeed,   as  shown in  \Figref{PT_resum_nunu-ZZ}(a), 
the  resummed amplitude, to be denoted by  $\overline{\mathcal T}_s (Z_{ L} Z_{ L})$, may
be constructed from ${\mathcal  T}_s (Z_{ L} Z_{ L})$ in Eq.\ (\ref{THsZZ}), if
$\Delta_{ H}  (s)$ is   replaced by the  resummed Higgs-boson  propagator
$\widehat{\Delta}_{ H}  (s)$,    and $\Gamma_{{HZZ}}^{\mu\nu}$   by  the
expression $\Gamma_{{HZZ}}^{\mu\nu} + \widehat{\Gamma}^{\mu\nu}_{{HZZ}}$,
where $\widehat{\Gamma}^{\mu\nu}_{{HZZ}}$  is the one-loop $HZZ$ vertex
calculated  within the PT.  It is then straightforward
to  show that  the   Higgs-mediated amplitude $\widetilde{\mathcal  T}_s
(Z_{ L}Z_{ L}) = 
\overline{\mathcal T}_s (Z_{ L}Z_{ L}) + {\mathcal T}_s^{ P}$ respects
the  equivalence theorem {\it individually}; to that  end we only   need to employ the
following tree-level-type PT WIs
\begin{eqnarray}
  \label{PTHZZ1}
k_{2\nu} \widehat{\Gamma}^{\mu\nu}_{{HZZ}}
(q,k_1,k_2) + i M_{ Z} \widehat{\Gamma}_{{HZ\chi}}^\mu (q,k_1,k_2)
&=& - \frac{g_w}{2c_w} \widehat{\Pi}_{{Z\chi}}^\mu (k_1),\nonumber\\
  \label{PTHZZ2}
k_{1\mu} \widehat{\Gamma}^\mu_{{HZ\chi}}
(q,k_1,k_2) + i M_{ Z} \widehat{\Gamma}_{{H\chi\chi}}(q,k_1,k_2)
&=& - \frac{g_w}{2c_w}\!\left[\widehat{\Pi}_{{HH}}(q^2) + 
\widehat{\Pi}_{{\chi\chi}}(k^2_2)\right],\nonumber\\
  \label{PTHZZ3}
k_{1\mu} k_{2\nu} \widehat{\Gamma}^{\mu\nu}_{{HZZ}}
(q,k_1,k_2) + M^2_{ Z}\widehat{\Gamma}_{{H\chi\chi}}(q,k_1,k_2)
&=& \frac{ig_w M^2_{ Z}}{2c_w}\!\left[\widehat{\Pi}_{{HH}}(q^2) + 
\widehat{\Pi}_{{\chi\chi}}(k^2_1) + \widehat{\Pi}_{{\chi\chi}}(k^2_2)\right]\!,\nonumber \\
\end{eqnarray}
where $\widehat{\Gamma}^{{HZ\chi}}_\mu$  and $\widehat{\Gamma}^{{H\chi\chi}}$
are  the one-loop PT $HZ\chi$ and  $H\chi\chi$ vertices, respectively. In addition,  
one should also make use  of the PT WI involving the
$Z\chi$- and  $\chi\chi$- self-energies, namely 
\be
k_\mu \widehat{\Pi}_{{Z\chi}}^\mu (k) = -i M_{ Z} \widehat{\Pi}_{{\chi\chi}}(k^2),
\ee
which is the exact analogue of Eq.~(\ref{PWI3}). 

\noindent
{\it vi}.\hspace{.1cm} {\it Renormalization group invariance}

Physical quantities, such as scattering amplitudes, must be invariant under the RG, \ie they should 
not depend on the renormalization point $\mu$ chosen to carry out the 
subtractions, nor the renormalization scheme 
($\overline{\rm MS}$, on-shell scheme, momentum subtraction, etc). 
 \newline
 \indent
Let us consider, for concreteness, a two-to-two amplitude, mediated 
(at tree-level) by a gauge boson (photon, gluon, $W$ or $Z$).
In QED the RG-invariance of such an amplitude is realized in a very particular way.
Due to the characteristic relations  (\ref{qed3}) and (\ref{qed4}),  
the amplitude may be decomposed in  a unique way into three parts that are 
individually RG-invariant: ({\it i}) a universal 
(process-independent)  part, corresponding to the effective charge defined in  (\ref{alphaqed}), 
[expanded to the given order], which is RG-invariant due to (\ref{qed4}); ({\it ii})  a process-dependent part 
composed by   the  vertex corrections and the wave-function renormalization of the external particles, 
which is RG-invariant due to (\ref{qed3});
({\it iii}) a process-dependent part, coming from 
UV finite  boxes; this is trivially  RG-invariant, since it is UV finite and does not get renormalized. 
\newline
\indent
In non-Abelian theories the  RG-invariance of scattering amplitudes is
enforced  order-by-order  in  perturbation  theory, regardless  of the PT
rearrangement, by  virtue of  equations such as (\ref{qed1}),  which hold
also in a non-Abelian context [but not (\ref{qed3}) and (\ref{qed4})].
There is  an important  difference, however, with  respect to  the QED
case: while  the entire amplitude is  RG-invariant, the identification
of a universal part corresponding  to an effective charge is no longer
possible. Since the PT rearrangement  restores 
relations of the type (\ref{qed3}) and (\ref{qed4}), 
the  three  individually RG-invariant
quantities introduced above for QED can also  be identified in a non-Abelian
context; in  particular, non-Abelian effective  charges constitute the
universal part of the amplitude.
\newline
\indent
It  should be  clear  from  the above  discussion,  together with  the
analysis of  the previous  subsections, that when  resonant amplitudes
are   regulated   following  the   PT   procedure  they are   automatically
RG-invariant.  In  fact, in the cases where  the resonating particles
are $W$, $Z$, or Higgs bosons, one can isolate a universal part, which, 
in turn, may be identified with the {\it lineshape} of the corresponding particle.

\noindent
{\it vii}.\hspace{.1cm} {\it  Good high energy behavior.}

On physical grounds   one  expects 
that     far from the   resonance   the
Born-improved   amplitude must behave exactly  as  its  tree-level
counterpart; in fact, 
a  self-consistent resummation formalism should  have
this property built in, \ie far from resonance one should 
recover the correct high energy behavior {\it without} having 
to  re-expand  the Born-improved amplitude   perturbatively. 
Recovering the correct asymptotic behavior is particularly tricky,
however, when the final particles are gauge bosons.
In order
to accomplish  this,  
in addition   to   the correct  one-loop
(running) width, the
appropriate  one-loop vertex corrections  must   be supplemented;
these vertex corrections and the width must be related 
by a crucial tree-level Ward identity.
In practice this WI
ensures   that
the massive  cancellations, which take place at  tree-level,  
will still go  through after the Born-amplitude has been ``dressed''.
The exact mechanism that 
enforces the correct high energy behavior of the
Born-improved amplitude, when the PT width and vertex are used,
has been studied in detail in~\cite{Papavassiliou:1999qn}, for the specific reference process  
$f(p_1)\bar{f}(p_2)\to  Z(k_1)Z(k_2)$ used  above.

\newpage


\section{\seclab{beyond_1l} Beyond one loop: from two loops to all orders}
\noindent
In this section we generalize the  PT to two loops and beyond; this is
a  rather more  difficult  problem  than pinching  at  one loop,  both
conceptually and operationally.  In the first part of  this section we
will establish the rules of the game beyond one-loop; the final upshot
of  these considerations is  simple but  cumbersome to  implement: one
must repeat exactly what one did  at one loop, only now with many more
diagrams.   There are essentially  two important  lessons that  can be
learned  from the two-loop  construction: ({\it i})  the PT  works perfectly
well  beyond   one  loop,  and  ({\it ii})   the  PT  must   evolve  into  a
non-diagrammatic procedure.   The second lesson is  taken seriously in
the second part  of this section, where the  all-order PT construction
is  presented.  There  we  show  that the  entire  pinching action  is
actually encoded  into the STI  satisfied by a very  special all-order
kernel  shared by  propagator and  vertex diagrams.   This fundamental
observation allows us to rise beyond the diagram-by-diagram treatment,
and eventually generalize the PT to all orders.

\subsection{The pinch technique at two loops}
\noindent
In this subsection we present the PT construction 
at two loops~\cite{Papavassiliou:1999az,Papavassiliou:1999bb}.  
Specifically, we will show that the PT can 
be generalized at two loops   
by resorting {\it exactly} 
to the same physical and field-theoretical principles 
as at one-loop. 
In addition, it will become clear 
that the correspondence between the PT and BFG, 
established at one-loop, persists also at the two-loop level.
\newline
\indent
Historically,  the  basic  conceptual  difficulty associated  with  the
generalization of  the PT  beyond one loop  has been to  determine the
origin of the  pinching momenta.  Let us assume  that, without loss of
generality, one  chooses from  the beginning the  conventional Feynman
gauge.  Then,  the only  sources of possible pinching  momenta are
the three-gluon  vertices. The question  is whether all  such vertices
must  be somehow forced  to pinch,  or, in  other words,  whether the
standard PT decomposition of  Eq.~(\ref{decomp}) should be carried out
to  all available  three-gluon  vertices.  The  problem  with such  an
operation, however,  is the following:  for the case of  a three-gluon
vertex nested  inside a Feynman diagram,  how does one  choose what is
the  ``special'' momentum?   Or,  in  other words,  which  way is  one
supposed to break the Bose-symmetry of the vertex?  It turns out that the
solution  to these  questions is  very simple:  one should  only apply
Eq.  (\ref{decomp}) to the  vertices that  have the  physical momentum
incoming (or  outgoing) in one of  their legs (not  mixed with virtual
momenta); the special  leg is precisely the one  carrying $q$. We will
call such  a vertex  ``external''.  All other  vertices are not  to be
touched, i.e.  they should not be decomposed in any way; such vertices
have virtual momenta in every one  of their three legs, and are called
``internal'' (see \Figref{TWL0}).
 A simple way  to understand  why the  decomposition of
internal  vertices would  lead to  inconsistencies is  to  consider the
special unitarity  properties that the PT  subamplitude must satisfy
(see corresponding subsection).
\newline
\indent
We   emphasize
that  throughout the entire two-loop analysis we will maintain  a
diagrammatic   interpretation  of  the    various  contributions.   In
particular, no sub-integrations
should to be carried out.  This additional feature renders the method all
the more  powerful, because unitarity  is manifest, and can  be easily
verified by means  of the Cutkosky  cuts.  
\newline
\indent
In addition to the result of Eq.~(\ref{dreg}), 
throughout this section we will employ the following formulas, valid
in dimensional regularization: 
\bea
&&\int_k \frac{k_{\alpha}k_{\beta}}{k^4}
=\left(\frac{1}{4-\epsilon}\right)\int_k \frac{1}{k^2} = 0,\label{dimreg-1}\nonumber\\
&&\int_k \frac{(2k+q)_{\alpha}}{k^2 (k+q)^2} = 0,\label{dimreg-2}\nonumber\\
&&\int_k \frac{\ln^n (k^2)} {k^2} = 0\qquad n=0,1,2,\dots 
\label{dimreg-3}
\eea
Finally, in order  
to make contact with the notation of~\cite{Papavassiliou:1999az,Papavassiliou:1999bb},  
we introduce the dimensionful projection operator
\be
t_{\mu\nu}(q) \equiv q^2 P_{\mu\nu}(q)\,.
\ee
\begin{figure}[!t]
\bce
\includegraphics[width=15cm]{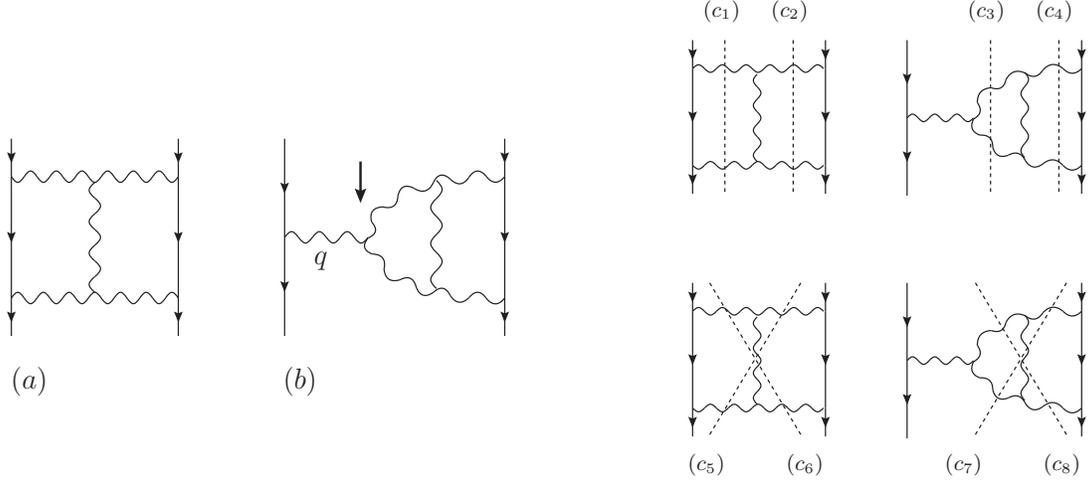}
\ece
\caption{\figlab{TWL0} Left panel:Some examples of external and 
internal vertices which appears in the two-loop graphs. Diagram $(a)$ has only 
internal three-gluon vertices, while diagram $(b)$ has two internal vertices 
and an external one (indicated by the arrow).
Right panel: The two- and three-particle Cutkosky cuts (cutting through gluons only).}
\end{figure}

\subsubsection{\label{1prg}The one-particle reducible graphs}
\noindent
We begin the two-loop construction by treating the 1PR graphs, 
collectively shown in \Figref{TWL1}.  
All such graphs are the product of two one-loop subgraphs, 
which may be individually converted into their 
 PT counterparts, following precisely the standard PT procedures 
established in section II. 
Note, in fact, that,  
as has been explained in detail in~\cite{Papavassiliou:1995fq,Papavassiliou:1995gs,Papavassiliou:1996zn}, the resummability of
the one-loop PT self-energy requires precisely this: the conversion of 
one-particle reducible (1PR) strings of 
conventional self-energies $\Pi^{(1)}$
into strings containing PT self-energies  $\widehat\Pi^{(1)}$.
\newline
\indent
There is an important point, however, that one must realize. 
The conversion of the 1PR graphs into the corresponding PT 
does not take place for free; instead, the process of the conversion 
gives rise to 
certain residual pieces, all of which have the crucial characteristic of being 
effectively 1PI.  The way these pieces are produced is ({\it i}) because  
there is a mismatch between the propagator-like terms obtained from the available  
quark-gluon vertices $\Gamma^{(1)}$, and those needed to convert the string of two $\Pi^{(1)}$s
into a string of two $\widehat\Pi^{(1)}$s
[this happens with graphs $(a)$,  $(b)$,  $(c)$  and $(d)$ in \Figref{TWL2}], 
or ({\it i})  terms that in the one-loop construction were vanishing, due to the on-shell 
conditions, now they do not vanish, because they do not communicate with the external 
quarks (i.e., the Dirac equation cannot be triggered) [see graph $(b)$ in \Figref{TWL2}] .
\newline
\indent
To see this in detail, 
let us return for a moment to the one-loop construction,  
and consider the conventional quark-gluon vertex at one-loop, 
to be denoted by $\Gamma^{(1)}_{\alpha}(p_1,p_2)$.  
We will repeat the calculation of subsection \ref{PT3}, but now we will 
not assume that the external quarks are on-shell; this is because, 
at two-loops, we can have the situation depicted in \Figref{TWL2}.  
In particular,
\be
\Gamma^{(1)}_{\alpha} = \widehat{\Gamma}_{\alpha}^{(1)}\, + \,
\frac{1}{2}\, V^{(1)}_{\mathrm{P}\alpha\sigma}(q) \gamma^{\sigma}
+X_{1\alpha}^{(1)}(p_1,p_2)(\not\! p_2 -m)
+   (\not\! p_1 -m) X_{2\alpha}^{(1)}(p_1,p_2)\, ,
\label{PTact}
\ee
where
\bea
X_{1\alpha}^{(1)}(p_1,p_2) &=&  g^2 C_A \int_k \frac{1}{k^2 (k+q)^{2}} \gamma_{\alpha} S^{(0)}(p_2+k), 
\nonumber\\
X_{2\alpha}^{(1)}(p_1,p_2) &=&  g^2 C_A \int_k \frac{1}{k^2 (k+q)^{2}} S^{(0)}(p_2+k)\gamma_{\alpha}.
\label{somedef2} 
\eea
By
$ \frac{1}{2}\, V_{{\rm P}\, \alpha\beta}^{(1)}(q)$ 
we denote the dimensionless propagator-like contribution to be alloted to $\Pi^{(1)}(q)$, i.e.
\be
V_{{\rm P}\, \alpha\beta}^{(1)}(q) = 2g^2 C_A P_{\alpha\beta}(q) \int_k \frac{1}{k^2(k+q)^2}\
\ee
Thus, Eq.~(\ref{proplikevert}) reads (completely equivalently)
\be
\Pi_{{\rm P}\, \alpha\beta}^{(1)}(q) = q^2 V_{{\rm P}\, \alpha\beta}^{(1)}(q)
\label{awow}
\ee 
or 
\be
\Pi_{{\rm P}\,\alpha\beta}^{(1)}(q) =  
V_{{\rm P}\,\alpha\sigma}^{(1)}(q)t_{\beta}^{\sigma}(q) \, .
\label{PP1}
\ee
Of course, on-shell ($\ps_1=\ps_2=m$) the $\Gamma^{(1)}_{\alpha}$ 
of (\ref{PTact}) collapses to that of (\ref{PTvert}). 
\begin{figure}[!t]
\bce
\includegraphics[width=15cm]{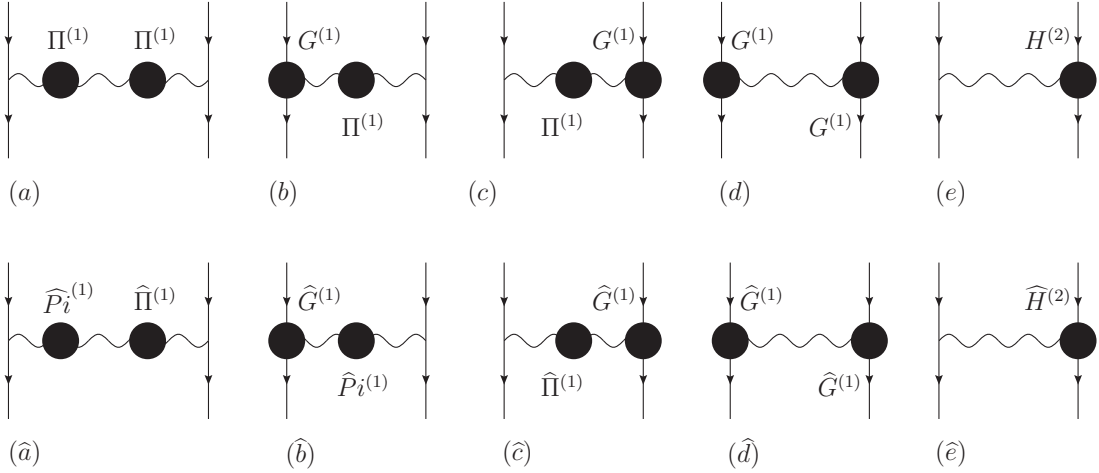}
\ece
\caption{\figlab{TWL1}The one-particle reducible graphs  
($a$), ($b$), ($c$), ($d$), and ($e$) before the PT rearrangement.
${G}^{(1)}$ denotes  the sum of the conventional one-loop 
quark-gluon vertex and  one-loop 
wave-function corrections to the external quarks.
${H}^{(2)}$ is the product of the  one-loop quark-gluon vertex 
times the one-loop wave-function corrections to the external quarks
(see also \Figref{TWL2}).
Note that ($a$), ($b$), ($c$), and ($d$) become disconnected  
by cutting a gluon line, whereas ($e$) by cutting an external quark line.
All these graphs are converted into their PT equivalent graphs, i.e. 
($\widehat{a}$), ($\widehat{b}$), ($\widehat{c}$), ($\widehat{d}$), and ($\widehat{e}$)
are produced, together with some residual terms that, due to the pinching action, 
are effectively 1PI.}  
\end{figure}
\newline
\indent
It is relatively straightforward to establish that 
\bea
(2a) &=& (2\widehat{a}) - d(q)\, R^{(2)}_{{\rm P}\,\alpha\beta}(q)\, d(q)\,,
\label{1prse} \nonumber\\
(2b)+(2c)+(2d)+(2e)\, &=& \,(2 \widehat{b})+(2\widehat{c})+(2\widehat{d}) + (2\widehat{e}) 
- F^{(2)}_{{\rm P}\,\alpha}(p_1,p_2)\,,
\label{RandF}
\eea
with
\bea
iR^{(2)}_{{\rm P}\,\alpha\beta}(q) &=& 
\Pi^{(1)}_{\alpha\rho}(q) V_{{\rm P}\beta}^{(1)\rho}(q) + 
\frac{3}{4} \, q^2 \,V_{{\rm P}\,\alpha\rho}^{(1)}(q)V_{{\rm P}\,\beta}^{(1)\rho}(q)\, ,
\label{RP} \nonumber\\
F^{(2)}_{{\rm P}\,\alpha}(p_1,p_2) &=&
\Pi_{{\rm P}\alpha}^{(1)\beta}(q)\, d(q)\,
\widehat{\Gamma}_{\beta}^{(1)}(p_1,p_2) +  
d(q)\, Y_{{\rm P}\alpha}^{(2)}(p_1,p_2),
\label{1prmixed}
\eea
with
\be
 Y_{{\rm P}\alpha}^{(2)}(p_1,p_2) \equiv
 X_{1\alpha}^{(1)}(p_1,p_2)\Sigma^{(1)}(p_1)+
 X_{2\alpha}^{(1)}(p_1,p_2)\Sigma^{(1)}(p_2).
\label{FP}
\ee
\indent
Notice that 
\be
 R^{(2)}_{\mathrm{P}\,\alpha\beta}(q)  = 
I_2 \left[ L_{\alpha\beta}(q,k)+ 3t_{\alpha\beta}(q)\right],
\label{RP2}
\ee
where 
\be
L_{\alpha\beta}(q,k) \equiv  
\Gamma_{\alpha}^{\sigma\rho}(q,k,-k-q)
\Gamma_{\beta\sigma\rho}(q,k,-k-q) - 2 k_{\alpha}(k+q)_{\beta}
\label{Tense1}
\ee
is simply the numerator of the conventional $\Pi_{\alpha\beta}^{(1)}(q)$ 
[\ie the terms in square brackets on the rhs of Eq.~(\ref{PTprop})]. 
 
\subsubsection{ Quark-gluon vertex and gluon self-energy at two loops}
\noindent
In this subsection we will first demonstrate the construction
of the two-loop PT quark-gluon vertex 
$\widehat{\Gamma}_{\alpha}^{(2)}(p_1,p_2)$, which turns out
to have the exact same properties as its one-loop counterpart 
$\widehat{\Gamma}_{\alpha}^{(1)}(p_1,p_2) $. At the same time
we will determine the two-loop propagator-like
contributions 
$V_{{\rm P}\alpha\sigma}^{(2)} \gamma_{\sigma}$,  which will be subsequently
converted into 
$\Pi_{{\rm P}\alpha\sigma}^{(2)}$, i.e. the
two-loop version of 
$\Pi_{{\rm P}\alpha\sigma}^{(1)}$
of Eq.~(\ref{PP1}). In addition, out of this procedure
the terms  $Y_{{\rm P}\alpha}^{(2)}(p_1,p_2)$
of Eq.~(\ref{FP}) will emerge again, but with opposite sign. Then, 
we will turn to the two-loop gluon self-energy, and we will outline the 
basic steps that must be followed in its construction.
\begin{figure}[!t]
\bce
\includegraphics[width=14cm]{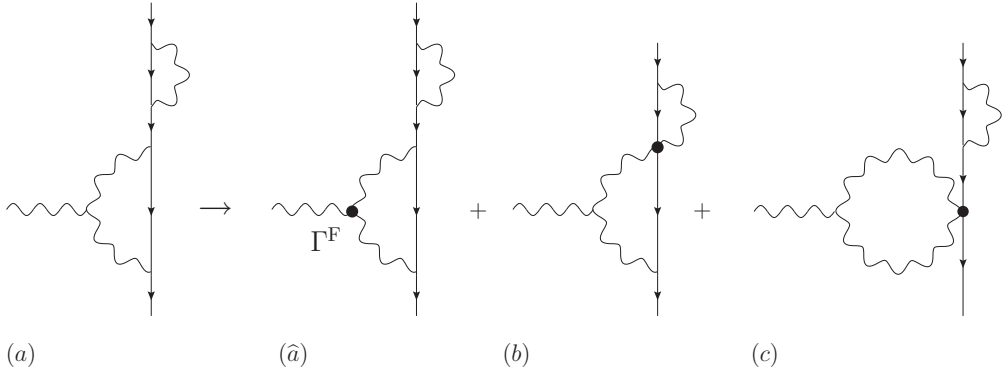}
\ece
\caption{\figlab{TWL2}The PT rearrangement of the non-Abelian part of ${\bf H}^{(2)}$ 
[graph $(a)$], giving rise  to its PT counterpart [graph $\widehat{a}$], 
and to additional contributions  [graphs $(b)$ and $(c)$]. Diagram 
 $(b)$ is effectively 1PI, and is alloted to  $Y_{\rm P}^{(2)}$, whereas 
 $(c)$ contributes to the first term of $F_{\rm P}^{(2)}$.}
\end{figure}
\begin{figure}[!t]
\bce
\includegraphics[width=14cm]{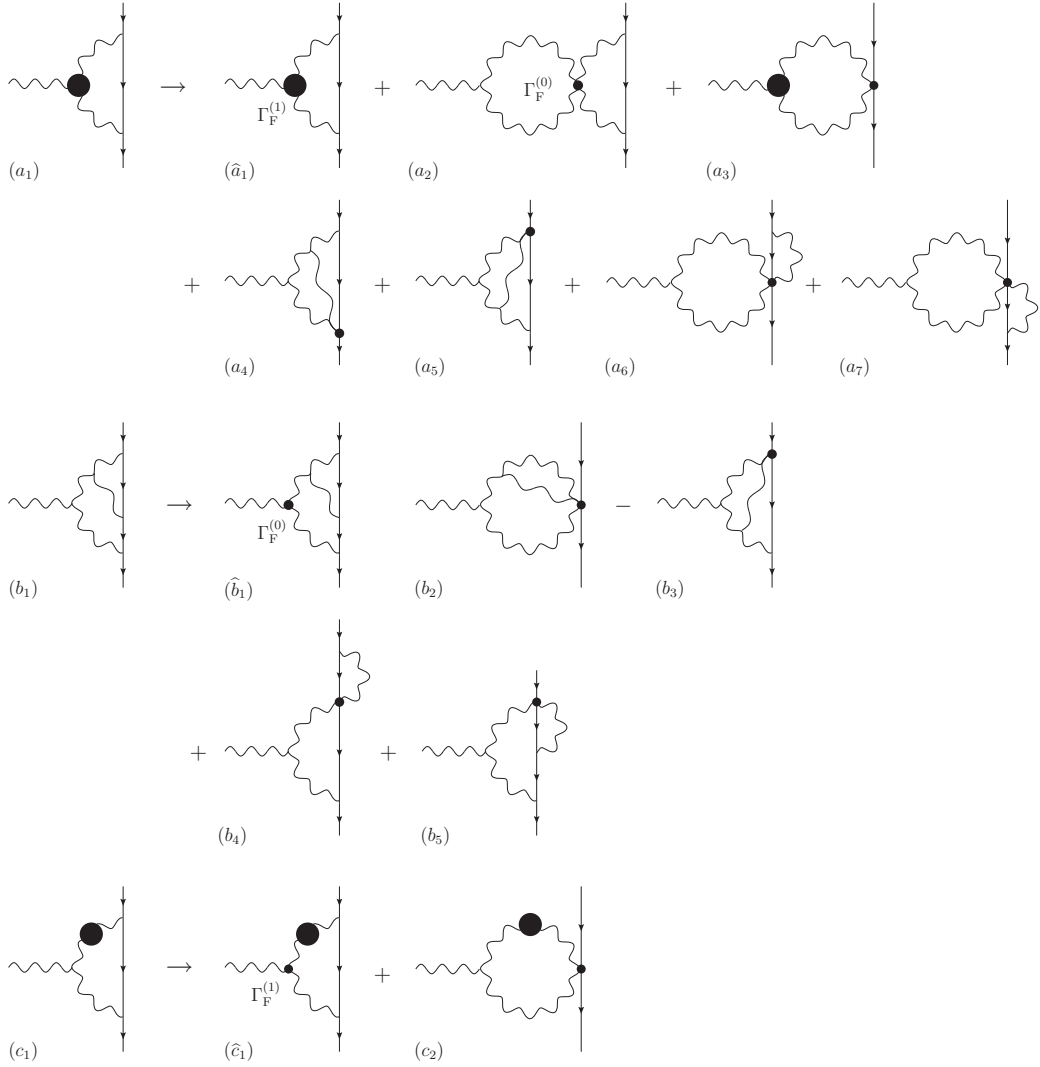}
\ece
\caption{\figlab{TWL3}The result of enforcing the PT decomposition on
the external vertices of some of the two-loop Feynman diagrams
contributing the conventional two-loop quark-gluon vertex
${\Gamma}_{\alpha}^{(2)}$}
\end{figure}
\begin{figure}[!t]
\bce
\includegraphics[width=14cm]{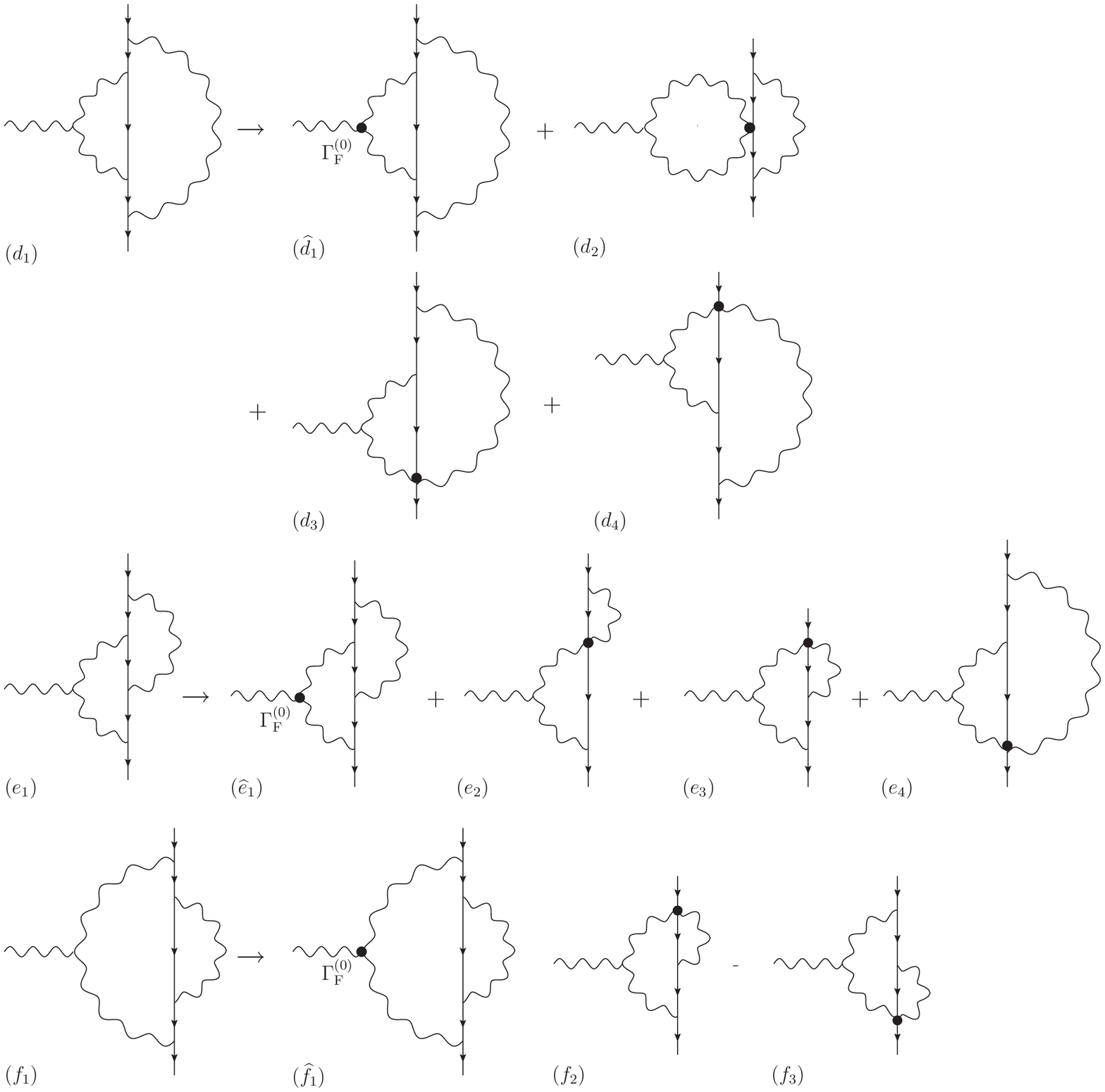}
\ece
\caption{\figlab{TWL4}The result of enforcing the PT decomposition on
the external vertices of some of the remaining two-loop vertex graphs. }
\end{figure}
\newline
\indent
The construction of the two-loop quark-gluon vertex proceeds as follows:
the Feynman graphs contributing to $\Gamma_{\alpha}^{(2)}(p_1,p_2)$
can be classified into two sets:
({\it i}) those containing an ``external'' three-gluon vertex 
i.e. a three-gluon vertex where the momentum $q$ is incoming,
as shown in Fig.s \ref{fig:TWL3} and~\ref{fig:TWL4}, and 
({\it ii}) those which do not have an ``external'' three-gluon vertex.
This latter set contains either graphs with no three gluon vertices
(Abelian-like), or graphs with
internal three-gluon vertices (all three
legs are irrigated by virtual momenta, i.e. $q$ never enters  
alone into any of the legs).  
Of course, all  three-gluon vertices appearing
in the computation of the one-loop $S$-matrix are external, and
so are those appearing in the 1PR part of the two-loop $S$-matrix
(see previous subsection).
Then one
carries out the decomposition of Eq. (\ref{decomp}) 
to the external three-gluon vertex   
of all graphs belonging to set ($i$), leaving {\it all} their other
vertices unchanged, and identifies the propagator-like pieces
generated at the end of this procedure.
\newline
\indent
Let us define the following quantities (the two loop integration symbol $\int_k\int_\ell$ will be suppressed throughout)
\bea
iI_{1} &=& g^4 C_A^2 
[\ell^2 (\ell-q)^2 k^2 (k+\ell)^2 (k+\ell-q)^2]^{-1},\nonumber\\
iI_{2}  &=& g^4 C_A^2 
[\ell^2 (\ell-q)^2 k^2 (k+q)^2]^{-1},\nonumber\\
iI_{3} &=& g^4 C_A^2 
[\ell^2 (\ell-q)^2 k^2 (k+\ell)^2]^{-1},\nonumber\\
iI_{4} &=& g^4 C_A^2 
[\ell^2 \ell^2 (\ell-q)^2 k^2 (k+\ell)^2]^{-1},\nonumber\\
iI_{5} &=& g^4 C_A^2 [\ell^2 k^2 (k+q)^2]^{-1},
\eea
which will be used extensively in what follows.
The calculation is straightforward, but lengthy [see also again Fig.s \ref{fig:TWL3} and~\ref{fig:TWL4}];
we find 
\be
\Gamma_{\alpha}^{(2)}(p_1,p_2)
=
\frac{1}{2} F^{(2)}_{{\rm P}\,\alpha} (p_1,p_2)
+\frac{1}{2}V_{{\rm P}\alpha\sigma}^{(2)} \gamma^{\sigma}
+\widehat{\Gamma}_{\alpha}^{(2)} (p_1,p_2),
\label{VPT2}
\ee
with
\be
V_{{\rm P}\alpha\sigma}^{(2)}(q) = I_4 L_{\alpha\sigma}(\ell,k)
+ (2 I_{2} + I_{3}) g_{\alpha\sigma}
-I_1 \left[k_{\sigma}g_{\alpha\rho}+  
\Gamma_{\rho\sigma\alpha}^{(0)}(-k,-\ell,k+\ell)\right] (\ell-q)^{\rho}.
\label{BRes1}
\ee
\indent
The interpretation of the three terms appearing on the rhs of
Eq.~(\ref{VPT2}) is as follows:

\begin{itemize}

\item[{\it i}.]  The term 
$\frac{1}{2} F^{(2)}_{{\rm P}\,\alpha}(p_1,p_2)$ 
is half of the 
vertex-like part necessary to cancel the
corresponding term appearing in Eq.~(\ref{RandF}), during
the conversion of 
conventional 1PR graphs into their PT counterparts. The other
half will come from the mirror vertex (not shown).
\newline
\item[{\it ii}.] $\frac{1}{2}V_{{\rm P}\alpha\sigma}^{(2)} \gamma^{\sigma} $
is  the total propagator-like term originating from the 
two-loop quark-gluon vertex; together with the
equal contribution  from the mirror set of two-loop vertex graphs (not shown) will
give rise to the self-energy term
\be
\Pi_{{\rm P}\alpha\beta}^{(2)}(q) = V_{{\rm P}\alpha\sigma}^{(2)}(q)t^{\sigma}_{\beta}(q) , 
\label{2LPinch}
\ee
which 
will be part of the 
effective two-loop PT gluon self-energy,
to be constructed below.
\newline
\item[{\it iii}.] $\widehat{\Gamma}_{\alpha}^{(2)}(p_1,p_2)$ is the PT two-loop
quark-gluon vertex; it {\it coincides} with the 
corresponding two-loop quark-gluon vertex computed
in the BFG, \ie
\be
\widehat{\Gamma}_{\alpha}^{(2)} (p_1,p_2) = 
\widetilde{\Gamma}_{\alpha}^{(2)}(p_1,p_2,\xi_{ Q} =1)
\ee
as happens in the one-loop case. 
Either by virtue of the above equality and 
the formal properties of the BFM, 
or by means of
an explicit diagrammatic calculation,  
where one acts with $q^{\alpha}$ on individual diagrams,
one can establish that 
$\widehat{\Gamma}_{\alpha}^{(2)}(p_1,p_2)$ satisfies the WI
\be
q^{\alpha}\widehat{\Gamma}_{\alpha}^{(2)}(p_1,p_2) =
\widehat{\Sigma}^{(2)}(p_1)-\widehat{\Sigma}^{(2)}(p_2),
\label{WI2}
\ee
which is the exact two-loop 
analogue of Eq.~(\ref{WIvert}). 
$\widehat{\Sigma}^{(2)}(p)$ is the two-loop PT fermion self-energy;
it is given by 
\be
\widehat{\Sigma}^{(2)}(p) = 
\widetilde{\Sigma}^{(2)}(p,\xi_{ Q}=1) 
 =\Sigma^{(2)}(p,\xi=1). 
\ee
Again, this is the precise generalization of the one-loop result
of Eq.~(\ref{Sigmahat}). 
At this point this result comes as no surprise, since 
all three gluon vertices
appearing in the Feynman graphs contributing to
$\Sigma^{(2)}(p,\xi)$ are internal; therefore, at $\xi=1$ there
will be no pinching.
\end{itemize}
\indent
Notice that, as happened in the one-loop case, 
in deriving the above results 
no integrations (or sub-integrations) over virtual momenta have been carried out.

The construction of 
${\widehat\Pi}^{(2)}_{\alpha\beta}(q)$ proceeds as follows:
To the conventional two-loop gluon self-energy 
$\Pi^{(2)}_{\alpha\beta}(q)$ (shown in \Figref{TWL5}) we add two additional terms;
({\it i}) the propagator-like term  
$\Pi_{\rm{P}\alpha\beta}^{(2)}(q)$ derived in the previous subsection,
 Eq.~(\ref{2LPinch}), and ({\it ii})
the propagator-like part \linebreak $- R^{(2)}_{{\rm P}\,\alpha\beta}(q)$ 
given in Eq.~(\ref{RP}),
stemming from the conversion of the conventional string
into a PT string; this term must be removed from the  
1PR reducible set and be alloted to
${\widehat\Pi}^{(2)}_{\alpha\beta}(q)$, as described in~\cite{Papavassiliou:1995fq,Papavassiliou:1995gs,Papavassiliou:1996zn}. 
Thus, ${\widehat\Pi}^{(2)}_{\alpha\beta}(q)$ reads
\be
{\widehat\Pi}^{(2)}_{\alpha\beta}(q) =
\Pi^{(2)}_{\alpha\beta}(q) + \Pi_{{\rm P}\alpha\beta}^{(2)}(q) -
R^{(2)}_{{\rm P}\,\alpha\beta}(q).
\label{FinalA}
\ee
\newline
\indent
It is a rather lengthy exercise to establish that the two-loop PT gluon self-energy 
coincides with that of the BFG, \ie 
\be
{\widehat\Pi}^{(2)}_{\alpha\beta}(q) = 
{\widetilde\Pi}^{(2)}_{\alpha\beta}(q,\xi_{ Q} =1).
\label{FinalB}
\ee
\begin{figure}[!t]
\bce
\includegraphics[width=12cm]{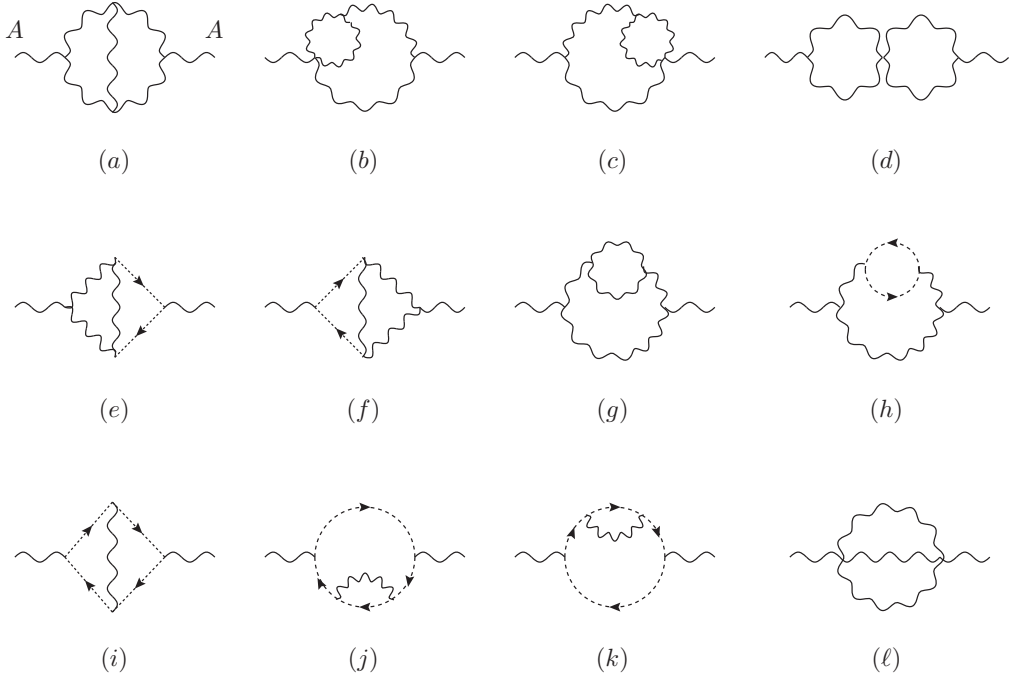}
\ece
\caption{\figlab{TWL5}The Feynman diagrams contributing to the conventional
two-loop gluon self-energy $\Pi_{\alpha\beta}^{(2)}$, 
in the $R_{\xi}$ gauges.}
\end{figure}
To verify this important result explicitly, we simply start out with 
the diagrams contributing to $\Pi^{(2)}_{\alpha\beta}(q)$, 
and convert them into the corresponding diagrams contributing
to ${\widetilde\Pi}^{(2)}_{\alpha\beta}(q,\xi_{ Q} =1)$, shown in \Figref{BFM_2l}, 
keeping track of the terms that are left over. 
In doing so we only need to carry out algebraic manipulations
in the numerators of individual Feynman diagrams,
and use judiciously the identity of Eq.~(\ref{INPTDEC1}). 
Again, no integrations over virtual momenta need be carried out,
except for identifying 
vanishing contributions by means of the formulas listed
in  Eqs~(\ref{dimreg-1}), (\ref{dimreg-2}) and~(\ref{dimreg-3}). 
It turns out that all terms left over after this 
diagram-by-diagram conversion 
cancel exactly against the terms  
$[\Pi_{{\rm P}\alpha\beta}^{(2)}(q) - R^{(2)}_{{\rm P}\,\alpha\beta}(q)]$, 
finally furnishing the crucial equality of Eq.~(\ref{FinalB}).
\newline
\indent
Since we have used dimensional regularization throughout,
and no integrations have been performed, 
the results of this section do not depend on the value $d$ of the
space-time; 
in particular they are valid for $d=3$, 
which is of additional field-theoretical interest
\cite{Cornwall:1984eu,Cornwall:1988ad,Cornwall:1997dc}. 
Clearly, when $d\to 4$ the renormalization program
should also be carried out (for details see~\cite{Papavassiliou:1999bb}).

\subsubsection{The two-loop absorptive construction}
\noindent
We now turn to the absorptive two-loop construction. 
Operationally the situation is significantly more involved 
compared to the one-loop case; the full details have been 
presented in \cite{Papavassiliou:1999bb}. Here we will give 
a brief qualitative description  
of how this construction proceeds.
\begin{figure}[!t]
\bce
\includegraphics[width=12cm]{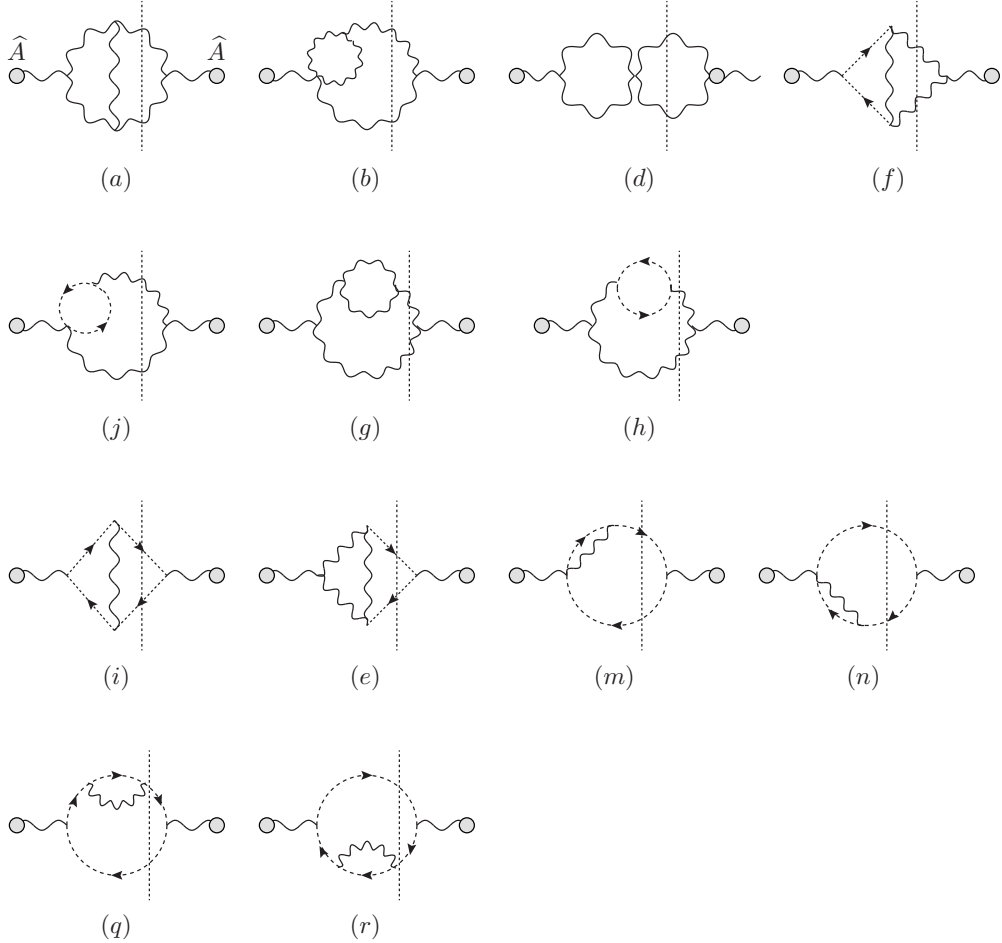}
\ece
\caption{\figlab{TWL8}The two-particle Cutkosky cuts of the PT (and BFG) 
two-loop gluon self-energy.
We have used the same labeling of individual diagrams as in 
\Figref{BFM_2l}. The two upper (lower) rows show graphs where two gluon 
(ghost) lines have been cut.}
\end{figure}
\newline
\indent
The central result  is that the strong version of the OT, 
given in (\ref{OTPTstrong}), holds also for the two-loop case. 
Specifically, the   imaginary parts of the
two-loop PT Green's  functions (under construction)
are related by  the OT  to precisely  identifiable and very  special  parts  of  two  different
squared amplitudes,  namely the  one-loop squared amplitude for the  process
$q\bar{q} \to  gg$ and the   tree-level squared amplitude for  the process
$q\bar{q} \to ggg$.  For example, the two-particle Cutkosky cuts of
the two-loop PT self-energy, \Figref{TWL8},
are related to the propagator-like part of the
PT-rearranged one-loop squared amplitude for $q\bar{q} \to gg$,  \Figref{TWL7},
while, at the same time, the three-particle Cutkosky  cuts of the same  quantity, 
\Figref{TWL10},
are related to the $s$-channel part of the 
PT-rearranged tree-level squared amplitude for $q\bar{q} \to
ggg$ (\Figref{TWL9}).  The same holds for vertex- and box-like contributions.
The aforementioned PT-rearranged squared amplitudes
are to be constructed in a way exactly
analogous to the PT-rearranged tree-level $q\bar{q} \to gg$
squared amplitude. 
\newline
\indent
In order to get a firmer understanding of the above statements, 
let  us   consider  the OT for the forward  scattering   process  
$q(p_1)  {\bar q}(p_2)\rightarrow q(p_1){\bar q}(p_2)$
at two loops. We have that 
\bea
\Im m 
\langle q\bar{q}| T^{[6]} | q\bar{q} \rangle &=&
\frac{1}{2} \left(\frac{1}{3!}\right)\int_{\mathrm{PS}_{3g}}\!
{\langle ggg | T^{[2]} |q\bar{q} \rangle}^{*}
\langle ggg | T^{[2]} |q\bar{q} \rangle
\nonumber\\
&+& \frac{1}{2} \left(\frac{1}{2!}\right)\int_{\mathrm{PS}_{2g}}\!
2 \,\Re e \left( {\langle gg | T^{[4]} |q\bar{q} \rangle}^{*}
\langle gg | T^{[2]} |q\bar{q} \rangle \right),
\label{abq}
\eea
where 
$\int_{\mathrm{PS}_{2g}}$ and $\int_{\mathrm{PS}_{3g}}$ stand for the two- and three-body
phase-space for massless gluons, respectively.
The superscripts $[n]$ denotes the order of the corresponding 
amplitude in the coupling constant $g$; when counting powers of 
$g$ remember the couplings from hooking up the external particles.
The lhs of (\ref{abq})
can be obtained by carrying out all possible 
Cutkosky cuts on the two-loop diagrams contributing to the 
process  $q(p_1)  {\bar q}(p_2)\rightarrow q(p_1){\bar q}(p_2)$. 
Specifically, one must sum the contributions of both 
two-particle and three-particle cuts; the cuts involve 
gluons and  ghosts.   
Equivalently, on the rhs one has 
intermediate physical process involving two and three gluons 
[$q\bar{q}\to gg$ and $q\bar{q} \to ggg$]; as usual 
one averages over the initial state polarizations and sums 
over the final state polarizations.
\newline
\indent
Let us now assume that the amplitude 
$q(p_1)  {\bar q}(p_2)\rightarrow q(p_1){\bar q}(p_2)$
has been cast in its PT form, as described in the previous subsection. 
That means that the 
subamplitudes ${\widehat T}^{[6]}_{\ell}$ have been 
constructed, where 
$\ell=1,2,3$ denotes (as in the one-loop case) the 
propagator-like subamplitudes ($\ell=1$), the vertex-like subamplitudes ($\ell=2$), 
and the  box-like ones ($\ell=3$). Thus ${\widehat T}^{[6]}_{1}$
is composed of the two-loop PT gluon self-energy 
${\widehat\Pi}^{(2)}_{\alpha\beta}(q)$, 
${\widehat T}^{[6]}_{2}$ of the two-loop PT vertex 
$\widehat{\Gamma}_{\alpha}^{(2)}(p_1,p_2)$ (and its mirror), 
and ${\widehat T}^{[6]}_{3}$ of the two-loop PT box 
(which coincides with the regular two-loop box in the Feynman gauge).
\newline
\indent
Then, one can show that 
\bea
\Im m 
\langle q\bar{q}|\widehat{T}^{[6]}_{\ell} | q\bar{q} \rangle &=&
\frac{1}{2} \left(\frac{1}{3!}\right) \int_{\mathrm{PS}_{3g}}\,
[{\langle ggg | \widehat{T}^{[2]} |q\bar{q} \rangle}^{*}
\langle ggg | \widehat{T}^{[2]} |q\bar{q} \rangle]_{\ell}
\nonumber\\
&+& \frac{1}{2} \left(\frac{1}{2!}\right)\int_{\mathrm{PS}_{2g}}\,
2 \,\Re e \left( {\langle gg | \widehat{T}^{[4]} |q\bar{q} \rangle}^{*}
\langle gg | \widehat{T}^{[2]} |q\bar{q} \rangle \right)_{\ell} \, .
\label{abqs}
\eea
\newline
\indent
\begin{figure}[!t]
\bce
\includegraphics[width=15cm]{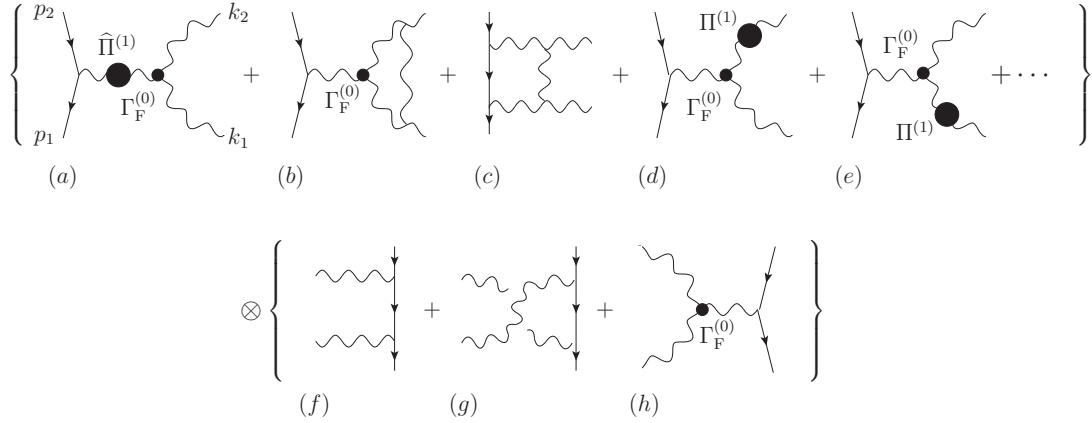}
\ece
\caption{\figlab{TWL7}
The product of the PT-rearranged amplitudes 
of the process $q\bar{q}\to q\bar{q}$ at one-loop (up) and tree-level (down). 
The longitudinal momenta from the polarization tensors will produce 
additional cancellations between $s$-channel and $t$-channel graphs
furnishing finally the first term on the rhs of Eq.~(\ref{abq}).}
\end{figure}
Lets us now see how one constructs the 
``PT-rearranged'' squared amplitudes appearing 
on the rhs of (\ref{abqs}); 
the construction is identical to what we did in the one-loop 
case, but is sufficiently more complicated at the technical level 
to merit some additional clarifying remarks.
\newline
\indent
A  PT-rearranged squared amplitude means the following.
Consider a normal squared amplitude, i.e. the product of two 
regular amplitudes [remember that 
``product'' means that they are also connected (multiplied) 
by the corresponding polarization 
tensors]. Then each amplitude must be first cast into its PT form; 
this is done following simply the PT rules for 
converting a on-shell amplitude into its PT form.
However, this is not the end of the story as far as 
the  PT-rearrangement of the square is concerned. 
One must go through the additional exercise  
of letting the longitudinal momenta coming from the 
polarization vectors trigger the $s$-$t$ cancellation 
at that order. That will identify the genuine propagator-like 
piece of the entire product. 
\begin{figure}[!t]
\bce
\includegraphics[width=12cm]{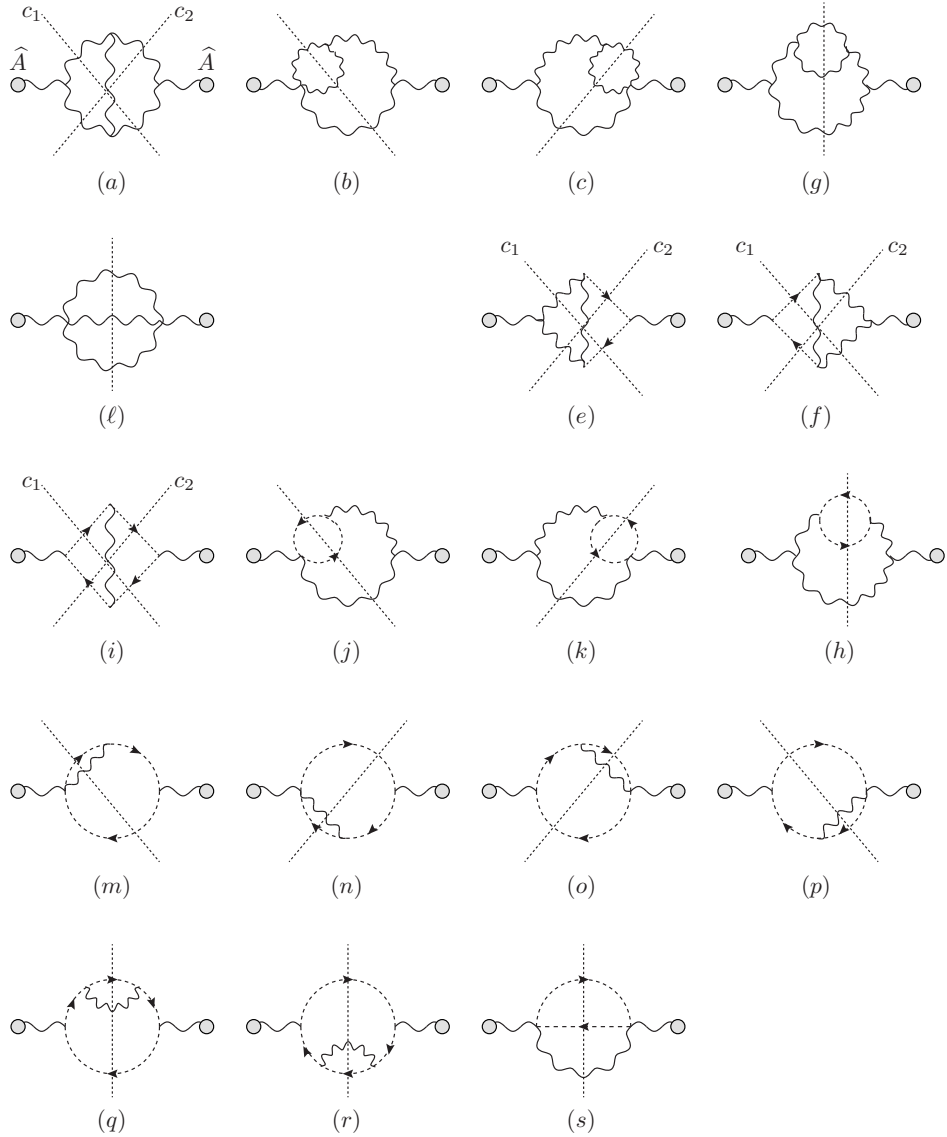}
\ece
\caption{\figlab{TWL10}The three-particle Cutkosky cuts of the PT two-loop
 gluon self-energy. 
The first five graphs have three-gluon cuts, 
the next two have two-gluon--one-ghost cuts,
while the remaining ones have one-gluon-two-ghost cuts.}
\end{figure}
\newline
\indent
The reader should recognize 
that this is exactly the procedure followed in the 
one-loop absorptive construction of \secref{QCD_one-loop} for 
arriving at the $\widehat{{\mathcal M}}_{\ell}$. First,  
the tree-level amplitude ${\mathcal T}_s$ was converted  
into ${{\mathcal T}_{s}}^{F}$; this is the conversion of 
 ${\mathcal T}_s$ into its  PT form. This was done  
simply by carrying out the PT decomposition of (\ref{decomp}) 
to the only available 
three-gluon vertex; note that 
 this vertex has a preferred momentum ($q$)
(employing the terminology introduce in this chapter, it is ``external''), 
because the momenta of the other two legs are 
integrated over all available phase-space.  
Then, the longitudinal momenta from the polarization tensors were 
used to trigger the $s$-$t$ cancellation, which, in turn, 
furnished the three $\widehat{{\mathcal M}}_{\ell}$.
\begin{figure}[!t]
\bce
\includegraphics[width=12cm]{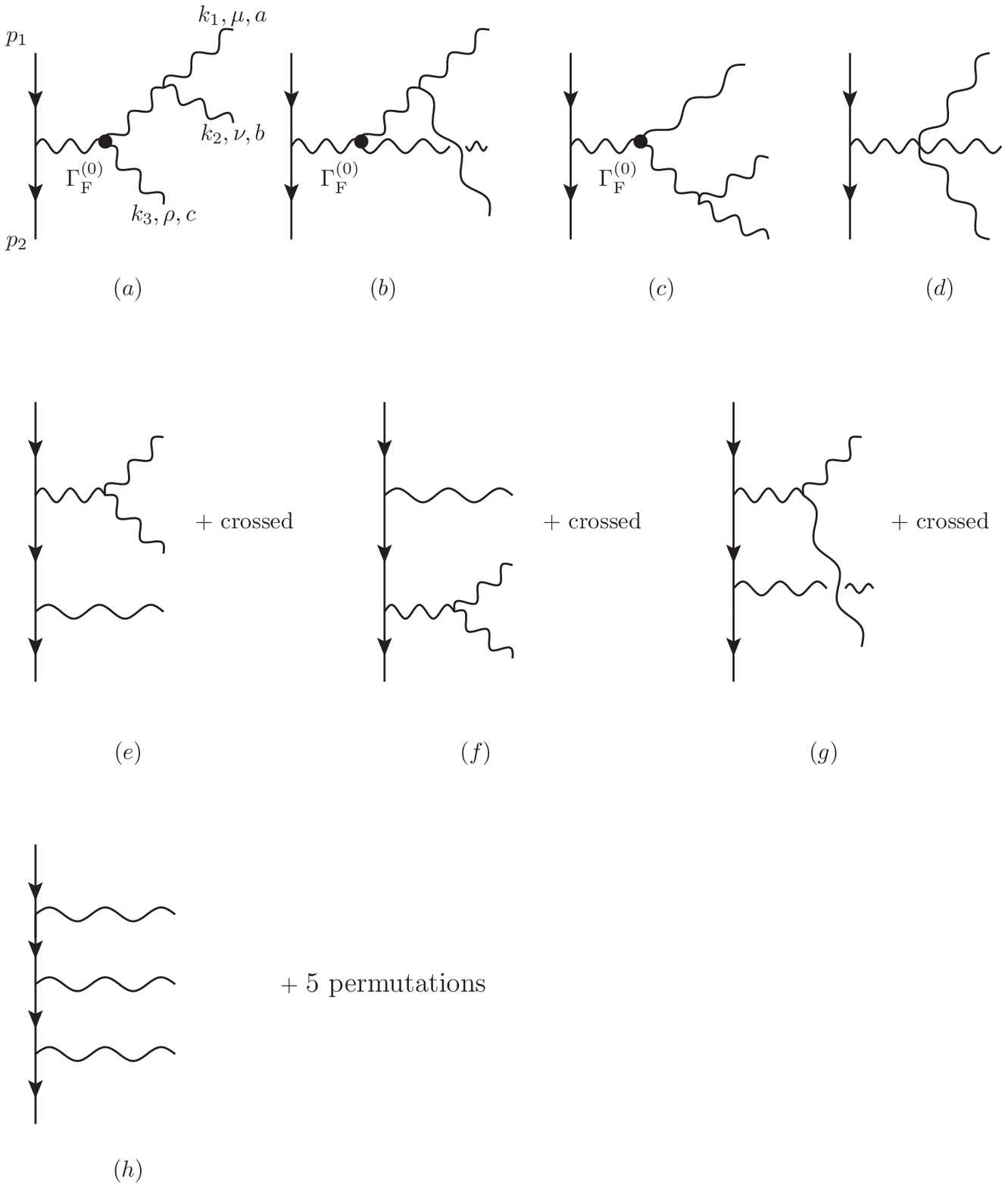}
\ece
\caption{\figlab{TWL9}The tree-level graphs contributing to the process
$q(p_1) \bar{q}(p_2) \to g(k_1) g(k_2) g(k_3)$, after the PT rearrangement.
The momenta $k_i$ are to be integrated over in the three-body phase-space integral.}
\end{figure}
\newline
\indent
For the two-loop case at hand
this procedure must be applied for both squared amplitudes appearing 
on the  rhs of (\ref{abqs}). 
For ${\langle ggg | \widehat{T}^{[2]} |q\bar{q} \rangle}^{*}
\langle ggg | \widehat{T}^{[2]} |q\bar{q} \rangle$ 
we cast the tree-level 
process $q(p_1)\bar{q}(p_2)\to g(k_1)g(k_2)g(k_3)$
into its PT form. This means to carry out 
(\ref{decomp}) only to the three-gluon vertices shown in 
\Figref{TWL9}; all other vertices are internal, because 
they are irrigated by momenta (the $k_i$) that are being integrated over 
in $\int_{\mathrm{PS}_{3g}}$. It turns out that 
the $\Gamma^{P}$ parts cancel completely, as expected; this 
takes some algebra to demonstrate. Then, the 
longitudinal momenta ($k_1^{\mu}$, $k_1^{\mu'}$, $k_2^{\nu}$, $k_2^{\nu'}$, $k_3^{\rho}$, $k_3^{\rho'}$) 
contained in the three polarization tensors will act, giving rise to the $s$-$t$ cancellations 
corresponding to this particular process; its realization is 
operationally more complicated, but otherwise completely analogous 
to the one showed in \secref{QCD_one-loop}.  
For ${\langle gg | \widehat{T}^{[4]} |q\bar{q} \rangle}^{*}
\langle gg | \widehat{T}^{[2]} |q\bar{q} \rangle$ 
things are more involved. 
Now the amplitude $q(p_1)\bar{q}(p_2)\to g(k_1)g(k_2)$
must be cast into its PT form both at tree-level [for  
$\langle gg | \widehat{T}^{[2]} |q\bar{q} \rangle$], 
and at one-loop  
[for $\langle gg | \widehat{T}^{[4]} |q\bar{q} \rangle^{*}$]. 
The tree-level PT amplitude is, of course, known from the absorptive construction of 
subsection \ref{abspt}. 
The construction of the one-loop PT  amplitude  $q(p_1)\bar{q}(p_2)\to g(k_1)g(k_2)$ 
has been described in great detail 
in the literature~\cite{Watson:1994tn,Binosi:2002ez}.
In fact, as mentioned in subsection 2.4.3, this process 
was the first explicit example~\cite{Watson:1994tn} 
where the universality (process-independence) 
of the PT gluon self-energy was explicitly demonstrated; the end result is shown in \Figref{TWL7}. 
As we see there, the procedure of the PT rearrangement 
leads to the appearance of $\widehat{\Pi}(q)$ and  
the conversion of the conventional one-loop three-gluon vertex  
 $\Gamma_{\,\alpha\mu\nu}^{(1)}(q,k_1,k_2)$
into $\Gamma_{F\,\alpha\mu\nu}^{(1)}(q,k_1,k_2)$, which
is the BFG  three-gluon vertex at one-loop, 
with one background gluon ($q$) and two quantum ones ($k_1$, $k_2$). 
It is straightforward to show that 
$\Gamma_{F\,\alpha\mu\nu}^{(1)}(q,k_1,k_2)$
satisfies the following WI
\be
q^{\alpha} \Gamma_{F\,\alpha\mu\nu}^{(1)}(q,k_1,k_2)
= \Pi_{\mu\nu}^{(1)}(k_1) - \Pi_{\mu\nu}^{(1)}(k_2) , 
\label{wigf1}
\ee
which is the exact one-loop analogue of 
the tree-level WI of Eq.~(\ref{WI2B}); indeed the rhs is the
difference of two one-loop self-energies computed in the 
conventional Feynman gauge.
Then, one must let the longitudinal momenta in the 
polarization tensors trigger the corresponding $s$-$t$
cancellation. When they hit on 
$\langle gg | \widehat{T}^{[2]} |q\bar{q} \rangle$ they will simply 
trigger the prototype tree-level $s$-$t$ cancellation of \secref{QCD_one-loop}; but 
when they hit on $\langle gg | \widehat{T}^{[4]} |q\bar{q} \rangle^{*}$
they will trigger the {\it one-loop} version of the 
same $s$-$t$ cancellation~\cite{Papavassiliou:1999az,Papavassiliou:1999bb}.
To demonstrate this latter cancellation one needs 
to know the result of the action of the longitudinal momenta $k_1^{\mu}$ and $k_2^{\nu}$
on $\Gamma_{F\,\alpha\mu\nu}^{(1)}(q,k_1,k_2)$, exactly as happened 
in the tree-level construction, see Eq.(\ref{WIGFLR-2}).
Given that these momenta correspond to quantum fields 
(as opposed to the $q^{\alpha}$, which corresponds to a background field) 
the result of their contraction with $\Gamma_{F\,\alpha\mu\nu}^{(1)}(q,k_1,k_2)$ 
is not a WI, as in (\ref{wigf1}), but rather an STI, which is given in Eq.(\ref{STI:mixed1}).
\newline
\indent
The reader should  appreciate at this point that 
any rearrangement of
the (internal) vertices of two-loop box-diagrams 
cannot be  reconciled   with  the   arguments   of  the
simultaneous two-  and    three-particle Cutkosky cuts  presented here. 
To see this with an example, 
let us return to the two representative two-loop diagrams of \Figref{TWL0}.
After their PT rearrangement, the two-particle Cutkosky cut on $(a)$, denoted by $(c_2)$ in \Figref{TWL0}, 
must reproduce 
$(c)\otimes [(f)+(g)]$ in  \Figref{TWL7}, and  cut $(c_4)$ on $(b)$ must reproduce  $(b)\otimes [(f)+(g)]$. 
Obviously, if we were to modify the internal three-gluon vertices of $(a)$ or $(b)$ in \Figref{TWL0}
in any way, this identification would not work: one must modify {\it only} the vertex injected with $q$ 
(turning it to $\Gamma^{{\rm F}}$).
This argument may be generalized to include all remaining two-particle and   
three-particle cuts, making  the above conclusion completely airtight.
\newline
\indent
We emphasize that the arguments presented here do not
postulate at any point the existence of any relation between the PT and the 
BFM. On the other hand, in hindsight, all conclusions drawn 
are in complete agreement with the known PT/BFM correspondence.
Specifically, switching now to the BFM language, 
the fact that internal vertices should not be touched is precisely what the BFM Feynman rules dictate:
since one cannot have background fields propagating inside loops, all internal vertices 
have three {\it quantum} gluons merging. 

Let us finally comment on an additional point. 
The philosophy of this subsection has been to establish that the two-loop 
PT Green's functions, constructed explicitly in the previous subsection, satisfy the 
strong version of the OT. 
Evidently, one could adopt an alternative approach, 
and consider the absorptive formulation as the defining construction for 
the PT Green's functions. Thus, one can start from the rhs of the OT,   
carry out the same analysis that we did here, and 
postulate  that the sum of the  propagator-like parts of the rhs furnish the imaginary part 
of the (now unknown) two-loop gluon self-energy  on the lhs, and similarly for vertices and boxes. 
Then, the real parts would have to be reconstructed from the 
corresponding dispersion relation, in the spirit of subsection \ref{abspt}.
The advantage of this strategy is that 
all the PT-rearranged (squared) amplitudes appearing on the rhs are at least one loop lower 
than the amplitude on the lhs. Therefore, one can actually reconstruct  
the lhs, by working directly on the rhs, 
because one knows already how to pinch at lower orders. 
Of course,  the downside is that, in order to obtain the final answer,   
one would have  to carry out complicated phase-space and dispersion integrations. 
\begin{figure}[t]
\bce
\includegraphics[width=12cm]{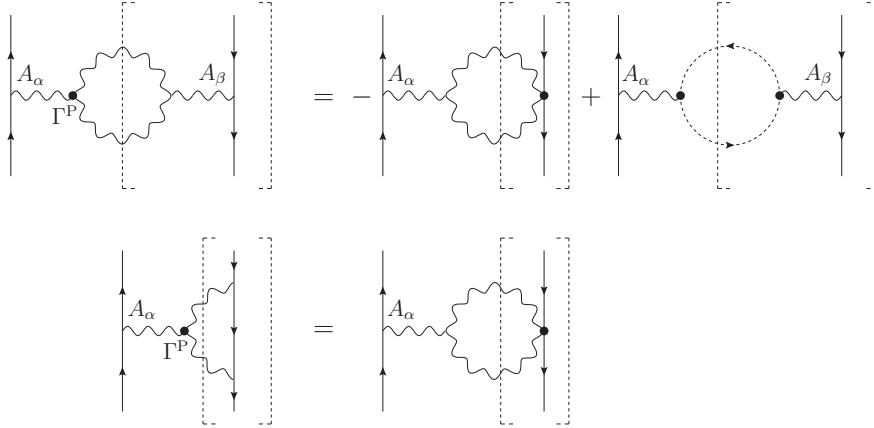}
\ece
\caption{\figlab{1l_fund_canc} The one-loop PT algorithm seen in terms of the fundamental $s$-$t$ cancellation. The self-energy like contribution coming from the vertex is exactly canceled by the one coming from the propagator. Notice that none of the vertices appearing in the rhs of the above diagrammatic identities is an elementary vertex of the theory, they are all {\it unphysical} vertices.}
\end{figure}

\subsection{The PT to all orders in perturbation theory}

\noindent
The two-loop construction of the PT quark-gluon vertex and gluon propagator 
outlined in the previous subsections follows 
to the letter the one-loop recipe: we have tracked down the 
rearrangements induced to individual Feynman diagrams 
when the longitudinal pinching momenta 
acting on their bare vertices trigger 
elementary WIs. Clearly such a method does not lend itself 
to an all-order generalization; realistically speaking, 
the two-loop construction is as far as one can push this diagrammatic procedure.
\newline
\indent
In order to generalize the PT to higher orders one must 
find an expeditious way of enforcing the crucial cancellations 
without manipulating individual diagrams, but, instead, through the collective  treatment 
of entire sets of diagrams.
It turns out that such a breakthrough is indeed possible~\cite{Binosi:2002ft,Binosi:2003rr}:
the  PT algorithm can  be successfully generalized to  {\it all orders}  in perturbation theory 
through the judicious use  of the STI satisfied  by a  special Green's
function,  which  serves  as  a  common kernel  to  all  higher  order
self-energy and vertex diagrams.
\newline
\indent
In fact, as was first realized in ~\cite{Binosi:2002ft,Binosi:2003rr}, 
the one- and two-loop PT rearrangements are but lower-order manifestations 
of a  fundamental
cancellation, taking place between  graphs of distinct kinematic nature
when computing the divergence  of the four-point function 
$AAq\bar q$ (with the gluons off-shell, and the quarks on-shell).
At tree-level, this is nothing else than 
the ``$s$-$t$ cancellation'' that we  
already encountered in the absorptive PT construction of 
\secref{QCD_one-loop} and~\ref{sec:SM_one-loop}. 
As for the one-loop PT rearrangement, it is actually encoded in the two graphs of \Figref{1l_fund_canc}: 
both graphs have the $\Gamma^\mathrm{P}$ terms in common, while their terms shown 
in dotted brackets are the tree-level $t$- and $s$-channel contributions, respectively, 
to the four-particle amplitude $AA\to q\bar q$. 
Dressing the above amplitude with higher order corrections, 
and exploiting its STI, will eventually provide a compact way of generalizing the PT to all orders. 

\subsubsection{The four-point kernel $AAq\bar q$ and its Slavnov-Taylor identity}

\begin{figure}[t]
\bce
\includegraphics[width=7cm]{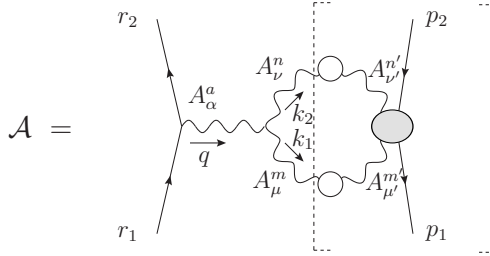}
\ece
\caption{\figlab{fund_ampl} The fundamental amplitude receiving the action of the longitudinal momenta stemming from $\Gamma^\mathrm{P}$. The gray blob represent the (connected) kernel corresponding to the process $AA\to q\bar q$.}
\end{figure}
\noindent
With this intention in mind, of all the diagrams contributing to the QCD amplitude under consideration 
we will focus on  the  subset that will receive  the action of
the     longitudinal    momenta     stemming     from    $\Gamma^\mathrm
{P}$, to be denoted by ${\mathcal A}$. It is given by (see \Figref{fund_ampl})
\bea
{\mathcal A}(r_1,-r_2,p_2,-p_1)&=&\bar u(r_2)g\gamma_\alpha t^au(r_1)\frac{-i}{q^2}\int_{k_1}i\Gamma^{am'n'}_{\alpha\mu'\nu'}(q,-k_1,-k_2){\mathcal T}^{\mu\nu}_{mn}(k_1,k_2,p_2,-p_1),\nonumber\\
{\mathcal T}^{\mu\nu}_{mn}(k_1,k_2,p_2,-p_1)&=&\bar u(p_1)\left[i\Delta_{mm'}^{\mu\mu'}(k_1)i\Delta_{nn}^{\nu\nu'}(k_2){\mathcal C}_{\mu'\nu'}^{m'n'}(k_1,k_2,p_2,-p_1)\right]u(p_2).
\label{T_ampl}
\eea
From the definition above we clearly see that ${\mathcal T}$ 
is the subamplitude $AA\to q\bar q$ with the gluon off-shell and the fermions on-shell. 
The next step is to carry out the usual PT vertex decomposition on the three-gluon vertex $\Gamma^{am'n'}_{\alpha\mu'\nu'}$
appearing inside the integrand on the rhs of (\ref{T_ampl}), and concentrate on the action of its $\Gamma^\mathrm{P}$ part 
on the amplitude ${\mathcal T}$. Therefore, we need to determine  the STI satisfied by the amplitude ${\mathcal T}$ of Eq.~(\ref{T_ampl}).
\newline
\indent
In order to get this STI, one starts from the trivial identity in configuration space (valid due to ghost charge conservation)
\be
\left\langle T[\bar c^m(x) A^n_\nu(y)q(z)\bar q(w)]\right\rangle=0,
\ee
where $T$ denotes the time-ordered product of fields. By rewriting the fields in terms of their BRST transformed counterparts, using the equations of motions and the equal time commutation relations~\cite{Pascual:1984zb}, and, finally, Fourier transforming the result to momentum space, one arrives at the following identity
\be
k_1^\mu C_{\mu\nu}^{mn}=k_{2\nu}G_1^{mn}
-igf^{nrs} Q_{1\nu}^{mrs}+gX_{1\nu}^{mn}+
g\bar X_{1\nu}^{mn}.
\label{BasSTI}
\ee
\begin{figure}
\bce
\includegraphics[width=13.5cm]{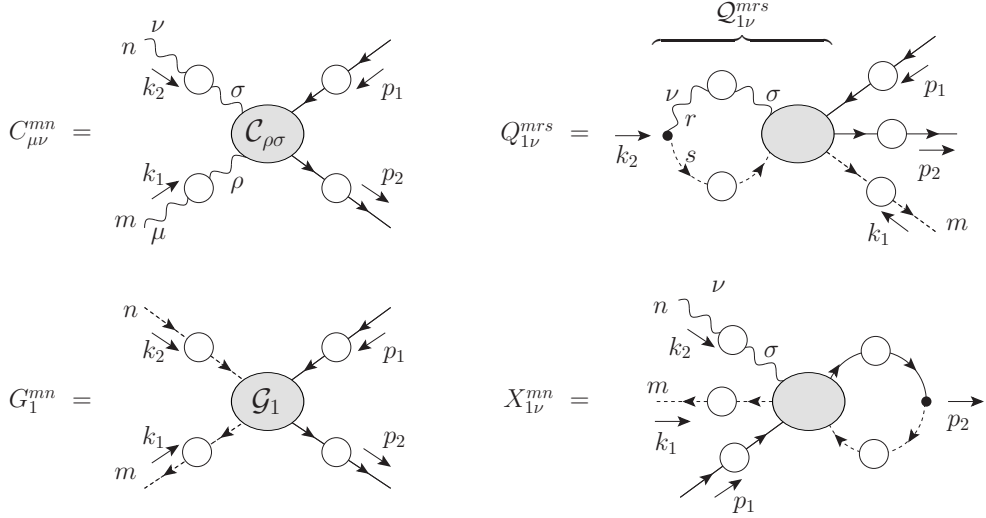}
\ece
\caption{\figlab{defaf}Diagrammatic representation of the GreenÕs functions appearing in the 
Slavnov-Taylor identity (\ref{BasSTI}). The function $\bar X_{1\nu}^{mn}$ (which coincides with  $X_{1\nu}^{mn}$ when the fermion arrows reversed) is not shown.}
\end{figure}
The various Green's functions that appears in the rhs of the equation above are shown in \Figref{defaf}.
The  terms
$X_{1\nu}$ and $\bar  X_{1\nu}$    
correspond  to contributions  that vanish when the external quarks are assumed to be  on-shell, 
since  they are  missing one full  quark propagator;  at  lowest   order  they  are  simply  the  terms
proportional  to  the  inverse  tree-level propagators  $(\psm_1-m)$  and
\mbox{$(\psm_2-m)$} appearing  in the PT calculations. Indeed,  if we multiply
both sides  of Eq.~(\ref{BasSTI}) by  the product $S^{-1}(p_2)S^{-1}(-p_1)$
of  the two  inverse propagators  of the  external quarks,  and then
sandwich the  resulting amplitude  between the on-shell  spinors $\bar
u(p_1)$ and $u(p_2)$,  the vanishing of the aforementioned
terms  follows by virtue  of the (all-order) Dirac equation. Thus we arrive at the on-shell STI
\bea
k_1^\mu {\mathcal T}_{\mu\nu}^{mn}(k_1,k_2,p_2,-p_1)&=&{\mathcal S}^{mn}_{1\nu}(k_1,k_2,p_2,-p_1),
\label{onshSTI_k1}\nonumber\\
{\mathcal S}^{mn}_{1\nu}(k_1,k_2,p_2,-p_1)&=&
\bar  u(p_2)\left[
k_{2\nu} iD^{mm'}(k_1)iD^{nn'}(k_2){\mathcal
G}_1^{m'n'}(k_1,k_2,p_2,-p_1)\right.\nonumber\\
&-&\left.igf^{nrs} iD^{mm'}(k_1){\mathcal
Q}_{1\nu}^{m'rs}(k_1,k_2,p_2,-p_1)\right]u(p_2),
\label{onshdef_k1}
\eea
where the auxiliary function ${\mathcal Q}_{1\nu}^{m'rs}$ is shown in \Figref{Q1_Q2_defs}. A similar (Bose-symmetric) relation will hold when hitting the kernel with the $k_2$ momentum.
\begin{figure}[t]
\bce
\includegraphics[width=11cm]{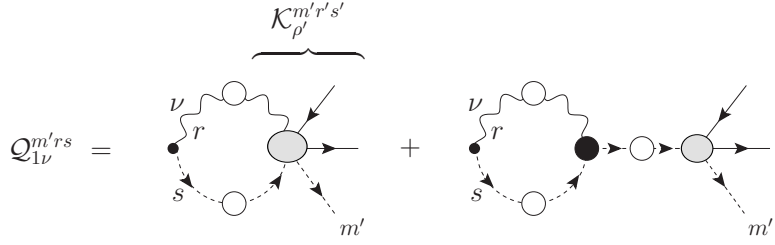}
\ece
\caption{\figlab{Q1_Q2_defs} Definition of the momentum space auxiliary function ${\mathcal Q}_{1\nu}^{m'rs}$. Gray blobs represent connected kernels, while black (respectively white) blobs represent 1PI (respectively connected) $n$-point Green's functions. 
Notice that the kernel ${\cal K}^{m'r's'}_{\rho'}$ is 1PI with respect to $s$-channel cuts.}
\end{figure}

\subsubsection{The fundamental all-order $s$-$t$ cancellation}
\noindent
Having established the STIs of Eq.~(\ref{onshSTI_k1}),  we now turn to
the main conceptual point related  to the all orders PT construction. 
The basic observation is
the   following.   In  perturbation   theory  the   quantities  ${\mathcal
T}_{\mu\nu}^{mn}$, ${\mathcal S}_{1\nu}^{mn}$, and
${\mathcal S}_{2\mu}^{mn}$ are given by Feynman  diagrams,
which can be separated
into  distinct classes,  depending on  their kinematic  dependence ($s$/$t$ separation) and
their geometrical properties (1PI/1PR separation).  
We recall that $s$ graphs are those which do not contain information
about  the  kinematic  details   of  the  incoming  test-quarks 
(self-energy graphs),  whereas $t$ graphs are those that  depend also on the
mass of the test quarks (vertex graphs).  
Thus,  the Feynman  graphs allow  the following
separation (see also \Figref{s_t_I_R_decomp}, where we show the decomposition below at an arbitrary perturbative order $n$)
\bea
{\mathcal T}_{\mu\nu}^{mn}&=&[{\mathcal
T}_{\mu\nu}^{mn}]_{s,\scriptscriptstyle{\rm I}}+ 
[{\mathcal
T}_{\mu\nu}^{mn}]_{s,\scriptscriptstyle{\rm R}}+
[{\mathcal
T}_{\mu\nu}^{mn}]_{t,\scriptscriptstyle{\rm I}}+ 
[{\mathcal
T}_{\mu\nu}^{mn}]_{t,\scriptscriptstyle{\rm R}}, 
\label{stsa-1}\nonumber\\
{\mathcal S}_{1\nu}^{mn}&=&[{\mathcal
S}_{1\nu}^{mn}]_{s,\scriptscriptstyle{\rm I}}+ 
[{\mathcal
S}_{1\nu}^{mn}]_{s,\scriptscriptstyle{\rm R}}+[{\mathcal
S}_{1\nu}^{mn}]_{t,\scriptscriptstyle{\rm I}}+ 
[{\mathcal
S}_{1\nu}^{mn}]_{t,\scriptscriptstyle{\rm R}},
\label{stsa-2}
\eea
\indent
On the other hand, we know by now that the action of the (longitudinal) momenta 
$k_1^\mu$ or $k_2^\nu$ on 
${\mathcal T}_{\mu\nu}^{mn}$ does not respect, in general,
the above separation;
their action  mixes propagator- with vertex-like terms, since the
latter generate (through pinching) effectively 
propagator-like terms. Similarly, the separation between 1PI and 1PR terms is no longer
unambiguous, since  1PR diagrams are 
converted (again through pinching) into effectively 1PI ones
(the opposite cannot happen).
An early example of this effect 
appeared in the construction of the PT three-gluon vertex in \secref{QCD_one-loop}, 
and was encountered  again in subsection \ref{1prg}.
\newline
\indent
Therefore, even though Eq~(\ref{onshSTI_k1}) holds for the entire amplitude 
${\mathcal T}_{\mu\nu}^{mn}$, 
it is not true for the individual subamplitudes
defined in  Eqs~(\ref{stsa-1}), \ie we have 
\be
k^\mu_1 [{\mathcal T}^{mn}_{\mu\nu}]_{x,\scriptscriptstyle{\rm Y}} \neq  
[{\mathcal S}_{1\nu}^{mn}]_{x,\scriptscriptstyle{\rm
Y}},  \qquad x=s,t;\ {\rm Y=I,R},
\label{neq-1}
\ee
where $I$ (respectively $R$) indicates the one-particle {\it
irreducible} (respectively {\it reducible}) parts of the amplitude
involved. The reason  for  this  inequality are  precisely  the
propagator-like  terms, such  as  those found  in  the one-  and
two-loop calculations (see \Figref{2l_R_terms}).
One should notice that the situation is exactly analogous 
to that encountered when studying the OT,  
which holds at the level of the full amplitude but not at the level of individual subamplitudes.
\begin{figure}[t]
\bce
\includegraphics[width=11cm]{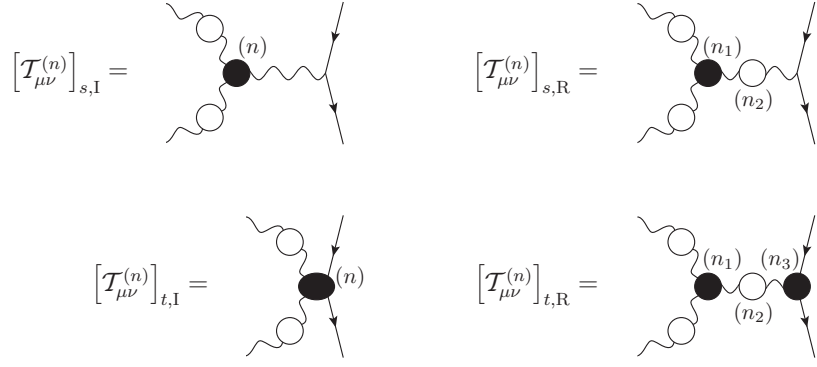}
\ece
\caption{\figlab{s_t_I_R_decomp} Decomposition at an arbitrary perturbative level $n$ of the fundamental amplitude ${\mathcal T}^{mn}_{\mu\nu}$ in terms of $s$ and $t$ channel, 1PI and 1PR components.}
\end{figure}
\newline
\indent
In particular, for individual subamplitudes we have that
\bea
k_1^\mu[{\mathcal T}^{mn}_{\mu\nu}]_{s,\scriptscriptstyle{\rm R}}
&=&[{\mathcal S}^{mn}_{1\nu}]_{s,\scriptscriptstyle{\rm R}} 
+[{\mathcal R}^{mn}_{1\nu}]_{s,\scriptscriptstyle{\rm
I}}^{\rm int},
\label{xYres-1}\nonumber\\
k_1^\mu[{\mathcal T}^{mn}_{\mu\nu}]_{s,\scriptscriptstyle{\rm I}}
&=& [{\mathcal S}^{mn}_{1\nu}]_{s,\scriptscriptstyle{\rm I}} 
-[{\mathcal R}^{mn}_{1\nu}]_{s,\scriptscriptstyle{\rm I}}^{\rm int}
+[{\mathcal R}^{mn}_{1\nu}]_{s,\scriptscriptstyle{\rm I}}^{\rm ext}, 
\label{xYres-2}\nonumber\\
k_1^\mu[{\mathcal T}^{mn}_{\mu\nu}]_{t,\scriptscriptstyle{\rm R}}
&=& [{\mathcal S}^{mn}_{1\nu}]_{t,\scriptscriptstyle{\rm R}} 
+[{\mathcal R}^{mn}_{1\nu}]_{t,\scriptscriptstyle{\rm I}}^{\rm int},
\label{xYres-3}\nonumber\\
k_1^\mu[{\mathcal T}^{mn}_{\mu\nu}]_{t,\scriptscriptstyle{\rm I}}
&=& [{\mathcal S}^{mn}_{1\nu}]_{t,\scriptscriptstyle{\rm I}} 
-[{\mathcal R}^{mn}_{1\nu}]_{t,\scriptscriptstyle{\rm I}}^{\rm int}
-[{\mathcal R}^{mn}_{1\nu}]_{s,\scriptscriptstyle{\rm I}}^{\rm ext},
\label{xYres-4}
\eea
with similar equations holding when acting with the momentum $k_2^\nu$. In the
above equations the superscript ``ext'' and ``int'' stands for
``external'' and ``internal'' respectively, and refers to whether or not
the pinching part of the diagram at hand has touched the external
on-shell fermion legs. At order $n$, some example of the ${\mathcal R}^{(n)}$ terms
introduced in the equations above are shown in \Figref{nl_R_terms}.
The structure of the $[{\mathcal
R}^{mn}_{1\nu}]_{s,{\rm Y}}$ terms is very characteristic:
kinematically they only depend on $s$; whereas this is obviously true
for the first two equations of (\ref{xYres-1})
 (since these terms originate from the action of 
$k_1^\mu$ on an $s$-dependent piece $[{\mathcal
T}^{mn}_{\mu\nu}]_{s,{\rm Y}}$), it is 
far less obvious for those appearing in the last two equations of ~(\ref{xYres-3}), 
since they stem from the action of 
$k_1^\mu$ on a $t$-dependent term $[{\mathcal
T}^{mn}_{\mu\nu}]_{t,{\rm Y}}$. 
In addition, all the ${\mathcal R}$ terms share a common feature, \ie they  
{\it cannot} be written in terms of conventional
Feynman rules; instead they are built out of unphysical vertices,  
which do not correspond to any term in the original
Lagrangian. It should be clear by now that the ${\mathcal R}$ terms are precisely 
the terms that cancel against each other when we carry out the PT procedure {\it diagrammatically}.
\begin{figure}[t]
\bce
\includegraphics[width=16cm]{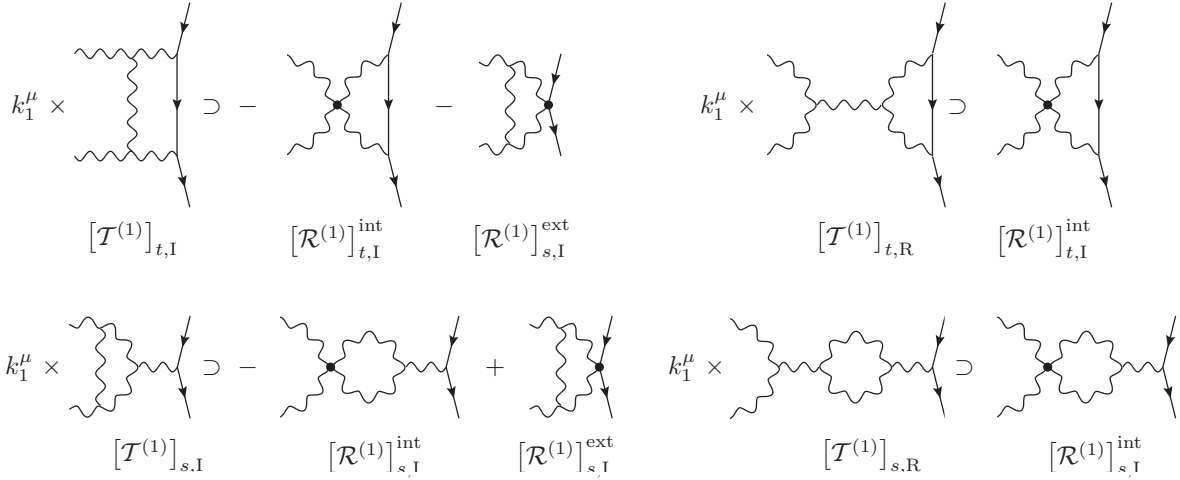}	
\ece
\caption{\figlab{2l_R_terms} Some two-loop examples of the ${\mathcal R}$ terms,
  together with the Feynman diagrams from which they originate.}
\end{figure}
\newline
\indent
Thus, after the PT cancellations have been enforced, we find that 
for the 1PI $t$-channel part of the amplitude we have the equality 
\be
[k^\mu_1 {\mathcal T}^{mn}_{\mu\nu}]^{\scriptscriptstyle{\rm PT}}_{t,\scriptscriptstyle{\rm I}} =  
[{\mathcal S}_{1\nu}^{mn}]_{t,\scriptscriptstyle{\rm
I}},  \label{eq-1} 
\ee
What is crucial in the above result is that it automatically
takes care of both the  $s$-$t$ as well as the 1PR and 1PI
cancellations of the ${\mathcal R}$ terms, which is  characteristic
of the PT, without having to actually trace them down. 
Thus, on hindsight, 
the PT algorithm as applied \eg in \secref{QCD_one-loop} and~\ref{sec:SM_one-loop}, 
and even in the two-loop generalization carried out in the initial part of this section,
has amounted to enforcing diagram-by-diagram the 
vast, BRST-driven $s$-$t$ channel cancellations, without    
making explicit use of the all-order STIs.   
Evidently, tracing down the action of the longitudinal momenta  
and the resulting exchanges  
between vertex and self-energy graphs, 
is equivalent to deriving (loop-by-loop) Eq.~(\ref{eq-1}). 
Therefore,
the non-trivial step for generalizing the PT to all orders is to 
recognize that the result obtained after 
the implementation of the PT algorithm 
on the  lhs of  Eq.~(\ref{eq-1}) is already encoded  
on the rhs!
Indeed, the rhs involves  only conventional (ghost) Green's functions,
expressed  in terms  of normal  Feynman  rules, with  no reference  to
unphysical  vertices. 
\newline
\indent
This last point merits some additional comments.  
It should be clear that no pinching is possible when looking at the
$t$-channel irreducible part of the rhs of Eq.~(\ref{onshSTI_k1}).
So, if we were to enforce  
the  PT cancellations on both sides of the $t$ irreducible part of 
these equations, on the 
rhs there would be nothing to pinch (all the vertices are
internal), and therefore,
there would be no unphysical vertices generated. Therefore,
on the lhs, where pinching is possible, 
all contributions containing unphysical vertices must cancel.
The only way to distort this equality is to violate the PT rules,
allowing, for example, the generation of additional 
longitudinal momenta by carrying out sub-integrations, or 
by decomposing internal vertices. Violating 
these rules, however, will result in undesirable consequences: 
in the first case the appearance of terms of the form 
$q\cdot k$ in the denominators will interfere with the 
manifest analyticity of 
the PT Green's functions constructed, whereas, in the second, the 
special unitarity properties of the  PT Green's functions will 
be inevitably compromised. 

\subsubsection{The PT to all orders: the quark-gluon vertex and the gluon propagator}
\noindent
The considerations just presented can be used in order to accomplish in a rather compact form 
the all-order generalization of the PT construction.
\newline
\indent
To  begin with,  it is  immediate  to recognize  that, in  the \Rxi Feynman gauge (RFG for short),  
box diagrams of arbitrary order $n$,  to be denoted by $B^{(n)}$, coincide
with the PT boxes ${\widehat B}^{(n)}$, since all three-gluon vertices
are  ``internal'',  \ie  they  do  not  provide  longitudinal
momenta. Thus, they coincide  with the BFG boxes, $\tilde{B}^{(n)}$,
{\it  i.e.},
\be
{\widehat B}^{(n)}  = B^{(n)}  =  \tilde{B}^{(n)}
\label{res2:boxes}
\ee
for every~$n$. The same is true for the PT quark self-energies;  
for exactly the same reason, they coincide with their RFG (and BFG)  
counterparts, \ie
\be
{\widehat \Sigma}^{ij \,(n)}  = \Sigma^{ij \,(n)}  =  
\tilde{\Sigma}^{ij \,(n)}. 
\label{res2:sigmas}
\ee
\begin{figure}[t]
\bce
\includegraphics[width=14.5cm]{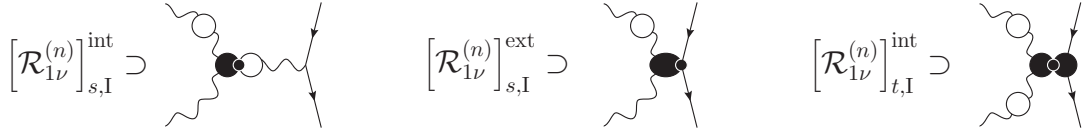}	
\ece
\caption{\figlab{nl_R_terms} Examples of the generic structures included in the ${\mathcal R}$ terms. 
A black dot indicates that a propagator has been removed through pinching.}
\end{figure}
The next step will be  the construction of the all-order PT
quark-gluon 1PI vertex 
$\widehat\Gamma^a_\alpha$. 
Now, of all the diagrams that contribute to this
vertex in the RFG (shown in \Figref{qg_Rxi_vertex}),  
the only one receiving the action of the longitudinal 
pinching momenta is diagram $(a)$; in fact, we know from the two-loop construction that (in the RFG) 
when the external gluon couples to a diagram through 
any type of interaction vertex other than a three-gluon vertex, then that  
diagram is inert, as far as pinching is concerned.  
Thus, we carry out the PT vertex decomposition of Eq.~(\ref{decomp}) in diagram $(a)$, and concentrate on the $\Gamma^\mathrm{P}$ part only; specifically
\be
(a)^\mathrm{P}=gf^{amn}\int_{k_1}\!(g^\nu_\alpha k_1^\mu-g^\mu_\alpha k_2^\nu)\left[{\mathcal T}^{mn}_{\mu\nu}(k_1,k_2,p_2,-p_1)\right]_{t,\mathrm{I}}.
\ee
Following the discussion presented in the previous subsection,
the  pinching   action  amounts  to  the   replacements
\bea
k_1^\mu\left[{\mathcal T}^{mn}_{\mu\nu}\right]_{t,\mathrm{I}}&\to&
\left[k_1^\mu{\mathcal T}^{mn}_{\mu\nu}\right]^\mathrm{PT}_{t,\mathrm{I}}=\left[{\mathcal
S}_{1\nu}^{mn}\right]_{t,{\rm  I}},
\label{r1}\nonumber\\
k_2^\nu\left[{\mathcal T}^{mn}_{\mu\nu}\right]_{t,\mathrm{I}}&\to&
\left[k_2^\nu{\mathcal T}^{mn}_{\mu\nu}\right]^\mathrm{PT}_{t,\mathrm{I}}=\left[{\mathcal
S}_{2\mu}^{mn}\right]_{t,{\rm  I}},
\label{r2}
\eea
or, equivalently,
\be
(a)^\mathrm{P}\to gf^{amn}\int_{k_1}\!\left\{\left[{\mathcal S}^{mn}_{1\alpha}(k_1,k_2)\right]_{t,{\rm  I}}-\left[{\mathcal S}^{mn}_{2\alpha}(k_1,k_2)\right]_{t,{\rm  I}}\right\}.
\ee
In the formula above, and in what follows, we indicate only the momenta of the gluons $k_i$, omitting the momenta of the external fermions $p_i$. 
\newline
\indent
At   this  point  the   construction  of   the  effective   PT  quark-gluon vertex
has been  completed, and we have 
\bea
\widehat\Gamma^a_\alpha(q,p_2,-p_1)&=&
(a)^\mathrm{F}+(b)+(b')+(c)+(c')+(d)\nonumber \\
&+& gf^{amn}\int_{k_1}\!\left\{\left[{\mathcal S}^{mn}_{1\alpha}(k_1,k_2)\right]_{t,{\rm  I}}-\left[{\mathcal S}^{mn}_{2\alpha}(k_1,k_2)\right]_{t,{\rm  I}}\right\}.
\eea
We emphasize that 
in the construction presented thus far we have
never resorted to the BFM   formalism, 
but have only used the BRST identities of Eq.~(\ref{onshSTI_k1}), 
together with~(\ref{eq-1}).
\newline
\indent
The next crucial question will be then to establish the
connection between the effective PT vertex and the
quark-gluon vertex 
$\widetilde\Gamma^a_\alpha$ written in the BFG.
To that end, we start be observing that all the inert terms
contained in the original RFG $\Gamma^a_\alpha$ vertex carry
over to the same sub-groups of BFG graphs.
In order to facilitate this identification  
we recall (see also~\appref{Frules}) that to lowest order one has the identities $\Gamma^\mathrm{F}=\Gamma_{\widehat{A}AA}$ , while $\Gamma_{Aq\bar q}=\Gamma_{\widehat{A}q\bar q}$ and $\Gamma_{AAAA}=\Gamma_{\widehat{A}AAA}$, so that 
\be
(a)^\mathrm{F}=(\widehat{a}), \qquad (b)=(\widehat{b})\quad (b')=(\widehat{b'}), \qquad (d)=(\widehat{d}),
\ee
where a ``hat'' on the corresponding diagram means 
that the (external) gluon $A^a_\alpha$ has been effectively converted into a  background gluon $\widehat{A}^a_\alpha$.
\newline
\indent 
As should be familiar by now, the only exception to this rule are 
the ghost diagrams $(d)$ and $(d')$, which need to be combined with the remaining terms from the PT construction
in order to arrive at (see \Figref{qg_BFM_topo}) both  the {\it  symmetric}
ghost gluon vertex $\Gamma_{\widehat{A}c\bar  c}$, characteristic
of the BFG, as well as at the
four-particle ghost vertex $\Gamma_{\widehat{A} Ac\bar c}$, totally absent in the conventional $R_\xi$  
diagrams.
\begin{figure}[t]
\bce
\includegraphics[width=15.5cm]{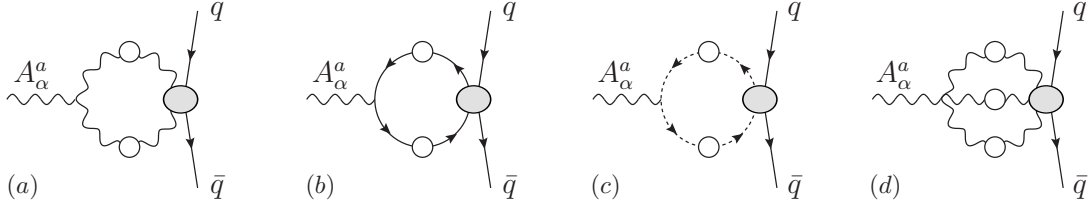}	
\ece
\caption{\figlab{qg_Rxi_vertex} The Feynman diagrams contributing to the quark-gluon vertex $\Gamma^a_\alpha$ in the $R_\xi$ gauge. Kernels appearing in these diagrams are $t$-channel and 1PI. Not shown are the contribution $(b')$ and $(c')$ corresponding to diagrams $(b)$ and $(c)$ with the fermion and ghost arrows reversed.}
\end{figure}
\indent
Indeed, using Eq.~(\ref{onshdef_k1}), we find (omitting the spinors)
\bea
gf^{amn}\int_{k_1}\!\left[{\mathcal S}^{mn}_{1\alpha}(k_1,k_2)\right]_{t,{\rm  I}}&=&-gf^{amn}\int_{k_1}\!k_{2\alpha}D^{mm'}(k_1)D^{nn'}(k_2)\!\left[{\mathcal G}^{m'n'}_1(k_1,k_2)\right]_{t,{\rm  I}}\nonumber \\
&+&g^2f^{amn}f^{nrs}\int_{k_1}\!D^{mm'}(k_1)\left[{\mathcal Q}^{m'rs}_{1\alpha}(k_1,k_2)\right]_{t,{\rm  I}},
\eea
with a similar relation holding for the ${\mathcal S}_2$ term. Then, we find
\bea
& & (c)-gf^{amn}\int_{k_1}\!k_{2\alpha}D^{mm'}(k_1)D^{nn'}(k_2)\left[{\mathcal G}^{m'n'}_1(k_1,k_2)\right]_{t,{\rm  I}}\nonumber \\
& &\hspace{1cm}=\ gf^{amn}\int_{k_1}\!(k_1-k_2)_\alpha D^{mm'}(k_1)D^{nn'}(k_2)\left[{\mathcal G}^{m'n'}_1(k_1,k_2)\right]_{t,{\rm  I}}=(\widehat{c}),
\eea
and, using the decomposition for the ${\mathcal Q}^{m'rs}_{1\nu}$ shown in \Figref{Q1_Q2_defs},
\bea
& & g^2f^{amn}f^{nrs}\int_{k_1}\!D^{mm'}(k_1)\left[{\mathcal Q}^{m'rs}_{1\alpha}(k_1,k_2)\right]_{t,{\rm  I}}\nonumber \\
& &\hspace{.5cm}=\ ig^2f^{amn}f^{nrs}g_{\alpha\rho}\int_{k_1}\!\int_{k_3}\!D^{mm'}(k_1)D^{ss'}(k_3)\Delta^{\rho\rho'}_{rr'}(k_4)\left\{\left[{\mathcal K}^{m'r's'}_{\rho'}(k_1,k_3,k_4)\right]_{t,{\rm  I}}\right.\nonumber \\
& &\hspace{.5cm}\left.+\
i\Gamma^{gr's'}_{\rho'}(-k_2,k_3,k_4)D^{gg'}(k_2)\left[{\mathcal G}^{m'g'}_1(k_1,k_2)\right]_{t,{\rm  I}}\right\}\nonumber \\
& &\hspace{.5cm}=\ (\widehat{e})+(\widehat{f}),
\eea
with ${\mathcal K}^{m'r's'}_{\sigma'}$ representing the 1PI five-particle kernel shown in \Figref{Q1_Q2_defs},  
while $\Gamma^{gr's'}_{\sigma'}$ is the usual ghost-gluon vertex.
\newline
\indent
In exactly the same way, the remaining ${\mathcal S}_2$, when added to the \Rxi ghost diagram $(c')$,  will 
give rise to graphs $(\widehat{c'})$, $(\widehat{e'})$, and $(\widehat{f'})$; so, we finally arrive at the relation
\be
\widehat{\Gamma}^a_\alpha(q,p_2,-p_1)=\widetilde{\Gamma}^a_\alpha(q,p_2,-p_1).
\label{res1:vertex}
\ee
\indent
The final  step is  to construct the  all-order PT  gluon self-energy
$\widehat\Pi_{\alpha\beta}^{ab}(q)$.  Notice  that at this  point one would
expect  that  it  too  coincides  with the  BFG  gluon
self-energy  $\widetilde\Pi^{ab}_{\alpha\beta}(q)$,  since  both the  boxes
${\widehat  B}$  and  the vertex  $\widehat\Gamma^a_\alpha$  do
coincide with  the corresponding BFG boxes ${\widetilde
B}$ and vertex  $\widetilde{\Gamma}^a_\alpha$, and the $S$-matrix
is unique. We will conclude this Section by furnishing an indirect proof of this statement by means of the 
``strong induction principle'', which states that a given 
predicate $P(n)$ on $\mathbb N$ is true $\forall\ n\in{\mathbb N}$, 
if $P(k)$ is true whenever $P(j)$ is true $\forall\  j\in{\mathbb N}$
with $j<k$.  In order to avoid notational clutter, we will use the schematic notation
where all the group, Lorentz, 
and momentum indices will be suppressed. 
\newline
\indent
At  one-  and  two-loop (\ie
$n=1,2$), we know that the result is true from our explicit calculations.  
Let  us then assume that
the PT construction has been  successfully carried out up to the order
$n-1$ (strong  induction hypothesis):  we will show  then that  the PT
gluon self-energy is equal to  the BFG gluon self-energy
at order~$n$, \ie $\widehat\Pi^{(n)}\equiv\widetilde\Pi^{(n)}$;  
hence,  by  the  strong
induction principle, this last result  will be valid at any given $n$,
showing finally that the PT construction holds true to all orders.
\newline
\indent
\begin{figure}[t]
\bce
\includegraphics[width=9.5cm]{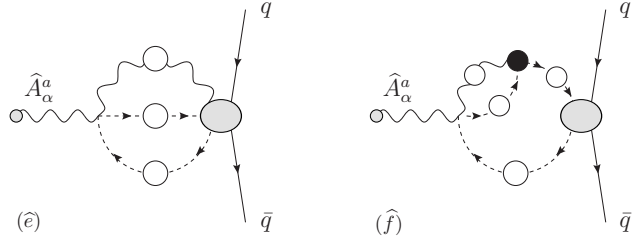}	
\ece
\caption{\figlab{qg_BFM_topo} Additional topologies present in the BFM quark-gluon vertex. As before, not shown are two similar contributions $(e')$ and $(f')$ with the ghost arrows reversed.}
\end{figure}
From the strong inductive hypothesis, we know that
\be
\widehat\Pi^{(\ell)} \equiv \widetilde\Pi^{(\ell)},\,\,\,\,
\widehat\Gamma^{(\ell)} \equiv \widetilde\Gamma^{(\ell)},\,\,\,\, 
\widehat B^{(\ell)} \equiv \widetilde B^{(\ell)}\equiv B^{(\ell)},
\label{hypo-3}
\ee
where $\ell=1,\dots,n-1$. 
\newline
\indent
Now, the $S$-matrix element of order $n$, to be denoted as $S^{(n)}$,
assumes the form 
\be
S^{(n)}=\left\{\Gamma\Delta\Gamma\right\}^{(n)}+B^{(n)}.
\ee 
Moreover, since it is unique, regardless if it is written in the Feynman
gauge, in the BFG, as well as before and after the PT
rearrangement, we have that \mbox{$S^{(n)}\equiv
\widehat S^{(n)}\equiv\widetilde 
S^{(n)}$}. Using then
Eq.~(\ref{res2:boxes}) (which holds to all orders, implying that 
Eq.~(\ref{hypo-3}) holds true also when $\ell=n$), we find that
\be
\left\{\Gamma\Delta\Gamma\right\}^{(n)}\equiv
\{\widehat\Gamma\widehat\Delta\widehat\Gamma\}^{(n)}\equiv
\{\widetilde\Gamma\widetilde\Delta\widetilde\Gamma\}^{(n)}.
\label{fun1}
\ee
\newline
\indent
The above amplitudes can then be split into 1PR and 1PI parts; in
particular, due to the strong inductive 
hypothesis of Eq.~(\ref{hypo-3}), the 1PR part after 
the PT rearrangement coincides with the 1PR part written in the 
BFG, since
\be
\left\{\Gamma\Delta\Gamma\right\}^{(n)}_{\scriptscriptstyle{\rm R}}=
\Gamma^{(n_1)}\Delta^{(n_2)}\Gamma^{(n_3)}, \qquad \left\{
\begin{array}{l}
n_1,n_2,n_3<n, \\
n_1+n_2+n_3=n. 
\end{array}
\right.
\ee
\newline
\indent
Then Eq.~(\ref{fun1}) state the equivalence of the 1PI parts, {\it i.e.},
\be
\{\widehat\Gamma\widehat\Delta\widehat\Gamma\}_{\scriptscriptstyle{\rm
I}}^{(n)}\equiv 
\{\widetilde\Gamma\widetilde\Delta\widetilde\Gamma\}_{\scriptscriptstyle{\rm
I}}^{(n)}, 
\ee
which implies
\be
\left[\widehat\Gamma^{(n)}-\widetilde\Gamma^{(n)}\right]
\Delta^{(0)}\Gamma^{(0)}+\Gamma^{(0)}\Delta^{(0)}\left[\widehat\Gamma^{(n)}
-\widetilde\Gamma^{(n)}\right]+\Gamma^{(0)}\Delta^{(0)}
\left[\widehat\Pi^{(n)}-\widetilde\Pi^{(n)}\right]
\Delta^{(0)}\Gamma^{(0)}=0.
\ee
At this point we do  not have the equality we want yet, but only that
\bea
\widehat\Gamma^{(n)}&=&\widetilde\Gamma^{(n)}+f^{(n)}\Gamma^{(0)},\nonumber\\
\widehat\Pi^{(n)}&=&\widetilde\Pi^{(n)}-2iq^2f^{(n)},
\eea
with $f^{(n)}$ an arbitrary function of $q^2$. However, by means of the
{\it explicit} construction of the PT vertex just presented, 
we know that 
$\widehat\Gamma^{(n)}\equiv\widetilde\Gamma^{(n)}$, and therefore $f=0$.
Then one immediately concludes that 
\be
\widehat\Pi^{(n)}\equiv\widetilde\Pi^{(n)}.
\ee
Hence, by strong induction, the above relation is true for any given
order $n$, \ie inserting back the Lorentz and gauge group structures,
we arrive at the announced result
\be
\widehat\Pi^{ab}_{\alpha\beta}(q)\equiv\widetilde\Pi^{ab}_{\alpha\beta}(q).
\ee
\indent
The same techniques described in this section have been used  in~\cite{Binosi:2004qe} 
to generalize the PT algorithm to all orders in the electroweak sector of the SM.

\newpage


\section{PT in the Batalin-Vilkovisky framework\seclab{PTBV}}
\noindent
In the previous section we demonstrated how the PT algorithm can be generalized 
beyond the one-loop level. In passing from 
the two-loop to the all-order construction 
we abandoned the diagrammatic treatment, and 
resorted instead to  a  more  formal procedure,  
where the longitudinal pinching momenta trigger  the  STIs satisfied  
by  specific subsets  of  fully dressed  vertices
appearing  in  the   ordinary  perturbative  expansion. Due to  the  non-linearity  of the  BRST
transformation, we have also seen that the Green's functions generated by this process involve
composite  operators of  the type  $\langle0\vert T[s\Phi(x)\cdots]\vert0\rangle$,  
with $s$  the BRST operator and $\Phi$ a generic QCD field.
\newline
\indent
While it is rather satisfactory that the PT could be generalized to all orders, 
the way this was actually accomplished leaves some important issues still open, 
mainly for the following reasons.
\begin{itemize}
\item[$\ast$]
The PT quark-gluon vertex was constructed explicitly, but was only partially off-shell, since 
the external fermions were assumed (as always up to now) to be on-shell; however, 
for the upcoming SD analysis (next section) we should be able to construct {\it fully off-shell} vertices.
To be sure, a fully off-shell quark-gluon vertex can be constructed through appropriate embedding (as was done for the 
completely off-shell three-gluon vertex in \secref{QCD_one-loop}), but this would take us further  
away from the non-diagrammatic formulation that we are striving for. 
\newline
\item[$\ast$]
The only fully off-shell Green's function considered was the PT gluon propagator; however, 
it was not {\it explicitly} constructed, but rather indirectly shown to coincide with the corresponding 
BFG gluon propagator.
It is clearly preferable to devise an {\it explicit} procedure for obtaining the PT gluon propagator to all orders.
\newline
\item[$\ast$]
During the all-order construction presented, 
no pinching contributions have actually been ever identified; 
rather, all cancellations of such terms were concealed in the crucial steps corresponding to Eqs~(\ref{r1}). 
However, the determination of closed expressions for the pinching contributions is instrumental 
for understanding the process-independence of the algorithm,  
as well as for determining crucial identities relating the \Rxi Green's functions with those of the BFG.
\end{itemize}  
\indent 
As it  turns out~\cite{Binosi:2002ez},  the most  efficient framework  for dealing
with  the type  of quantities  appearing in  the PT  procedure  is the
so-called  Batalin-Vilkovisky   (BV)  formalism~\cite{Batalin:1984jr}.  
This formalism not only 
streamlines  the  derivation  of  the  STIs satisfied  by  1PI  Green's
functions and connected kernels, but has two additional important  
key features~\cite{Binosi:2002ez,Grassi:1999tp}: it allows for the construction of
the auxiliary (ghost) Green's functions in terms of a well-defined set
of Feynman  rules, and furnishes  a set  of useful
identities, the  so-called Background-Quantum Identities  (BQIs), which 
relate  Green's  functions  involving  background  fields  to  Green's
functions  involving quantum  fields.  As we  will  see, these  latter
identities will allow for the effortless 
identification of the PT Green's functions
with  those of the  BFG,  and  constitute  one  of  the  main
advantages gained from employing the BV formalism~\cite{Binosi:2002ez}.
\newline
\indent  
In  this  section  we  present  an  introduction  to  the  BV
formalism,  providing the  minimum  amount of  information needed to
establish notation,  and  arrive at  the relevant  generating
functionals  and master equations,  together with  the differentiation
rules needed  to generate from  the latter useful identities.  We will
then concentrate on  putting to work all the  machinery introduced, by
casting in the BV language  the one- and two-loop results presented in
earlier sections.
\newline
\indent
Thus, this introductory section sets up the stage  
for an alternative, and more efficient, way of 
generalizing the PT to all orders. In fact, as will be shown 
in the next section, casting the PT in the BV language permits to 
go on step further, allowing  for the generalization of the 
PT algorithm to the Schwinger-Dyson equations of QCD.

\subsection{Green's functions: conventions}

\begin{figure}[!t]
\begin{center}
\includegraphics[width=5cm]{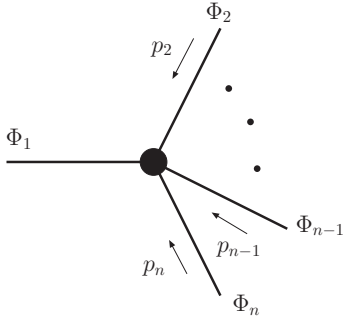}
\end{center}
\caption{Our conventions for the (1PI) Green's functions $\Gamma_{\Phi_1\cdots\Phi_n}(p_1,\dots,p_n)$. All momenta $p_2,\dots,p_n$ are assumed to be incoming, and are assigned to the corresponding fields starting from the rightmost one. The momentum of the leftmost field $\Phi_1$ is determined through momentum conservation ($\sum_i p_i=0$) and will be suppressed.}
\label{fig:Green_conv}
\end{figure}
\noindent
The Green's functions of the theory can be constructed in terms of time-ordered products of free fields $\Phi^0_1\cdots\Phi^0_n$ and vertices of the interaction Lagrangian ${\mathcal L}_\mathrm{int}$ (constructed from the pieces of ${\mathcal L}$ which are not bilinear in the fields) through the standard Gell-Man-Low formula for the 1PI truncated Green's functions
\begin{eqnarray}
\Gamma_{\Phi_1\cdots\Phi_n}(x_1,\dots,x_n)&=&\langle T[\Phi_1(x_1)\cdots\Phi_n(x_n)]\rangle^\mathrm{1PI}\nonumber \\
&=& \langle T[\Phi^0_1(x_1)\cdots\Phi^0_n(x_n)]\exp(-i\int\!d^4x\,{\mathcal L}_\mathrm{int})\rangle^\mathrm{1PI}.
\label{gml}
\end{eqnarray}
The complete set of Green's functions can be handled most efficiently by introducing a generating functional (see also \secref{PTBFM}), which in Fourier space reads
\begin{equation}
\Gamma[\Phi]=\sum_{n=0}^\infty\frac{(-i)^n}{n!}\int\!\prod_{i=0}^nd^4p_i\ \delta^4(\sum_{j=1}^np_j)\Phi_1(p_1)\cdots\Phi_n(p_n)\Gamma_{\Phi_1\cdots\Phi_n}(p_1,\dots,p_n),
\end{equation}
with $p_i$ the (in-going) momentum of the $\Phi_i$ field. 
Since in perturbation theory $\Gamma_{\Phi_1\cdots\Phi_n}$ is a formal power series in $\hbar$, 
we will denote its  $m$-loop contribution as $\Gamma^{(m)}_{\Phi_1\cdots\Phi_n}$. 
Then, the Green's functions of the theory may be obtained from the 
generating functional $\Gamma[\Phi]$ by means of functional differentiation:
\begin{equation}  
\Gamma_{\Phi_1\cdots\Phi_n}(p_1,\dots,p_n) = i^n\left.\frac{\delta^{n}\Gamma}{\delta\Phi_1(p_1)\delta\Phi_2(p_2)\cdots\delta\Phi_n(p_n)}\right|_{\Phi_i=0},
\label{greenfunc}
\end{equation} 
where $\Phi(p)$ denotes the Fourier transform of $\Phi(x)$,  and our convention 
on the external momenta is  summarized in Fig.~\ref{fig:Green_conv}.
From the definition given in Eq.~(\ref{greenfunc}) it follows that the Green's functions $i^{-n}\Gamma_{\Phi_1\cdots\Phi_n}$ are simply given by the corresponding Feynman diagrams in Minkowsky space. Finally, notice that upon inversion of two (adjacent) fields we have
\begin{equation}
\Gamma_{\Phi_1\cdots\Phi_i\Phi_{i+1}\cdots\Phi_n}(p_1,\dots,p_i,p_{i+1},\dots,p_n)=\pm
\Gamma_{\Phi_1\cdots\Phi_{i+1}\Phi_i\cdots\Phi_n}(p_1,\dots,p_{i+1},p_i,\dots,p_n),
\end{equation}
with the minus appearing only when both fields $\Phi_i$ and $\Phi_{i+1}$ obey Fermi statistics.
\newline
\indent
The Green's functions defined so far are sufficient for building 
all possible amplitudes involved in the $S$-matrix computation; however, due to the non-linearity of the BRST transformations, they do not cover the complete set of Green's functions appearing in the STIs of the theory (and therefore needed for its renormalization, as well as the PT construction). 

\subsection{The Batalin-Vilkovisky formalism for pedestrians \label{BV}}
\noindent
Let us now introduce  for each field  $\Phi$ appearing in
the  theory\footnote{In the rest of this report, quark fields will be denoted by the generic letter $\psi$} 
  a  corresponding  anti-field, to be denoted by $\Phi^*$. 
The anti-field $\Phi^*$
has opposite  statistics   with  respect   to  $\Phi$; its ghost charge, ${\rm gh}(\Phi^*)$,
is related to the  ghost  charge ${\rm gh}(\Phi)$ of the field $\Phi$ by 
${\rm gh}(\Phi^*)=-1- {\rm gh}(\Phi)$.
For convenience, we summarize the ghost charges and statistics of the various QCD fields and anti-fields 
in Table~\ref{tableI}.
\begin{table}
	\centering
		\begin{tabular}{|c|c|c|c|c|c||c|c|c|c|c||c|c|}
		\hline
		$\Phi$ & $A_\mu$ & $\psi$ & $\bar\psi$ & $c$ & $\bar c$ & $A^*_\mu$ & $\psi^*$ & $\bar\psi^*$ & $c^*$ & $\bar c^*$ & $\Omega_\mu^a$\\
		\hline\hline
		${\rm gh}(\Phi)$ & 0 & 0 & 0 & 1 & -1 & -1 & -1 &  -1 & -2 & 0 & 1\\
		\hline
		${\rm st}(\Phi)$ & B & F & F & F & F & F & B & B & B & B & F\\
		\hline
		\end{tabular}
		\caption{Ghost charges and statistics (B for Bose, F for Fermi) of the QCD fields, anti-fields and BFM sources.\label{tableI}}
\end{table}
Next, we add to the original gauge invariant Lagrangian ${\mathcal L}_{\mathrm{I}}$ of Eq.~(\ref{Linv}) a 
term coupling the anti-fields with the BRST variation of the corresponding fields, to get
\bsub
\begin{eqnarray}
{\mathcal L}_{\mathrm{BV}}&=&{\mathcal L}_{\mathrm I}+{\mathcal L}_{\mathrm{BRST}}, \label{BRST_Lag-1} \\
{\mathcal L}_{\mathrm{BRST}}&=& \sum_{\Phi}\Phi^*s\Phi\nonumber \\
&=&A^{*a}_\mu(\partial^\mu c^a+gf^{abc}A^b_\mu c^c)-\frac12gf^{abc}c^{*a}c^bc^c+ig\bar\psi^*c^at^a\psi-igc^a\bar\psi t^a\psi^*.
\label{BRST_Lag}
\end{eqnarray}
\esub
Notice that we have suppressed all spinor indices (both flavor and color), since they play no role in the 
ensuing construction.
\newline
\indent
Then, the action $\Gamma^{(0)}[\Phi,\Phi^*]$ constructed from ${\mathcal L}_{\mathrm{BV}}$, 
will satisfy the master equation
\begin{equation}
\int\!\!d^4x\sum\frac{\delta\Gamma^{(0)}}{\delta\Phi^*}\frac{\delta\Gamma^{(0)}}{\delta\Phi}=0.
\label{master_eq}
\end{equation}
To verify this, observe that, on one hand, the terms in $\delta\Gamma^{(0)}/{\delta\Phi}$ 
that are independent of the anti-fields $\Phi^*$ are zero, due the BRST (actually the gauge) invariance of the action
\begin{equation}
\int\!\!d^4x\sum s\Phi\frac{\delta\Gamma^{(0)}_{\mathrm{I}}}{\delta\Phi}=\int\!\!d^4x (s\Gamma^{(0)}_{\mathrm{I}}[\Phi])=0\,;
\end{equation} 
on the other hand, terms in $\delta\Gamma^{(0)}/{\delta\Phi}$ that are linear in the anti-fields vanish, due to the nihilpotency of the BRST operator
\begin{equation}
\int\!\!d^4x\sum s\Phi'\frac{\delta (s\Phi)}{\delta\Phi'}=\int\!\!d^4x \sum s^2\Phi'=0.
\end{equation}
Now, since the anti-fields are external  sources, we must constrain them to
suitable values before the action $\Gamma^{(0)}$ can be used for calculating the $S$-matrix
elements of the theory~\cite{Weinberg:1996kr}. 
To  that end, we  introduce an  arbitrary
fermionic  functional, $\Psi[\Phi]$, with ghost charge -1, and set for all the anti-fields $\Phi^*$
\begin{equation}
\Phi^*=\frac{\delta\Psi[\Phi]}{\delta\Phi}.
\end{equation}
Then the action becomes
\begin{eqnarray}
\Gamma^{(0)}[\Phi,\delta\Psi/\delta\Phi]&=&\Gamma^{(0)}[\Phi]+(s\Phi)\frac{\delta\Psi[\Phi]}{\delta\Phi}\nonumber \\
&=& \Gamma^{(0)}[\Phi]+s\Psi[\Phi],
\end{eqnarray}
and therefore, choosing the functional $\Psi$ to satisfy the relation
\begin{equation}
s\Psi=\int\!d^4x\left({\mathcal L}_{\mathrm{GF}}+{\mathcal L}_{\mathrm{FPG}}\right),
\end{equation}  
we see that the action $\Gamma^{(0)}$ (obtained from ${\mathcal L}_{\mathrm{BV}}$) 
is equivalent to the gauge-fixed action obtained from the original Lagrangian ${\mathcal L}$ of Eq.~(\ref{QCD_lag}). The functional $\Psi$ is often referred to  as the ``gauge fixing fermion''.
\newline
\indent
It is well-known that the BRST symmetry is crucial for endowing a (gauge) theory  with a unitary $S$-matrix 
and gauge-independent physical observables; therefore, it must be implemented to all orders. 
For doing so we establish the quantum corrected version of the master equation (\ref{master_eq}) in the form of the STI functional
\begin{eqnarray}
{\mathcal S}(\Gamma)[\Phi]&=&\int\!d^4x\sum\frac{\delta\Gamma}{\delta\Phi^*}\frac{\delta\Gamma}{\delta\Phi}\nonumber \\
&=&\int\!d^4x\left\{\frac{\delta\Gamma}{\delta A^{*\mu}_m}\frac{\delta\Gamma}{\delta A^{m}_\mu}+\frac{\delta\Gamma}{\delta c^{*m}}\frac{\delta\Gamma}{\delta c^{m}}+\frac{\delta\Gamma}{\delta \psi^{*}}\frac{\delta\Gamma}{\delta\bar \psi}+\frac{\delta\Gamma}{\delta \psi}\frac{\delta\Gamma}{\delta\bar \psi^*}+B^m\frac{\delta\Gamma}{\delta\bar c^m}\right\}\nonumber \\
&=&0,
\label{STIfunc_nm}
\end{eqnarray}
where $\Gamma[\Phi,\Phi^*]$ is now the effective action. 
\newline
\indent
In order to simplify the structure of the STI generating functional of Eq.~(\ref{STIfunc_nm}), 
let us notice that the anti-ghost  $\bar c^{a}$ and the multiplier $B^a$ have {\it linear} BRST transformations; 
therefore they do not present the usual complications (due to non-linearity) of the other QCD fields. 
Together with their corresponding anti-field, they enter bi-linearly in the action, and one can write the complete action (which we now explicitly indicate with a C subscript) as a sum of a minimal and non-minimal sector
\begin{equation}
\Gamma_{\mathrm{C}}^{(0)}[\Phi,\Phi^*]=\Gamma^{(0)}[A^a_\mu,A^{*a}_\mu,\psi,\psi^*,\bar\psi,\bar\psi^*,c^a,c^{*a}]+\bar c^{*a}B^a.
\end{equation}
The last term has no effect on the master equation (\ref{master_eq}), which is satisfied by
$\Gamma^{(0)}$ alone; the fields $\{A^a_\mu,A^{*a}_\mu,\psi,\psi^*,\bar\psi,\bar\psi^*,c^a,c^{*a}\}$ are usually called {\it minimal variables} while  $\bar c^{a}$ and $B^a$ are referred to as non-minimal variables or ``trivial pairs''. 
Equivalently, one can introduce the minimal (or reduced) action by subtracting from the complete one the local term corresponding to the gauge-fixing Lagrangian, \ie
\be
\Gamma=\Gamma_{\mathrm{C}}-\int\!d^4x\,{\mathcal L}_\mathrm{GF}.
\ee 
In either cases, the result is that the STI functional is now written as
\be
{\mathcal S}(\Gamma)[\Phi]
=\int\!d^4x\left\{\frac{\delta\Gamma}{\delta A^{*\mu}_m}\frac{\delta\Gamma}{\delta A^{m}_\mu}+\frac{\delta\Gamma}{\delta c^{*m}}\frac{\delta\Gamma}{\delta c^{m}}+\frac{\delta\Gamma}{\delta \psi^{*}}\frac{\delta\Gamma}{\delta\bar \psi}+\frac{\delta\Gamma}{\delta \psi}\frac{\delta\Gamma}{\delta\bar \psi^*}\right\}=0.
\label{STIfunc}
\ee
In practice, the STIs of Eq.~(\ref{STIfunc}), generated from the reduced functional $\Gamma$,  
coincide with those  obtained from $\Gamma_{\mathrm{C}}$ after the implementation of 
the Faddeev-Popov equation, described in the next subsection~\cite{Itzykson:1980rh}. 
One should also keep in mind that the Green's functions involving unphysical fields that are
generated by $\Gamma$ coincide with those generated by $\Gamma_{\mathrm{C}}$ 
only up to constant terms proportional to the gauge fixing parameter, \eg $\Gamma_{A_\mu A_\nu}(q)=\Gamma^{\mathrm{C}}_{A_\mu A_\nu}(q)-i\xi^{-1}q_\mu q_\nu$. 
The differences between employing  $\Gamma$ or $\Gamma_{\mathrm{C}}$ 
is further explored in \appref{FPEs-STIs-BQIs}; there, all relevant identities 
needed for the generalization of the PT procedure to the SDEs of QCD (next section) are derived and discussed.
\newline
\indent
Taking functional derivatives
of  ${\mathcal S}(\Gamma)[\Phi]$  and  setting afterwards  all fields  and
anti-fields to  zero will generate  the complete set of  the all-order
STIs of the theory; this is in exact analogy to what happens with  the effective action,
where  taking  functional derivatives  of  $\Gamma[\Phi]$ and  setting
afterwards all fields  to zero generates the Green's  functions of the
theory, see Eq.~(\ref{greenfunc}).  
However, in order to reach meaningful expressions, one needs to keep in mind that:
\begin{enumerate} 
\item[$\ast$] ${\mathcal S}(\Gamma)$ has ghost charge 1; 
\newline
\item[$\ast$] functions with non-zero ghost charge vanish, since the ghost charge is a conserved quantity.
\end{enumerate}
Thus, in order to extract
non-trivial identities from Eq.~(\ref{STIfunc}) one needs to differentiate the latter with respect to a  combination of fields, containing either one ghost field, or two ghost fields and one anti-field. The only exception to this rule is when differentiating with respect to a ghost anti-field, which needs to be compensated by three ghost fields. In particular, identities 
involving one or more gauge fields are obtained by differentiating Eq.~(\ref{STIfunc})
with respect to the set of fields in which one gauge boson has been replaced by the corresponding ghost field. This is due to the fact that  the linear part of the BRST transformation of the gauge field is proportional to the ghost field: $sA^a_\mu|_\mathrm{linear}=\partial_\mu c^a$. For completeness we also notice that, for obtaining STIs involving Green's functions that contain ghost fields, one ghost field must be replaced by two ghost fields, due to the non linearity of the BRST ghost field transformation [$sc^a\propto f^{abc}c^bc^c$, see Eq.~(\ref{BRSTtrans})].
The last technical point to be clarified is  the dependence of the STIs on the (external) momenta. 
One should notice that the integral over $d^4x$ present in Eq.~(\ref{STIfunc}), together with the conservation of momentum flow of the Green's functions, 
implies that no momentum integration is left over; as a result, the STIs will be expressed as a sum of products of (at most two) Green's functions. 
\newline
\indent
An advantage of working with the BV formalism is the fact that the STI functional~(\ref{STIfunc}) is valid in any gauge, \ie
it will not be affected when switching from one gauge to another. In particular, if we want to consider the BFM gauge, the only additional step we need to take is to implement the equations of motion for the background fields at the quantum level. 
This is achieved most efficiently by extending the BRST symmetry to the background gluon field, through the relations
\begin{equation}
s\widehat{A}_\mu^m=\Omega^m_\mu, \qquad s\Omega^m_\mu=0,
\label{extBRST}
\end{equation}
where $\Omega_\mu^m$ represents a (classical) vector field, with the same quantum numbers as the gluon, ghost charge $+1$, and Fermi statistics (see also Table~\ref{tableI}). The dependence of the Green's  functions on the background fields is then controlled by  the modified STI functional
\be
{\mathcal S}'(\Gamma')[\Phi]={\mathcal S}(\Gamma')[\Phi]+\int\!d^4x\
\Omega_m^\mu\left(\frac{\delta\Gamma'}{\delta\widehat{A}^m_\mu}-\frac{\delta\Gamma'}{\delta A^m_\mu}\right)=0,
\label{STIfunc_BFM}
\ee
where $\Gamma'$ denotes the effective action that depends on the background sources $\Omega^m_\mu$ (with $\Gamma\equiv\Gamma'\vert_{\Omega=0}$), and ${\mathcal S}(\Gamma')[\Phi]$ is the STI functional of Eq.~(\ref{STIfunc}). 
Differentiation of the STI functional (\ref{STIfunc_BFM}) with respect to the background source and background or quantum fields will 
then provide the BQIs, which relate 1PI Green's functions involving background fields with the ones involving quantum fields. 
The BQIs are particularly useful in the PT context, since they  allow for a direct  comparison between PT  and BFM Green's functions.
\newline
\indent
Finally, the background gauge invariance of the BFM effective action implies that Green's functions involving background fields satisfy 
linear WIs when contracted with the momentum corresponding to a background leg. 
They are generated by taking functional differentiations of the WI functional
\be
{\mathcal W}_{\vartheta}[\Gamma']=\int\!d^4x\,\sum_{\Phi,\Phi^*}\left(\delta_{\vartheta(x)}\Phi\right)\frac{\delta\Gamma'}{\delta\Phi}=0,
\label{WI_gen_funct}
\ee 
where $\vartheta^a(x)$ are the local infinitesimal parameters corresponding to the $SU(3)$ generators $t^a$, which now play the role of the ghost field. The transformations $\delta_{\vartheta}\Phi$ are thus given by
\bsub
\bea
\delta_{\vartheta}A^a_\mu=gf^{abc}A^b_\mu \vartheta^c &\qquad& \delta_{\vartheta}\widehat{A}^a_\mu=\partial_\mu \vartheta^a+gf^{abc}\widehat{A}^b_\mu \vartheta^c,\label{theta_trans-1} \nonumber \\
\delta_{\vartheta} c^a=-g f^{abc}c^b\vartheta^c &\qquad& \delta_{\vartheta} \bar c^a=-g f^{abc}\bar c^b\vartheta^c,
\label{theta_trans-2} \nonumber\\
\delta_{\vartheta}\psi^i_\mathrm{f}=ig \vartheta^a(t^a)_{ij}\psi^j_\mathrm{f} &\qquad& 
\delta_{\vartheta}\bar\psi^i_\mathrm{f}=-ig \vartheta^a\bar\psi^j_\mathrm{f} (t^a)_{ji}, 
\label{theta_trans-3}
\eea
\esub
and the background transformations of the anti-fields  $\delta_{\vartheta}\Phi^*$ coincide with the gauge transformations 
of the corresponding quantum gauge fields according to their specific representation. 
Notice that, in order to obtain the WIs satisfied by the Green's functions involving background gluons $\widehat{A}$, one has to differentiate the functional (\ref{WI_gen_funct}) with respect to the corresponding parameter $ \vartheta$. 
\newline
\indent
The  STIs and BQIs needed  for the PT construction have been derived in~\cite{Binosi:2008qk}, 
together with the  method  for constructing  the
auxiliary functions  appearing in  these identities; 
they are reported for convenience in Appendix~\ref{Appendix:STIs}   and~\ref{Appendix:BQIs}.

\subsection{Faddeev-Popov equation(s)\label{FPEs}}
\noindent
The Faddev-Popov equation is a highly non-trivial identity,  which is extremely 
useful in the PT context, since it determines the result of the contraction of longitudinal momenta on auxiliary Green's functions. 
The FPE is also instrumental in proving the equivalence between the STIs and BQIs derived 
using $\Gamma$ or $\Gamma_{\mathrm{C}}$ (see Appendix).
Since the FPE depends crucially on the form of the ghost Lagrangian, which, in turn, depends on the gauge fixing function 
[see Eq.~(\ref{FPG_Lag})], we will first present the corresponding derivation in the $R_\xi$ gauges, and then generalize it to the BFM.
\newline
\indent
To derive the FPE in the $R_\xi$ gauges, one observes that in the QCD action the only term proportional to the anti-ghost fields comes from the 
Faddeev-Popov Lagrangian density, which can be rewritten as
\begin{equation}
{\mathcal L}^{R_\xi}_{\mathrm{FPG}}=-\bar c^m\partial^\mu(sA^m_\mu)=-\bar c^m\partial^\mu\frac{\delta\Gamma}{\delta A^{*m}_\mu}.
\label{Rxi}
\end{equation}
Differentiation of the action with respect to $\bar c^a$ yields the FPE in the form of the identity
\begin{equation}
\frac{\delta\Gamma}{\delta \bar c^m}+\partial^\mu\frac{\delta\Gamma}{\delta A^{*m}_\mu}
=0,
\label{FPpos}
\end{equation}
so that, taking the Fourier transform, we arrive at 
\begin{equation}
\frac{\delta\Gamma}{\delta \bar c^m}+iq^\mu\frac{\delta\Gamma}{\delta A^{*m}_\mu}
=0.
\label{FPeqRxi}
\end{equation}
Thus,  in the $R_\xi$ case, the FPE amounts to the simple statement that the contraction 
of a leg corresponding to a 
gluon anti-field ($A^{*m}_\mu$)  by its own momentum ($q^\mu$) converts it to an anti-ghost leg ($\bar c^m$). 
Functional differentiation of this identity with respect to QCD fields (but not background sources and fields, see below) 
furnishes useful identities, that will be used extensively in our construction. 
\newline
\indent
For obtaining FPEs for Green's functions involving background gluons and sources, 
one has to modify Eq.~(\ref{FPeqRxi}), in order to account for the presence of extra terms in the BFM gauge fixing function (and therefore in the BFM Faddeev-Popov ghost Lagrangian). Eq.~(\ref{FPpos}) gets then generalized to
\be
\frac{\delta\Gamma'}{\delta \bar c^m}+\left(\widehat{\cal D}^\mu\frac{\delta\Gamma'}{\delta A^*_\mu}\right)^m-\left({\cal D}^\mu\Omega_\mu\right)^m=0.
\label{FPeqBFM}
\ee
\newline
\indent
Once again, the specific FPEs needed for the PT construction have been derived in~\cite{Binosi:2008qk}, and are reported in Appendix~\ref{Appendix:FPEs}.

\subsection{The (one-loop) PT algorithm in the BV language} 
\noindent
After introducing all this formal machinery, it would be important to
make contact  with the PT algorithm. 
This is clearly best done at  the one-loop level, since in  this case all calculations
are  rather straightforward,  and  it  is relatively  easy  to compare  the
standard  diagrammatic  results  with  those coming  from  the  BV
formalism.  This comparison will ({\it i})  help us identify the pieces that
will be generated  when applying the PT algorithm,  and ({\it ii}) establish
the rules for distributing the  pieces obtained in ({\it i})
among the different Green's functions appearing in the calculation.
\begin{figure}[!t]
\begin{center}
\includegraphics[width=13.5cm]{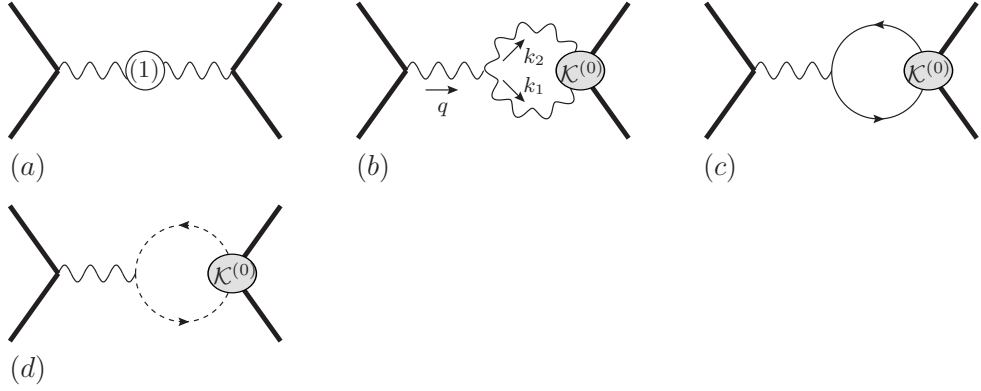}
\end{center}
\caption{The $S$-matrix PT setup for constructing the gluon propagator at one loop. The external particles are left unspecified since, due to the process-independence of the algorithm, they can equally well be quarks or gluons. 
When the external particles are, \eg gluons, the kernel ${\mathcal K}^{(0)}$ appearing in diagram $(b)$  is the tree-level version of the one shown in Fig.~\ref{fig:gggg_SDker}. Mirror diagrams having the kernels on the opposite side are not shown.}
\label{fig:1l_PT}
\end{figure}
\newline
\indent
The starting point is the usual embedding of the (one-loop) gluon propagator 
into an $S$-matrix element (Fig.~\ref{fig:1l_PT}).  Choosing the external legs to be, \eg gluons, and 
carrying out the  PT decomposition $\Gamma=\Gamma^{\rm P}+\Gamma^{\rm F}$ 
on the tree-level three-gluon vertex of diagram $(b)$
[see Eqs~(\ref{GF})], we find
\bea
(b)&=&(b)^{\rm F}+(b)^{\rm P},\nonumber \\
(b)^{\rm P}&=&-\frac12gf^{am'n'}\int_{k_1}(g_{\alpha\nu'}k_{1\mu'}-g_{\alpha\mu'}k_{2\nu'})\Delta_{m'm}^{(0)\mu'\mu}(k_1)\Delta_{n'n}^{(0)\nu'\nu}(k_2){\mathcal K}^{(0)}_{A^m_\mu A^n_\nu A^r_\rho A^s_\sigma}(k_2,p_2,-p_1).\nonumber \\
&=&-gf^{amn}g_\alpha^\nu\int_{k_1}\frac1{k_1^2}\frac1{k_2^2}k_1^\mu{\mathcal K}^{(0)}_{A^m_\mu A^n_\nu A^r_\rho A^s_\sigma}(k_2,p_2,-p_1).
\eea
The $1/2$ symmetry factor carried by the kernel ${\mathcal K}$ is due to the explicit presence of the crossing term.
Then, expanding at tree-level the STI~(\ref{STISDgggg}), dropping the terms proportional to the inverse gluon propagator $\Gamma_{AA}^{(0)}(p_i)$, 
since they vanish for ``on-shell'' gluons, and finally making use of Eq.~(\ref{BQI:auxOmAs}), we get
\bea
(b)^{\rm P}&=& -\Gamma^{(1)}_{\Omega^a_\alpha A^{*\gamma}_d}(-q)\Gamma^{(0)}_{A^d_\gamma A^r_\rho A^s_\sigma}(p_2,-p_1)
+ (b')\nonumber \\
(b')&=&-gf^{amn}g_\alpha^\nu\!\int_{k_1}\!\frac1{k_1^2}\frac1{k_2^2}{\mathcal K}^{(0)}_{ c^{m}A^r_\rho A^s_\sigma A^{*\gamma}_d}(p_2,-p_1,k_2)\Gamma^{(0)}_{A^d_\gamma A^n_\nu}(k_2).
\label{PT_contr_ggg}
\eea
\indent
At this point the PT calculation is over, and one needs to reshuffle the pieces generated.
Since $(b')$ is a vertex-like contribution, the PT vertex $\widehat{\Gamma}_{AAA}$, which is obtained by considering the corresponding Green's function embedded in the $S$-matrix element, will be given by
\be
(b)^{\mathrm F}+(b')+(c)+(d)=(b)+(c)+(d)+(b')-(b)^{\mathrm P},
\ee
or
\be
i\widehat{\Gamma}^{(1)}_{A^a_\alpha A^r_\rho A^s_\sigma}(p_2,-p_1)
=i\Gamma^{(1)}_{A^a_\alpha A^r_\rho A^s_\sigma}(p_2,-p_1)+\Gamma^{(1)}_{\Omega^a_\alpha A^{*\gamma}_d}(-q)\Gamma^{(0)}_{A^d_\gamma A^r_\rho A^s_\sigma}(p_2,-p_1).
\label{pippo1}
\ee
The PT self-energy will be given instead by the combination $(a)+2(b)^{\mathrm P}$ [the factor of 2 coming from the mirror diagram of $(b)$], \ie
\be
\widehat{\Pi}^{(1)}_{\alpha\beta}(q)=\Pi^{(1)}_{\alpha\beta}(q)+\Pi^{{\rm P}(1)}_{\alpha\beta}(q), 
\ee
where we have defined, with the aid of  Eq.~(\ref{ga_tree_lev}),
\be
\Pi^{{\rm P}(1)}_{\alpha\beta}(q)
=2i\Gamma^{(1)}_{\Omega_\alpha A^{*\gamma}}(q)\Gamma^{(0)}_{A_\gamma A_\beta}(q).
\ee
\newline
\indent
We can now proceed to the comparison of the PT Green's functions with those of the BFG, by resorting to the BQIs. Clearly,
Eq.~(\ref{pippo1}) represents the one-loop expansion of the BQI~(\ref{BQI:ggg}), and we immediately conclude that
\be
\widehat{\Gamma}^{(1)}_{A^a_\alpha A^r_\rho A^s_\sigma}(p_2,-p_1)=\Gamma^{(1)}_{\widehat{A}^a_\alpha A^r_\rho A^s_\sigma}(p_2,-p_1).
\ee
For the self-energy we have instead [recall that in our conventions $-\Gamma_{AA}(q)=\Pi(q)$]
\be
\delta^{ab}\widehat{\Pi}^{(1)}_{\alpha\beta}(q)=-\Gamma^{(1)}_{A^a_\alpha A^b_\beta}(q)+
2i\Gamma^{(1)}_{\Omega^a_\alpha A^{*\gamma}_d}(q)\Gamma^{(0)}_{A^d_\gamma A^b_\beta}(q)\,,
\label{BQI_1l}
\ee
which represents the one-loop version of the BQI of Eq.~(\ref{BQI:gg}), \ie we have
\be
\delta^{ab}\widehat{\Pi}^{(1)}_{\alpha\beta}(q)=-\Gamma^{(1)}_{\widehat{A}^a_\alpha\widehat{A}^b_\beta}(q).
\ee

\subsection{The two-loop case}
\noindent
As can  be inferred on  the basis of  the analysis carried out  in the
previous section, at the two-loop  level all the diagrams appearing in
the skeleton expansion of  the propagator and vertex are participating
in the  PT construction. Thus,  it would be  useless to carry  out the
same comparison between the diagrammatic  and the BV formulation as we
did at the  one-loop level: while the level of difficulty would  be the same as
that of the all-order analysis, no deeper understanding would be gained in the process.
\newline
\indent 
However, what we can still show is how the BV formalism enormously facilitates the comparison between the final PT result and the corresponding BFG Green's function, drastically reducing the additional effort required in order to compare the PT self-energy to the corresponding  BFG one. This is particularly relevant in the SM case, where the use of the BQIs has been instrumental for generalizing the PT algorithm to two loops~\cite{Binosi:2002bs}.  
Let us start with the two-loop gluon self-energy, which is given by [see Eq.~(\ref{FinalA})]
\be
{\widehat\Pi}^{(2)}_{\alpha\beta}(q) =
\Pi^{(2)}_{\alpha\beta}(q) + \Pi_{\alpha\beta}^{{\rm P}\,(2)}(q) -
R^{{\rm P}\,(2)}_{\alpha\beta}(q),
\label{2lPTtwopf}
\ee
with 
\bea
\Pi^{{\rm P}\,(2)}_{\alpha\beta}(q)-
R^{{\rm P}\,(2)}_{\alpha\beta}(q) & = & q^2P_\beta^\gamma(q)
\bigg\{I_4 L_{\alpha\gamma}(\ell,k)
+ I_{3}g_{\alpha\rho}  \nonumber \\
& - & 
I_1 \left[k_{\gamma}g_{\alpha\sigma}+  
\Gamma_{\sigma\gamma\alpha}^{(0)}(-k,-\ell,k+\ell)\right](\ell-q)^{\sigma}\bigg\}
\nonumber \\
&+& iV^{{\rm
P}\,(1)\,\gamma}_\beta(q)\Pi^{(1)}_{\gamma\alpha}(q)-q^2I_2P_{\alpha\beta}(q).
\label{ta}
\eea
On the other hand, the two loop version of the BQI~(\ref{BQI:gg}) satisfied by the gluon two-point function gives
\bea
{\widehat\Pi}^{(2)}_{\alpha\beta}(q) &=&
\Pi^{(2)}_{\alpha\beta}(q)+2i\Gamma^{(2)}_{\Omega_\alpha A^{*\gamma}}(q)\Gamma^{(0)}_{A_\gamma A_\beta}(q)
+2i\Gamma^{(1)}_{\Omega_\alpha A^{*\gamma}}(q)\Gamma^{(1)}_{A_\gamma A_\beta}(q)\nonumber \\
&+&\Gamma^{(1)}_{\Omega_\alpha A^{*\gamma}}(q)\Gamma^{(0)}_{A_\gamma A_\epsilon}(q)
\Gamma^{(1)}_{\Omega_\beta A^{*\epsilon}}(q),
\eea
where we have used the fact that $-\Gamma_{A_\alpha A_\beta}=\Pi_{\alpha\beta}$.
From the one-loop results of the previous subsections it is immediate to establish the identity
\be
2i\Gamma^{(1)}_{\Omega_\alpha A^{*\gamma}}(q)\Gamma^{(1)}_{A_\gamma A_\beta}(q)+\Gamma^{(1)}_{\Omega_\alpha A^{*\gamma}}(q)\Gamma^{(0)}_{A_\gamma A_\epsilon}(q)= iV^{{\rm
P}\,(1)\,\gamma}_\beta(q)\Pi^{(1)}_{\gamma\alpha}(q)-q^2I_2P_{\alpha\beta}(q).
\ee
Finally, the perturbative expansion at the two-loop level of the auxiliary function $\Gamma_{\Omega A^*}$ is shown in \Figref{2l_aux_exp}; 
one has the results
\bea
& & {(a)}=\frac12I_1\Gamma^{(0)}_{\sigma\gamma\alpha}(-k,-\ell,k+\ell)
(\ell-q)^\sigma, \qquad
{(b)}=\frac12I_1(\ell-q)_\alpha k_\gamma, \nonumber\\
& & {(c)}=-\frac12I_4L_{\alpha\gamma}(\ell,k), \hspace{3.85cm}
{(d)}=-\frac12I_3g_{\alpha\gamma},
\label{li}
\eea
or
\bea
2i\Gamma^{(2)}_{\Omega_\alpha A^{*\gamma}}(q)\Gamma^{(0)}_{A_\gamma A_\beta}(q)&=&q^2P_\beta^\gamma(q)
\bigg\{I_4 L_{\alpha\gamma}(\ell,k)
+ I_{3}g_{\alpha\gamma}  \nonumber \\
& - & 
I_1 \left[k_{\gamma}g_{\alpha\sigma}+  
\Gamma_{\sigma\gamma\alpha}^{(0)}(-k,-\ell,k+\ell)\right](\ell-q)^{\sigma}\bigg\}.
\eea
Putting everything together, we see that 
\bea
\Pi^{{\rm P}\,(2)}_{\alpha\beta}(q)-
R^{{\rm P}\,(2)}_{\alpha\beta}(q) & = &2i\Gamma^{(2)}_{\Omega_\alpha A^{*\gamma}}(q)\Gamma^{(0)}_{A_\gamma A_\beta}(q)
+2i\Gamma^{(1)}_{\Omega_\alpha A^{*\gamma}}(q)\Gamma^{(1)}_{A_\gamma A_\beta}(q)\nonumber \\
&+&\Gamma^{(1)}_{\Omega_\alpha A^{*\gamma}}(q)\Gamma^{(0)}_{A_\gamma A_\epsilon}(q),
\eea
which thus implies
\be
\widehat{\Pi}^{(2)}_{\alpha\beta}(q)=\widetilde{\Pi}^{(2)}_{\alpha\beta}(q),
\ee
where $\widetilde{\Pi}^{(2)}_{\alpha\beta}(q)=-\Gamma_{A_\alpha A_\beta}$ is the BFG two-loop gluon self-energy. 
\begin{figure}[!t]
\bce
\includegraphics[width=12cm]{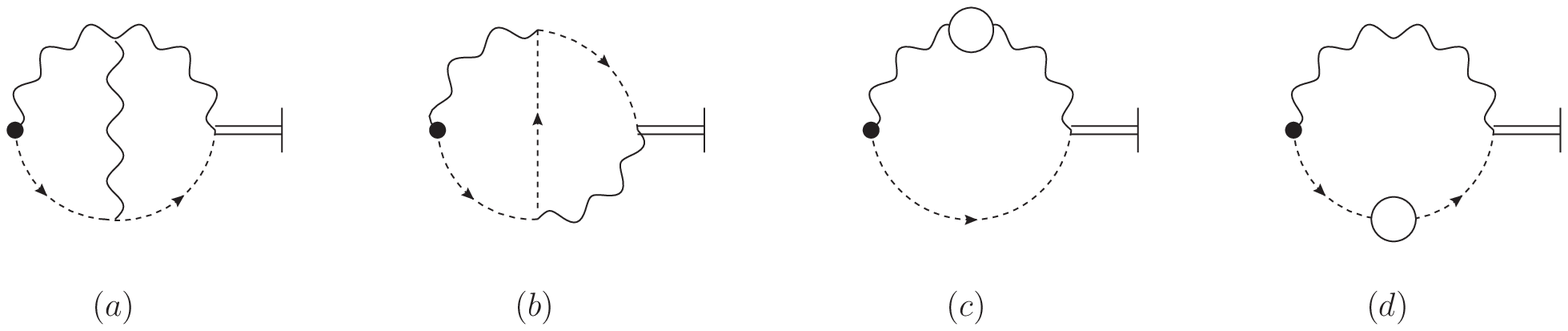}
\caption{\figlab{2l_aux_exp}The two-loop diagrams contributing to the auxiliary function $\Gamma_{\Omega A^*}$.}
\ece
\end{figure}
\indent
The proof in the case of the PT two-loop quark-gluon vertex is even easier; after using the one-loop result, the two-loop PT vertex can be cast in the form
\bea
\widehat\Gamma^{(2)}_{\alpha}(p_1,p_2)  & = &
\Gamma^{(2)}_{\alpha}(p_1,p_2)+
\frac i2V_\alpha^{{\rm P}\,(1)\,\rho}(q)\Gamma_{\rho}^{(1)}(p_1,p_2)
-\frac12\bigg\{I_4 L_{\alpha\rho}(\ell,k)
+ I_{3}g_{\alpha\rho} \nonumber \\
& - & I_1 \left[k_{\rho}g_{\alpha\sigma}+  
\Gamma_{\sigma\rho\alpha}^{(0)}(-k,-\ell,k+\ell)\right](\ell-q)^{\sigma}\bigg\}
\gamma^\rho.
\eea
The two-loop BQI~(\ref{BQI:gff}) for the gluon-quark vertex reads
\bea
i\Gamma^{(2)}_{\widehat{A}^a_\alpha \psi\bar\psi}(p_2,-p_1)&=&i\Gamma^{(2)}_{\widehat{A}^a_\alpha \psi\bar\psi}(p_2,-p_1)+\Gamma^{(2)}_{\Omega_\alpha A^{*\gamma}}(q)\Gamma^{(0)}_{A^a_\gamma \psi\bar\psi}(p_2,-p_1)\nonumber \\
&+&\Gamma^{(1)}_{\Omega_\alpha A^{*\gamma}}(q)\Gamma^{(1)}_{A^a_\gamma \psi\bar\psi}(p_2,-p_1),
\eea
where we have neglected pieces that vanish on-shell. Then one has
\bsub
\bea
\Gamma^{(1)}_{\Omega_\alpha A^{*\rho}}(q)\Gamma^{(1)}_{A^a_\rho \psi\bar\psi}(p_2,-p_1)&=&\frac i2 V^{(1)\,\rm{P}\,\rho}_{\alpha}(q)\Gamma^{(1)}_{A^a_\rho \psi\bar\psi}(p_2,-p_1), \\
\Gamma^{(2)}_{\Omega_\alpha A^{*\gamma}}(q)\Gamma^{(0)}_{A_\gamma \psi\bar\psi}(p_2,-p_1)&=&-\frac12\bigg\{I_4 L_{\alpha\rho}(\ell,k)
+ I_{3}g_{\alpha\rho} \nonumber \\
& - & I_1 \left[k_{\rho}g_{\alpha\sigma}+  
\Gamma_{\sigma\rho\alpha}^{(0)}(-k,-\ell,k+\ell)\right](\ell-q)^{\sigma}\bigg\}
\gamma^\rho,
\label{p2}
\eea
\esub
where in  (\ref{p2}) we have suppressed a factor $gt^a$. 
Thus we get the equality between the two-loop PT and BFG quark-gluon vertex
[one needs to take into account that, due to the different conventions used,  $\Gamma_\alpha(p_1,p_2)=i\Gamma_{A_\alpha\psi\bar\psi}(p_2,-p_1)$].
\newline
\indent
 
\newpage


\section{\seclab{PT_SDEs}The PT Schwinger-Dyson Equations for QCD Green's functions}
\noindent
After recasting the PT algorithm in the BV language, and making contact with the original diagrammatic formulation,
we are now ready to face the final challenge:
apply the PT program to the (non-perturbative) Schwinger-Dyson equations
(SDEs) of QCD. In fact, historically, this was the main motivation for introducing the method in the 
first place~\cite{Cornwall:1976hg,Cornwall:1981ru,Cornwall:1981zr,Cornwall:1989gv}.
\newline
\indent
In this section, after briefly discussing the difficulties encountered 
when attempting to truncate the 
SDEs within the conventional formalism, we will scrutinize, 
in detail how the PT algorithm can be extended to the construction 
of new SDEs for the QCD Green's functions. Next we proceed to the actual construction of  
the new SDEs for the gluon two- and three-point functions. We will finally discuss 
in depth the theoretical and practical advantages of the new SDE series.

\subsection{SDEs for non-Abelian gauge theories: difficulties with the conventional formulation}
\noindent
The SDEs provide  a formal framework
for tackling physics  problems requiring a non perturbative treatment. 
In fact,  even though  these  equations  are   derived  from  an  expansion  about  the
free-field  vacuum,  they finally  make  no  reference  to it,  or  to
perturbation theory,  and can be  used to address problems  related to
chiral  symmetry  breaking, dynamical  mass  generation, formation  of
bound      states,     and     other      non-perturbative     
effects~\cite{Cornwall:1974vz,Marciano:1977su}.  
\newline
\indent
In practice, however, their usefulness hinges crucially on one's ability to devise
a  self-\linebreak consistent truncation  scheme  that would  select a  tractable
subset of these equations,  without compromising the physics one hopes
to describe.  Inventing such a scheme for the SDE of gauge theories is
a highly non-trivial proposition. The problem originates from the fact
that the SDEs are built out of unphysical Green's functions; thus, the
extraction  of  reliable  physical  information depends  critically  on
the delicate all-order cancellations we have been describing in this review,  
which may be inadvertently distorted
in the  process of the truncation.   For example, several of the  issues related to
the truncation of  the SDEs of QED have been addressed  in~\cite{Curtis:1990zs,Curtis:1990zr,Curtis:1993py,Bashir:1994az,Bashir:1995qr,Bashir:1997qt,Bashir:2002dz,Bashir:2004yt,Kondo:1988md,Sauli:2002tk}; it goes without saying that   
the  situation  becomes even  more
complicated for  strongly coupled non-Abelian gauge  theories, such as
QCD~\cite{Mandelstam:1979xd}, mainly because  of the following two reasons.

\begin{itemize}

\item[{\it i}.]  As we have seen, the  complications  caused  by  the  dependence  of  the  Green's
functions on the gfp are more acute in non-Abelian
gauge-theories. For example, recall that in  QED  the photon
self-energy (vacuum  polarization) is independent of the 
gfp, both perturbatively (to all orders)
and  non-perturbatively; when 
multiplied by $e^2$ it forms  a physical observable,
the QED effective charge.  In  contradistinction, the gluon self-energy is
gfp-dependent  already at one loop;  depending on  the gauge-fixing
scheme employed,  this dependence may  be more or less  virulent.  This
difference is clearly of little practical importance when computing $S$-matrix
elements  at a fixed order in  perturbation theory, but has far-reaching
consequences  when  attempting  to  truncate  the  corresponding  SDEs,
written in some gauge. Moreover, contrary to what happens in the perturbative calculation,
even if one were to put together 
the non-perturbative expressions from these truncated SDEs to form
a physical observable, the gauge-cancellations may 
not go through completely, because the process of the truncation 
might have distorted them. Thus, there is a high probability of ending up
with a residual gauge-dependence infesting one's non-perturbative
prediction for a physical observable.  
\newline
\item[{\it ii}.] In Abelian gauge theories the Green's functions satisfy linear 
WIs, which in non-Abelian theories are replaced by the non-linear STIs, involving,
in addition to the basic Green's functions of the theory, various composite ghost operators. 
In order to appreciate how the fact that the Green's functions 
satisfy STIs may complicate the truncation procedure of the SDEs,
let us consider the simplest STI (and WI in this case)
satisfied by the photon and gluon self-energies alike, namely
\be
q^{\alpha} \Pi_{\alpha\beta}(q) = 0.
\label{fundtrans}
\ee
Eq.~(\ref{fundtrans}) is without a doubt the most fundamental 
statement at the level of Green's functions that one can obtain
from the BRST symmetry: it affirms the transversality of the 
gauge-boson self-energy, be it a photon or a gluon, and is valid both  
perturbatively to all orders as well as non-perturbatively. 
The problem stems from the fact that 
in the SDE of $\Pi_{\alpha\beta}$ enter higher order 
Green's functions, 
namely the fully-dressed fundamental vertices of the theory.  
 It is these latter Green's functions that in the Abelian context 
satisfy WIs, whereas in the non-Abelian context satisfy STIs.
Thus, whereas in QED the validity of Eq.~(\ref{fundtrans}) 
can be easily seen at  
the level of the SDE, simply because 
$q^{\alpha}\Gamma_{\alpha}(p,p+q) =e\left[ S^{-1}(p+q)-S^{-1}(p)\right]$, in QCD 
proving Eq.~(\ref{fundtrans}) explicitly, \ie 
by contracting with $q^{\alpha}$ the SDE 
of the gluon self-energy, requires a subtle 
conspiracy of all the (fully-dressed) vertices appearing in it.
Truncating the SDE naively usually amounts to leaving out some of these vertices, 
and, as a result, Eq.~(\ref{fundtrans}) is compromised.

\end{itemize}
\begin{figure}[!t]
\includegraphics[width=14cm]{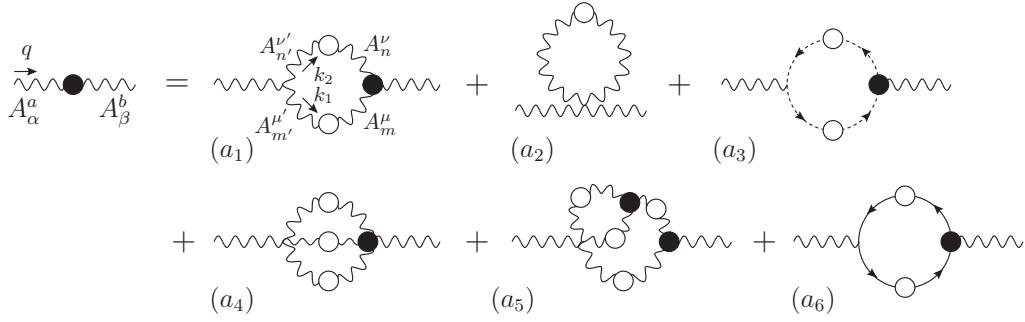}
\caption{\figlab{gg_SDE} The SDE satisfied by the gluon self-energy $-\Gamma_{AA}$. The symmetry factors of the diagrams are $s_{a_1}=s_{a_2}=s_{a_6}=1/2$, $s_{a_3}=s_{a_4}=-1$, $s_{a_5}=1/6$.}
\end{figure}
To be concrete, consider the SDE for the gluon propagator shown in \Figref{gg_SDE}; 
then, Eq.~(\ref{fundtrans}) translates at the level of the SDE to
the statement
\be
q^{\alpha} \sum_{i=1}^{6}(a_i)_{\alpha\beta} =0.
\label{convSD}
\ee
The diagrammatic verification 
of (\ref{convSD}), \ie through contraction of the individual graphs by $q^{\alpha}$, 
is practically very difficult, essentially due to the 
complicated STIs satisfied by the vertices involved. The most typical example
of such an STI is the one satisfied by the 
conventional three-gluon vertex of Eq.~(\ref{sti3gv}).
In addition, some of the pertinent STIs are either too complicated,
such as that of the conventional four-gluon vertex, or they 
cannot be cast in a particularly convenient  form.
For instance, in the case of the conventional gluon-ghost vertex, $\Gamma_{\mu}(q,p)$, 
the STI that one may obtain formally for  $q^{\mu}\Gamma_{\mu}(q,p)$ 
is the sum of two terms, one of which is $p^{\mu}\Gamma_{\mu}(q,p)$; this
clearly limits its usefulness in applications.
\newline
\indent
The main practical consequence of these complicated STIs  
is that one cannot truncate (\ref{convSD}) in any obvious way 
without violating the transversality of the resulting $\Pi_{\alpha\beta}(q)$.
For example, keeping only graphs $(a_1)$ and $(a_2)$ is not correct even perturbatively,
since the ghost loop is crucial for the transversality of  $\Pi_{\alpha\beta}$
already at one-loop;
adding $(a_3)$ is still not sufficient for a SD analysis, because
(beyond one-loop) $q^{\alpha}[(a_1)+(a_2) + (a_3)]_{\alpha\beta} \neq 0$.
\newline
\indent
Last but not least, these complications  are often compounded by 
additional problems related to the loss 
of multiplicative renormalizability and the inability to 
form renormalization-group invariant quantities.
\newline
\indent
It should be clear by now that the difficulties pointed out are exactly of the type that 
can be circumvented using a PT approach (in fact, a gauge invariant truncation scheme for SDEs has been the original motivation 
for introducing it).
In particular, the way \eg point ({\it i}) is resolved,
for the  prototype case  of the gluon  self-energy, is  the following.
The  BFG  is  a privileged gauge,  
in  the  sense that  it  is  selected  {\it
dynamically} when  the gluon self-energy  is embedded into  a physical
observable (such as an on-shell test amplitude).   Specifically, the BFG
captures the net propagator-like subamplitude emerging after QED-like
conditions have been replicated inside the test-amplitude, by means of
the PT  procedure. Thus, once  the PT rearrangements have  taken place,
the  propagator  is removed  from  the  amplitude  and is  studied  in
isolation: one considers the SDE for the background gluon self-energy,
$\widehat{\Pi}_{\alpha\beta}$,  at $\xiQ=1$.   Solving the  SDE in
the BFG  eliminates any gauge-related exchanges  between the solutions
obtained   for   $\widehat{\Pi}_{\alpha\beta}$   and  other   Green's
functions, when put together  to form observables; thus, the solutions
are free of gauge artifacts.
Regarding point ({\it ii}), all full vertices appearing in the new SDE  
satisfy now Abelian WIs; as a result, 
gluonic   and  ghost   contributions   are  {\it
separately}   transverse,    within   {\it   each}    order   in   the
``dressed-loop'' expansion (as was already noticed in our discussion of the BFM two-loop 
gluon self-energy, see subsection~\ref{2lBFM}). Thus, as we will see explicitly in the next section,
it is much easier to devise truncation schemes that manifestly preserve the validity of
Eq.~(\ref{fundtrans}). Let us for now turn to the problem of how the PT algorithm can be generalized 
to a non-perturbative setting.

\subsection{The PT algorithm for Schwinger-Dyson equations\label{SDE}}
\noindent
The BV (re)formulation of the PT algorithm reveals its true power when
dealing  with  the  problem  of constructing  (off-shell)  PT  Green's
functions without resorting to fixed  order calculations, as it is the
case                 when                 dealing                 with
SDEs~\cite{Binosi:2006da,Binosi:2007pi,Binosi:2008qk}.  In particular,
it is immediate to realize  that the (one-loop) procedure described at
the end of the previous  section for the various QCD Green's functions
carries  over  practically  unaltered   to  the  construction  of  the
corresponding (non-perturbative)  SDEs.  This is basically  due to the
following                         three                        crucial
observations~\cite{Binosi:2006da,Binosi:2007pi,Binosi:2008qk}:
\begin{itemize}
\item[$\ast$] the pinching  momenta will be always  determined from
the  tree-level decomposition given in Eq.~(\ref{decomp}); 
\newline
\item[$\ast$] their
action is completely  fixed by the structure of  the STIs they trigger; 
\newline
\item[$\ast$] the kernels encountered in these STIs  are those 
appearing in  the corresponding BQIs;  thus, it is always  possible to
write  the  result of  the  action of  the pinching  momenta  in terms  of the 
auxiliary Green's functions of the BQIs.
\end{itemize}
\indent
The only operational difference is that, in the case of the SDEs for vertices,  
{\it all} external legs  will be {\it off-shell}. 
This is of course unavoidable, because the (fully dressed) vertices 
are nested inside, for example, the SDE of the off-shell gluon self-energy, 
(see \Figref{gg_SDE}); thus, all legs are off-shell (the external leg because the physical 
four-momentum ($q^2$) is off-shell, 
and the legs inside the diagrams because they are irrigated by the virtual off-shell momenta).
As a result, 
the equations of motion employed in the previous section in order to drop some of the resulting terms
cannot be used in this case; therefore, 
the corresponding contributions, proportional to inverse 
self-energies, do not vanish, and form part of the resulting BQI.
\newline
\indent
Thus the  PT rules for the  construction of SDEs may be summarized as follows: 
\begin{itemize}
\item[$\ast$]
   For the SDEs  of vertices, with {\it all  three} external legs {\it
  off-shell},  the pinching  momenta,  coming from  the only  external
  three-gluon  vertex   undergoing  the  decomposition  (\ref{decomp}),
  generate four types of terms: one of them, corresponding to the term
  $(b')$  in   Eq.~(\ref{PT_contr_ggg}),  is  a   genuine  vertex-like
  contribution that  must be included in  the final PT  answer for the
  vertex under construction, while the remaining three-terms will form
  part  of  the  emerging  BQI (and thus would be discarded from the PT vertex).     
  These  latter  terms  have  a  very
  characteristic structure, which facilitates their identification in
  the calculation.   Specifically, one of them  is always proportional
  to the auxiliary function $\Gamma_{\Omega A^*}$, while the other two
  are proportional  to the inverse propagators of  the fields entering
  into  the  two  legs  that  did {\it not}  undergo  the  decomposition  of
  Eq.~(\ref{decomp}).
\newline
\item[$\ast$] In the case of the SDE for the gluon propagator, 
the  pinching  momenta  will  only  generate  pieces  proportional  to
$\Gamma_{\Omega A^*}$;  these terms should be  discarded from the  PT answer
for the  gluon two-point function  (since they  are exactly those  that cancel against
the contribution coming from the corresponding vertices), and will contribute instead 
to the corresponding BQI.
\end{itemize}
\indent
Thus, using these rules, and starting  from the corresponding SDEs written in the Feynman gauge of the $R_\xi$,
we  will first derive the new SDEs  for the $\widehat{\Gamma}_{AAA}$ vertex (the quark-gluon vertex $\widehat{\Gamma}_{A\psi\bar\psi})$ has been also explicitly constructed in~\cite{Binosi:2008qk}), and will then address the more
complicated case of the SDE for the PT gluon propagator $\widehat{\Gamma}_{AA}$. 

\subsubsection{Three-gluon vertex\label{tgv}}

\begin{figure}[!t]
\bce
\includegraphics[width=14.5cm]{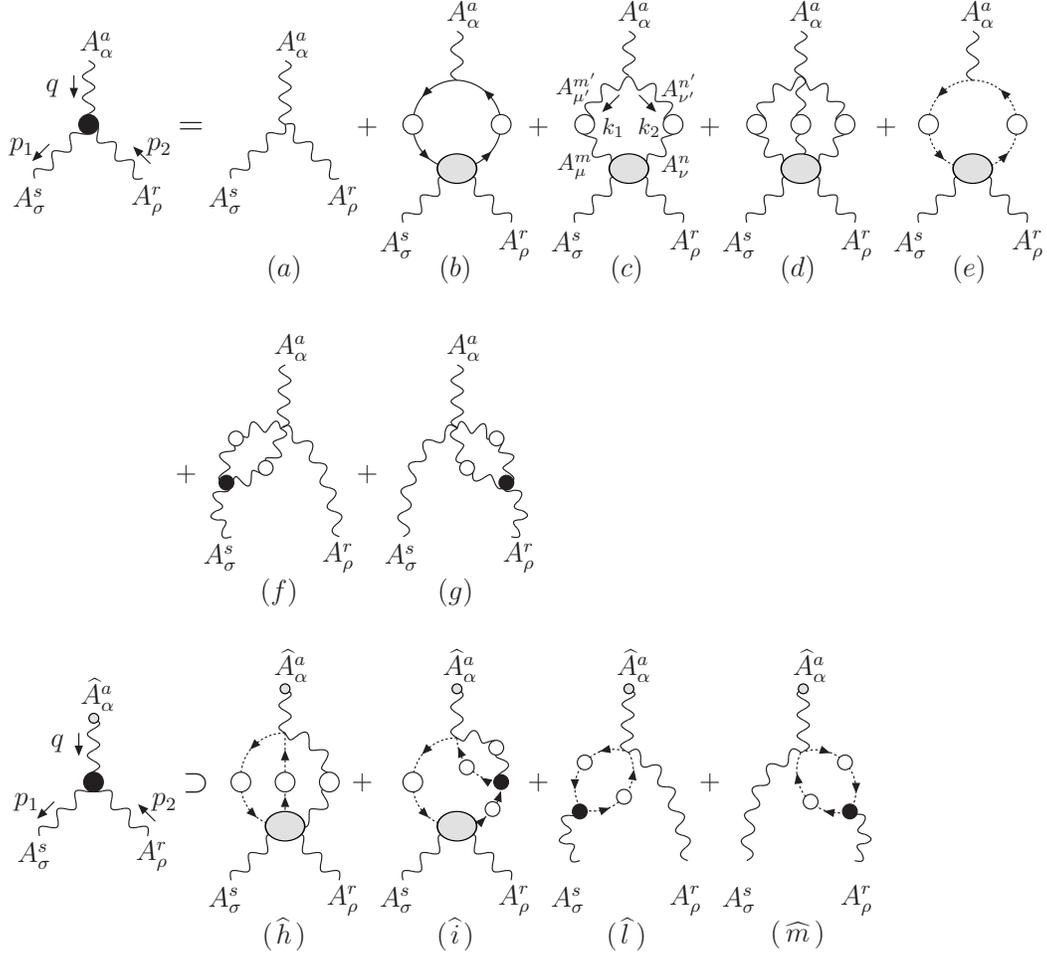}
\ece
\caption{\figlab{ggg_SDE}The SDE of the  three-gluon
vertex. The symmetry factors of  the $R_\xi$ (first and second line of
the   figure)  diagrams   are   $s(a,b)=1$.  $s(c)=1/2$,   $s(d)=1/6$,
$s(e)=-1$,  $s(f,g)=1/2$. In  the third  line we  show  the additional
topologies  present  in the  BFM  version  of  the equation 
[$s(\widehat{h},\widehat{i},\widehat{l},\widehat{m})=-1$], 
generated during the PT procedure.}
\end{figure}
\noindent
The SDE for the conventional three-gluon vertex is shown in \Figref{ggg_SDE}. Before starting the calculation, let us emphasize that the purpose of this exercise is to generate dynamically the vertex $\Gamma_{\widehat{A}AA}$ and {\it not} the fully Bose-symmetric vertex~\cite{Cornwall:1989gv,Binger:2006sj} $\Gamma_{\widehat{A}\widehat{A}\widehat{A}}$ studied in \secref{QCD_one-loop}. 
The reason for this is the fact that it is the former vertex that appears in the SDEs for the gluon propagator 
[see, {\it e.g.}, diagram $(d_1)$ in \Figref{ghatghat_SDE}], making it the relevant vertex to be studied at this level.
\newline
\indent
We then begin by carrying out the  decomposition of Eq.~(\ref{decomp}) to the  tree-level vertex appearing in diagram $(c)$, 
which will be the only one modified in our construction. Concentrating on the $\Gamma^\mathrm{P}$ part,  
we find
\be
(c)^\mathrm{P}= gf^{amn'}g_{\alpha\nu'}\int_{k_1}\frac1{k_1^2}\Delta_{n'n}^{\nu'\nu}(k_2)k_1^{\mu}{\mathcal K}_{A^m_\mu A^n_\nu A^r_\rho A^s_\sigma}(k_2,p_2,-p_1),
\label{neqref}
\ee
where the kernel $K_{AAAA}$ is shown in \Figref{gggg_SDker}.
\newline
\indent
Using the STI~(\ref{STISDgggg}) satisfied by this kernel, we obtain from Eq.~(\ref{neqref}) four terms, namely \mbox{$(c)^\mathrm{P}=(s_1)+(s_2)+(s_3)+(s_4)$}. Then, using Eqs~(\ref{BQI:auxOmAs}) and~(\ref{BQI:auxOmAbarc}), it is fairly straightforward to prove that
\begin{eqnarray}
(s_1)&=& -\Gamma_{\Omega^a_\alpha A^{*\gamma}_d}(-q)\Gamma_{A^d_\gamma A^r_\rho A^s_\sigma}(p_2,-p_1),\nonumber\\
(s_2) & = &-\Gamma_{\Omega^a_\alpha A^s_\sigma A^{*\gamma}_d}(-p_1,p_2)\Gamma_{A^d_\gamma A^r_\rho}(p_2),\nonumber\\
(s_3)& = & -\Gamma_{\Omega^a_\alpha A^r_\rho A^{*\gamma}_d}(p_2,-p_1)\Gamma_{A^d_\gamma A^s_\sigma}(p_1).
\label{gggs3-3}
\end{eqnarray}
\begin{figure}[!t]
\bce
\includegraphics[width=15cm]{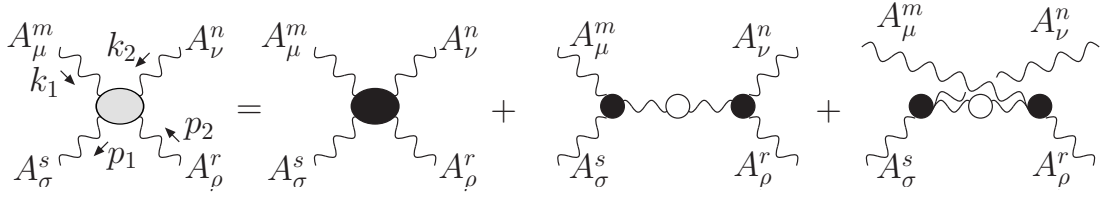}
\ece
\caption{\figlab{gggg_SDker}Skeleton expansion of the kernel appearing in the SDE for the three-gluon vertex [diagram $(c)$ of \Figref{ggg_SDE}]}
\end{figure}
\newline
\indent
Evidently, the term $(s_1)$ gives rise to  the  propagator-like
contribution,~\footnote{Note that this term  is identical   
to the one found in the construction of the SDE for the quark-gluon vertex
case~\cite{Binosi:2008qk}. This is the (all-order) manifestation  of the PT 
process-independence (see 2.4.3): the propagator-like 
contributions do not depend on the details of the external (embedding) particles.} 
which, in the $S$-matrix PT, would  be allotted  to  the new two-point  function, $\widehat{\Gamma}_{AA}$.  
As for $(s_2)$  and $(s_3)$, they correspond to terms that would vanish on-shell, but now, 
due to the off-shell condition of the external legs, must be retained in the final answer.
\newline
\indent
Finally, let us consider  the term $(s_4)$, given by
\begin{eqnarray}
(s_4)&=& gf^{am'n'}g_{\alpha\nu'}\int_{k_1}D^{m'm}(k_1)\Delta
_{n'n}^{\nu'\nu}(k_2){\mathcal K}_{c^{m} A_d^{*\gamma}A^r_\rho A^s_\sigma }(k_2,p_2,-p_1)\Gamma_{A^d_\gamma A^n_\nu}(k_2),\hspace{1cm}
\end{eqnarray}
and show how it combines with the remaining $R_\xi$ diagrams to generate the BFG vertex $\Gamma_{\widehat{A}AA}$. To this end, using Eq.~(\ref{gainvprop}) and the FPE satisfied by the kernel ${\mathcal K}_{c A^*A A}$, 
we can write $(s_4)=(s_{4a})+(s_{4b})$,  with 
\begin{eqnarray}
(s_{4a})&=&-igf^{am'd}g_{\alpha\gamma}\int_{k_1}D^{m'm}(k_1){\mathcal K}_{c^{m} A_d^{*\gamma}A^r_\rho A^s_\sigma}(k_2,p_2,-p_1), \nonumber \\
(s_{4b})&=&-gf^{am'n'}g_{\alpha\nu'}\int_{k_1}\delta^{dn'}\frac{k_2^{\nu'}}{k_2^2}D^{m'm}(k_1){\mathcal K}_{c^{m}\bar c^dA^r_\rho A^s_\sigma}(k_2,p_2,-p_1).
\end{eqnarray}
The kernel ${\mathcal K}_{c\bar c AA}$ is defined by 
replacing in Eq.~(\ref{1PRker:gggg3}) every anti-field leg $A^*$ by 
the corresponding anti-ghost field $\bar c$.
\begin{figure}[t]
\bce
\includegraphics[width=11cm]{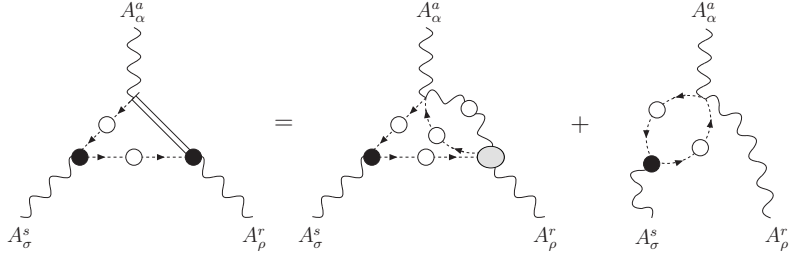}
\ece
\caption{\figlab{1PR_treelevel}The 1PR terms appearing in ${\mathcal K}_{c^mA^{*\gamma}_dA^r_\rho A^s_\sigma}$ 
contain a tree-level contribution  
generating the missing BFM topologies. 
Here we show  the case for $(\widehat{l})$;
 a symmetric term  generates $(\widehat{m})$. 
The first term on the rhs is part of the skeleton 
expansion of diagram $(\widehat{h})$ of \Figref{ggg_SDE}. 
Notice that the lhs is simply a pictorial representation of 
the rhs: the anti-fields are static sources and do not propagate.}
\end{figure}
First of all, notice that the apparently missing topologies $(\widehat{l})$ and $(\widehat{m})$ of \Figref{ggg_SDE} will be  
generated by the tree-level contribution appearing in the SDE of the auxiliary function $\Gamma_{cAA^*}$. To prove this, let us write
\begin{eqnarray}
{\mathcal K}_{c^{m} A_d^{*\gamma} A^r_\rho A^s_\sigma}(k_2,p_2,-p_1)
&=&
{\mathcal K}'_{c^{m} A_d^{*\gamma} A^r_\rho A^s_\sigma}(k_2,p_2,-p_1)-igf^{dre'}g_\rho^\gamma\Gamma_{c^{m}A^s_\sigma\bar c^{e'}}(-p_1,\ell)D^{ee'}(\ell)\nonumber \\
&-&igf^{dse}g^\gamma_\sigma D^{ee'}(\ell')\Gamma_{c^{m}A^r_\rho \bar c^{e'}}(p_2,-\ell'),
\end{eqnarray}
where the prime denotes that the $\Gamma_{cAA^*}$ appearing in the corresponding 1PR terms starts at one-loop. We then find (see also~\Figref{1PR_treelevel})
\begin{eqnarray}
(s_{4a})&=&(s'_{4a})+(\widehat{l})+(\widehat{m}),\nonumber \\
(s'_{4a})&=&-igf^{am'd}g_{\alpha\gamma}\int_{k_1}D^{m'm}(k_1){\mathcal K}'_{c^{m}A_d^{*\gamma}A^r_\rho A^s_\sigma }(k_2,p_2,-p_1).
\end{eqnarray}
Consider then the terms $(s'_{4a})$ and $(s_{4b})$. Their general structure suggests that $(s'_{4a})$ 
should give rise to the ghost quadrilinear vertex, while  $(s_{4b})$, when added to diagram $(e)$, should symmetrize the trilinear ghost gluon coupling. It turns out that this expectation is essentially correct, but its realization is not immediate, mainly due to  the fact that the $(s_{4b})$ contains a tree-level instead of a full ghost propagator [$(k_2^2)^{-1}$ instead of  $D(k_2)$], while
$(s'_{4a})$ can  reproduce, at most,  diagram $(\widehat{h})$ of \Figref{ggg_SDE}, but not  $(\widehat{i})$. 
The solution to this apparent mismatch is rather subtle: one must employ the SDE satisfied by the {\it ghost propagator}. This SDE   
is common to both the $R_{\xi}$-gauge and the BFM, given that there are no background ghosts.
\newline
\indent
To show how this works in detail,  we
add and subtract to the sum $(s'_{4a})+(s_{4b})$ the missing term  (see \Figref{cggAstar_SDaux}), obtaining
\begin{eqnarray}
(s'_{4a})&=&-igf^{am'd}g_{\alpha\gamma}\int_{k_1}D^{m'm}(k_1)\left[{\mathcal K}'_{c^{m}A_d^{*\gamma}A^r_\rho A^s_\sigma }(k_2,p_2,-p_1)\right.\nonumber \\
&-&\left.\Gamma'_{c^gA_d^{*\gamma}}(k_2)iD^{gg'}(k_2){\mathcal K}_{c^{m}\bar c^{g'}A^r_\rho A^s_\sigma}(k_2,p_2,-p_1)\right]\nonumber\\
&=&-igf^{am'd}g_{\alpha\gamma}\int_{k_1}D^{m'm}(k_1){\mathcal K}^{\mathrm{full}}_{c^{m}A^{*\gamma}_d\psi\bar\psi}(k_2,p_2,-p_1)\label{K_full}\nonumber \\
(s_{4b})&=&-gf^{am'n'}g^{\nu'}_\alpha\int_{k_1}\left[\delta^{dn'}\frac{k_{2\nu'}}{k_2^2}
-\Gamma'_{c^eA_{\nu'}^{*n'}}(k_2)D^{ed}(k_2)
\right]\times\nonumber \\
&\times&D^{m'm}(k_1){\mathcal K}_{c^{m}\bar c^dA^r_\rho A^s_\sigma}(k_2,p_2,-p_1).
\end{eqnarray}
\begin{figure}[!t]
\bce
\includegraphics[width=12cm]{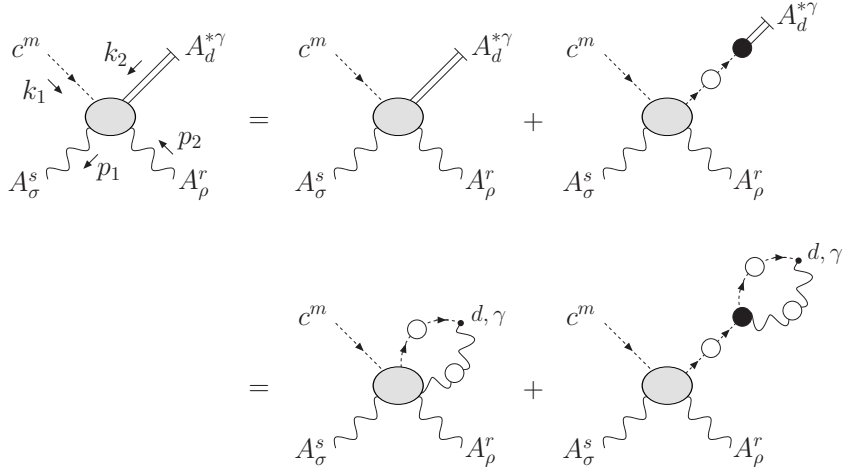}
\ece
\caption{\figlab{cggAstar_SDaux}Diagrammatic decomposition of the kernel 
${\mathcal K}^{\mathrm{full}}_{c^{m}A^{*\gamma}_d A^r_\rho A^s_\sigma}$ introduced in Eq.~(\ref{K_full}). 
The first term represents the kernel ${\mathcal K}'_{c^{m} A_d^{*\gamma} A^r_\rho A^s_\sigma}$; 
therefore the $\Gamma_{cAA^*}$ appearing in the corresponding 1PR terms start at one-loop. 
The second term is the one added (and subtracted) to the original sum $(s'_{4a})+(s_{4b})$.
After replacing the gluon anti-field  $A^{*\gamma}_d$ with the corresponding composite operator (second line),  
this kernel generates the BFM terms $(\widehat{h})+(\widehat{i})$.}
\end{figure}
Then, using Eq~(\ref{Astrick}) (which can be safely done now, since tree-level contribution
has been already taken into account) 
\be
(s'_{4a})=(\widehat{h})+(\widehat{i}).
\ee
We finally turn to $(s_{4b})$ and consider the ghost SD equation of \Figref{ghost_SDE}. One has
\begin{equation}
iD^{dn'}(k_2)=i\frac{\delta^{dn'}}{k_2^2}+i\frac{\delta^{dg}}{k_2^2}\left[-\Gamma'_{c^g \bar c^{g'}}(k_2)\right]iD^{g'n'}(k_2),
\label{ghost_SDE}
\end{equation}
where $\Gamma'_{c^g \bar c^{g'}}$ is given by  $\Gamma_{c^g \bar c^{g'}}$ minus its tree-level part.
Multiplying the above equation by $k_2^2$, using the FPE~(\ref{FPE:ghprop}), 
and factoring out a $k_{2\nu'}$ we get the relation 
\be
k_{2\nu'}D^{dn'}(k_2)=\delta^{dn'}\frac{k_{2\nu'}}{k_2^2}-\Gamma'_{c^gA_{\nu'}^{*n'}}(k_2)D^{gd}(k_2).
\label{ghSDE2}
\ee
Therefore, we obtain 
\be
(s_{4b})=-gf^{am'n'}\int_{k_1}k_{2\alpha}D^{m'm}(k_1)D^{n'n}(k_2){\mathcal K}_{c^{m}\bar c^nA^r_\rho A^s_\sigma}(k_2,p_2,-p_1),
\ee
so that
\begin{equation}
(s_{4b})+(e)=(\widehat{e}).
\end{equation}
\newline
\indent
Using the tree-level Feynman rules (see \appref{Frules}),
it is straightforward  to establish that the graphs $(b)$, $(d)$, $(f)$, and $(g)$
can be converted to hatted ones automatically, and that 
$(c)^\mathrm{F}=(\widehat{c})$.
Thus, collecting all the pieces we have, and using the standard PT decomposition (\ref{decomp}) on the tree-level contribution $(a)$, we get 
\begin{eqnarray}
i\Gamma_{A^a_\alpha A^r_\rho A^s_\sigma}(p_2,-p_1)&=&-\Gamma_{\Omega^a_\alpha A^{*\gamma}_d}(-q)\Gamma_{A^d_\gamma A^r_\rho A^s_\sigma}(p_2,-p_1)-\Gamma_{\Omega^a_\alpha A^s_\sigma A^{*\gamma}_d}(-p_1,p_2)\Gamma_{A^d_\gamma A^{r}_\rho}(p_2)\nonumber \\
&-&\Gamma_{\Omega^a_\alpha A^r_\rho A^{*\gamma}_d}(p_2,-p_1)\Gamma_{A^d_\gamma A^{s}_\sigma}(p_1)+[(\widehat{a})+(\widehat{b})
+ (\widehat{c})
+(\widehat{d})+(\widehat{e})\nonumber\\
&+&(\widehat{f})+(\widehat{g})+
(\widehat{h})+(\widehat{i})+
(\widehat{l})+(\widehat{m})]^{ars}_{\alpha\rho\sigma}-igf^{ars}\Gamma^\mathrm{P}(p_2,-p_1).\hspace{.5cm}
\label{PT-ggg-res}
\end{eqnarray}
As in the previous case, the sum of diagrams in the brackets is nothing but the kernel expansion of the SDE governing the vertex $\Gamma_{\widehat{A}AA}$, {\it i.e.}, 
\bea
i\Gamma_{\widehat{A}^a_\alpha A^r_\rho A^s_\sigma}(p_2,-p_1)&=&[(\widehat{a})+(\widehat{b})
+ (\widehat{c})
+(\widehat{d})+(\widehat{e}))\nonumber \\
&+&(\widehat{f})+(\widehat{g})+(\widehat{h})+(\widehat{i})+
(\widehat{l})+(\widehat{m})]^{ars}_{\alpha\rho\sigma}.
\label{final_ggg}
\eea This, in turn, implies that Eq.~(\ref{PT-ggg-res}) represents the
BQI of  Eq.~(\ref{BQI:ggg}) up  to the last  (tree-level) term  on the
rhs.  Of  course, this tree-level  discrepancy is to be  expected, since
the PT  algorithm cannot possibly  work at tree-level if  the external
legs are amputated, as is the  case in the SDEs we are considering. To
be sure,  if we start  from the tree-level  $\Gamma^{(0)}_{AAA}$ only,
\ie without hooking (two of) the external legs to (conserved) external
currents,   we   can   still    carry   out   the   decomposition   of
Eq.~(\ref{decomp}), but the $\Gamma^\mathrm{P}$ term will have nothing
to act upon.
\begin{figure}[!t]
\bce
\includegraphics[width=12cm]{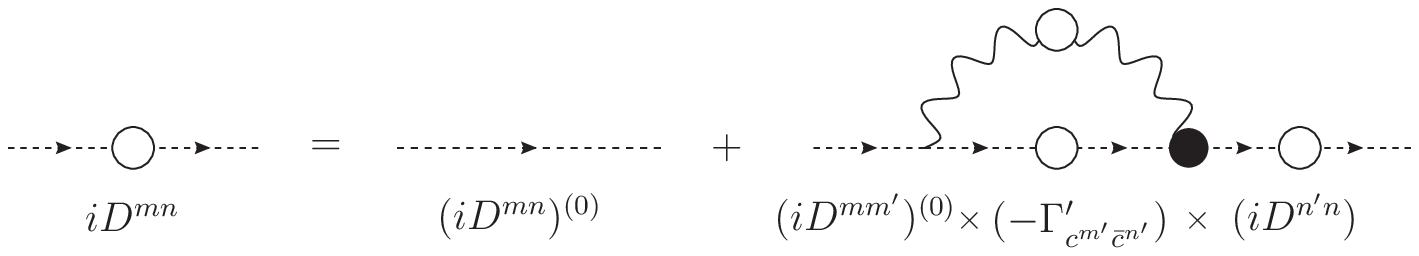}
\ece
\caption{\figlab{ghost_SDE}The SDE~(\ref{ghost_SDE}) satisfied by the ghost propagator.}
\end{figure}
\newline
\indent In summary, the application  of the PT to the conventional SDE
for the  trilinear gluon  vertex ({\it i})  has converted  the initial
kernel   expansion   [graphs   ({\it   a})  through   ({\it   g})   in
\Figref{ggg_SDE}]  into   the  graphs  corresponding   to  the  kernel
expansion of the vertex $\Gamma_{\widehat{A}AA}$; ({\it ii}) all other
pinching terms extracted from the original diagram $(c)$ are precisely
the combinations  of auxiliary Green's functions appearing  in the BQI
that relates the two vertices.
\newline
\indent
Notice at this point that the skeleton expansion of the multi-particle kernels appearing in the SDE for $\Gamma_{\widehat{A}AA}$ is still written in terms of the conventional fully dressed vertices and propagators (involving only quantum fields). 
Thus, Eq.~(\ref{final_ggg}) is not manifestly dynamical, \ie it does not involve the same unknown quantities on the right and left hand side. Specifically, in order to convert (\ref{final_ggg}) into a genuine SDE, one has two possibilities, both involving the use of the above BQI: ({\it i}) substitute the lhs of Eq.~(\ref{PT-ggg-res}) into the rhs of Eq.~(\ref{final_ggg}) and solve for the conventional $\Gamma_{AAA}$ vertex, or ({\it ii}) invert Eq.~(\ref{PT-ggg-res}) and use it to convert every $\Gamma_{AAA}$ vertex appearing in the rhs of Eq.~(\ref{final_ggg}) into a $\Gamma_{\widehat{A}AA}$ vertex. 
It would seem that the latter option is operationally more cumbersome, especially taking into account that a similar procedure has to be followed for all the Green's functions that appear in the coupled system of SDEs that one considers.

\subsubsection{The gluon propagator}
\noindent
In  this  section  we  turn to the SDE of the  gluon
self-energy. From a 
technical point of view, the construction is somewhat more involved  
compared to that of the the vertices, simply   
because the PT decomposition of Eq.~(\ref{decomp}) must be carried out 
on both sides of the self-energy diagram. Put in a different way,  
now we must convert not one, but two external gluons to background gluons. 
This is achieved through a procedure  consisting of the following three  
basic steps~\cite{Binosi:2007pi,Binosi:2008qk}: ({\it i})  
Start with the conventional SDE for $\Gamma_{A A}$ and 
convert (through pinching) $A$ to $\widehat{A}$ ; this generates the SDE for $\Gamma_{\widehat{A} A}$.  
({\it ii}) Use the symmetry of $\Gamma_{\widehat{A} A}$ to interchange legs: 
$\Gamma_{\widehat{A} A} = \Gamma_{A\widehat{A}}$; this saves a lot of algebra in the next step.
({\it iii}) In the SDE for $\Gamma_{A\widehat{A}}$, convert (through pinching) $A$ to $\widehat{A}$; 
this generates the SDE for $\Gamma_{\widehat{A}\widehat{A}}$. Let us now go over these steps in detail.  

\noindent
$\ast$\hspace{.4cm}{\it First step}

The starting point is diagram $(a_1)$ of \Figref{gg_SDE}.
Following the PT procedure, we decompose the tree-level 
three-gluon vertex according to (\ref{decomp}), 
and concentrate on the pinching part, 
\be
(a_1)^\mathrm{P}= igf^{amn'}g_{\alpha\nu'}\int_{k_1}\frac1{k_1^2}\Delta_{n'n}^{\nu'\nu}(k_2)k_1^{\mu}\Gamma_{A^m_\mu A^n_\nu A^b_\beta}(k_2,-q).
\ee
The application of the STI of Eq.~(\ref{STI:ggg}), together with Eq.~(\ref{gainvprop}) and the FPE~(\ref{FP:gcc}), results in the following terms
\begin{eqnarray}
(a_1)^\mathrm{P}&=&igf^{am'n'}g_{\alpha\nu'}\int_{k_1}D^{m'm}(k_1)\Delta_{n'n}^{\nu'\nu}(k_2)\Gamma_{c^m A^n_\nu A^{*\gamma}_d}(k_2,-q)\Gamma_{A^d_\gamma A^b_\beta}(q)\nonumber \\
&+&gf^{am'd}g_{\alpha\gamma}\int_{k_1}D^{m'm}(k_1)\Gamma_{c^m A^b_\beta A^{*\gamma}_d}(-q,k_2)\nonumber \\
&-&igf^{am'n'}g_{\alpha\nu'}\int_{k_1}\delta^{dn'}\frac{k_2^{\nu'}}{k_2^2}D^{m'm}(k_1)\Gamma_{c^m A^b_\beta\bar c^d}(-q,k_2)\nonumber \\
&=& (s_1)+(s_2)+(s_3).
\end{eqnarray}
Using the SDE of the auxiliary function $\Gamma_{\Omega A^*}$, shown in Eq.~(\ref{BQI:auxOmAs}), one has immediately that
\begin{equation}
(s_1)=-i\Gamma_{\Omega^a_\alpha A^{*\gamma}_d}(q)\Gamma_{A^d_\gamma A^b_\beta}(q).
\end{equation}
This corresponds to half of the pinching contribution coming from the vertex in the $S$-matrix PT.
\begin{figure}[!t]
\bce
\includegraphics[width=16cm]{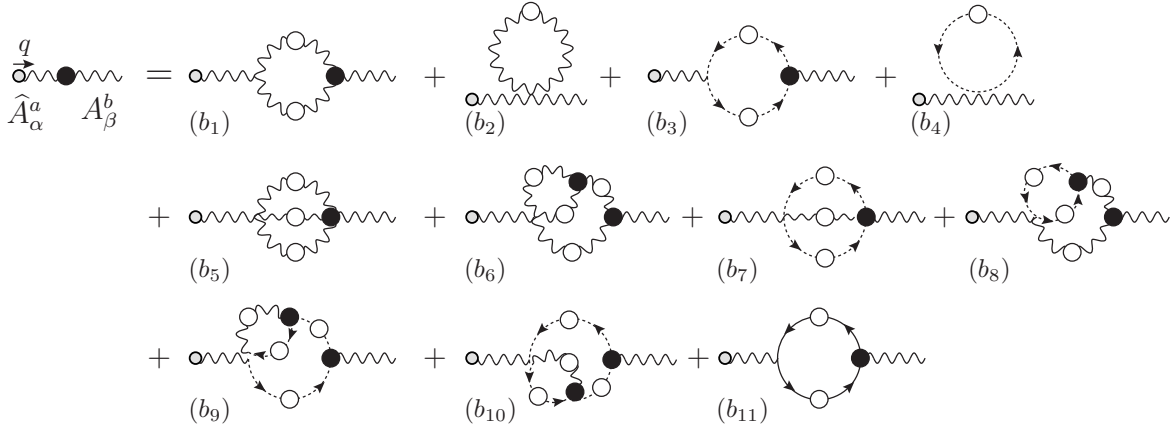}
\ece
\caption{\figlab{ghatg_SDE} The SDE satisfied by the gluon self-energy $-\Gamma_{\widehat{A} A}$. The symmetry factors of the diagrams are $s(b_1,b_2,b_6)=1/2$, $s(b_5)=1/6$, and all the remaining diagrams have $s=-1$.}
\end{figure}
\newline
\indent
As far as the $(s_2)$ and $(s_3)$ terms are concerned, 
let us start by adding and subtracting to them the expression needed to convert 
the tree-level ghost propagator of $(s_{3})$ into a full one; making use of the ghost SDE~(\ref{ghSDE2}), we obtain 
\begin{eqnarray}
(s_{2})&=&-gf^{am'd}g_{\alpha\gamma}\int_{k_1}iD^{m'm}(k_1)\left[i\Gamma_{c^m A^b_\beta A^{*\gamma}_d}(-q,k_2)\right.\nonumber \\
&+&\left.\Gamma'_{c^{g'}A^{*\gamma}_d}(k_2)D^{g'g}(k_2)\Gamma_{c^m A^b_\beta\bar c^g}(-q,k_2)\right],\nonumber \\
(s_{3})&=&-igf^{am'n'}\int_{k_1}k_{2\alpha}D^{m'm}(k_1)D^{n'n}(k_2)\Gamma_{c^mA^b_\beta\bar c^n}(-q,k_2).
\label{s4}
\end{eqnarray}
The second term symmetrizes the  trilinear ghost-gluon coupling, and one has
\begin{equation}
(s_{3})+(a_3)=(b_3),
\end{equation}
where $(b_3)$ is shown in \Figref{ghatg_SDE}. The term $(s_{2})$ will finally generate all the remaining terms.
To see how this happens, we denote by $(s_{2a})$ and $(s_{2b})$ 
the two terms appearing in the square brackets of $(s_2)$, and concentrate on the first one. 
Making use of the SDE~(\ref{BQI:auxcAAs}) satisfied by the auxiliary function $\Gamma_{cAA^*}$, 
and the decomposition~(\ref{SDE_kercAAbarc}) of the kernel 
appearing in the latter, we get
\begin{eqnarray}
(s_{2a}) &=&g^2f^{am'd}f^{mdb}g_{\alpha\beta}\int_{k_1}D^{m'm}(k_1)\nonumber \\
&+&g^2f^{am'd}f^{dn's'}g_{\alpha\sigma'}\int_{k_1}\int_{k_3}D^{m'm}(k_1)\Delta_{s's}^{\sigma'\sigma}(k_3)D^{n'n}(k_4){\mathcal K}_{c^mA^b_\beta A^s_\sigma\bar c^n}(-q,k_3,k_4)\nonumber \\
&=&(b_4)+(b_7)+(b_8)+(b_{10}).
\end{eqnarray}
\begin{figure}[!t]
\bce
\includegraphics[width=16cm]{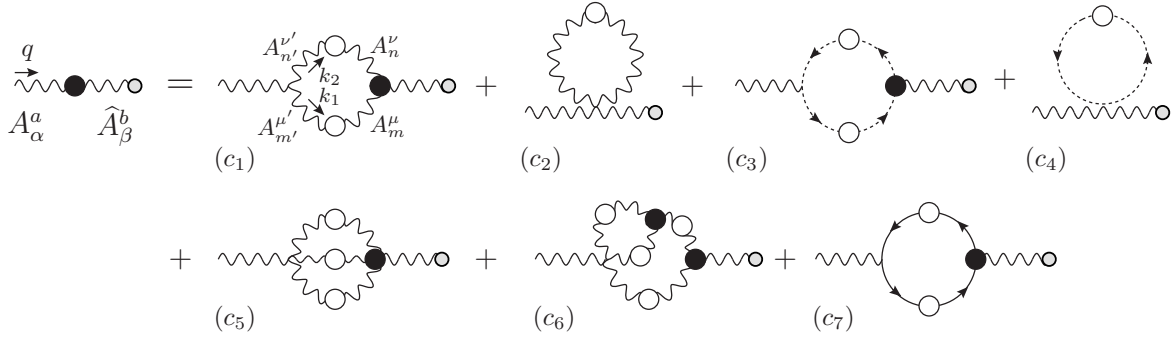}
\ece
\caption{\figlab{gghat_SDE} The SDE satisfied by the gluon self-energy $-\Gamma_{A\widehat{A}}$. The symmetry factors of the diagrams are $s(c_1,c_2,c_6)=1/2$, $s(c_3,c_4,c_7)=-1$, $s(c_5)=1/6$.}
\end{figure}
For the second term, using the SDE satisfied by $\Gamma_{cA^*}$, shown in Eq.~(\ref{BQI:auxcAs}), we obtain
\begin{eqnarray}
(s_{2b}) &=&ig^2f^{am'd}f^{dse}g_\alpha^\sigma\int_{k_1}\int_{k_3}D^{m'm}(k_1)\Delta^{\sigma\sigma'}_{ss'}(k_3)D^{ee'}(k_4)\Gamma_{c^{g'} A_{s'}^{\sigma'}\bar c^{e'}}(k_3,k_4)D^{g'g}(k_2)\times\nonumber\\
&\times&\Gamma_{c^m A^b_\beta\bar c^g}(-q,k_2)\nonumber\\
&=&(b_9).
\end{eqnarray}
Finally, since the diagrams $(a_2)$, $(a_4)$ $(a_5)$, and $(a_6)$ 
carry over directly to the corresponding 
BFM ones $(b_2)$, $(b_5)$, $(b_6)$, and $(b_{11})$, and 
given that $(a_1)^\mathrm{F}=(b_1)$, we have the final identity
\begin{equation}
(s_{2})+(s_{3})+\left[(a_1)^\mathrm{F}+\sum_{i=2}^6(a_i)\right]
=\sum_{i=1}^{11}(b_i),
\end{equation}
and therefore 
\begin{equation}
-\Gamma_{A^a_\alpha A^b_\beta}(q)=-i\Gamma_{\Omega^a_\alpha A^{*\gamma}_d}(q)\Gamma_{A^d_\gamma A^b_\beta}(q)-\Gamma_{\widehat{A}^a_\alpha A^b_\beta}(q),
\end{equation}
which is the BQI of Eq.~(\ref{twoBQI1}).

\noindent
$\ast$\hspace{.4cm}{\it Second step\label{step_2}}

The second step in the propagator construction is to employ the obvious relation
\begin{equation}
\Gamma_{\widehat{A}^a_\alpha A^b_\beta}(q)=\Gamma_{A^a_\alpha \widehat{A}^b_\beta}(q),
\end{equation}
that is to  interchange the  background  and quantum
legs (the SDE for the self-energy  $-\Gamma_{A\widehat{A}}$ is shown in \Figref{gghat_SDE}).
This apparently trivial operation introduces a considerable simplification. First of all, 
it allows for the identification of the  pinching  momenta from the usual PT decomposition of the (tree-level) 
$\Gamma$ appearing in diagram $(c_1)$ of \Figref{gghat_SDE} 
[something not directly possible from diagram $(b_1)$]; thus, from the operational 
point of view, we remain on familiar ground.
In addition,  it  avoids the need to employ the 
(formidably complicated) BQI for the four-gluon vertex;
 indeed, the equality between diagrams $(c_5)$, $(c_6)$, $(c_7)$  
of \Figref{gghat_SDE}, and $(d_5)$, $(d_6)$, $(d_{11})$ 
of \Figref{ghatghat_SDE}, 
respectively, is now immediate [as it was before, 
between the diagrams $(a_4)$, $(a_5)$, $(a_6)$ and $(b_5)$, $(b_6)$, $(b_{11})$, respectively].

\noindent
$\ast$\hspace{.4cm}{\it Third step}

\begin{figure}[!t]
\includegraphics[width=16cm]{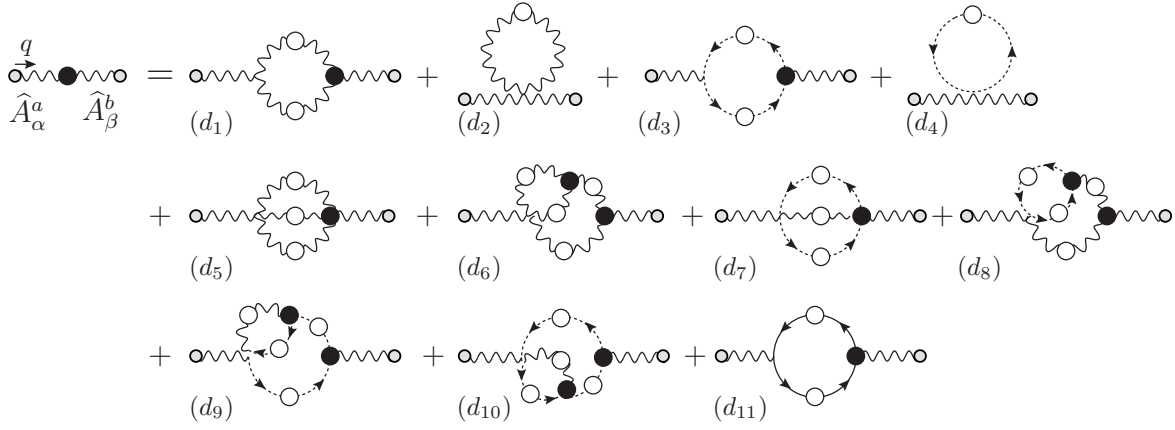}
\caption{\figlab{ghatghat_SDE} The SDE satisfied by the gluon self-energy $-\Gamma_{\widehat{A}\widehat{A}}$. The symmetry factors are the same as the one described in \Figref{ghatg_SDE}.}
\end{figure}

We now turn to diagram $(c_1)$ and concentrate on its pinching part, given by
\begin{equation}
(c_1)^\mathrm{P}= igf^{amn'}g_{\alpha\nu'}\int_{k_1}\frac1{k_1^2}\Delta_{n'n}^{\nu'\nu}(k_2)k_1^{\mu}\Gamma_{A^m_\mu A^n_\nu \widehat{A}^b_\beta}(k_2,-q).
\end{equation}
Notice the appearance of  
the full BFM vertex $\Gamma_{AA\widehat A}$
instead of the standard $\Gamma_{AAA}$ (in the $R_{\xi}$). 
The STI satisfied by the former vertex has been derived in Eq.~(\ref{STI:mixed1}). 
Now, the first three terms, $(s_1)$, $(s_2)$ and $(s_3)$, appearing in this STI, 
will give rise to PT contributions exactly equal to those encountered 
in first step described above, the only difference being that 
the $A^b_\beta$ field appearing there is now a background field $\widehat{A}^b_\beta$. 
Thus, following exactly the reasoning described before, 
we find [see again \Figref{ghatghat_SDE} for the diagrams corresponding to each $(d_i)$]
\bea
(s_1)&\to&-i\Gamma_{\Omega_\alpha^aA^{*e}_\epsilon}(q)\Gamma_{ A^e_\epsilon\widehat{A}^b_\beta}(q),\nonumber \\
(s_2)+(s_3)+(c_3)&=&(d_3)+(d_4)+(d_7)+(d_8)+(d_9)+(d_{10}).
\eea
For the term $(s_4)$ we have instead
\be
(s_4)\to g^2f^{am'e}f^{ebm}g_{\alpha\mu'}g_{\beta\mu}\int_{k_1}\Delta_{m'm}^{\mu'\mu}(k_1).
\ee
Clearly this has a seagull-like structure; in particular, it is immediate to prove 
that when added to $(c_2)$ it will convert it into $(d_2)$
\begin{equation}
(s_4)+(c_2)=(d_2).
\end{equation}
Thus, since as always $(c_1)^\mathrm{F}=(d_1)$ we get
\be
(s_{2})+(s_{3})+(s_4)+\left[(c_1)^\mathrm{F}+\sum_{i=2}^7(c_i)\right]
=\sum_{i=1}^{11}(d_i),
\ee
and therefore
\begin{equation}
-\Gamma_{A^a_\alpha \widehat{A}^b_\beta}(q)=-i\Gamma_{\Omega_\alpha^aA^{*e}_\epsilon}(q)\Gamma_{A^e_\epsilon \widehat{A}^b_\beta}(q)-\Gamma_{\widehat{A}^a_\alpha \widehat{A}^b_\beta}(q),
\end{equation}
which is the BQI of Eq.~(\ref{twoBQI2}). This concludes our proof.
\newline
\indent
In \Figref{4steps} we summarize the steps that allowed the successful construction of the 
SDE for the PT propagator. 

\begin{figure}[!t]
\bce
\includegraphics[width=8cm]{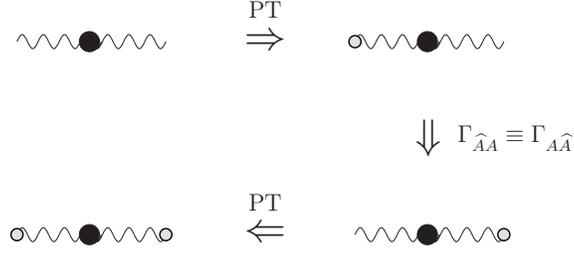}
\ece
\caption{\figlab{4steps}Summary of the PT procedure employed for the construction of the new SDE describing
 the gluon propagator.}
\end{figure}

\subsection{The new Schwinger-Dyson series}
\noindent
We will now have a closer look at the structure and physical consequences of the new SDEs  
obtained in the previous subsections. 
We focus,  for concreteness, 
on the SDE for the gluon propagator; notice, however, 
that the following analysis applies with minimal modification to the three-point functions SDEs. 
\newline
\indent
The PT rearrangement gives rise dynamically to the new SD series (shown in \Figref{PT_newSDE}) which has the following characteristics: 
\begin{itemize}
\item[$\ast$] On the rhs, it is as if the external gluons had been converted dynamically into background gluons, since we have graphs that are made out of new vertices, which coincide precisely with the BFG ones. 
Notice, however, an important point: the graphs contain inside them the same gluon propagator as before, namely $\Delta$.
\newline
\item[$\ast$] On the lhs, we have the sum of three terms:  
in addition to the term  $\Delta^{-1}(q^2) P_{\alpha\beta}(q)$, present there from the 
beginning, we have two additional contributions, 
$ 2 G(q^2) \Delta^{-1}(q^2) P_{\alpha\beta}(q)$ and $G^2(q^2) \Delta^{-1}(q^2) P_{\alpha\beta}(q)$,
which appear during the PT rearrangement of the rhs (and are subsequently 
carried to the lhs). Thus, the term appearing on the lhs of the new SDE is 
$\Delta^{-1}(q^2)[1+G(q^2)]^2 P_{\alpha\beta}(q)$, with the function $G(q^2)$ defined through 
\be
\Gamma_{\Omega_\alpha A^*_\beta}(q)=ig_{\alpha\beta}G(q^2)+\dots ,
\ee
where the omitted terms are proportional to $q_\alpha q_\beta$ (and therefore irrelevant due to transversality).
\end{itemize}

Summarizing, one may write schematically
\be
\Delta^{-1}(q^2)[1+G(q^2)]^2 P_{\alpha\beta}(q) = q^2 P_{\alpha\beta}(q) + 
i \sum_{1=1}^{11}(d_i)_{\alpha\beta}, 
\label{newSDa}
\ee
which is nothing but a rewriting of the BQI relating the background and the conventional gluon two-point function [see Eq.~(\ref{BQI:gg})].
Equivalently, one can cast Eq.~(\ref{newSDa}) into a more conventional form by isolating on the lhs the inverse of the unknown quantity, thus writing 
\be
\Delta^{-1}(q^2)P_{\alpha\beta}(q)= 
\frac{q^2 P_{\alpha\beta}(q) + i \sum_{1=1}^{11}(d_i)_{\alpha\beta}}{[1+G(q^2)]^2}. 
\label{newSDb}
\ee
\begin{figure}[!t]
\bce
\includegraphics[width=16cm]{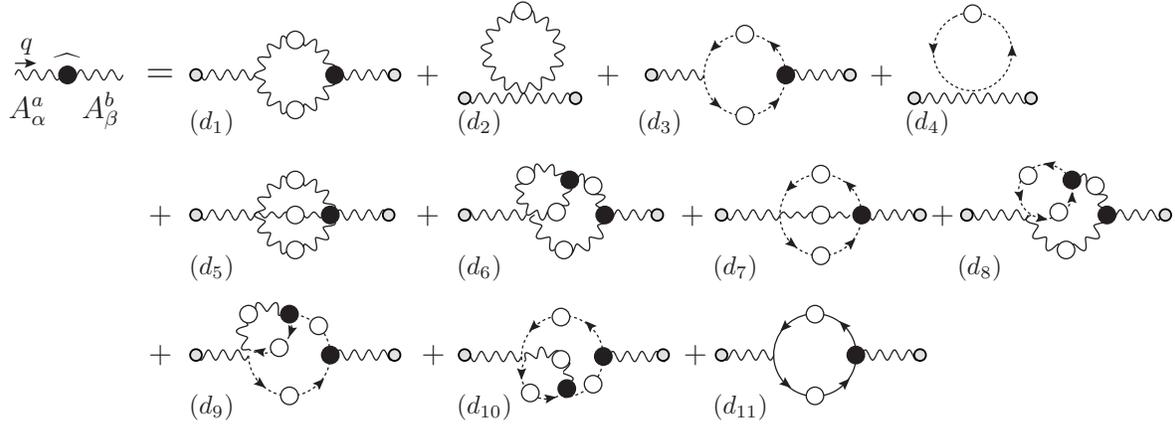}
\ece
\caption{\figlab{PT_newSDE}The new SD series projected out dynamically by the PT algorithm.}
\end{figure}

\subsubsection{The PT as a gauge-invariant truncation scheme: advantages over the conventional SDEs}
\noindent
The new SD series of Eqs~(\ref{newSDa}) and (\ref{newSDb}) has a very special structure. 
In order to gain a deeper understanding of the situation at hand, let us first step back and  
consider the one-loop case again, 
in which the application of the intrinsic PT algorithm to the (one-loop) 
gluon self-energy, amounts to carry out the PT rearrangement of Eq.~(\ref{INPTDEC1}) 
of the two elementary three-gluon vertices appearing in diagram $(a)$ of \Figref{1l_PT_gprop}, 
thus getting the result
\be
\widehat{\Pi}^{(1)}_{\alpha\beta}(q) = \frac12g^2C_{A}\left\{ 
\int_k\!\frac{\Gamma^{{\rm F}}_{\alpha \mu \nu}\Gamma^{{\rm F}\,\mu \nu}_{\beta}}{k^2 (k+q)^2}-\int_k\!\frac{2 (2k+q)_{\alpha}(2k+q)_{\beta}}{k^2 (k+q)^2}\right\}- 
2g^2C_{A} \int_k\!\frac{q^2 P_{\alpha\beta}(q)}{k^2 (k+q)^2}.
\label{prop_exp}
\ee
It is elementary to verify that each of the two terms in braces on the rhs of 
the equation above are transverse; 
thus,  the PT rearrangement has created 
three manifestly transverse structures. That in itself might not be so important, 
if it were not for the fact that, as we know, these structures 
admit a special diagrammatic representation and a unique field-theoretic interpretation. 
Specifically, the two terms in the square bracket correspond precisely  
to diagrams $(\widehat{a})$ and $(\widehat{b})$ of \Figref{PT_1loop},  
defining the one-loop gluon self-energy in the BFG,
while the third term on the rhs of Eq.~(\ref{prop_exp}) 
is the one-loop 
expression of the special auxiliary Green's function $\Gamma_{\Omega A^*}$,  
identified in the BV formulation of the PT (or, more precisely, the $g_{\alpha\beta}$ part of this latter function);
it corresponds to diagram $(c)$ in  \Figref{PT_1loop}.
Since the one-loop PT self-energy
is obtained by simply dropping this last term from the rhs of Eq.~(\ref{prop_exp}), 
one has
\be
\Pi_{\alpha\beta}^{(1)}(q) = \widehat\Pi_{\alpha\beta}^{(1)}(q) + (c)_{\alpha\beta},
\label{BQI_1l-2}
\ee
which, as we know from \secref{PTBV} [see Eq.~(\ref{BQI_1l})], is nothing but the  one-loop version 
of the BQI of Eq.~(\ref{BQI:gg}) [and, therefore, also of the SDE (\ref{newSDa})]. 
Let us now focus on $\Pi_{\alpha\beta}^{(1)}(q)$, and  
imagine for a moment that no ghost loops may be considered when computing it, 
\ie the graphs $({\widehat b})_{\alpha\beta}$ must be omitted; 
in a SDE context  this ``omission'' would amount to a ``truncation'' of the series.
One may still obtain a {\it transverse} approximation  for 
$\Pi_{\alpha\beta}^{(1)}(q)$ with no ghost-loop, given by
\be
\Pi_{\alpha\beta}^{(1)} (q) = ({\widehat a})_{\alpha\beta} + (c)_{\alpha\beta} 
= 4  q^2 f (q^2) P_{\alpha\beta}(q),
\ee
where the function $f(q^2)$ has been defined in Eq.~(\ref{f_qsquare}). 
Interestingly enough, the PT rearrangement offers already at one-loop   
the ability to truncate gauge-invariantly, \ie preserving  
the transversality of the truncated answer. 
\begin{figure}[t]
\bce
\includegraphics[width=12.5cm]{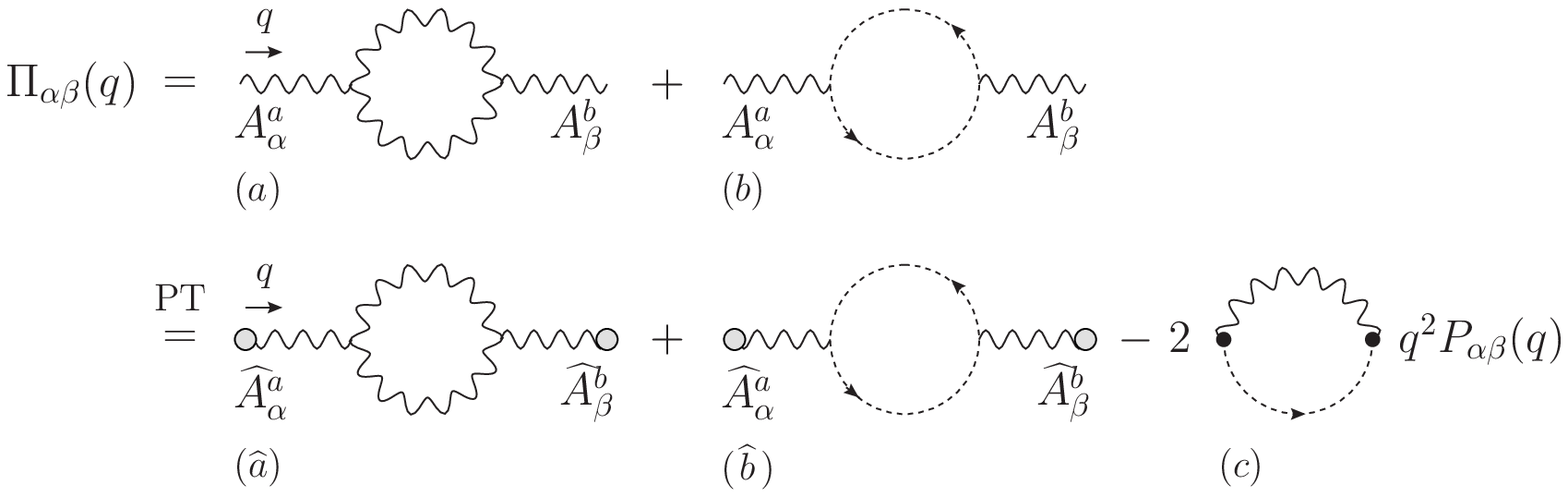}
\ece
\caption{\figlab{PT_1loop}The conventional one-loop gluon self-energy before (first line) and after (second line) the PT rearrangement.}
\end{figure}
\newline
\indent
Turning now to the full SDE (\ref{newSDb}), one can prove that the special transversality property found above 
holds true non-perturbatively, with gluonic and ghost contributions separately transverse, and, in addition, 
no mixing between the one- and two-loop dressed diagrams. In the BFM context this property has been already 
seen explicitly for the divergent parts of the two-loop gluon self-energy diagrams (\secref{PTBFM}); 
here we will show that this is in fact a consequence of the all-orders WIs satisfied by the 
full vertices appearing in the diagrams defining the PT self-energy (\Figref{ghatghat_SDE}).
There are four fully dressed vertices appearing in $\widehat\Pi$, whose WIs we need: 
$\Gamma_{\widehat{A}AA}$, $\Gamma_{c\widehat{A}\bar c}$, $\Gamma_{\widehat{A}AAA}$ and, finally, $\Gamma_{c\widehat{A}\bar c A}$. 
One way to derive their WIs is to differentiate the WI functional of Eq.~(\ref{WI_gen_funct}) with respect 
to the corresponding field combination where the background field has been replaced by the corresponding gauge parameter $\vartheta$. 
On the other hand, a much more expeditious way is to derive the corresponding tree-level WIs, 
and then use linearity to generalize them to all orders, as we do in QED. Either way, one obtains the following results
\bea
q^\alpha\Gamma_{\widehat{A}^a_\alpha A^m_\mu A^n_\nu}(k_1,k_2)&=& gf^{amn}\left[\Delta^{-1}_{\mu\nu}(k_1)-\Delta^{-1}_{\mu\nu}(k_2)\right],\label{ghatgg_WI}\nonumber\\
q^\alpha\Gamma_{c^n\widehat{A}^a_\alpha \bar c^m}(q,-k_1)&=&
igf^{amn}\left[D^{-1}(k_2)-D^{-1}(k_1)\right],\label{ghatcc_WI}\nonumber\\
q^\alpha\Gamma_{\widehat{A}^a_\alpha A^b_\beta A^m_\mu A^n_\nu}(k_1,k_2,k_3)&=& gf^{adb}\Gamma_{A^d_\beta A^m_\mu A^n_\nu}(k_2,k_3)+gf^{adm}\Gamma_{A^d_\mu A^b_\beta A^n_\nu}(k_1,k_3)\nonumber \\
&+& gf^{adn}\Gamma_{A^d_\nu A^b_\beta A^m_\mu}(k_1,k_2),\label{ghatggg_WI}\nonumber\\
q^\alpha\Gamma_{c^n\widehat{A}^a_\alpha A^b_\beta \bar c^m}(q,k_3,-k_1)&=&gf^{adb}\Gamma_{c^n A^d_\beta\bar c^m}(q+k_3,-k_1)+gf^{adm}\Gamma_{c^n A^b_\beta\bar c^d}(k_3,q-k_1)\nonumber\\
&+&gf^{adn}\Gamma_{c^d A^b_\beta\bar c^m}(k_3,-k_1).\label{ghatgcc_WI}
\eea
Armed with these WIs, we can now prove that the four groups identified (at two-loop level) in Eq.~(\ref{1ld_g}) 
are in fact independently transverse even non-perturbatively. 
\newline
\indent
Let's start from the one-loop dressed gluonic contributions given by the combination of \mbox{$(d_1)+(d_2)$} of \Figref{ghatghat_SDE}. Using
the first WI of Eq.~(\ref{ghatgg_WI}) we get 
\be
q^\beta(d_1)^{ab}_{\alpha\beta}=-g^2C_A\delta^{ab}q_\alpha\int_k\!\Delta^\mu_\mu(k),
\ee
while, by simply computing the divergence of the 
tree-level vertex $\Gamma_{\widehat A\widehat A AA}$ given in \appref{Frules}, we get
\be
q^\beta(d_2)^{ab}_{\alpha\beta}=g^2C_A\delta^{ab}q_\alpha\int_k\!\Delta^\mu_\mu(k),
\ee
so that clearly 
\be
q^\beta\left[(d_1)+(d_2)\right]^{ab}_{\alpha\beta}=0.
\label{4tr}
\ee
Exactly the same procedure yields for the one-loop dressed ghost contributions 
\bea
q^\beta(d_3)^{ab}_{\alpha\beta}&=& -2g^2C_A\delta^{ab}q_\alpha\int_k\!D(k),\nonumber\\
q^\beta(d_4)^{ab}_{\alpha\beta}&=& 2g^2C_A\delta^{ab}q_\alpha\int_k\!D(k),
\eea
and therefore
\be
q^\beta\left[(d_3)+(d_4)\right]^{ab}_{\alpha\beta}=0.
\ee
\indent
For the two-loop dressed contributions the proof is only slightly more involved. 
We begin with the gluonic contributions. Using the third WI of Eq.~(\ref{ghatggg_WI})  
in diagram $(d_5)$, after appropriate manipulation of the the terms produced, 
and taking into account the symmetry factor of $1/6$, we obtain
\be
q^\beta(d_5)^{ab}_{\alpha\beta}=\frac i2gf^{bmn}\Gamma^{(0)}_{\widehat{A}^a_\alpha A^m_{\mu'}A^g_{\gamma'}A^e_{\epsilon'} }\int_k\!\int_\ell\!\Delta^{\epsilon'\epsilon}(k)\Delta^{\gamma'\gamma}(\ell+k)\Gamma_{A^g_\gamma A^e_\epsilon A^n_\mu}(k,\ell)\Delta^{\mu'\mu}(\ell+q).
\label{q_d5}
\ee
Similarly, after making use the full Bose symmetry of the three-gluon vertex, graph $(d_6)$ gives 
\bea
q^\beta(d_6)^{ab}_{\alpha\beta}&=&\frac i2gf^{bmn}\Gamma^{(0)}_{\widehat{A}^a_\alpha A^m_{\mu'}A^g_{\gamma'}A^e_{\epsilon'} }\int_k\!\int_\ell\!\Delta^{\epsilon'\epsilon}(k)\Delta^{\gamma'\gamma}(\ell+k)\Gamma_{A^g_\gamma A^e_\epsilon A^{n'}_{\nu'}}(k,\ell)\times\nonumber\\
&\times&\left[\Delta^{\nu'\nu}(\ell)g^{\mu'}_\nu-\Delta^{\mu'\mu}(\ell+q)g^{\nu'}_\mu\right].
\eea
The first term in the square brackets vanishes (the integral is independent of $q$, and therefore the free Lorentz index $\beta$ cannot be saturated). 
Furthermore, the  second term is exactly equal but opposite in sign to the one appearing in Eq.~(\ref{q_d5}), so that we obtain 
\be
q^\beta\left[(d_5)+(d_6)\right]^{ab}_{\alpha\beta}=0.
\ee
Finally, we turn to the two-loop dressed ghost contributions. Using the last WI of Eq.~(\ref{ghatgcc_WI}), we see that the divergence of diagram $(d_7)$ gives us three terms, namely
\bea
q^\beta(d_7)^{ab}_{\alpha\beta}&=&-i\Gamma^{(0)}_{c^{m'}\widehat{A}^a_\alpha A^{r'}_{\rho'}\bar c^{n'}}\int_k\!\int_\ell\!D^{m'm}(\ell+k)D^{n'n}(\ell+q)\Delta^{\rho'\rho}_{r'r}(k)\times\nonumber \\
&\times&\left[gf^{ber}\Gamma_{c^nA^e_\rho\bar c^e}(k-q,-\ell-k)+gf^{ben}\Gamma_{c^e A^r_\rho\bar c^m}(k,-\ell-k)\right.\nonumber\\
&+&\left.gf^{ben}\Gamma_{c^n A^r_\rho\bar c^n}(k,-q-\ell-k)\right].
\label{q7_div}
\eea
Each one of these three terms can be easily shown to cancel exactly against the individual 
divergences of the remaining three graphs. 
To see this in detail, let us consider for example diagram $(d_{10})$ and use the WI~(\ref{ghatcc_WI}) to obtain
\bea
q^\beta(d_{10})^{ab}_{\alpha\beta}&=&-igf^{bmn}\Gamma^{(0)}_{c^{m'}\widehat{A}^a_\alpha A^{r'}_{\rho'}\bar c^{n'}}\int_k\!\int_\ell\!D^{m'm}(\ell+k)D^{n'd}(\ell+q)D^{d'n}(\ell+k+q)\Delta^{\rho'\rho}_{r'r}(k)\!\times\nonumber \\
&\times&\Gamma_{c^d A^r_\rho\bar c^{d'}}(k,-q-\ell-k)\left[D^{-1}(\ell+k+q)-D^{(-1)}(\ell+k)\right].
\eea
Then, we see that the second inverse propagator in the square brackets will give rise to a $q$-independent integral that will integrate to zero, while the first term will cancel exactly the third term appearing in the square brackets of Eq.~(\ref{q7_div}). It is not difficult to realize that the same pattern will be encountered when calculating the divergence of diagrams $(d_8)$ and $(d_9)$, so that one has the identity
\be
q^\beta\left[(d_7)+(d_8)+(d_9)+(d_{10})\right]^{ab}_{\alpha\beta}=0.
\ee
This concludes the proof of the special transversality property of $\widehat{\Pi}^{ab}_{\alpha\beta}(q)$,  
showing that gluon and ghost loops are separately transverse, and that dressing loops of different orders do not mix.
\newline
\indent
This last property has far-reaching practical consequences 
for the  treatment of the SD series~\cite{Binosi:2007pi,Binosi:2008qk}. Specifically, it furnishes a systematic gauge-invariant truncation scheme that preserves the transversality of the answer. 
In fact, we can drastically reduce the number of coupled SDEs that must be included 
in order to maintain the gauge (or BRST) symmetry  
of the theory intact, as reflected, for example, in the validity of Eq.(\ref{fundtrans}). For example, keeping only the 
diagrams in the first group, we obtain the truncated SDE
\be
\Delta^{-1}(q^2) P_{\alpha\beta}(q) = 
\frac{q^2 P_{\alpha\beta}(q) + i[(d_1)+(d_2)]_{\alpha\beta}}{[1+G(q^2)]^2}, 
\label{trua}
\ee
and from Eq.~(\ref{4tr}) we know that 
$[(d_1)+(d_2)]_{\alpha\beta}$ is transverse, \ie 
\be
[(d_1)+(d_2)]_{\alpha\beta}= {(d-1)}^{-1} [(d_1)+(d_2)]^{\mu}_{\mu}P_{\alpha\beta}(q).
\ee
Thus,  the transverse projector $P_{\alpha\beta}(q)$ appears 
{\it exactly}
on both sides of (\ref{trua}); one may subsequently 
isolate the scalar cofactors on both sides, 
obtaining a scalar equation of the form
\be
\Delta^{-1}(q^2) = 
\frac{q^2 + i[(d_1)+(d_2)]^{\mu}_{\mu}}{[1+G(q^2)]^2}. 
\ee
A truncated equation similar to (\ref{trua}) may be written for any other 
of the four groups previously isolated, or for sums of these groups, 
without compromising the transversality of the answer. 
The price one has to pay for this advantageous situation 
is that one must consider, in  addition, the    
equation determining the scalar function $G(q^2)$. This price is, however, rather modest, 
since one can approximate this function via a dressed-loop expansion [see, \eg  \Figref{Composite_ope} together with Eq.~(\ref{BQI:auxOmAs})], without 
jeopardizing the transversality of $\Pi_{\alpha\beta}(q)$, given that 
$[1+G(q^2)]^2$ affects only the size of the scalar prefactor.
\newline
\indent
Thus, in the case of pure Yang-Mills, within this new formulation,   
the minimum number of 
equations that one must consider is only two: The SDE for the gluon self-energy, given by the 
first gauge-invariant subset {\it only} ({\it i.e.},  
$\left[ (d_1) + (d_2) \right]_{\alpha\beta}$ in \Figref{PT_newSDE}), and 
the SDE for the full three-gluon vertex, shown in \Figref{ggg_SDE} 
(which is instrumental in assuring the gauge invariance of the subset chosen).  
This is to be contrasted to what happens within the conventional formulation: there the SDEs for {\it all} 
vertices must be considered, or else  Eq.~(\ref{fundtrans}) is violated (which is what usually happens).

\subsubsection{Some important theoretical and practical issues}
\noindent
We now turn to some additional points that, due to their theoretical and practical relevance, 
deserve further elaboration.  
\newline
\indent
It is important to emphasize that 
the analysis presented here does {\it not} furnish a simple diagrammatic 
truncation, analogous to that of the gluon self-energy,
for the SDE of the 
three gluon  vertex $\Gamma_{\widehat{A}AA}$, shown in \Figref{ggg_SDE}. 
Thus, if one were to truncate the SDE for the three-gluon vertex by keeping any subset of the graphs appearing 
in Fig.\ref{fig:ggg_SDE}, 
one would violate the validity of the {\it all-order} WI of $\Gamma_{\widehat{A}AA}$ [Eq.~(\ref{ghatgg_WI}) first line]; 
this, in turn, would lead immediately to the violation of Eq.~(\ref{fundtrans}), thus 
making the entire truncation scheme collapse. 
\newline
\indent
The strategy one should adopt is instead the following (see also the discussion in the next section). Given that the proposed 
truncation scheme hinges crucially on the validity of  the WI of $\Gamma_{\widehat{A}AA}$, one should 
start out with an approximation that manifestly preserves it. 
The way to enforce this, familiar to the SDE practitioners already from the QED era, is to resort 
to the ``gauge-technique''~\cite{Salam:1963sa}, namely ``solve'' the WI. 
Specifically, one must  express  the three-gluon vertex 
as a functional of the corresponding self-energies, in such a way that (by construction) 
its WI is automatically satisfied. For example, an Ansatz with this property would be 
\be
\Gamma_{\widehat{A}_\alpha A_\mu A_\nu}(k_1,k_2)=\Gamma^{(0)}_{\widehat{A}_\alpha A_\mu A_\nu}(k_1,k_2) - 
i\frac{(k_2-k_1)_\alpha}{k_2^2-k_1^2}\left[\Pi_{\mu\nu}(k_2)-\Pi_{\mu\nu}(k_1)\right].
\label{GTA}
\ee
Contracting the rhs with $q_{\alpha}=(k_1+k_2)_{\alpha}$ yields automatically the first WI of Eq.~(\ref{ghatgg_WI}). 
Thus, the minimum amount of ingredients for initiating a {\it self-consistent} non-perturbative 
treatment is the SD for the gluon self-energy, consisting of $\left[ (d_1) + (d_2) \right]_{\alpha\beta}$, 
supplemented by an Ansatz for the three-gluon vertex like the one given in (\ref{GTA}). 
Note that the  ``gauge-technique'' leaves the transverse (\ie automatically conserved) part 
of the vertex undetermined. This is where the SDE for the vertex enters; 
it is used precisely to determine the transverse parts. 
Specifically, following standard techniques~\cite{Ball:1980ax,Binger:2006sj}, 
one must expand the vertex into a suitable tensorial basis, consisting of fourteen independent 
tensors, and then isolate the transverse subset. This procedure will lead to a large number 
of coupled integral equations, one for each of the form-factors multiplying 
the corresponding tensorial structures, which may or may not be tractable.
However, at this point, one may simplify the resulting equations (\eg linearize, etc)  
without jeopardizing the transversality of $\Pi_{\alpha\beta}$, which only depends on the 
``longitudinal'' part of the vertex, \ie the one determined by (\ref{GTA}).
Thus, the transverse parts will be approximately determined, but gauge invariance, 
as captured by $q^{\alpha}\Pi_{\alpha\beta}=0$, will remain exact. 
\newline
\indent
Note by the way that the methodology described above constitutes, even to date,  
the standard procedure  even in the context of QED, where the structure of the SDE is much simpler, given 
that the SDE for the photon contains one single graph [diagram $(a_6)$ in \Figref{gg_SDE}], 
and  the photon-electron vertex satisfies automatically a naive all-order WI.
Thus, while  the PT approach described here 
replicates QED-like properties at the level of the SDEs of QCD, admittedly a striking fact in itself, 
does not make QCD easier to solve than QED.  
\newline
\indent
One should appreciate an additional point: any attempt to apply the approach described above 
in the context of the conventional SDE is bound to lead 
to the violation of the transversality of $\Pi_{\alpha\beta}$, because ({\it i}) 
the vertices satisfy complicated STI's instead of the WIs of Eq.~(\ref{ghatgg_WI}), 
a fact that makes the application of the  ``gauge-technique'' impractical, and ({\it ii})  
even if one came up with the analogue of Eq.~(\ref{GTA}) for all vertices, one should still 
keep all self-energy diagrams in \Figref{gg_SDE} to guarantee that $q^{\alpha}\Pi_{\alpha\beta}=0$. 
From this point of view, the improvement of the PT approach over the standard formulation becomes evident.
\newline
\indent
Finally, one should be aware of the fact that there is no a-priori 
guarantee that the gauge-invariant subset kept ({\it i.e.},  $\left[ (d_1) + (d_2) \right]_{\alpha\beta}$) 
capture necessarily most of the dynamics, or, in other words, that 
they represent the numerically dominant contributions  
(however, for a variety of cases it seems to be true, see next section). 
But, the point is that one can {\it systematically} improve the picture by including more terms, 
without worrying that the initial approximation is plagued 
with artifacts, originating from the  violation of the gauge invariance or of the BRST symmetry.   
\newline
\indent
Now, in going from Eq.~(\ref{newSDa}) to Eq.~(\ref{newSDb})
one essentially chooses to retain the original propagator $\Delta(q)$ as the  
unknown quantity, to be dynamically determined from the SDE.
There is, of course, an alternative strategy: one may define 
a new ``variable'' from the quantity appearing on the lhs (\ref{newSDa}),  
namely 
\be
\widehat{\Delta} (q)\equiv \left[1+G(q^2)\right]^{-2} {\Delta}(q),
\label{BQI}
\ee
which leads to  a new form for (\ref{newSDa}),
\be
\widehat\Delta^{-1}(q^2)P_{\alpha\beta}(q) = q^2 P_{\alpha\beta}(q) + i
\sum_{i=1}^{11}(d_i)_{\alpha\beta} \,.
\label{newSDa2}
\ee
Obviously, the special transversality properties established above hold as well for 
Eq.~(\ref{newSDa2}); for example, one may truncate it gauge-invariantly as
\be
\widehat\Delta^{-1}(q^2) P_{\alpha\beta}(q) = 
q^2 P_{\alpha\beta}(q) + i[(d_1)+(d_2)]_{\alpha\beta}.
\label{trub}
\ee
\newline
\indent
Should one opt for treating  $\widehat{\Delta} (q)$ as the 
new unknown quantity, then  an additional step must be carried out:
one must 
use (\ref{BQI}) to rewrite  the entire rhs of (\ref{newSDa2}) 
in terms of $\widehat{\Delta}$ instead of $\Delta$, \ie
carry out the replacement  
\mbox{$\Delta \to \left[1+G\right]^2 \,\widehat{\Delta}$}
 {\it inside} every diagram on the rhs of Eq.~(\ref{newSDa2})
that contains $\Delta$'s.
\newline
\indent
Thus, while Eq.~(\ref{trua}) furnishes a gauge-invariant approximation
for the conventional gluon self-energy $\Delta(q)$, 
Eq.~(\ref{newSDa2}) is the gauge-invariant approximation for the effective 
PT self-energy $\widehat{\Delta}$. The crucial point is that
one may switch from one to the other by means of Eq.~(\ref{BQI}).
For practical purposes this means, for example, that one
may reach a gauge-invariant approximation not just for the PT quantity (BFG)
but also for the 
{\it conventional} self-energy computed in the Feynman gauge (RFG).
Eq.~(\ref{BQI}), which is the all-order generalization of the one-loop 
relation given in Eq.~(\ref{BQI_1l-2}),  
plays an instrumental
role in this entire construction, allowing one 
to convert the SDE series into a dynamical equation for either $\widehat{\Delta} (q)$ 
or $\Delta(q)$.
\newline
\indent
Let us end this section by observing that the new SDEs constructed in the previous subsections have been dynamically projected out in the BFG, which captures, as usual, the net gauge-independent and   
universal (\ie process-independent) contribution 
contained in any physical quantity. In practice, however, one would like to be able to truncate  
gauge-invariantly (\ie maintaining transversality) sets of 
SDEs written in different gauges.
This becomes particularly relevant, for example,  
when one attempts to compare SDE predictions with 
lattice simulations, carried out usually in the Landau gauge, as we do in the next section.
\newline
\indent
This can be achieved by using the GPT algorithm described in subection~\ref{GPT}.
As described there, the GPT 
modifies the  starting point of the PT algorithm distributing  
differently the longitudinal momenta between $\Gamma_{\alpha \mu \nu}^{{\rm F}}$
and $\Gamma_{\alpha \mu \nu}^{{\rm P}}$.
Specifically, the non-pinching part, \ie the analogue of 
$\Gamma_{\alpha \mu \nu}^{{\rm F}}$, 
must satisfy, instead of (\ref{WI2B}), 
a WI whose rhs is the difference of two inverse tree-level propagators 
in the gauge one wishes to consider.
In the context of SDEs, one starts out with the conventional SDE in the chosen gauge, 
carrying out the generalized  PT vertex decomposition.
Then, the action of the corresponding $\Gamma_{\alpha \mu \nu}^{{\rm P}}$
projects one to the corresponding BFM gauge.
This new SD series contains full vertices 
that, even though they are in a different gauge,
satisfy the same QED-like WIs given in Eq.~(\ref{ghatgcc_WI}).
Therefore, the truncation properties of this SDE 
are the same as those just discussed 
for the case of the Feynman gauge.
The analogy is completed by realizing that 
the BQIs in the corresponding gauge allow 
one to switch back and forth 
from the conventional to the BFM Green's function.
Thus, one may obtain, for example, 
transverse approximations for
the gluon propagator 
in the conventional Landau gauge by studying the SDE written 
in the BFM Landau gauge, computing the 
$[1+G(q^2)]^2$ in the same gauge, \ie by employing 
 Eq.~(\ref{newSDb}) and using for the diagrams on its rhs 
the BFM Feynman rules in the Landau gauge.

\newpage


\section{\seclab{Applications-II}Applications part II: 
Infrared properties of QCD Green's functions and dynamically generated gluon mass}

\noindent
The generation  of mass  gaps in  QCD is one  of the  most fundamental
problems in particle physics. In  part the difficulty lies in the fact
that  the  symmetries  governing   the  QCD  Lagrangian  prohibit  the
appearance of  mass terms  at tree-level 
for all fundamental  degrees of  freedom and,  
provided  that  these symmetries  are  not  violated
through the procedure of regularization, this masslessness persists to
all orders in  perturbation theory.  Thus, mass  generation in QCD
becomes  an   inherently  non-perturbative  problem,   whose  tackling
requires the use of rather  sophisticated calculational tools
and approximation schemes \cite{Marciano:1977su}.  
\newline
\indent
Whereas the  generation of quark  masses is intimately  connected with
the breaking of chiral symmetry~\cite{Lane:1974he}, 
Cornwall argued long
ago~\cite{Cornwall:1981zr} that an  effective gluon  mass 
can be generated {\it dynamically}, while   
preserving   the   local  $SU(3)_c$   invariance   of
QCD,  in   close  analogy  to  what   happens  in  QED$_2$
(Schwinger model)~\cite{Schwinger:1962tn}, 
where the  photon acquires a mass without violating
the  Abelian  gauge  symmetry (see discussion below).   
The gluon mass furnishes,  at  least  in
principle, a regulator for all infrared (IR) divergences of QCD.
It must be emphasized that the gluon mass is  not  a  directly
measurable  quantity,  and  that its  value  is  determined
by  relating it  to  other dimensionful  non-perturbative
parameters,  such  as  the  string  tension,  glueball  masses,  gluon
condensates, and the vacuum energy of QCD~\cite{Shifman:1978by}.
\newline
\indent
Since gluon  mass generation is a purely  non-perturbative effect, the
most  standard way for  studying it  in the  continuum is  through the
SDEs governing  the   relevant  Green's functions, and most importantly the 
gluon self-energy.
One of the cornerstones in the original analysis of~\cite{Cornwall:1981zr}
was  the insistence  on  preserving, at  every level  of
approximation,   crucial    properties   such   as   gauge-invariance,
gauge-independence,  and  invariance  under the renormalization group.  
With this motivation, a physical gluon propagator, $\widehat{\Delta}_{\mu\nu}$,
was derived through  the systematic  rearrangement of  Feynman graphs,  
which led to the birth of the PT.
As the reader knows very well by now, the  self-energy 
$\widehat{\Pi}_{\mu\nu}$ of  this
propagator is gauge-independent, and captures the leading logarithms of
the theory,  exactly as happens  with the vacuum polarization  in QED.
The central  result of~\cite{Cornwall:1981zr} was  that, when solving
a simplified (one-loop inspired) SDE 
governing  the  PT propagator, 
one  finds (under  special assumptions  for  the  form  of  the  three-gluon  vertex)
solutions that are free of  the Landau singularity, and reach a finite
(non-vanishing) value in the deep IR.  
These solutions may
be  successfully   fitted  by  a   ``massive''  propagator  of   the 
form $\Delta^{-1}(q^2)  =  q^2  +  m^2(q^2)$; the  crucial  characteristic,
enforced by  the SDE itself, is  that $m^2(q^2)$ is  not ``hard'', but
depends non-trivially  on the momentum  transfer $q^2$.  Specifically,
$m^2(q^2)$  is  a monotonically  decreasing  function,  starting at  a
non-zero   value  in   the  IR  ($m^2(0)   >0$)   and  dropping
``sufficiently  fast'' in  the  deep UV.
\newline
\indent
Arguments based on the Operator Product Expansion (OPE) suggest that 
$m^2(q^2)$ should display power-law running,    
of the type  $m^2(q^2)\sim  \langle G^2 \rangle/q^2$,
where $\langle G^2 \rangle$ is 
 the  gauge-invariant
gluon condensate of  dimension  four: $\langle  G^2 \rangle = \langle 0|\!:\!G_{\mu\nu}^{a}
G^{\mu\nu}_{a}\!:\!|0  \rangle$.  
The unambiguous connection between the gluon mass and the gluon condensate 
 established by Lavelle ~\cite{Lavelle:1988eg}
merits further comments, because it 
constitutes another important success of the PT (for early calculations relating 
the gluon condensate and the effective gluon mass, see \cite{Graziani:1984cs}). 
Specifically, 
the OPE was used to find the contribution of the  
$\langle G^2 \rangle$
condensate to the {\it conventional} gluon propagator,
but the results turned out to be of limited usefulness: 
in addition to $\langle G^2 \rangle$, 
gauge-dependent condensates involving the
ghost fields $c$ and $\bar{c}$ also appeared \cite{Lavelle:1988eg}.  
This calculation amply demonstrates that no physical
results can be obtained from the OPE for a gauge-dependent quantity, such as
the usual gluon propagator.  
Then, the  same calculation was repeated 
for the PT propagator with very different results~\cite{Lavelle:1991ve}: 
{\it only} the gauge-invariant condensate $\langle G^2 \rangle$
appeared, and in just such a way 
that it could be interpreted as contributing to a running mass.  This result
is equivalent to saying that, at large 
Euclidean momentum, $\widehat{\Delta}$ in  $SU(N)$ behaves as
\begin{equation}
\label{lavelle}
\widehat{\Delta}^{-1}(q)\rightarrow q^2 + \frac{17N}{18(N^2-1)}
\frac{\langle G^2 \rangle }{q^2}.
\end{equation}
Note that the multiplicative constant is positive, so this OPE correction has the
right sign to represent a running mass, since the condensate $\langle G^2 \rangle$  is also
positive (actually, powers of logarithms of $q^2$ can also occur, but we ignore them here).
We emphasize also that this kind of power-law 
running has also been obtained from independent SDE studies
~\cite{Cornwall:1981zr,Cornwall:1985bg,Aguilar:2007ie}.
\newline
\indent
An effective low-energy field 
theory for describing the gluon mass   
is  the  gauged non-linear sigma model  known  as ``massive
gauge-invariant Yang-Mills''~\cite{Cornwall:1979hz}, with    
Lagrangian density  
\begin{equation}
{\cal L}_{\mathrm{MYM}}= \frac{1}{2} G_{\mu\nu}^2 - 
m^2 {\rm Tr} \left[A_{\mu} - {g}^{-1} U(\theta)\partial_{\mu} U^{-1}(\theta) \right]^2\,,
\label{nlsm}
\end{equation}
where 
$A_{\mu}= \frac{1}{2i}\sum_{a} \lambda_a A^{a}_{\mu}$, the $\lambda_a$ are the SU(3) generators
(with  ${\rm Tr} \lambda_a  \lambda_b=2\delta_{ab}$), 
and the $N\times N$
unitary matrix $U(\theta) = \exp\left[i\frac{1}{2}\lambda_a\theta^{a}\right]$ 
describes the scalar fields $\theta_a$.  
Note that ${\cal L}_{\mathrm{MYM}}$ is locally gauge-invariant under the combined gauge transformation 
\be
A^{\prime}_{\mu} = V A_{\mu} V^{-1} - {g}^{-1} \left[\partial_{\mu}V \right]V^{-1}\,, 
\qquad
U^{\,\prime} = U(\theta^{\,\prime}) = V U(\theta)\,,
\label{gtransfb}
\ee
for any group matrix $V= \exp\left[i\frac{1}{2}\lambda_a\omega^{a}(x)\right]$, where 
$\omega^{a}(x)$ are the group parameters. 
One might think that, by employing (\ref{gtransfb}), the fields  $\theta_a$ can always 
be transformed to zero, but this is not so if the $\theta_a$ contain vortices.
To use the ${\cal L}_{\mathrm{MYM}}$ in (\ref{nlsm}), one solves the equations of motion for $U$ 
in terms of the gauge potentials and substitutes the result in the equations for the gauge potential.  
One then finds Goldstone-like massless modes, that, as we will see later in this section, 
are instrumental for enforcing gauge-invariance. 
This model admits  vortex
solutions~\cite{Cornwall:1979hz},  with a  long-range pure  gauge term  in  their potentials,
which endows  them with a topological quantum  number corresponding to
the center  of the gauge group  [$Z_N$ for $SU(N)$], and  is, in turn,
responsible for quark  confinement and gluon screening~\cite{Cornwall:1979hz,Bernard:1982my}. 
Specifically, center vortices of  thickness $\sim m^{-1}$,  where $m$ is
the induced mass of the gluon, form a condensate because their entropy
(per  unit  size) is  larger  than  their  action.  This  condensation
furnishes an  area law to  the fundamental representation  
Wilson loop, thus confining quarks. 
 On the  other hand, the adjoint potential shows a
roughly linear  regime followed by string breaking  when the potential
energy is about $2m$, corresponding to gluon screening~\cite{Cornwall:1981zr,Cornwall:1979hz}. 
Of course, ${\cal L}_{\mathrm{MYM}}$ is not renormalizable, and breaks down in the ultraviolet.  
This breakdown simply reflects the fact that the gluon mass $m$ in (\ref{nlsm})
is assumed to be constant, while, as commented above, both the OPE and the SDEs  
furnish a momentum-dependent gluon mass, vanishing at large $q^2$.  
\begin{figure}[!t]
\bce
\includegraphics[width=6.5cm,angle=-90]{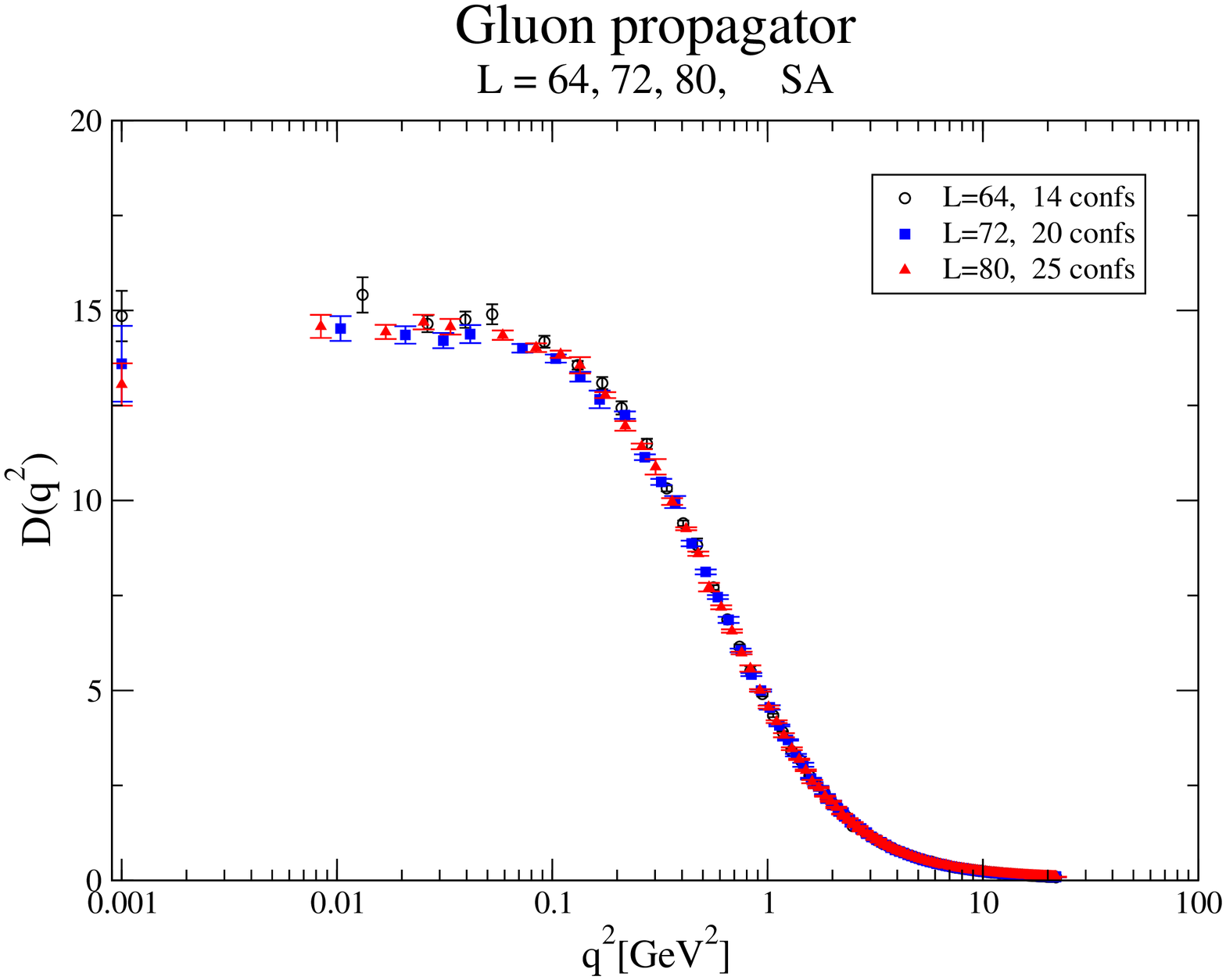}\hspace{-1cm}
\includegraphics[width=6.5cm,angle=-90]{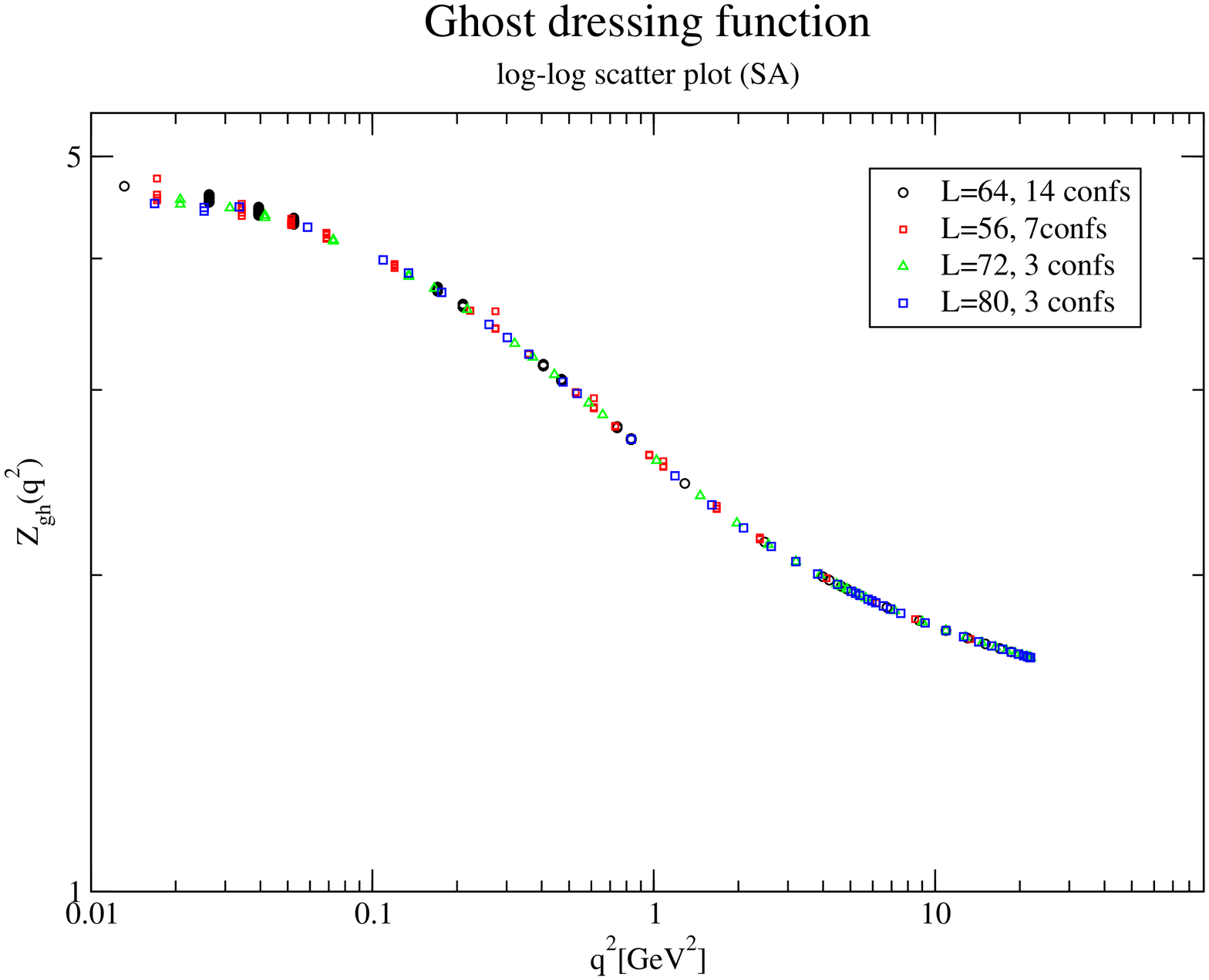}
\ece
\caption{\figlab{lat_res_gluon&ghost}({\it Left panel}) The gluon propagator calculated for different lattice sizes at $\beta=5.7$, from Ref.~\cite{Bogolubsky:2007ud}. The data points drawn at $q^2=0.001$ represent the zero-momentum gluon propagator $\Delta(0)$. ({\it Right panel}) The ghost dressing function $Z_{\mathrm{gh}}(q^2)=q^2 D(q^2)$  for the same value of $\beta$. Notice that no power-law enhancement is observed for this quantity (see discussion in subsection~\ref{latcomp}).} 
\end{figure}
\newline
\indent
The main theoretical tool for quantitative calculations in the infrared region of QCD, 
aside from the SDEs,  is the lattice. In this framework, QCD is approximated 
by a lattice gauge theory with a non-zero 
lattice spacing and a finite space-time volume. 
In this way, one reduces the infinite functional integrals to a finite number of finite integrations, 
thus allowing the computation of correlation functions 
by numerical evaluations of these integrals via Monte-Carlo methods.
The gluon and ghost propagators (in various gauges) 
have been studied extensively 
on the lattice~\cite{Alexandrou:2000ja,Alexandrou:2001fh,Alexandrou:2002gs}.
To be sure, lattice simulations  of gauge-dependent quantities are known to 
suffer from the problem of the Gribov copies, especially in the infrared 
regime, but it is generally believed 
that the effects are  quantitative rather than qualitative. 
The effects of the Gribov ambiguity on the ghost propagator 
become more pronounced in the infrared, 
while their impact on the gluon propagator 
usually stay within the statistical error of the simulation~\cite{Williams:2003du,Sternbeck:2004qk,Silva:2004bv}.
It turns out that a large body of lattice 
data, produced over several years, confirm that the 
gluon propagator  reaches indeed a finite (non-vanishing) value in the deep 
IR, as predicted by Cornwall. This  rather characteristic  behavior 
was already suggested by early studies, and 
has been  firmly established  recently using
large-volume lattices, for pure  Yang-Mills (no quarks included), for
both $SU(2)$~\cite{Cucchieri:2007md} and $SU(3)$~\cite{Bogolubsky:2007ud} (see Fig.~\fref{lat_res_gluon&ghost}). 

\subsection{PT Schwinger-Dyson equations for the gluon and ghost propagators}

\noindent
As mentioned above, in the original analysis of gluon mass generation 
a simplified (and linearized)  SDE was considered, that involved only 
the gluon self-energy, with no ghost loops included~\cite{Cornwall:1981zr}. 
In this section we go one step further: we will 
exploit the powerful machinery offered by the 
SDE truncation scheme introduced in the previous section, in order to study 
gauge-invariantly the gluon-ghost system. In particular, we will show how  
to obtain self-consistently an infrared finite gluon propagator and a divergent (but non-enhanced) 
ghost propagator, in qualitative agreement with recent lattice data~\cite{Bogolubsky:2007ud}.
It is worth emphasizing that this behavior has also been 
confirmed within the Gribov-Zwanziger formalism~\cite{Dudal:2008sp}. 
\newline
\indent
In the previous sections 
we have employed the standard notation of the BV formalism, where all Green's functions
are denoted by the letter $\Gamma$, and  
all incoming fields (together with their color and Lorentz indices) are explicitly displayed as subscripts.
This notation is completely unambiguous, and is particularly suited 
for formal manipulations carried out so far, 
but is rather cumbersome for actual applications. Therefore, in this section we will 
switch to a simplified notation, that will result much more familiar to the SDE practitioners.
\begin{figure}
\bce
\hspace{-.8cm}\includegraphics[width=16cm]{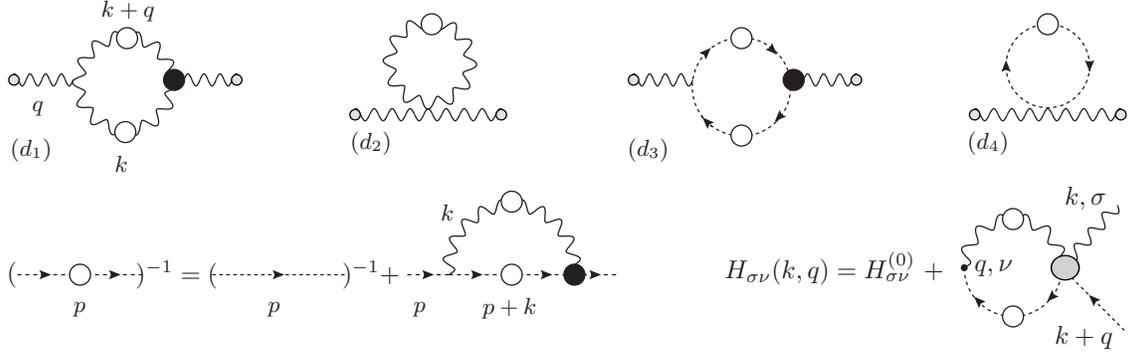}
\ece
\caption{\figlab{SDeqs}The new SDE for the gluon-ghost system.
Wavy lines with white blobs are 
full gluon propagators, dashed lines with 
white blobs are full-ghost propagators, black blobs are full vertices, and 
the gray blob denotes the scattering kernel. The circles attached 
to the external gluons denote that, 
from the point of view of Feynman rules, they are treated as background fields.}
\end{figure}
\newline
\indent 
The SDEs for the gluon-ghost system are shown in \Figref{SDeqs}. Evidently, 
we only consider the ``one-loop dressed'' contributions, leaving out 
(gauge-invariantly!) two-loop dressed diagrams. Indeed, 
as we know from the previous analysis 
this truncation preserves gauge-invariance, in the sense that 
it does {\it not} compromise the transversality of the gluon self-energy. 
In the case of pure (quark-less) QCD these SDEs read
\bea
&&\Delta^{-1}(q^2){P}_{\mu\nu}(q) =
\frac{q^2 {P}_{\mu\nu}(q) + i\sum_{i=1}^{4}(d_i)_{\mu\nu}}{[1+G(q^2)]^2},
\label{newSDb-1}\nonumber\\
&& i D^{-1}(p^2) = p^2 + i \lambda
\int_k \Gamma^{\mu}\Delta_{\mu\nu}(k)\gb^{\nu}(p,k) D(p+k),
\label{newSDb-2}\nonumber\\
&& i\Lambda_{\mu \nu}(q) = \lambda
\int_k H^{(0)}_{\mu\rho} D(k+q)\Delta^{\rho\sigma}(k)\, H_{\sigma\nu}(k,q),
\label{newSDb-3}
\eea
where \mbox{$\lambda=g^2 C_{\rm {A}}$}. 
$\Gamma_{\mu}$ is the standard (asymmetric) gluon-ghost vertex at tree-level,
and $\gb^{\nu}$ the fully-dressed one.
$G(q^2)$ is the $g_{\mu\nu}$ component
of the auxiliary two-point function $\Lambda_{\mu \nu}(q)$, and 
the function $H_{\sigma\nu}$ is 
defined diagrammatically in \Figref{SDeqs}.
$H_{\sigma\nu}$, and  
is related to the full gluon-ghost vertex by 
\mbox{$q^{\sigma} H_{\sigma\nu}(p,r,q) = -i{\gb}_{\nu}(p,r,q)$}; 
at tree-level, $H_{\sigma\nu}^{(0)} = ig_{\sigma\nu}$.
\newline
\indent
Using the BFM rules of \appref{Frules} to evaluate the diagrams $(d_i)$,
we find
\bea
(d_1)&=&
\frac{\lambda}{2}\int_k \widetilde{\Gamma}_{\mu\alpha\beta}
\Delta^{\alpha\rho}(k) {\g}_{\nu\rho\sigma} \Delta^{\beta\sigma}(k+q),
\label{d-1}\nonumber\\
(d_2)&=&
-i\lambda g_{\mu\nu}\int_k \!\Delta^{\rho}_{\rho}(k)
-i \lambda \left(\frac{1}{\xi}-1\right) \int_k \!\Delta_{\mu\nu}(k),
\label{d-2}\nonumber\\
(d_3)&=& -\lambda \int_k  \widetilde{\Gamma}_{\mu} D(k)  D(k+q) {\g}_{\nu},
\nonumber\\
(d_4)&=&   2i \lambda g_{\mu\nu} \int_k D(k).
\eea
In the above formulas, 
\mbox{$\widetilde{\Gamma}_{\mu\alpha\beta} (q,p_1,p_2)\! =\!
\Gamma_{\mu\alpha\beta} (q,p_1,p_2)\! + \xi^{-1}\Gamma^\mathrm{P}_{\mu\alpha\beta}(q,p_1,p_2)$}, 
where  
$\Gamma_{\mu\alpha\beta}$ is the standard QCD three-gluon vertex, and  ${\g}_{\mu\alpha\beta}$ its
the fully-dressed version. Similarly, 
$\widetilde{\Gamma}_{\mu}$ denotes the {\it symmetric} gluon-ghost vertex at tree-level 
and ${\g}_{\mu}$ its fully-dressed counterpart. Due to the Abelian all-order WIs that these two full 
vertices  satisfy (for all $\xi$), namely
\bea
q^{\mu}{\g}_{\mu\alpha\beta}=i\Delta^{-1}_{\alpha\beta}(k+q)
-i\Delta^{-1}_{\alpha\beta}(k)\,,
\label{SDEWIa}\nonumber\\
q^{\mu}{\g}_{\mu} = iD^{-1}(k+q) - iD^{-1}(k)\,, 
\label{SDEWIb}
\eea
one can easily demonstrate that 
$q^{\mu}[(d_1)+(d_2)]_{\mu\nu} =0$
and \mbox{$q^{\mu}[(d_3)+(d_4)]_{\mu\nu} =0$}~ \cite{Aguilar:2006gr}. 
\newline
\indent
In order to make contact with the lattice results of~\cite{Bogolubsky:2007ud,Cucchieri:2007md} 
shown in Fig.~\fref{lat_res_gluon&ghost}, 
we will have to project the above system of coupled SDEs 
in the LG ($\xi =0$). This is a subtle exercise, 
because one cannot set directly $\xi =0$ in the integrals on the 
rhs of Eqs~(\ref{d-1}), due to the terms proportional to  $\xi^{-1}$. 
Instead, one has to use the expressions for general $\xi$, 
carry out explicitly the set of cancellations produced 
when the terms proportional to $\xi$ generated by the identity
\be
k^{\mu} \Delta_{\mu\nu}(k)= -i \xi k_{\nu}/k^2,
\label{fund_id}
\ee are used to cancel $\xi^{-1}$  terms, and set $\xi =0$ only at the
very end (this exercise is very similar to the BFM
pinching    we   have    carried    at   the    one-loop   level    in
subsection~\ref{PT/BFG:conceptual}).  
\newline
\indent
Let us focus on  graph $(d_1)$.   
First of all, it is relatively straightforward
to establish that only the bare part of the full ${\g}$ 
may furnish contributions proportional to $\xi^{-1}$.  
To see why this is so,  consider the  SDE of  ${\g}$  shown   in
\Figref{ggg_SDE}. Evidently, the $\xi^{-1}$ parts  of the bare 
vertex are longitudinal; therefore, by virtue of  Eq.~(\ref{fund_id}), 
they cancel when contracted with an internal gluon propagator. 
Moreover, all kernels appearing 
in this SDE are regular in the LG,  since all the fields entering 
are  quantum ones (it is like computing the kernels in the normal LG, 
where no  $\xi^{-1}$ terms exist).
Therefore, 
writing ${\g}_{\nu\alpha\beta} = \widetilde{\Gamma}_{\nu\alpha\beta} + {\k}_{\nu\alpha\beta}$,
we have that ${\k}_{\nu\alpha\beta}$ 
is regular in the limit $\xi\to 0$; 
we will denote by ${\kb}_{\nu\alpha\beta}$ its value at $\xi =0$. 
Thus, the only divergent contributions contained in  $(d_1)$  
reside in the product $\widetilde{\Gamma}_{\mu\alpha\beta} \widetilde{\Gamma}_{\nu\rho\sigma}$.  
It is then a simple algebraic exercise to 
demonstrate that they  cancel exactly against the part of the seagull diagram $(d_2)$ 
proportional to $\xi^{-1}$.  
Thus, taking the LG limit we obtain 
\bea
\sum_{i=1}^2(d_i)_{\mu\nu}&=&\lambda\Bigg\{\frac{1}{2}\int_k
\Gamma_\mu^{\alpha\beta}\Delta_{\alpha\rho}^\mathrm{t}(k)\Delta_{\beta\sigma}^{t}(k+q)\l_\nu^{\rho\sigma}
-\frac{9}{4}g_{\mu\nu}\int_k \Delta(k) \nonumber \\
&+&\int_k \!\!\Delta_{\alpha\mu}^\mathrm{t}(k) 
\frac{(k+q)_{\beta}}{(k+q)^2}[\Gamma+ {\l}]_\nu^{\alpha\beta}
+\int_k \frac{k_{\mu}(k+q)_{\nu}}{k^2 (k+q)^2}\Bigg\},
\label{contr}
\eea
where $\Delta^\mathrm{t}_{\mu\nu}(q)=P_{\mu\nu}(q)\Delta(q^2)$, and 
$\l_{\mu\alpha\beta} \equiv \Gamma_{\mu\alpha\beta}+\kb_{\mu\alpha\beta}$.  
Note that $\l_{\mu\alpha\beta}$ satisfies the WI
\be
q^{\mu} \l_{\mu\alpha\beta} = {\rm P}_{\alpha\beta}(k+q)\Delta^{-1}(k+q)
-{\rm P}_{\alpha\beta}(k)\Delta^{-1}(k)\,,  
\ee
as a direct consequence of (\ref{SDEWIa}).
Therefore, one can verify that the 
lhs of (\ref{contr}) vanishes when contracted by $q^{\mu}$, 
thus proving the announced transversality of this subset of graphs.
\newline
\indent
In order to proceed with our analysis, we need to furnish some information 
about the all-order vertices $\l_{\mu\alpha\beta}$, ${\g}_{\mu}$, $\gb_{\mu}$, and $H_{\mu\nu}$, 
entering in our system of SDEs. Of course, these vertices satisfy their own SDEs; 
so, in principle, one should couple all these equations together, 
to form an even more extended (and more intractable) system of coupled integral equations. 
As mention at the end of the previous section, 
given the 
practical difficulties of such a task, 
one resorts to  the ``gauge technique''~\cite{Salam:1963sa,Salam:1964zk,Delbourgo:1977jc},  
expressing the vertices as functionals of the various self-energies 
involved, in such a way as to satisfy by construction the correct WIs\footnote{As already mentioned, this method leaves the transverse 
(\ie identically conserved) part of the vertex undetermined. The 
transverse parts are known to be subleading in the IR, when a mass gap is formed, 
but must be supplied in the UV, because they are instrumental 
for enforcing the cancellation of the overlapping divergences and the 
correct renormalization group properties.}.
The Ansatz we will use for ${\l}_{\mu\alpha\beta}$ and ${\g}_{\mu}$ is 
\bea
{\l}_{\mu\alpha\beta}&=& \Gamma_{\mu\alpha\beta} + i\frac{q_{\mu}}{q^2}
\left[\Pi_{\alpha\beta}(k+q)-\Pi_{\alpha\beta}(k)\right],
\label{gluonv}\nonumber\\
{\g}_{\mu}&=& \widetilde{\Gamma}_{\mu} -i\frac{q_{\mu}}{q^2}
\left[L(k+q)-L(k)\right].
\label{ghostv}
\eea
where $L$ denotes the ghost self-energy, $D^{-1}(p^2)=p^2-iL(p^2)$. 
As announced, the corresponding WIs, (\ref{SDEWIa}) and (\ref{SDEWIa}), are 
identically satisfied.
On the other hand, for the conventional ghost-gluon vertex ${\gb}_{\mu}$,
appearing in the SDE of (\ref{newSDb-2}) we will use its
tree-level expression, {\it i.e.}, ${\gb}_{\nu}\to \Gamma_{\mu} = -p_{\mu}$.
Note that, unlike ${\g}_{\mu}$,  the conventional ${\gb}_{\mu}$ 
satisfies a STI of rather limited usefulness; the ability to 
employ such a different treatment for ${\g}_{\mu}$ and  ${\gb}_{\mu}$  
without compromising gauge-invariance 
is indicative of the versatility of the new SD formalism used here.
Finally, for $H_{\mu\nu}$ we use its tree-level value, $H_{\mu\nu}^{(0)}= i g_{\mu\nu}$.
\newline
\indent
The above Ansatz for the vertices needs further explaining, 
in view of the fact that it  contains a 
 longitudinally coupled pole $1/q^2$.
 We hasten to emphasize that the origin of this 
pole is not kinematic but rather dynamical, {\it i.e.}, it is a  composite (bound-state) pole, 
whose presence is instrumental for the realization of 
the Schwinger mechanism in $d=4$, leading to 
\mbox{$\Delta^{-1}(0)\neq 0$}. Given the importance of this mechanism to our approach, 
and the subtlety of the various concepts invoked, in the next subsection we 
present a brief overview of the Schwinger mechanism and its connection with 
the Goldstone phenomenon and the Higgs mechanism~\cite{Farhi:1982vt,Jackiw:1973ha}.

\subsection{Schwinger mechanism, dynamical gauge-boson mass generation, and bound-state poles}
\begin{figure}
\bce
\includegraphics[width=14cm]{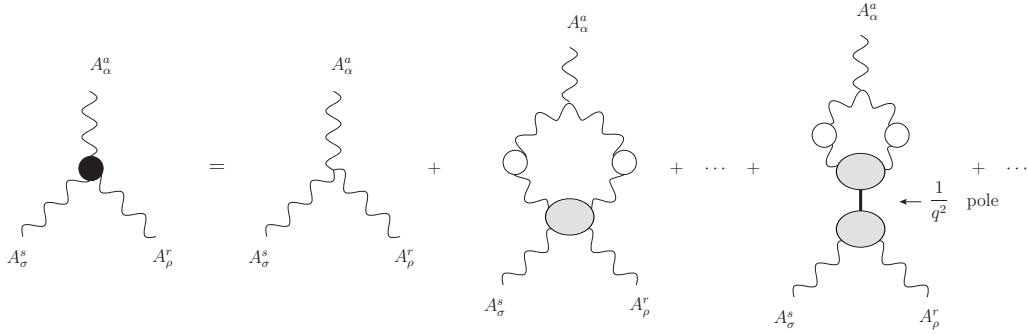}
\ece
\caption{\figlab{SDE_ggg_poles} The new SDE for the three-gluon vertex. Note that, as always, the kernels 
are one-particle irreducible; thus, the $1/q^2$ pole is not kinematic but dynamical, and 
is completely non-perturbative, \ie it vanishes to all orders in perturbation theory.
Physically it corresponds to a  
longitudinally coupled massless composite excitation, acting as the (composite) Goldstone mode 
necessary for maintaining the local gauge invariance~\cite{Jackiw:1973tr}.}
\end{figure}
\noindent
According to Goldstone's theorem~\cite{Nambu:1960tm,Goldstone:1961eq,Goldstone:1962es}, 
the spontaneous breaking 
of a continuous global symmetry is accompanied by massless excitations, known 
as ``Goldstone'' particles\footnote{In most cases these particles are scalars, 
and are hence  referred to as bosons, even though  
spin $\frac{1}{2}$ Goldstone particles may exist, as for 
example when supersymmetry is spontaneously broken.}.
Even though Goldstone's theorem is most frequently 
realized by a scalar field acquiring a vacuum expectation value,
Nambu and Jona-Lasinio \cite{Nambu:1961tp,Nambu:1961fr} 
introduced the notion of a dynamical 
Goldstone boson, demonstrating that the Goldstone mechanism can take place 
even when the Lagrangian does not include scalar fields. 
Thus, when chiral symmetry is broken, 
the associated Goldstone boson (the pion)
is not an elementary excitation of a fundamental scalar field, but it is rather formed 
as a quark-antiquark bound state.
\newline
\indent
Goldstone's theorem, however, does not apply 
when the symmetry that is spontaneously broken is not global but local, 
accompanied by the associated massless gauge boson mediating the interaction (equivalently, one may say that 
the theorem  does not apply 
in the presence of gauge-invariant long range forces).
In such a case  
the Goldstone phenomenon is replaced by the Higgs phenomenon: 
no massless excitations appear in the spectrum, because the 
would-be Goldstone bosons
combine with the transverse gauge boson to furnish the third helicity state 
of a massive spin one particle 
\cite{Higgs:1964pj,Higgs:1964ia,Englert:1964et,Guralnik:1964eu,Higgs:1966ev}. 
In addition, a massive scalar particle, 
the Higgs boson, also appears in the spectrum, and is 
instrumental for the renormalizability of the theory.
Of course, both mechanisms may co-exist;
if a local and a global symmetry are broken, 
one may have both phenomena. For example, the 
spontaneous breaking of a gauge symmetry through the Higgs mechanism 
gives mass to the gauge bosons; if, in addition, a global symmetry 
(e.g. chiral symmetry) is broken, then 
massless Goldstone bosons are present in the spectrum.
As we will see in a moment, the opposite may also happen:
the breaking of a global symmetry may furnish the Goldstone bosons  
that (in the absence of elementary scalar fields) will be 
absorbed by the gauge bosons, which will thus become massive.
\newline
\indent
Independently of the above consideration, and before the 
Higgs mechanism was even invented, 
Schwinger argued  
that the gauge invariance of a vector field does not necessarily 
imply zero mass for the associated particle, if the 
current vector coupling is sufficiently strong \cite{Schwinger:1962tn}. 
Schwinger's fundamental observation 
was that if (for some reason) the vacuum polarization 
tensor $\Pi_{\mu\nu}(q)$ acquires a pole at zero momentum transfer, then the 
 vector meson becomes massive, even if the gauge symmetry 
forbids a mass at the level of the 
fundamental Lagrangian \cite{Schwinger:1962tp}.
Indeed, casting the self-energy in the familiar form 
\be
\Delta^{\mu\nu}(q^2) = \left(g^{\mu\nu} - \frac{q^{\mu}q^{\nu}}{q^2} \right) 
\frac{-i}{q^2[1+{\bf \Pi}(q^2)]},
\label{JS}
\ee
it is clear that if ${\bf \Pi}(q^2)$ has a pole at  $q^2=0$ with positive 
residue $\mu^2$, then 
the vector meson 
is massive, even though it is massless in the absence of interactions 
($g=0$, ${\bf \Pi} =0$). 
\newline
\indent
Indeed, there is {\it no} physical principle which would preclude ${\bf \Pi}(q^2)$ from 
acquiring a pole.  
Since bound states are expected to exist in most physical systems, and to produce poles 
in $\Pi_{\mu\nu}$ at time-like momenta, one may suppose that for sufficiently 
strong binding, the mass of such a bound state will be reduced to zero, thus generating a mass
for the vector meson without interfering with gauge invariance. 
Schwinger demonstrated his general ideas in two-dimensional massless spinor QED 
(``Schwinger model''), which, by virtue of the special properties 
of the Dirac algebra at $d=2$, is explicitly solvable: 
$\Pi_{\mu\nu}$ {\it does} indeed have a pole at $q^2=0$, and the photon acquires a mass,   
$\mu^2 = e^2/\pi$ (in $d=2$, $e$ has dimensions of mass)\footnote{The generation of the pole in the Schwinger model is 
related to the anomaly of the axial-vector current, which 
is the reason why Godstone's theorem is evaded in this case. 
This is in fact consistent with Coleman's theorem \cite{Coleman:1973ci}, 
stating the absence of Goldstone bosons in $d=2$.}.
\newline
\indent
Perhaps the most appealing feature of the Schwinger mechanism is that  
the appearance of the required pole may happen for purely dynamical reasons, 
and, in particular, without the 
need to introduce fundamental scalar field in the Lagrangian.
In fact, the Higgs mechanism can be viewed as just 
a {\it very special realization} of the Schwinger mechanism: 
the vacuum expectation value $v$ of a canonical 
scalar field coupled to the vector meson
gives rise to tadpole contributions to ${\bf \Pi}(q^2)$, which produce a pole.
Thus, the Higgs mechanism corresponds to the special case where 
the residue of the pole is saturated by $v^2$, furnishing 
a gauge-boson mass $\mu^2 = 2 g^2 v^2$; the pole required is provided  
by the would-be Goldstone particles, which decouple from the spectrum.
\newline
\indent
In order to employ the Schwinger mechanism in realistic field theories,  
ultimately one must be able to demonstrate that a  pole is generated somehow.
 Of course, the Higgs mechanism guarantees 
that, at the classical or semi-classical level. In the absence of 
fundamental scalar fields the realization of the mechanism is more subtle, 
but, at the same time, conceptually superior, because 
one does not have to assume the existence  of fundamental scalars 
(not observed in Nature, to date).
Therefore, in the seventies,  
an appealing alternative to the Higgs mechanism 
was extensively considered. 
The idea was to combine the {\it dynamical} Goldstone mechanism 
(\ie the Nambu--Jona-Lasino mechanism 
with composite Goldstone particles)
with a gauge theory,  in order to give  
{\it dynamical gauge-invariant masses} to the vector mesons. 
In this two-step scenario, mass generation is proceeded  
by the spontaneous breaking of a global (chiral) symmetry;
 the required pole in $\Pi_{\mu\nu}$ 
is provided by composite Goldstone bosons 
(fermion-antifermion bound states).
Gauge invariance (long range forces), on the other hand,    
ensures that these massless excitations will decouple from the 
complete physical scattering amplitude. 
The actual realization of the decoupling 
goes through a rather subtle 
mechanism, as has been 
demonstrated explicitly in the toy models considered
~\cite{Jackiw:1973tr,Cornwall:1973ts}.
\newline
\indent
Even though this activity was mainly directed towards 
an alternative descriptions (\ie without resorting to tree-level Higgs mechanism and elementary scalars)
of the electroweak sector, Grand Unified Theories, and general model-building, 
these profound ideas invariably influenced our approach to the strong interactions.
The general philosophy adopted when 
applying some of the dynamical concepts described above
to pure Yang-Mills theories 
(without matter fields), such as quarkless QCD, 
is the following~\cite{Eichten:1974et}.
One assumes that, 
in a strongly-coupled gauge theory,   
longitudinally coupled,   
zero-mass bound-state 
excitations are dynamically produced. To be sure, the demonstration of  
the existence of a bound state, and in particular a zero-mass bound state, 
in realistic field theories is a difficult dynamical problem, usually  
studied by means of integral equations, known as Bethe-Salpeter equations
(see, e.g.,~\cite{Poggio:1974qs}).
Thus, it is clear that a vital ingredient for this scenario 
is strong coupling, which can only come from the infrared
instabilities of a non-Abelian gauge theories. 
The aforementioned excitations are {\it like} 
dynamical Nambu-Goldstone bosons, in the sense that
they are massless, composite, and longitudinally coupled; 
but, at the same time, they differ from Nambu-Goldstone bosons 
as far as their origin is concerned: they do {\it not}
originate from the spontaneous breaking of any global symmetry.
The main role of these  excitations is to trigger the Schwinger mechanism, 
i.e. to provide the required pole in the gluon self-energy  
[specifically, the gauge-independent ${\bf {\widehat\Pi}}(q^2)$
obtained with the PT] thus furnishing 
(gauge-invariantly) a dynamical 
mass for the gluons.  The additional 
important step is then to demonstrate that every such 
Goldstone-like scalar that is ``eaten" by
a gluon to give it mass is actually canceled out of the $S$-matrix 
by other massless poles, or by current conservation.    
\newline
\indent
Exactly how should the idea of a composite excitation 
be incorporated  
at the level of Green's functions 
and the corresponding SDEs?
A composite excitation is represented as a pole in an off-shell Green's function
representing a field that does not exist in the classical action, but that
occurs in the solution of the SDE for that Green's
function, as a sort of bound state.
To make contact with our starting point, i.e. the Ansatz of (\ref{gluonv})-(\ref{ghostv}),
we turn to the following simplified situation. 
Consider the  WI of Eq.~(\ref{new_WI3g_PT}),   
and ask the following question: supposing that $\Delta_{\mu\nu}^{(0)}$ develops  
a mass, how must one modify $\tilde{\Gamma}_{\alpha\mu\nu}$
in order for the WI to continue been valid,  
which is tantamount to saying that the gauge-invariance remains intact?
Thus, replace in $\Delta_{\mu\nu}^{0}$ of (\ref{GluProp}) the 
$d^{-1}(k)= k^2$ by $d^{-1}_m(k) =  k^2+m^2$, and substitute 
the resulting $\Delta_{m\,\mu\nu}^{(0)}$ in the rhs of Eq.~(\ref{new_WI3g_PT}).
In order to maintain the validity of the WI one must simultaneously replace  
$\tilde{\Gamma}_{\alpha\mu\nu}$ on the lhs by $\tilde{\Gamma}_{\alpha\mu\nu}^{m}$
given by 
\be
\tilde{\Gamma}_{\alpha\mu\nu}^{m}(q,k_1,k_2) = 
\tilde{\Gamma}_{\alpha\mu\nu}(q,k_1,k_2) -\left[\frac{m^2}{2}
\frac{q_{\alpha}k_{1\mu}(q-k_1)_{\nu}}{q^2 k_1^2} + {\rm c.p.}\right],
\label{massvertex}
\ee
where ``c.p.'' stands for cyclic permutations.
The new vertex $\tilde{\Gamma}_{\alpha\mu\nu}^{m}(q,k_1,k_2)$
has, as we mentioned above, terms with
longitudinally-coupled massless poles, whose residue is $m^2$.  
If the propagator is
transverse and has mass, this is the
only way that the original WI can be satisfied, 
just as the only way a massive gauge-boson
propagator can be transverse is if it has similar poles in the transverse
projector $P_{\mu\nu}$. 

\subsection{\label{latcomp}Results and comparison with the lattice}
\noindent
Substituting into the system of SDEs 
the expressions for the various vertices, as discussed above,  
and carrying out elementary but lengthy algebraic manipulations, 
we arrive at the final form of the SDEs, presented in ~\cite{Aguilar:2008xm}.
\newline
\indent
The crucial point is the behavior of 
(\ref{newSDb-1}) as $q^2\to 0$, where the ``freezing'' of the 
gluon propagator is observed.
In this limit, Eq.(\ref{newSDb-1}) yields
\be
\Delta^{-1}(0) \sim
15 \int_k \Delta(k) - 6 \int_k k^2 \Delta^2(k).
\label{tad}
\ee
The rhs  of~Eq.(\ref{tad}) vanishes perturbatively, by virtue  of the
dimensional regularization  result 
\be 
\int_k  \frac{\ln^{n}\! k^2}{k^2}=0,\qquad n={0,1,2,}\dots 
\label{dimreg}
\ee  
which ensures  the masslessness  of the
gluon to all orders. However, $\Delta^{-1}(0)$ does
{\it not} have to vanish non-perturbatively,  provided that the 
quadratically divergent integrals defining it 
can be properly regulated and made finite, {\it without} 
introducing counterterms of the 
form $ m^2_0 (\Lambda^2_{\chic{\mathrm{UV}}}) A^2_{\mu}$, 
which are forbidden by the local gauge invariance 
of the  fundamental  QCD Lagrangian.
It turns out that this is indeed possible:
the divergent integrals can be regulated by 
subtracting appropriate combinations of ``dimensional regularization zeros''.
Specifically, for large enough $k^2$  
the $\Delta(k^2)$ goes over to its 
perturbative expression, to be denoted by 
$\Delta_{\rm pert}(k^2)$; it has the form  
\be
\Delta_{\rm pert}(k^2)=\sum_{n=0}^N a_n\frac{\ln^{n} k^2}{k^2}, 
\ee
where the coefficient $a_n$ are known from the perturbative expansion. 
Thus, we obtain for the regularized $\Delta^{-1}_{\rm reg}(0)$ 
\be 
\Delta^{-1}_{\rm reg}(0) \sim 15 \int_0^s\!\!dy\ y
\left[   \Delta(y)    -\Delta_{\rm pert}(y)\right]
- 6 \int_0^s\!\!dy\ y^2\left[\Delta^2(y)-\Delta^2_{\rm pert}(y)\right].
\label{regtad}
\ee
\indent
The obvious ambiguity of the regularization described above is 
the choice of the point $s$, past which the two 
curves, $\Delta(y)$ and $\Delta_{\rm pert}(y)$, are 
assumed to coincide. Due to this ambiguity 
one cannot pin down $\Delta_{\rm reg}(0)$ completely, which must be treated, at this level, 
as an arbitrary initial (boundary) condition.
In~\cite{Aguilar:2008xm} $\Delta_{\rm reg}(0)$ was fixed by  
resorting to the lattice data 
of ~\cite{Bogolubsky:2007ud}; specifically, 
when solving the system of SDEs numerically, 
$\Delta^{-1}_{\rm reg}(0)$ was chosen to have the same value as the lattice data at the origin
\mbox{$\Delta_{\rm reg}(0) = 7.3 \,\mbox{GeV}^{-2}$}.
Once this boundary condition is imposed, the system of SDE is solved for the 
entire range of (Euclidean) momenta, from the deep IR to the deep UV.
The solutions obtained are shown in \Figref{figb}, where we show the numerical results for  $\Delta(q^2)$  
and the ghost dressing function renormalized at $\mu=M_b=4.5\,\mbox{GeV}$, and the comparison 
with the corresponding lattice data of Ref.\cite{Bogolubsky:2007ud}.
\begin{figure}
\mbox{}\hspace{-2cm}\includegraphics[width=20cm]{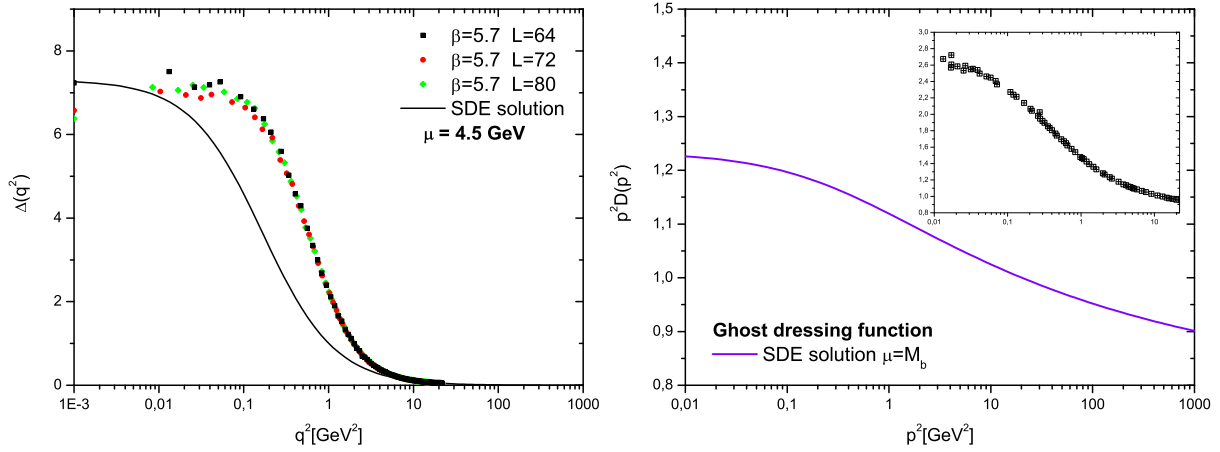}
\caption{\figlab{figb} ({\it Left panel}) The numerical solution for the gluon propagator from the SDE
(black continuous line) compared to the 
lattice data of Ref. \cite{Bogolubsky:2007ud}. ({\it Right panel}) The ghost dressing function $p^2D(p^2)$ 
obtained from the SDE. In the insert we show  the lattice data for the same quantity; notice that there is no 
``infrared enhancement''~\cite{Boucaud:2008ky}; instead, the dressing function saturates at a finite value}
\end{figure}
\newline
\indent
It is interesting to mention that      
the non-perturbative transverse gluon propagator, being finite in the IR, is 
automatically less singular than a simple pole, thus satisfying 
the corresponding Kugo-Ojima confinement criterion~\cite{Kugo:1979gm,Fischer:2006ub}.
Note that for \mbox{$q^2\le 10\, \mbox{GeV}^2$}
both gluon propagators (lattice and SDE)
shown in \Figref{figb} 
may be fitted very accurately using a unique functional form,  
given by \mbox{$\Delta^{-1}(q^2)= a+b\,(q^{2})^{c-1}$}. Specifically, 
measuring $q^2$ in $\mbox{GeV}^2$,
the lattice data are 
fitted by \mbox{$a=0.14$}, \mbox{$b=0.31$},
and \mbox{$c=2.51$}, 
while the SDE solution  is 
described setting \mbox{$a=0.14$}, \mbox{$b=0.86$}, and 
\mbox{$c=2.02$} .
\newline
\indent 
Let us now take a closer look at the ghost propagator, $D(p^2)$. 
First of all, 
the $D(p^2)$ obtained from the ghost SDE diverges
at the origin, in qualitative agreement with the lattice data. 
From the SDE point of view, 
this divergent behavior is due to 
the fact  
that the vertex ${\gb}_{\mu}$ employed does not contain $1/p^{2}$ 
poles, as suggested 
by previous lattice studies~\cite{Cucchieri:2004sq}; actually, 
 ${\gb}_{\mu}$ was fixed at its tree-level value. 
Notice, however, that away from the LG one may obtain an IR finite 
(massive-like) $D(p^2)$, due to the contribution of the 
longitudinal form factor of the vertex ${\gb}_{\mu}$
(i.e. proportional to the gluon momentum $k_{\mu}$); the latter  
gets annihilated in the LG when contracted with the gluon propagator,    
but contributes away from it~\cite{Aguilar:2007nf}.
\newline
\indent 
Of particular theoretical interest is 
the IR behavior of the ghost dressing function $p^2 D(p^2)$,  
because it is intimately 
related to the Kugo-Ojima confinement criterion
for the ghost propagator. This criterion would be satisfied if  
the non-perturbative ghost propagator (in the LG) 
were more singular in the IR than a simple pole; this type of behaviour is 
usually referred to as IR ``enhancement''.
Thus, if we were to fit the 
dressing function \mbox{$p^2 D(p^2)$} [obtained either from the lattice data or the corresponding SDE] 
with a function of the form form \mbox{$p^2 D(p^2)=c_1 (p^{2})^{-\gamma}$}, a positive $\gamma$ 
would indicate that the aforementioned criterion is satisfied. 
\newline
\indent 
It is relatively obvious from both the lattice data and our SDE solutions that 
no such ``enhancement'' is observed; $p^2 D(p^2)$ reaches a finite (positive) 
value as $p^2\to 0$; a detailed fitting exercise confirms this point. 
Indeed, a IR-finite fit of the form  
\mbox{$p^2 D(p^2)=\kappa_1 -\kappa_2\ln(p^2+\kappa_3)$},  
[with $\kappa_3$ acting as an effective gluon mass; 
$p^2$ in $\mbox{GeV}^2$, $k_1=1.12$, $k_2=0.04$, and $k_3=0.08$, valid for the range $p^2<10$ ]
is far superior to any power-law fit that has been tried ~\cite{Aguilar:2008xm}. 
The absence of power-law enhancement has also been 
verified in alternative SDE studies~\cite{Boucaud:2008ky}. 
The fact that the Kugo-Ojima criterion is not satisfied 
is by no means an inconsistency, since a criterion is a sufficient but not a 
necessary condition. In that sense, the old confinement criterion 
associated with a gluon propagator going like 
$1/k^4$ in the IR is not valid either; this simply indicates that 
the real relation between the gluon propagator and confinement is more sophisticated 
than simply calculating a Fourier transform 
(see, for example, \cite{Greensite:2003bk}).  
\newline
\indent
Comparing  the PT solution for the gluon propagator 
with the lattice data we notice that, 
whereas their asymptotic  behavior coincides  (perturbative limits),
there is  a discrepancy of about  a factor of 1.5--2  in the intermediate
region of  momenta, especially around the  fundamental QCD mass-scale
[reflected also in the different values of the two sets of fitting parameters $(a,b,c)$].
In the case of the ghost dressing function,  
a relative difference of similar size is observed. 
These discrepancies may  be accounted for by extending  
the gluon SDE to include the ``two-loop dressed'' graphs, omitted 
(gauge-invariantly) from the original analysis presented in~\cite{Aguilar:2008xm},  
and/or by supplying to the vertex given in (\ref{gluonv}) the missing transverse parts.

\subsection{The non-perturbative effective charge of QCD}
\noindent
As we have seen in detail in subsection~\ref{QCDech}, the PT permits the  
generalization of the prototype QED construction of an effective charge 
in the case of non-Abelian gauge theories, and in particular QCD.  
To remind the reader of the basic steps, we recall that, 
due to the Abelian WIs satisfied by the PT effective Green's functions,    
the PT self-energy $\widehat\Delta^{-1}(q^2)$ absorbs all  
the RG-logs, exactly as happens in QED with the photon self-energy;
specifically, in the deep UV, 
\be 
\widehat\Delta^{-1}(q^2)= q^2\left[1+ b g^2\ln\left(\frac{q^2}{\mu^2}\right)\right],
\label{rightRG}
\ee
where  $b = 11 C_A/48\pi^2$  is the first coefficient of the QCD $\beta$-function. 
Equivalently, since $Z_{g}$ and ${\widehat Z}_{A}$, the renormalization constants 
of the gauge-coupling and the effective self-energy, respectively, 
satisfy the QED relation ${Z}_{g} = {\widehat Z}^{-1/2}_{A}$,  
the product 
${\widehat d}(q^2) = g^2 \widehat\Delta(q^2)$ forms a RG-invariant 
($\mu$-independent) quantity~\cite{Cornwall:1981zr};
for large momenta $q^2$,
\be
{\widehat d}(q^2) = \frac{\overline{g}^2(q^2)}{q^2}\,,
\label{ddef1}
\ee
where $\overline{g}^2(q^2)$ is the RG-invariant effective charge of QCD,
\be
\overline{g}^2(q^2) = \frac{g^2}{1+  b g^2\ln\left(q^2/\mu^2\right)}
= \frac{1}{b\ln\left(q^2/\Lambda^2\right)}\,.
\label{effch}
\ee
\begin{figure}
\hspace{-1.8cm}\includegraphics[width=19.5cm]{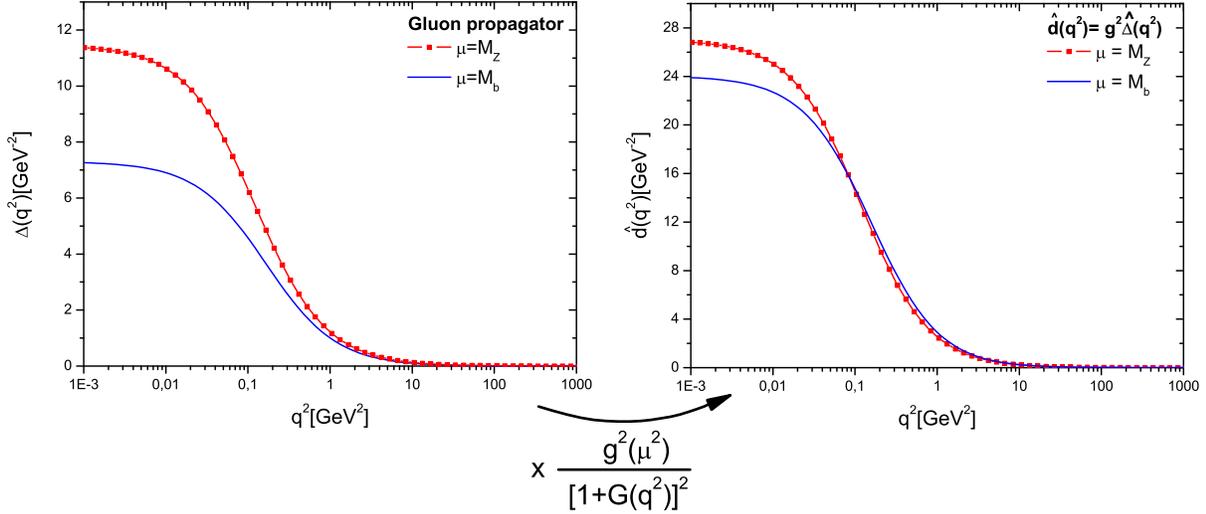}
\caption{\figlab{rgi_fig} ({\it Left panel}) The solution of SDE renormalized at \mbox{$\mu=M_b=4.5\,\mbox{GeV}$}
(continuous blue curve) and  $\mu=M_Z=91\,\mbox{GeV}$ (red line plus square curve). ({\it Right panel}) The corresponding PT-BFM $\widehat{\Delta}(q^2)$ obtained as the convolution of $\Delta(q^2)$  and the function $g^2(\mu^2)/[1+G(q^2)]^2$. }
\end{figure}
\indent
Let us now address the following question: assuming that one has non-perturbative 
information about the IR behavior of the {\it conventional} gluon propagator $\Delta(q^2)$, 
how should one extract an effective charge, which, perturbatively, 
will go over to Eq.~(\ref{effch})?  To accomplish this, one must use 
an additional field-theoretic ingredient, namely the BQI of Eq.~(\ref{BQI:gg}), relating the  conventional $\Delta(q^2)$ and the PT-BFM $\widehat{\Delta}(q^2)$ in any $R_{\xi}$-like gauge, \ie 
\be
\Delta(q^2) = 
\left[1+G(q^2)\right]^2 \widehat{\Delta}(q^2).
\label{bqi}
\ee
Note that the  $G(q^2)$  already appears in  Eq.~(\ref{newSDb-1}) and \Figref{SDeqs}.
The dynamical equation for  $G(q^2)$ is obtained from the $g_{\mu\nu}$ part of 
$\Lambda_{\mu\nu}$ in (\ref{newSDb-3});
after using some of the aforementioned approximations for the vertices, we obtain 
\be
G(q^2) = - \frac{\lambda}{3}\int_k\,\left[
2+\frac{(k\cdot q)^2}{k^2q^2}\right]\Delta(k)D(k+q).
\label{gg}
\ee
First of all, it is easy to verify that, at lowest order, 
the $G(q^2)$ obtained from  Eq.~(\ref{gg})  restores the $\beta$ function coefficient  
in front of ultraviolet logarithm. In that limit 
\bea
1+G(q^2) &=& 1 +\frac{9}{4}
\frac{C_{\rm {A}}g^2}{48\pi^2}\ln(\frac{q^2}{\mu^2}),\nonumber\\
\Delta^{-1}(q^2)&=& q^2 \left[1+\frac{13}{2}
\frac{C_{\rm {A}}g^2}{48\pi^2}\ln(\frac{q^2}{\mu^2})\right]. 
\eea
Thus, using  Eq.~(\ref{bqi}) we recover the $\widehat{\Delta}^{-1}(q^2)$ of Eq.~(\ref{rightRG}), as we should. 
Then, non-perturbati- vely, one substitutes into  Eq.~(\ref{bqi}) the $G(q^2)$ and $\Delta(q^2)$ 
obtained from solving the system in  Eq.~(\ref{newSDb}), to obtain $\widehat{\Delta}(q^2)$.  
This latter quantity is the non-perturbative generalization of Eq.~(\ref{rightRG}); 
for the same reasons explained above, when multiplied by $g^2$ it should form an RG-invariant quantity, \eg the non-perturbative 
generalization of ${\widehat d}(q^2)$. In \Figref{rgi_fig} we present the combined result of the above steps: 
${\widehat d}(q^2)$ is obtained from two different sets of solutions of the system of Eqs~(\ref{newSDb-1}) one  
renormalized at $\mu=M_b=4.5\,\mbox{GeV}$ and one at $\mu=M_Z=91\,\mbox{GeV}$. Ideally
the two curves of ${\widehat d}(q^2)$ should be identical; even though this does not happen, due to the approximations employed 
when solving the SDE system, the two curves are fairly close, indicating that ${\widehat d}(q^2)$ is to a very good approximation an RG-invariant quantity, as it should.
\begin{figure}
\bce
\includegraphics[width=9cm]{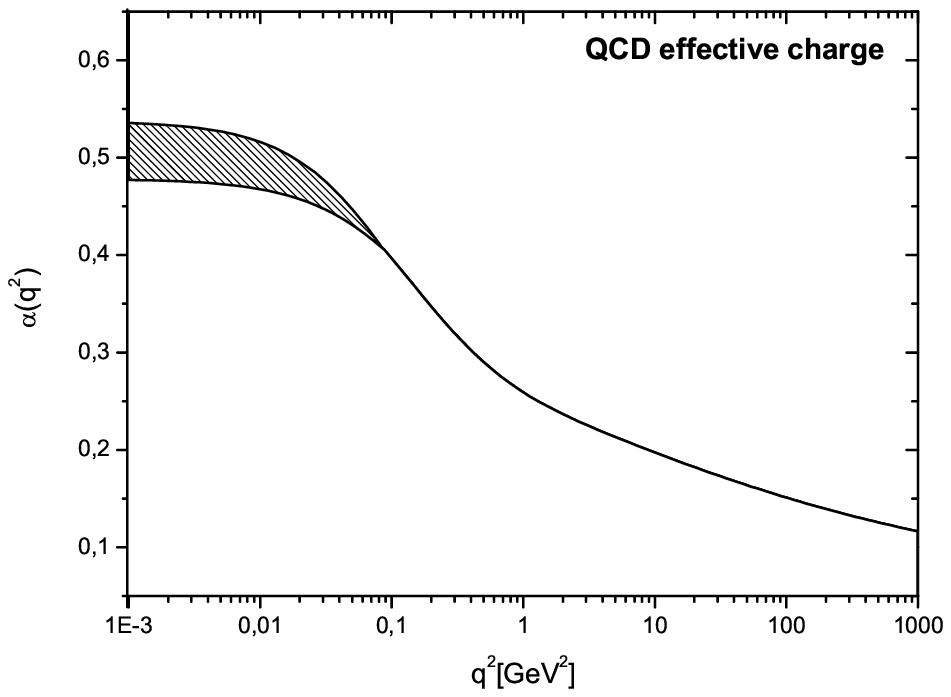}
\ece
\caption{\figlab{effcharge} The QCD effective charge, \mbox{$\alpha(q^2)=\overline{g}^2(q^2)/4\pi$},
extracted from \Figref{rgi_fig} by factoring out $(q^2+m^2(0))^{-1}$, with a  
gluon mass of $m(0)$=500\, \mbox{MeV}.}
\end{figure}
\newline
\indent
We are now in the position to define the non-perturbative 
QCD effective charge from the RG-invariant quantity ${\widehat d}(q^2)$. Of course, 
as already mentioned in subsection~\ref{QCDech}, 
given that ${\widehat d}(q^2)$ reaches a finite value in the deep infrared, 
it would be unwise to define the effective charge 
by  factoring out of  ${\widehat d}(q^2)$ a factor of $1/q^2$, 
because this would give rise to 
the unphysical situation where the strong QCD coupling vanishes in the deep IR. 
This is wrong not only operationally, \ie 
forcing the coupling to vanish when it does not want to, 
but also conceptually, 
because it suggests that QCD in the presence of a gluon mass in non-interactive\footnote{To see how unphysical this procedure is, imagine applying it to the electroweak sector.
Specifically, given that the propagators of the $W$ and $Z$ are  finite in the IR, 
(due to the standard Higgs mechanism),  
pulling a factor $1/q^2$ out of them, instead of $(q^2_E+M^2)$ 
[{\it viz}. (\ref{aw}), with $q^2 \to - q^2_E$], would 
give rise to an electroweak coupling that vanishes in the IR. Does that mean 
that the electroweak theory is non-interacting in the IR?  Does Fermi's constant vanish all of a sudden?
Or is $\beta$ decay no longer observed?}.
Of course, nothing could be further from the physical reality. First of all, a 
multitude of phenomenological studies find, with virtually no exception, 
that the QCD effective charge freezes at a non-zero 
value~\cite{Dokshitzer:1995qm,Brodsky:2002nb,Aguilar:2002tc,Aguilar:2004td,Prosperi:2006hx,Albacete:2009fh}. 
Second, a finite QCD effective charge constitutes a central assumption of the QCD/CFT correspondence~\cite{deTeramond:2005su,Brodsky:2008be}.  
Third, as we mentioned in detail at the beginning of this section 
[see discussion following (\ref{nlsm})], 
the dynamical gluon mass is responsible for a  
very rich dynamical structure, being intimately connected, 
among other things, to both quark confinement and gluon screening. 
The correct procedure corresponds to factor out a ``massive'' propagator, \ie write 
\be
\widehat{d}(q^2) = \frac{\overline{g}^2(q^2)}{q^2 + m^2(q^2)}.
\label{ddef}
\ee
\indent
Of course, as we have emphasized,  $m^2(q^2)$ itself is running, which must also 
be taken into account in a more sophisticated treatment.
For the purposes of this report, however, we assume that 
$m^2(q^2)$ is constant, $m^2(q^2)=m^2(0)$, and use for $m(0)$ the value of $500\, \mbox{MeV}$ 
favored by phenomenology~\cite{Aguilar:2001zy}. 
The $\alpha(q^2)$ obtained is shown in \Figref{effcharge}; as announced, 
at low energies it freezes to a finite value, indicating the appearance of an infrared fixed point of QCD.

\newpage


\section{\seclab{conclrem}Concluding remarks}
\noindent
In this report we have given a detailed account of the 
pinch technique and some of its most characteristic applications.
The present work may be separated in a natural way, into two large parts.
The first part, comprising of sections 2 to 5, contains 
practically the entire one-loop PT, both in QCD and the electroweak sector,
and the connection with the BFM formalism.
The second part contains the developments 
following the non-diagrammatic formulation of the PT, the 
streamlining achieved by resorting to the Batalin-Vilkovisky 
formalism, and finally the non-perturbative QCD applications, with particular 
highlights the gauge-invariant truncation of the SDE series and the 
dynamical generation of an effective gluon mass. 
\newline
\indent
We hope to have conveyed to the reader the underlying unity of  
the multitude of topics covered, and to have succeeded in demonstrating  
the versatility of the PT formalism and the wide range of its applicability.
The reader should be able to appreciate, for example, 
how a simple one-loop calculation contains the 
seed of a non-trivial truncation of the SDEs, accomplished a quarter of a century later; 
or how a seemingly innocuous (or even redundant, according to some) rearrangement of graphs  
is able to give rise to a resummation formalism for resonant transition amplitudes 
that satisfies such a plethora of tightly interwoven physical constraints.
\newline
\indent
Throughout this report we have attempted to maintain a balance between the technical presentation 
(how to pinch) and the physical motivations and phenomenological 
applications (when to pinch and why). 
Even so, there is a considerable number of additional important applications that 
we could not possibly cover. Let us mention a few.
There have been several application in the area of finite temperature field theory, 
starting with the early work by Cornwall and collaborators~\cite{Cornwall:1984eu,Cornwall:1985bg}, 
the calculation of the plasmon decay constant by Nadkarni~\cite{Nadkarni:1988ti}, 
the gauge-independent thermal $\beta$ function computed 
by Sasaki~\cite{Sasaki:1996jf,Sasaki:1995ck},
and the work on magnetic screening for the quark-gluon plasma
by Alexanian and Nair~\cite{Alexanian:1995rp}. 
In addition, the explicit one-loop 
PT calculations in the context  of the Coulomb and temporal axial
gauges have been presented by Passera and Sasaki in \cite{Passera:1996jc}.
Moreover, Pilaftsis applied the PT to the 
resonant CP violation a decade ago~\cite{Pilaftsis:1996ac,Pilaftsis:1997dr}, 
and recently  to resonant leptogenesis~\cite{Pilaftsis:2003gt,Pilaftsis:2008qt}.
The PT has also been used
in order to obtain scale and gauge-independent mixing angles for scalar particles
\cite{Yamada:2001px,Espinosa:2002cd} and \cite{Pilaftsis:2002nc}.
In addition, Caporaso and Pasquetti applied 
to  the non-commutative QED~\cite{Caporaso:2005xf}, and 
non-commutative (softly broken) supersymmetric Chern-Simons theory~\cite{Caporaso:2006ij}. 
\newline
\indent
To be sure, there are still many things one would like to know about the PT and 
its field-theoretic origin. Most importantly, as mentioned in the Introduction, 
a formal definition of the PT Green's functions 
in terms of fundamental fields, encoding ``ab initio'' their special properties, still eludes us. 
Ideally, one would like to find that particular combination of fields or operators, which, when 
appropriately combined, will furnish the PT answer without pinching, \ie regardless of whether 
or not one tracks down the various cancellations explicitly, and in any gauge-fixing scheme considered.
Such a situation 
would be, of course, far  superior than what happens now with the BFG, where 
there is no pinching only because of a kinematic accident, 
namely the lack of pinching momenta in that particular gauge. 
\newline
\indent
One basic and rather obvious question, that, perhaps surprisingly, has not been addressed to date, is the following.
It is well known that one can construct a gauge-invariant operator
out of a gauge-variant one 
by means of a path-order exponential containing the gauge field $A$ \cite{Wilson:1973jj}.
For example, in the case of the fermion propagator 
\mbox{$S(x,y) = \langle 0|\psi(x) \bar{\psi}(y)|0\rangle$}
the corresponding
gauge-invariant propagator $S_{PO}$ reads 
(``PO'' stands for 
``path-ordered'')
\begin{equation}
S_{PO} (x,y) = 
\langle 0|\psi(x) P \exp\left(i \int_{x}^{y} dz \cdot A(z)  
\right)\bar{\psi}(y)|0\rangle.
\end{equation}
Is this gauge-invariant propagator related in any way to the PT fermion propagator, constructed in 
\secref{QCD_one-loop}?  Of course, completely 
related to this question is the construction presented in \cite{Catani:1989fe}; in fact, 
the distinction made there between the Wightman  and the causal two-point function might be worth pursuing 
from the PT point of view.
\newline
\indent
Anyone remotely familiar with the PT gets the tantalizing feeling that, 
in addition to the BRST symmetry, some other powerful 
(yet undiscovered)  mechanism must be at work, enforcing  the PT properties. 
The remarkable supersymmetric relations discovered by Binger and Brodsky~\cite{Binger:2006sj}
(see \secref{QCD_one-loop}) intensify this impression;   
their results indeed beg the question of whether one has actually stumbled into something bigger.
Could it be, for example, that the PT rearrangements end up exposing 
some sort of hidden symmetry?  Such a possibility is not unprecedented;     
an  interesting  3-d   example  of  a (topological) 
field-theory, which,  when formulated  in the background  Landau gauge
($\xi_Q  =  0$),  displays  an  additional  (non-BRST  related)  rigid
supersymmetry, is given in ~\cite{Birmingham:1988ap}. 
\newline
\indent
Finally, it would be most interesting to explore possible 
connections with other field-theoretic methods~\cite{DAttanasio:1996jd,Arnone:2005fb,Morris:2005tv},
~\cite{Schubert:1996jj},~\cite{Nielsen:1975ph,Nielsen:1975fs},~\cite{Feng:1995vg},~\cite{Simonov:1993kt}, 
or string-inspired approaches~\cite{Bern:1994zx,Bern:1994cg,Bern:1991an}, \cite{Feng:1994ji},~\cite{DiVecchia:1996uq}, 
in order to either acquire a more formal understanding of the PT,
or to encompass various related approaches into a unique coherent framework.


\section*{Acknowledgments}
\noindent
\noindent
We are very happy to acknowledge numerous conversations over the years on the Pinch Technique 
and related topics with 
A. C. Aguilar,
L. Alvarez-Gaume, 
G. Barnich,
J. Bernab\`eu,
J-P. Blaizot,
S. Brodsky,
L. G. Cabral-Rosetti
J. M. Cornwall,
E. de Rafael,
V. Mathieu,
N. E. Mavromatos,
P. Minkowski,
M. Passera,  
N. Petropoulos,
K. Philippides, 
A. Pilaftsis, 
A. Santamaria,
A. Sirlin, 
R. Stora,
V. Vento,
and J. N. Watson.
Some of them have made important contributions to several of the topics discussed in this review, while others
were just happy to share ideas with us.
\newline
\noindent
The research of J.~P. is supported by the European FEDER and  Spanish MICINN under grant FPA2008-02878, and the Fundaci\`on General of the UV.
\newline
\noindent
Feynman diagrams have been drawn using \verb+JaxoDraw+ \cite{Binosi:2003yf,Binosi:2008ig}.

\newpage


\begin{appendix}


\section{\seclab{SU(N)ids}$SU(N)$ group theoretical identities}

In this Appendix we collect some useful group theoretical identities for the $SU(N)$ gauge group.
\newline
\indent
For any representation of $SU(N)$ 
the generators $t^{a}$ ($a =1,2 ... N^2-1$)
are hermitian, traceless matrices, generating the closed 
algebra
\be
[t^{a}, t^{b}]  = i f^{abc} t^{c},
\label{comrel}
\ee
where $f^{abc}$ are the (totally antisymmetric) structure constants, which satisfy the Jacobi identity
\be
f^{abx} f^{cdx}+f^{acx} f^{dbx}+f^{adx} f^{bcx} = 0. 
\label{Jacobi}
\ee
\indent
The fundamental representation $t_f^{a}$ is $N$-dimensional, with the normalization 
\be
{\rm Tr}(t_f^{a} t_f^{b}) = \frac{1}{2}\delta^{ab}
\label{norm}
\ee
In the case of QCD, the fundamental ${t}_f^{a} = \lambda^{a}/2$, where $\lambda^{a}$ are the Gell-Mann  matrices.
\newline
\indent
The adjoint representation has dimension $N^2-1$, and its generators ${t}_A^{a}$ have matrix elements given by  the relation
\be
({t}_A^{a})_{bc} = -i f^{abc}.
\ee
\indent 
The Casimir eigenvalue $C_r$ of a representation $r$ is defined as 
\be
t_r^{a}t_r^{a} = C_r \, 1
\ee
while the Dynkin index $d_r$ is defined as 
\be
{\rm Tr}(t_r^{a} t_r^{b}) = d_r \delta^{ab}
\ee
The Casimir eigenvalue and the Dynkin index of a representation of a group $G$ are related by the general formula
\be
C_r = \frac{\mathrm{dim}(G)}{\mathrm{dim}(r)} d_r
\ee
where $\mathrm{dim}(G)$ is the dimension of the group and $\mathrm{dim}(r)$ the dimension of the representation.
Thus, for the adjoint representation $r=A$, $\mathrm{dim}(G)=\mathrm{dim}(A)$, and 
therefore $C_A= d_A$. Specializing to the $SU(N)$ case, one has $C_A=N$, and from Eq.~(\ref{norm}) we have that  $d_f= \frac{1}{2}$, and thus $C_f=(N^2-1)/2N$; for QCD, $C_f=4/3$.   
\newline
\indent
We conclude by quoting some identities involving the structure functions
\bsub
\bea
& & f^{aex}f^{bex} = C_A \delta^{ab}, \label{ff}\\
& & f^{axm}f^{bmn}f^{cnx}  = \frac{1}{2} C_A f^{abc},\label{fff}\\
& & f^{alm}f^{bmn}f^{cne}f^{del}-
f^{alm}f^{bmn}f^{dne}f^{cel}=-\frac12C_Af^{abx}f^{cdx}.\label{ffff}
\eea
\esub

\newpage


\section{Feynman rules \seclab{Frules}}

\subsection{$R_\xi$ and BFM gauges \label{RxiBFMFR}}
\begin{figure}[!t]
\bce
\includegraphics[width=15.5cm]{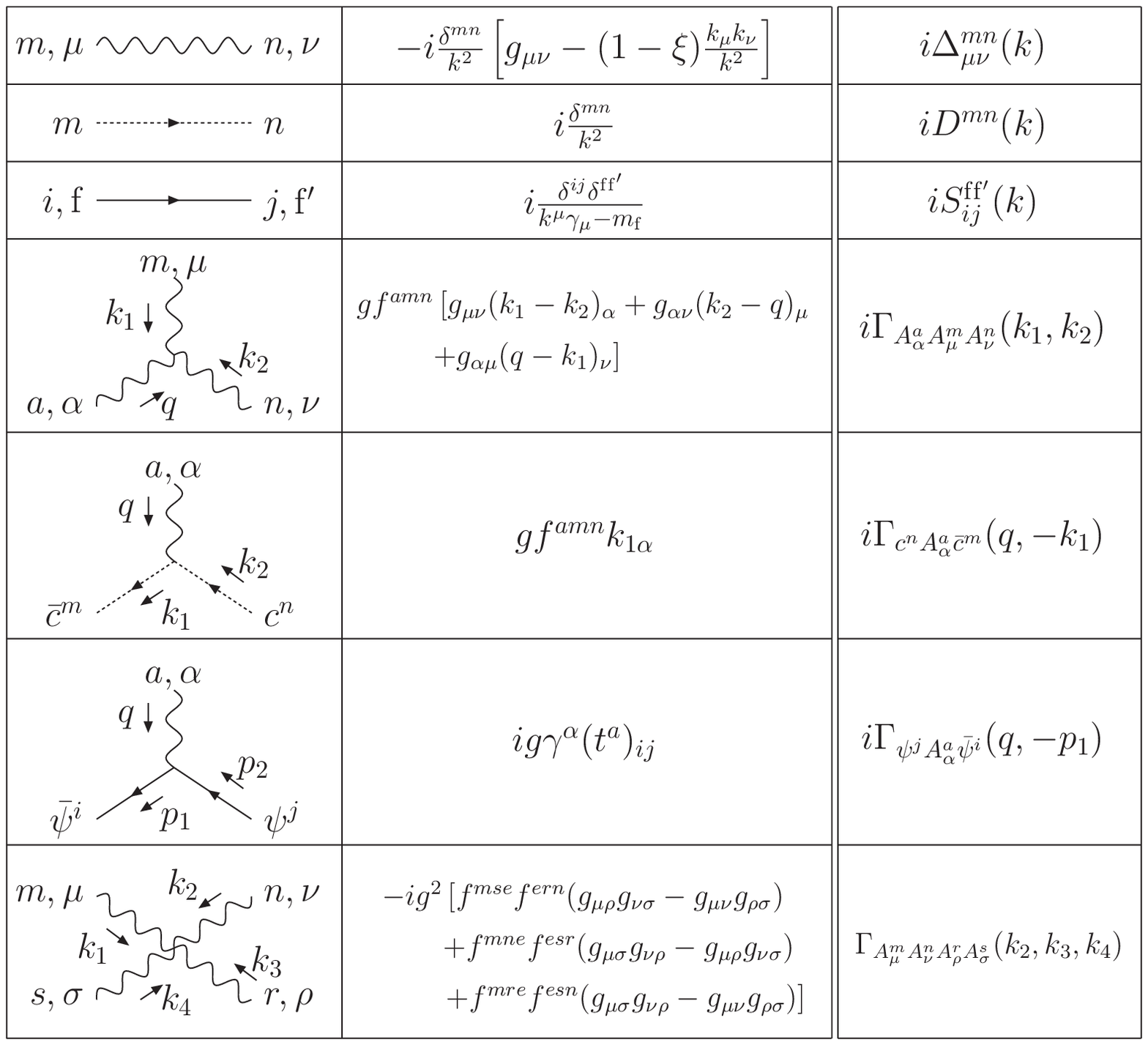}
\ece
\caption{\figlab{Frules_Rxi}Feynman rules for QCD in the $R_\xi$ gauges. The first two columns show the lowest order Feynman diagrams and rule respectively, while the last one shows the corresponding all-order Green's function according to the conventions of Eq.(\ref{greenfunc}).}
\end{figure}
\noindent
The Feynman rules for QCD in $R_\xi$ gauges are given in \Figref{Frules_Rxi}. In the case of the BFM gauge, since the gauge fixing
term is quadratic in the quantum fields, apart from vertices involving ghost fields
only vertices containing exactly two quantum fields might differ from the conventional ones. Thus, 
the vertices $\Gamma_{\widehat A\psi\bar\psi}$ and $\Gamma_{\widehat AAAA}$ have to lowest order the same expression as the corresponding $R_\xi$ ones $\Gamma_{A\psi\bar\psi}$ and $\Gamma_{AAAA}$ (to higher order their relation is described by the corresponding BQIs).
\begin{figure}[!t]
\includegraphics[width=15.8cm]{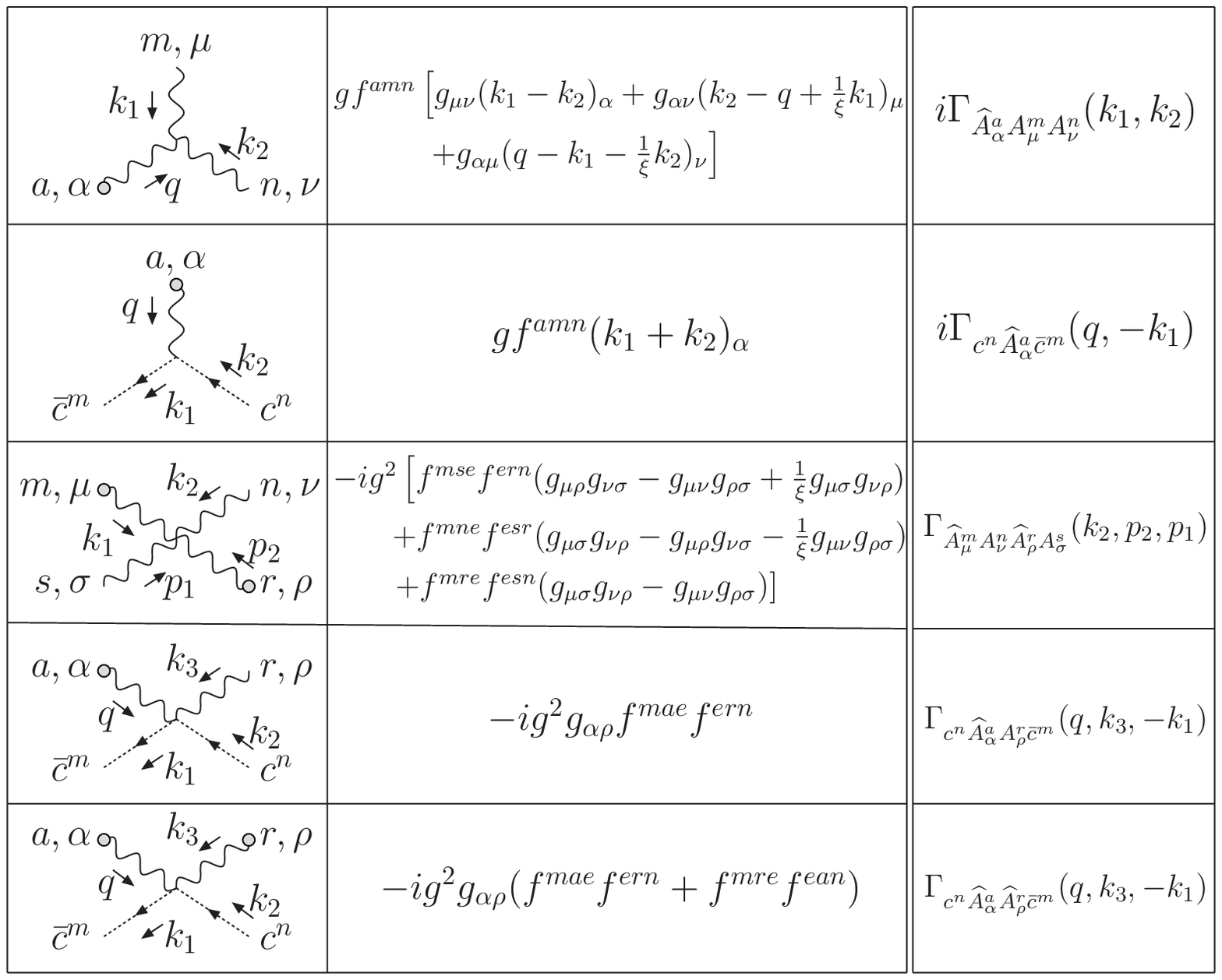}
\caption{\figlab{Frules_BFM}Feynman rules for QCD in the BFM gauge. We only include those rules which are different from the $R_\xi$ ones to lowest order. A gray circle on a gluon line indicates a background field.}
\end{figure}

\subsection{Anti-fields \label{afFR}}
\noindent
The couplings of the anti-fields $\Phi^*$ with fields is entirely encoded in the BRST Lagrangian of Eq.~(\ref{BRST_Lag}). 
When choosing the BFM gauge the additional coupling $gf^{amn}A^{*m}_\mu \widehat{A}^n_\nu c^a$ will arise in the BRST Lagrangian ${\mathcal L}_\mathrm{BRST}$ as a consequence of the BFM splitting $A\to\widehat{A}+A$.
One then gets the Feynman rules given in \Figref{Frules_anti}.
\begin{figure}[!t]
\bce
\includegraphics[width=16cm]{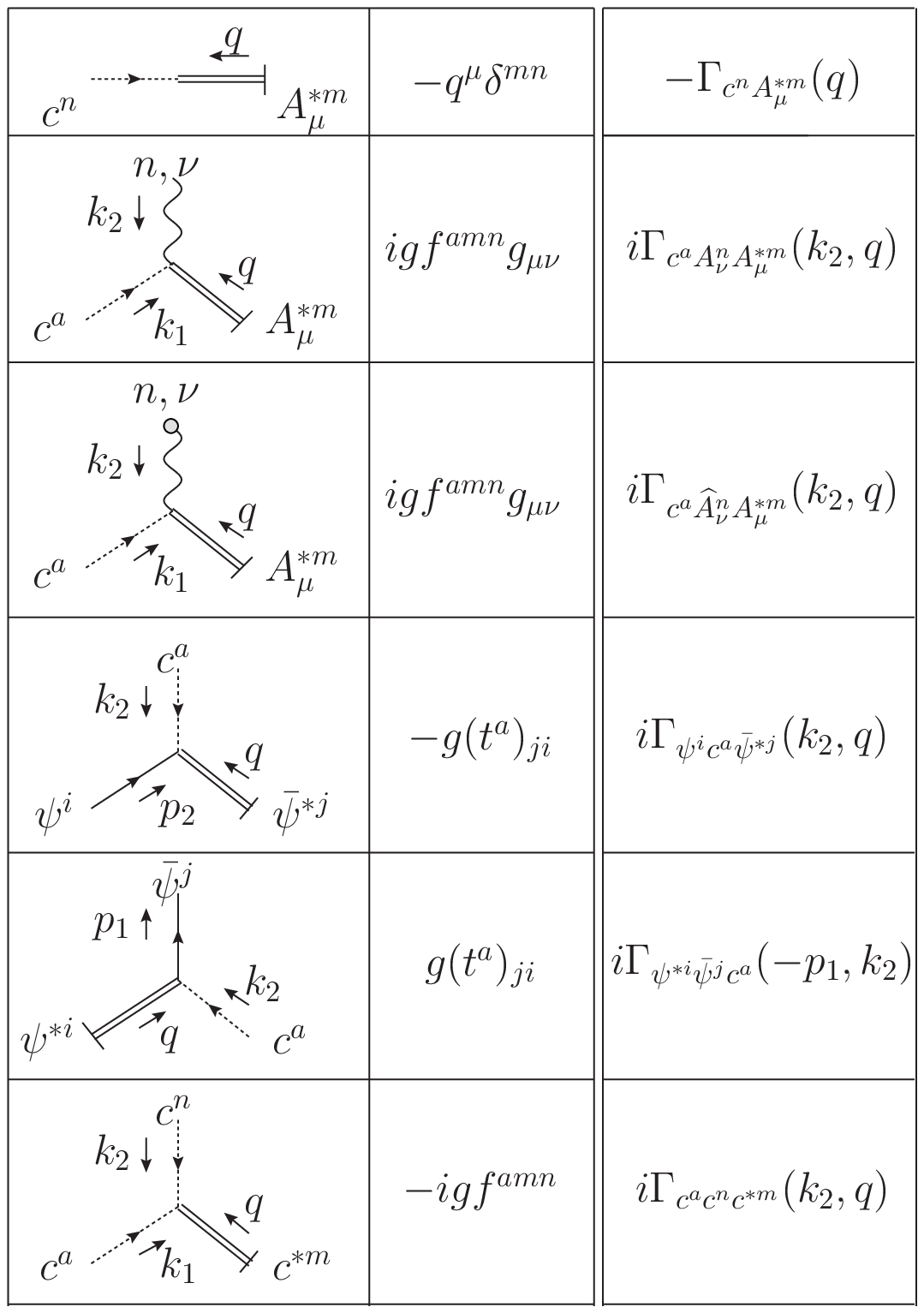}
\ece
\caption{\figlab{Frules_anti}Feynman rules for QCD anti-fields.}
\end{figure}

\subsection{BFM sources \label{BFMsFR}}
\begin{figure}[!h]
\bce
\includegraphics[width=16cm]{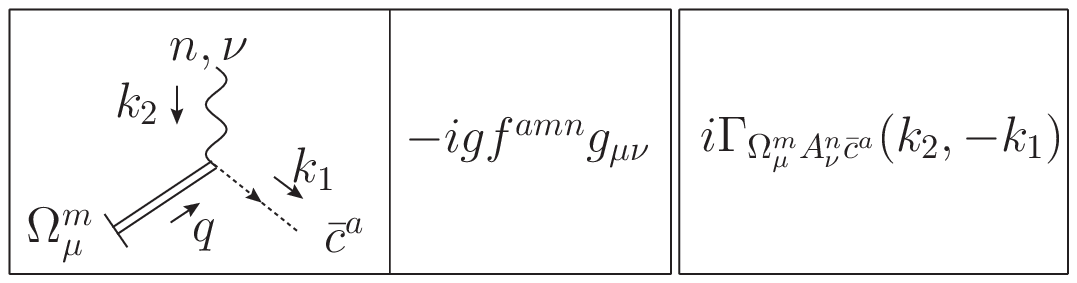}
\ece
\caption{\figlab{Frules_BFMsources}Feynman rule for the BFM gluon source $\Omega^m_\mu$.}
\end{figure}
\noindent
The coupling of the BFM source $\Omega^m_\mu$ with the ghost and gluon fields can be derived from the Faddeev-Popov ghost Lagrangian, since making use of the BRST transformation of Eq.~(\ref{extBRST}) we get
\begin{equation}
{\mathcal L}_\mathrm{FPG}=-\bar c^a s{\mathcal F}^a_\mathrm{BFM}
\supset-\bar c^a g f^{amn}(s\widehat{A}^m_\mu) A^\mu_n
=-g f^{amn}\bar c^a\Omega^m_\mu A^\mu_n.
\end{equation}
The corresponding Feynman rule is given in \Figref{Frules_BFMsources}. In general Feynman rules involving the BFM source $\Omega$ can be derived by the one involving the (gluon) anti-field $A^*$ through the replacements $A^*\to\Omega$ and $c\to\bar c$.

\newpage


\section{\seclab{FPEs-STIs-BQIs}Faddev-Popov equations, Slavnov-Taylor Identities and Background Quantum Identities for QCD}

\subsection{Faddeev-Popov Equations\label{Appendix:FPEs}}
\noindent
As a first example of the use of the FPE introduced in Section~\ref{FPEs}, let us 
differentiate the functional equation (\ref{FPeqRxi}) with respect to the ghost field $c^b$; after setting the fields/anti-fields to zero we get (relabeling the color and Lorentz indices)
\begin{equation}
\Gamma_{c^m\bar c^n}(q)+iq^\nu\Gamma_{c^mA^{*n}_\nu}(q)=0,
\label{FPE:ghprop}
\end{equation}
which can be used to relate the auxiliary function $\Gamma_{c^mA^{*n}_\nu}(q)$ 
with the full ghost propagator $D^{ab}(q)$. Due to Lorentz invariance, we can in fact write $\Gamma_{c^mA^{*n}_\nu}(q)=q_\nu\Gamma_{c^mA^{*n}}(q)$, and therefore
\begin{equation}
\Gamma_{c^m\bar c^n}(q)=-iq^\nu\Gamma_{c^mA^{*n}_\nu}(q)=-iq^2\Gamma_{c^mA^{*n}}(q).
\end{equation}
On the other hand, due to our definition of the Green's functions [see Eq.~(\ref{greenfunc})], one has that 
\begin{equation}
iD^{mr}(q)\Gamma_{c^r\bar c^n}(q)=\delta^{mn}, 
\label{twop_inverse}
\end{equation}
and therefore we get the announced relation:
\begin{eqnarray}
\Gamma_{c^mA^{*n}_\nu}(q)&=&q_\nu\Gamma_{c^mA^{*n}}(q)\nonumber \\
&=&q_\nu[q^2 D^{mn}(q)]^{-1}.
\label{gaprop}
\end{eqnarray}
\newline
\indent
As a second example, let us differentiate  Eq.~(\ref{FPeqRxi}) twice, 
once with respect to $A^n_\nu$ and once with respect to $c^r$,
and then set the fields/anti-fields to zero; in this way we get the identity
\begin{equation}
\Gamma_{c^rA^n_\nu\bar c^m}(k,q)+iq^\mu\Gamma_{c^rA^n_\nu A^{*m}_\mu}(k,q)=0,
\label{FP:gcc}
\end{equation}
which is particularly useful for the PT construction.  
All these identities can be easily checked at tree-level; for example, using the Feynman rules of \appref{Frules}, we have
\begin{equation}
iq^\mu\Gamma^{(0)}_{c^rA^n_\nu A^{*m}_\mu}(k,q)=igf^{mnr}q_\nu=-\Gamma^{(0)}_{c^rA^n_\nu\bar c^m}(k,q).
\end{equation}
\newline
\indent
Differentiation of the functional (\ref{FPeqBFM}) with respect to a BFM source $\Omega$ 
and a quantum gluon field $A$ or a ghost field $c$ and a background gluon $\widehat{A}$, 
provides instead the identities ($k_1+k+q=0$)
\begin{eqnarray}
\Gamma_{\Omega^r_\rho A^n_\nu\bar c^m}(k,q)+iq^\mu\Gamma_{\Omega^r_\rho A^n_\nu A^{*m}_\mu}(k,q)&=&
gf^{mnr}g_{\nu\rho}, \label{FPE_Om1} \\
\Gamma_{c^r_\rho \widehat{A}^n_\nu\bar c^m}(k,q)+iq^\mu\Gamma_{c^r\widehat{A}^n_\nu A^{*m}_\mu}(k,q)&=&
-igf^{mne}\Gamma_{c^rA^{*e}_\nu}(-k_1), \label{FPE_Om2}
\end{eqnarray}
that can be easily checked at tree-level.

\subsection{Slavnov-Taylor Identities\label{Appendix:STIs}}
\noindent
STIs are obtained by functional differentiation of the STI functional of Eq.~(\ref{STIfunc}) with respect to suitable combinations of fields chosen following the rules discussed in \secref{PTBV}.

\subsubsection{STIs for gluon proper vertices}
\noindent
Let us start by deriving the well-known STI for the trilinear 
gluon vertex which in the conventional formalism has been introduced in Eq.~(\ref{sti3gv}). 
By considering the functional differentiation
\begin{equation}
\left.\frac{\delta^3{\mathcal S}(\Gamma)}{\delta c^a(q)\delta A^m_\mu(k_1)\delta A^n_\nu(k_2)}\right|_{\Phi,\Phi^*=0}=0 \qquad q+k_1+k_2=0,
\label{3g_diff}
\end{equation}
and using Eq.~(\ref{gaprop}) one obtains
\bea
q^\alpha\Gamma_{A^a_\alpha  A^m_\mu A^n_\nu}(k_1,k_2)&=&[q^2D^{aa'}(q)]\left\{\Gamma_{c^{a'} A^n_\nu A^{*\gamma}_d}(k_2,k_1)\Gamma_{A^d_\gamma A^m_\mu}(k_1)\right.\nonumber \\
&+&\left.\Gamma_{c^{a'} A^m_\mu A^{*\gamma}_d}(k_1,k_2)\Gamma_{A^d_\gamma A^n_\nu}(k_2)\right\}.
\label{STI:ggg}
\eea
At this point one would need to find out the relation between the (full) gluon propagator and the two point function $\Gamma_{A^a_\alpha A_\beta^b}$. First of all let us notice that since we are working in the Feynman gauge [see also Eq.~(\ref{prop_cov})]
\begin{equation}
i\Delta^{ab\ (0)}_{\alpha\beta}(q)=-\frac i{k^2}\left\{P_{\alpha\beta}(q)+\frac{k_\alpha k_\beta}{k^2}\right\}\delta^{ab} \qquad P_{\alpha\beta}(q)=g_{\alpha\beta}-\frac{q_\alpha q_\beta}{q^2},
\end{equation}
which translates to the all order formula
\begin{equation}
i\Delta^{ab}_{\alpha\beta}(q)=-i\delta^{ab}\left\{P_{\alpha\beta}(q)\Delta(q^2)+\frac{q_\alpha q_\beta}{q^4}\right\},
\label{fullgprop}
\end{equation}
with
\begin{equation}
\Delta(q^2)=\frac1{q^2+i\Pi(q^2)}, \qquad \Pi_{\alpha\beta}(q)=P_{\alpha\beta}(q)\Pi(q^2).
\end{equation}
Notice that the way the gluon self-energy $\Pi_{\alpha\beta}(q)$ has been defined in the above equation, {\it i.e.}, with the imaginary $i$ factor in front, implies that it is given simply by the corresponding Feynman diagrams in Minkowski space.
Imposing then the condition
\begin{equation}
i\Delta^{am}_{\alpha\mu}(q)(\Delta^{-1})^{\mu\beta}_{mb}(q)=\delta^{ab}g^\alpha_\beta,
\end{equation}
we get
\begin{equation}
(\Delta^{-1})^{\mu\beta}_{mb}(q)=i\delta^{mb}\left\{P^{\mu\beta}(q)\Delta^{-1}(q^2)+q^\mu q^\beta\right\}.
\label{invprop}
\end{equation}
On the other hand, recall that we are working with minimal variables, and thus with the reduced functional $\Gamma$; in the case of linear gauge fixings (as the $R_\xi$ and the BFM are) the latter is equivalent to the complete one after subtracting the local term $\int\!d^4x\,{\mathcal L}_\mathrm{GF}$. This implies in turn that Green's functions involving unphysical fields generated by the reduced functional coincide with the ones generated by the  complete one only up to constant terms. In our case this affects only the two point function of the gluon field, for which one has the tree level expression
\begin{equation}
\Gamma^{(0)}_{A^a_\alpha A^b_\beta}(q)=iq^2\delta^{ab}P_{\alpha\beta}(q),
\label{ga_tree_lev}
\end{equation}
which furnishes the sought-for all-order formula
\bea
\Gamma_{A^a_\alpha A^b_\beta}(q)&=&(\Delta^{-1})^{ab}_{\alpha\beta}(q)-i\delta^{ab}q_\alpha q_\beta \nonumber \\
&=& i\delta^{ab} P_{\alpha\beta}(q) \Delta^{-1}(q^2).
\label{gainvprop}
\eea
\newline
\indent
Using the above relation, we can now check the identity at tree-level; we get
\begin{eqnarray}
q^\alpha\Gamma^{(0)}_{A^a_\alpha  A^m_\mu A^n_\nu}(k_1,k_2)&=&
\left\{\Gamma^{(0)}_{c^{a} A^n_\nu A^{*\gamma}_d}(k_2,k_1)\Gamma^{(0)}_{A^d_\gamma A^m_\mu}(k_1)+\Gamma^{(0)}_{c^{a} A^m_\mu A^{*\gamma}_d}(k_1,k_2)\Gamma^{(0)}_{A^d_\gamma A^n_\nu}(k_2)\right\}\nonumber \\
&=&igf^{amn}\left[(g_{\mu\nu}k_1^2-k_{1\mu}k_{1\nu})-(g_{\mu\nu}k_2^2-k_{2\mu}k_{2\nu})\right].
\end{eqnarray}
Notice also that Eq.~(\ref{gainvprop}) allows us to compare the STI of Eq.~(\ref{STI:ggg}) with the one written 
in the conventional formalism of Eq.~(\ref{sti3gv}). Factoring out the color structure, one arrive at the following identification 
\be
H_{\mu\gamma}(k_1,k_2)=\Gamma_{cA_\mu A^*_\gamma}(k_1,k_2),
\ee
which also shows that the FPE (\ref{FP:gcc}) corresponds to the well-known relation existing between the auxiliary function
$H_{\alpha\beta}$ and the conventional gluon-ghost vertex $\Gamma_\beta$ shown in Eq.~(\ref{H_ghost_rel}). 
\newline
\indent
We pause here to show what would have happened had we worked with the complete generating functional. In this case, due to the extra term appearing in the master equation (\ref{STIfunc_nm}) satisfied by the complete action, the differentiation carried out in Eq.~(\ref{3g_diff}) would generate two more terms with respect to the ones already appearing in Eq.~(\ref{STI:ggg}),  namely
\be
\delta^{dn}k_{2\nu}\Gamma_{c^a A^m_\mu\bar c^d}(k_1,k_2)+\delta^{dm}k_{2\mu}\Gamma_{c^a A^n_\nu\bar c^d}(k_2,k_1).
\ee
To get to the terms above we have used the equation of motion of the Nakanishi-Lautrup multiplier $B$ eliminating the latter in favor of the corresponding gauge-fixing function ${\mathcal F}$. Then, making use of the FPE (\ref{FP:gcc}), we get  
\be
-i\delta^{dn}k_{2\nu}k_{2\gamma}\Gamma_{c^a A^m_\mu A^{*\gamma}_d}(k_1,k_2)-i\delta^{dm}k_{1\mu}k_{1\gamma}\Gamma_{c^a A^n_\nu A^{*\gamma}_d}(k_2,k_1),
\ee
so that we finally would get the STI
\bea
q^\alpha\Gamma_{A^a_\alpha  A^m_\mu A^n_\nu}(k_1,k_2)&=&[q^2D^{aa'}(q)]\left\{\Gamma_{c^{a'} A^n_\nu A^{*\gamma}_d}(k_2,k_1)\left[\Gamma^\mathrm{C}_{A^d_\gamma A^m_\mu}(k_1)-i\delta^{dm}k_{1\mu}k_{1\gamma}\right]\right.\nonumber \\
&+&\left.\Gamma_{c^{a'} A^m_\mu A^{*\gamma}_d}(k_1,k_2)\left[\Gamma^\mathrm{C}_{A^d_\gamma A^n_\nu}(k_2)-i\delta^{dn}k_{2\gamma}k_{2\nu}\right]\right\},
\eea
where we have indicated explicitly that the two-point functions are to be evaluated from the completed functional (for the three point functions appearing in the STI above there is no difference). We then see that the difference amounts to a tree-level piece appearing in the two-point function, as has been anticipated in our general discussion of subsection~\ref{BV} (recall that we are using the Feynman gauge $\xi=1$). In particular notice that we correctly find the relation $\Gamma^\mathrm{C}_{A^a_\alpha A^b_\beta}(q)= (\Delta^{-1})^{ab}_{\alpha\beta}(q)$.
\newline
\indent
Another STI  that will  be needed  in the PT  construction is  the one
involving the  quadrilinear gluon vertex; carrying  out the functional
differentiation
\begin{equation}
\left.\frac{\delta^4{\mathcal S}(\Gamma)}{\delta c^m(k_1)\delta A^n_\nu(k_2)\delta A^r_\rho(p_2)\delta A^s_\sigma(-p_1)}\right|_{\Phi,\Phi^*=0}=0 \qquad k_1+k_2+p_2=p_1,
\end{equation}
and using Eq.~(\ref{gaprop}), we arrive at the result
\begin{eqnarray}
& & k_1^\mu\Gamma_{A_\mu^mA_\nu^n A^r_\rho A^s_\sigma}(k_2,p_2,-p_1)=[k_1^2D^{mm'}(k_1)]\bigg\{\Gamma_{c^{m'} A^s_\sigma A^{*\gamma}_d}(-p_1,k_2+p_2)\Gamma_{A^d_\gamma A^n_\nu A^r_\rho}(k_2,p_2)\nonumber \\
&&\hspace{.3cm}+\Gamma_{c^{m'} A^r_\rho A^{*\gamma}_d}(p_2,k_2-p_1)\Gamma_{A^d_\gamma A^n_\nu A^s_\sigma}(k_2,-p_1)+\Gamma_{c^{m'} A^n_\nu A^{*\gamma}_d}(k_2,p_2-p_1)\Gamma_{A^d_\gamma A^r_\rho  A^s_\sigma}(p_2,-p_1)\nonumber\\
&&\hspace{.3cm}+\Gamma_{c^{m'} A^r_\rho A^s_\sigma A^{*\gamma}_d}(p_2,-p_1,k_2)\Gamma_{A^d_\gamma A^n_\nu}(k_2)+\Gamma_{c^{m'} A^n_\nu A^s_\sigma A^{*\gamma}_d}(k_2,-p_1,p_2)\Gamma_{A^d_\gamma A^r_\rho}(p_2)\nonumber \\
&&\hspace{.3cm}+ \Gamma_{c^{m'} A^n_\nu A^r_\rho A^{*\gamma}_d}(k_2,p_2,-p_1)\Gamma_{A^d_\gamma A^s_\sigma}(p_1)\bigg\}.
\label{STI:gggg}
\end{eqnarray}

\subsubsection{STIs for mixed quantum/background Green's functions}
\noindent
Let us consider a Green's function  involving background as  well as
quantum fields. Clearly, when contracting such a function with  the momentum
corresponding to a background leg it will satisfy a linear WI [such as the ones presented in Eq.s~(\ref{ghatgg_WI}), (\ref{ghatcc_WI}), (\ref{ghatggg_WI}), and~(\ref{ghatgcc_WI})],
whereas  
when contracting it with  the momentum corresponding to a 
quantum leg it will satisfy  a non-linear STI.
Let us then study the particularly  interesting case of
the  STI   satisfied  by  the   vertex  $\Gamma_{\widehat{A}AA}$  when
contracted  with  the momentum  of  one  of  the quantum  fields. Taking the functional differentiation
\be
\left.\frac{\delta^3{\mathcal S'}(\Gamma')}{\delta c^m(k_1)\delta\widehat{A}^a_\alpha(q)\delta A^n_\nu(k_2)}\right|_{\Phi,\Phi^*,\Omega=0}=0 \qquad q+k_1+k_2=0,
\ee
we get
\bea
k_1^\mu\Gamma_{\widehat{A}^a_\alpha A^m_\mu A^n_\nu}(k_1,k_2)&=&[k_1^2D^{mm'}(k_1)]\left\{\Gamma_{c^{m'}A^n_\nu A^{*\epsilon}_e}(k_2,q)\Gamma_{\widehat{A}^a_\alpha A^e_\epsilon}(q)\right.\nonumber \\
&+&\left.\Gamma_{c^{m'}\widehat{A}^a_\alpha A^{*\epsilon}_e}(q,k_2)\Gamma_{A^e_\epsilon A^n_\nu}(k_2)\right\}.
\label{STI:mixed_wtl}
\eea
Notice that the same result can be achieved by contracting directly the BQI of Eq.~(\ref{BQI:ggg_tlc}) with the momentum of one of the quantum fields and then using the STI of Eq.~(\ref{STI:ggg}) together with the BQIs of Eq.s(\ref{twoBQI1}) and~(\ref{BQI:cgbarc}) to bring the result in the above form.
\newline
\indent
It is particularly important to correctly identify, in the above identity, the missing tree-level contributions (due to the use of the reduced functional, see also the discussion in Section~\ref{BQI:3p}). In order to do that, one can either work with the complete functional and use the FPE (\ref{FPE_Om2}), or add them by hand using Eq.~(\ref{BQI:ggg_tlc}), obtaining in either cases the STI 
\bea
k_1^\mu\Gamma_{\widehat{A}^a_\alpha A^m_\mu A^n_\nu}(k_1,k_2)&=&[k_1^2D^{mm'}(k_1)]\left\{\Gamma_{c^{m'}A^n_\nu A^{*\epsilon}_e}(k_2,q)\Gamma_{\widehat{A}^a_\alpha A^e_\epsilon}(q)\right.\nonumber \\
&+&\left.\Gamma_{c^{m'}\widehat{A}^a_\alpha A^{*\epsilon}_e}(q,k_2)\Gamma_{A^e_\epsilon A^n_\nu}(k_2)\right\}-igf^{amn}(k_1^2g_{\alpha\nu}-k_{1\alpha}k_{2\nu}).
\label{STI:mixed}
\eea
\indent
This STI can be further manipulate by using Eq.~(\ref{gainvprop}) and the FPE~(\ref{FPE_Om2}) 
for rewriting the term proportional to $\Gamma_{AA}(k_2)$ as
\bea
\Gamma_{c^{m'}\widehat{A}^a_\alpha A^{*\epsilon}_e}(q,k_2)\Gamma_{A^e_\epsilon A^n_\nu}(k_2)&=&\Gamma_{c^{m'}\widehat{A}^a_\alpha A^{*\epsilon}_e}(q,k_2)(\Delta^{-1})^{en}_{\epsilon\nu}(k_2)+k_{2\nu}\Gamma_{c^{m'}\widehat{A}^a_\alpha\bar c^n}(q,k_2)\nonumber \\
&+&igf^{aen} k_{2\nu}\Gamma_{c^{m'}A^{*e}_\alpha}(-k_1).
\label{fmassage}
\eea
On the other hand, employing Eq.~(\ref{gaprop}) we find
\be
[k_1^2D^{mm'}(k_1)](igf^{nae}k_{2\nu})\Gamma_{c^{m'}A^{*e}_\alpha}(-k_1)=-igf^{amn}k_{1\alpha}k_{2\nu};
\ee
so, inserting Eq.~(\ref{fmassage}) back into Eq.~(\ref{STI:mixed}) we see 
that the term above partially cancels the tree level contribution, thus leaving us with the STI
\bea
k_1^\mu\Gamma_{\widehat{A}^a_\alpha A^m_\mu A^s_\nu}(k_1,k_2)&=&[k_1^2D^{mm'}(k_1)]\left\{\Gamma_{c^{m'}A^n_\nu A^{*\epsilon}_e}(k_2,q)\Gamma_{\widehat{A}^a_\alpha A^e_\epsilon}(q)\right.\nonumber \\
&+&\left.\Gamma_{c^{m'}\widehat{A}^a_\alpha A^{*\epsilon}_e}(q,k_2)(\Delta^{-1})^{en}_{\epsilon\nu}(k_2)+k_{2\nu}\Gamma_{c^{m'}\widehat{A}^a_\alpha\bar c^n}(q,k_2)
\right\}\!-\!igf^{amn}k_1^2g_{\alpha\nu}.\nonumber \\
\label{STI:mixed1}
\eea

\subsubsection{STIs for the gluon SD kernel}
\noindent
In the construction of the SDEs for the gluon self-energy and three-gluon vertex, one 
needs the knowledge of the STI satisfied by the kernel (see Fig.~\ref{fig:gggg_SDker})
\begin{eqnarray}
{\mathcal K}_{A^m_\mu A^n_\nu A^r_\rho A^s_\sigma}(k_2,p_2,-p_1)&=&
\Gamma_{A^m_\mu A^n_\nu A^r_\rho A^s_\sigma}(k_2,p_2,-p_1)\nonumber\\
&+&i\Gamma_{A^s_\sigma A^m_\mu A^e_\epsilon}(k_1,\ell)i\Delta_{ee'}^{\epsilon\epsilon'}(\ell)i\Gamma_{A^{e'}_{\epsilon'}A^n_\nu A^r_\rho}(k_2,p_2)\nonumber \\
&+&i\Gamma_{A^s_\sigma A^n_\nu A^e_\epsilon}(k_2,\ell')i\Delta_{ee'}^{\epsilon\epsilon'}(\ell')i\Gamma_{A^{e'}_{\epsilon'}A^m_\mu A^r_\rho}(k_1,p_2).
\end{eqnarray}
\indent
Using the above relation, together with STI of Eq.~(\ref{STI:ggg}), we find the following result
\begin{eqnarray}
&&k_1^\mu i\Gamma_{A^s_\sigma A^m_\mu A^e_\epsilon}(k_1,\ell)i\Delta_{ee'}^{\epsilon\epsilon'}(\ell)i\Gamma_{A^{e'}_{\epsilon'}A^n_\nu A^r_\rho}(k_2,p_2)=-[k_1^2D^{mm'}(k_1)]\Gamma_{A^{e'}_{\epsilon'} A^n_\nu A^r_\rho}(k_2,p_2)\times\nonumber \\
&&\hspace{1cm}\times\left\{\Gamma_{c^{m'} A^s_\sigma A^{*e'}_\epsilon}(-p_1,\ell)P^{\epsilon\epsilon'}(\ell)+i\Gamma_{c^{m'} A^e_\epsilon A^{*\gamma}_d}(\ell,-p_1)\Gamma_{A^d_\gamma A^s_\sigma}(p_1)\Delta_{ee'}^{\epsilon\epsilon'}(\ell)
\right\}.
\end{eqnarray}
In this case this is, however, not the end of the story, since the 
first term in the equation above still contains (virtual) longitudinal momenta, 
which will trigger the STI of Eq.~(\ref{STI:ggg}) together with the FPE~(\ref{FP:gcc}). 
After taking this into account, we obtain
\begin{eqnarray}
&&k_1^\mu i\Gamma_{A^s_\sigma A^m_\mu A^e_\epsilon}(k_1,\ell)i\Delta_{ee'}^{\epsilon\epsilon'}(\ell)i\Gamma_{A^{e'}_{\epsilon'}A^n_\nu A^r_\rho}(k_2,p_2)=-[k_1^2D^{mm'}(k_1)]\times\nonumber \\
&&\hspace{.5cm}\times\left\{
\left[\Gamma_{c^{m'} A^s_\sigma A^{*\epsilon'}_{e'}}(-p_1,\ell)+
i\Gamma_{c^{m'} A^e_{\epsilon}A^{*\gamma}_d}(\ell,-p_1)\Gamma_{A^d_\gamma A^s_\sigma}(p_1)\Delta_{ee'}^{\epsilon\epsilon'}(\ell)
\right]\Gamma_{A^{e'}_{\epsilon'} A^n_\nu A^r_\rho}(k_2,p_2)\right.\nonumber \\
&&\hspace{.5cm}+i\Gamma_{c^{m'} A^s_\sigma \bar c^e}(-p_1,\ell)D^{ee'}(\ell)\left[\Gamma_{c^{e'}A^r_\rho A^{*\gamma}_d}(p_2,k_2)\Gamma_{A^d_\gamma A^n_\nu}(k_2)\right.\nonumber \\
&&\left.\left.\hspace{.5cm}+\Gamma_{c^{e'}A^n_\nu A^{*\gamma}_d}(k_2,p_2)\Gamma_{A^d_\gamma A^r_\rho}(p_2)\right]\right\}.\nonumber \\
\end{eqnarray}
Similarly we find
\begin{eqnarray}
&&k_1^\mu i\Gamma_{A^s_\sigma A^n_\nu A^e_\epsilon}(k_2,\ell')i\Delta_{ee'}^{\epsilon\epsilon'}(\ell')i\Gamma_{A^{e'}_{\epsilon'}A^m_\mu A^r_\rho}(k_1,p_2)=-[k_1^2D^{mm'}(k_1)]\times\nonumber \\
&&\hspace{.5cm}\times\bigg\{\Gamma_{A^s_\sigma  A^n_\nu A^{e}_{\epsilon}}(k_2,\ell')
\left[\Gamma_{c^{m'} A^r_\rho A^{*\epsilon}_e}(p_2,-\ell')+i\Delta_{ee'}^{\epsilon\epsilon'}(\ell')\Gamma_{c^{m'}A^{e'}_{\epsilon'}A^{*\gamma}_d}(-\ell',p_2)\Gamma_{A^d_\gamma A^r_\rho}(p_2)\right]\nonumber \\
&&\hspace{.5cm}+iD^{ee'}(\ell')\left[
\Gamma_{c^{e}A^n_\nu A^{*\gamma}_d}(k_2,-p_1)
\Gamma_{A^d_\gamma A^s_\sigma}(p_1)+\Gamma_{c^{e}A^s_\sigma A^{*\gamma}_d}(-p_1,k_2)\Gamma_{A^d_\gamma A^n_\nu}(k_2)\right]\times\nonumber\\
&&\hspace{.5cm}\times\Gamma_{c^{m'} A^r_\rho \bar c^{e'}}(p_2,-\ell')\bigg\}.
\end{eqnarray}
\indent
As before, after combining these results with the four-gluon 1PI vertex STI of  Eq.~(\ref{STI:gggg}) 
we arrive  at the needed STI for the four-gluon SD kernel, namely
\begin{eqnarray}
k_1^\mu{\mathcal K}_{A^m_\mu A^n_\nu A^r_\rho A^s_\sigma}(k_2,p_2,-p_1)&=&[k_1^2D^{mm'}(k_1)]\bigg\{\Gamma_{c^{m'} A^n_\nu A^{*\gamma}_d}(k_2,-k_1-k_2)\Gamma_{A^d_\gamma A^r_\rho  A^s_\sigma}(p_2,-p_1)\nonumber \\
&+&{\mathcal K}_{c^{m'} A^n_\nu A^s_\sigma A^{*\gamma}_d}(k_2,-p_1,p_2)\Gamma_{A^d_\gamma A^r_\rho}(p_2)\nonumber \\
&+&{\mathcal K}_{c^{m'}A^n_\nu A^r_\rho  A^{*\gamma}_d}(k_2,p_2,-p_1)\Gamma_{A^d_\gamma A^s_\sigma}(p_1)\nonumber \\
&+&{\mathcal K}_{ c^{m'}A^r_\rho A^s_\sigma A^{*\gamma}_d}(p_2,-p_1,k_2)\Gamma_{A^d_\gamma A^n_\nu}(k_2)\bigg\},
\label{STISDgggg}
\end{eqnarray}
where the following auxiliary kernels have been defined
\begin{eqnarray}
{\mathcal K}_{c^{m'} A^n_\nu A^s_\sigma A^{*\gamma}_d}(k_2,-p_1,p_2)&=&
\Gamma_{c^{m'} A^n_\nu A^s_\sigma A^{*\gamma}_d}(k_2,-p_1,p_2)\nonumber\\
&+&i\Gamma_{A^s_\sigma A^n_\nu A^{e}_\epsilon}(k_2,\ell')i\Delta^{\epsilon\epsilon'}_{ee'}(\ell')i\Gamma_{c^{m'} A^{e'}_{\epsilon'} A^{*\gamma}_d}(-\ell',p_2)\nonumber \\
&+&i\Gamma_{c^{m'} A^s_\sigma\bar c^e}(-p_1,\ell)iD^{ee'}(\ell)i\Gamma_{c^{e'}A^n_\nu A^{*\gamma}_d}(k_2,p_2),
\label{1PRker:gggg1} \nonumber \\
{\mathcal K}_{c^{m'}A^n_\nu A^r_\rho  A^{*\gamma}_d}(k_2,p_2,-p_1)&=&
\Gamma_{c^{m'}A^n_\nu A^r_\rho  A^{*\gamma}_d}(k_2,p_2,-p_1)\nonumber \\
&+&i\Gamma_{c^{m'} A^{e}_{\epsilon}A^{*\gamma}_d}(\ell,-p_1)i\Delta^{\epsilon\epsilon'}_{ee'}(\ell)i\Gamma_{A^{e'}_{\epsilon'} A^r_\rho A^n_\nu}(k_2,p_2)\nonumber \\
&+&i\Gamma_{c^e A^n_\nu A^{*\gamma}_d}(k_2,-p_1)iD^{ee'}(\ell')i\Gamma_{c^{m'}A^r_\rho\bar c^{e'}}(p_2,-\ell'),
\label{1PRker:gggg2} \nonumber \\
{\mathcal K}_{ c^{m'}A^r_\rho A^s_\sigma A^{*\gamma}_d}(p_2,-p_1,k_2)&=&\Gamma_{ c^{m'}A^r_\rho A^s_\sigma A^{*\gamma}_d}(p_2,-p_1,k_2)\nonumber \nonumber \\
&+&i\Gamma_{c^{m'}A^s_\sigma\bar c^{e'}}(-p_1,\ell)iD^{ee'}(\ell)i\Gamma_{c^{e'}A^r_\rho A^{*\gamma}_d}(p_2,k_2)\nonumber \\
&+&i\Gamma_{c^eA^s_\sigma A^{*\gamma}_d}(-p_1,k_2)iD^{ee'}(\ell')i\Gamma_{c^{m'}A^r_\rho \bar c^{e'}}(p_2,-\ell').\label{1PRker:gggg3} 
\end{eqnarray}

\subsection{Background-Quantum Identities\label{Appendix:BQIs}}
\noindent
BQIs are obtained by functional differentiation of the STI functional of Eq.~(\ref{STIfunc_BFM}) with respect to combinations of background fields, quantum fields and background sources.

\subsubsection{BQIs for two-point functions}
\noindent
The  first BQI  we can  construct is  the one  relating  the 
conventional with the BFM gluon self-energies. 
To this
end,   consider   the   following  functional   differentiation 
\bea
\left.\frac{\delta^2{\mathcal               S'}\left(\Gamma'\right)}{\delta
\Omega_\alpha^a(p)\delta A_\beta^b(q)}\right|_{\Phi,\Phi^*,\Omega=0}=0
&\qquad&       q+p=0,     \nonumber  \\     
 \left.\frac{\delta^2{\mathcal
S'}\left(\Gamma'\right)}{\delta
\Omega_\alpha^a(p)\delta\widehat{A}_\beta^b(q)}\right|_{\Phi,\Phi^*,\Omega=0}=0
&\qquad&   q+p=0,   
\eea   
which   will  give   the   relations  
 \bea
i\Gamma_{\widehat{A}_\alpha^a    A_\beta^b}(q)&=&\left[ig_\alpha^\gamma
\delta^{ad}+                                    \Gamma_{\Omega_\alpha^a
A^{*\gamma}_d}(q)\right]\Gamma_{A^d_\gamma A^b_\beta}(q),
\label{twoBQI1}\nonumber \\
i\Gamma_{\widehat{A}_\alpha^a\widehat{A}_\beta^b}(q)&=&\left[i g_\alpha^\gamma
\delta^{ad}+ 
\Gamma_{\Omega_\alpha^a A^{*\gamma}_d}(q)\right]\Gamma_{A^d_\gamma
\widehat A^b_\beta}(q).
\label{twoBQI2}
\eea
\begin{figure}[!t]
\includegraphics[width=15cm]{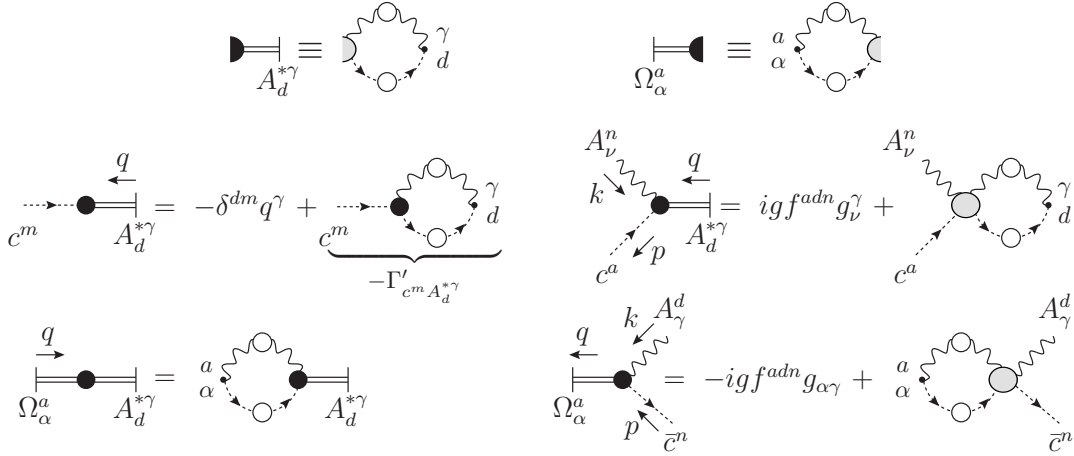}
\caption{\figlab{Composite_ope}Expansions of the gluon anti-field and BFM source in terms of
the corresponding  composite operators. Notice that  if the anti-field
or the BFM sources are attached to  a 1PI vertex, as shown in the first
line, such an expansion will, in  general, convert the 1PI vertex into a
(connected) SD kernel. The equivalence shown is therefore not valid at
tree-level ({\it e.g.},  in the case of three-point  functions such an
equivalence would imply that the  kernels shown on the rhs
of the  corresponding expansions would be  disconnected); when present,
the tree-level needs  to be added by hand, as  explicitly shown in the
two expansions of the second line  and the last one of the third line.
This type of expansion allows one to express the terms appearing in
the BQIs  in a  form that reveals kernels  appearing in  the STIs
[see,   {\it   e.g.},   Eq.s  
(\ref{Aux-ggg:1}) and (\ref{Aux-ggg:2})]}
\end{figure}
%
\indent
We can now combine Eq.s~(\ref{twoBQI1}) and~(\ref{twoBQI2}) such that
the two-point function mixing background and quantum fields drops out, to get the BQI
\bea
i\Gamma_{\widehat{A}_\alpha^a\widehat{A}_\beta^b}(q)&=&i\Gamma_{A^a_\alpha A^b_\beta}(q)+\Gamma_{\Omega_\alpha^a A^{*\gamma}_d}(q)\Gamma_{A^d_\gamma A^b_\beta}(q)+\Gamma_{\Omega_\beta^b A^{*\gamma}_d}(q)\Gamma_{A^a_\alpha A^d_\gamma}(q)\nonumber \\
&-&i\Gamma_{\Omega_\alpha^a A^{*\gamma}_d}(q)\Gamma_{A^d_\gamma A^{e}_{\epsilon}}(q)\Gamma_{\Omega_\beta^b A^{*\epsilon}_{e}}(q)\nonumber \\
&=&i\Gamma_{A^a_\alpha A^b_\beta}(q)+2\Gamma_{\Omega_\alpha^a A^{*\gamma}_d}(q)\Gamma_{A^d_\gamma A^b_\beta}(q)-i\Gamma_{\Omega_\alpha^a A^{*\gamma}_d}(q)\Gamma_{A^d_\gamma A^{e}_{\epsilon}}(q)\Gamma_{\Omega_\beta^b A^{*\epsilon}_{e}},
\label{BQI:gg}
\eea
where the last identity is due to the transversality of the $\Gamma_{AA}$ two-point function.
\newline
\indent
In order for our PT procedure to be self-contained,
it is important to express  
the  1PI auxiliary Green's function involved 
in the various STIs  and the BQIs in terms of kernels  
that also appear in  the relevant STIs. 
The key observation that  makes this possible
is  that  one may always  replace  an
anti-field  or  BFM  source  with  its  corresponding  BRST  composite
operator. Thus, for example, one has (see \Figref{Composite_ope})
\begin{eqnarray}
A^{*\gamma}_d(q)&\to&i\Gamma^{(0)}_{c^{e'} A^{n'}_{\nu'} A^{*\gamma}_d}\int_{k_1}i\Delta_{n'n}^{\nu'\nu}(k_2)iD^{e'e}(k_1), \label{Astrick}\\
\Omega^a_\alpha(q)&\to&i\Gamma^{(0)}_{\Omega^{a}_{\alpha} A^{n'}_{\nu'} \bar c^{e'}}
\int_{k_1}i\Delta_{n'n}^{\nu'\nu}(k_2)iD^{e'e}(k_1), 
\label{Omtrick}
\end{eqnarray}
where $k_1$ and $k_2$ are related through $k_2=q-k_1$.
In this way we get the following SDEs (see again \Figref{Composite_ope})
\begin{eqnarray}
-\Gamma_{c^mA^{*\gamma}_d}(q)&=&-\delta^{dm}q_\gamma-\Gamma'_{c^mA^{*\gamma}_d}(q)\nonumber \\
&=&-\delta^{dm}q_\gamma+gf^{dn'e'}g^\gamma_{\nu'}\int_{k_1}D^{e'e}(k_1)\Delta_{\nu'\nu}^{nn'}(k_2)\Gamma_{c^mA^{n}_{\nu}\bar c^{e}}(k_2,k_1),
\label{BQI:auxcAs} \\
i\Gamma_{c^{a} A^n_\nu A^{*\gamma}_d}(k,q)&=&igf^{adn}g^\gamma_\nu-igf^{e'ds'}g^\gamma_{\sigma'}\int_{k_1}D^{ee'}(k_1)\Delta_{ss'}^{\sigma\sigma'}(k_2){\mathcal K}_{c^a A^n_\nu A^s_\sigma\bar c^e}(k,k_2,k_1),
\label{BQI:auxcAAs} \\
- \Gamma_{\Omega_\alpha^a A^{*\gamma}_d}(q)&=&gf^{ae'n'}g_{\alpha\nu'}\int_{k_1}D^{e'e}(k_1)\Delta_{n'n}^{\nu'\nu}(k_2)\Gamma_{c^e A^n_\nu A^{*\gamma}_d}(k_2,-q),
\label{BQI:auxOmAs}\\
i\Gamma_{\Omega_\alpha^a A^d_\gamma\bar c^n}(k,p)&=&-igf^{adn}g_{\alpha\gamma}-igf^{ae'n'}g_{\alpha\nu'}\int_{k_1}D^{e'e}(k_1)\Delta_{n'n}^{\nu'\nu}(k_2){\mathcal K}_{c^e A^n_\nu A^d_\gamma\bar c^n}(k_2,k,p).\nonumber \\
\label{BQI:auxOmAbarc}
\end{eqnarray}
The kernel ${\mathcal K}_{cAA\bar c}$ appearing in the SDEs~(\ref{BQI:auxcAAs}) and~(\ref{BQI:auxOmAbarc})  is shown in \Figref{cggc_SDker} and reads
\begin{eqnarray}
{\mathcal K}_{c^a A^n_\nu A^s_\sigma\bar c^e}(k,k_2,k_1)&=&\Gamma_{c^a A^n_\nu A^s_\sigma\bar c^e}(k,k_2,k_1)\nonumber\\
&+&i\Gamma_{A^n_\nu A^s_\sigma A^r_\rho}(k_2,-k-k_2)i\Delta^{\rho\rho'}_{rr'}(k+k_2)i\Gamma_{c^mA^{r'}_{\rho'}\bar c^e}(k+k_2,k_1)\nonumber \\
&+&i\Gamma_{c^a A^s_\sigma\bar c^r}(k_2,-k_1-k_2)iD^{rr'}(k_1+k_2)i\Gamma_{c^{r'} A^n_\nu\bar c^e}(k,k_1).
\label{SDE_kercAAbarc}
\end{eqnarray}
\begin{figure}[!t]
\includegraphics[width=15cm]{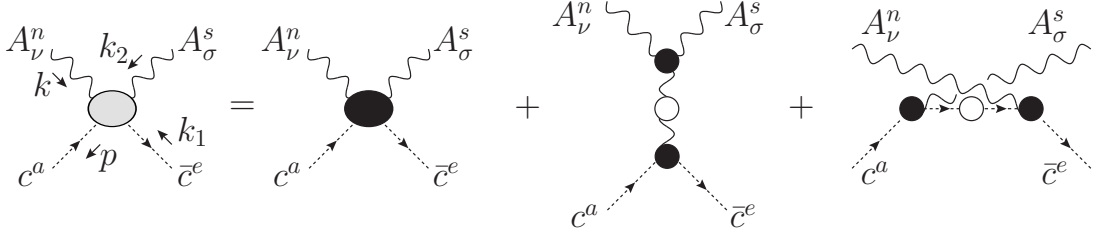}
\caption{\figlab{cggc_SDker}Skeleton expansion of the kernel appearing in the SDE for the auxiliary function $\Gamma_{cAA^*}$.}
\end{figure}

\subsubsection{BQIs for three-point functions \label{BQI:3p}}

\noindent
The relation between the trilinear gluon vertex 
and the trilinear background gluon vertex, 
can be obtained by considering the following functional differentiation
\begin{equation}
\left.\frac{\delta^3{\mathcal S'}(\Gamma')}{\delta \Omega^a_\alpha(q)\delta A^r_\rho(p_2)\delta A^s_\sigma(-p_1)}\right|_{\Phi,\Phi^*,\Omega=0}=0 \qquad q+p_2=p_1.
\label{hggg_diff}
\end{equation}
We then get
\begin{eqnarray}
i\Gamma_{\widehat{A}^a_\alpha A^r_\rho A^s_\sigma}(p_2,-p_1)&=&[ig^\gamma_\alpha\delta^{ad}+\Gamma_{\Omega^a_\alpha A^{*\gamma}_d}(-q)]\Gamma_{A^d_\gamma A^r_\rho A^s_\sigma}(p_2,-p_1)\nonumber \\
&+&\Gamma_{\Omega^a_\alpha A^s_\sigma A^{*\gamma}_d}(-p_1,p_2)\Gamma_{A^d_\gamma A^{r}_\rho}(p_2)+\Gamma_{\Omega^a_\alpha A^r_\rho A^{*\gamma}_d}(p_2,-p_1)\Gamma_{A^d_\gamma A^{s}_\sigma}(p_1).\nonumber \\
\label{BQI:ggg}
\end{eqnarray}
In order to explore further the all-order structure of these two  auxiliary Green's functions, replace the BFM source 
with the corresponding composite operator using Eq.~(\ref{Omtrick}),  thus obtaining
\begin{eqnarray}
i\Gamma_{\Omega^a_\alpha A^s_\sigma A^{*\gamma}_d}(-p_1,p_2)&=&i\Gamma^{(0)}_{\Omega^a_\alpha A^{n'}_{\nu'}\bar c^{m'}}\int_{k_1}iD^{m'm}(k_1)i\Delta_{n'n}^{\nu'\nu}(k_2){\mathcal K}_{c^m A^n_\nu A^s_\sigma A^{*\gamma}_d}(k_2,-p_1,p_2),\nonumber\\
\label{Aux-ggg:1}\\
i\Gamma_{\Omega^a_\alpha A^r_\rho A^{*\gamma}_d}(p_2,-p_1)&=&i\Gamma^{(0)}_{\Omega^a_\alpha A^{n'}_{\nu'}\bar c^{m'}}\int_{k_1}iD^{m'm}(k_1)i\Delta_{n'n}^{\nu'\nu}(k_2){\mathcal K}_{c^m A^n_\nu A^r_\rho A^{*\gamma}_d}(k_2,p_2,-p_1),\nonumber\\
\label{Aux-ggg:2}
\end{eqnarray}
with the corresponding kernels defined
in Eq.s~(\ref{1PRker:gggg1})  and~(\ref{1PRker:gggg2}). 
Notice the  emergence  of the  pattern exploited in the application of the PT to the SDEs of QCD: namely that  the  auxiliary
functions  appearing in  the  BQI satisfied  by  a particular  Green's
function  can be written  in terms  of kernels  appearing in  the STIs
triggered when the  PT procedure is applied to  that same Green's
function.
\newline
\indent
Now, the BQI of Eq.~(\ref{BQI:ggg}) gives at tree-level the result
\be
\Gamma_{\widehat{A}^a_\alpha A^r_\rho A^s_\sigma}(p_2,-p_1)=\Gamma_{A^a_\alpha A^r_\rho A^s_\sigma}(p_2,-p_1).
\ee
This is once again due to the use of the reduced functional: in fact in such case the two (tree-level) vertices need to coincide, since the  difference between them is proportional to the inverse of the gauge fixing parameter (see \appref{Frules}) and therefore entirely due to the gauge fixing Lagrangian. To restore the correct tree-level terms one would have to use the complete functional; in that case the differentiation of Eq.~(\ref{hggg_diff}) shows the two additional terms 
\be
-\delta^{ds}p_{1\sigma}\Gamma_{\Omega^a_\alpha A^r_\rho \bar c^d}(p_2,-p_1)+\delta^{dr}p_{2\rho}\Gamma_{\Omega^a_\alpha A^s_\sigma  \bar c^d}(-p_1,p_2),
\ee
which, with the help of Eq.~(\ref{FPE_Om1}) become
\be
-i\delta^{ds}p_{1\sigma}p_{1\gamma}\Gamma_{\Omega^a_\alpha A^r_\rho A^{*\gamma}_d}(p_2,-p_1)-i\delta^{dr}p_{2\rho}p_{2\gamma}\Gamma_{\Omega^a_\alpha A^s_\sigma A^{*\gamma}_d}(-p_1,p_2)+gf^{ars}(q_{\alpha\rho}p_{1\sigma}+g_{\alpha\sigma}p_{2\rho}).
\ee
Therefore we get the final identity
\bea
i\Gamma^\mathrm{C}_{\widehat{A}^a_\alpha A^r_\rho A^s_\sigma}(p_2,-p_1)&=&[ig^\gamma_\alpha\delta^{ad}+\Gamma_{\Omega^a_\alpha A^{*\gamma}_d}(-q)]\Gamma_{A^d_\gamma A^r_\rho A^s_\sigma}(p_2,-p_1)+gf^{ars}(q_{\alpha\rho}p_{1\sigma}+g_{\alpha\sigma}p_{2\rho})\nonumber \\
&+&\Gamma_{\Omega^a_\alpha A^s_\sigma A^{*\gamma}_d}(-p_1,p_2)\left[\Gamma^\mathrm{C}_{A^d_\gamma A^{r}_\rho}(p_2)-i\delta^{dr}p_{2\rho}p_{2\gamma}\right]\nonumber \\
&+&\Gamma_{\Omega^a_\alpha A^r_\rho A^{*\gamma}_d}(p_2,-p_1)\left[\Gamma^\mathrm{C}_{A^d_\gamma A^{s}_\sigma}(p_1)
-i\delta^{ds}p_{1\sigma}p_{1\gamma}\right],
\eea
which gives the expected tree-level result. Once again we see that the difference between working with the reduced and complete functional lies in some constant (tree-level) terms that one recovers after applying the FPE for writing the STI/BQI at hand in the same form using $\Gamma$ or $\Gamma_\mathrm{C}$. Thus, opting for the fast way of deriving the STI/BQI  with the reduced functional and adding the correct tree-level term, we write the BQI in its final form
\begin{eqnarray}
i\Gamma_{\widehat{A}^a_\alpha A^r_\rho A^s_\sigma}(p_2,-p_1)&=&[ig^\gamma_\alpha\delta^{ad}+\Gamma_{\Omega^a_\alpha A^{*\gamma}_d}(-q)]\Gamma_{A^d_\gamma A^r_\rho A^s_\sigma}(p_2,-p_1)\nonumber \\
&+&\Gamma_{\Omega^a_\alpha A^s_\sigma A^{*\gamma}_d}(-p_1,p_2)\Gamma_{A^d_\gamma A^{r}_\rho}(p_2)+\Gamma_{\Omega^a_\alpha A^r_\rho A^{*\gamma}_d}(p_2,-p_1)\Gamma_{A^d_\gamma A^{s}_\sigma}(p_1)\nonumber \\
&+&gf^{ars}\left(p_{2\rho}g_{\alpha\sigma}+p_{1\sigma}g_{\alpha\rho}\right).
\label{BQI:ggg_tlc}
\end{eqnarray}
\indent
We conclude by giving the relation between the trilinear quantum gluon-quark vertex 
and the trilinear background gluon-quark vertex; this
can be obtained by considering the following functional differentiation
\begin{equation}
\left.\frac{\delta^3{\mathcal S'}(\Gamma')}{\delta \Omega^a_\alpha(q)\delta\psi(p_2)\delta\bar\psi(-p_1)}\right|_{\Phi,\Phi^*,\Omega=0}=0 \qquad q+p_2=p_1.
\end{equation}
We then get
\begin{eqnarray}
i\Gamma_{\widehat{A}^a_\alpha\psi\bar\psi}(p_2,-p_1)&=&[ig^\gamma_\alpha\delta^{ad}+\Gamma_{\Omega^a_\alpha A^{*\gamma}_d}(-q)]\Gamma_{A^d_\gamma\psi\bar\psi}(p_2,-p_1)\nonumber \\
&+&\Gamma_{\psi^*\bar\psi\Omega^a_\alpha}(-p_1,q)\Gamma_{\psi\bar\psi}(p_2)+\Gamma_{\psi\bar\psi}(p_1)\Gamma_{\psi\Omega^a_\alpha\bar\psi^*}(q,-p_1).
\label{BQI:gff}
\end{eqnarray}

\subsubsection{BQI for the ghost-gluon trilinear vertex}
\noindent
In this section we are going to derive the BQIs relating the $R_\xi$ ghost sector with the BFM ones.
We start from the trilinear ghost-gluon coupling, for which we choose the following functional differentiation 
\begin{equation}
\left.\frac{\delta^3{\mathcal S'}(\Gamma')}{\delta \Omega^a_\alpha(-q)\delta c^m(k_1)\delta \bar c^n(k_2)}\right|_{\Phi,\Phi^*,\Omega=0}=0 \qquad k_1+k_2=q,
\end{equation}
thus getting the result
\begin{eqnarray}
i\Gamma_{c^m\widehat{A}^a_\alpha\bar c^n}(-q,k_2)&=&[i\delta^{da}g^\gamma_\alpha+\Gamma_{\Omega^a_\alpha A^{*\gamma}_d}(q)]
\Gamma_{c^m A^d_\gamma\bar c^n}(-q,k_2)\nonumber\\
&-&\Gamma_{c^m A^{*\gamma}_d}(-k_1)\Gamma_{\Omega^a_\alpha A^d_\gamma\bar c^n }(k_1,k_2)-
\Gamma_{\Omega^a_\alpha c^m c^{*d}}(k_1,k_2)\Gamma_{c^d\bar c^n}(k_2).
\label{BQI:cgbarc}
\end{eqnarray}

\end{appendix}

\newpage


\end{document}